\documentclass[11pt]{book}
\usepackage[utf8]{inputenc}
\pdfoutput=1
\usepackage[T1]{fontenc}
\usepackage{lmodern}
\usepackage[protrusion=true,expansion=true]{microtype}
\usepackage{amsmath,amssymb,amsfonts,amsthm}
\usepackage{subcaption}
\usepackage{graphicx}
\usepackage{fullpage}
\usepackage{setspace}
\usepackage[backref=page]{hyperref}
\usepackage{color}
\usepackage{wrapfig}
\usepackage{tikz}
\usetikzlibrary{decorations.pathreplacing,patterns}
\usepackage{algorithm}
\usepackage[noend]{algpseudocode}
\usepackage[framemethod=tikz]{mdframed}
\usepackage{xspace}
\usepackage{pgfplots}
\usepackage{framed}
\usepackage{thmtools}
\usepackage{thm-restate}
\usepackage{tabu}
\usepackage{fancyhdr}
\pgfplotsset{compat=1.5}
\usepackage{bbm}
\usepackage[most]{tcolorbox}
\usepackage{titlesec}
\usepackage{xcolor}
\usepackage{lmodern}
\usepackage{fontawesome5}
\usepackage{placeins}
\usepackage{titling}
\usepackage{chngcntr}

\newtheorem{theorem}{Theorem}[section]
\newtheorem{corollary}[theorem]{Corollary}
\newtheorem{lemma}[theorem]{Lemma}

\newtheorem{definition}[theorem]{Definition}
\newtheorem{remark}[theorem]{Remark}
\newtheorem{claim}[theorem]{Claim}
\newtheorem{invariant}[theorem]{Invariant}
\newtheorem{observation}[theorem]{Observation}

\newtheorem{fact}[theorem]{Fact}

\newtheorem{question}[theorem]{Question}
\newtheorem{assumption}[theorem]{Assumption}

\newenvironment{proofof}[1]{\begin{trivlist} \item {\bf Proof
#1:~~}}
  {\qed\end{trivlist}}

\newcommand{\namedref}[2]{\hyperref[#2]{#1~\ref*{#2}}}
\newcommand{\thmlab}[1]{\label{thm:#1}}
\newcommand{\thmref}[1]{\namedref{Theorem}{thm:#1}}
\newcommand{\lemlab}[1]{\label{lem:#1}}
\newcommand{\lemref}[1]{\namedref{Lemma}{lem:#1}}
\newcommand{\claimlab}[1]{\label{claim:#1}}
\newcommand{\claimref}[1]{\namedref{Claim}{claim:#1}}
\newcommand{\corlab}[1]{\label{cor:#1}}
\newcommand{\corref}[1]{\namedref{Corollary}{cor:#1}}
\newcommand{\seclab}[1]{\label{sec:#1}}
\newcommand{\secref}[1]{\namedref{Section}{sec:#1}}

\newcommand{\factlab}[1]{\label{fact:#1}}
\newcommand{\factref}[1]{\namedref{Fact}{fact:#1}}
\newcommand{\invlab}[1]{\label{inv:#1}}
\newcommand{\invref}[1]{\namedref{Invariant}{inv:#1}}

\newcommand{\figlab}[1]{\label{fig:#1}}
\newcommand{\figref}[1]{\namedref{Figure}{fig:#1}}
\newcommand{\alglab}[1]{\label{alg:#1}}
\renewcommand{\algref}[1]{\namedref{Algorithm}{alg:#1}}

\newcommand{\deflab}[1]{\label{def:#1}}
\newcommand{\defref}[1]{\namedref{Definition}{def:#1}}
\newcommand{\eqnlab}[1]{\label{eq:#1}}
\newcommand{\eqnref}[1]{\namedref{Equation}{eq:#1}}

\newcommand{\obslab}[1]{\label{obs:#1}}
\newcommand{\obsref}[1]{\namedref{Observation}{obs:#1}}
\newcommand{\linlab}[1]{\label{line:#1}}

\newcommand{\chaplab}[1]{\label{chap:#1}}
\newcommand{\chapref}[1]{\namedref{Chapter}{chap:#1}}

\newcommand{\invarlab}[1]{\label{invar:#1}}
\newcommand{\invarref}[1]{\namedref{Invariant}{invar:#1}}
\newcommand{\assumref}[1]{\namedref{Assumption}{assum:#1}}
\newcommand{\assumlab}[1]{\label{assum:#1}}

\def \Adv    {\mdef{\mathsf{Adv}}}
\def \Alg    {\mdef{\mathsf{Alg}}}
\def \AMS    {\mdef{\mathsf{AMS}}}

\def \MaintainIter    {\mdef{\mathsf{MaintainIter}}}

\def \EstLevel    {\mdef{\mathsf{EstLevel}}}

\def \HAM    {\mdef{\mathsf{HAM}}}

\def \detgapeq    {\mdef{\textsc{DetGapEQ}}}
\def \PrivMed    {\mdef{\textsc{PrivMed}}}
\def \FLAG    {\mdef{\mathsf{STATE}}}
\def \SPARSE    {\mdef{\mathsf{SPARSE}}}
\def \sampler    {\mdef{\textsc{Sampler}}}
\def \fracpart    {\mdef{\textsc{Frac}}}

\def \countsketch    {\mdef{\textsc{CountSketch}}}
\def \bptree   {\mdef{\textsc{BPTree}}}
\def \heavyhitters    {\mdef{\textsc{HeavyHitters}}}
\def \zeroestimate    {\mdef{\textsc{F0Estimate}}}
\def \DENSE    {\mdef{\mathsf{DENSE}}}
\def \LZeroEst    {\mdef{\textsc{LZeroEst}}}
\def \FPEst    {\mdef{\textsc{FPEst}}}
\def \SparseRecover    {\mdef{\textsc{SparseRecover}}}

\def \GapNorm    {\mdef{\textsc{GapNorm}}}
\def \PRG    {\mdef{\textsc{PRG}}}

\def \trunc {\textup{trunc}}

\def \accgame    {\mdef{\mathsf{Acc}}}

\newcommand{\PPr}[1]{\ensuremath{\mathbf{Pr}\left[#1\right]}}
\newcommand{\PPPr}[2]{\ensuremath{\underset{#1}{\mathbf{Pr}}\left[#2\right]}}
\newcommand{\Ex}[1]{\ensuremath{\mathbb{E}\left[#1\right]}}
\newcommand{\EEx}[2]{\ensuremath{\underset{#1}{\mathbb{E}}\left[#2\right]}}
\renewcommand{\O}[1]{\ensuremath{\mathcal{O}\left(#1\right)}}
\newcommand{\tO}[1]{\ensuremath{\tilde{\mathcal{O}}\left(#1\right)}}
\newcommand{\eps}{\varepsilon}
\newcommand{\TVD}{d_{\mathrm{tv}}}
\newcommand{\KLD}{d_{\mathrm{KL}}}

\newcommand{\coloralg}{\textsc{color}\xspace}
\newcommand{\avoid}{\textsc{avoid}\xspace}
\newcommand{\kavoid}{\ensuremath{\textsc{avoid}^k}\xspace}

\def \calA    {\mdef{\mathcal{A}}}
\def \calB    {\mdef{\mathcal{B}}}
\def \calC    {\mdef{\mathcal{C}}}
\def \calD    {\mdef{\mathcal{D}}}
\def \calE    {\mdef{\mathcal{E}}}
\def \calF    {\mdef{\mathcal{F}}}
\def \calG    {\mdef{\mathcal{G}}}
\def \calH    {\mdef{\mathcal{H}}}
\def \calI    {\mdef{\mathcal{I}}}

\def \calL    {\mdef{\mathcal{L}}}
\def \calM    {\mdef{\mathcal{M}}}
\def \calN    {\mdef{\mathcal{N}}}

\def \calP    {\mdef{\mathcal{P}}}
\def \calQ    {\mdef{\mathcal{Q}}}
\def \calR    {\mdef{\mathcal{R}}}
\def \calS    {\mdef{\mathcal{S}}}
\def \calT    {\mdef{\mathcal{T}}}
\def \calU    {\mdef{\mathcal{U}}}

\def \calX    {\mdef{\mathcal{X}}}

\def \bA    {\mdef{\mathbf{A}}}
\def \bB    {\mdef{\mathbf{B}}}
\def \bD    {\mdef{\mathbf{D}}}
\def \bG    {\mdef{\mathbf{G}}}
\def \bH    {\mdef{\mathbf{H}}}

\def \bL    {\mdef{\mathbf{L}}}
\def \bM    {\mdef{\mathbf{M}}}
\def \bP    {\mdef{\mathbf{P}}}
\def \bR    {\mdef{\mathbf{R}}}
\def \bS    {\mdef{\mathbf{S}}}

\def \bV    {\mdef{\mathbf{V}}}
\def \bU    {\mdef{\mathbf{U}}}
\def \bSigma  {\mdef{\mathbf{\Sigma}}}
\def \bmu    {\mdef{\mathbf{\mu}}}
\def \bfEta    {\mdef{\mathbf{\eta}}}
\def \bX    {\mdef{\mathbf{X}}}
\def \bY    {\mdef{\mathbf{Y}}}
\def \bZ    {\mdef{\mathbf{Z}}}
\def \bW    {\mdef{\mathbf{W}}}
\def \ba    {\mdef{\mathbf{a}}}
\def \bb    {\mdef{\mathbf{b}}}
\def \bc    {\mdef{\mathbf{c}}}
\def \be    {\mdef{\mathbf{e}}}

\def \bq    {\mdef{\mathbf{q}}}

\def \bs    {\mdef{\mathbf{s}}}
\def \bu    {\mdef{\mathbf{u}}}
\def \bw    {\mdef{\mathbf{w}}}
\def \bv    {\mdef{\mathbf{v}}}
\def \bx    {\mdef{\mathbf{x}}}
\def \by    {\mdef{\mathbf{y}}}
\def \bz    {\mdef{\mathbf{z}}}
\def \bg    {\mdef{\mathbf{g}}}
\def \bh    {\mdef{\mathbf{h}}}

\newcommand{\mdef}[1]{{\ensuremath{#1}}\xspace}  
\newcommand{\myfunc}[1]{\mdef{\mathsf{#1}}}      

\DeclareMathOperator*{\Lap}{Lap}
\DeclareMathOperator*{\argmin}{argmin}
\DeclareMathOperator*{\argmax}{argmax}
\DeclareMathOperator*{\polylog}{polylog}
\DeclareMathOperator*{\poly}{poly}

\DeclareMathOperator*{\Bin}{Bin}

\DeclareMathOperator*{\dist}{dist}

\DeclareMathOperator*{\Var}{Var}
\DeclareMathOperator*{\rank}{rank}
\DeclareMathOperator*{\nnz}{nnz}
\DeclareMathOperator*{\Span}{Span}
\DeclareMathOperator*{\sgn}{sign}
\DeclareMathOperator*{\Trace}{Tr}
\DeclareMathOperator*{\Bern}{Bern}

\def \negl     {\mdef{\myfunc{negl}}}                
\newcommand{\flr}[1]{\mdef{\left\lfloor#1\right\rfloor}}              
\newcommand{\ceil}[1]{\mdef{\left\lceil#1\right\rceil}}               

\newcommand{\ignore}[1]{}

\newif\ifnotes\notestrue 
\ifnotes
\newcommand{\samson}[1]{\textcolor{blue}{{\bf (Samson:} {#1}{\bf ) }} \marginpar{\tiny\bf
             \begin{minipage}[t]{0.5in}
               \raggedright S:
            \end{minipage}}}
\newcommand{\david}[1]{\textcolor{purple}{{\bf (David:} {#1}{\bf ) }} \marginpar{\tiny\bf
             \begin{minipage}[t]{0.5in}
               \raggedright D:
            \end{minipage}}} 
\else
\newcommand{\samson}[1]{}
\newcommand{\david}[1]{}
\fi

\makeatletter
\renewcommand*{\@fnsymbol}[1]{\textcolor{mahogany}{\ensuremath{\ifcase#1\or *\or \dagger\or \ddagger\or
 \mathsection\or \triangledown\or \mathparagraph\or \|\or **\or \dagger\dagger
   \or \ddagger\ddagger \else\@ctrerr\fi}}}
\makeatother

\providecommand{\email}[1]{\href{mailto:#1}{\nolinkurl{#1}\xspace}}

\definecolor{mahogany}{rgb}{0.75, 0.25, 0.0}
\definecolor{darkblue}{rgb}{0.0, 0.0, 0.55}
\definecolor{darkpastelgreen}{rgb}{0.01, 0.75, 0.24}
\definecolor{darkgreen}{rgb}{0.0, 0.2, 0.13}
\definecolor{darkgoldenrod}{rgb}{0.72, 0.53, 0.04}
\definecolor{darkred}{rgb}{0.55, 0.0, 0.0}
\definecolor{forestgreenweb}{rgb}{0.13, 0.55, 0.13}
\definecolor{greencss}{rgb}{0.0, 0.5, 0.0}
\definecolor{bleudefrance}{rgb}{0.19, 0.55, 0.91}
\definecolor{darkslateblue}{rgb}{0.22, 0.36, 0.52}

\hypersetup{
     colorlinks   = true,
     citecolor    = mahogany,
     linkcolor	  = mahogany
}

\fancypagestyle{pg}
{
\lhead{}
\rhead{}
\cfoot{--\ \thepage\ --}

}

\AtBeginDocument{%
  \DeclareFontShape{T1}{lmr}{m}{scit}{<->ssub*lmr/m/scsl}{}%
}

\newtcolorbox{tallchapterbannerbox}{
  colback=white,
  colframe=violet!80!black,
  width=\textwidth,
  arc=10pt,
  boxrule=1pt,
  sharp corners=south,
  drop shadow southeast,
  left=20pt, right=20pt, top=6pt, bottom=6pt,
  fontupper=\small\itshape,
  enhanced,
  overlay unbroken={
    \node[anchor=north west, xshift=8pt, yshift=-14pt, font=\small] at (frame.north west) {\faExclamationCircle};
    \node[anchor=north east, xshift=-8pt, yshift=-14pt, font=\small] at (frame.north east) {\faExclamationCircle};
  }
}

\newtcolorbox{chapterbannerbox}{
  colback=white,
  colframe=violet!80!black,
  width=\textwidth,
  arc=10pt,
  boxrule=1pt,
  sharp corners=south,
  drop shadow southeast,
  left=20pt, right=20pt, top=6pt, bottom=6pt,
  fontupper=\small\itshape,
  enhanced,
  overlay unbroken={
    \node[anchor=north west, xshift=8pt, yshift=-10pt, font=\small] at (frame.north west) {\faExclamationCircle};
    \node[anchor=north east, xshift=-8pt, yshift=-10pt, font=\small] at (frame.north east) {\faExclamationCircle};
  }
}

\newtcolorbox{shortbannerbox}{
  colback=white,
  colframe=violet!80!black,
  width=\textwidth,
  arc=10pt,
  boxrule=1pt,
  sharp corners=south,
  drop shadow southeast,
  left=8pt, right=8pt, top=6pt, bottom=6pt,
  fontupper=\small\itshape,
  enhanced,
  overlay unbroken={
    \node[anchor=north west, xshift=8pt, yshift=-6pt, font=\small] at (frame.north west) {\faStar};
    \node[anchor=north east, xshift=-8pt, yshift=-6pt, font=\small] at (frame.north east) {\faStar};
  }
}

\makeatletter
\let\ps@plain\ps@pg
\makeatother

\renewcommand{\footnoterule}{%
  \kern -3pt
  \hrule width 0.4\columnwidth height 0.4pt
  \kern 2.6pt
}

\usepackage{xpatch}

\makeatletter
\xpatchcmd{\maketitle}
  {\@thanks}
  {%
    \par\vfill
    \noindent\rule{\textwidth}{0.4pt}\par
    \vspace{-1.5em}
    \@thanks
  }
  {}
  {\PackageWarning{mypatch}{Could not patch maketitle}}
\makeatother
\allowdisplaybreaks

\thanksmarkseries{fnsymbol}

\title{
\Huge
The Adversarial Robustness of Sketching and Streaming Algorithms\thanks{\normalsize Version to appear as a monograph in NOW Publishers \emph{Foundations and Trends in Theoretical Computer Science} series.}
\normalsize
\vspace{0.5in}
}

\author{
\LARGE
David P. Woodruff\thanks{Carnegie Mellon University and Google Research.
E-mail: \email{dwoodruf@andrew.cmu.edu}.}
\and
\LARGE
Samson Zhou\thanks{Texas A\&M University.
E-mail: \email{samsonzhou@gmail.com}.}
}

\date{
\vspace{0.5in}
\LARGE
\today
\normalsize
}

\begin{document}
\maketitle

\pagestyle{pg}
\setcounter{page}{1}

\chapter*{Abstract}
Sketching and streaming algorithms are vital for handling massive datasets. While classical methods guarantee correctness on fixed inputs, they often fail with \emph{adaptive inputs}, where future data depends on past algorithm outputs. This is common in settings such as  optimization, databases, finance, and network monitoring. This monograph surveys recent advances in adversarial robustness, including techniques for insertion-only streams, connections to differential privacy, and cryptographic methods that achieve adversarial robustness. We also discuss fundamental limitations, especially for linear sketches and streams with insertions and deletions, where robustness often requires polynomial space or sketching dimension. Throughout, we explore core problems like adaptively answering queries for optimization problems, norm estimation, frequency moments, and heavy hitters, and highlight emerging tools and open challenges at the intersection of streaming, sketching, privacy, and adversarial robustness.

\tableofcontents

\chapter{Introduction}
\chaplab{chap:intro}
\begin{chapterbannerbox}
\centering
It is not always adversaries who bring algorithms to failure;\\ sometimes, it is our own unintended misuse.
\end{chapterbannerbox}
\vspace{0.4in}
Suppose you are managing the traffic control system for an internet service provider. 
Each second, routers across the country send updates. 
You cannot afford to store all the traffic coming through, so you rely on an efficient randomized data structure, or {\it sketch}, that estimates statistics of packet flows to detect anomalies, which is cheap, fast, and accurate.

But there's a twist: when the system flags suspicious activity, it re-routes traffic, rate-limits IP addresses, or updates firewall rules. 
In other words, your algorithm not only observes the network, but also changes and defines it. 

And the moment it acts, adversaries adapt. 
An adversary that was spiking a few flows now switches its strategy, spreading out its traffic just enough to evade detection. 
Your sketch was designed for the old distribution, but the distribution just changed in response to your output. 
This makes the problem \emph{adaptive}. 
If your algorithm does not account for this feedback loop, then it may fail to detect future attacks. 
This failure highlights a fundamental theoretical vulnerability: the violation of independence between an algorithm’s internal randomness and its input stream.

\paragraph{A technical perspective.}
The example above captures a typical scenario in the \emph{streaming model of computation}, a standard and well-studied model for massive datasets, where the goal is to perform computation using space sublinear in the size of the dataset. 

Consider the seminal algorithm~\cite{AlonMS99} of Alon, Matias, and Szegedy (AMS), which is a sketch for estimating the Euclidean norm ($L_2$ norm) of a vector that is so simple that we cannot resist describing it here: generate a random sign vector $\bs\in\{-1,+1\}^n$ and for a vector $\bx\in\mathbb{R}^n$ defined by a data stream, maintain $\langle\bs,\bx\rangle$, e.g., by updating a counter by $\Delta\cdot s_i$ each time $x_i$ changes by a value $\Delta$. 
Then $\langle\bs,\bx\rangle^2$ is an unbiased estimate of $\|\bx\|_2^2$. 
This algorithm just uses $\O{\log n}$ bits of space to store a single counter $\langle\bs,\bx\rangle$\footnote{The random sign vector $\bs$ can be stored as a pseudorandom seed, since it just needs to be $4$-wise independent; we refer to \cite{AlonMS99} for the full details.}.

The above analysis crucially requires that $\bs$ is a random sign vector that is independent of $\bx$, as the value $\langle\bs,\bx\rangle^2$ is clearly \emph{not} an unbiased estimate to $\|\bx\|_2^2$ if $s_i=\sgn(x_i)$, i.e., if the sign of $x_i$ equals the random sign $s_i$ for each $i\in[n]$. 
This example illustrates a problem with all analyses of classical streaming algorithms beginning with the work of \cite{AlonMS99} up until the work of \cite{Ben-EliezerY20,Ben-EliezerJWY22} more than twenty years later\footnote{There are a few exceptions, such as \cite{HardtW13}, but they are for other models rather than streaming, we discuss this later.} on adaptively chosen inputs. 
The algorithm's randomness is not independent of its input, and therefore there is no guarantee that the algorithm will be correct!

Fortunately, this example cannot happen if all entries of $x$ are positive, and although there are still specific attacks against the AMS algorithm, it turns out that if the stream contains only positive updates to an underlying vector, referred to as an {\it insertion-stream}, then there are algorithms~\cite{Ben-EliezerJWY22} that can leverage the number of times the $L_2$ norm (or whichever function of interest) can change by a ``significant'' amount to achieve adversarial robustness, i.e., correctness even with adaptively-chosen inputs. 
In fact, using a tree-like structure and the concept of so-called difference estimators~\cite{WoodruffZ21}, it is possible to obtain adversarially robust algorithms with almost no overhead of space complexity in terms of the accuracy parameter over classical streaming algorithms\footnote{This is up to logarithmic factors, we omit the details in this introduction.}.

The key to these results is that an approximation to the $L_2$ norm, or other function of interest, need not change too many times in an insertion-only stream. 
Indeed, once the norm is non-zero, then for integer-valued vectors, the norm is at least $1$, and if one is only interested in a $2$-approximation, then there are only $\O{\log n}$ powers of $2$ between $1$ and the maximal possible value, assuming the stream length is bounded by a polynomial in $n$, denoted $\poly(n)$. 
One way to exploit this fact is by taking a classical non-robust streaming algorithm and setting its failure probability $\delta$ to be sufficiently small so that one can union bound over all possible times in the stream that the sketch might increase its output by a factor of $2$. 
As classical algorithms typically have a $\log\frac{1}{\delta}$ dependence in their memory usage, and one can show that $\frac{1}{\delta}$ can be upper bounded by $\left(\poly(n)\right)^{\O{\log n}}$, one obtains an adversarially robust streaming algorithm with only a logarithmic factor overhead \cite{Ben-EliezerJWY22}. 
Other general frameworks for obtaining adversarially robust algorithms when the function value cannot change too often include the sketch-switching framework of \cite{Ben-EliezerJWY22}, which achieves similar though incomparable bounds; we provide more details on these frameworks in later sections. 

But even if the function value can change more frequently, \cite{HassidimKMMS22} showed a surprising connection to differential privacy in this context. 
Namely, they showed that differential privacy can be used to hide the internal randomness of an algorithm, and using such techniques they achieved a quadratic improvement in space complexity upon the approaches of \cite{Ben-EliezerJWY22}!
Beyond its utility in hiding randomness, the connection to differential privacy provides a rigorous theoretical framework for stability; by treating robustness as a byproduct of information-theoretic privacy, we can quantify the ``leakage'' of the internal state of a sketch and derive new bounds for adaptive queries. 

However, for streams of length $\poly(n)$, for which updates to an underlying vector may be positive or negative, all of the above approaches for approximating the Euclidean norm, or most statistics of interest, would require space polynomial in $n$. 
This is because the approximate function value may change many times, e.g., if the underlying vector in a stream switches $\poly(n)$ times between being $0$ and non-zero, then any relative approximation to its norm must change $\poly(n)$ times. 
Such streams with insertions and deletions are called turnstile streams, and while the AMS algorithm uses polylogarithmic space, the above technique achieving adversarial robustness would use polynomial space. 
For long enough turnstile streams, the best known algorithms we have for any problem are so-called linear sketches; namely, they maintain $\bA \cdot \bx$ as their data structure for an $m \times n$ matrix $\bA$, $m \ll n$, given updates to an underlying vector $\bx$ in a stream. 
Typically the sketching matrix $\bA$ itself can be represented implicitly, and so such algorithms compress $n$ words of memory to $m$ words of memory and enjoy other advantages such as being easy to update when the coordinates of $\bx$ undergo positive or negative updates in a turnstile stream. 
There is also some evidence that the optimal non-robust algorithm for any problem in a turnstile stream is in fact a linear sketch 
\cite{LiNW14,AiHLW16}. 
Perhaps surprisingly, it turns out that one can show that no linear sketch based streaming algorithm for turnstile streams can be adversarially robust, that is, any such algorithm must use $\poly(n)$ memory \cite{HardtW13,GribelyukLWYZ24,GribelyukSWY25}.
There is a saving grace here though, in that if the stream has a large number of +1/-1 updates, then the Euclidean norm and other common statistics take too long to change by a large amount and \cite{Ben-EliezerEO22,WoodruffZ24} showed there are further optimizations upon the differential privacy-based framework, though still the memory required is polynomial in the length of the stream.

Finally, in some settings the adversary may in fact be much stronger and not only interacts with the output of the algorithm, but may also observe parts of the underlying data structure itself, e.g., if the data structure is distributed across devices and its contents are shared over a communication channel.  
\cite{AjtaiBJSSWZ22} showed that norm estimation cannot be performed in sublinear space in general, but surprisingly, if we assume the adversary is computationally bounded, then we can use \emph{cryptographic} techniques to perform sparse recovery and estimate a number of statistics securely. 
For example, one can perform $L_2$ estimation within a multiplicative factor of $M\sqrt{\frac{n}{r}}$ for $r$-dimensional sketches with input vectors having integer entries with magnitude at most $M$; see also \cite{BogdanovRVV26} for connections to property-preserving hashing. 

The above leads to a large number of open questions. 
For example, are non-linear sketches helpful for robust algorithms in turnstile streams for $L_2$ estimation? What about other statistics and other models? 
Throughout this monograph, we study other functions such as frequency moments, heavy hitters, as well as more involved optimization problems. 
We also study other models such as adaptive data analysis and linear sketching over the reals, and so on.

\section{Why Adversarial Robustness?}
In the previous section, we presented a motivating example illustrating how classical streaming algorithms can fail when their inputs adapt based on the algorithm’s outputs. 
This feedback loop creates a fundamental challenge: the very act of processing data can change the data itself, leading to potential failures if the algorithm is not designed to handle such adaptive scenarios.

In this section, we broaden the perspective to explain why adversarial robustness is a critical property for algorithms deployed in modern, data-driven systems. 
As these technologies increasingly support critical infrastructure, it is essential that algorithms remain reliable and secure even when facing inputs that may be manipulated or influenced by adversaries.

The relentless expansion of data-driven technologies into critical infrastructure necessitates algorithms designed with properties beyond mere computational efficiency. 
In particular, adversarial robustness plays a crucial role in ensuring the dependability, security, and correctness of algorithmic systems operating under perturbation or manipulation. 
Data in contemporary applications, ranging from high-frequency financial arbitrage to large-scale adaptive recommendation systems, frequently manifests as a continuous, high-volume stream. 
The intrinsic memory and pass constraints characteristic of the streaming model typically assume benign, oblivious data sources. 
However, in practice, this idealized assumption often breaks down: input sequences may be shaped by external observers or even deliberately crafted by adversaries aiming to exploit subtle algorithmic vulnerabilities. 
This monograph presents a comprehensive study of the fundamental principles underlying adversarial robustness in dynamic streaming environments.

We focus on designing streaming algorithms that remain accurate, even when faced with adaptive or adversarial inputs. 
Unlike classical analyses that assume a fixed or independent input stream, we consider settings where future data can depend on the algorithm’s previous outputs or internal state. 
Importantly, this form of ``adaptivity'' need not be malicious; inputs may simply reflect natural dependencies resulting from interactive systems. 

In fact, these dependencies may even be unintentional, such as using the same subroutine across many steps of an interactive procedure such as in convex optimization. 
Yet, even benign forms of feedback can undermine standard correctness guarantees and render traditional analysis techniques ineffective. 
Indeed, in many practical systems, a sequence of seemingly innocuous interactions can gradually erode the statistical guarantees of streaming algorithms. 
Each interaction, taken in isolation, may appear benign; however, when the algorithm’s internal randomness influences its outputs, and those outputs in turn shape future inputs, subtle dependencies begin to accumulate, c.f., \secref{sec:intro:attacks}.  
Over time, these dependencies compromise the independence assumptions that underpin standard correctness analyses. 
This phenomenon is not limited to overtly adversarial scenarios. 
Rather, it can emerge organically in settings such as interactive data exploration, recommendation systems, or collaborative analytics platforms, where users iteratively refine their inputs in response to prior outputs. 
As a result, even well-designed algorithms may exhibit degraded performance or biased outcomes if not explicitly robust to such adaptivity. 
We discuss such examples in more detail in \secref{sec:intro:applications}. 

To illustrate a more fundamental vulnerability, consider the widely used technique of random sampling. 
In static streams, methods like Bernoulli or reservoir sampling typically provide ``representative'' subsets. 
However, even simple adaptive strategies can exploit these techniques, so that the resulting samples are heavily skewed. 
For example, an adversary can always ensure that a collection of $k$ samples from a dataset of $n$ items are always the $k$ smallest or $k$ largest items, distorting estimates such as the median; for more details, see \secref{sec:uniform:sampling:attack}. 
A similar phenomenon arises in more sophisticated algorithms like the AMS sketch~\cite{AlonMS99}, a classic tool for estimating $L_2$ norms in data streams. 
While effective under random or fixed inputs, the AMS sketch can be misled by inputs that depend on prior outputs, leading to large errors despite their theoretical guarantees. 
We provide more details on these attacks in \secref{sec:intro:attacks}, as well as throughout the monograph. 
More generally, these examples underscore a broader point: adaptivity, even when mild or unintentional, can severely undermine the reliability of randomized streaming algorithms.

The need for robust streaming algorithms becomes even more pressing in environments where adversaries have greater access to the internal state of a system. 
For example, many successful attacks on machine learning models utilized knowledge of internal parameters and training weights to minimize loss functions near the original input, and recent advancements~\cite{BiggioCMNSLGR13,SzegedyZSBEGF14,GoodfellowSS14} have allowed imperceptible modifications to generate adversarial inputs in images~\cite{SzegedyZSBEGF14,HuangPGDA17} and the physical world~\cite{SharifBBR16,KurakinGB17a,AthalyeEIK18} that lead to incorrect classifications. 
Similar attacks on neural networks in streaming settings have also been studied~\cite{MladenovicBBHLV22}. 
In sensor networks, localized decisions may be influenced by global summaries, allowing malicious nodes to alter their readings in response to aggregate feedback. 
Crowd-sourced platforms face manipulation when contributors adapt to dynamic scoring or reputation metrics, subtly gaming the system through repeated interactions. 
Across these domains, the common thread is clear: when algorithmic behavior is exposed or influences future inputs, traditional guarantees quickly break down, necessitating new frameworks for ensuring robustness.

This monograph consolidates a broad array of recent research to present a systematic overview of adversarially robust streaming algorithms. 
We formally introduce a number of distinct adversarial models, ranging from the \emph{black-box} adversary, who observes only the algorithm’s outputs, to the significantly more powerful \emph{white-box} adversary, who has complete access to the algorithm’s internal state and randomness. 
For each scenario, we outline the key challenges and survey the range of algorithmic tools that have been designed to achieve  provable guarantees of robustness.

\section{Motivating Applications}
\seclab{sec:intro:applications}
\paragraph{Recommendation systems.}
You open your favorite streaming platform, and a tailored list of movie recommendations appears, reflecting your past preferences, current trends, and maybe even a touch of serendipity. 
You skip over titles that do not appeal to you, hide ones you’ve already seen, and eventually pick something to watch. 
This interaction feels seamless, but behind the scenes, your feedback—what you skip, hide, or select—is being used to reshape the system’s understanding of your preferences in real time.

Such platforms do not just process static data; they operate in dynamic, interactive environments where users continuously shape the input stream. 
Each recommendation list is crafted not only from an underlying catalog (or stream) of content, but also from prior algorithmic outputs and user actions—many of which depend on internal randomness or learned heuristics. 
This creates a feedback loop: the algorithm influences the user, and the user influences the algorithm. 
Over time, this blurs the line between inputs and outputs, making the data itself a function of the system’s past behavior.

This adaptivity, while crucial for personalization, introduces fundamental challenges. 
The input stream is no longer independent or fixed in advance; it evolves in response to the system’s own outputs. 
Without robust algorithmic design, such feedback can lead to performance degradation—manifesting as overfitting, narrowing of diversity, or amplification of minor preferences into persistent biases. 
What starts as personalization can become stagnation.

To counter this, recommendation systems must exhibit adaptive robustness—the ability to maintain reliable performance even when the data they process has been shaped by their own prior outputs.
This need aligns closely with the goals of adversarially robust submodular maximization~\cite{krause2008robust,MitrovicBNTC17,KazemiZK18,orlin2018robust,AvdiukhinMYZ19}, which studies how to make sequential decisions that remain effective under adaptively chosen inputs. 
Ensuring robustness in this setting is critical not only for sustained accuracy, but also for fairness, diversity, and long-term user trust in the recommendation process.

\paragraph{Database queries.}
A natural application of adaptive robustness arises in interactive database systems operating in the streaming model, where a user issues a sequence of queries over a dynamic dataset. 
In many real-world settings, such as financial monitoring systems, network logs, or user analytics platforms, data arrives continuously, and queries must be answered on-the-fly with sublinear memory. 
Crucially, each response returned by the streaming algorithm may influence the user’s subsequent queries: a user may refine or alter future queries based on information gleaned from earlier outputs. 
This adaptivity implies that the query sequence is not fixed in advance and may be adversarially or adaptively chosen based on the algorithm’s previous behavior, including its internal randomness. 
As a result, traditional streaming guarantees that assume independence between the data stream and the algorithm’s randomness may no longer hold. 
Robust streaming algorithms in this setting must ensure consistent and accurate responses, even when the queries are adaptively correlated with prior outputs, preventing information leakage, unaccounted feedback loops, or exploitation via carefully crafted query sequences.

\paragraph{Compressed sensing and control feedback.}
An illustrative example of adaptive adversarial influence given by \cite{HardtW13} originates from real-time signal processing and control applications, such as radar-based navigation systems. 
In the compressed sensing radar example introduced in~\cite{GilbertHRSW12,GilbertHSWW12}, a ship receives high-dimensional measurements modeled as $\bA\bx$, where $\bA$ is a fixed sensing matrix and $\bx$ is the underlying sparse signal, e.g., positions of incoming threats. 
Based on the sketch $\bA\bx$, the ship executes a sequence of evasive maneuvers. 
However, if an attacker observes the ship’s movements, they can potentially infer information about $\bA\bx$ and alter their attack strategy in subsequent rounds. 
Since the sensing matrix $\bA$ must remain unchanged across interactions due to efficiency and calibration constraints, the system must rely on robust sketching algorithms that maintain correctness even when the underlying signal $\bx$ is chosen adaptively based on prior outputs. 
This motivates the study of sketching under adaptively chosen inputs, in which the adversary may exploit correlations across repeated queries to amplify error or breach guarantees.

\paragraph{Financial analytics, search engine optimization, and strategic manipulation.}
Another important setting is high-frequency trading, where streaming algorithms are employed to analyze rapidly evolving financial data in real time. 
Consider an algorithm that computes low-memory sketches of order flows to detect market trends or anomalies. 
A trader may monitor such summaries to guide trading decisions. 
However, a competitor may observe the trades and strategically manipulate the incoming data stream—such as by introducing noise, bursts of transactions, or misleading signals—to alter the sketch's internal state and induce suboptimal decisions. 
This creates a feedback loop in which the sketch’s outputs influence future inputs in a potentially adversarial manner. 
To guard against such manipulation, streaming algorithms must provide adversarial robustness guarantees that ensure statistical accuracy even under adaptively chosen and possibly adversarially corrupted inputs. 
Such guarantees go beyond traditional worst-case analyses and call for more refined algorithmic techniques that are stable under input-output correlations.

Similarly, consider the setting of search engine optimization. 
Here, web search engines run some webpage ranking algorithm that is known, but whose parameters are hidden to general users and may also change from time to time, e.g., due to periodic maintenance or manual adjustment. 
In this case, website owners may wish to improve the ranking of their website, such as through Google bombing. 
Here, the search engine optimization manipulators are the adversary, as they wish to learn the parameters in the sketch. 
Intentionally or unintentionally, this manipulation may be at the expense of other website owners. 

\paragraph{Multiparty sketching and correlated data sources.}
Adaptive robustness is also critical in distributed and multiparty settings, as originally discussed in~\cite{MironovNS11}. 
In these scenarios, different parties (e.g., institutions or sensors) contribute to a global sketch maintained by a central aggregator. 
Suppose the aggregator responds to each party’s queries using shared random coins or stateful mechanisms. 
Then, subsequent contributions or queries may become correlated with prior outputs—either intentionally or due to structural feedback in the system. 
For example, in collaborative analytics between hospitals or data providers, each party might adjust their data contributions or interpretation strategies based on the aggregator’s intermediate responses. 
Even in non-adversarial settings, this breaks the independence assumptions underlying traditional sketch analyses. 
Therefore, ensuring robustness against such adaptive interactions becomes essential for sketching algorithms deployed in real-world multi-agent systems. 
Addressing this challenge requires new frameworks that allow for input adaptivity while still providing strong approximation and privacy guarantees.

\section{Motivating Attacks}
\seclab{sec:intro:attacks}
In this section, we describe a number of simple attacks on standard algorithms as motivation for the necessity to study adversarial robustness. 
We first discuss a model where the goal is to sample a small representative subset from an underlying and evolving dataset. 
We then describe a scenario where the goal is to estimate the norm of an underlying vector defined by an evolving data stream. 

\subsection{Attack on Random Sampling}
\seclab{sec:intro:sampling:attack}
Random sampling is a fundamental and versatile technique for handling massive datasets across various scientific domains. 
Random sampling is used in numerous applications, including statistics, databases, networking, data mining, approximation algorithms, randomized algorithms, machine learning, and more. 
For a more formal discussion, we defer to \secref{sec:random:sampling}. 

The main goal is to select a small yet representative subset of the data, conduct the analysis on this subset, and extrapolate approximate conclusions for the full dataset. 
Traditionally, the analysis of sampling algorithms has focused on the non-adaptive (or static) setting, where the dataset is fixed in advance and the sampling algorithm operates over this fixed input. 
However, in many real-world scenarios, the assumption that the dataset remains unchanged during sampling is not realistic. 

\paragraph{Sampling in an adaptive environment.} 
Consider a setting where the set of sampled items can impact future updates to the dataset. 
One way to model this is as a two-player game between a sampler and an adversary. 
In each round, the adversary first presents an element to the sampler. 
The element may be influenced, possibly in a probabilistic or adversarial manner, by all previously submitted elements and any information the adversary has gathered about the samples so far. 
After receiving the element, the sampler may choose to update both its internal state and the set of samples. 
The adversary's ultimate goal is to manipulate the sample so that it no longer accurately represents the underlying data stream, thereby causing the sampler to produce misleading or incorrect conclusions. 
Informally, we model the adversary as a sequence of functions that map the previous outputs of an algorithm up to some time $t$ to the next stream element $x_t$, effectively making the input dependent on the internal coins of the sampler. 
We defer the formal model to \secref{sec:random:sampling}. 

Rather than formalizing the notion of representative samples, consider the following illustrative examples given by \cite{Ben-EliezerY20}, which demonstrate how an adaptive adversary can compromise the performance of both Bernoulli and reservoir sampling algorithms.
Consider a stream consisting of $n$ real-valued points $x_1, \ldots, x_n$ drawn from the interval $[0, 1]$. 
The Bernoulli sampling algorithm processes each point in the stream and includes the point in the sample independently with probability $p\in(0,1)$. 
In a static context, and assuming $p$ is sufficiently large, the sampled subset generally captures the entire dataset well under various definitions of representativeness. 
For example, the median of the sample will, with high probability, be within additive $\eps$ to the median of the full stream when $p = \frac{c}{\eps^2 n}$ for some sufficiently large constant $c>0$. 
This property extends to other quantiles as well.

Now consider how this behavior changes under an adaptive adversary. 
In this scenario, the adversary maintains a ``working range'' throughout the game, beginning with the entire interval $[0, 1]$. 
At the first step, the adversary submits $x_1 = 0.5$ to the sampler. 
If $x_1$ is included in the sample, the adversary restricts the range to $[0.5, 1]$; otherwise, the adversary narrows the focus to $[0, 0.5]$. 
For each subsequent round $i$, the adversary submits the midpoint of the current range. 
Formally, let $a_1 = 0$ and $b_1 = 1$. 
In round $i$, the adversary submits $x_i = \frac{a_i + b_i}{2}$. 
If $x_i$ is sampled, then the next range becomes $[x_i, b_i]$; otherwise, it becomes $[a_i, x_i]$. 
This process continues for $n$ rounds, yielding a stream $x_1, \ldots, x_n$.

This strategy ensures that, at each step, the newly submitted element lies above all previously sampled elements and below all elements not yet sampled. 
Consequently, the $k$ elements that eventually comprise the sample will, with probability $1$, be exactly the smallest $k$ elements in the entire stream. 
This subset is highly unrepresentative and, in fact, might be considered the most skewed possible with respect to the full dataset's distribution. 
In such a case, the median of the sample deviates significantly from the true median of the stream.
\cite{Ben-EliezerY20} also showed that a similar attack applies to the reservoir sampling algorithm. 
Although the mechanics differ slightly, the adversary’s strategy can still force all $k$ sampled elements to fall within the first $\O{k\log n}$ elements of the stream, with high probability. 
Additional details regarding this attack and its implications are discussed further in \secref{sec:uniform:sampling:attack}.

\subsection{Attack on Norm Estimation}
\seclab{sec:intro:attack:ams}
In this section, we present an informal sketch of an attack on the classical Alon-Matias-Szegedy (AMS) streaming algorithm~\cite{AlonMS99}, which is widely used to estimate the squared $L_2$ norm of a frequency vector. 
The attack is designed to target the AMS sketch and force it to return an inaccurate estimate of the true value $\|f\|_2^2$, where $f$ is the underlying frequency vector defined by an adversarial data stream. 
Intuitively, the attacker aims to ``query'' the sketch with specific updates to observe how the output changes; this reveals the random signs in the underlying sketch matrix, allowing the adversary to subsequently adaptively query a vector that is perfectly correlated with those signs, so that the algorithm fails. 

\paragraph{AMS sketch description.}
The AMS sketch implicitly generates a random matrix $\bA \in \mathbb{R}^{t \times n}$, where each entry $A_{i,j}$ is independently drawn from the Rademacher distribution, i.e., uniformly chosen from $\{-1, 1\}$. 
At time $j\in[m]$ in the stream, the algorithm maintains the vector $\bA\cdot\bx^{(j)}\in\mathbb{R}^t$, where $\bx^{(j)}$ denotes the frequency vector after the first $j$ updates. 
Note that due to the linearity of $\bA$, each stream update can be efficiently incorporated by
\[\bA\bx^{(j+1)} = \bA\bx^{(j)} + \bA\be_{i_{j+1}}\cdot\Delta_{j+1},\]
where the $(j+1)$-th update is represented by the pair $(i_{j+1}, \Delta_{j+1})$, which means that the $i_{j+1}$-th coordinate of the frequency vector should be changed by $\Delta_{j+1}$. 
In our current setting, we allow $\Delta_{j+1}$ to be either positive or negative. 

The squared $L_2$ norm at time $j$ is approximated using $\frac{1}{t}\|\bA\bx^{(j)}\|_2^2$. 
In the standard non-adaptive setting, this estimate is guaranteed to approximate the true norm within a $(1 \pm \eps)$ factor with high probability, provided the sketch uses $t = \Theta\left(\frac{1}{\eps^2}\right)$ rows.
However, the analysis crucially uses the fact that $A$ is a matrix consisting of scaled random signs. 
In particular, the random signs are independent of the frequency vector $\bx^{(j)}$, so for any row $\bv$ of $\bA$, we have $\Ex{\langle\bv,\bx^{(j)}\rangle}=0$. 
In an adaptive setting where $\bx^{(j)}$ may not be independent of $\bA$, this property may no longer hold. 
Indeed, we describe a simple attack that exploits this property. 

\paragraph{Norm estimation in an adaptive environment.}
Initialize $\bx$ to the all zeros vector $\bx=\mathbf{0}^n$ and initialize the counter $i=1$. 
Consider an attack that does the following. 
For each $i\in[n]$, if $\|\bA(\bx+\be_i)\|_2^2=\|\bA\bx\|_2^2+\Omega(\sqrt{i})$, then we add $\be_i$ to $\bx$. 
Otherwise, we do not change $\bx$ and move onto the next $i$, i.e., we increment the counter $i$. 

We emphasize that the sketch matrix $\bA$ is not observed directly by the attack; instead, it tries to add $\be_i$ and observes the new estimate. 
The intuition is that if $\bA$ has a large correlation with coordinate $i$, then we add $i$ to the frequency vector; otherwise we ignore $i$. 
In essence, the frequency vector is designed specifically to exploit the randomness of $\bA$. 
It can be shown, e.g., by Khintchine's inequality, that for each new $\be_i$, we have that with constant probability $\|\bA(\bx+\be_i)\|_2^2=\|\bA\bx\|_2^2+\Omega(\sqrt{i})$. 
Then by a standard concentration inequality, i.e., Chernoff bound, we will add $\Theta(n)$ different values of $i$ into $\bx$ with high probability. 
In that case, we have $\|\bx\|_2^2\le n$ but by induction 
\[\|\bA\bx\|_2^2\ge\Theta(n)\cdot\Theta(\sqrt{n})=\Theta(n^{3/2}).\] 
As a result, the AMS fails to output a constant-factor approximation to $\|\bx\|_2^2$. 
We formalize this analysis, albeit with a different attack on AMS by \cite{Ben-EliezerJWY22}, in \secref{sec:ams:attack}.

\section{Organization of this Monograph}
The exposition of this monograph is organized as follows:
\begin{itemize}
\item 
\textbf{Foundational concepts and preliminaries (\chapref{chap:prelims}):} 
Before exploring the nuances of adversarial robustness, we begin by establishing a precise common ground. 
This chapter provides a concise review of essential mathematical notation, key probability distributions, and foundational concepts from communication complexity, information theory, and differential privacy. 
This foundational material provides unified notation and preliminaries for more technical discussions in subsequent sections. 
\item 
\textbf{The black-box adversarial model for insertion-only streams (\chapref{chap:black:box}):} 
We begin our discussion with the black-box model, where the adversary observes only the outputs of the streaming algorithm, without access to its internal state or randomness. 
This chapter focuses on insertion-only streams, a simplified yet foundational setting for understanding robust algorithm design. 
We introduce key conceptual tools such as \emph{sketch-switching} and \emph{bounded computation paths}, which enable the systematic transformation of non-robust algorithms into robust ones by carefully regulating the exposure of internal randomness. 
In addition, we present advanced techniques like \emph{difference estimators}, which attain near-optimal space bounds for core streaming problems including $F_p$ norm estimation, distinct element estimation, and identifying heavy hitters. 
These results demonstrate that, in the insertion-only model, the cost of achieving adversarial robustness can be remarkably low. 
We further extend our framework to applications in entropy estimation and sampling, covering both uniform and importance-based approaches.
\item 
\textbf{Connecting robustness to differential privacy and adaptive data analysis (\chapref{chap:dp:ada}):} 
This section reveals deep, and at times counterintuitive, connections between adversarial robustness and core areas of theoretical computer science. 
We show how techniques from \emph{differential privacy}, originally developed for safeguarding individual data in statistical analyses, can be effectively adapted to enhance the robustness of streaming algorithms by deliberately obscuring aspects of their internal state. 
In fact, the differential privacy framework can be isolated as a primitive to robustly answer a number of adaptive queries in other settings as well, such as the dynamic model or sketching for optimization problems, e.g., approximate nearest neighbors, linear regression, or half-space queries. 
In the opposite direction, we show that tools from \emph{adaptive data analysis} offer a powerful lens through which to establish formal separations between the performance of oblivious and adaptive streaming algorithms. 
These separations serve to quantify the inherent computational overhead required to ensure correctness in the presence of adaptively chosen inputs.
\item
\textbf{Navigating turnstile (insertion-deletion) streams (\chapref{chap:turnstile}):} 
The introduction of explicit deletions in the data stream, transitioning from insertion-only to turnstile models, markedly increases the challenge of ensuring adversarial robustness. 
Many of the structural properties and algorithmic tools that enable efficient robustness in simpler settings no longer apply. 
This chapter undertakes a rigorous analysis of techniques specifically designed for turnstile streams, including the use of \emph{dense-sparse decomposition}, which achieves improved space complexity over na\"{i}ve extensions of black-box frameworks. 
Importantly, we also establish strong lower bounds and present targeted attacks on linear sketches, highlighting fundamental vulnerabilities. 
These results show that even widely deployed sketching algorithms can be compromised by carefully crafted adversaries when deletions are allowed—particularly for core problems such as $F_p$ and $F_0$ estimation.
\item 
\textbf{The white-box adversarial model (\chapref{chap:white:box}):} 
We then address the most formidable adversarial model: the white-box adversary, who has full access to the algorithm’s internal state and all its random bits. 
This chapter investigates the fundamental limitations of achieving robustness against such powerful opponents, often drawing deep connections to communication complexity theory. 
Despite the significant challenges, we present robust algorithms for problems such as sparse vector recovery, low-rank matrix recovery, and tensor recovery. 
These algorithms provide provable robustness guarantees against computationally bounded white-box adversaries, relying critically on contemporary cryptographic hardness assumptions—most notably, the Short Integer Solution (SIS) problem.
\item 
\textbf{Adversarially robust algorithms on turnstile streams (\chapref{chap:turnstile:algs}):} 
At this point, the impression created by adaptive attacks, especially those targeting linear sketches, might be that adversarial robustness is impossible in turnstile streams. 
Contrary to this view, we show in this chapter that for a broad class of problems satisfying an approximate triangle inequality, there exist streaming algorithms that are robust to adaptive adversaries while supporting both insertions and deletions. 
As a concrete result, we present an adversarially robust algorithm that achieves a $(1+\eps)$-approximation for $F_2$ estimation using $\poly\left(\frac{1}{\eps},\log n\right)$ space. 
\item 
\textbf{Conclusion and future directions (\chapref{chap:conclusion}):} 
Finally, we bring together the insights gained from our study of the various adversarial models and algorithmic techniques.
This concluding chapter offers a clear summary of the main theoretical contributions, highlights the outstanding open problems, and sketches potential directions for future research in the continually advancing field of adversarial robustness in streaming algorithms.
\end{itemize}

By unifying these diverse lines of inquiry and rigorously analyzing their theoretical underpinnings, this monograph aims to provide a holistic understanding of adversarial robustness in streaming algorithms. 
It is our hope that it can serve as a foundational resource, illuminating both the theoretical principles and practical implications relevant to the design of robust algorithmic systems under increasingly adversarial models motivated by modern applications.

\chapter{Notation/Preliminaries}
\chaplab{chap:prelims}
We denote the set of integers from $1$ to $n$ (for any positive integer $n > 0$) by $[n] = \{1, 2, \ldots, n\}$. 
The notation $\poly(n)$ refers to an arbitrary fixed polynomial in $n$, where the degree may depend on relevant constants.
Similarly, we use $\polylog(n)$ to refer to an arbitrary fixed polynomial in $\log n$, again with degree possibly depending on relevant parameters. 
For a (possibly multivariate) function $f$, we use $\tO{f}$ to denote $f\cdot\polylog(f)$. 
An event is said to occur \emph{with high probability} if it happens with probability at least $1 - \frac{1}{\poly(n)}$. 

For a vector $\bv$, we define
\[\|\bv\|_p = \left(|v_1|^p + \ldots + |v_n|^p\right)^{1/p}\] 
to be its entrywise $L_p$ norm for $p\ge 1$ and the entrywise $L_p$ quasi-norm for $p\in(0,1)$.  
We similarly define $\|\bv\|_0=|\{i\in[n]: v_i\neq 0\}|$. 
For a matrix $\bA \in \mathbb{R}^{n \times d}$, the \emph{operator norm} is defined as $\|\bA\|_2 = \max_{\bx \in \mathbb{R}^d} \frac{\|\bA \bx\|_2}{\|\bx\|_2}$ and \emph{Frobenius norm} is defined as $\|\bA\|_F = \left( \sum_{i,j} A_{i,j}^2 \right)^{1/2}$. 
{
\renewcommand{\thetheorem}{\thechapter.\arabic{theorem}}
\renewcommand{\thedefinition}{\thechapter.\arabic{theorem}}
\begin{definition}[Leverage scores and sensitivities]
\deflab{def:leverage:score}
Let $\ba_1,\ldots,\ba_n\in\mathbb{R}^d$ be the rows of matrix $\bA\in\mathbb{R}^{n\times d}$. 
Given $p>0$, the $L_p$ sensitivity of a row $\ba_t$ is defined to be
\[\max_{\substack{\bx\in\mathbb{R}^d\\ \|\bA\bx\|_p>0}}\frac{|\langle\ba_t,\bx\rangle|^p}{\|\bA\bx\|_p^p}.\]
For $p=2$, we also call this quantity the leverage score of $\ba_t$.
\end{definition}
}

For two vectors $\bu,\bv\in\mathbb{R}^n$, we define their Hamming distance by $\HAM(\bu,\bv)=\|\bu-\bv\|_0$, though we note the more intuitive definition is the number of coordinates on which two vectors $\bu$ and $\bv$ differ. 

Given two probability mass functions $P$ and $Q$, their \emph{total variation distance} is defined as
\[\TVD(P, Q) = \frac{1}{2} \cdot \|P - Q\|_1.\]

\section{Common Probability Distributions}
To begin, we first recall the probability distribution for standard Gaussian random variables. 
\begin{definition}[Continuous Gaussian]
\deflab{def:continuous:gaussian}
For any $s>0$, the spherical Gaussian function with center $\mu \in \mathbb{R}^n$ is defined by
\[\rho_s(\bx) = \exp\left(-\frac{\|\bx - \mu\|_2^2}{2s^2}\right),\]
for all $\bx \in \mathbb{R}^n$. 
Similarly, given any full-rank matrix $\bS \in \mathbb{R}^{m \times n}$, the ellipsoidal Gaussian function with mean $\mathbf{\mu} \in \mathbb{R}^n$ and covariance $\bSigma = \bS^{\top} \bS$ is given by
\[\rho_{\bS}(\bx) = \exp\left( -\frac{1}{2}(\bx - \mathbf{\mu})^{\top} \bSigma^{-1} (\bx - \mathbf{\mu}) \right)\]
for all $\bx \in \mathbb{R}^n$.
For $n=1$, we use the notation $\calN(\mathbf{\mu},\sigma^2)$ to denote the normal distribution, i.e., the univariate Gaussian with mean $\mu$ and variance $\sigma^2$. 
\end{definition}
Recall that a lattice $\calL\subset\mathbb{R}^n$ is a discrete additive subgroup of $\mathbb{R}^n$, equivalently the set of all integer linear combinations of a finite set of linearly independent basis vectors in $\mathbb{R}^n$.
\begin{definition}[Discrete Gaussian]
\deflab{def:discrete:Gaussian}
Given a lattice $\calL \subset \mathbb{R}^n$, a shift vector $\textbf{c} \in \mathbb{R}^n$, and a point $\bx \in \calL + \textbf{c}$, the discrete Gaussian distribution over $\calL + \textbf{c}$ with mean $\bmu \in \mathbb{R}^n$ and covariance matrix $\bSigma = \bS^{\top} \bS \in \mathbb{R}^{n \times n}$ is defined by the probability mass function
\[\mathcal{D}_{\calL + \textbf{c}, \bS}(\bx) = \frac{\rho_{\bS}(\bx)}{\rho_{\bS}(\calL + \textbf{c})},\]
for all $\bx \in \calL + \mathbf{c}$, where $\rho_{\bS}(A) = \sum_{\bx \in A} \rho_{\bS}(\bx)$ for any subset $A \subseteq \mathbb{R}^n$. 
We write $\mathcal{D}(0,\sigma^2\mathbb{I}_n)$ to refer to the discrete Gaussian distribution over $\calL = \mathbb{Z}^n$ with mean $0^n$ and covariance $\bSigma = \sigma^2 \cdot\mathbb{I}_n$.
\end{definition}
We recall the following result on the normalization constant for the probability mass function of discrete Gaussians. 
\begin{fact}[Normalization constant]
\factlab{fact:normalization}
\cite{CanonneKS20}
For any $\sigma > 0$ with $\sigma \in \mathbb{R}$, the following inequality holds:
\[\max\left\{ \sqrt{2\pi \sigma^2},\, 1 \right\} \le \sum_{z \in \mathbb{Z}} e^{-z^2 / 2\sigma^2} \le \sqrt{2\pi \sigma^2} + 1.\]
\end{fact}
We also note the following connection between the value of the discrete Gaussian probability mass function at a point $\bv$ and the value of the continuous Gaussian probability density function evaluated at $\bv$. 
The proof is entirely routine and is provided solely for completeness.
\begin{lemma}
\lemlab{lem:pmf}
Suppose that $\sigma> n^{C+1}$ is sufficiently large. 
Let $p$ denote the probability mass function of the discrete Gaussian distribution $\calD(0, \sigma^2 \mathbb{I}_n)$, and let $q$ be the probability mass function obtained by sampling $x \sim \calN(0, \sigma^2 \mathbb{I}_n)$ and truncating each coordinate of $x$ to the nearest integer. 
Then, for every $\bv \in \mathbb{Z}^n$ such that $\|\bv\|_1 \le \frac{\sigma^2}{n^{C+1}}$, we have
\[\frac{p(\bv)}{q(\bv)} \in \left[1 - \frac{1}{n^C},\, 1 + \frac{1}{n^C} \right].\]
\end{lemma}
\begin{proof}
We have the following probability mass function for the distribution $q$:
\[q(\bv) = \frac{1}{(2\pi \sigma^2)^{n/2}} \int_{v_1}^{v_1 + 1} \cdots \int_{v_n}^{v_n + 1} e^{-\|\bx\|_2^2 / 2\sigma^2} \, dx_1 \cdots dx_n.\]
Note that since $\|\bv\|_1 \le \frac{\sigma^2}{n^{C+1}}$, then for each coordinate $i$, the term $e^{-(x_i+1)^2 / 2\sigma^2}$ lies within the interval
\[\left[e^{-x_i^2 / 2\sigma^2} \cdot e^{-1 / n^{C+1}}, e^{-x_i^2 / 2\sigma^2} \cdot e^{1 / n^{C+1}} \right].\]
Using this to approximate the integral above, it follows that
\[q(\bv) \in \left[\frac{e^{-\|\bv\|_2^2 / 2\sigma^2}}{(2\pi \sigma^2)^{n/2}} \cdot e^{-1 / n^{C}}, \frac{e^{-\|\bv\|_2^2 / 2\sigma^2}}{(2\pi \sigma^2)^{n/2}} \cdot e^{1 / n^{C}}\right].\]
Next, recall that for any $\bv \in \mathbb{Z}^n$, the probability mass function of $\calD(0, \sigma^2 \mathbb{I}_n)$ is
\[p(\bv) = \frac{e^{-\|\bv\|_2^2 / 2\sigma^2}}{\left( \sum_{z \in \mathbb{Z}} e^{-z^2 / 2\sigma^2} \right)^n}.\]
By \factref{fact:normalization}, we know
\[p(\bv) \in \left[\frac{e^{-\|\bv\|_2^2 / 2\sigma^2}}{(\sqrt{2\pi \sigma^2} + 1)^n}, \frac{e^{-\|\bv\|_2^2 / 2\sigma^2}}{(\sqrt{2\pi \sigma^2})^n}\right].\]
Finally, the desired result follows by combining these two bounds and using the assumption $\sigma>n^{C+1}$.
\end{proof}

\section{Preliminaries for Communication Complexity and Information Theory}
In this section, we recall a number of preliminaries from communication complexity and information theory. 
\begin{definition}[Entropy and conditional entropy]
Let $X$ be a random variable taking on possible values in a finite domain $\Omega$ and probability mass function $p(x)=\PPr{X=x}$. 
Then the \emph{entropy} of $X$ is defined as 
\[H(X):=\sum_{x\in\Omega} p(x)\log\frac{1}{p(x)}.\]  
The \emph{conditional entropy} of $X$ with respect to a random variable $Y$ is defined as 
\[H(X|Y)=\mathbb{E}_{y}{H(X|Y=y)},\]
where $H(X|Y=y):=\sum_{x\in\Omega} p(x|y)\log\frac{1}{p(x|y)}$, for the conditional probability mass function $p(x|y)$. 
\end{definition}

\begin{definition}[Mutual information and conditional mutual information]
Given random variables $X$ and $Y$, we define their \emph{mutual information} by
\[I(X;Y)=H(X)-H(X|Y)=H(Y)-H(Y|X)=I(Y;X).\]
We define the \emph{conditional mutual information} between $X$ and $Y$ conditioned on a random variable $Z$ by 
\[I(X;Y|Z)=H(X|Z)-H(X|Y,Z).\]
\end{definition}

\begin{theorem}[Data-processing inequality]
\cite{cover1999elements}
Let $X, Y, Z$ be random variables such that $X \rightarrow Y \rightarrow Z$ forms a Markov Chain, i.e., $X$ and $Z$ are conditionally independent given $Y$. 
Then 
\[I(X; Y) \ge I(X;Z).\]
\end{theorem}

\begin{theorem}[Chain rule for mutual information]
\thmlab{thm:chain:information}
\cite{cover1999elements}
Let $X_1,\ldots,X_n,Z$ be random variables. 
Then
\[I(X_1,\ldots,X_n;Z)= \sum_{i = 1}^n I(X_i; Z | X_1, ..., X_{i-1}).\]
\end{theorem}

\section{Preliminaries for Differential Privacy}
\seclab{sec:prelims:dp}
We first define differential privacy~\cite{DworkMNS06}, which informally demands that changing one data point in the input does not significantly change the algorithm's output distribution, making it difficult to tell if any individual was in the dataset.
\begin{restatable}[Differential privacy, \cite{DworkMNS06,DworkKMMN06}]{definition}{defdp}
\deflab{def:dp}
Given a privacy parameter $\eps>0$ and an additive parameter $\delta\in(0,1)$, a randomized algorithm $\calA$ is $(\eps,\delta)$-differentially private if for all datasets $D_1$ and $D_2$ that differ on a single element, and all subsets $S$ of the range of $\calA$, 
\[\PPr{\calA(D_1)\in S}\le e^{\eps}\cdot\PPr{\calA(D_2)\in S}+\delta.\]
\end{restatable}
We next define the Laplace distribution, which is a bell-shaped distribution centered at $0$, often used to add noise to protect privacy in data analysis.
\begin{definition}[Laplace distribution]
For $x\in\mathbb{R}$, the probability density function $f(x)$ of a random variable drawn from the Laplace distribution $\Lap(b)$ is $f(x)=\frac{1}{2b}\cdot\exp\left(-\frac{|x|}{b}\right)$.
\end{definition}
We next define sensitivity, which measures how much a function’s output can change when one item in the dataset is changed.
\begin{definition}[Sensitivity]
The sensitivity of a function $f:X\to\mathbb{R}$ is the maximum value of $|f(S)-f(S')|$, taken across all pairs $S,S'\in X$ of datasets that differ on a single item. 
\end{definition}
It is known that one method to achieving differential privacy for a numeric function is through the Laplace mechanism, which adds Laplace noise scaled to the sensitivity of the function.
\begin{theorem}[Laplace mechanism, \cite{DworkMNS06}]
Let $f:X\to\mathbb{R}$ be a function with sensitivity $\ell$. 
Then the mechanism that takes input $S\in X$ and outputs $f(S)+\Lap\left(\frac{\ell}{\eps}\right)$ is $(\eps,0)$-differentially private. 
\end{theorem}

\paragraph{Composition of differential privacy.}
Composition theorems in differential privacy quantify the accumulation of privacy loss across the application of multiple differentially private mechanisms on the same dataset. 
These theorems provide formal guarantees on the overall privacy budget, enabling practitioners to analyze and control cumulative privacy leakage over multiple queries or computations. 
The basic composition theorem states that the privacy loss over multiple mechanisms is roughly additive:
\begin{theorem}[Basic composition of differential privacy]
\cite{DworkMNS06}
\thmlab{thm:dp:basic:comp}
Let $\calA_i:X\to\mathbb{R}$ be an $(\eps_i,\delta_i)$-differentially private algorithm for each $i\in[k]$. 
Then $\calA_{[k]}(x)=(\calA_1(x),\ldots,\calA_k(x))$ is $\left(\sum_{i=1}^k\eps_i,\sum_{i=1}^k\delta_i\right)$-differentially private. 
\end{theorem}
Advanced composition is a more sophisticated technique to show that the privacy loss can be sublinear in the sum of the privacy losses across each of the mechanisms.
\begin{theorem}[Advanced composition of differential privacy]
\cite{DworkRV10}
\thmlab{thm:dp:adv:comp}
Let $\eps,\delta'\in(0,1]$ and $\delta\in[0,1]$. 
Any mechanism that permits $k$ adaptive interactions with mechanisms that preserve $(\eps,\delta)$-differential privacy guarantees $(\eps',k\delta+\delta')$-differential privacy, where $\eps'=\sqrt{2k\ln\frac{1}{\delta'}}\cdot\eps+2k\eps^2$. 
\end{theorem}
Here, an adaptive interaction means that each query to the differentially private mechanism can depend on the answers received from all previous queries. 
In other words, the choice of the $i$-th query is allowed to be a function of the outputs of the first $i-1$ queries, rather than being fixed in advance. 
This models scenarios where an analyst uses prior results to determine future queries.

\paragraph{Sparse vector technique.}
The sparse vector technique is an approach introduced by Dwork, Naor, Reingold, Rothblum, and Vadhan~\cite{DworkNRRV09} to efficiently and privately answer a sequence of queries while minimizing noise. 
In particular, it only adds noise to a small subset of queries that exceed a certain threshold. 
Formally, a data analyst holding a dataset $S$ is given a sequence of functions $f_1,f_2,\ldots$ with sensitivity $1$. 
The sparse vector technique privately reports the first index $i$ such that $f_i(S)$ exceeds a threshold $t$. 
Given in \algref{alg:svt}, the sparse vector technique offers the following guarantees:
\begin{theorem}[Sparse vector technique]
\thmlab{thm:svt}
\cite{DworkNRRV09}
There exists an algorithm $\textsc{AboveThreshold}$ that is $\eps$-differentially private and identifies the first among a sequence $f_1,f_2,\ldots$ of sensitivity $1$ queries that exceeds a noisy threshold $t$.
\end{theorem}

\begin{algorithm}[!htb]
\caption{Algorithm $\textsc{AboveThreshold}$, i.e., the sparse vector technique~\cite{DworkNRRV09}}
\alglab{alg:svt}
\begin{algorithmic}[1]
\Require{Database $S\in X$, privacy parameter $\eps$, threshold $t$, functions $f_1,f_2,\ldots:X\to\mathbb{R}$ with sensitivity $1$}
\State{$\tau\gets t+\Lap\left(\frac{2}{\eps}\right)$}
\For{each round $i$}
\State{$\widehat{f_i}\gets f_i(S)+\Lap\left(\frac{4}{\eps}\right)$}
\If{$\widehat{f_i}\ge\tau$}
\State{Output $\top$ and halt}
\Else
\State{Output $\bot$ and continue to next iteration}
\EndIf
\EndFor
\end{algorithmic}
\end{algorithm}

\paragraph{Private median.}
The goal of an algorithm that privately computes a median of a dataset $S\subset\mathbb{R}$ is to output $x\in\mathbb{R}$ such that with high probability, there are at least $\frac{|S|}{2}-k$ elements in $S$ that are at least $x$, and at least $\frac{|S|}{2}-k$ elements in $S$ that are at most $x$. 
A standard approach is to use the exponential mechanism~\cite{McSherryT07} to sample some number, where the score function is the number of elements in $S$ between $x$ and the median. 
Because the range of the samples is $\mathbb{R}$, then the sampling process can be efficiently simulated, thereby achieving the following guarantees:

\begin{theorem}[Private median, e.g.,~\cite{HassidimKMMS22}]
\thmlab{thm:dp:median}
Given a database $S\in X$, there exists a parameter $k=\O{\frac{1}{\eps}\log\frac{|X|}{\delta}}$ and an $(\eps,0)$-differentially private algorithm $\PrivMed$ that outputs an element $x\in X$ such that with probability at least $1-\delta$, there are at least $\frac{|S|}{2}-k$ elements in $S$ that are at least $x$, and at least $\frac{|S|}{2}-k$ elements in $S$ that are at most $x$. 
\end{theorem}

\paragraph{Generalization of differential privacy.}
In the context of differential privacy, generalization seeks guarantees on the properties of a randomly sampled subset of a dataset or distribution, compared to the overall dataset or distribution. 
For example, \cite{DworkFHPRR15,BassilyNSSSU21} showed that for the value of a predicate $h$ computed on a random sample of size $n$ from a distribution $\calD$ in a differentially private manner, the empirical mean of $h$ over the same samples is ``close'' to the true population mean of $h$ under the distribution $\calD$:

\begin{theorem}[Generalization of differential privacy, e.g.,~\cite{DworkFHPRR15,BassilyNSSSU21}]
\thmlab{thm:generalization}
Let $\eps\in(0,1/3)$ be a privacy parameter, $\delta\in(0,\eps/4)$ be an additive parameter, and $n\ge\frac{1}{\eps^2}\log\frac{2\eps}{\delta}$. 
Let $\calA:X^n\to 2^X$ be an $(\eps,\delta)$-differentially private algorithm that processes a database of size $n$ and produces a predicate $f:X\to\{0,1\}$. 
Suppose $\calD$ is a distribution over $X$ and $S$ is a dataset of $n$ elements drawn independently and identically distributed from $\calD$. 
Then
\[\PPPr{S\sim\calD,f\gets\calA(S)}{\left|\frac{1}{|S|}\sum_{x\in S}f(x)-\EEx{x\sim\calD}{f(x)}\right|\ge10\eps}<\frac{\delta}{\eps}.\]
\end{theorem}
We remark that unlike a standard Chernoff bound, which only applies to a fixed predicate independent of the sample, this result applies to predicates $f$ that are chosen based on the dataset $S$ itself. 
Thus, differential privacy ensures that even though $f$ may depend on $S$, the empirical average over the sample remains close to the true expectation over the distribution.

\chapter{Black-Box Model}
\chaplab{chap:black:box}

\begin{tallchapterbannerbox}
\centering
For some central problems on insertion-only streams, one can obtain adversarially robust algorithms that come close to matching the performance of their non-robust counterparts.
\end{tallchapterbannerbox}
\vspace{0.4in}

In this chapter, we introduce the \emph{black-box} adversarial model, where an adversary $\Adv$ has repeated interactions with outputs of an algorithm $\Alg$ through a data stream that represents the queries of $\Adv$. 
First introduced by \cite{Ben-EliezerJWY22}, the model can be summarized as the following two-player game between a streaming algorithm $\Alg$ and a source $\Adv$ of adaptive or adversarial input to $\Alg$. 
Prior to the game, a fixed query function $\calQ$ is determined. 
The game then proceeds over $m$ rounds, so that in the $t$-th round:
\begin{enumerate}
\item
$\Adv$ computes an update $s_t$ for the stream, which possibly depends on all previous stream updates and all previous outputs from $\Alg$. 
\item
$\Alg$ updates its internal data structures $\calD_t$ with $s_t$, possibly drawing a fresh batch $R_t$ of random bits, and outputs a response $Z_t$.
\item
$\Adv$ observes and records the response $Z_t$.
\end{enumerate}
The goal of $\Alg$ is to produce a correct answer $Z_t$ to the query function $\calQ$ on the dataset $\{s_1,\ldots,s_t\}$ across all times $t\in[m]$. 
Conversely, the goal of the adversary $\Adv$ is to compel an incorrect response $Z_t$ to the query $\calQ$ at some time $t\in[m]$ throughout the stream through its choices of $s_1,\ldots,s_m$. 
By the nature of the game, the algorithm $\Alg$ is permitted space sublinear in the size of the input $m$ and only a single pass over the stream. 

Throughout this section, we focus on the setting where data stream updates can only be inserted, meaning that once an element enters the stream, it cannot be deleted or modified. 
In \secref{sec:random:sampling}, we first suppose that all possible stream updates are elements from some underlying universe and the goal is to acquire a representative sample of the dataset. 

In the latter sections, we assume that the underlying universe is the set of integers $[n]=\{1,2,\ldots,n\}$ and we focus on frequency vectors, so that the underlying dataset is the number of times each item appears (or the sum of the updates to each item). 
Then the data stream defines an underlying frequency vector $x\in\mathbb{R}^n$ so that each stream update $s_t=(a_t,\Delta_t)$ increases coordinate $a_t\in[n]$ of $x$ by some $\Delta_t>0$.  
In other words, the frequency vector $x$ at the end of the stream of length $m$ is defined so that for all $i\in[n]$,
\[x_i=\sum_{t: a_t=i}\Delta_t.\]
In our setting, we assume that each increase $\Delta_t$ is a positive integer upper bounded by a fixed polynomial in $n$, so that the update can be encoded using $\O{\log n}$ bits. 

\paragraph{Chapter organization.}
We first show in \secref{sec:random:sampling} that random sampling approaches such as Bernoulli sampling and reservoir sampling can be adversarially robust with a sufficient number of samples, in the sense that the density of the samples accurately captures the distribution of the dataset. 

For a number of subsequent sections, it would be instructive to recall that the standard AMS algorithm~\cite{AlonMS99} can be adapted to use $\tO{\frac{1}{\eps^2}\log n}$ bits of space and provide $(1+\eps)$-approximation to the $F_2$ moment at all times over the course of an insertion-only stream~\cite{BravermanCIW16}. 
In \secref{sec:ams:attack}, we give an attack on the AMS algorithm, showing it is not adversarially robust on insertion-only streams. 
We then present two generic frameworks in the insertion-only model that were first introduced by \cite{Ben-EliezerJWY22} and use $\tO{\frac{1}{\eps^3}{\log^2 n}}$ bits of space. 
These frameworks efficiently transform non-robust streaming algorithms into adversarially robust streaming algorithms. 
The first approach is called sketch switching and is presented in \secref{sec:sketch:switch}. 
The second approach is called bounded computation paths and is presented in \secref{sec:bounded:computations}. 
Both approaches utilize the fact that specific functions of interest have values that do not change too many times over the duration of the stream. 
However, they both incur an extraneous $\tO{\frac{1}{\eps}\log n}$ multiplicative factor over the optimal algorithm for $F_2$ estimation in the non-adaptive setting. 

This overhead was shown to be unnecessary by a difference estimator approach by \cite{WoodruffZ21}, which we detail in \secref{sec:diff:est}. 
Namely, they show that there exists an adversarially robust algorithm on insertion-only streams that uses $\tO{\frac{1}{\eps^2}\log n}$ and outputs a $(1+\eps)$-approximation to the $F_2$ moment at all times. 
We remark that chronologically speaking, difference estimators were preceded by an approach that used differential privacy and achieved $\tO{\frac{1}{\eps^{2.5}}}\cdot\polylog(n)$ bits of space~\cite{HassidimKMMS20}; however, we defer discussion of this approach to \chapref{chap:dp:ada} for the purposes of a unified presentation. 

In light of these approaches, one may ask whether there exist any problems that admit a separation between adaptive and non-adaptive insertion-only streams. 
To this end, \cite{KaplanMNS21} introduced the streaming adaptive data analysis problem and showed such a separation for this problem; again for the purposes of presentation, we defer a discussion of this problem to \chapref{chap:dp:ada}. 
Subsequently, \cite{ChakrabartiGS22} showed that such a separation exists for the problem of graph coloring; we discuss this result in \secref{sec:graph:color}. 

\section{Random Sampling}
\seclab{sec:random:sampling}
Random sampling is a simple, general-purpose technique that plays a foundational role in processing large-scale data across numerous scientific and engineering disciplines. 
Its effectiveness spans applications in statistics, databases, networking, data mining, approximation algorithms, randomized algorithms, and machine learning, among others. 
The power of random sampling lies in its ability to yield accurate, high-probability approximations by selecting only a small subset of a massive dataset, thus circumventing the need for expensive or impractical computations on the full data. 
By analyzing a randomly chosen, representative subset, one can efficiently generalize certain properties to the entire dataset. 
In this section, we study the adversarial robustness of random sampling.

\subsection{Uniform Sampling}
We first discuss algorithmic design for producing a representative sample from an adaptively evolving dataset. 
A more systematic investigation of the power of uniform sampling against adaptive adversaries was conducted by \cite{AlonBDMNY21}, who showed that proving sample complexity results in this setting is equivalent to proving regret bounds in an online learning setting. 
As it turns out, the \emph{Littlestone dimension}~\cite{Littlestone87} captures the complexity in both cases. 
In any case, we follow here the exposition of \cite{Ben-EliezerY20}, which is simpler but yields weaker bounds. 

A widely accepted notion of representativeness in data sampling is the concept of an $\eps$-approximation, introduced by Vapnik and Chervonenkis~\cite{VapnikC71} and further explored in the context of discrepancy theory~\cite{Chazelle01, MustafaV17}. 
This notion is intricately connected to the theory of VC-dimension~\cite{VapnikC71}, and it captures a key property desired in representative samples: the ability to approximate distributions over subsets of the universe.

Let $X = (x_1, \ldots, x_n)$ be a sequence of (possibly repeated) elements from a universe $U$, and let $R \subseteq U$ be any subset. 
The \emph{density} of $R$ in $X$ is defined as the proportion of elements in $X$ that belong to $R$, i.e., $d_R(X) = \mathbf{\Pr}_{i \in [n]}[x_i \in R]$.

A \emph{set system} is a pair $(U, \calR)$, where $\calR \subseteq 2^U$ is a collection of subsets of $U$. 
A (non-empty) subsequence $S$ of $X$ is said to be an \emph{$\eps$-approximation} of $X$ with respect to the set system $(U, \calR)$ if it approximates the density of every set $R \in \calR$ within additive error $\eps$:

\begin{definition}[$\eps$-approximation]
A sample $S$ is an $\eps$-\emph{approximation} of $X$ with respect to $\calR$ if for every $R \in \calR$,
\[\left| d_R(X) - d_R(S) \right| \le \eps.\]
\end{definition}

In settings where the universe $U$ is well-ordered, a common and meaningful choice for $\calR$ is the family of all intervals of the form $[a, b] \subseteq U$ (including degenerate intervals of the form $[a, a]$, i.e., single points). 
With this set system, an $\eps$-approximation provides a powerful and intuitive notion of representativeness for streaming algorithms, with deep connections to a number of classical research problems, such as approximate median and more generally, quantile estimation~\cite{GreenwaldK01, WangLYC13,KarninLL16,GribelyukSWY24,GuptaSW24,GribelyukSWY25}, as well as range queries~\cite{BagchiCEG07}. 
Specifically, if $S$ is an $\eps$-approximation of $X$ with respect to $(U, \calR)$, then for any quantile rank $q \in [0,1]$, the $q$-quantile of $S$ is guaranteed to be within $\eps$ of the corresponding $q$-quantile in $X$. 

For adversarial robustness, the model can be described as the following two-player game between a streaming sampler $\Alg$ and an adaptive adversary $\Adv$. 
The game unfolds over $n$ rounds, and proceeds as follows:

\begin{enumerate}
\item 
At each round $i \in [n]$, the adversary $\Adv$ selects an element $x_i$ from a fixed universe $U$, possibly using a randomized strategy that depends on the entire history $(x_1, \ldots, x_{i-1})$ and the sampler’s internal state $\sigma_{i-1}$.
\item 
The sampler $\Alg$ receives the next stream element $x_i$, performs arbitrary (possibly unbounded) computation, updates its internal state to $\sigma_i \gets \Alg(\sigma_{i-1}, x_i)$, and optionally records information from the stream. 
The sampler does not need to know $n$ in advance.
\item 
After $n$ rounds, the sampler outputs a final state $\sigma_n$. 
In the sampling algorithms considered in this section, this final state defines a sample $S \subseteq (x_1, \ldots, x_n)$, typically a (possibly non-consecutive) subsequence of the input stream.
\end{enumerate}
We remark that this is actually a white-box model, since the adversary has full access to the internal parameters of the sampling algorithm, c.f., \chapref{chap:white:box}. 
Nevertheless, using the game defined above, we now describe what it means for a sampling algorithm to be adversarially robust. 
\begin{definition}[Robust sampling algorithm]
We say that a sampling algorithm $\Alg$ is $(\eps,\delta)$-robust with respect to the set system $(U, \mathcal{R})$ and the stream length $n$ if for any (even unbounded) strategy of $\Adv$, it holds that with probability at least $1-\delta$, the final sample $S$ produced by $\Alg$ is $\eps$-representative of $X=(x_1,\ldots,x_n)$. 
The memory size used by $\Alg$ is defined to be the maximal size of $\sigma_1,\ldots,\sigma_n$. 
\end{definition}

The analysis of adversarial strategies relies heavily on concentration inequalities for martingales.

\begin{definition}[Martingale]
A \emph{martingale} is a sequence of random variables $X = (X_0, \ldots, X_m)$ with finite expectations such that for all $0 \le i < m$, the following holds:
\[\Ex{X_{i+1} \mid X_0, \ldots, X_i} = X_i.\]
\end{definition}

Rather than the most common formulations for concentration inequalities for martingales such as Azuma's inequality, which applies to martingales with bounded differences $|X_{i+1} - X_i|$, the analysis of \cite{Ben-EliezerY20} requires a more general result that incorporates both bounded differences and bounded conditional variance, originally proven by \cite{Freedman75,McDiarmid98} and refined as follows: 

\begin{lemma}[\cite{ChungL06}, Theorem 6.1]
\lemlab{lem:martingale-with-var}
Let $X = (X_0, \ldots, X_n)$ be a martingale. 
Suppose that for all $1 \le i \le n$, the conditional variance satisfies $\Var(X_i \mid X_0, \ldots, X_{i-1}) \le \sigma_i^2$ for some $\sigma_1, \ldots, \sigma_n \ge 0$, and that $|X_i - X_{i-1}| \le M$ for some constant $M \ge 0$. 
Then, for any $\lambda \ge 0$,
\[\PPr{|X - X_0| \ge \lambda} \le 2\exp\left(-\frac{\lambda^2}{2 \sum_{i=1}^{n} \sigma_i^2 + M\lambda/3}\right).\]
In particular,
\[\PPr{X - X_0 \ge \lambda} \le \exp\left(-\frac{\lambda^2}{2 \sum_{i=1}^{n} \sigma_i^2 + M\lambda/3}\right).\]
\end{lemma}

Unlike Azuma’s inequality, \lemref{lem:martingale-with-var} is especially useful in cases where the possible step size $M$ is large but only rarely attained, resulting in significantly smaller variances. 
The martingales analyzed by \cite{Ben-EliezerY20} exhibit precisely this behavior, making this inequality particularly suited to our setting.

\subsubsection{Bernoulli Sampling}
We first show that Bernoulli sampling is robust against adaptive adversaries, following the presentation of \cite{Ben-EliezerY20}. 
Recall that in Bernoulli sampling, each element is independently included in the sample with a fixed probability $p$. 
For each step $0 \leq i \leq n$ in the process, let $X_i = (x_1, \ldots, x_i)$ denote the sequence of elements selected by the adversary up to round $i$, and let $S_i \subseteq X_i$ be the corresponding subsequence of elements that were sampled. 
Note that $X_n = X$ and $S_n = S$, so to prove the lemma, it suffices to show that $|d_R(X_n) - d_R(S_n)| \leq \eps$.

A natural first approach is to analyze the sequence of random variables $(Y_0, Y_1, \ldots, Y_n)$, where we define $Y_i = d_R(X_i) - d_R(S_i)$, and attempt to apply a martingale concentration inequality. 
Since our objective is to bound the probability that $Y_n$ deviates significantly from zero, this seems promising. 
However, a direct computation reveals that $(Y_i)$ is not a martingale in general, as the condition $\mathbb{E}[Y_i \mid Y_0, \ldots, Y_{i-1}] = Y_{i-1}$ fails to hold.

To address this issue, we instead define an alternate sequence of random variables that do form a martingale. 
For any fixed $R \subseteq U$ and $0 \leq i \leq n$, define:
\begin{align*}
A^R_i &= \frac{i}{n} \cdot d_R(X_i) = \frac{|R \cap X_i|}{n}, \\
B^R_i &= \frac{|R \cap S_i|}{np}, \\
Z^R_i &= B^R_i - A^R_i.
\end{align*}
Here, as before, $R \cap X_i$ denotes the subsequence of $X_i$ consisting of elements that belong to $R$.

Crucially, as established in the following claim, the sequence $Z^R = (Z^R_0, \ldots, Z^R_n)$ forms a martingale. 
Moreover, the claim provides useful bounds on the conditional variance and step size of the process, which will be instrumental in conjunction with \lemref{lem:martingale-with-var}.

\begin{lemma}
\cite{Ben-EliezerY20}
\lemlab{lem:uniform-martingale}
The sequence $(Z^R_0, \ldots, Z^R_n)$ is a martingale. 
In addition, for each $i$, the conditional variance satisfies $\Var(Z^R_i \mid Z^R_0, \ldots, Z^R_{i-1}) \leq\frac{1}{n^2p}$, and the difference between successive terms is bounded as $|Z^R_i - Z^R_{i-1}| \leq\frac{1}{np}$.
\end{lemma}
\begin{proof}
We first verify that the sequence $(Z^R_0, Z^R_1, \ldots, Z^R_n)$ indeed forms a martingale. 
Fix an index $i\in[n]$, and suppose that the first $i-1$ rounds of interaction between the adversary and the sampling algorithm have passed. 
That is, the values $Z^R_0, \ldots, Z^R_{i-1}$ are determined, and the adversary now selects the element $x_i$ to present in round $i$.

First consider the case where $x_i \notin R$. In this case, both $A^R_i = A^R_{i-1}$ and $B^R_i = B^R_{i-1}$, which implies that $Z^R_i = Z^R_{i-1}$. 
Hence, the conditional expectation satisfies:
\[\mathbb{E}[Z^R_i \mid Z^R_0, \ldots, Z^R_{i-1};\ x_i \notin R] = Z^R_{i-1},\]
as required.

Now consider the case where $x_i \in R$. Then:
\[A^R_i = A^R_{i-1} + \frac{1}{n},\]
and depending on whether $x_i$ is sampled:
\[B^R_i = 
\begin{cases}
B^R_{i-1}, & \text{if } x_i \text{ is not sampled}, \\
B^R_{i-1} + \frac{1}{np}, & \text{if } x_i \text{ is sampled}.
\end{cases}\]
It follows that:
\[Z^R_i = 
\begin{cases}
Z^R_{i-1} - \frac{1}{n}, & \text{if } x_i \text{ is not sampled}, \\
Z^R_{i-1} + \frac{1}{np} - \frac{1}{n}, & \text{if } x_i \text{ is sampled}.
\end{cases}\]

Since each element is independently sampled with probability $p$ (regardless of previous rounds), we compute the conditional expectation as:
\begin{align*}
\mathbb{E}[Z^R_i \mid Z^R_0, \ldots, Z^R_{i-1};\ x_i \in R] &= Z^R_{i-1} + p \cdot \left( \frac{1}{np} - \frac{1}{n} \right) + (1 - p) \cdot \left( -\frac{1}{n} \right) \\
&= Z^R_{i-1}.
\end{align*}

Combining the two cases ($x_i \in R$ and $x_i \notin R$), we conclude that
\[\mathbb{E}[Z^R_i \mid Z^R_0, \ldots, Z^R_{i-1}] = Z^R_{i-1},\]
which confirms the martingale property.

We now prove the remaining two properties claimed in the statement, upper bounding the conditional variance and the difference between successive terms. 
The maximal change in $Z^R_i$ across a single round occurs when $x_i \in R$, and is given by:
\[\left| Z^R_i - Z^R_{i-1} \right| \leq \max \left\{ \frac{1}{n},\ \frac{1}{np} - \frac{1}{n} \right\} \leq \frac{1}{np}.\]

For the variance bound, observe that if $x_i \notin R$, then $Z^R_i = Z^R_{i-1}$ deterministically, and the conditional variance is zero. If $x_i \in R$, then:
\begin{align*}
\Var(Z^R_i \mid Z^R_0, \ldots, Z^R_{i-1};\ x_i \in R) &= (1 - p) \cdot \left( \frac{1}{n} \right)^2 + 
p \cdot \left( \frac{1}{np} - \frac{1}{n} \right)^2\\
&= \frac{1}{n^2} \left( \frac{1}{p} - 1 \right) \leq \frac{1}{n^2 p}.
\end{align*}
Thus, in all cases,
\[\Var(Z^R_i \mid Z^R_0, \ldots, Z^R_{i-1}) \leq \frac{1}{n^2 p},\]
completing the proof.
\end{proof}

\begin{lemma}
\cite{Ben-EliezerY20}
Given $\eps,\delta\in(0,1)$, consider a fixed universe $U$ and a subset $R\subseteq U$, and let $X = (x_1, x_2, \ldots, x_n)$ be the sequence chosen by $\Adv$. 
Then for a sample $S$ produced by Bernoulli sampling with probability $p \geq 10 \cdot \frac{\ln(4/ \delta)}{\eps^2 n}$, we have 
\[\PPr{|d_R(X) - d_R(S)| \geq \eps} \leq \delta.\]
\end{lemma}
\begin{proof}
We prove the following two bounds for any sampling probability $p$ that satisfies the conditions stated in the lemma for Bernoulli sampling:
\begin{align}
\label{eq:union-bound-uniform}
\PPr{|A^R_n - B^R_n| \geq\frac{\eps}{2}}\leq \frac{\delta}{2}, \qquad \PPr{|B^R_n - d_R(S_n)| \geq\frac{\eps}{2}} \leq\frac{\delta}{2}.
\end{align}
Applying a union bound to these two inequalities, and using the triangle inequality, we deduce that
\[\PPr{|d_R(X_n) - d_R(S_n)| \geq \eps} \leq \delta,\]
since $A^R_n = d_R(X_n)$, which is exactly the desired result.

We begin with the first inequality. 
From \lemref{lem:uniform-martingale} and \lemref{lem:martingale-with-var}, we are justified in applying the concentration bound for martingales to the sequence $(Z^R_0, \ldots, Z^R_n)$, using the parameters $\lambda = \eps/2$, variance bound $\sigma_i^2 =\frac{1}{n^2 p}$, and maximum step size $M = \frac{1}{np}$. 
Since $Z^R_0 = 0$, we note that $|A^R_n - B^R_n| = |Z^R_n - Z^R_0|$, and thus:
\[\PPr{|A^R_n - B^R_n| \geq \eps / 2} \leq 2 \exp\left(-\frac{(\eps/2)^2}{2n \cdot \frac{1}{n^2 p} + \frac{\eps}{6np}}\right) < 2 \exp\left(-\frac{\eps^2 np}{9}\right).\]
This upper bound is at most $\delta/2$ provided that $np \geq \frac{9}{\eps^2} \ln\frac{4}{\delta}$, which establishes the first inequality in~\eqref{eq:union-bound-uniform}.

We now address the second inequality. Observe that:
\[B^R_n = d_R(S_n) \cdot \frac{|S_n|}{np}.\]
Since elements are sampled independently with probability $p$, the total number of sampled elements $|S_n|$ follows a binomial distribution $\Bin(n, p)$, regardless of the adversary’s choices. Applying a Chernoff bound with deviation $\eps/2$, we obtain:
\[\PPr{\left||S_n| - np\right| \geq \eps np / 2} \leq 2 \exp\left(- \frac{(\eps/2)^2 np}{2 + \eps/3} \right) < 2 \exp\left(- \frac{\eps^2 np}{10} \right).\]
This probability is at most $\frac{\delta}{2}$ whenever $np \geq \frac{10 \ln(4/\delta)}{\eps^2}$.

Now, conditioning on the event that $|S_n|$ lies within $\eps np/2$ of its expectation, we compute:
\[\left|d_R(S_n) - B^R_n\right| = \left|1 - \frac{|S_n|}{np}\right| \cdot d_R(S_n) \leq \left|1 - \frac{|S_n|}{np}\right| \leq \frac{\eps}{2},\]
where the first inequality uses the fact that $d_R(S_n) \leq 1$, and the second follows from the bound on $|S_n|$. This completes the proof of the second inequality in~\eqref{eq:union-bound-uniform}.
\end{proof}

\subsubsection{Reservoir Sampling}
Next, we show that reservoir sampling is robust against adaptive adversaries, along the results of \cite{Ben-EliezerY20}. 
Recall that reservoir sampling maintains a sample of the stream, so that at each time $t$, the sample is replaced with the new item $x_t$ with probability $\frac{1}{t}$. 
Reservoir sampling with memory $k$ is then initialized by storing the first $k$ items of the data stream, and then for each time $t>k$, replacing a uniformly random sample with the new item $x_t$ with probability $\frac{k}{t}$. 
Note that for reservoir sampling, the sample size is fixed, whereas the sample size is a random variable in Bernoulli sampling. 
The high-level approach is similar to the analysis for Bernoulli sampling, except that a different martingale is used. 
In particular, we define for $i\in(k,n]$:
\begin{align*}
A^R_i &= i \cdot d_R(X_i) = |R \cap X_i|, \\ B^R_i &= i \cdot d_R(S_i)
= \frac{i}{k} \cdot |R \cap S_i|, \\
 Z^R_i &= B^R_i - A^R_i. 
\end{align*} 
For indices $i \in[k]$, we simply set $A^R_i = B^R_i = |R \cap X_i|$. 
This choice is a natural continuation of the definitions given for $i > k$: indeed, due to the structure of $B^R_i$, when the stream contains no more than $k$ elements, the reservoir retains all of them without replacement.
We now state the counterpart of \lemref{lem:uniform-martingale} adapted to the reservoir sampling context.

\begin{lemma}
\cite{Ben-EliezerY20}
\lemlab{lem:reservoir-martingale}
The sequence $(Z^R_0, \ldots, Z^R_n)$ forms a martingale. 
Additionally, for all $i$, the conditional variance $\Var(Z^R_i \mid Z^R_0, \ldots, Z^R_{i-1})$ is at most $\frac{i}{k}$, and the absolute step size satisfies $|Z^R_i - Z^R_{i-1}| \leq \frac{i}{k}$.
\end{lemma}
\begin{proof}
The proof mirrors the reasoning used in \lemref{lem:uniform-martingale}. 
First, note that for $i\le k$, the claim holds trivially. 
Next, consider a fixed $i>k$ and suppose the first $i-1$ rounds have completed, so the values $Z^R_0, \ldots, Z^R_{i-1}$ are fixed. 
Let $x_i$ denote the element generated by the adversary in round $i$. 
We start by noting that
\[A^R_i = \begin{cases}
A^R_{i-1} & \text{if } x_i \notin R, \\
A^R_{i-1} + 1 & \text{if } x_i \in R.
\end{cases}\]
The computation of $B^R_i$ is more delicate, as it depends on three aspects: (i) whether $x_i \in R$, (ii) whether $x_i$ is sampled, and (iii) if sampled, whether the item it replaces, denoted $r_i$, belongs to $R$.

\paragraph{Case 1: $x_i \notin R$.} 
If $x_i$ is not sampled, or if it is sampled and replaces an element $r_i \notin R$, then $R \cap S_i = R \cap S_{i-1}$, meaning no elements from $R$ enter or leave the sample. 
Thus,
\[B^R_i = \frac{i}{k} \cdot |R \cap S_i| = \frac{i-1}{k} \cdot |R \cap S_{i-1}| + \frac{1}{k} \cdot |R \cap S_{i-1}| = B^R_{i-1} + d_R(S_{i-1}),\]
where the last equality uses the fact that $|S_{i-1}| = k$ for $i > k$.

Now consider the case where $x_i$ is sampled and replaces $r_i \in R$. 
This event occurs with probability $\frac{k}{i} \cdot d_R(S_{i-1})$, since the sampling probability is $\frac{k}{i}$ and the removed element belongs to $R$ with probability $d_R(S_{i-1})$. 
In this case, $|R \cap S_i| = |R \cap S_{i-1}| - 1$, so
\[B^R_i = \frac{i}{k} \cdot |R \cap S_i| = \frac{i}{k} \cdot |R \cap S_{i-1}| - \frac{i}{k} = B^R_{i-1} + d_R(S_{i-1}) - \frac{i}{k}.\]

Taking the expectation over the possible outcomes when $x_i \notin R$, we obtain:
\begin{align*}
\mathbb{E}[B^R_i \mid Z^R_0, \ldots, Z^R_{i-1};&\ x_i \notin R]\\
& = \left(1 - \frac{k}{i} \cdot d_R(S_{i-1})\right) \cdot (B^R_{i-1} + d_R(S_{i-1})) \\
&\quad + \frac{k}{i} \cdot d_R(S_{i-1}) \cdot \left(B^R_{i-1} + d_R(S_{i-1}) - \frac{i}{k} \right) \\
&= B^R_{i-1}.
\end{align*}
Since $A^R_i = A^R_{i-1}$ in this case, we conclude that
\[\Ex{Z^R_i \mid Z^R_0, \ldots, Z^R_{i-1};\ x_i \notin R} = Z^R_{i-1}.\]

\paragraph{Case 2: $x_i \in R$.}  
In this case, if we have $S_i = S_{i-1}$, we again have $B^R_i = B^R_{i-1} + d_R(S_{i-1})$. 
On the other hand, the only time $S_i\neq S_{i-1}$ can occur is when $x_i$ is sampled and $r_i \notin R$, so that $R\cap S_i\neq R\cap S_{i-1}$. 
This occurs with probability $\frac{k}{i} \cdot (1 - d_R(S_{i-1}))$ and as a result, the size of $R \cap S_i$ increases by one. 
Thus, 
\[B^R_i = \frac{i}{k} \cdot |R \cap S_i| = \frac{i}{k} \cdot (|R \cap S_{i-1}| + 1) = B^R_{i-1} + d_R(S_{i-1}) + \frac{i}{k}.\]
Combining both cases, the expected value of $B^R_i$ conditioned on $x_i \in R$ becomes
\begin{align*}
\Ex{B^R_i \mid x_i \in R} &= B^R_{i-1} + d_R(S_{i-1}) + \left(\frac{k}{i} \cdot (1 - d_R(S_{i-1}))\right) \cdot \frac{i}{k} \\
&= B^R_{i-1} + 1.
\end{align*}
Since $x_i \in R$ also implies $A^R_i = A^R_{i-1} + 1$, we conclude that
\[\Ex{Z^R_i \mid Z^R_0, \ldots, Z^R_{i-1};\ x_i \in R} = Z^R_{i-1}.\]
Together with Case 1, this shows that the sequence $(Z^R_0, \ldots, Z^R_n)$ forms a martingale.

We now derive bounds on the absolute difference $|Z^R_i - Z^R_{i-1}|$ and on the conditional variance of $Z^R_i$ given $Z^R_0, \ldots, Z^R_{i-1}$. 
These follow from the previous analysis and the fact that $d_R$ takes values in $[0, 1]$. 
Indeed, observe that if $x_i \notin R$, then $A^R_i = A^R_{i-1}$ and 
\[B^R_i \in [B^R_{i-1} - i/k,\ B^R_{i-1} + 1].\]
If $x_i \in R$, then $A^R_i = A^R_{i-1} + 1$ and 
\[B^R_i \in [B^R_{i-1},\ B^R_{i-1} + 1 + i/k].\]
In both scenarios, the change satisfies
\[|Z^R_i - Z^R_{i-1}| \leq \frac{i}{k}.\]

Next, we consider the conditional variance $\Var(Z^R_i \mid Z^R_0, \ldots, Z^R_{i-1})$. 
This calculation assumes a fixed value of $d_R(S_{i-1})$, but the bound we obtain will hold without conditioning on it.
We first consider the case where $x_i \notin R$ and observe that with probability $\frac{k}{i} \cdot d_R(S_{i-1})$, the deviation below the expectation is $\frac{i}{k} - d_R(S_{i-1})$; otherwise, the deviation above the expectation is $d_R(S_{i-1})$. 
Therefore:
\begin{align*}
\Var(Z^R_i \mid Z^R_0, \ldots, Z^R_{i-1},&\ x_i \notin R,\ d_R(S_{i-1})) \\
&= \frac{k}{i} \cdot d_R(S_{i-1}) \cdot \left( \frac{i}{k} - d_R(S_{i-1}) \right)^2 \\
&\quad + \left(1 - \frac{k}{i} \cdot d_R(S_{i-1})\right) \cdot \left(d_R(S_{i-1})\right)^2 \\
&= \frac{i}{k} \cdot d_R(S_{i-1}) - \left(d_R(S_{i-1})\right)^2 \leq \frac{i}{k}.
\end{align*}

Otherwise, if $x_i\in R$, then observe that with probability $\frac{k}{i} \cdot (1 - d_R(S_{i-1}))$, the deviation above the expectation is $\frac{i}{k} + d_R(S_{i-1}) - 1$; otherwise, the deviation below is $1 - d_R(S_{i-1})$. 
Therefore:
\begin{align*}
\Var(Z^R_i \mid Z^R_0, \ldots, Z^R_{i-1},&\ x_i \in R,\ d_R(S_{i-1})) \\
&= \frac{k}{i} \cdot (1 - d_R(S_{i-1})) \cdot \left(\frac{i}{k} + d_R(S_{i-1}) - 1\right)^2 \\
&\quad + \left(1 - \frac{k}{i} \cdot (1 - d_R(S_{i-1}))\right) \cdot \left(1 - d_R(S_{i-1})\right)^2 \\
&= \frac{i}{k} \cdot (1 - d_R(S_{i-1})) - (1 - d_R(S_{i-1}))^2 \leq \frac{i}{k}.
\end{align*}
In both cases, it follows that the variance is bounded by $\frac{i}{k}$ regardless of the value of $d_R(S_{i-1})$ or whether $x_i \in R$ or not. Hence,
\[\Var(Z^R_i \mid Z^R_0, \ldots, Z^R_{i-1}) \leq \frac{i}{k},\]
which concludes the proof.
\end{proof}

We now have the following guarantees for adversarial robustness for reservoir sampling.

\begin{lemma}
\cite{Ben-EliezerY20}
Given $\eps,\delta\in(0,1)$, consider a fixed universe $U$ and a subset $R\subseteq U$, and let $X = (x_1, x_2, \ldots, x_n)$ be the sequence chosen by $\Adv$. 
Then for a sample $S$ produced by reservoir sampling with memory $k\ge 2\cdot\frac{\ln(2/ \delta)}{\eps^2}$, we have 
\[\PPr{|d_R(X) - d_R(S)| \geq \eps} \leq \delta.\]
\end{lemma}
\begin{proof}
Observe the following equivalence:
\begin{align*}
\PPr{|d_R(X) - d_R(S)| \geq \eps} 
&= \PPr{|B^R_n - A^R_n| \geq \eps n} \\
&= \PPr{Z^R_n - Z^R_0| \geq \eps n}.
\end{align*}
Given \lemref{lem:reservoir-martingale}, we can apply \lemref{lem:martingale-with-var} to the martingale sequence $Z^R = (Z^R_0, \ldots, Z^R_n)$ using parameters $\lambda = \eps n$, $\sigma^2_i = i/k$ for $i \geq k$ (and $\sigma^2_i = 0$ for $i < k$), and a maximal step size of $M = \frac{n}{k}$. 
This yields the bound:
\begin{align*}
\PPr{|Z^R_n - Z^R_0| \geq \lambda} & \leq 2 \exp\left( -\frac{\lambda^2}{2 \sum_{i=1}^n \sigma_i^2 + M\lambda/3} \right) \\
&= 2 \exp\left( -\frac{\eps^2 n^2}{2 \sum_{i=1}^n (i/k) + (n/k)(\eps n)/3} \right) \\
&= 2 \exp\left( -\frac{\eps^2 k n^2}{n(n+1) + (\eps n^2)/3} \right) \\
&\leq 2 \exp\left( -\frac{\eps^2 k n^2}{2n^2} \right) = 2 \exp\left( -\frac{\eps^2 k}{2} \right),
\end{align*}
where the final inequality uses the fact that $n \geq 2$.

Thus, to ensure that the failure probability $\PPr{|d_R(X) - d_R(S)| \geq \eps}$ is at most $\delta$, it suffices to choose
\[k \geq \frac{2}{\eps^2} \ln\left(\frac{2}{\delta}\right).\]
\end{proof}

\subsubsection{Attack on Uniform Sampling}
\seclab{sec:uniform:sampling:attack}
In this section, we formalize the attack on uniform sampling discussed in \secref{sec:intro:sampling:attack} and previously presented by \cite{Ben-EliezerY20}. 
Specifically, \cite{Ben-EliezerY20} shows that in the adversarial setting, the required sample size for obtaining a representative subset cannot depend solely on the VC-dimension; rather, it must also take into account the cardinality of the domain. 
To illustrate this, \cite{Ben-EliezerY20} constructs a set system $(U, \mathcal{R})$ with a universe $U$ of large size and a VC-dimension of just one, and design an adversarial strategy that causes the sampling algorithm to produce a set that is not an $\epsilon$-approximation of $(U, \mathcal{R})$ with high probability. 
This contrasts with the static case, where the same sample size would suffice to guarantee an $\eps$-approximation with high probability. 
In particular uniform sampling algorithms such as Bernoulli sampling or reservoir sampling return an extremely unrepresentative sample: it consists exactly of the $k$ smallest elements in the stream, where $k$ is the final sample size.

\begin{algorithm}[!htb]
\caption{Adversary's strategy for producing an unrepresentative sample}
\alglab{alg:attack:uniform:sampling}
\begin{algorithmic}[1]
\State{Initialize interval endpoints: $a \gets 1$, $b \gets N$}
\State{Define effective sampling probability: $p' \gets \max\{p, \ln n / n\}$}
\For{$i = 1$ to $n$}
\State{Compute next stream element: $x_i \gets \lfloor a + (1 - p')(b - a) \rfloor$}
\If{$x_i$ is sampled}
\State{Update interval: $a \gets x_i$, $b \gets b$}
\Else
\State{Update interval: $a \gets a$, $b \gets x_i$}
\EndIf
\EndFor
\State{Output stream: $X = \{x_1, x_2, \ldots, x_n\}$}
\end{algorithmic}
\end{algorithm}
We show that the attack in \algref{alg:attack:uniform:sampling} breaks both Bernoulli sampling and reservoir sampling. 
\begin{theorem}
\cite{Ben-EliezerY20}
\thmlab{thm:uniform:sampling:attack}
There exists a constant $c > 0$ and a set system $(U, \mathcal{R})$ with VC-dimension 1 such that, for any $0 < \eps, \delta < \frac{1}{2}$, the following holds:
\begin{enumerate}
\item 
The Bernoulli sampling algorithm with sampling probability $p < \frac{c \cdot \ln |\mathcal{R}|}{n \ln n}$ is not $(\eps, \delta)$-robust.
\item 
The reservoir sampling algorithm with sample size $k < \frac{c \cdot \ln |\mathcal{R}|}{\ln n}$ is not $(\eps, \delta)$-robust.
\end{enumerate}
Moreover, for any $n^{6 \ln n} \le N \le 2^{n/2}$, there exists such a set system $(U, \mathcal{R})$ with $|U| = |\mathcal{R}| = N$.
\end{theorem}
\begin{proof}
For a universe $U = \{1, 2, \ldots, N\}$, consider the attack in \algref{alg:attack:uniform:sampling}, where $N$ satisfies $n^{6 \ln n} \leq N \leq 2^{n/2}$, and defines $\mathcal{R} = \{[1, b] : b \in U\}$. 
This set system clearly has VC-dimension 1. 
The adversary’s strategy is the same against both Bernoulli sampling and reservoir sampling; it proceeds iteratively by maintaining a working interval $[a_i, b_i]$, initialized to $[1, N]$. 
At each step $i$, the adversary selects the point $x_i = \lfloor a_i + (1 - p')(b_i - a_i) \rfloor$, where $p' = \max\left(p,\frac{\ln n}{n}\right)$, and updates the interval based on whether $x_i$ was sampled: if it is sampled, the adversary narrows the interval to $[x_i, b_i]$; otherwise, the adversary narrows the interval to $[a_i, x_i]$. 
This process continues for $n$ rounds, generating the final stream $x_1, \ldots, x_n$.

Let $S$ denote the subset of sampled elements produced by the sampling algorithm. 
Since the expected size of $S$ is at most $np'$, an application of Markov's inequality shows that the probability $|S| \geq 2np'$ is less than $\frac{1}{2}$. 
Hence, with constant probability at least $\frac{1}{2}$, $|S| < 2np'$. 
\cite{Ben-EliezerY20} shows that under this condition, the adversary’s strategy is well-defined and does not terminate prematurely; in particular, the invariant $a_i < b_i$ holds throughout the entire stream construction.

\begin{claim}
\claimlab{clm:big-range}
\cite{Ben-EliezerY20}
If the sample size $|S|$ satisfies $|S| < 2np'$, then for every round $i \in [n]$, the maintained interval satisfies $b_i - a_i \geq n$.
\end{claim}
\begin{proof}
Define $\ell_i = b_i - a_i$. 
We prove by induction that $\ell_i \geq n$ for all $i$.

At each step $i$, if the inserted element $x_i$ is sampled, then the interval shrinks by a factor of at most $p'$, so $\ell_{i+1} \geq p'\ell_i$. 
Otherwise if $x_i$ is not sampled, then the interval becomes smaller as $\ell_{i+1} \geq (1 - p')\ell_i - 2 \geq (1 - 2p')\ell_i$, where the final inequality uses the inductive assumption.

Since $|S| < 2np'$, we can bound $\ell_i$ from below as follows:
\begin{align*}
\ell_i &\ge p'^{|S|}(1 - 2p')^{n - |S|} \cdot N \\
&\ge p'^{|S|}(1 - 2p')^n \cdot N \\
&= e^{-\left(|S| \ln\frac{1}{p'} + n \ln\frac{1}{1 - 2p'}\right)} \cdot N \\
&> e^{- (2np' \ln\frac{1}{p'} + 3np')} \cdot N \\
&\ge e^{\ln n - \ln N} \cdot N = n~,
\end{align*}
where we used the inequality $\ln\frac{1}{1 - 2p'} \le 3p'$ for sufficiently small $p'$, and the fact that $p' \le \frac{\ln N}{6n \ln n}$ and $p' \ge \frac{\ln n}{n}$ to conclude:
\[\ln N \ge 6np' \ln n \ge 2np' \ln\frac{1}{p'} + 3np' + \ln n.\]
This establishes the inductive step and completes the proof.
\end{proof}

The claim implies that under the assumption $|S| < 2np'$, the adversary's strategy in \algref{alg:attack:uniform:sampling} successfully constructs a stream of $n$ elements. 
We now argue that the resulting sample is not an $\eps$-approximation. 
We begin with the case of the Bernoulli sampling algorithm.

\begin{claim}
\cite{Ben-EliezerY20}
Under the attack on Bernoulli sampling, the following properties hold at each round $i$:
\begin{itemize}
\item 
All elements sampled before round $i$ are at most $a_i$. 
\item 
All non-sampled elements submitted before round $i$ are at least $b_i$.
\item 
The element $x_i$ submitted at round $i$ lies strictly within the interval $(a_i, b_i)$.
\end{itemize}
\end{claim}
\begin{proof}
We prove the claim by induction, noting that the base case $i = 1$ is trivial.

Suppose the claim holds for round $i - 1$. 
Then, due to the adversary’s update rules and \claimref{clm:big-range}, we know:
\[a_{i-1} \le a_i < b_i \le b_{i-1}.\]
Thus, for any previously submitted element $x_j$ with $j < i - 1$, the inductive assumption implies $x_j$ satisfies the stated conditions, i.e., $x_j\le a_i$ if $x_j$ was sampled and $x_j\ge b_i$ if $x_j$ was not sampled. 
For $j = i - 1$, if $x_{i-1}$ was sampled, the adversary sets $a_i = x_{i-1}$, so $x_{i-1} \le a_i$. 
Otherwise, $x_{i-1}$ was not sampled, and the adversary sets $b_i = x_{i-1}$, hence $x_{i-1} \ge b_i$.
Lastly, by construction of $x_i$, it always holds that $a_i<x_i<b_i$. 
Therefore, the claim holds at round $i$.
\end{proof}

As established in the previous claim, at every point in the stream, all \emph{sampled} elements are strictly smaller than all \emph{non-sampled} elements. 
This property alone suffices to ensure that the sample set is not an $\eps$-approximation of $(U, \mathcal{R})$.

Let $S$ denote the set of sampled elements, and let $s$ be the largest element in $S$ (if $S$ is empty, the approximation clearly fails). 
Now consider the range $[1, s] \in \mathcal{R}$. 
Within the sample, this range contains all elements in $S$, so its empirical density is
\[d_{[1,s]}(S) = 1~.\]
In contrast, the density of this range in the full stream is
\[d_{[1,s]}(X) = \frac{|S|}{n}~.\]
Therefore, the absolute difference between these densities is
\[|d_{[1,s]}(S) - d_{[1,s]}(X)| = 1 - \frac{|S|}{n} \ge 1 - 2p' > \frac{1}{2} \ge \eps~.\]
This shows that the sampled set $S$ is not an $\eps$-approximation whenever $|S| < 2np'$, which, as previously shown, occurs with probability at least $\frac{1}{2}$. 
Hence, the Bernoulli sampling algorithm with the specified sampling probability $p$ is not $(\eps, \delta)$-robust.

The same argument extends naturally to the reservoir sampling algorithm. 
Recall that $k$ is the final sample size, and define $k'$ as the total number of elements that were ever sampled during the reservoir sampling process, including those that were later evicted. 
We can bound $k'$ as follows:
\[\Ex{k'} = k + \sum_{i=1}^n \frac{k}{i} \le 2k \ln n~.\]
By Markov's inequality, with probability at least $\frac{1}{2}$, we have $k' \le 4k \ln n$.
According to the adversary’s strategy, the first $k'$ elements of the stream are the smallest ones, and the final sample $S$ is a subset of these. 
That is, while $S$ may not contain the $k$ absolute smallest elements, it only includes elements from among the $k'$ smallest.
Let $s$ be the largest element among the $k'$ smallest ones. 
Then, in the sample $S$, all elements fall within the interval $[1, s]$, so:
\[d_{[1,s]}(S) = \frac{k}{k} = 1~.\]
Meanwhile, the density of this range in the full stream satisfies:
\[d_{[1,s]}(X) = \frac{k'}{n} \le \frac{4k \ln n}{n} \le \frac{\ln N}{n} \le \frac{1}{2}~.\]
Therefore, the density gap is:
\[|d_{[1,s]}(S) - d_{[1,s]}(X)| > 1 - \frac{1}{2} \ge \eps~,\]
demonstrating that reservoir sampling with sample size $k$ (as specified in the theorem) also fails to be $(\eps, \delta)$-robust.
\end{proof}

\subsection{Coreset Construction}
The previously discussed sampling methods are uniform, meaning that each update in the stream is sampled with a fixed probability, independent of the item's identity. 
In this section, we focus on robustness for a range of algorithms that utilize non-uniform sampling strategies, where the probability of sampling each item is approximately proportional to its ``importance''.

\subsubsection{Merge and Reduce}
\seclab{sec:mr}
We first demonstrate that the general merge-and-reduce framework ensures adversarial robustness. 
This paradigm is commonly used in the design of coresets, which are compact, weighted summaries of large datasets that preserve essential properties for downstream tasks, thus enabling more efficient algorithmic processing.
Recall that informally, a query space $(P,\dist,Q)$ consists of a dataset $P$, a distance function $\dist$, and a family of candidate solutions $Q$; the quality of a solution is measured by aggregating the distances $\dist(p,Q)$ over all data points $p\in P$. 
\begin{definition}[$\eps$-coreset]
Let $P$ be a dataset, let $z \ge 0$, $\eps \in (0,1)$, and let $(P, \dist, Q)$ denote a query space. 
A subset $C \subseteq P$, equipped with a weight function $w: P \rightarrow \mathbb{R}$, is called an \emph{$\eps$-coreset} for $(P, \dist, Q)$ if
\[(1 - \eps) \sum_{p \in P} \dist(p, Q)^z \le \sum_{p \in C} w(p) \dist(p, Q)^z \le (1 + \eps) \sum_{p \in P} \dist(p, Q)^z.\]
\end{definition}
Here, $P$ should intuitively be interpreted as the input dataset, $Q$ should intuitively be interpreted as an underlying space (such as the location of possible centers for clustering), and $\dist$ is the underlying metric. 

The construction of efficient offline coresets has been extensively studied for a variety of problems in computational geometry~\cite{FeldmanMSW10,FeldmanL11,BravermanFL16,BachemLK17,SohlerW18, BravermanLUZ19,HuangV20,Feldman20}, linear algebra~\cite{BravermanDMMUWZ20}, machine learning~\cite{MunteanuSSW18,BaykalLGFR19,MussayOBZF20,TukanBFR20,TukanZMRBF23}. 
These problems include linear regression, low-rank approximation, $L_1$ subspace embeddings, $k$-means and $k$-median clustering, $k$-center, support vector machines, Gaussian mixture models, $M$-estimators, Bregman clustering, projective clustering, PCA, $k$-line center, and $j$-subspace approximation. 
For these problems, merge-and-reduce often translates to an oblivious streaming algorithm that is optimal in space up to polylogarithmic factors. 
\cite{BravermanHMSSZ21} showed that the merge-and-reduce approach can be leveraged to transform these offline constructions into adversarially robust streaming algorithms. 
As a result, it follows that there exist adversarially robust streaming algorithms for these problems that are space optimal up to polylogarithmic factors.  

The merge-and-reduce method proceeds as follows: suppose the input stream consists of $n = 2^k$ elements, denoted $p_1, \ldots, p_n$ (if $n$ is not a power of two, it can be padded appropriately). 
At level $0$, we define $C_{0,j} = p_j$ for all $j \in [n]$. 
The process proceeds through $k$ hierarchical levels. 
At each level $i \in [k]$, we construct $\frac{n}{2^i}$ coresets $C_{i,1}, \ldots, C_{i,n/2^i}$, where each $C_{i,j}$ is an $\frac{\eps}{2k}$-coreset for the union of the two coresets $C_{i-1,2j-1}$ and $C_{i-1,2j}$ from the previous level.

This strategy is well-suited to the streaming model: once both $C_{i-1,2j-1}$ and $C_{i-1,2j}$ are available, $C_{i,j}$ can be computed immediately. 
Afterward, the two lower-level coresets can be safely discarded, ensuring memory efficiency throughout the process.

We recall the following approach for constructing coresets based on sensitivity sampling. 
Informally, the sensitivity of a point quantifies the ``importance'' of the point with respect to the dataset, e.g., \defref{def:leverage:score} for subspace embeddings. 

\begin{lemma}[Adapted from Lemma 2.3 in \cite{BachemLK17}]
\lemlab{lem:gen:sensitivity}
Let $\eps > 0$ and $\delta \in (0,1)$, and consider a weighted point set $P$ with non-negative weight function $\mu: P \to \mathbb{R}_{\geq 0}$. 
Suppose $s: P \to \mathbb{R}_{\geq 0}$ provides an upper bound on the sensitivity of each point in $P$. Define $S = \sum_{p \in P} \mu(p) s(p)$. 
Let 
\[m = \Omega\left(\frac{S^2}{\eps^2} \left( d' + \log \frac{1}{\delta} \right)\right),\]
where $d'$ is the pseudo-dimension of the query space.

Construct a sample $C$ by selecting $m$ points from $P$ independently with replacement, where each point $p$ is chosen with probability 
\[q(p) = \frac{\mu(p) s(p)}{S},\]
and if selected, assigned weight 
\[\frac{\mu(p)}{m \cdot q(p)}.\]
Then, with probability at least $1 - \delta$, the weighted sample $C$ forms an $\eps$-coreset of $P$.
\end{lemma}
For completeness, we recall that here, the pseudo-dimension is defined as follows:
\begin{definition}[Pseudo-dimension]
Let $\calQ$ be a class of functions $q:X\to\mathbb{R}$.  
The \emph{pseudo-dimension} is the largest integer $d$ such that there exist $x_1,\ldots,x_d\in X$ and $r_1,\ldots,r_d\in\mathbb{R}$ with the property that for every $T\subseteq[d]$ there is a $q\in\calQ$ satisfying
\[q(x_i)\ge r_i \ (i\in T), \qquad q(x_i)< r_i \ (i\notin T).\]
\end{definition}

We begin by observing that any streaming algorithm with linear memory usage is inherently adversarially robust, as it can recompute exact or approximate solutions incrementally at each step.

\begin{restatable}{lemma}{lemadvcoreset}
\lemlab{lem:adv:coreset}
\cite{BravermanHMSSZ21}
Given a set of points $P$, there exists an offline coreset construction that, with probability at least $1 - \delta$, outputs an $\eps$-coreset of $P$ that is robust to adaptive adversaries.
\end{restatable}
\begin{proof}
Let adversary $\Adv$ generate a sequence of points $P = (p_1, \ldots, p_n)$ such that each point $p_i$ may depend on previous points $p_1, \ldots, p_{i-1}$. 
Let $s(p)$ denote an upper bound on the sensitivity of each point in $P$, and consider using the sensitivity sampling method from \lemref{lem:gen:sensitivity}. 
The goal is to sample each point $p$ with probability $q(p)$, proportional to its sensitivity. 
However, if the sampling relies on internal randomness that an adversary can infer or correlate with (e.g., the seed of a pseudorandom generator), then the sampling process may no longer be independent across points, and coreset correctness is not guaranteed. 
To avoid this, we observe that the algorithm uses fresh, public randomness at each time step. 
That is, the randomness used at time $i$ is independent of the adversary’s prior choices $p_1, \ldots, p_{i-1}$. 
Thus, regardless of the adversarial input, the sampling decisions themselves remain independent and so by \lemref{lem:gen:sensitivity}, the resulting set $C$ is an $\eps$-coreset for $P$ with probability at least $1 - \delta$.
\end{proof}

This result establishes that any offline coreset construction using independent randomness remains robust in adversarial settings. 
The sensitivity sampling approach here is particularly relevant for our use of the merge-and-reduce technique in clustering. 

We now proceed to prove the main result.

\begin{theorem}[Merge and reduce is adversarially robust]
\thmlab{thm:merge:reduce}
\cite{BravermanHMSSZ21}
Given an offline $\eps$-coreset construction, the merge and reduce framework gives an adversarially robust streaming construction for an $\eps$-coreset with high probability.
\end{theorem}
\begin{proof}[Informal]
Let $\delta = \frac{1}{\poly(n)}$, and consider an $\eps$-coreset construction with failure probability at most $\delta$. 
We show that the merge-and-reduce framework produces an $\eps$-coreset that is robust to adversarial inputs with probability at least $1 - (2n\log n)\delta$.

We prove this by induction on the number of input points $n$. 
Let $k$ be the largest integer such that $2^k\le n$. 
At level $0$, we define $C_{0,j} = p_j$ for $j \in [n]$, which are trivially $\eps$-coresets (of one point each). 
At each subsequent level $i \in [k]$, define the event $\calE_i$ to be the event that for all $j \in \left[\frac{n}{2^i}\right]$, the coreset $C_{i,j}$ is an $\frac{\eps}{2k}$-coreset for the union of $C_{i-1,2j-1}$ and $C_{i-1,2j}$.

By \lemref{lem:gen:sensitivity}, each such coreset $C_{i,j}$ succeeds with probability at least $1 - \delta$, and applying a union bound over the $\frac{n}{2^i}$ such pairs, we have
\[\PPr{\calE_i}\ge 1 - \frac{n\delta}{2^i}.\]
Crucially, we note that we construct each coreset only after the stream of adaptive updates for the corresponding point set has completed. 
Moreover, the sampling process uses fresh randomness that is independent of the previous randomness of the algorithm and thus also independent of the adversary. 

Define $\calE = \bigcap_{i=1}^{k} \calE_i$ to be the event that all merge steps succeed. Then
\[\PPr{\calE} \ge 1 - \sum_{i=1}^{k} (1 - \PPr{\calE_i}) \ge 1 - \sum_{i=1}^{k} \frac{n\delta}{2^i} \ge 1 - 2n\delta.\]
When $\calE$ holds, the final coreset $C_{k,1}$ approximates the cost of the first $2^k$ points $P_{2^k}$ up to a multiplicative factor of
\[\left(1 + \frac{\eps}{2k}\right)^k \le e^{\eps/2} \le 1 + \eps.\]
Thus, with probability at least $1 - 2n\delta$, $C_{k,1}$ is an $\eps$-coreset of $P_{2^k}$. 

Now, consider a dyadic decomposition of $n$, i.e., $n=2^{k_1}+2^{k_2}+\ldots+2^{k_\ell}$, where $k_1>k_2>\ldots>k_\ell$ are non-negative integers. 
Partition the stream into consecutive blocks $Q_i$ of length $2^{k_i}$ for $i\in[\ell]$. 
By a similar reasoning as above, we have an $\eps$-coreset for each set of points $Q_i$ with probability $1-2n\delta$. 
Since $\ell\le\log n$, then by a union bound, we have coresets for all $Q_i$ with probability $1-(2n\log n)\delta$. 
In particular, it should be noted that the coreset for each block $Q_i$ is constructed only after all of the adaptive updates to $Q_i$ are complete. 
Then it follows that the union of these coresets is also an $\eps$-coreset for $P$. 
\end{proof}

We now give a number of applications for \thmref{thm:merge:reduce}. 

\begin{definition}[Spectral Approximation]
\deflab{def:spectral_approx}
Given a matrix $\bA \in \mathbb{R}^{n \times d}$ and an approximation parameter $\eps > 0$, the goal is to compute a smaller matrix $\bM \in \mathbb{R}^{m \times d}$ with $m \ll n$ such that for all vectors $\bx \in \mathbb{R}^d$, the following holds:
\[(1 - \eps)\|\bA\bx\|_2 \le \|\bM\bx\|_2 \le (1 + \eps)\|\bA\bx\|_2,\]
which, up to a scaling factor of $\eps$, is equivalent to ensuring the matrix inequality:
\[(1 - \eps)\bA^\top \bA \preceq \bM^\top \bM \preceq (1 + \eps)\bA^\top \bA.\]
\end{definition}

Spectral approximation serves as a foundation for several applications, including linear regression.

\begin{definition}[Projection-Cost Preservation]
Given a matrix $\bA \in \mathbb{R}^{n \times d}$, a target rank $k > 0$, and an approximation parameter $\eps > 0$, the objective is to compute a matrix $\bM \in \mathbb{R}^{m \times d}$ with $m \ll n$, such that for any rank-$k$ orthogonal projection matrix $\bP \in \mathbb{R}^{d \times d}$, we have:
\[(1 - \eps)\|\bA - \bA\bP\|_F^2 \le \|\bM - \bM\bP\|_F^2 \le (1 + \eps)\|\bA - \bA\bP\|_F^2.\]
\end{definition}

If $\bM$ preserves the projection cost of $\bA$, then the best low-rank approximation of $\bM$ yields a projection matrix that approximately preserves the low-rank structure of $\bA$.

\begin{definition}[Low-Rank Approximation]
Given a matrix $\bA \in \mathbb{R}^{n \times d}$, a target rank $k > 0$, and an approximation parameter $\eps > 0$, the task is to find a rank-$k$ matrix $\bM \in \mathbb{R}^{n \times d}$ such that:
\[\|\bA - \bA_{(k)}\|_F^2 \le \|\bA - \bM\|_F^2 \le (1 + \eps)\|\bA - \bA_{(k)}\|_F^2,\]
where $\bA_{(k)}$ denotes the best possible rank-$k$ approximation to $\bA$.
\end{definition}

\begin{restatable}[$L_p$ subspace embedding]{definition}{deflpsubspace}
\deflab{def:lp:subspace}
Given a parameter $p>0$, a matrix $\bA \in \mathbb{R}^{n \times d}$ and an approximation parameter $\eps > 0$, the goal is to construct a smaller matrix $\bM \in \mathbb{R}^{m \times d}$ with $m \ll n$ such that for all $\bx \in \mathbb{R}^d$, the following $L_p$ norm condition holds:
\[(1 - \eps)\|\bA\bx\|_p \le \|\bM\bx\|_p \le (1 + \eps)\|\bA\bx\|_p.\]
\end{restatable}

Using coreset constructions from \cite{DrineasMM06,DrineasMMW12,ClarksonW13,CohenP15,CohenMM17} along with \thmref{thm:merge:reduce}, we obtain the following result:

\begin{theorem}
\thmlab{thm:nla:coreset}
\cite{BravermanHMSSZ21}
There exist adversarially robust streaming algorithms, based on merge-and-reduce, for the row-arrival model such that with probability $1-\frac{1}{\poly(n)}$ at each time $t\in[n]$:
\begin{enumerate}
\item
The algorithm outputs a matrix $\bM_t$ that satisfies the spectral approximation guarantee:
\[(1 - \eps) \bA_t^\top \bA_t \preceq \bM_t^\top \bM_t \preceq (1 + \eps) \bA_t^\top \bA_t,\]
using $\O{\frac{d}{\eps^2} \log^4 n}$ sampled rows. 
This result applies to spectral approximation, subspace embedding, linear regression, and generalized regression.
\item
The algorithm outputs a matrix $\bM_t$ such that for every rank-$k$ orthogonal projection matrix $\bP \in \mathbb{R}^{d \times d}$,
\[(1 - \eps)\|\bA_t - \bA_t \bP\|_F^2 \le \|\bM_t - \bM_t \bP\|_F^2 \le (1 + \eps)\|\bA_t - \bA_t \bP\|_F^2,\]
while using $\O{\frac{k}{\eps^2} \log^4 n}$ sampled rows. 
This gives both projection-cost preservation and low-rank approximation.
\item
The algorithm outputs a matrix $\bM_t$ that satisfies the $L_1$ subspace embedding guarantee:
\[(1 - \eps)\|\bA_t \bx\|_1 \le \|\bM_t \bx\|_1 \le (1 + \eps)\|\bA_t \bx\|_1 \quad \text{for all } \bx \in \mathbb{R}^d,\]
using $\O{\frac{d}{\eps^2} \log^4 n}$ sampled rows.
\end{enumerate}
\end{theorem}

By applying the coreset constructions from \cite{Cohen-AddadSS21,Cohen-AddadLSS22,CohenLNSSS22,Cohen-AddadWZ23}, we can also use \thmref{thm:merge:reduce} to achieve adversarial robustness for $(k, z)$-clustering problems, such as $k$-median clustering when $z = 1$ and $k$-means clustering when $z = 2$. 
Informally, the goal of the $(k,z)$-clustering problem is to find $k$ centers to minimize the total cost incurred by aggregating, across all points $x$ in an input dataset $X$, the $z$-th power of the distance between $x$ and its closest center. 
Additionally, as observed in \cite{BachemLK17}, the constructions of \cite{FeldmanL11} yield coresets for Bregman clustering, which handles $\mu$-similar Bregman divergences including the Itakura-Saito distance, KL-divergence, Mahalanobis distance, and others.

\begin{theorem}
\thmlab{thm:merge:reduce:clustering:coreset}
\cite{BravermanHMSSZ21}
There exist adversarially robust streaming algorithms, based on merge-and-reduce, for the insertion-only model such that with probability $1-\frac{1}{\poly(n)}$ at each time $t\in[n]$:
\begin{enumerate}
\item
The algorithm outputs an $\eps$ coreset for the optimal $(k,z)$-clustering problem, including $k$-means ($z=2$) and $k$-median ($z=1$), with $\tO{\min\left(\frac{1}{\eps^2}\cdot k^{2-\frac{z}{z+2}},\frac{1}{\min(\eps^4,\eps^{2+z})}\cdot k\right)}\cdot\polylog(n)$ points.
\item
The algorithm outputs an $\eps$ coreset for optimal Bregman $k$-clustering, with $\O{\frac{1}{\eps^2}\,dk^3\log^3 n}$ points.
\end{enumerate}
\end{theorem}

Furthermore, by leveraging the sensitivity bounds from \cite{VaradarajanX12a,VaradarajanX12b} and the coreset construction framework of \cite{BravermanFL16}, \thmref{thm:merge:reduce} can be applied to a variety of shape fitting problems:

\begin{theorem}
\thmlab{thm:merge:reduce:shape:coreset}
\cite{BravermanHMSSZ21}
There exist adversarially robust streaming algorithms, based on merge-and-reduce, for the insertion-only model such that with probability $1-\frac{1}{\poly(n)}$ at each time $t\in[n]$:
\begin{enumerate}
\item
The algorithm outputs an $\eps$ coreset for the $k$-line clustering problem, with $\O{\frac{d}{\eps^2}\,f(d,k)k^{f(d,k)}\log^4 n}$ points in $\mathbb{R}^d$, where $f(d,k)$ is a fixed function.
\item
The algorithm outputs an $\eps$ coreset for the dimension-$j$ subspace fitting problem, storing $\O{\frac{d}{\eps^2}\,g(d,j)\log^4 n}$ points in $\mathbb{R}^d$, for some fixed function $g(d,j)$.
\item
The algorithm outputs an $\eps$ coreset for the  $(j, k)$-projective clustering problem, storing at most $\O{\frac{d}{\eps^2}\,h(d,j,k)\log^3 n(\log n)^{h(d,j,k)}}$ points in $\mathbb{R}^d$, for some fixed function $h(d,j,k)$, under the assumption that the input points have integer coordinates.
\end{enumerate}
\end{theorem}

\subsubsection{Importance Sampling}
Next, we show adversarial robustness for algorithms based on importance sampling, i.e., each item of the stream is sampled with probability proportional to some quantity that measures its ``importance''.
Although there are various problems that can be solved using importance sampling, in this section we focus on the problem of $L_p$ subspace embeddings, which we recall is defined as follows:
\deflpsubspace*
We remark that there exist constructions of $L_p$ subspace embeddings that use $m\approx\poly(d)$ rows, e.g.,~\cite{DrineasMM06,DrineasMMW12,ClarksonW13,CohenP15,CohenMM17}. 

As before, we consider the row-arrival model, so that the rows $\ba_1,\ldots,\ba_n$ of $\bA$ arrive sequentially, generated by an adversary. 
It is known that in the non-adaptive setting, the following definition of the online $L_p$ sensitivities provides a method to perform importance sampling that gives a subspace embedding:

\begin{definition}[Online $L_p$ sensitivities]
Consider a matrix $\bA = \ba_1 \circ \ldots \circ \ba_n \in \mathbb{R}^{n \times d}$.  
The online sensitivity of the $i$-th row $\ba_i$ (for each $i \in [n]$) is defined as 
\[\max_{\bx \in \mathbb{R}^d} \frac{|\langle \ba_i, \bx \rangle|^p}{\|\bA_{i-1} \bx\|_p^p},\]
where $\bA_{i-1} = \ba_1 \circ \ldots \circ \ba_{i-1}$ denotes the submatrix formed by the first $i-1$ rows.
\end{definition}

We remark that constant-factor approximations to the online $L_p$ sensitivities can be efficiently computed and suffice for the purposes of row sampling, c.f., \algref{alg:robust:row:sampling}. 

\begin{algorithm}[!htb]
\caption{Row sampling algorithm}
\alglab{alg:robust:row:sampling}
\begin{algorithmic}[1]
\Require{A stream of rows $\ba_1, \ldots, \ba_n \in \mathbb{R}^d$, parameter $p > 0$, and accuracy $\eps > 0$}
\Ensure{A $(1+\eps)$-approximate $L_p$ subspace embedding}
\State{$\bM \gets \emptyset$}
\State{$\alpha \gets \frac{C d}{\eps^2} \log(n\kappa)$ for a sufficiently large parameter $C > \kappa^2$}
\For{each row $\ba_i$, $i \in [n]$}
\If{$\ba_i \in \Span(\bM)$}
\State{$\tau_i \gets 2 \cdot \max_{\substack{\bx \in \mathbb{R}^d \\ \bx \in \Span(\bM)}} \frac{|\langle \ba_i, \bx \rangle|^p}{\|\bM \bx\|_p^p + |\langle \ba_i, \bx \rangle|^p}$}
\Else
\State{$\tau_i \gets 1$}
\EndIf
\State{$p_i \gets \min(1, \alpha \tau_i)$}
\State{With probability $p_i$, update $\bM \gets \bM \circ \frac{\ba_i}{p_i^{1/p}}$}
\Comment{Sampling based on online sensitivity}
\EndFor
\State{\Return $\bM$}
\end{algorithmic}
\end{algorithm}

We first show correctness in a non-adaptive setting. 
To that end, we use the following concentration inequality for the sum of a number of random variables, possibly non-independent. 
\begin{theorem}[Freedman's Inequality] 
\thmlab{thm:scalar:freedman}
\cite{Freedman75}
Consider a scalar martingale $Y_0, Y_1, \ldots, Y_n$ with difference sequence $X_1, \ldots, X_n$, where $Y_0 = 0$ and $Y_i = Y_{i-1} + X_i$ for all $i \in [n]$. 
Suppose $|X_t| \le R$ for all $t \in [n]$ with high probability. 
Define the predictable quadratic variation by
\[w_k := \sum_{t=1}^k \underset{t-1}{\mathbb{E}}\left[X_t^2\right], \quad \text{for } k \in [n].\]
Then for any $\eps \ge 0$, $\sigma^2 > 0$, and $k \in [n]$,
\[\PPr{\max_{t \in [k]} |Y_t| > \eps \text{ and } w_k \le \sigma^2} \le 2\exp\left(-\frac{\eps^2/2}{\sigma^2 + R\eps/3}\right).\]
\end{theorem}

We now recall that $L_p$ subspace embeddings can be achieved by importance sampling. 
Here, we define the condition number of the data stream as the maximum, over all intermediate matrices induced by prefixes of the stream, of the ratio between the largest and smallest nonzero singular values. 
\begin{lemma}[$L_p$ subspace embedding by row sampling]
\lemlab{lem:lp:subspace}
\cite{BravermanDMMUWZ20,BravermanHMSSZ21}
Let $\eps > \frac{1}{n}$, $p\le 2$, and $C > \kappa^p$, where $\kappa$ is the condition number of the data stream. 
Then \algref{alg:robust:row:sampling} returns a matrix $\bM$ such that, for all $\bx \in \mathbb{R}^d$,
\[\left|\|\bM\bx\|_p - \|\bA\bx\|_p\right| \le \eps\|\bA\bx\|_p\]
with high probability.
\end{lemma}
\begin{proof}
By a constant rescaling of $\eps$, we would like to prove that with high probability, 
\[\left| \|\bM \bx\|_p^p - \|\bA \bx\|_p^p \right| \le \eps \|\bA\bx\|_p^p,\]
for all $\bx \in \mathbb{R}^d$. 
Fix any vector $\bx \in \mathbb{R}^d$ and let $\eps \in\left(0,\frac{1}{2}\right)$ satisfy $\eps > \frac{1}{n}$. 
We prove by induction the stronger claim that for all $j\in[n]$, with high probability, 
\[\left| \|\bM_j \bx\|_p^p - \|\bA_j \bx\|_p^p \right| \le \eps \|\bA_j \bx\|_p^p,\]
where $\bA_j = \ba_1 \circ \ldots \circ \ba_j$ is the matrix formed by the first $j$ rows of $\bA$, and $\bM_j$ contains those rows sampled into $\bM$ by time $j$.

For the base case $j=1$, since either $\ba_1=0$ or $p_1=1$, it follows that $\bM_1=\bA_1$, so the claim holds trivially.
Now, suppose the claim holds for all $j<n$.  
Define a martingale sequence $Y_0, \ldots, Y_n$ with differences $X_1, \ldots, X_n$, where for each $j \geq 1$,
\begin{align*}
X_j = 
\begin{cases}
0 & \text{if } Y_{j-1} > \eps \|\bA_{j-1}\bx\|_p^p, \\
\left(\frac{1}{p_j} - 1\right) |\ba_j^\top \bx|^p & \text{if } Y_{j-1} \le \eps \|\bA_{j-1}\bx\|_p^p \text{ and } \ba_j \text{ sampled}, \\
-|\ba_j^\top \bx|^p & \text{if } Y_{j-1} \le \eps \|\bA_{j-1}\bx\|_p^p \text{ and } \ba_j \text{ not sampled}.
\end{cases}
\end{align*}
This construction ensures $\{Y_j\}$ is a martingale and satisfies
\[Y_j = \|\bM_j \bx\|_p^p - \|\bA_j \bx\|_p^p\]
while $X_j\neq 0$. 

We now upper bound the variance. 
If $p_j < 1$, write $p_j = \alpha \tau_j$. 
Using the inductive hypothesis and the definition of $\tau_j$, one can verify
\[\tau_j \ge \frac{|\ba_j^\top \bx|^p}{\|\bA_j \bx\|_p^p},\]
since $\eps\in(0,1)$. 
Therefore,
\[\Ex{X_j^2 \mid \mathcal{F}_{j-1}} \le \frac{\|\bA_j\bx\|_p^p}{\alpha}\cdot|\ba_j^\top\bx|^p.\]
Summing over $i\in[j]$ yields
\[\sum_{i=1}^j \Ex{X_i^2 \mid \mathcal{F}_{i-1}} \le \frac{\|\bA_j\bx\|_p^{2p}}{\alpha}.\]
To reduce notational clutter, we use $\bA$ to refer to $\bA_j$ in the remainder of the proof. 
Let $\kappa_1=\sigma_{\min}(\bA)$ be the smallest singular value of $\bA$ and $\kappa_2=\sigma_{\max}(\bA)$ be the largest singular value of $\bA$, so that $\frac{\kappa_2}{\kappa_1}\le\kappa$, by definition of the condition number $\kappa$ of the data stream. 
Note that for $p$-norms, we use $\sigma_{\max}(\bA)$ to denote $\sup_{\bx\neq\bf{0}}\frac{\|\bA\bx\|_p^p}{\|\bx\|_p^p}$ and similarly, for $\sigma_{\min}(\bA)$. 

Before applying Freedman's inequality, it remains to upper bound the increments. 
To that end, we have
\[|X_j| \le \frac{1}{p_j} |\ba_j^\top \bx|^p \le \frac{2\eps^2}{C d \log n} \|\bA \bx\|_p^p,\]
for a suitable parameter $C>\kappa^p$, using the bounds on $p_j$ and $\tau_j$. 

With variance and increment bounded as above, Freedman's inequality in \thmref{thm:scalar:freedman} implies
\[\PPr{|Y_n| > \eps \kappa_1^p \|\bx\|_p^p} \le \frac{1}{2^d \poly(n)},\]
for the choice $C>\kappa^p$ and assuming we have $\kappa_1\le\|\bA\|_p\le\kappa_2$. 
Since $\kappa_1^p \|\bx\|_p^p \le \|\bA \bx\|_p^p$, we obtain
\[\PPr{\left| \|\bM \bx\|_p^p - \|\bA \bx\|_p^p \right| > \eps \|\bA \bx\|_p^p} \le \frac{1}{2^d \poly(n)}.\]
Adjusting $\eps$ as needed (since $p \le 2$) yields
\[\left|\|\bM \bx\|_p - \|\bA \bx\|_p \right| \le\eps \|\bA \bx\|_p,\]
with high probability.

Finally, it remains to show correctness for all $\bx\in\mathbb{R}^d$ by a standard net argument. 
Define the unit ball
\[B = \{\bA \by \in \mathbb{R}^n \mid \|\bA \by\|_p = 1 \},\]
and let $\calN$ be an $\eps$-net of $B$ with respect to the $L_p$ norm, with size at most $\left(\O{\frac{1}{\eps}}\right)^d$. 
By a union bound, with probability at least $1-\frac{1}{\poly(n)}$,
\[\left| \|\bM \by\|_p - \|\bA \by\|_p \right| \le \eps \quad \text{for all } \bA \by \in \mathcal{N}.\]

For any $\bz$ with $\|\bA \bz\|_p = 1$, construct a sequence $\bA \by_1, \bA \by_2, \ldots$ with
\[\left\| \bA \bz - \sum_{j=1}^i \bA \by_j \right\|_p \le \eps^i,\]
and scaling factors $\gamma_i \le \eps^{i-1}$ so that $\frac{1}{\gamma_i} \bA \by_i \in \mathcal{N}$.

Then,
\[\left| \|\bM \bz\|_p - \|\bA \bz\|_p \right| \le \sum_{i=1}^\infty \left| \|\bM \by_i\|_p - \|\bA \by_i\|_p \right| \le \sum_{i=1}^\infty \eps^i = \O{\eps},
\]
completing the induction and proof.
\end{proof}

\paragraph{Adversarially robust subspace embedding.}
We now show that the algorithm corresponding to \lemref{lem:lp:subspace} can be adjusted to achieve adversarial robustness.
\begin{lemma}[Adversarially robust subspace embedding]
\cite{BravermanHMSSZ21}
\lemlab{lem:lp:downsamplerobust}
\algref{alg:robust:row:sampling} is adversarially robust.
\end{lemma}
\begin{proof}
We follow the proof of \lemref{lem:lp:subspace}, now accounting for an adversary that observes previous rows and randomness.  
At step $i$, let $H_{i-1}$ denote the \emph{history} of all previously sampled rows $\ba_1, \ldots, \ba_{i-1}$, indicators $B_1, \ldots, B_{i-1}$, and any randomness used by the algorithm or adversary up to that point. 
The row $\ba_i$ and sampling indicator $B_i$ may depend on $H_{i-1}$ as well as the adversary's strategy at step $i$.
Define the incremental contribution
\begin{align*}
X_j = \left(\left(\frac{1}{p_j} - 1\right) |\ba_j^\top \bx|^p B_j - |\ba_j^\top \bx|^p (1 - B_j)\right) \mathbb{I}[{Y_{j-1} < \eps \|\bA_{j-1}\bx\|_p^p]},
\end{align*}
so that $Y_i = Y_{i-1} + X_i$. 
It is straightforward to verify that conditioned on $X_j\neq 0$, 
\[\mathbb{E}[X_j \mid H_{j-1}] = \left(\left(\frac{1}{p_j} - 1\right) |\ba_j^\top \bx|^p p_j - |\ba_j^\top \bx|^p (1 - p_j)\right) \mathbb{I}[Y_{j-1} < \eps \|\bA_{j-1}\bx\|_p^p] = 0,\]
because $B_i$ is a Bernoulli random variable with parameter $p_i$ conditioned on the history, and all other terms are fixed given $H_{i-1}$. 
Hence, $(Y_i)_{i=0}^n$ forms a martingale with respect to the history $H_0 \subset H_1 \subset \cdots \subset H_n$.  
The rest of the proof of \lemref{lem:lp:subspace} relies only on the martingale property and the boundedness of the increments $X_i$, so it applies unchanged in the adversarial setting. 
Therefore, \algref{alg:robust:row:sampling} is robust to adaptive adversaries. 
\end{proof}

We recall known upper bounds on the total sum of online $L_p$ sensitivities, such as those stated in Theorem 2.2 of~\cite{CohenMP20} and Lemmas 2.2 and 4.7 of~\cite{BravermanDMMUWZ20}.

\begin{lemma}[Bound on the sum of online $L_p$ sensitivities]
\cite{CohenMP20,BravermanDMMUWZ20,WoodruffY23}
\lemlab{lem:online:space}
Consider a stream of rows $\bA = \ba_1 \circ \ldots \circ \ba_n \in \mathbb{R}^{n \times d}$ with condition number at most $\kappa$. 
For $p\in[1,2]$, let $s_i$ denote the online $L_p$ sensitivity of the row $\ba_i$.  
Then
\[\sum_{i=1}^n s_i = \O{d \log(n\kappa)}.\]
\end{lemma}

Note that the parameter $\kappa$ can be chosen adversarially since the rows of the input matrix $\bA$ may be generated by an adversary.  
Instead, we assume $\kappa$ is an input parameter that serves as an \emph{upper bound} on the condition number of the data stream. 

\begin{lemma}[Adversarially robust $L_p$ subspace embedding]
\lemlab{lem:sample:robust:lp}
\cite{BravermanHMSSZ21}
For any $\eps > 0$, $p \in[1,2]$, and a matrix $\bA \in \mathbb{R}^{n \times d}$ whose rows $\ba_1, \ldots, \ba_n$ arrive sequentially in a streaming fashion with condition number bounded by $\kappa$, there exists an adversarially robust streaming algorithm that outputs a $(1+\eps)$-subspace embedding with high probability. 
This algorithm samples at most 
$\O{\frac{d^2 \kappa^2}{\eps^2} \log^2(n\kappa)}$, with high probability, where $\kappa$ denotes the ratio between upper and lower bounds on $\|\bA\|_p$.
\end{lemma}
\begin{proof}
Consider \algref{alg:robust:row:sampling}, which is adversarially robust by virtue of \lemref{lem:lp:downsamplerobust}. 
What remains is to bound the space complexity of \algref{alg:robust:row:sampling}. 

Applying \lemref{lem:lp:downsamplerobust} and taking a union bound over all $n$ rows in the stream, we see that each row $\ba_i$ is sampled with probability at most $4\alpha \tau_i$, where $\tau_i$ denotes the online leverage score of $\ba_i$.  
By \lemref{lem:online:space}, we have the bound $\sum_{i=1}^n \tau_i = \O{d \log(n\kappa)}$, and recall that $\alpha = \O{\frac{d \kappa^2}{\eps^2} \log(n\kappa)}$. 
Let $\gamma > 0$ be a sufficiently large constant such that
\[\sum_{i=1}^n \alpha \tau_i \leq \frac{d^2 \gamma \kappa \log \kappa}{\eps^2} \log n.\]
We use a martingale argument to upper bound the total number of sampled rows. 
Define a martingale sequence $U_0, U_1, \ldots, U_n$ with differences $W_1, \ldots, W_n$, where for each $j \geq 1$, 
\[W_j = \begin{cases}
1 - p_j, & \text{if } \ba_j \text{ is sampled in } \bM, \text{ and } U_{j-1} \le \frac{d^2 \gamma \kappa \log(n\kappa)}{\eps^2} \log n, \\
-p_j, & \text{if } \ba_j \text{ is not sampled and } U_{j-1} \le \frac{d^2 \gamma \kappa \log(n\kappa)}{\eps^2} \log n, \\
0, & \text{otherwise}.
\end{cases}
\]
This construction ensures that $\Ex{W_j \mid U_1, \ldots, U_{j-1}} = 0$ and therefore $(U_j)_{j=0}^n$ forms a martingale. 
Intuitively, $U_n$ measures the deviation between the actual number of sampled rows and the expected sum $\sum_{j=1}^n p_j$.
Since each $p_j \in [0,1]$, we have the variance bound
\[\Ex{W_j^2 \mid U_1, \ldots, U_{j-1}} \le p_j \le \alpha \tau_j,\]
and also $\Ex{|W_j| \mid U_1, \ldots, U_{j-1}} \le 1$.

Applying Freedman's inequality, c.f., \thmref{thm:scalar:freedman}, with variance parameter $\sigma^2 = \sum_{j=1}^n \alpha \tau_j \le \frac{d^2 \gamma \kappa^2 \log^2(n\kappa)}{\eps^2}$ and uniform bound $R \le 1$, we obtain
\begin{align*}
\PPr{|U_n| > \frac{d^2 \gamma \kappa^2 \log^2(n\kappa)}{\eps^2}}&\le 2 \exp\left(-\frac{\frac{d^4 \gamma^2 \kappa^4 \log^4(n\kappa)}{2 \eps^4}}{\sigma^2 + \frac{R d^2 \gamma \kappa^2 \log^2(n \kappa)}{3 \eps^2}}\right)\\
&\le \frac{1}{\poly(n)}.
\end{align*}
Consequently, with high probability, the total number of sampled rows is bounded by
\[\O{\frac{d^2 \kappa^2 \log^2(n\kappa)}{\eps^2}}.\]
\end{proof}

Along the lines of the proof techniques presented in this section, \cite{BravermanHMSSZ21} also showed adversarial robustness of importance sampling techniques for the purposes of low-rank approximation and graph sparsification. 
However, it should be noted that those results also suffer from an extra multiplicative factor of $d$ compared to the best offline results due to the union bound over an $\eps$-net, as well as extra multiplicative factors of $\kappa$ due to the gap in the largest and smallest possible values of $\|\bA\bx\|_p$ across different unit vectors $\bx\in\mathbb{R}^d$. 
By considering correctness in the periods where this ratio increases geometrically, \cite{JiangPW23} is able to improve the dependencies on $\kappa$ to be polylogarithmic, i.e., $\polylog(\kappa)$. 

\section{Attacking the AMS Sketch in Insertion-Only Streams}
\seclab{sec:ams:attack}
In this section, we describe an attack by \cite{Ben-EliezerJWY22} on the well-known Alon-Matias-Szegedy (AMS) streaming algorithm~\cite{AlonMS99} used to estimate the $L_2$ norm of a frequency vector. 
The attack targets an AMS sketch and causes it to output an inaccurate approximation of the true norm $\|\bx\|_2^2$. 
In fact, \cite{Ben-EliezerJWY22} proves a stronger claim: for any $r \geq 1$, and for an AMS sketch with $\frac{r}{\eps^2}$ rows, an adversary can adaptively craft just $\O{r}$ stream updates to cause a significant error in the sketch's estimate.

\paragraph{AMS algorithm overview.} 
We recall the description of the AMS sketching algorithm presented in \secref{sec:intro:attack:ams}. 
The AMS sketch uses a (typically implicit) random matrix $\bA \in \mathbb{R}^{t \times n}$ whose entries $A_{i,j}$ are i.i.d.\ Rademacher variables, i.e., uniformly sampled from $\{-1,1\}$. 
The sketch maintains the vector $\bA\cdot\bx^{(j)}$ at step $j$ in the stream, where $\bx^{(j)}$ is the frequency vector. 
Since $\bA$ is linear, updates can be applied incrementally: 
\[\bA\bx^{(j+1)} = \bA\bx^{(j)} + \bA\be_{i_{j+1}} \Delta_{j+1},\]
for an update $(i_{j+1}, \Delta_{j+1})$. 
The estimated $L_2$ norm at step $j$ is then given by $\frac{1}{t} \|\bA\bx^{(j)}\|_2^2$, which, under non-adaptive conditions, is a $(1 \pm \eps)$ approximation with high probability for $t = \Theta\left(\frac{1}{\eps^2}\right)$.

\paragraph{Attack strategy.} 
Let $\bA$ denote the $t \times n$ AMS sketch matrix, where each $A_{i,j}$ is sampled uniformly from $\{ -t^{-1/2}, t^{-1/2} \}$. 
The estimated norm at step $j$ is $\|\bA\cdot\bx^{(j)}\|_2^2$. 
Let $\be_i \in \mathbb{R}^n$ be the standard basis vector with $1$ in the $i$-th position. 

In this attack, the vector $\bw$ always represents the current frequency vector $\bx^{(j)}$ of the stream. 
Notably, the adversary only observes the scalar value $\|\bA\bx^{(j)}\|_2^2$ after each step and does not require access to the internal structure of $\bA$. 
Initially, the adversary inserts the update $(1, C \cdot \sqrt{t})$, corresponding to the operation that increases the first coordinate of the frequency vector by $C\cdot\sqrt{t}$, where $C$ is a sufficiently large constant. 
In other words, the frequency vector is initialized to $C\cdot\sqrt{t}\cdot\be_1$, for the elementary vector $\be_1$. 
Then, for each $i = 2, \ldots, n$, the adversary probes the effect of inserting item $i$ once. 
If the increase in the AMS estimate is less than $1$, the adversary adds the item again. 
If the increase is exactly $1$, the adversary inserts the item once more with probability $\frac{1}{2}$.
The main intuition is that the adversary tries to increase the frequency vector in directions that are less-correlated with the sketch. 
The full attack is presented in \algref{alg:attack:AMS}. 

\begin{algorithm}[!htb]
\caption{Adversary against AMS sketch \cite{Ben-EliezerJWY22}}
\alglab{alg:attack:AMS}
\begin{algorithmic}
\State{$\bw \gets C \cdot \sqrt{t} \cdot\be_1$}
\For{$i = 2$ to $m$}
\State{$\text{old} \gets \|\bA\bw\|_2^2$}
\State{$\bw \gets \bw + \be_i$}
\State{$\text{new} \gets \|\bA\bw\|_2^2$}
\If{$\text{new} - \text{old} < 1$}
\State{$\bw \gets \bw + \be_i$}
\ElsIf{$\text{new} - \text{old} = 1$}
\State{With probability $\frac{1}{2}$: $\bw \gets \bw + \be_i$}
\EndIf
\EndFor
\end{algorithmic}
\end{algorithm}

We claim that, with high probability, after a stream of $m = \O{t}$ updates, the AMS sketch produces an estimate $\|\bA\bx^{(m)}\|_2^2$ that lies outside the $(1 \pm \eps)$ approximation range of the true value $\|\bx^{(m)}\|_2^2$, noting that at the end of the stream, we have $\bw = \bx^{(m)}$. 
In fact, \cite{Ben-EliezerJWY22} proves that the adversary can force the sketch to output an estimate that is not even a $2$-approximation, regardless of the number of rows $t$ in the AMS sketch.

\begin{theorem}
\thmlab{thm:attack:AMS}
\cite{Ben-EliezerJWY22}
Let $\bA \in \mathbb{R}^{t \times n}$ be an AMS sketch where each entry is an independent Rademacher random variable scaled by $t^{-1/2}$, and suppose $1 \leq t < \frac{n}{c}$ for some absolute constant $c$. 
Then there exists an adversary that creates an adaptive stream of length $m=\O{t}$, so that with probability at least $\frac{9}{10}$, the AMS sketch fails to provide a $\left(1 \pm \frac{1}{2}\right)$-approximation of the squared $L_2$ norm of the frequency vector $\bx^{(m)}$. 
Specifically:
\[\|\bA\bx^{(m)}\|_2^2 < \frac{1}{2} \|\bx^{(m)}\|_2^2.\]
\end{theorem}
\begin{proof}
For $j = 2, 3, \ldots$, we define the $j$-th \textit{step} of \algref{alg:attack:AMS} as the iteration of the \texttt{for} loop where the loop variable $i$ equals $j$. 
The \emph{first} step is defined to be the state of the stream immediately after executing line 1 of \algref{alg:attack:AMS}.

Let $\bw^i$ denote the frequency vector at the end of the $i$-th step, and let $\by^i = \bA\bw^i$ be the corresponding AMS sketch.
Define $s_i = \|\by^i\|_2^2 = \|\bA\bw^i\|_2^2$ as the sketch's estimate of the squared norm at step $i$. 
Initially, we have $\bw^1 = C \cdot \sqrt{t} \cdot \be_1$ for a sufficiently large constant $C$, which implies $s_1 = C^2 t$. 
Since the stream only allows insertions, the true squared norm $\|\bw^i\|_2^2$ is non-decreasing, so it remains at least $C^2 t$ for all $i$. 
Therefore, to show the sketch is inaccurate, it suffices to demonstrate that with high probability, $s_i < \frac{C^2t}{2}$ for some $i \geq 2$.

At each step $i \geq 1$, consider the effect of inserting $\be_{i+1}$:
\begin{itemize}
\item
If we insert $\be_{i+1}$ once, the updated estimate becomes:
\[s_{i+1} = \|\by^i + \bA\be_{i+1}\|_2^2 = s_i + 1 + 2 \sum_{j=1}^t y_j^i A_{j,i+1}.\]
\item
If we insert $\be_{i+1}$ twice, we have:
\[s_{i+1} = \|\by^i + 2\bA\be_{i+1}\|_2^2 = s_i + 4 + 4 \sum_{j=1}^t y_j^i A_{j,i+1}.\]
\end{itemize}
By construction, the insertion decision is based on the sign of the inner product $\sum_{j=1}^t y_j^i A_{j,i+1}$:
\begin{itemize}
\item
If the sum is negative, we insert $\be_{i+1}$ twice.
\item
If the sum is positive, we insert $\be_{i+1}$ once.
\item 
If the sum is zero, we insert $\be_{i+1}$ either once or twice with equal probability, i.e., flip a fair coin.
\end{itemize}
The key observation is that $\sum_{j=1}^t y_j^i A_{j,i+1}$ is a symmetric random variable because each $A_{j,i+1}$ is an independent Rademacher random variable, and $\by^i$ is fixed. 
Consequently, we have:
\begin{equation*}
\begin{split}
\Ex{\left| \sum_{j=1}^t y_j^i A_{j,i+1} \right|} &= \Ex{\sum_{j=1}^t y_j^i A_{j,i+1} \; \middle| \; \be_{i+1} \text{ inserted once}} \\
&= -\Ex{\sum_{j=1}^t y_j^i A_{j,i+1} \; \middle| \; \be_{i+1} \text{ inserted twice}}.
\end{split}
\end{equation*}
Now recall that the vector $A_{*,i+1}$, which is the $(i+1)$-th column of the matrix $\bA$, consists of independent Rademacher entries scaled by $\frac{1}{\sqrt{t}}$. 
By Khintchine’s inequality~\cite{haagerup1981best}, we have:
\[\Ex{\left| \sum_{j=1}^t y_j^i A_{j,i+1} \right|} = \frac{\alpha}{\sqrt{t}} \cdot \|\by^i\|_2 = \alpha \cdot \frac{\sqrt{s_i}}{\sqrt{t}},\]
for some absolute constant $\alpha > 0$. 
In fact, Theorem 1.1 of~\cite{haagerup1981best} shows that $\alpha \geq\frac{1}{\sqrt{2}}$ suffices. 

By a similar reasoning, it follows that whether $\be_{i+1}$ is inserted once or twice is a symmetric random variable, depending on the vector $A_{*,i+1}$. 
Thus, $\be_{i+1}$ is inserted once with probability $\frac{1}{2}$ and otherwise inserted twice with probability $\frac{1}{2}$. 
Using this, we can compute the expectation of the AMS sketch at step $i+1$:
\begin{equation*}
\begin{split} 
\Ex{s_{i+1}} &= \frac{1}{2}\left(s_i + 1 + 2\alpha \cdot \frac{\sqrt{s_i}}{\sqrt{t}}\right) + \frac{1}{2}\left(s_i + 4 - 4\alpha \cdot \frac{\sqrt{s_i}}{\sqrt{t}}\right) \\
&= s_i + \frac{5}{2} - \alpha \cdot \frac{\sqrt{s_i}}{\sqrt{t}} \\
&\leq s_i + \frac{5}{2} - \sqrt{\frac{s_i}{2t}},
\end{split}
\end{equation*}
where the last inequality uses $\alpha \geq\frac{1}{\sqrt{2}}$. 
This recurrence implies:
\[\Ex{s_{i+1}} \le \Ex{s_i} + \frac{5}{2} - \Ex{\sqrt{\frac{s_i}{2t}}}.\]

Now suppose there exists some $i \leq C^2 t + 2$ such that $\Ex{\sqrt{s_i}} < C \cdot \sqrt{\frac{t}{200}}$. 
Then, by the definition of expectation,
\[\sum_j \sqrt{j} \cdot \PPr{s_i = j} < C \cdot \sqrt{\frac{t}{200}}.\]
In particular, we have:
\[\sqrt{\frac{C^2 t}{2}} \cdot \PPr{s_i \geq\frac{C^2 t}{2}} \leq \sum_{j \geq\frac{C^2 t}{2}} \sqrt{j} \cdot \PPr{s_i = j} < \sqrt{\frac{C^2 t}{200}}.\]
This yields:
\[\PPr{s_i \geq\frac{C^2 t}{2}} \leq \frac{1}{10},\]
which implies $\PPr{s_i < \frac{C^2 t}{2}} \geq \frac{9}{10}$. 
Hence, at this step $i$, the AMS sketch underestimates the squared norm with a high constant probability. 

Suppose now that no such $i \leq C^2 t + 2$ exists. 
That is, suppose $\Ex{\sqrt{s_i}} \geq C \cdot \sqrt{\frac{t}{200}}$ for all $i = 2, 3, \ldots, C^2 t + 2$. 
Then by the recurrence above,
\[\Ex{s_{i+1}} < \Ex{s_i} - 1.\]
Since $s_1 = C^2 t$, this would imply:
\[\Ex{s_{C^2 t + 2}} < C^2 t - (C^2 t + 1) = -1,\]
which is a contradiction, as $s_i$ is always non-negative being a squared norm.

Therefore, such an index $i \leq C^2 t + 2$ must exist for which 
\[\PPr{s_i < \frac{C^2 t}{2}} \geq \frac{9}{10},\]
i.e., the AMS sketch is fooled by this point with high probability, completing the proof.
\end{proof}

\section{Sketch Switching}
\seclab{sec:sketch:switch}
In this section, we describe the sketch switching framework introduced by \cite{Ben-EliezerJWY22}. 
The framework maintains a relatively small number of independent instances $A_1,\ldots,A_\lambda$ of a non-robust strong tracking streaming algorithm. 
The main intuition is to change the output of the algorithm very rarely, so that these small number of independent instances can collectively achieve a ``good'' approximation over the entire course of the stream. 

Specifically, at each time in the stream, only one of these independent instances is ``active''.  
As long as the current output by instance $A_c$ is a sufficiently accurate multiplicative approximation of output by the active instance, then the overall framework outputs the same value and the active instance remains the same. 
However, when the current output by instance $A_c$ is no longer a sufficiently accurate multiplicative approximation of output by the active instance, we update the output of the overall framework and replace the active instance $A_c$ with the next independent instance $A_{c+1}$ of the algorithm. 

The main point is that the overall framework carefully exposes the randomness of the independent instances $A_1,\ldots,A_\lambda$. 
By the strong tracking guarantee of each independent instance, we have correctness of the active algorithm until the output of the overall framework changes. 
However, at that point, the active instance is replaced with the next independent instance anyway, so we retain correctness throughout the course of the stream. 
The main intuition is that many functions of interest can only increase a ``small'' number of times, so we do not need too many instances $\lambda$. 
To capture this property, we first define the following quantity:
\begin{definition}[Flip number]
\deflab{def:flipno}
Given an accuracy parameter $\eps\ge 0$ and a stream length $m$, let $y=(y_0,y_1,\ldots,y_m)$ be a sequence of real numbers. 
We say that the \emph{$\eps$-flip number} $\lambda_\eps(y)$ of the sequence $y$ is the maximum integer $k$ such that there exist $k$ subindices $0\le t_1<\ldots<t_k\le m$ such that $y_{t_{j-1}}\notin(1\pm\eps) y_{t_j}$, i.e., either $y_{t_{j-1}}<(1-\eps) y_{t_j}$ or $y_{t_{j-1}}>(1+\eps) y_{t_j}$. 

For a fixed function $f:\mathbb{R}^n\to\mathbb{R}$ and a class $\calC\subseteq([n]\times\mathbb{Z})^m$ of stream updates, the $(\eps,m)$-flip number $\lambda_{\eps,m}(f)$ of $f$ over $\calC$ is the maximum, over all sequences $((a_1,\Delta_1),\ldots,(a_m,\Delta_m))\in\calC$ of the $\eps$-flip number of the sequence $y=(y_0,\ldots,y_m)$ defined by $y_t=f(x^{(t)})$ for all $0\le t\le m$, where $x^{(t)}$ denotes the underlying frequency vector defined by the first $t$ updates of the stream. 
\end{definition}
Observe that the flip number is monotonic in $\eps$, so that $\lambda_{\eps',m}(f)\ge\lambda_{\eps,m}(f)$ if $\eps'<\eps$. 
Moreover, the flip number is cleanly preserved under approximations:
\begin{lemma}
\lemlab{lem:flip:acc}
\cite{Ben-EliezerJWY22}
For a fixed $\eps\in(0,1)$, let $u=(u_0,\ldots,u_m)$, $v=(v_1,\ldots,v_m)$, and $w=(w_0,\ldots,w_m)$ be three sequences of real numbers such that:
\begin{enumerate}
\item 
For any $0\le i\le m$, we have $v_i=\left(1\pm\frac{\eps}{8}\right)\cdot u_i$.
\item 
$w_0=v_0$ and for any $i>0$, if $w_{i-1}=\left(1\pm\frac{\eps}{2}\right)\cdot v_i$, then $w_i=w_{i-1}$. 
Otherwise, we have $w_i=v_i$. 
\end{enumerate}
Then $w_i=(1\pm\eps)\cdot u_i$ for any $0\le i\le m$. 
Moreover, $\lambda_0(w)\le\lambda_{\eps/8}(u)$. 
\end{lemma}

\begin{definition}[Strong tracking]
For each $t\in[m]$, let $\bx^{(t)}$ be the frequency vector defined after the first $t$ updates of an arbitrary but fixed data stream. 
Let $f:\mathbb{R}^n\to\mathbb{R}$ be a function on frequency vectors. 
Then we say an algorithm $\calA$ achieves $(\eps,\delta)$-strong tracking for $f$ if at each step $t\in[m]$, the algorithm outputs an estimate $Z_t$ such that
\[|Z_t-f(\bx^{(t)})|\le\eps\cdot|f(\bx^{(t)}|\]
for all $t\in[m]$ simultaneously, with probability at least $1-\delta$. 
\end{definition}

\begin{algorithm}[!htb]
\caption{Sketch Switching Framework}
\alglab{alg:sketch:switch}
\begin{algorithmic}[1]
\Require{Accuracy parameter $\eps\in(0,1)$, an adaptive stream $(a_t,\Delta_t)$ for $t\in[m]$, and an $(\eps,\delta)$-strong tracker for a given function}
\Ensure{Adversarially robust $(1+\eps)$-streaming algorithm}
\State{$\lambda\gets\lambda_{\eps/8,m}(f)$}
\State{Initialize $A_1,\ldots,A_\lambda$ as independent instances of $\left(\frac{\eps}{8},\frac{\delta}{\lambda}\right)$-strong tracking algorithm for $f$}
\State{$Z\gets f(\mathbf{0}^n)$}
\State{$c\gets 1$}
\For{each stream update $(a_t,\Delta_t)$}
\State{Update $A_1,\ldots,A_\lambda$ with $(a_t,\Delta_t)$}
\State{Let $y$ be the current output of $A_c$}
\If{$y\notin\left(1\pm\frac{\eps}{2}\right)\cdot Z$}
\State{$Z\gets y$}
\State{$c\gets c+1$}
\EndIf
\State{\Return $Z$ for the estimate at time $t$}
\EndFor
\end{algorithmic}
\end{algorithm}

\begin{theorem}[Sketch switching]
\thmlab{thm:sketch:switch}
For a fixed function $f:\mathbb{R}^n\to\mathbb{R}$, let $\calA$ be an $(\eps,\delta)$-strong tracking streaming algorithm that uses space $S(\eps,\delta)$ for any accuracy parameter $\eps\in(0,1)$ and failure probability $\delta\in(0,1)$. 
Then \algref{alg:sketch:switch} is an adversarially robust streaming algorithm that outputs a $(1+\eps)$-approximation for $f(x^{(t)})$ at every time $t\in[m]$ with probability $1-\delta$. 
Moreover, the algorithm uses $\O{S(\eps/8,\delta/\lambda)\cdot\lambda}$ space, where $\lambda=\lambda_{\eps/8,m}(f)$. 
\end{theorem}
\begin{proof}
We first observe that without loss of generality, we can assume the adversary $\Adv$ against \algref{alg:sketch:switch} is deterministic because for any randomized adversary that succeeds with probability greater than $\delta$, then by an averaging argument, there exists a fixing of the random bits of $\Adv$ that also succeeds with probability greater than $\delta$, across the coin flips of \algref{alg:sketch:switch}. 
Hence, we fix the string of randomness to the adversary, which results in a deterministic adversary. 
Then given a sequence $Z_1,\ldots,Z_t$ of outputs by the streaming algorithm in \algref{alg:sketch:switch} and the sequence $(a_1,\Delta_1),\ldots,(a_t,\Delta_t)$ of stream updates up to time $t$, then the next update $(a_{t+1},\Delta_{t+1})$ is deterministically fixed. 

We first fix the string of randomness for the first instance $A_1$. 
Let $s_1^1,s_2^1,\ldots,s_m^1$ be the updates that the adversary would make if $Z_0=f(\mathbf{0}^n)$ were output at every step of the stream. 
Let $\bx^{(t),1}$ denote the underlying frequency vector after stream updates $s_1^1,\ldots,s_t^1$. 
Let $t_1\in[m]$ be the first step at which $Z_0\notin\left(1\pm\frac{\eps}{2}\right)\cdot A_1(t_1)$, if such a step exists. 
Otherwise, we can set $t_1=m+1$. 
Note that at time $t_1$, the output of the algorithm is changed to $A_1(t_1)$. 
conditioning on the event that $A_1$ provides strong tracking for $f$ with accuracy parameter $\frac{\eps}{8}$ over the stream updates $s_1^1,\ldots,s_m^1$, then we have 
\[A_1(t)=\left(1\pm\frac{\eps}{8}\right)\cdot f(\bx^{(t)}),\]
for all steps $t<t_1$. 
This event happens with probability $1-\frac{\delta}{\lambda}$. 
Thus with probability $1-\frac{\delta}{\lambda}$, we have that $Z_0=\left(1\pm\eps\right)\cdot f(\bx^{(t)})$ for all steps $t<t_1$. 
Moreover, we further have $A_1(t_1)$ is a $\left(1\pm\frac{\eps}{8}\right)$-approximation to $f(\bx^{(t_1)})$. 

Now, \algref{alg:sketch:switch} switches to instance $A_2$. 
Let $o_1=A_1(t_1)$ so that the algorithm outputs $o_1$ provided that $o_1=\left(1\pm\frac{\eps}{2}\right)\cdot A_2(t)$. 
Since the randomness of the adversary is already fixed, then there exists a sequence $u_{t_1+1}^2,\ldots,u_m^2$ of updates that the adversary would send upon both the previous history of inputs $u_1^1,\ldots,u_{t_1}^1$ and the corresponding outputs, as well as if the algorithm would always output $o_1$ from step $t_1$ onwards until time $m$. 
Conditioning on the strong tracking guarantee of $A_2$ to provide a $\left(1+\frac{\eps}{8}\right)$-approximation on this fixed sequence of updates, we set $t_2\in[m]$ to be the first step at which $o_1\notin\left(1\pm\frac{\eps}{2}\right)\cdot A_2(t_2)$, if such a step exists. 
Then by the same above reasoning, we have that $o_1=\left(1\pm\eps\right)\cdot f(\bx^{(t)})$ for all $t\in[t_1,t_2)$. 
Moreover, by setting $o_2=A_2(t_2)$, we have $o_2=\left(1\pm\frac{\eps}{8}\right)\cdot f(\bx^{(t_2)})$. 

We can then apply this reasoning inductively for each instance $A_c$, across all $c\in[\lambda]$, so that the output is a $(1+\eps)$-approximation to the correct value $f(\bx^{(t)})$ for all $t\in[t_c,t_{c+1}-1)$. 
Moreover, since each instance fails with probability $\frac{\delta}{\lambda}$, then by a union bound, all $\lambda$ instances provide strong tracking with probability at least $1-\delta$. 

Toward correctness, it remains to show that all updates of the stream can be handled before the $\lambda$ independent instances $A_1,\ldots,A_{\lambda}$ are exhausted. 
To that end, we apply \lemref{lem:flip:acc} with the sequence $u=(f(\bx^{(0)}),\ldots,f(\bx^{(m)}))$, $v=(f(\bx^{(0)}),A_1(1),\ldots,A_1(t_1),A_2(t_1+1),\ldots)$ and $w$ being the output of the \algref{alg:sketch:switch}. 
Since $w$ is generated precisely according to $v$ in the statement of \lemref{lem:flip:acc}, then it follows that $\lambda=\lambda_{\eps/8,m}(f)$ independent instances $A_1,\ldots,A_{\lambda}$ suffice. 

Finally, it remains to analyze the space complexity of \algref{alg:sketch:switch}. 
There are $\lambda$ instances, each which use $S(\eps/8,\delta/\lambda)$ space. 
Thus, the algorithm uses $\O{S(\eps/8,\delta/\lambda)\cdot\lambda}$ space in total. 
\end{proof}

We now discuss a number of applications for the sketch switching technique, and particularly for \thmref{thm:sketch:switch}. 

\subsection{Applications to Moment Estimation}
We first consider the problem of moment estimation, defined as follows:
\begin{definition}[Moment estimation]
Given a vector $\bv\in\mathbb{R}^n$ and a parameter $p>0$, the $F_p$ moment of $\bv$ is defined as 
\[F_p(\bv)=\|\bv\|_p^p=\sum_{i\in[n]}|v_i|^p=|v_1|^p+\ldots+|v_n|^p.\]
Given an accuracy parameter $\eps\in(0,1)$, the goal is to output a $(1+\eps)$-approximation to $F_p(\bv)$. 
\end{definition}
We remark that contained within the above definition of the $F_p$ moment is the definition of the $L_p$ norm $\|\bv\|_p$, defined as follows:
\begin{definition}[Norm estimation]
Given a vector $\bv\in\mathbb{R}^n$ and a parameter $p>0$, the $L_p$ norm of $\bv$ is defined as 
\[\|\bv\|_p=\left(\sum_{i\in[n]}|v_i|^p=|v_1|^p+\ldots+|v_n|^p\right)^{1/p}.\]
Given an accuracy parameter $\eps\in(0,1)$, the goal is to output a $(1+\eps)$-approximation to $\|\bv\|_p$. 
\end{definition}
The $F_p$-moment estimation problem is frequently referred to as the norm estimation problem and vice versa. 
In particular, an algorithm that produces a $(1+\eps)$-approximation for one can be adapted to approximate the other within a $(1+\eps)$ factor by appropriately adjusting $\eps$ (for constant $p$). 
Therefore, we refer to these problems interchangeably. 

We first show that the flip number of these problems is bounded.

\begin{lemma}
\lemlab{lem:monfpflip}
\cite{Ben-EliezerJWY22}
Let $g:\mathbb{R}^n \to \mathbb{R}$ be a monotone function, so that $g(x) \geq g(y)$ whenever $x_i \geq y_i$ for all $i \in [n]$. 
Suppose further that $g(x) \geq T^{-1}$ for all non-zero $x$, and $g(M \cdot \vec{1}) \leq T$, where $M$ is an upper bound on the magnitude of the entries in the frequency vector and $\vec{1}$ is the all-ones vector. 
Then the $(\eps,m)$-flip number of $g$ in the insertion-only streaming model satisfies $\lambda_{\eps,m}(g) = \O{\frac{1}{\eps}\log T}$.
\end{lemma}
\begin{proof}
Observe that $g(f^{(1)}) \geq T^{-1}$ and $g(f^{(m)}) \leq g(M \cdot \vec{1}) \leq T$. 
Since the stream includes only insertions, the function values form a non-decreasing sequence: $g(f^{(0)}) \leq g(f^{(1)}) \leq \ldots \leq g(f^{(m)})$.

Now consider any maximal increasing subsequence of time steps $1 \leq y_1 < y_2 < \ldots < y_k \leq m$ such that for each $i \in [k-1]$, we have $g(f^{(y_i)}) < (1 - \eps)g(f^{(y_{i+1})})$. 
Since we ignore the $0$-th step, the flip number is at most $k+1$. 

After each $y_i$, the function value increases by a factor of at least $\frac{1}{1 - \eps}$. 
The number of such multiplicative increases needed to go from $T^{-1}$ to $T$ is $\O{\frac{1}{\eps}\log T}$. 
Hence, if $k$ exceeds this bound, then by the pigeonhole principle, two function values must lie within the same multiplicative interval, contradicting the definition of the sequence. 
This completes the proof.
\end{proof}

A special case of this proposition includes the $F_p$ moments of data streams. 
We also define $\|\bx\|_0 = |\{ i: x_i \neq 0 \}|$ to be the number of non-zero entries in the vector $\bx$, c.f., \defref{def:distinct:elements}.  
Then we have the following upper bound on the flip number for a stream length of $m=\poly(n)$.

\begin{corollary}
\corlab{cor:fpflip}
\cite{Ben-EliezerJWY22}
Let $p \geq 0$. 
Then, in the insertion-only streaming model, the $(\eps,m)$-flip number of $\|\bx\|_p^p$ satisfies:
\begin{itemize}
\item 
$\lambda_{\eps,m}(\| \cdot \|_p^p) = \O{\frac{1}{\eps}\log n}$ for $p \leq 2$,
\item 
$\lambda_{\eps,m}(\| \cdot \|_p^p) = \O{\frac{p}{\eps}\log n}$ for $p > 2$,
\item 
$\lambda_{\eps,m}(\| \cdot \|_0) = \O{\frac{1}{\eps}\log m}$ for $p = 0$.
\end{itemize}
\end{corollary}
\begin{proof}
We note that $\|\mathbf{0}^n\|_p^p = 0$ and $\|\bx\|_p^p \geq 1$ for any nonzero integral vector $\bx$. 
Moreover, $\|\bx^{(m)}\|_p^p \leq M^p n \leq n^{1 + cp}$ for some constant $c$, since we assume $\|\bx\|_\infty \leq M$ where $M = \poly(n)$. 
Applying \lemref{lem:monfpflip} with $T = n^{c \cdot \max\{p, 1\}}$ yields the result.

For $p = 0$, observe that $\|\bx^{(m)}\|_0$ either stays the same or increases by one with each insertion, so the maximum number of multiplicative increases is $\O{\log m}$.
\end{proof}

For the insertion-only setting, there is a number of streaming algorithms that achieve $(1+\eps)$-approximation to the $F_p$ moment across various regimes~\cite{AlonMS99,IndykW05,Indyk06,Li08,KaneNW10a,AndoniKO11,Ganguly11,GangulyW18,BlasiokDN17}. 
We use the following strong tracker for $F_2$ estimation:
\begin{restatable}[Oblivious $F_2$ strong tracking]{theorem}{thmstrongFtwo}
\thmlab{thm:strong:F2}
\cite{BlasiokDN17}
Given an accuracy parameter $\eps>0$ and a failure probability $\delta\in(0,1)$, there exists an insertion-only streaming algorithm that provides $(\eps,\delta)$-strong $F_2$ tracking and uses 
\[\O{\frac{\log n}{\eps^2}\left(\log\frac{1}{\eps}+\log\frac{1}{\delta}+\log\log n\right)}\]
bits of space. 
\end{restatable}
For $p\in(0,2)$, we use the following strong tracker:
\begin{restatable}[Oblivious $F_p$ strong tracking for $0<p<2$]{theorem}{thmstrongFpsmallp}
\thmlab{thm:strong:Fp:smallp}
\cite{BlasiokDN17}
For $0<p<2$, there exists an insertion-only streaming algorithm that provides $(\eps,\delta)$-strong $F_p$ tracking, using 
\[\O{\frac{\log n}{\eps^2}\left(\log\log n+\log\frac{1}{\eps}+\log\frac{1}{\delta}\right)}\]
bits of space. 
\end{restatable}
Hence by combining the guarantees of the framework in \thmref{thm:sketch:switch} with the flip number bounds in \corref{cor:fpflip} and the strong-tracking bounds in either \thmref{thm:strong:F2} or \thmref{thm:strong:Fp:smallp}, we have the following:
\begin{theorem}
\thmlab{thm:fp:small:sketch:switching}
\cite{Ben-EliezerJWY22}
Given $p\in(0,2]$ and $\eps\in(0,1)$, there exists an adversarially robust algorithm on insertion-only streams of length $m=\poly(n)$ that uses $\tO{\frac{1}{\eps^3}\log^3 n}$ bits of space and with probability at least $\frac{2}{3}$, outputs a $(1+\eps)$-approximation to the $F_p$ moment at all times.
\end{theorem}

For $p>2$, we utilize the following strong tracker:
\begin{restatable}[\cite{Ganguly11}, Theorem 22 in \cite{GangulyW18}]{theorem}{thmFpbigp}
\thmlab{thm:Fp:bigp}
For any $p>2$, there exists an insertion-only streaming algorithm that uses $\O{\frac{1}{\eps^2}n^{1-2/p}\log^2 n\log\frac{1}{\delta}}$ bits of space and outputs a $(1+\eps)$-approximation to the $F_p$ moment with probability at least $1-\delta$. 
\end{restatable}

By combining the guarantees of the framework in \thmref{thm:sketch:switch} with the flip number bounds in \corref{cor:fpflip} and the strong-tracker of \thmref{thm:Fp:bigp}, we have the following:

\begin{theorem}
\thmlab{thm:fp:big:sketch:switching}
\cite{Ben-EliezerJWY22}
Given $p>2$ and $\eps\in(0,1)$, there exists an adversarially robust algorithm on insertion-only streams of length $m=\poly(n)$ that uses $\tO{\frac{1}{\eps^3}n^{1-2/p}}$ bits of space and with probability at least $\frac{2}{3}$, outputs a $(1+\eps)$-approximation to the $F_p$ moment at all times.
\end{theorem}

\subsection{Applications to Distinct Element Estimation}
In this section, we consider the problem of distinct element estimation, also known as $F_0$ estimation or even $L_0$ estimation, though it is not quite a norm as it does not satisfy homogeneity. 
The problem is defined as follows.

\begin{definition}[Distinct element estimation]
\deflab{def:distinct:elements}
Given a vector $\bv\in\mathbb{R}^n$, we define the number of nonzero coordinates in $\bv$ to be 
\[F_0(\bv)=\|\bv\|_0 = |\{i:v_i \neq 0 \}|.\]
Given an accuracy parameter $\eps\in(0,1)$, the goal is to output a $(1+\eps)$-approximation to $F_0(\bv)$. 
We remark that when $\bv$ is defined by an insertion-only data stream, $F_0(\bv)$ is the number of distinct elements in the stream. 
\end{definition}
We state the following strong tracker for distinct element estimation. 
\begin{restatable}[Oblivious $F_0$ strong tracking]{theorem}{thmstrongFzero}
\thmlab{thm:strong:F0}
\cite{Blasiok20}
There exists an insertion-only streaming algorithm $\zeroestimate$ that uses $\O{\frac{1}{\eps^2}\log\frac{1}{\delta}+\log n}$ bits of space and provides $(\eps,\delta)$-strong $F_0$ tracking. 
\end{restatable}

Putting together the guarantees of the framework in \thmref{thm:sketch:switch} with the flip number bounds in \corref{cor:fpflip} and the strong-tracker of \thmref{thm:strong:F0}, we have the following:

\begin{theorem}
\thmlab{thm:fzero:sketch:switching}
\cite{Ben-EliezerJWY22}
Given $\eps\in(0,1)$, there exists an adversarially robust algorithm on insertion-only streams of length $m=\poly(n)$ that uses $\tO{\frac{1}{\eps^3}\log^2 n}$ bits of space and with probability at least $\frac{2}{3}$, outputs a $(1+\eps)$-approximation to the number of distinct elements at all times.
\end{theorem}

\section{Bounded Computation Paths}
\seclab{sec:bounded:computations}
Whereas the sketch switching framework in \secref{sec:sketch:switch} created a number of independent instances of non-robust streaming algorithms to handle each time the function value $f$ of the data stream increased, in this section, we describe an approach that just uses a single instance of a non-robust streaming algorithm to achieve adversarial robustness. 

For a fixed function $f:\mathbb{R}^n\to\mathbb{R}$, let $\calA$ be an $(\eps,\delta)$-strong tracking algorithm for $f$ that uses space $S(\eps,\delta)$. 
Consider an algorithm $\calA'$ that simply runs $\calA$ with a finer accuracy $\eps'$ and a smaller probability of failure $\delta'$. 
Moreover, suppose we round the output as in the sketch switching technique in \secref{sec:sketch:switch}. 

Specifically, we set $\eps'=\frac{\eps}{8}$ and $\delta'=\frac{\delta}{\binom{m}{\lambda}T^{\O{\lambda}}}$ for $\lambda=\lambda_{\eps/8,m}(f)$ to union bound over all possible ``computation paths'' the adversary might force, c.f., \figref{fig:computation:paths}. 
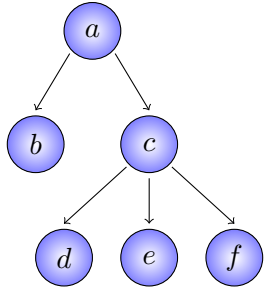
\begin{figure*}[!htb]
\centering
\begin{tikzpicture}[scale=0.75]
\filldraw[shading=radial, inner color = white, outer color = blue!50!, opacity=1] (0,0) circle (0.5);
\node at (0,0){$a$};

\filldraw[shading=radial, inner color = white, outer color = blue!50!, opacity=1] (-1,-2) circle (0.5);
\node at (-1,-2){$b$};
\draw [->] (-0.4,-0.4) -- (-1,-1.4);

\filldraw[shading=radial, inner color = white, outer color = blue!50!, opacity=1] (1,-2) circle (0.5);
\node at (1,-2){$c$};
\draw [->] (0.4,-0.4) -- (1,-1.4);

\filldraw[shading=radial, inner color = white, outer color = blue!50!, opacity=1] (-0.5,-4) circle (0.5);
\node at (-0.5,-4){$d$};
\draw [->] (0.6,-2.4) -- (-0.5,-3.4);

\filldraw[shading=radial, inner color = white, outer color = blue!50!, opacity=1] (1,-4) circle (0.5);
\node at (1,-4){$e$};
\draw [->] (1,-2.6) -- (1,-3.4);

\filldraw[shading=radial, inner color = white, outer color = blue!50!, opacity=1] (2.5,-4) circle (0.5);
\node at (2.5,-4){$f$};
\draw [->] (1.4,-2.4) -- (2.5,-3.4);
\end{tikzpicture}
\caption{Example of possible computation paths by adversary to union bound for correctness. 
Each node represents a set of possible inputs to the data stream by the adversary, based on the previous outputs by the algorithm.}
\figlab{fig:computation:paths}
\end{figure*}

Then for a sequence $v_0,\ldots,v_t$ of outputs by $\calA$ with these parameters up to time $t$, we generate the sequence $w_t$ of outputs of $\calA'$ by first setting $w_0=v_0$, in the manner of \lemref{lem:flip:acc}.  
For $t>0$, we set $w_t=w_{t-1}$ if $w_{t-1}\in\left(1\pm\frac{\eps}{2}\right)\cdot v_t$. 
Otherwise, we set $w_t=v_t$. 

\begin{theorem}
\thmlab{thm:comp:paths}
Let $f:\mathbb{R}^n\to\mathbb{R}$ be a fixed function whose output uses $\log T$ bits of precision. 
Let $\calA$ be an $(\eps,\delta)$-strong tracking algorithm for a stream of length $m$ that uses space $S(\eps,\delta)$.  
Then there exists a streaming algorithm $\calA'$ such that:
\begin{enumerate}
\item
$\calA'$ is an adversarially robust streaming algorithm that outputs a $(1+\eps)$-approximation to $f(\bx^{(t)})$ for all steps $t\in[m]$, with probability at least $1-\delta$.
\item 
$\calA'$ uses space $\O{S(\eps',\delta')}$ for $\eps'=\frac{\eps}{8}$ and $\delta'=\frac{d}{\binom{m}{\lambda}T^{\O{\lambda}}}$ for $\lambda=\lambda_{\eps/8,m}(f)$. 
\end{enumerate}
\end{theorem}
\begin{proof}
As in the proof of \thmref{thm:sketch:switch}, we can assume without loss of generality that the adversary $\Adv$ is deterministic. 
Therefore, any fixed sequence of outputs fully determines the stream of updates $(a_1,\Delta_1),\ldots,(a_m,\Delta_m)$.

Let $\lambda=\lambda_{\eps/8,m}(f)$ and let $\calC$ be the collection of all possible output sequences with $\log T$ bits of precision with $0$-flip number at most $\lambda$. 
Observe that $|\calC|\le\binom{m}{\lambda}T^{\O{\lambda}}$, as the data stream has at most $\lambda$ times across the $m$ updates where the previous answer is incorrect, forcing the algorithm to output a new answer that can only be one of $T$ possible values, due to the encoding in $\log T$ bits. 
Since the adversary is deterministic, each possible stream of updates generated by the adversary corresponds to one of these possible output sequences. 

For $\delta'\le\frac{\delta}{|\calC|}$, by a union bound over all possible such sequences, we have that algorithm $\calA$ with accuracy parameter $\eps'=\frac{\eps}{8}$ and failure probability $\delta'$ provides an $\frac{\eps}{8}$-strong tracking guarantee for all possible input streams. 
Correctness then follows by applying \lemref{lem:flip:acc} to each possible input stream to show correctness after the rounding procedure. 
\end{proof}

\subsection{Applications to High Probability Regimes}
While for most values of failure probability $\delta$, the sketch switching technique has better space complexity for $F_p$ estimation, the computation paths technique has better space complexity for the regime of very small failure probability.

\begin{theorem}[$F_p$-estimation for small $\delta$]
\thmlab{thm:fp:small:comp:paths}
Given $\eps\in(0,1)$, $p\in(0,2]$, $\delta < n^{-C \frac{1}{\eps}\log n}$ for a sufficiently large constant $C>1$, there exists an adversarially robust algorithm on insertion-only streams of length $m=\poly(n)$ that uses $\O{\frac{1}{\eps^2}\log^2 n\log\frac{1}{\delta}}$ bits of space and with probability at least $1-\delta$, outputs a $(1+\eps)$-approximation to the $F_p$ moment at all times.
\end{theorem}
\begin{proof}
The proof is a direct application of the guarantees of the computation paths framework in \thmref{thm:comp:paths} with the flip number bounds in \corref{cor:fpflip} and the strong-tracker of \thmref{thm:strong:Fp:smallp}. 
In particular, the flip number satisfies $\lambda = \O{\frac{\log n}{\eps}}$ and for the sufficiently small regime of $\delta$ in the assumption, we have $\log\frac{m^\lambda}{\delta} = \Theta\left(\log\frac{1}{\delta}\right)$.
\end{proof}

In contrast to \thmref{thm:fp:small:comp:paths}, adapting the sketch-switching paths technique would require space $\tO{\frac{1}{\eps^3}\log^2 n\log\frac{1}{\delta}}$. 
We can also achieve similar guarantees for $p>2$, c.f., \cite{Ben-EliezerJWY22}. 

\subsection{Applications to Subspace Embeddings}
Recall the following definition of $L_2$ subspace embeddings from \defref{def:lp:subspace}:
\deflpsubspace*
In the non-adaptive setting, we have already discussed sampling rows of $\bA$ with probability proportional to their $L_2$ leverage scores (or equivalently $L_2$ sensitivities) as one way to achieve $L_2$ subspace embeddings. 
Another way is as follows:
\begin{theorem}
\cite{Woodruff14}
Let $\eps,\delta\in(0,1)$, and define $\bS = \frac{1}{\sqrt{k}}\cdot \bR \in \mathbb{R}^{k \times n}$, where each entry $R_{i,j}$ of $\bR$ is drawn independently from the standard normal distribution. 
Suppose $k = \Theta\left( \frac{d + \log(1/\delta)}{\eps^2} \right)$. 
Then for any fixed matrix $\bA \in \mathbb{R}^{n \times d}$, with probability at least $1 - \delta$, the matrix $\bS\bA$ serves as a $(1 \pm \eps)$-$L_2$-subspace embedding for $\bA$. 
\end{theorem}
Now suppose the rows of $\bA\in\mathbb{R}^{n\times d}$ arrive sequentially and adversarially. 
Moreover, suppose we only output a subspace embedding of $\bA^{(t)}$ when one of the singular values of $\bA^{(t)}$ has increased by $(1+\eps)$. 
Then there are $\O{\frac{d}{\eps}\log n}$ possible times this can happen, assuming that the entries of $\bA$ are bounded in magnitude by at most $\poly(n)$. 
Then by directly applying the computation paths framework in \thmref{thm:comp:paths}, we have the following guarantee: 
\begin{theorem}[$L_2$ subspace embeddings]
\thmlab{thm:subspace:embed:comp:paths}
Given $\eps\in(0,1)$, there exists an adversarially robust algorithm on row-arrival streams of length $n$ with entries bounded in magnitude by $M=\poly(n)$ that uses $\O{\frac{d}{\eps^3}\log^2 n}$ bits of space and with probability at least $0.99$, outputs a $(1+\eps)$-$L_2$ subspace embedding at all times.
\end{theorem}

\section{Difference Estimators}
\seclab{sec:diff:est}
Although both sketching switching and bounded computation paths offer simple means to achieve adversarially robust streaming algorithms from existing oblivious streaming algorithms, the resulting space complexity is not tight with the optimal results known for the insertion-only model. 
Namely, these approaches incur extra factors in both $\frac{1}{\eps}$ and $\log n$ due to the dependency on the flip number. 
In this section, we give a framework by \cite{WoodruffZ21} that does not lose these factors and are therefore tight with the optimal (non-adaptive) insertion-only results up to $\polylog\left(\frac{1}{\eps}\right)$ factors. 

The main takeaway from the sketch switching (and also bounded computations paths) technique is that each time an underlying function $F$ increases by $(1+\eps)$, we need to run an additional independent instance of a non-adaptive streaming algorithm, which generally results in $\O{\frac{1}{\eps}\log n}$ overall copies, translating to $\O{\frac{1}{\eps}\log n}$ multiplicative overhead. 
However, this seems wasteful because the function has only increased additively by $\eps\cdot F$ and so we should intuitively only need to roughly estimate the increase, say to a constant-factor approximation. 
The following algorithmic building block, called a difference estimator, offers a means to achieve this. 
Intuitively, a difference estimator computes (or approximates) how much a function changes between two different times throughout the course of a (non-adaptive) data stream. 
\begin{definition}[Difference Estimator]
\deflab{def:diff:est}
Let $\bu$ and $\bv$ be frequency vectors, $\eps\in\left(0,\frac{1}{2}\right)$ be an accuracy parameter, $\delta\in(0,1)$ be a failure parameter, and $\gamma\in(0,1]$ be a ratio parameter. 
We say an algorithm is a $(\gamma,\eps,\delta)$-difference estimator for a function $F$ if the algorithm outputs an additive $\eps\cdot F(\bu)$ approximation to $F(\bu+\bv)-F(\bu)$ with probability at least $1-\delta$, given $F(\bu+\bv)-F(\bu)\le\gamma\cdot F(\bu)$.  
\end{definition}
For the purposes of adversarial robustness, one should interpret $\bu$ in \defref{def:diff:est} as a frequency vector that arrives in some prefix of the stream and $\bv$ as a frequency vector that arrives afterwards. 
Then the goal is to estimate the difference $F(\bu+\bv)-F(\bu)$ caused by the arrival of $\bv$ as it continuously evolves, while $\bu$ is fixed. 
Due to this reason, \cite{WoodruffZ21} defines this notion as a fixed-prefix difference estimator. 
In particular, \cite{WoodruffZ21} also defines a fixed-suffix difference estimator, where $\bv$ is fixed and $\bu$ is changing, and shows such a primitive can be used to improve upon algorithms in the sliding window model, where only the most recent $W$ updates in the data stream are considered to be the dataset of interest. 
Although the sliding window model has a rich history of research~\cite{LeeT06a,LeeT06b,BravermanO07,BravermanOZ12,CrouchMS13,BravermanGO14,BravermanLLM15,BravermanLLM16,Cohen-AddadSS16,BravermanGLWZ18,BravermanWZ21,WoodruffZ21,AjtaiBJSSWZ22,EpastoMMZ22,JayaramWZ22,BlockiLMZ23,WoodruffZZ23,Cohen-AddadJYZZ25,BravermanG0WZ26,BravermanWWZ26,Cohen-AddadWXZ26,NagawanshiPWWZ26}, our goal is to focus on adversarial robustness in the insertion-only model in this section. 
Thus whenever we refer to difference estimators, we mean the fixed-prefix difference estimators in the context of \cite{WoodruffZ21}. 

\subsection{Framework}
In this section, we describe the framework for adversarial robustness using difference estimators. 
We defer the discussion of the implementation of difference estimators for various functions to subsequent sections. 

\paragraph{Sketch stitching.}
Let $F$ be a function defined over frequency vectors, and let $\bv$ be a frequency vector implicitly defined via a data stream. 
Suppose there exists a $(\gamma, \eps, \delta)$-difference estimator $\calB$ for $F$, and a streaming algorithm $\calA$ that provides a $(1+\eps)$-approximation to $F(\bu)$ for any frequency vector $\bu$. 
To build intuition, assume that both $\calA$ and $\calB$ use space $\tO{\frac{1}{\eps^2}}$.

\cite{WoodruffZ21} introduces a new framework called \emph{sketch stitching}, which partitions the stream into contiguous blocks. 
Let $\bu_1, \ldots, \bu_\beta$ be frequency vectors induced by stream prefixes of increasing lengths $t_1 < \ldots < t_\beta$, such that $\bv = \bu_{\beta+1}$. 
Then:
\[F(\bv) = F(\bu_1) + \sum_{k=1}^{\beta} \left(F(\bu_{k+1}) - F(\bu_k)\right).\]
If we maintain a $(1+\eps)$-multiplicative approximation to $F(\bu_1)$ and a $(1+\eps)$-multiplicative approximation to each difference $F(\bu_{k+1})-F(\bu_k)$, then we can approximate $F(\bv)$ to within a $(1+\eps)$ factor. 
In this construction, $\calA$ estimates $F(\bu_1)$, while $\calB$ estimates each difference. 
These estimates can be summed to obtain a final approximation, allowing us to stitch together contributions from individual blocks. 
Importantly, the vectors $\bu_1, \ldots, \bu_\beta$ are unconstrained, and we next describe a principled way to define them.

\paragraph{Granularity changing.}
Assume $\bv \succeq \bu$, meaning $v_j \ge u_j \ge 0$ for all $j$. 
Let $i$ be the smallest integer such that $F(\bv) \le 2^i$, and let $t$ be the first time at which the prefix frequency vector $\bu$ satisfies $F(\bu) \ge 2^{i-1}$. 
Then the difference $F(\bv) - F(\bu) \le \frac{1}{2}F(\bv)$, so combining a $(1 + \frac{\eps}{2})$-approximation to $F(\bu)$ with a $(1 + \eps)$-approximation to $F(\bv) - F(\bu)$ yields a $(1+\eps)$-approximation to $F(\bv)$.

Generalizing this, instead of performing a single granularity change at scale $2^{i-1}$, we introduce multiple granularity changes at progressively finer dyadic scales. 
Fix the reference prefix frequency vector $\bu_0 := \bu$ at time $t$, and let $\bv$ denote the final frequency vector. 
For each $k$, define $t_k$ to be the last time before the increase $F(\bu_k)-F(\bu_{k-1})$ exceeds $\tau_k:=\frac{F(\bu)}{2^k}$, if such a time exists. 
Otherwise, we say that level $k$ is inactive, omit it from the decomposition, and set $t_k=t_{k-1}$. 
Then we define each frequency vector $\bu_k$ as the vector defined by the updates in the time interval $(t_{k-1},t_k]$. 
Observe that if we compute a $\left(1 + \frac{2^k\eps}{\beta}\right)$-approximation to each difference $F(\bu_{k+1}) - F(\bu_k)$, with $\bu_{\beta+1} = \bv$, then their sum is a $(1+\eps)$-approximation to $F(\bv)$. 
The key idea is that since each difference $F(\bu_{k+1}) - F(\bu_k) \le \frac{1}{2^k}\cdot F(\bv)$, a lower-accuracy estimator suffices for later blocks. 
We refer to the estimator used at level $k$ as a \emph{level $k$ difference estimator}.

By choosing $\beta = \Theta\left(\log\frac{1}{\eps}\right)$, we ensure that the contribution of the final block is $\O{\eps}\cdot F(\bv)$, so even omitting it adds only a small additive error. 
Thus, it suffices to sketch only the first $\beta$ blocks with increasing coarseness in estimation accuracy.

We emphasize that this is not a hierarchical data structure in the traditional sense. 
The sketches are not applied to differences of frequency vectors such as $\bv - \bu$, which correspond to $F(\bv - \bu)$. 
Rather, we use sketches of $\bu$ and $\bv$ to estimate $F(\bv) - F(\bu)$ directly, a technique requiring specialized difference estimators. 
This construction enables sketching with stronger guarantees than what is achievable using only $\calA$.

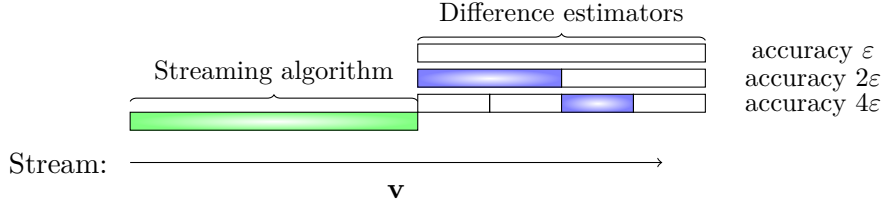
\begin{figure*}[!htb]
\centering
\begin{tikzpicture}[scale=0.95]
\draw [->] (0,-0.45) -- (7.4,-0.45);
\node at (-1,-0.45){Stream:};
\node at (3.7,-0.85){$\bv$};

\filldraw[shading=radial, inner color = white, outer color = green!50!, opacity=1] (0,0) rectangle+(4,0.25);
\draw (4,0.25) rectangle+(1,0.25);

\draw [decorate,decoration={brace}] (0,0.4) -- (4,0.4);
\node at (2,0.8){\small{Streaming algorithm}};

\draw (5,0.25) rectangle+(1,0.25);
\filldraw[shading=radial, inner color = white, outer color = blue!50!, opacity=1] (6,0.25) rectangle+(1,0.25);
\draw (7,0.25) rectangle+(1,0.25);
\node at (9.5,0.25+0.1){\small{accuracy $4\eps$}};

\filldraw[shading=radial, inner color = white, outer color = blue!50!, opacity=1] (4,0.6) rectangle+(2,0.25);
\draw (6,0.6) rectangle+(2,0.25);
\node at (9.5,0.6+0.1){\small{accuracy $2\eps$}};

\draw (4,0.95) rectangle+(4,0.25);
\node at (9.5,0.95+0.1){\small{accuracy $\eps$}};

\draw [decorate,decoration={brace}] (4,1.25) -- (8,1.25);
\node at (6,1.65){\small{Difference estimators}};

\end{tikzpicture}
\caption{Difference estimator outputs (in blue) are stitched together with streaming algorithm output (in green) as estimate estimate for $F(\bv)$. }
\figlab{fig:tree}
\end{figure*}

\paragraph{Adversarial robustness.}
We now describe how to implement the framework in the adversarially robust streaming model. 
Recall that sketch-switching only reveals outputs at geometrically increasing intervals—specifically, when internal estimates grow by a $(1+\eps)$ factor. 
This requires maintaining $\O{\frac{1}{\eps}\log n}$ independent algorithms and switching to fresh, unused sketches each time an output is revealed. 

The approach of \cite{WoodruffZ21} instead computes a $(1+\O{\eps})$-approximation of $F(\bu)$, where $\bu$ is the frequency vector at the earliest time for which $F(\bu)\ge 2^{i-1}$, for some integer $i$, and the suffix vector $\bv$ (representing the rest of the stream) satisfies $F(\bu+\bv)\ge 2^i$; we shall use $\O{\log n}$ instances of this approach to handle the geometric scales $2^i$. 
Before $\bu$ can grow to a vector $\bu_1$ such that $F(\bu_1)\ge \frac{3}{2}\cdot F(\bu)$, it must first pass through a vector $\bw_1$ where $F(\bw_1) \approx (1+\eps)\cdot F(\bu)$. 
Rather than maintaining a high-accuracy level-1 estimator for $\bw_1$, we instead use a low-accuracy level-$\beta$ estimator for the small difference $F(\bw_1)-F(\bu)$, with $\beta=\O{\log\frac{1}{\eps}}$. 
Once the stream reaches $\bw_1$, we estimate $F(\bw_1)$ by stitching together the estimates of $F(\bu)$ and $F(\bw_1)-F(\bu)$, then discard the estimator for $F(\bw_1)-F(\bu)$.
As the stream progresses to a vector $\bw_2$ for which $F(\bw_2)\approx (1+2\eps)\cdot F(\bu)$, we maintain a level-$(\beta-1)$ estimator for $F(\bw_2)-F(\bu)$, and again reveal the output once $\bw_2$ is reached, discarding the sketch afterward.

It is important to track progress using \emph{additive} steps of $\eps$ rather than geometric ones. 
For instance, if we used a single sketch to track an increase from $(1+2\eps)\cdot F(\bu)$ to $(1+4\eps)\cdot F(\bu)$, then exposing the estimate at $(1+3\eps)\cdot F(\bu)$ could compromise the sketch’s randomness, and if we withheld the output, we could lose the ability to form a $(1+\eps)$ approximation to $F$ at that point.

Instead, at each increment, we define $\bw_k$ such that $F(\bw_k)\approx(1+k\eps)\cdot F(\bu)$ and recursively estimate differences $F(\bw_k)-F(\bw_{k-1})$, each using a level-$\beta$ estimator if the difference is small.
Because the difference $F(\bw_k)-F(\bw_{k-1})$ is roughly $\eps F(\bu)$, we only need low-accuracy sketches.

To manage overlapping estimators, we leverage the binary representation of $k$ to determine which level-$j$ estimators to use. 
This effectively encodes a path through a binary tree over the stream, c.f., \figref{fig:tree}, where each estimator corresponds to a different block whose internal randomness is only revealed once. 
Thus, we require $\O{2^k}$ instances of level-$k$ estimators for each $k\in[\beta]$. 
Since each level-$k$ estimator only needs accuracy $(1+2^k\eps)$, the total space does not grow too rapidly. 
The resulting space dependence remains bounded by $\tO{\frac{1}{\eps^2}}$, achieving the desired efficiency. 
Finally, we recall that we require $\O{\log n}$ instances of this approach, to handle the geometric scales $2^i$, i.e., when the value of the function doubles. 

\paragraph{Space optimization.}
We can further optimize our framework using standard streaming techniques as in \cite{Ben-EliezerJWY22}:
\begin{enumerate}
\item 
Instead of storing $\O{\log n}$ instances of $\calA$, it suffices to retain only $\O{\log\frac{1}{\eps}}$ at any time, since we can discard prefixes contributing just a $\poly(\eps)$ fraction of $F(\bv)$.
\item 
If $\calA$ and $\calB$ satisfy the \emph{strong-tracking} property, then we only need to union bound over $\poly\left(\log n,\log\frac{1}{\eps}\right)$ sketches, instead of all $m$ possible prefixes (where $m$ is the length of the data stream).
\end{enumerate}
Note that our framework does not rely on existing sketches, so we must establish strong tracking for our difference estimators to fully utilize the second optimization.

For the ease of presentation, we present a slightly sub-optimal framework for adversarially robust streaming algorithms in \algref{alg:framework}. 
We clarify that within the context of \algref{alg:framework}, $\calA(s,t,\eta,\delta)$ is a stronger tracker for a function $F$ between a start time $s$, an end time $t$, with an accuracy parameter $\eta$, and a failure probability $\delta$. 
Similarly, $\calB(s,t_0,t,\gamma,\eta,\delta)$ is a difference estimator where the prefix vector is defined by the updates between start time $s$ and end time $t_0$, and the suffix vector is defined by the updates between start time $t_0+1$ and end time $t$. 
Moreover, the accuracy parameter is $\eta$ and the failure probability is $\delta$, while the difference is guaranteed to have a value that is at most a $\gamma$-factor of the value of the prefix. 
Finally, we remark that we shall show that $a=\O{2^b}$ and $b=\O{\log\frac{1}{\eps}}$. 

\begin{algorithm}[!htb]
\caption{Framework for Robust Algorithms on Insertion-Only Streams}
\alglab{alg:framework}
\begin{algorithmic}[1]
\Require{Stream $s_1,\ldots,s_t\in[n]$, accuracy parameter $\eps\in(0,1)$, oblivious strong tracker $\calA$ for a function $F$, $(\gamma,\eps,\delta)$-difference estimator $\calB$ for $F$}
\Ensure{Robust $(1+\eps)$-approximation to $F$}
\State{$\delta\gets\frac{1}{\poly\left(\frac{1}{\eps}, \log n\right)}$, $\eta\gets\frac{\eps}{1000\log\frac{1}{\eps}}$, $\beta\gets64\ceil{\log\frac{1}{\eps}}$}
\State{$c\gets 0$, $a\gets 0$, $\gamma_j\gets 2^{j-3}\eps$ for $j\in[\beta]$}
\For{each update $s_t\in[n]$, $t\in[m]$}
\If{there exists $b$ such that $\calB_{c,b}(1,t_{c,b},t,\gamma_b,\eta,\delta)>\frac{1}{16}\cdot\gamma_b\cdot\calA_c(1,t_c,\eta,\delta)$}
\State{Let $b$ be the largest such index}
\State{$a\gets\flr{\frac{a}{2^{b-1}}}\cdot 2^{b-1}+2^{b-1}$}
\For{all $i\in[b]$}
\State{$t'_{c,i}\gets t_{c,i}$, $t_{c,i}\gets t$}
\Comment{Update difference estimator times}
\EndFor
\EndIf
\If{$\calA_{c+1}(1,t,\eta,\delta)\ge 2\cdot\calA_{c}(1,t_c,\eta,\delta)$}
\linlab{lin:frame:init}
\Comment{Switch sketch at top layer}
\State{$c\gets c+1$, $t_c\gets t$, $a\gets 0$, $t_{c,i}\gets t$ for $i\in[\beta]$}
\EndIf
\State{Let $z_1<\ldots<z_k$ be the nonzero bits in the binary representation of $a$}
\Comment{Compile previous frozen components}
\State{\Return $\calA_{c}(1,t_c,\eta,\delta)+\sum_{i\in[k-1]}\calB_{c,z_i}(1,t'_{c,z_{i+1}},t'_{c,z_i},\eta,\delta)$}
\Comment{Compute estimator $X$ for $F$ using revealed sketches}
\EndFor
\end{algorithmic}
\end{algorithm}

We begin by establishing the correctness of \algref{alg:framework} at the time steps where the counter~$a$ increments, which intuitively corresponds to times between which the function doubles. 
This, in turn, allows us to restrict our attention to time steps $t\in(t_i,t_{i+1})$, that is, the intervals between successive increments of the counter.
\begin{lemma}[Correctness when $F$ doubles]
\lemlab{lem:constant}
\cite{WoodruffZ21}
Let $\eps\in\left(0,\frac{1}{2}\right)$ and let $F$ be a monotonic function with $(\eps,m)$-flip number $\lambda=\O{\frac{1}{\eps}\log n}$ for $\log m=\O{\log n}$. 
For any integer $i>0$, let $t_i\in[m]$ be the step at which the counter $c$ in \algref{alg:framework} is first set to $i$. 
Then with probability at least $1-\lambda\delta$, \algref{alg:framework} outputs a $\left(1+\frac{\eps}{100\log\frac{1}{\eps}}\right)$-approximation to $F$ at all times $t_i$. 
\end{lemma}
\begin{proof}
Let $\calE$ be the event that all $\lambda$ instances of $\calA$ are correct at all times $t\in[m]$ for an oblivious stream. 
By the correctness of $\calA$, we have that $\PPr{\calE}\ge1-\lambda\delta$. 

Since the counter $c$ in \algref{alg:framework} is first set to $i$ at step $t_i$, then all previous outputs of \algref{alg:framework} have not used the subroutine $\calA_i$, so that the input at time $t_i$ is independent of the randomness of $\calA_i$. 
Conditioned on $\calE$, $\calA_i$ outputs a $(1+\eta)$-approximation to $F$ at time $t_i$, with probability $1-\delta$, where $\eta=\frac{\eps}{100\log\frac{1}{\eps}}$. 
Moreover, for $c$ to increase, the output of $\calA_i$ has to be at least $2^c$. 
Hence conditioned on $\calE$, each time the value of the counter $c$ has increased, the value of $F$ must have increased by at least $\frac{3}{2}$. 

Because $F$ has flip number $\lambda=\O{\frac{1}{\eps}\log n}$ for $\eps<\frac{1}{2}$, then there are at most $\lambda$ update times $t_i$ in which the counter $c$ is increased. 
By a union bound, \algref{alg:framework} outputs a $\left(1+\frac{\eps}{100\log\frac{1}{\eps}}\right)$-approximation to $F$ at all times $t_i$, conditioned on $\calE$. 
Since we can recall that $\PPr{\calE}\ge1-\lambda\delta$, then the desired claim follows. 
\end{proof}
We next upper and lower bound the change in the value of the function between when a difference estimator is initialized and when its output is incorporated into the global estimate. 
\begin{lemma}[Bounds on inputs to difference estimator]
\lemlab{lem:diff:est:bounds}
\cite{WoodruffZ21}
With probability at least $1-\delta\cdot\poly\left(\frac{1}{\eps},\log n\right)$, we have $\frac{\gamma_{z_i}}{256}\cdot F(1,t'_{c,z_{i+1}})\le F(1,t'_{c,z_i})-F(1,t'_{c,z_{i+1}})\le\gamma_{z_i}\cdot F(1,t'_{c,z_{i+1}})$ for all defined indices $c,i$. 
\end{lemma}
\begin{proof}
Let $c$ and $a$ be fixed in \algref{alg:framework}, so that $z_i$ corresponds to a nonzero bit in the binary representation of $a$. 
Let $\calE_1$ be the event that all instances of $\calA$ succeed, so that by \lemref{lem:constant}, $\calE_1$ holds with probability at least $1-\lambda\delta$. 
Let $\calE_2$ be the event that all instances of $\calB$ succeed, provided the input satisfies the conditions of the difference estimators, so that $\PPr{\calE_2}\ge1-\delta\cdot\poly\left(\frac{1}{\eps},\log n\right)$ by a union bound. 
Let $\calE$ be the event that both $\calE_1$ and $\calE_2$ occur so that $\PPr{\calE}\ge1-\delta\cdot\poly\left(\frac{1}{\eps},\log n\right)$ by a union bound over $\calE_1$ and $\calE_2$. 
We thus condition on $\calE$. 

Observe that a bit $b=z_i$ in the binary representation of $a$ can only be set to $1$ if at some time $\calB_{c,b}(1,t_{c,b},t,\gamma_b,\eta,\delta)>\frac{1}{16}\cdot\gamma_b\cdot\calA_c(1,t_c,\eta,\delta)$ but $\calB_{c,b+1}(1,t_{c,b+1},t,\gamma_{b+1},\eta,\delta)\le\frac{1}{16}\cdot\gamma_{b+1}\cdot\calA_c(1,t_c,\eta,\delta)$. 
Since $t_{c,b+1}\le t_{c,b}$, then $F(1,t_{c,b+1})\ge F(1,t_{c,b})$. 
Because $\eta\le\gamma_b$ for all $b$, then the resulting additive error implies that each algorithm $\calB$ is a $2$-approximation to the difference. 
Thus the above conditions imply we must have $\frac{1}{16}\cdot\gamma_b\cdot\calA_c(1,t_c,\eta,\delta)<\calB_{c,b}(1,t_{c,b},t,\gamma_b,\eta,\delta)\le\frac{1}{4}\cdot\gamma_b\cdot\calA_c(1,t_c,\eta,\delta)$. 

Observe that the algorithm then sets $t'_{c,b}$ to be the previous $t_{c,b}$ and updates the new value of $t_{c,b}$ to be the time $t$. 
Thus, we have $t'_{c,b}=t'_{c,z_i}$ and moreover, from the binary representation of $a$, we have $t_{c,b}$ is not further updated. 
Hence, $t'_{c,z_{i+1}}=t_{c,b}$. 
By the correctness of $\calB$, we have $F(1,t'_{c,z_i})-F(1,t'_{c,z_{i+1}})\le 2\calB_{c,b}(1,t_{c,b},t_{c,b},\gamma_b,\eta,\delta)$. 
Similarly, by the correctness of $\calA$, we have $\calA_c(1,t_c,\eta,\delta)\le 2F(1,t_c)\le2F(1,t'_{c,z_{i+1}})$. 
Moreover by the correctness of $\calA$, we have $F(1,t'_{c,z_i})-F(1,t'_{c,z_{i+1}})\le F(1,t)$ since $\calA_c(1,t_c,\eta,\delta)$
Putting these inequalities together, we have
\[F(1,t'_{c,z_i})-F(1,t'_{c,z_{i+1}})\le\gamma_{z_i}\cdot F(1,t'_{c,z_{i+1}}),\]
which proves the upper bound.

For the lower bound, we have from above that $\frac{1}{16}\cdot\gamma_b\cdot\calA_c(1,t_c,\eta,\delta)<\calB_{c,b}(1,t_{c,b},t,\gamma_b,\eta,\delta)$. 
By the correctness of $\calB$, we have $F(1,t'_{c,z_i})-F(1,t'_{c,z_{i+1}})\ge\frac{1}{2}\calB_{c,b}(1,t_{c,b},t_{c,b},\gamma_b,\eta,\delta)$. 
By the conditions of the algorithm, we also must have $\calA_{c+1}(1,t,\eta,\delta)\le 2\calA_c(1,t_c,\eta,\delta)$. 
Similarly, by the correctness of $\calA$, we have $\calA_{c+1}(1,t,\eta,\delta)\ge 2F(1,t)\ge 2F(1,t'_{c,z_{i+1}})$. 
Putting these inequalities together, we have
\[F(1,t'_{c,z_i})-F(1,t'_{c,z_{i+1}})\ge\frac{\gamma_{z_i}}{16}\cdot F(1,t'_{c,z_{i+1}}),\]
which proves the lower bound.
\end{proof}
We now show correctness of the framework on non-adaptive streams. 
\begin{lemma}[Correctness on non-adaptive streams]
\lemlab{lem:frame:correct}
\cite{WoodruffZ21}
With probability at least $1-\delta\cdot\poly\left(\frac{1}{\eps},\log n\right)$, \algref{alg:framework} outputs a $(1+\eps)$-approximation to $F$ at all times.
\end{lemma}
\begin{proof}
Consider a fixed value of $c$ in  \algref{alg:framework}. 
As before, let $\calE_1$ be the event that all instances of $\calA$ succeed, so that by \lemref{lem:constant}, $\calE_1$ holds with probability at least $1-\lambda\delta$. 
Similarly, let $\calE_2$ be the event that all instances of $\calB$ succeed, provided the input satisfies the conditions of the difference estimators, so that $\PPr{\calE_2}\ge1-\delta\cdot\poly\left(\frac{1}{\eps},\log n\right)$ by a union bound. 
Let $\calE$ be the event that both $\calE_1$ and $\calE_2$ occur, so that $\PPr{\calE}\ge1-\delta\cdot\poly\left(\frac{1}{\eps},\log n\right)$ by a union bound over $\calE_1$ and $\calE_2$. 
We thus condition on $\calE$ and show correctness between $t_c$ and $t_{c+1}$. 

We fix a value of $a$ and consider its binary representation $z_1<\ldots<z_k$. 
By \lemref{lem:diff:est:bounds}, we have for all $i\in[k]$, $F(1,t'_{c,z_i})-F(1,t'_{c,z_{i+1}})\le\gamma_{z_i}\cdot F(1,t'_{c,z_{i+1}})$ with probability at least $1-\delta\cdot\poly\left(\frac{1}{\eps},\log n\right)$. 
We also have $t'_{c,z_1}=t_{c,1}$ and $\calB_{c,1}(1,t_{c,1},t,\gamma_1,\eta,\delta)\le\frac{1}{16}\cdot\gamma_1\cdot\calA_c(1,t_c,\eta,\delta)$ by construction of the algorithm. 
We decompose 
\[F(1,t)=\left(F(1,t)-F(1,t'_{c,z_1})\right)+\sum_{i=1}^k\left(F(1,t'_{c,z_i})-F(1,t'_{c,z_{i+1}})\right).\]
Observe that the input up to time $t'_{c,z_i}$ for a difference estimator $\calB_{c,z_i}(1,t'_{c,z_{i+1}},t'_{c,z_i},\eta,\delta)$ is independent of the internal randomness of the difference estimator, since $t'_{c,z_i}$ is the first time that the output of the algorithm is revealed. 
Thus by an argument similar to the sketch-switching framework, each estimator $\calB_{c,z_i}(1,t'_{c,z_{i+1}},t'_{c,z_i},\eta,\delta)$ reports additive error $\eta\cdot F(1,t'_{c,z_{i+1}})\le2\eta F(1,t)$, since $F(1,t'_{c,z_i})-F(1,t'_{c,z_{i+1}})\le\gamma_{z_i}\cdot F(1,t'_{c,z_{i+1}})$ for all $i\in[k]$. 
Therefore, the total error is at most
\[\gamma_1\cdot\calA_c(1,t_c,\eta,\delta)+\sum_{i=1}^k\left(2\eta F(1,t)\right).\]
Observe that $\calA_c(1,t_c,\eta,\delta)\le 2F(1,t)$ and $\gamma_1\le\frac{\eps}{4}$. 
Moreover, we have $k\le64\log\frac{1}{\eps}$ and $\eta=\frac{\eps}{1000\log\frac{1}{\eps}}$. 
Therefore, the total error is at most $\eps\cdot F(1,t)$. 
In other words, \algref{alg:framework} outputs a $(1+\eps)$-approximation to $F$ at all times.
\end{proof}
Given the correctness on non-adaptive streams, it remains to show correctness on adaptive inputs by proving that each time the internal randomness of a subroutine is revealed to the adversary, the subroutine is never used again. 
\begin{theorem}[Framework for adversarially robust algorithms on insertion-only streams]
\thmlab{thm:framework}
\cite{WoodruffZ21}
Let $\eps\in\left(0,\frac{1}{2}\right)$ be an accuracy parameter, $\delta\in(0,1)$ be a failure probability, and $F$ be a monotonic function with $(\eps,m)$-flip number $\lambda=\O{\frac{\log n}{\eps}}$ on a stream of length $m$, with $\log m=\O{\log n}$. 
Suppose there exists a $(\gamma,\eps,\delta)$-difference estimator for $F$ that uses $\frac{\gamma}{\eps^2}\cdot S(n,\delta,\eps)$ bits of space and a strong tracker for $F$ that use $\O{\frac{1}{\eps^2}\cdot S(n,\delta,\eps)}$ bits of space. 
Then there exists an adversarially robust streaming algorithm that outputs a $(1+\eps)$-approximation for $F$ that succeeds with constant probability, using $\frac{1}{\eps^2}\cdot S(n,\delta',\eps')\cdot\log n\cdot\log\frac{1}{\eps}$ bits of space, where $\eps'=\O{\frac{\eps}{\log\frac{1}{\eps}}}$ and $\delta'=\O{\frac{1}{\poly\left(\frac{1}{\eps},\,\log n\right)}}$. 
\end{theorem}
\begin{proof}
Consider \algref{alg:framework} and observe that \lemref{lem:frame:correct} proves correctness for the framework. 
Thus it remains to analyze the space complexity of \algref{alg:framework}. 
By assumption, each $(\gamma,\eps,\delta)$ difference estimator $\calB$ uses $\frac{\gamma}{\eps^2}\cdot S(n,\delta,\eps)$ bits of space. 

Consider a fixed $c$. 
By \lemref{lem:diff:est:bounds}, it follows that there are at most $\O{\frac{1}{\gamma_i}}$ instances of $(\gamma_i,\eta,\delta')$ difference estimator $\calB$ between times $t_c$ and $t_{c+1}$. 
Therefore, the total space of the difference estimators between times $t_c$ and $t_{c+1}$ is
\[\sum_{i\in[\beta]}\O{\frac{1}{\gamma_i}}\cdot\frac{\gamma_i}{\eps^2}\cdot S(n,\delta',\eta)=\O{\frac{\beta}{\eps^2}}\cdot S(n,\delta',\eta).\]
Similarly, the space used by $\calA_c$ is $\O{\frac{1}{\eps^2}\cdot S(n,\delta',\eta)}$, so the total space used between times $t_c$ and $t_{c+1}$ is $\O{\frac{\beta}{\eps^2}}\cdot S(n,\delta',\eta)$. 

Finally, note that the value of $\calA_c$ increases by at least a factor of two each time the counter $c$ increases. 
Since the $(\eps,m)$-flip number of $F$ is $\O{\frac{1}{\eps}\log n}$, then the value of the counter $c$ satisfies $c=\O{\log n}$ at the end of the stream for $\log m=\O{\log n}$. 
Hence, the total space usage is $\O{\frac{\beta}{\eps^2}}\cdot S(n,\delta',\eta)\cdot\log n$.
We have $\eps'=\eta$ and $\beta=\O{\log\frac{1}{\eps}}$. 
Therefore, the total space is $\frac{1}{\eps^2}\cdot S(n,\delta',\eps')\cdot\log n\cdot\log\frac{1}{\eps}$. 
\end{proof}

\subsection{Difference Estimator for \texorpdfstring{$F_2$}{F2} Estimation}
\seclab{sec:diff:est:F2}
In this section, we describe the construction of a difference estimator for $F_2$ estimation by \cite{WoodruffZ21}. 
To that end, we estimate $F_2(\bv) - F_2(\bu)$ by expressing $F_2(\bv)$ as $F_2((\bv - \bu) + \bu) = F_2(\bv - \bu) + 2\langle \bv - \bu, \bu \rangle + F_2(\bu)$. 
Consequently, when $F_2(\bv) - F_2(\bu) \le \gamma F_2(\bu)$, a multiplicative $\left(1 + \frac{\eps}{\gamma}\right)$ approximation to the difference $F_2(\bv) - F_2(\bu)$ implies an additive approximation with error at most $\eps \cdot F_2(\bu)$. 
We next argue that if a streaming algorithm $\calA$ produces a $(1 + \eps)$ approximation to $F_2(\bu)$ and a $\left(1 + \frac{\eps}{\sqrt{\gamma}}\right)$ approximation to $F_2(\bv - \bu)$, then it can approximate the inner product $\langle \bv - \bu, \bu \rangle$ up to additive error $\frac{\eps}{\sqrt{\gamma}} \cdot \|\bv - \bu\|_2 \|\bu\|_2 \le \eps \cdot F_2(\bu)$, assuming $F_2(\bv - \bu) \le \gamma \cdot F_2(\bu)$ and noting that $F_2(\bu) = \|\bu\|_2^2$.
Therefore, for the case $p = 2$, algorithm~$\calB$ can be implemented directly using the sketches produced by $\calA$, with space complexity scaling as $\frac{\gamma}{\eps^2}$.

We first show that the expected product of two sketched dot products is exactly the dot product. 
\begin{lemma}
\lemlab{lem:ams:ex}
\cite{WoodruffZ21}
Let $\bs\in\{-1,+1\}^n$ be a random sign vector with pair-wise independent entries. 
For any vectors $\bu,\bv\in\mathbb{R}^n$, 
\[\Ex{\langle \bs,\bu\rangle\cdot\langle \bs,\bv\rangle}=\langle \bu,\bv\rangle.\]
\end{lemma}
\begin{proof}
By linearity of expectation, 
\begin{align*}
\Ex{\langle \bs,\bu\rangle\cdot\langle \bs,\bv\rangle}&=\Ex{\left(\sum_{i\in[n]}s_iu_i\right)\left(\sum_{i\in[n]}s_iv_i\right)}\\
&=\Ex{\sum_{i\in[n]}\sum_{j\in[n]} s_iu_is_jv_j}=\sum_{i\in[n]}\sum_{j\in[n]}\Ex{s_iu_is_jv_j}.
\end{align*}
The random variables $s_i\in\{-1,+1\}$ are $4$-wise independent. 
Therefore, $\Ex{s_is_j}=1$ for $i=j$ and $\Ex{s_is_j}=0$ for $i\neq j$. 
Thus, 
\[\Ex{\langle \bs,\bu\rangle\cdot\langle \bs,\bv\rangle}=\sum_{i\in[n]}u_iv_i=\langle \bu,\bv\rangle.\]
\end{proof}
Next, we upper bound the variance of the product of two sketched dot products by roughly the square of the dot product. 
\begin{lemma}
\lemlab{lem:ams:var}
\cite{WoodruffZ21}
Let $\bs\in\{-1,+1\}^n$ be a random sign vector with four-wise independent entries.
For any $\bu,\bv\in\mathbb{R}^n$, 
\[\Var\left(\langle \bs,\bu\rangle\cdot\langle \bs,\bv\rangle\right)\le2\|\bu\|_2^2\|\bv\|_2^2.\]
\end{lemma}
\begin{proof}
Since $\Var{\langle \bs,\bu\rangle\cdot\langle \bs,\bv\rangle}\le\Ex{\left(\langle \bs,\bu\rangle\cdot\langle \bs,\bv\rangle\right)^2}$, it follows by linearity of expectation that
\begin{align*}
\Var\left(\langle \bs,\bu\rangle\cdot\langle \bs,\bv\rangle\right)&\le\Ex{\left(\sum_{i\in[n]}s_iu_i\right)^2\left(\sum_{i\in[n]}s_iv_i\right)^2}\\
&=\Ex{\sum_{i\in[n]}\sum_{j\in[n]}\sum_{k=1}^n\sum_{\ell=1}^n s_is_js_ks_\ell u_iv_ju_kv_\ell}\\
&=\sum_{i\in[n]}\sum_{j\in[n]}\sum_{k\in[n]}\sum_{\ell\in[n]}\Ex{s_is_js_ks_\ell u_iv_ju_kv_\ell}.
\end{align*}
The random variables $s_i\in\{-1,+1\}$ are $4$-wise independent. 
Hence, $\Ex{s_is_js_ks_\ell}=1$ if $i,j,k,\ell$ consists of two (possibly not distinct) pairs of indices. 
Otherwise, $\Ex{s_is_js_ks_\ell}=0$. 
Thus, 
\[\Var\left(\langle \bs,\bu\rangle\cdot\langle \bs,\bv\rangle\right)\le\sum_{i\in[n]}\sum_{j\in[n]} u_iv_iu_jv_j\le2\|\bu\|_2^2\|\bv\|_2^2.\]
\end{proof}
Putting these together, we have the following guarantees for our difference estimator at a particular time. 
Here, ``four-wise independent'' means that any subset of four entries in the matrix is independent, i.e., it is not required that rows are fully independent of one another.
\begin{lemma}[AMS $F_2$ approximation gives inner product approximation]
\cite{WoodruffZ21}
\lemlab{lem:approx:ip}
Let vectors $\bu,\bv\in\mathbb{R}^n$ and $\bM\in\mathbb{R}^{d\times n}$ be a sketching matrix so that each entry $M_{i,j}$ of $\bM$ is a four-wise independent random sign scaled by $\frac{1}{\sqrt{d}}$ for $d=\O{\frac{1}{\eps^2}}$. 
Then
\[\Pr[|\langle \bu,\bv\rangle-\langle \bM\bu, \bM\bv\rangle|\le\eps\|\bu\|_2\|\bv\|_2]\ge\frac{2}{3}.\]
\end{lemma}
\begin{proof}
The proof is a standard variance reduction argument, which follows from Chebyshev's inequality, given the expectationa and variance calculations in  \lemref{lem:approx:ip:all} and \lemref{lem:ams:ex}. 
In particular, $\bM$ can be viewed as taking the arithmetic mean of $d$ scaled sign vectors $\bs_1,\ldots,\bs_d$. 
\end{proof}

We also recall the following $F_2$ strong tracker. 
\thmstrongFtwo*

\begin{lemma}[Strong tracker for inner product]
\lemlab{lem:approx:ip:all}
\cite{WoodruffZ21}
Let $\mathbf{0}^n\preceq \bv_1\preceq \bv_2\preceq\ldots\preceq \bv_m\in\mathbb{R}^n$ be vectors with polynomially bounded entries. 
There exists an algorithm that uses a sketching matrix $\bM\in\mathbb{R}^{d\times n}$ with $d=\O{\frac{1}{\eps^2}\left(\log\frac{1}{\eps}+\log\frac{1}{\delta}+\log\log n\right)}$ such that for $m=\poly(n)$ and a fixed $\bu\in\mathbb{R}^n$ with $\bu\succeq\mathbf{0}^n$, 
\[\left\lvert\langle \bu,\bv_i\rangle-\langle \bM\bu,\bM\bv_i\rangle\right\rvert|\le\eps\|\bu\|_2\|\bv_i\|_2,\]
simultaneously for all $i\in[m]$ with probability at least $1-\delta$. 
\end{lemma}
\begin{proof}
Let $t_0$ be the start of the data stream that defines the sequence of vectors $\bv_1,\bv_2,\ldots$. 
Let $t_1<t_2<\ldots<t_q$ be a sequence of indices so that $t_{i+1}$ is the minimal index with $\langle \bu,\bv_{t_{i+1}}\rangle\ge\left(1+\frac{\eps^4}{100}\right)\langle \bu,\bv_{t_i}\rangle$. 
Note that $q=\poly\left(\frac{1}{\eps},\log n\right)$ for $m=\poly(n)$, since each vector $\bv_i$ has polynomially bounded entries. 
Observe that both $\|\bv_1\|_2\le\|\bv_2\|_2\le\ldots\le\|\bv_m\|_2$ and $\langle \bu,\bv_1\rangle\le\langle \bu,\bv_2\rangle\le\ldots\le\langle \bu,\bv_m\rangle$ holds, since $\mathbf{0}^n\preceq \bv_1\preceq \bv_2\preceq\ldots\preceq \bv_m\in\mathbb{R}^n$. 
Therefore, it suffices to show 
\[\left\lvert\langle \bu,\bv_{t_i}\rangle-\langle \bM\bu,\bM\bv_{t_i}\rangle\right\rvert\le\eps\|\bu\|_2\|\bv_{t_i}\|_2,\]
for all $i\in[q]$. 
By \lemref{lem:ams:ex} and \lemref{lem:ams:var} and Chebyshev's inequality, for a sketching matrix $\bM$ with $\O{\frac{1}{\eps^2}}$ rows, we have
\[\Pr[\left\lvert\langle \bu,\bv_{t_i}\rangle-\langle \bM\bu,\bM\bv_{t_i}\rangle\right\rvert\le\|\bu\|_2\eps\|\bv_{t_i}\|_2]\ge\frac{2}{3}.\]
Thus if $\bM$ has $d=\O{\frac{1}{\eps^2}\left(\log\frac{1}{\eps}+\log\frac{1}{\delta}+\log\log n\right)}$ rows, we can use a standard median-of-means approach and take a union bound over all $i\in[q]$ to get 
\[\left\lvert\langle \bu,\bv_i\rangle-\langle \bM\bu,\bM\bv_i\rangle\right\rvert\le\eps\|\bu\|_2\|\bv_i\|_2,\]
simultaneously for all $i\in[m]$ with probability at least $1-\delta$. 
\end{proof}

\begin{lemma}[Corollary 17 in \cite{BravermanCIW16}]
\lemlab{lem:f2:chaining}
Let $\bs\in\{-1,+1\}^n$ be a random sign vector with $8$-wise independent entries. 
Let $\mathbf{0}^n\preceq \bx^{(1)}\preceq\ldots\preceq \bx^{(m)}=\bx$. 
There exists a universal constant $C$ such that $\Ex{\sup_t|\langle \bs,\bx^{(t)}\rangle}\le C\cdot\|\bx\|_2$. 
\end{lemma}

We now give the $F_2$ difference estimator via the inner product approximation property. 
\begin{lemma}[$F_2$ difference estimator]
\lemlab{lem:diff:est:F2}
\cite{WoodruffZ21}
There exists a $(\gamma,\eps,\delta)$-\emph{difference estimator} for $F_2$ that uses 
\[\O{\frac{\gamma\log n}{\eps^2}\left(\log\log n+\log\frac{1}{\eps}+\log\frac{1}{\delta}\right)}\]
bits of space.
\end{lemma}
\begin{proof}
Given an oblivious stream $S$, let $\bu$ be the frequency vector induced by the updates of $S$ from time $t_1$ to $t_2$, and $\bv$ be the frequency vector induced by updates from time $t_2+1$ to $t$, and suppose $F_2(\bv)\le\gamma\cdot F_2(\bu)$. 
We define vectors $\bv^{(1)},\bv^{(2)},\ldots$ so that $\bv^{(i)}$ is the first time at which the difference exceeds $\frac{Ci\eps^4}{100}\cdot F_2(\bu)$, i.e., $F_2(\bu+\bv^{(i)})-F_2(\bu)\ge \frac{Ci\eps^4}{100}\cdot F_2(\bu)$, where $C$ is a sufficiently small constant to be fixed. 
Note that we have at most $\poly\left(\frac{\gamma}{\eps}\right)$ such times. 
We first show correctness at these times. 

Let $\bM\in\{-1,+1\}^{d\times n}$ be a sketching matrix, where each entry is a random sign with $4$-wise independence, with 
\[d=\O{\frac{\gamma}{\eps^2}\left(\log\log n+\log\frac{1}{\eps}+\log\frac{1}{\delta}\right)}.\] 
For a fixed time $t^{(i)}$, let $\bv$ be the frequency vector formed by the updates from time $t_2+1$ to $t^{(i)}$. 

By \lemref{lem:approx:ip} for sufficiently large constant in the dimension $d$, with probability at least $1-\frac{\delta}{2}$,
\[|\langle \bu,\bv\rangle-\langle \bM\bu,\bM\bv\rangle|\le\frac{\eps}{8\sqrt{\gamma}}\|\bu\|_2\|\bv\|_2.\]
Recall the identity 
\[F_2(\bu+\bv)-F_2(\bu)=2\langle \bu,\bv\rangle+\|\bv\|_2^2.\]
Therefore,
\begin{align*}
\big|\big[F_2(\bu+\bv)&-F_2(\bu)\big]-\left[2\langle \bM\bu,\bM\bv\rangle+\|\bM\bv\|_2^2\right]\big|\\
&=\left\lvert\left[2\langle \bu,\bv\rangle-2\langle \bM\bu,\bM\bv\rangle\right]+\left[\|\bv\|_2^2-\|\bM\bv\|_2^2\right]\right\rvert.
\end{align*}
Since $\bM$ also serves as a sketching matrix for $F_2$, then for sufficiently large dimension $d$, we have $\left\lvert\|\bv\|_2^2-\|\bM\bv\|_2^2\right\rvert\le\frac{\eps}{8}\cdot\|\bv\|_2^2$. 
Hence,
\begin{align*}
\big|\lvert\big[F_2(\bu+\bv)&-F_2(\bu)\big]-\left[2\langle \bM\bu,\bM\bv\rangle+\|\bM\bv\|_2^2\right]\big|\\
&\le\frac{\eps}{4\sqrt{\gamma}}\|\bu\|_2\|\bv\|_2+\frac{\eps}{8}\|\bv\|_2^2.
\end{align*}
Because $F(\bv)\le\gamma\cdot F(\bu)$, then $\|\bv\|_2\le\sqrt{\gamma}\cdot\|\bu\|_2$. 
Thus for $\gamma\le 2$, the total error of the estimator is at most 
\[\frac{\eps}{4}\|\bu\|_2^2+\frac{\eps}{8}\|\bv\|_2^2\le\frac{\eps}{4}\|\bu\|_2^2+\frac{\eps}{4}\|\bu\|_2^2\le\eps\cdot F_2(\bu).\]
Correctness at all times $t^{(1)},t^{(2)},\ldots$ then follows from setting the failure probability in each matrix $M$ to be $\frac{\delta}{\poly\left(\frac{\gamma}{\eps}\right)}$, since there are at most $\poly\left(\frac{\gamma}{\eps}\right)$ such times. 

It remains to consider the correctness of the difference estimator at times $t\in(t^{(i)},t^{(i+1)})$. 
To that end, note that the difference can increase by at most $C\eps^4\cdot F_2(\bu)$ at such times. 
Let the vector $\bw^{(t)}$ denote the vector induced by the updates between $(t^{(i)},t]$ for any $t<t^{(i+1)}$, so that $\|\bw^{(t)}\|_2\le\sqrt{C}\eps^2\cdot\|\bu\|_2$. 
By \lemref{lem:f2:chaining}, we have that 
\[\Ex{\sup_t\left\lvert\langle \bs,\bw^{(t)}\rangle\right\rvert}\le\O{\sqrt{C}\eps^2\cdot\|\bu\|_2}.\]
Hence for sufficiently small $C$, we have by Markov's inequality,
\[\PPr{\sup_t\left\lvert\langle \bs,\bw^{(t)}\rangle\right\rvert\le\frac{\eps^2}{100}\cdot\|\bu\|_2}\ge\frac{2}{3}.\]
Now taking the median of $\O{\log\frac{\gamma}{\eps\delta}}$ such instances suffices to obtain correctness over the entire stream. 
In particular, the above argument shows that we have correctness at the $\poly\left(\frac{\gamma}{\eps}\right)$ times $t^{(1)},t^{(2)},\ldots$ while a subsequent union bound gives correctness at all times between each $t^{(i)}$ and $t^{(i+1)}$.   

It remains to analyze the space complexity. 
By \thmref{thm:strong:F2}, each of the two instances of the $F_2$-estimation algorithm uses
\[\O{\frac{\gamma\log n}{\eps^2}\left(\log\frac{1}{\eps}+\log\frac{1}{\delta}+\log\log n\right)}\]
bits of space. 
The matrix $\bM$ must also maintain 
\[d=\O{\frac{\gamma}{\eps^2}\left(\log\log n+\log\frac{1}{\eps}+\log\frac{1}{\delta}\right)}\]
rows. 
Hence, the overall space complexity of the $(\gamma,\eps,\delta)$-difference estimator for $F_2$ is 
\[\O{\frac{\gamma\log n}{\eps^2}\left(\log\log n+\log\frac{1}{\eps}+\log\frac{1}{\delta}\right)}.\]
\end{proof}

\begin{theorem}
\cite{WoodruffZ21}
Let $\eps\in\left(0,\frac{1}{2}\right)$ be an accuracy parameter. 
There exists an adversarially robust streaming algorithm that outputs a $(1+\eps)$-approximation for $F_2$ moment estimation that succeeds with probability at least $\frac{2}{3}$ and uses $\tO{\frac{1}{\eps^2}\log^2 n}$ bits of space. 
\end{theorem}
\begin{proof}
Observe that $F_2$ is a monotonic function with $(\eps,m)$-flip number $\lambda=\O{\frac{1}{\eps}\log n}$. 
By \lemref{lem:diff:est:F2} and \thmref{thm:strong:F2}, there exists a $(\gamma,\eps,\delta)$-difference estimator that uses 
\[\O{\frac{\gamma\log n}{\eps^2}\left(\log\log n+\log\frac{1}{\eps}+\log\frac{1}{\delta}\right)}\]
bits of space, as well as an oblivious strong tracker that uses space
\[\O{\frac{\log n}{\eps^2}\left(\log\frac{1}{\eps}+\log\frac{1}{\delta}+\log\log n\right)}.\]
Hence, the framework of \algref{alg:framework} implies through \thmref{thm:framework} that there exists an adversarially robust streaming algorithm that outputs a $(1+\eps)$-approximation for the $F_2$ moment, while using $\frac{1}{\eps^2}\log^2 n\cdot\polylog\left(\log n,\frac{1}{\eps}\right)$ bits of space, and succeeds with probability at least $\frac{2}{3}$.    
\end{proof}

\paragraph{Optimized $F_2$ algorithm.}
We now briefly describe how to optimize the adversarially robust $F_2$ algorithm. 
The framework in \algref{alg:framework} tracks the active instances $\calA_c$ and $\calB_{c,b}$ for a counter $c$. 
Instead of maintaining all sketches $\calA_c$ and $\calB_{c,b}$ simultaneously, observe that it suffices to maintain sketches $\calA_c$ and $\calB_{c,b}$ for $\O{\log\frac{1}{\eps}}$ values of $c$ at a time. 

Since the output increases by a constant factor each time the counter $c$ increments, it suffices to maintain the sketches $\calA_i$ and $\calB_{i,b}$ for only the smallest $\O{\log\frac{1}{\eps}}$ values of $i$ that are at least $c$. 
In particular, any larger index will have only missed $\O{\eps}$ fraction of the $F_2$ moment of the stream prior to the time when the algorithms are initialized. 
Therefore, the resulting output is still a $(1+\eps)$-approximation. 

\begin{theorem}[Adversarially robust $F_2$ streaming algorithm]
\thmlab{thm:robust:opt:F2}
\cite{WoodruffZ21}
Given $\eps>0$, there exists an adversarially robust streaming algorithm that, with probability at least $\frac{2}{3}$, returns a $(1+\eps)$-approximation to $F_2$ using $\O{\frac{1}{\eps^2} \log n \log^4 \frac{1}{\eps} \left( \log \frac{1}{\eps} + \log \log n \right)}$ bits of space.
\end{theorem}
\begin{proof}
At any moment during the stream, there are $\O{\log \frac{1}{\eps}}$ active indices $a$ and $\O{1}$ active indices $c$, which correspond to the sketches $\calB_a$ and $\calA_{a,c}$, respectively. 
From \thmref{thm:framework}, recall that for any fixed index $a$, the total space used by the sketches $\calA_{a,j}$ across the $\beta$ granularities is bounded by
\[\O{\frac{1}{\eps^2} \log^3 \frac{1}{\eps} \cdot S_1(n, \delta', \eps) + \frac{1}{\eps} \log \frac{1}{\eps} \cdot S_2(n, \delta', \eps)},\]
where $S_1(n, \delta', \eps) = \log n \left( \log \frac{1}{\eps} + \log \frac{1}{\delta'} + \log \log n \right)$, and $S_2 = 0$ for our $F_2$ difference estimator and strong tracker.

While \thmref{thm:framework} assumes we maintain up to $\O{\log n}$ instances indexed by $a$ (thus requiring $\O{\log n}$ simultaneous $\calA_{i,j}$ instances), it turns out that maintaining only $\O{\log \frac{1}{\eps}}$ active $i$ values at any time suffices. 
Since there are still $\O{\log n}$ total values of $a$ over the stream’s duration, we require each sketch to have failure probability at most $\frac{\delta}{\poly\left(\log n, \frac{1}{\eps}\right)}$ to ensure that the overall failure probability remains bounded by $\delta = \frac{2}{3}$. 

In total, we have $\O{\log \frac{1}{\eps}}$ active $a$ indices, which amount to $\O{\frac{1}{\eps}}$ running subroutines. 
Therefore, storing the splitting times for each of these $\O{\frac{1}{\eps} \log \frac{1}{\eps}}$ subroutines across a stream of length $m$ (where $\log m = \O{\log n}$) requires an additional $\O{\frac{1}{\eps} \log n \log \frac{1}{\eps}}$ bits of space. 

Combining these bounds, the total space usage is 
\[\O{\frac{1}{\eps^2} \log n \log^4 \frac{1}{\eps} \left( \log \frac{1}{\eps} + \log \log n \right)}.\]
\end{proof}

\paragraph{Heavy-hitters.}
As a straightforward corollary, we observe that our framework also applies to the $L_2$-heavy hitters problem, where the goal is to identify the ``frequent'' items of a data stream. 
Formally, the problem is defined as follows:

\begin{definition}[$L_2$-heavy hitters]
\deflab{def:L2:HH}
Given a frequency vector $\bx\in\mathbb{R}^n$, the goal is to output a set $S\subset[n]$ such that with probability $1-\delta$, $i\in S$ for every $i \in [n]$ such that $|x_i| \geq \eps \|\bx\|_2$ and $i\notin S$ for every $i\in[n]$ such that $|x_i|\le\frac{\eps}{2}\cdot\|\bx\|_2$. 
\end{definition}
We remark that the problem is also often stated as producing a frequency vector $\widehat{\bx}\in\mathbb{R}^n$ such that $|\bx-\widehat{\bx}|_\infty\le\eps\cdot\|\bx\|_2$, since by scaling the error parameter $\eps$ to say $\frac{\eps}{4}$, then there is additive error $\frac{\eps}{4}\cdot\|\bx\|_2$ to each coordinate $i\in[n]$, which suffices to differentiate whether $x_i\ge\eps\cdot\|\bx\|_2$ or $x_i\le\frac{\eps}{2}\cdot\|\bx\|_2$. 
By executing independent $L_2$-heavy hitter streaming algorithms for each difference estimator $\calA$ and strong tracker $\calB$, where the threshold for identifying heavy hitters is aligned with the accuracy guarantee of the corresponding algorithm, we can recover a list that includes all potential heavy hitters along with an estimated count for each item.
We use the following $L_2$-heavy hitter streaming algorithm.
\begin{theorem}
\cite{BravermanCINWW17}
\thmlab{thm:bptree}
For any $\eps > 0$ and $\delta \in [0,1)$, there exists a streaming algorithm, denoted $(\eps,\delta)$-$\bptree$, which with probability at least $1-\delta$ outputs a set of $\frac{\eps}{2}$-heavy hitters that includes all $\eps$-heavy hitters, along with approximate frequencies for each reported item up to an additive error of $\frac{\eps}{4} \cdot L_2$. 
The algorithm uses $\O{\frac{1}{\eps^2}\left(\log\frac{1}{\delta\eps}\right)(\log n+\log m)}$ bits of space.
\end{theorem}

\begin{theorem}[Adversarially robust $L_2$-heavy hitters streaming algorithm]
\thmlab{thm:robust:opt:HH}
\cite{WoodruffZ21}
Given $\eps>0$, there exists an adversarially robust streaming algorithm $\heavyhitters$ that, with probability at least $\frac{2}{3}$, solves the $L_2$-heavy hitters problem using 
\[\O{\frac{1}{\eps^2}\log n\log^4\frac{1}{\eps}\left(\log\frac{1}{\eps}+\log\log n\right)}\]
bits of space. 
\end{theorem}
\begin{proof}
For each integer $j \ge 0$ and fixed $i$, define $u_{i,j}$ as the final round where $Z_{i,j}$ in \algref{alg:framework} is defined. 
Let $t_i$ be the update time when the counter $a$ is first set to $i$ for any $i > 0$, and define $t^{(b)}_i$ as the first round when counter $b$ is set to $j$, with the convention that $t^{(0)}_i = t_i$.

Let $\calE_1$ denote the event that $X_a$ provides a $\left(1+\frac{\eps}{32}\right)$-approximation to $F_2(1,t_a)$, and let $\calE_2$ denote the event that $F_2(1,u_{a,j}) - F_2(1,t_{a,j}) \le \frac{1}{2^{\beta-j-3}} F_2(1,t_a)$ holds for each integer $j > 1$. 
By \lemref{lem:constant} and \lemref{lem:diff:est:bounds}, we have $\PPr{\calE_1 \wedge \calE_2} \ge 1 - \O{\frac{\delta \log n}{\eps}}$.

Assume there exists $r \in [n]$ such that $(x_r)^2 \ge \eps^2 F_2(1,m)$. 
Then $r$ must be either a $\frac{\eps^2}{32^2}$-heavy coordinate with respect to $F_2(1,t_a)$, or a $\frac{2^k \cdot \eps^2}{32^2 \beta^2}$-heavy coordinate with respect to $F_2(t_{a,k}, t_{a,k-1})$ for some $k \in [\beta]$, where we define $t_{a,0} = t_a$.

Therefore, the $\bptree$ algorithm with threshold $\frac{2^k \cdot \eps^2}{32^2 \beta^2}$ over the interval $[t_{a,k}, t_{a,k-1}]$ will identify $r$ as heavy. 
Furthermore, even if $r$ is only flagged at level $k$, a significant portion, $\O{\eps^2} \cdot F_2$, of its total contribution still remains after detection. 
Hence, by tracking $r$’s frequency after it is identified, we can approximate it up to an additive error of $\O{\eps} \cdot\|\bx\|_2$. 
Since this matches the accuracy of $\calA_{a,k}$, the overall space bound follows accordingly.
\end{proof}

\subsection{Difference Estimator for \texorpdfstring{$F_p$}{Fp} Estimation, \texorpdfstring{$0<p<2$}{0<p<2}}
In this section, we use the framework of \algref{alg:framework} to give an adversarially robust streaming algorithm for $F_p$ moment estimation, with $p\in(0,2)$. 
To apply \algref{alg:framework}, we require an $F_p$ strong tracker and an $F_p$ difference estimator. 
There exist constructions for $F_p$ strong trackers, so we devote the majority of this section toward the development of an $F_p$ difference estimator. 
Finally, we also again present an optimization of \algref{alg:framework} to achieve near-optimal space guarantees, i.e., matching the best known $F_p$ algorithm on insertion-only streams, up to $\polylog\left(\log n,\frac{1}{\eps}\right)$ terms. 

We first recall the following definition for $p$-stable distributions. 
\begin{definition}[$p$-stable distribution]
\deflab{def:pstable}
A distribution $\calD_p$ over $\mathbb{R}$ is called \emph{$p$-stable} if for any vector $\bx\in\mathbb{R}^n$ and independent random variables $Z_1,\ldots,Z_n \sim \calD_p$, the linear combination $\sum_{i=1}^n Z_i x_i$ has the same distribution as $\|\bx\|_p \cdot Z$ for $Z \sim \calD_p$.
\end{definition}

\begin{theorem}[Existence of $p$-stable distributions]
\cite{Zolotarev89}
For every $0 < p \le 2$, there exists a $p$-stable distribution $\calD_p$.
\end{theorem}

The probability density function $f(x)$ of a $p$-stable is defined by $f(x)=\Theta\left(\frac{1}{1+|x|^{1+p}}\right)$ for $p<2$ and corresponds to the normal distribution for $p=2$. 
\cite{Nolan03} describes various approaches for generating $p$-stable random variables, such as to draw $\theta$ uniformly at random from the interval $\left[-\frac{\pi}{2},\frac{\pi}{2}\right]$, $r$ uniformly at random from the interval $[0,1]$, and generating the $p$-stable random variable 
\[X=f(r,\theta)=\frac{\sin(p\theta)}{\cos^{1/p}(\theta)}\cdot\left(\frac{\cos(\theta(1-p))}{\log\frac{1}{r}}\right)^{\frac{1}{p}-1}.\]
These $p$-stable random variables are crucial to obtaining a strong $F_p$ tracking algorithm.

\thmstrongFpsmallp*
Unfortunately, the algorithm corresponding to \thmref{thm:strong:Fp:smallp} is based on the $p$-stable sketch of~\cite{Indyk06}, which is a median-based estimator that seems difficult to adapt for the purposes of achieving a difference estimator.

Instead, we use ingredients from Li's geometric mean estimator~\cite{Li08}, which also provides a streaming algorithm for $F_p$, but was not previously analyzed to guarantee strong tracking. 
Let $q\ge 3$ be a positive integer, let $d$ be a multiple of $q$, and let $\bA\in\mathbb{R}^{d\times n}$ be a matrix whose entries are independent $p$-stable random variables. 
For a vector $\bx\in\mathbb{R}^n$, let $\by=\bA\bx$ so that each entry of $\by$ is the inner product of a $p$-stable random vector with the vector $\bx$. 
Let 
\[C_{q,p}=\left[\frac{2}{\pi}\cdot\Gamma\left(1-\frac{1}{q}\right)\cdot\Gamma\left(\frac{p}{q}\right)\cdot\sin\left(\frac{\pi p}{2q}\right)\right]^{-q}\]
be a fixed constant and let
\[z_i:=C_{q,p}\cdot\left(\prod_{j=q(i-1)+1}^{qi}|y_j|^{p/q}\right)\]
be the geometric mean of the inner products of $q$ random $p$-stable vectors with the vector $\bx$. 
It is known that the asymptotic behavior of $C_{q,p}$ can be characterized as follows:
\begin{observation}[Characterization of $C_{q,p}$]
\obslab{obs:c:behave}
\cite{Li08}
$C_{q,p}=\O{\exp(\gamma_e(q-1))}$, for the Euler-Mascheroni constant $\gamma_e\approx 0.57721$. 
\end{observation}
Intuitively, the value of $C_{q,p}$ is chosen so that each random variable $z_i$ is an unbiased estimate of $\|\bx\|_p^p$. 
Specifically, we have the following statements about the expectation and the variance of each random variable $z_i$.
\begin{lemma}[Expectation and variance of Li's geometric mean estimator, Lemma 2.2 in \cite{Li08}]
\lemlab{ligeoest:var}
For $p\in(0,2]$, let $q\ge 3$ be an integer, let $d$ be a multiple of $q$, and let $\xi:=\frac{C_{\frac{q}{2},p}}{C_{q,p}}$
Let $\bA\in\mathbb{R}^{d\times n}$ be a matrix whose entries are independent $p$-stable random variables. 
For a vector $\bx\in\mathbb{R}^n$ and $\by=\bA\bx$, let $z_i:=C_{q,p}\cdot\left(\prod_{j=q(i-1)+1}^{qi}|y_j|^{p/q}\right)$. 
Then $\Ex{z_i}=\|\bx\|_p^p$ and $\Ex{z_i^2}\le\left(\xi^2-1\right)\cdot\|\bx\|_p^{2p}$.
\end{lemma}
Hence, we obtain a $(1+\eps)$-approximation to $\|\bx\|_p^p$ with constant probability by Chebyshev's inequality by taking the arithmetic mean of $\O{\frac{1}{\eps^2}}$ variables $z_i$. 

To obtain our $F_p$ difference estimator and estimate $F_p(\bu+\bv)-F_p(\bu)$, we adapt Li's geometric mean estimator and maintain $\bA(\bu+\bv)$ and $\bA\bu$, where $\bA$ is the sketching matrix for Li's geometric mean estimator, and $\bu$ and $\bv$ are frequency vectors. 
Observe it does not suffice to use $\bA\bv=\bA(\bu+\bv)-\bA\bu$ to recover $F_p(\bv)$, since we seek an estimate of $F_p(\bu+\bv)-F_p(\bu)$.  
Instead, we use the sketches $\bA(\bu+\bv)$ and $\bA\bu$ to compute terms $z_1,z_2,\ldots,z'_1,z'_2,\ldots$, where each $z_i$ is the geometric mean of $q$ consecutive entries in $\bA(\bu+\bv)$ and $z'_i$ are the corresponding terms for $\bA\bu$. 
By \lemref{ligeoest:var}, $z_i$ is an unbiased estimator of $F_p(\bu+\bv)$ and $z'_i$ is an unbiased estimator of $F_p(\bu)$. 
Hence, $z_i-z'_i$ is an unbiased estimator of $F_p(\bu+\bv)-F_p(\bu)$. 
We can then take the arithmetic mean of the values $z_i-z'_i$ across roughly $\O{\frac{\gamma}{\eps^2}}$ indices of $i$ to obtain a single estimate, where $F_p(\bv)\le \gamma(F_p(\bu+\bv)-F_p(\bu))$. 
We can also perform a standard median-of-means approach to further boost the probability of success to union bound across the stream. 

The main analytic challenge is to obtain a sufficiently strong upper bound on the variance of $z_i-z'_i$ and then obtain the strong tracking property. 
Toward upper bounding the variance, we expand $z_i-z'_i$ as a sum of $2^q-1$ geometric means of $q$ terms, each with at least one term $|\langle \bA_j, \bv\rangle|^{p/q}$. 
Since $\bA_j$ is a vector consisting of independent $p$-stable random variables, then $(\langle \bA_j, \bv\rangle)^{p/q}$ has the same distribution as $(\|\bv\|_p\cdot X)^{p/q}$ for a $p$-stable random variable $X$. 
Hence, if $F_p(\bv)\le\gamma\cdot F_p(\bu)$, then we can upper bound the probability that $|\langle \bA_j,\bv\rangle|^{p/q}\ge\|\bu\|^{p/q}_p$. 
Note that for $p\ge 1$, the condition $F_p(\bu+\bv)-F_p(\bu)\le\gamma\cdot F_p(\bu)$ implies $F_p(\bv)\le\gamma\cdot F_p(\bu)$ since $F_p(\bu+\bv)-F_p(\bu)\ge F_p(\bv)$ by convexity. 
For $p<1$, we shall need to slightly modify the framework in \algref{alg:framework} to achieve this condition. 
We give the difference estimator in full in \figref{fig:diff:est:smallp}. 
\begin{figure*}
\begin{mdframed}
\begin{enumerate}
\item
Let $\bA$ be a $d\times n$ random matrix whose entries are independent $p$-stable random variables, for $d=\O{\frac{\gamma^{2/p}}{\eps^2}\left(\log\frac{1}{\eps}+\log\log n\right)}$
\item
For any integer $q\ge 3$ and for any integer $i\in\left\lfloor\frac{d}{q}\right\rfloor$, let $z_i=\prod_{j=q(i-1)+1}^{qi}\left\lvert(\bA\bu+\bA\bv)_j\right\rvert^{p/q}$ and $z'_i=\prod_{j=q(i-1)+1}^{qi}\left\lvert(\bA\bu)_j\right\rvert^{p/q}$. 
\item
Output the arithmetic mean of $(z_1-z'_1),(z_2-z'_2),\ldots,(z_{d/q}-z'_{d/q})$. 
\end{enumerate}
\end{mdframed}
\caption{Difference estimator for $F_p(\bu+\bv)-F_p(\bu)$ with $0<p<2$}
\figlab{fig:diff:est:smallp}
\end{figure*}

We first recall the following inequality. 
\begin{claim}
\claimlab{clm:ab}
For $a,b\ge 0$ and $p\le 2\le q$, it follows that $(a+b)^{p/q}\le a^{p/q}+2b^{p/q}$.
\end{claim}
\begin{proof}
Let $p\le 2\le q$. 
Then for $0\le a\le b$,  
\[(a+b)^{p/q}\le(2b)^{p/q}\le 2b^{p/q}\le a^{p/q}+2b^{p/q}.\]
On the other hand for $0\le b\le a$, then by Bernoulli's inequality,
\[(a+b)^{p/q}=a^{p/q}\left(1+\frac{b}{a}\right)^{p/q}\le a^{p/q}\left(1+\frac{pb}{qa}\right)\le a^{p/q}+b^{p/q}.\]
\end{proof}

We now compute the expectation and upper bound the variance of our difference estimator. 
\begin{lemma}[Expectation and variance of difference estimator terms]
\lemlab{lem:exp:var:smallp}
\cite{WoodruffZ21}
Let $\bu,\bv\succeq\mathbf{0}^n$ be frequency vectors such that $F_p(\bu+\bv)-F_p(\bu)\le\gamma F_p(\bu)$ and $F_p(\bv)\le\gamma F_p(\bu)$. 
Then for each $i\in[d/q]$, $z_i$, and $z'_i$ as defined in \figref{fig:diff:est:smallp}, we have $\Ex{z_i-z'_i}=F_p(\bu+\bv)-F_p(\bu)$ and 
\[\Var(z_i-z'_i)\le\O{2^{2q}\gamma^{2/p}\|\bu\|_p^{2p}}=\O{2^{2q}\gamma^{2/p}(F_p(\bu))^2}.\]
\end{lemma}
\begin{proof}
Consider \figref{fig:diff:est:smallp} and let $q\ge 3$ be an integer. 
For any integer $d=\O{\frac{\gamma^{2/p}}{\eps^2}\left(\log\frac{1}{\eps}+\log\log n\right)}$, such that $d$ is a multiple of $q$, let $A\in\mathbb{R}^{d\times n}$ be a random matrix whose entries are independent $p$-stable random variables. 

For $i\in\left[\frac{d}{q}\right]$, each random variable $z_i$ is a scaled geometric mean of $q$ separate inner products, so that
\begin{align*}
z_i:&=C_{q,p}\cdot\prod_{j=q(i-1)+1}^{qi}\left\lvert\langle \bA_j,\bu+\bv\rangle\right\rvert^{p/q}\\
&=C_{q,p}\cdot\left(\prod_{j=q(i-1)+1}^{qi}\left\lvert1+\frac{\langle \bA_j,\bv\rangle}{\langle \bA_j,\bu\rangle}\right\rvert^{p/q}\right)\cdot\left(\prod_{j=q(i-1)+1}^{qi}\left\lvert\langle \bA_j,\bu\rangle\right\rvert^{p/q}\right),
\end{align*}
where we use $\bA_j$ to denote the $j$-th column of $A$. 
Similarly, each variable $z'_i$ satisfies
\[z'_i:=C_{q,p}\cdot\prod_{j=q(i-1)+1}^{qi}\left\lvert\langle \bA_j,\bu\rangle\right\rvert^{p/q}.\]
By \lemref{ligeoest:var}, $z_i$ and $z'_i$ are unbiased estimators for $F_p(\bu+\bv)$ and $F_p(\bu)$ respectively. 
Hence, $z_i-z'_i$ is an unbiased estimator for $F_p(\bu+\bv)-F_p(\bu)$, so that
\[\Ex{z_i-z'_i}=F_p(\bu+\bv)-F_p(\bu).\]
Let
\[T_i:=\prod_{j=q(i-1)+1}^{qi}\left\lvert1+\frac{\langle \bA_j,\bv\rangle}{\langle \bA_j,\bu\rangle}\right\rvert^{p/q}.\]
Then we can write
\[z_i-z'_i=C_{q,p}\cdot(T_i-1)\cdot\prod_{j=q(i-1)+1}^{qi}\left\lvert\langle \bA_j,\bu\rangle\right\rvert^{p/q}.\]
By \claimref{clm:ab}, 
\[T_i\le\prod_{j=q(i-1)+1}^{qi}\left(1+2\left\lvert\frac{\langle \bA_j,\bv\rangle}{\langle \bA_j,\bu\rangle}\right\rvert^{p/q}\right).\]
Therefore, $T_i-1$ is a sum of $2^q-1$ terms. 
Similarly, $z_i-z'_i$ is a sum of $2^q-1$ products of $q$ terms, at least one of which is $\lvert\langle \bA_j,\bv\rangle\rvert^{p/q}$. 

By \lemref{ligeoest:var}, each of the products of $q$ terms has variance at most $(\xi^2-1)\gamma^{2/p}\|\bu\|_p^{2p}$, where $\xi=\frac{C_{\frac{q}{2},p}}{C_{q,p}}$.  
Therefore, $z_i-z'_i$ is the sum of the $2^q-1$ terms and satisfies
\[\Var(z_i-z'_i)\le(2^q)^2(\xi^2-1)\gamma^{2/p}\|\bu\|_p^{2p}=\O{2^{2q}\gamma^{2/p}\|\bu\|_p^{2p}}.\]
\end{proof}

\begin{corollary}[Pointwise $F_p$ difference estimator for $0<p<2$]
\corlab{cor:diff:est:Fp:smallp:once}
\cite{WoodruffZ21}
Let $\bu,\bv\succeq\mathbf{0}^n$ be frequency vectors such that $F_p(\bu+\bv)-F_p(\bu)\le\gamma F_p(\bu)$ and $F_p(\bv)\le\gamma F_p(\bu)$. 
Then there exists an algorithm that uses a sketch of dimension $d=\O{\frac{\gamma^{2/p}}{\eps^2}\left(\log\frac{1}{\delta}\right)}$ and outputs an additive $\eps\cdot F_p(\bu)$ approximation to $F_p(\bu+\bv)-F_p(\bu)$ with probability at least $1-\delta$. 
\end{corollary}
\begin{proof}
Let $i\in[d/q]$ be fixed and let $z_i$ and $z'_i$ be defined as in \figref{fig:diff:est:smallp}. 
By \lemref{lem:exp:var:smallp},
\[\Ex{z_i-z'_i}=F_p(\bu+\bv)-F_p(\bu),\qquad\Var(z_i-z'_i)\le\O{2^{2q}\gamma^{2/p}(F_p(\bu))^2}.\]
Thus by Chebyshev's inequality, the arithmetic mean of $\O{\frac{2^{2q}\gamma^{2/p}}{\eps^2}}$ differences $z_i-z'_i$ achieves an additive $\eps F_p(\bu)$ error of the difference $F_p(\bu+\bv)-F_p(\bu)$ with probability at least $\frac{2}{3}$. 
We can then boost the probability of success to $1-\delta$ by taking the median of $\O{\log\frac{1}{\delta}}$ such instances.  
\end{proof}

We next prove the strong tracking property. 
However, if we consider each inner product $(\langle \bA_j,\bv\rangle)^{p/q}$ separately over the evolution of $\bv$ over the stream of length $m$ and then take a union bound, we will incur extraneous $\log n$ factors, for $m=\poly(n)$. 
Instead, we interleave our previous argument with carefully chosen values of $\bv$ and bound the supremum of $(\langle \bA_j,\bv\rangle)^{p/q}$ across all values of $v$.  
Namely, we split the stream into $\O{\frac{1}{\eps^{16q/p^2}}}$ times $t_1,t_2,\ldots$ between which the difference increases by roughly $\eps^{16q/p^2}\cdot F_p(\bu)$. 
We first apply a union bound over these times to argue correctness at these times, which incurs a $\log\frac{1}{\eps}$ term. 
To analyze the difference estimator between times $t_i$ and $t_{i+1}$ for a fixed $i$, observe that the difference estimator only increases by roughly $\eps^{16q/p^2}\cdot F_p(\bu)$ from $t_i$ to $t_{i+1}$. 
Thus, even if our approximation to the difference is a multiplicative $\eps^{1-16q/p^2}$, we can still achieve an additive $\eps\cdot F_p(\bu)$ approximation to the difference $F_p(\bu+\bv)-F_p(\bu)$, for a frequency vector $\bv$ induced by updates between $t_i$ and $t_{i+1}$. 
We then use chaining results from \cite{BravermanCIW16, BravermanCINWW17, BlasiokDN17} to show that the supremum of the multiplicative error between times $t_i$ and $t_{i+1}$ is upper bounded by $\eps^{1-16q/p}$ with good probability. 

We first recall the following upper bound on the supremum of the inner product of a random process with a vector of independent $p$-stable random variables. 
\begin{lemma}
\lemlab{lem:chain:lp}
\cite{BlasiokDN17}
Let $\bx^{(1)},\bx^{(2)},\ldots,\bx^{(m)}\in\mathbb{R}^n$ satisfy $\mathbf{0}^n\preceq \bx^{(1)}\preceq\ldots\preceq \bx^{(m)}$. 
Let $\bZ\in\mathbb{R}^n$ be a vector of independent $p$-stable random variables. 
Then there exists a constant $C_p$ such that
\[\PPr{\underset{k\le m}{\sup}|\langle \bZ,\bx^{(k)}\rangle|\ge\lambda\|\bx^{(m)}\|_p}\le C_p\left(\frac{1}{\lambda^{2p/(2+p)}}+n^{-1/p}\right).\]
\end{lemma}

\begin{lemma}[$F_p$ difference estimator for $0<p<2$]
\lemlab{lem:sketching:Fp:smallp}
\cite{WoodruffZ21}
Let $p\in(0,2)$ and $d=\O{\frac{\gamma^{2/p}}{\eps^2}\left(\log\frac{1}{\eps}+\log\frac{1}{\delta}\right)}$. 
Then there exists a $(\gamma,\eps,\delta)$-difference estimator for $F_p$ that uses a sketching matrix $\bA\in\mathbb{R}^{d\times n}$ with entries that are independent $p$-stable random variables.  
\end{lemma}
\begin{proof}
Given an oblivious stream $S$, let $\bu$ be the frequency vector induced by the updates of $S$ from time $t_1$ to $t_2$, and $\bv$ be the frequency vector induced by updates from time $t_2+1$ to $t$, and suppose $F_p(\bv)\le\gamma\cdot F_p(\bu)$. 
For $p\in[1,2)$, we define vectors $\bv^{(1)},\bv^{(2)},\ldots$ so that $\bv^{(i)}$ is the vector defined by the last time at which the difference does not exceed $Ci\eps^{16q/p^2}{2^{q^2}}\cdot F_p(\bu)$, i.e., $F_p(\bu+\bv^{(i)})-F_p(\bu)\le Ci\eps^{16q/p^2}{2^{q^2}}\cdot F_p(\bu)$, where $C$ is a sufficiently small constant to be fixed. 
Observe that there are at most $\O{\frac{1}{\eps^{16q/p^2}}}$ such times. 
For $p\in(0,1]$, we define vectors $\bv^{(1)},\bv^{(2)},\ldots$ so that $\bv^{(i)}$ is the vector defined by the last time at which the $F_p$ moment does not exceed $Ci\eps^{16q/p^2}{2^{q^2}}\cdot F_p(\bu)$, i.e., $F_p(\bv^{(i)})\le Ci\eps^{16q/p^2}{2^{q^2}}\cdot F_p(\bu)$. 
Since $(u_a+v_a)^p-u_a^p\ge p(u_a+v_a)^{p-1}v_a\ge 2^{p-1}pv_a^p$ for $u_a\ge v_a$ and $(u_a+v_a)^p-u_a^p\ge (2^p-1)v_a^p$ for $v_a\ge u_a$, then it again follows that there are at most $\O{\frac{1}{\eps^{16q/p^2}}}$ such times. 
We first show correctness at these times. 

For $d=\O{\frac{\gamma^{2/p}}{\eps^2}\left(\log\frac{1}{\eps}+\log\frac{1}{\delta}\right)}$ and a sketching matrix $\bA\in\mathbb{R}^{d\times n}$ with entries that are independent $p$-stable random variables,  \corref{cor:diff:est:Fp:smallp:once} implies that we have additive $\O{\eps}\cdot F_p(\bu)$ error at all times $t_i$. 
Let $\bv$ be a frequency vector induced by updates from time $t_i$ and $t_{i+1}$, so that $F_p(\bv)\le\frac{1}{2^{q^2}}\eps^{16q/p^2}\cdot F_p(\bu)$ by convexity for $p\in[1,2)$ and by definition for $p\in(0,1]$. 
Then it suffices to show that the change in the output of the difference estimator is at most $\O{\eps}\cdot F_p(\bu)$ for all values of $\bv$ between $t_i$ and $t_{i+1}$. 

For $i\in\left[\frac{d}{q}\right]$, each random variable $z_i$ in the difference estimator is a geometric mean of $q$ separate inner products, so that 
\begin{align*}
z_i:&=C_{q,p}\cdot\prod_{j=q(i-1)+1}^{qi}\left(\langle \bA_j, \bu+\bv\rangle\right)^{p/q}\\
&=C_{q,p}\cdot\left(\prod_{j=q(i-1)+1}^{qi}\left(1+\frac{\langle \bA_j,\bv\rangle}{\langle \bA_j,\bu\rangle}\right)^{p/q}\right)\cdot\left(\prod_{j=q(i-1)+1}^{qi}\left(\langle \bA_j,\bu\rangle\right)^{p/q}\right),
\end{align*}
where $\bA_j$ is the $j$-th column of $\bA$. 
Similarly, each random variable $z'_i$ satisfies
\[z'_i:=C_{q,p}\cdot\prod_{j=q(i-1)+1}^{qi}\left(\langle \bA_j,\bu\rangle\right)^{p/q}.\]

We write $z_i-z'_i$ as a sum of $2^q-1$ geometric means of $q$ terms, at least one of which is $(\langle \bA_j,\bv\rangle)^{p/q}$. 
By \lemref{lem:chain:lp} with $\lambda=\frac{1}{\eps^{8/p}}$, 
\[\PPr{\underset{t\le t_2}{\sup}|\langle \bA_j,\bv\rangle|\ge\lambda\|\bv\|_p}\le C_p\left(\eps^{16/(2+p)}+n^{-1/p}\right)\]
and similarly 
\[\PPr{\underset{t\le t_2}{\sup}|\langle \bA_j,\bu\rangle|\ge\lambda\|\bu\|_p}\le C_p\left(\eps^{16/(2+p)}+n^{-1/p}\right).\]
Thus with probability at least $1-\O{\eps^4}$, none of the $2^q-1$ terms exceeds  
\[\lambda^p\|\bv\|_p^{p/q}\|\bu\|^{(q-1)p/q}_p\le\frac{1}{\eps^8}\|\bv\|_p^{p/q}\|\bu\|^{(q-1)p/q}_p\le\frac{\eps^8}{2^q}\cdot F_p(\bu),\]
since $F_p(\bv)\le\frac{1}{2^{q^2}}\eps^{64q/p^2}\cdot F_p(\bu)$. 
Therefore, with probability at least $1-\O{2^q\eps^4}$, the sum of the $2^q-1$ terms is at most $\eps^8\cdot F_p(\bu)$. 
That is, the difference $z_i-z'_i$ is at most $\eps^8\cdot F_p(\bu)$ with probability at least $1-\O{2^q\eps^4}$. 
By a union bound over $\O{\frac{1}{\eps^2}}$ geometric means, then output of the difference estimator $F_p(\bu+\bv)-F_p(\bu)$ changes by at most $\eps^8\cdot F(u)$ over the course of $t\in(t_i,t_{i+1})$, with constant probability, for sufficiently small $\eps$. 
Moreover, by taking the median of $\O{\log\frac{1}{\eps}+\log\frac{1}{\delta}+\log\log n}$ such estimators, then the probability of success increases to $1-\frac{\delta}{\poly\left(\log n,\frac{1}{\eps}\right)}$ by applying standard Chernoff bounds. 
By first taking a union bound over all $\O{\log n}$ times when the value of the stream doubles, we have correctness of the difference estimator at those times. 
Next, we take a union bound over all $\O{\frac{1}{\eps^{16q/p^2}}}$ such times corresponding to the vectors $\bv^{(1)},\bv^{(2)},\ldots$. 
Then it follows both that the difference estimator is correct at these times and that the difference estimator provides strong tracking between these times, and thus over the entire stream with probability at least $1-\delta$. 
\end{proof}

\paragraph{Derandomization of $p$-stable random variables.} 
Observe that the previous analysis assumes that the independent $p$-stable random entries of $\bA$ can be generated and stored. 
Hence, it remains to derandomize $\bA$
To that end, we first require the following property:
\begin{lemma}[Lemma 8 in \cite{JayaramW18}]
\lemlab{lem:derandom:once}
Let $\calA$ be any streaming algorithm that stores only a linear sketch $\bA\cdot \bx$ on a vector $\bx\in\mathbb{Z}^n$ with entries bounded by $M=\poly(n)$, where the entries of $\bA\in\mathbb{R}^{k\times n}$ are i.i.d., and can be sampled using $\O{\log n}$ bits. 
Then for any fixed constant $c\ge 1$, $\calA$ can be implemented using a random matrix $\bA'$ using $\O{k\log n(\log\log n)^2}$ bits of space, such that for all $\by\in\mathbb{R}^k$ with entry-wise bit complexity of $\O{\log n}$,
\[\left\lvert\PPr{\bA\cdot \bx=\by}-\PPr{\bA'\cdot \bx=\by}\right\rvert<n^{-ck}.\]
\end{lemma}
Unfortunately, our difference estimator actually stores both $\bA\cdot(\bu+\bv)$ and $\bA\cdot \bu$, so we cannot immediately apply \lemref{lem:derandom:once}. 
Hence, we require the following generalization:
\begin{corollary}
\corlab{lem:derandom:many}
\cite{WoodruffZ21}
For a constant $q\ge 1$, let $\bx_1,\ldots,\bx_q\in\mathbb{Z}^n$ with entries bounded by $M=\poly(n)$ be vectors defined by a stream $S$, such that for each $i\in[q]$, $\bx_i$ is defined by the updates of $S$ between given times $t_{i,1}$ and $t_{i,2}$. 
Let $\calA$ be any streaming algorithm that stores linear sketches $\bA\cdot\bx_1,\ldots,\bA\cdot\bx_q$, such that the entries of $\bA\in\mathbb{R}^{k\times n}$ are i.i.d. and can be sampled using $\O{\log n}$ bits, and outputs $g(\bA\cdot\bx_1,\ldots,\bA\cdot\bx_q)$ for some composition function $g:\mathbb{R}^q\to\mathbb{R}$. 
Then for any fixed constant $c\ge 1$, $\calA$ can be implemented using a random matrix $\bA'$ using $\O{k\log n(\log\log n)^2}$ bits of space, such that for all $\by\in\mathbb{R}^k$ with entry-wise bit complexity of $\O{\log n}$,
\[|\Pr[g(\bA\cdot\bx_1,\ldots,\bA\cdot\bx_q)=\by]-\Pr[g(\bA'\cdot\bx_1,\ldots,\bA'\cdot\bx_q) = \by]|<n^{-ck}.\]
\end{corollary}
\begin{proof}
Without loss of generality, we assume that all entries $\bA\cdot\bx_i$ are integers bounded by $\poly(n)$ for each $i\in[q]$, due to the bit complexity of $\bA$ and $\bx_i$. 
Let $\bw_i=\bA\cdot\bx_i$ for each $i\in[q]$ and suppose the maximum entry of $\bA$ satisfies $\|\bA\|_\infty\le n^\alpha$ for some constant $\alpha$. 
Let $N=M\cdot n^\alpha$ so that by 
\[\|\bA\cdot\bx_i\|_\infty\le\|\bA\|_\infty\cdot\|\bx_i\|_\infty\le n^\alpha\cdot M=N,\]
for all $i\in[q]$. 
We define the vector $\bv=\sum_{i=1}^q N^{2i} \bw_i$, so that $\|\bv\|_{\infty}<q\cdot N^{2q}\le N'$ for $N'=N^{3q}$. 
Since $N'=\poly(n)$, then all entries of $\bv$ can be stored in $\O{\log n}$ bits. 
Moreover, each coordinate $v_j$ with $j\in[n]$ has a unique $N^2$-ary representation, due to the magnitude of $N$ and the multiplication with $N^{2j}$. 
Specifically, if $v_j=\sum_{i=1}^q N^{2i} w_{i,j}$ where $w_{i,j}$ represents the $j$-th coordinate of vector $\bw_i$, then there is a unique solution to the system $v_j=\sum_{i=1}^q N^{2i}\alpha_i$ constrained to the condition that $|\alpha_i|\le N$. 
Therefore, the vectors $\bw_i=\bA\cdot\bx_i$ can be computed from the vector $\bv$. 

Observe that by \lemref{lem:derandom:once}, it suffices to use a random matrix $\bA'$ with $\O{k\log n(\log\log n)^2}$ bits of space, so that for all $\by \in \mathbb{R}^k$ with entry-wise bit complexity of $\O{\log n}$: 
\[\left\lvert\PPr{\bA\cdot \bv=\by}-\PPr{\bA'\cdot \bv=\by}\right\rvert<n^{-ck}.\]
Because a streaming algorithm $\calA$ can extract the vectors $\bA\cdot\bx_1,\ldots,\bA\cdot\bx_q$ from the vector $\bA\cdot\bv$ by the above argument, then $\calA$ can subsequently compute the composition $g(\bA\cdot\bx_1,\ldots, \bA\cdot\bx_q)$. 
Therefore,
\[\left\lvert\PPr{g(\bA\cdot \bx_1,\ldots,\bA\cdot \bx_q)=y}-\PPr{g(\bA'\cdot\bx_1,\ldots,\bA'\cdot \bx_q)}\right\rvert<n^{-ck}.\]
\end{proof}
It thus remains to show that the pseudorandom generator of \corref{lem:derandom:many} can be used to derandomize the correctness guarantees of the difference estimator. 
\begin{lemma}[$F_p$ difference estimator for $0<p<2$]
\lemlab{lem:diff:est:Fp:smallp}
\cite{WoodruffZ21}
For $0<p<2$, there exists a $(\gamma,\eps,\delta)$-difference estimator for $F_p$ that uses space
\[\O{\frac{\gamma^{2/p}\log n}{\eps^{2}}(\log\log n)^2\left(\log\frac{1}{\eps}+\log\frac{1}{\delta}\right)}.\]
\end{lemma}
\begin{proof}
The difference estimator for $F_p(\bu+\bv)-F_p(\bu)$ requires an input splitting time $t_1$ at which the frequency vector $\bu$ concludes and the frequency vector $\bv$ begins. 
We must then argue correctness over all possible stopping times $t$ with $t>t_1$, provided $F_p(\bu+\bv)-F_p(\bu)\le\gamma F_p(\bu)$ for $p<1$ or $F_p(\bv)\le\gamma F_p(\bu)$ for $p\ge 1$. 
We apply \corref{lem:derandom:many} with $q = 2$, so that there exists a pseudorandom generator that succeeds with high probability $1-\frac{1}{\poly(n)}$ for any fixed value of $t$. 
Now, since the stream has length $m=\poly(n)$, by taking a union bound over all $m$ possible values of the stopping times $t$, then the pseudorandom generator is correct with high probability over all possible stopping times $t$. 
Therefore we obtain the claimed guarantees for the difference estimator, from \lemref{lem:sketching:Fp:smallp}. 

In particular, recall that the difference estimator is used to choose a final stopping time for the purposes of the framework in \thmref{thm:framework}, i.e., when the estimated difference is sufficiently large. 
Therefore, the framework only requires that the marginal distribution of the difference estimator is correct at all times, rather than requiring the joint distribution to be correct. 
Specifically, the framework uses a sequence of outputs $s_{t_1+1},\ldots,s_t$ from the difference estimator over the course of the stream to choose a final stopping time $t_2$, based on the first output that exceeds a certain threshold $T$. 
Therefore, we do not fool the final stopping time $t_2$ chosen by the framework, as that requires a conditional statement on the output sequence $s_{t_1+1},\ldots,s_t$ of the difference estimator not exceeding the threshold $T$. 

For example, suppose the probability over the distribution of the independent $p$-stable random variables that $s_{t_1+1}$ exceeds the threshold $T$ is $\frac{1}{2}$, i.e., $\PPr{s_{t_1+1}\ge T}=\frac{1}{2}$. 
Similarly, suppose the probability that $s_{t_1+2}$ exceeds the threshold $T$ is $\frac{1}{2}$, i.e., $\PPr{s_{t_1+2}\ge T} = \frac{1}{2}$. 
Furthermore, suppose that we have that $s_{t_1+2}$ cannot exceed $T$ conditioned on the event that $s_{t_1+1}<T$, so that $\PPr{s_{t_1+2}\ge T\,\mid\,s_{t_1+1}<T}=0$. 
Then since \thmref{thm:framework} chooses the first time that exceeds $T$, the framework can never choose $t_2$ to be $t_1+2$, i.e., $\PPr{t_2=t_1+2}=0$. 
On the other hand, our derandomization does not use independent $p$-stable random variables and only fools the marginal probabilities (of a pair of times $(t, t_1)$ for each $t>t_1$). 
Hence, it is possible for \thmref{thm:framework} to select $t_2=t_1+2$,  using our derandomized difference estimator. 
On the other hand, $t_1+2$ is still a valid stopping time if the difference estimator is correct at time $t_1+1$, and the output does not exceed $T$. 
That is, we need not fool the joint distribution, and thus not fool the choice of stopping time. 
Therefore, even though the difference estimator derandomized using the pseudorandom generator could induce a different distribution on the final stopping time, the final stopping time $t'_2$ output by our derandomized algorithm can still be used in our framework because $t'_2$ corresponds to the first output of the algorithm that exceeds the threshold $T$. 
Moreover, the derandomization retains correctness at all intermediate times between $t_1$ and $t'_2$.  
\end{proof}

\paragraph{Bit complexity and rounding of $p$-stable random variables.}
Finally, we describe how to store each inner product in \lemref{lem:diff:est:Fp:smallp} using $\O{\log n}$ bits of space, by rounding the $p$-stable random variable to $\O{\log n}$ bits of precision. 
Observe that due to the rounding, each summand changes additively by $\frac{1}{\poly(n)}$. 
Hence, the total estimate also changes by an additive $\frac{1}{\poly(n)}$. 
Then the total error of the estimate of the difference $F_p(\bu+\bv)-F_p(\bu)$ is $\eps\cdot F_p(\bu)+\frac{1}{\poly(n)}$. 
Now, as long as $F_p(\bu)\neq 0$, then the additive $\frac{1}{\poly(n)}$ can be absorbed into the $\eps\cdot F(u)$ term with a rescaling of $\eps$. 
On the other hand, in the case $F_p(\bu)=0$, then our estimator will output $0$ regardless, so that the rounding of the $p$-stable random variables will not affect the overall guarantees of our algorithm, i.e., the difference estimator still achieves additive error $\eps\cdot F_p(\bu)$. 

\begin{theorem}
\cite{WoodruffZ21}
Let $\eps\in\left(0,\frac{1}{2}\right)$ be an accuracy parameter and $p\in(0,2)$. 
There exists an adversarially robust streaming algorithm that outputs a $(1+\eps)$-approximation for $F_p$ moment estimation that succeeds with probability at least $\frac{2}{3}$ and uses $\tO{\frac{1}{\eps^2}\log^2 n}$ bits of space. 
\end{theorem}
\begin{proof}
Observe that $F_p$ is a monotonic function with $(\eps,m)$-flip number $\lambda=\O{\frac{1}{\eps}\log n}$. 
By \lemref{lem:diff:est:Fp:smallp} and \thmref{thm:strong:Fp:smallp}, there exists a $(\gamma,\eps,\delta)$-difference estimator that uses 
\[\O{\frac{\gamma^{2/p}\log n}{\eps^{2}}(\log\log n)^2\left(\log\frac{1}{\eps}+\log\frac{1}{\delta}\right)}\]
bits of space, as well as an oblivious strong tracker that uses space
\[\O{\frac{\log n}{\eps^2}\left(\log\log n+\log\frac{1}{\eps}+\log\frac{1}{\delta}\right)}.\]
Hence, by using the framework of \algref{alg:framework}, then by \thmref{thm:framework}, there exists an adversarially robust streaming algorithm that outputs a $(1+\eps)$-approximation for the $F_p$ moment while using $\tO{\frac{1}{\eps^2}\log^2 n}$ bits of space and succeeds with probability at least $\frac{2}{3}$. 
\end{proof}

\paragraph{Optimized $F_p$ algorithm for $p\in(0,2)$.}
To reduce the space complexity, we adopt the same optimization strategy as described in \secref{sec:diff:est:F2}. 
Recall that the counter $a$ in \algref{alg:framework} keeps track of the currently active algorithm instances $\calA_a$ and $\calB_{a,c}$.
Rather than maintaining $\O{\log n}$ total sketches, we only retain those corresponding to the smallest $\O{\log\frac{1}{\eps}}$ indices $i \ge a$. 
Since each increment of $a$ corresponds to the output increasing by a factor of $2$, sketches associated with larger indices will only miss a small $\O{\eps}$ portion of the total $F_p$ value, and thus still yield a valid $(1+\eps)$-approximation.

\begin{theorem}[Adversarially robust $F_p$ streaming algorithm for $p \in (0,2)$]
\thmlab{thm:robust:opt:Fp:smallp}
\cite{WoodruffZ21}
For any $\eps > 0$ and $p \in (0,2)$, there exists an adversarially robust streaming algorithm that outputs a $(1+\eps)$ approximation to $F_p$ using 
\[\O{\frac{1}{\eps^{2}}\log n(\log\log n)^2\log\frac{1}{\eps}\left(\log\log n+\log\frac{1}{\eps}\right)}\]
bits of space, and succeeds with probability at least $\frac{2}{3}$.
\end{theorem}
\begin{proof}
Under the aforementioned optimization, the number of simultaneously active indices $a$ is $\O{\log\frac{1}{\eps}}$, and for each such $a$, there are $\O{1}$ active $c$ indices corresponding to sketches $\calA_a$ and $\calB_{a,c}$. 
From \thmref{thm:framework}, the total space used by each $\calB_{a,j}$ across the $\beta$ granularities is:
\[\O{\frac{1}{\eps^2} \cdot S_1(n,\delta',\eps) + \frac{1}{\eps} \log\frac{1}{\eps} \cdot S_2(n,\delta',\eps)},\]
where for our application, $S_1(n,\delta',\eps) = \log n (\log\log n)^2 (\log\frac{1}{\eps} + \log\frac{1}{\delta'} + \log\log n)$, and $S_2 = 0$ since both the strong tracker and the difference estimator for $F_p$ do not contribute to this term.

Over the full stream, there are $\O{\log n}$ total possible values of $a$, so the failure probability for each sketch must be bounded by $\delta / \poly(\log n, 1/\eps)$ to ensure the total failure probability remains below $\delta = \frac{2}{3}$. 
Since only $\O{\log\frac{1}{\eps}}$ values of $a$ are active at once, the overall sketching and granularity-changing framework requires 
\[\O{\frac{1}{\eps^2} \log n (\log\log n)^2 \log\frac{1}{\eps} \left( \log\log n + \log\frac{1}{\eps} \right)}\]
bits of space.

Furthermore, with $\O{\log\frac{1}{\eps}}$ active $a$ indices, there are $\O{\frac{1}{\eps}}$ subroutines in total, each requiring $\O{\log n \log\frac{1}{\eps}}$ bits to track their update times in a stream of length $m$, where $\log m = \O{\log n}$. 
This adds another 
\[\O{\frac{1}{\eps} \log n \log\frac{1}{\eps}}\]
bits to the space usage. 
Combining both contributions, the total space is 
\[\O{\frac{1}{\eps^{2}}\log n(\log\log n)^2\log\frac{1}{\eps}\left(\log\log n+\log\frac{1}{\eps}\right)}.\]
\end{proof}

We also obtain the following for the case $p\in(0,1]$:

\begin{theorem}
[Adversarially robust $F_p$ streaming algorithm for $p\in(0,1\text{]}$]
\thmlab{thm:robust:opt:Fp:tinyp}
\cite{WoodruffZ21}
For any $\eps>0$ and $p\in(0,1]$, there exists an adversarially robust streaming algorithm that outputs a $(1+\eps)$ approximation to $F_p$ using
\[\O{\frac{1}{\eps^{2}} \log\frac{1}{\eps} \left(\log\log n + \log\frac{1}{\eps} \right) + \frac{1}{\eps} \log\frac{1}{\eps} \log n}\]
bits of space, with success probability at least $\frac{2}{3}$.
\end{theorem}

\paragraph{Entropy estimation.}
We conclude with an application of our results to estimating Shannon entropy in adversarially robust streaming settings.
For a frequency vector $\bv \in \mathbb{R}^n$, the Shannon entropy is defined as
\[H(\bv) = -\sum_{i=1}^n v_i \log v_i.\]

\begin{observation}
\obslab{obs:entropy:addmult}
\cite{HarveyNO08}
An algorithm that provides an $\eps$-additive approximation to the Shannon entropy $H(\bv)$ also yields a $(1+\eps)$-multiplicative approximation to the quantity $h(\bv) := 2^{H(\bv)}$, and vice versa.
\end{observation}

As a result, we focus on designing algorithms that compute $(1+\eps)$-multiplicative approximations to $h(\bv) = 2^{H(\bv)}$.

\begin{lemma}[Section 3.3 in~\cite{HarveyNO08}]
\lemlab{lem:entropy:reduction}
Let $k = \log\frac{1}{\eps} + \log\log m$ and define $\eps' = \frac{\eps}{12(k+1)^3 \log m}$. 
There exists a set $\{y_0, \ldots, y_k\} \subset (0,2)$, computable in linear time, and a deterministic post-processing function such that, given $(1+\eps')$-approximations to $F_{y_i}(\bv)$ for all $i$, it outputs a $(1+\eps)$-approximation to $h(\bv) = 2^{H(\bv)}$.
\end{lemma}

We describe the construction of the exponents $\{y_0, \ldots, y_k\}$ used in \lemref{lem:entropy:reduction} due to~\cite{HarveyNO08}. 
Define 
\[\ell = \frac{1}{2(k+1) \log m}, \quad f(z) = \frac{(k^2 \ell) z - \ell(k^2 + 1)}{2k^2 + 1}.\]
Then for each $i = 0, \ldots, k$, we set 
\[y_i = 1 + f\left(\cos\left(\frac{i\pi}{k}\right)\right),\]
ensuring each $y_i \in (0,2)$ and enabling linear-time computation of the full set.

Finally, a $(1+\eps)$-multiplicative approximation to $h(\bv) = 2^{H(\bv)}$ can be recovered by evaluating $2^{P(0)}$, where $P(x)$ is the degree-$k$ polynomial that interpolates the values $F_{y_i}(\bv)$ at the points $y_i$.
Hence, we can utilize our adversarially robust streaming algorithms for estimating $F_p(\bv)$ for $p\in\{y_1,y_2,\ldots\}$:

\begin{theorem}[Adversarially robust entropy streaming algorithm]
\thmlab{thm:robust:entropy}
\cite{WoodruffZ21}
For any $\eps > 0$, there exists an adversarially robust streaming algorithm that outputs an additive $\eps$-approximation to the Shannon entropy using $\tO{\frac{1}{\eps^2} \log^3 n}$ bits of space, and succeeds with probability at least $\frac{2}{3}$.
\end{theorem}
\begin{proof}
From \obsref{obs:entropy:addmult} and \lemref{lem:entropy:reduction}, it suffices to compute $(1+\eps')$-approximations to $F_{y_i}(\bv)$ for a collection of exponents $y_i \in (0,2)$ drawn from the set $\{y_0, \ldots, y_k\}$, where $k = \log\frac{1}{\eps} + \log\log m$ and $\eps' = \frac{\eps}{12(k+1)^3 \log m}$.

Using \thmref{thm:robust:opt:Fp:smallp} with parameter $\eps'$, we can compute adversarially robust approximations to each $F_{y_i}(\bv)$ in space
\[\tO{\frac{1}{(\eps')^2} \log n (\log\log n)^2 \log\frac{1}{\eps} \left( \log\log n + \log\frac{1}{\eps} \right)}.\]

Each algorithm has failure probability at most $1 - \poly\left(\eps, \frac{1}{\log n}\right)$, achieved by setting the internal failure rate $\delta' = \frac{\delta}{\poly(1/\eps, \log n)}$ in both \lemref{lem:diff:est:Fp:smallp} and \thmref{thm:strong:Fp:smallp}.

Since we run $\O{k}$ such algorithms and $\log m = \O{\log n}$, the total space used is bounded by $\tO{\frac{1}{\eps^2} \log^3 n}$.
\end{proof}

\subsection{Difference Estimator for \texorpdfstring{$F_p$}{Fp} Estimation, Integer \texorpdfstring{$p>2$}{p>2}}
In this section, we use the framework of \algref{alg:framework} to give an adversarially robust streaming algorithm for $F_p$ moment estimation, for integral $p>2$. 
We again require both an $F_p$ strong tracker and an $F_p$ difference estimator to use \thmref{thm:framework}. 
We note that for integer $p>2$, the dominant space factor is $n^{1-2/p}$~\cite{ChakrabartiKS03,Ganguly12,WoodruffZ12,WoodruffZ21b}. 
Hence, we will not focus on optimizing the $\polylog(n)$ factors. 
We recall the following $F_p$ moment approximation algorithm. 

\thmFpbigp*

By setting the failure probability $\delta'=\frac{\delta}{\poly(n)}$, we can apply \thmref{thm:Fp:bigp} across all $m=\poly(n)$ times on a stream to obtain the following strong tracker:
\begin{theorem}[Oblivious $F_p$ strong tracking for integer $p>2$]
\thmlab{thm:strong:Fp:bigp}
\cite{Ganguly11, GangulyW18}
For $p>2$, there exists an insertion-only streaming algorithm that uses $\O{\frac{1}{\eps^2}n^{1-2/p}\log\frac{n}{\delta}\log^2 n}$ bits of space and provides $(\eps,\delta)$-strong tracking for the $F_p$ moment. 
\end{theorem}
We first introduce the following definition of perfect $L_p$ sampling. 
\begin{definition}[$L_p$ sampling]
Let $\bx\in\mathbb{R}^n$, $\delta\in(0,1]$ be a failure probability, and $c>0$ be any input constant.  
A \emph{perfect $L_p$ sampler} is an algorithm that either outputs a failure symbol $\bot$ with probability at most $\delta$ or an index $i^*\in[n]$ such that for each $i\in[n]$,
\[\PPr{i^*=i}=\frac{x_i^p}{\|\bx\|_p^p}+\O{n^{-c}}.\]
\end{definition}
Since $p>2$ is an integer, we can write $F_p(\bu+\bv)-F_p(\bu)=\sum_{k=1}^{p}\binom{p}{k}\langle \bu^{p-k},\bv^k\rangle$, where we abuse notation so that $\bv^k$ denotes the coordinate-wise $k$-th power of $\bv$. 
Consider a coordinate $a\in[n]$ acquired from a perfect $L_k$ sampler and let $Z$ be the $i$-th coordinate of the frequency vector $\bv^k$. 
Since the vector $\bv$ arrives after the frequency vector $\bu$, then we can sample $i\in[n]$ after $\bu$ arrives and then explicitly compute $Z$ as $\bv$ arrives. 
We can also obtain an unbiased estimate $Y$ to $\|\bu\|_k^k$ with low variance, so that the expected value of $YZ$ would be roughly $\langle \bu^{p-k},\bv^k\rangle$ and moreover, the variance of the estimator is small. 

Unfortunately, one issue with this approach is that perfect $L_p$ samplers for $p>2$ are not known. 
Indeed, while perfect $L_p$-samplers are known for $p\le 2$~\cite{JayaramW18,SwartworthWZ25}, their constructions are based on duplicating each stream update $\poly(n)$ times, so that adapting these constructions to build perfect $L_p$-samplers for $p>2$ would require $(\poly(n))^{1-2/p})$ space rather than $\polylog(\poly(n))$, though this gap has since been resolved by \cite{WoodruffXZ25}. 
In particular, observe that the former requires space larger than $n$ while the latter remains $\polylog(n)$. 
An alternative approach would be to use approximate $L_p$ samplers~\cite{MonemizadehW10,JowhariST11,AndoniKO11, MahabadiRWZ20,MahabadiWZ22}, but these samplers use prohibitively large $\frac{1}{\eps^2}$ space dependency. 
Instead, we use the following perfect $L_2$-sampler: 
\begin{theorem}[Perfect $L_2$ sampler]
\thmlab{thm:perfect:sampler}
\cite{JayaramW18,SwartworthWZ25}
Given failure probability $\delta\in(0,1]$, there exists a one-pass streaming algorithm $\sampler$ that is a perfect $L_2$ sampler and uses $\O{\log^3 n\log\frac{1}{\delta}}$ bits of space.
\end{theorem}
The general outline of our difference estimator is as follows. 
We first use the perfect $L_2$-sampler to return a coordinate $i\in[n]$ with probability $\frac{u_a^2}{\|\bu\|_2^2} \pm \frac{1}{\poly(n)}$. 
The perfect $L_2$ sampler can be used to obtain an unbiased estimate $X$ of $u_i^{k-2}$ with small variance. 
We can also obtain an unbiased estimate $Y$ of $\|\bu\|_2^2$, respectively. 
Subsequently, we can track the $i$-th coordinate of $\bv$ exactly and then compute $v_i^{p-k}$. 
We show that the product of $X$, $Y$, and $v_i^{p-k}$ is nearly an unbiased estimate of $\langle \bu^{p-k},\bv^k\rangle$ up to additive $\frac{1}{\poly(n)}$ factors.  
Moreover, we show that the mean of enough repetitions gives a sufficiently small variance to achieve a $(1+\eps)$-approximation to $\langle \bu^{p-k},\bv^k\rangle$. 
Hence by repeating the estimator for each summand in $\sum_{k=1}^{p}\binom{p}{k}\langle \bu^{p-k},\bv^k\rangle$, we can obtain a $(\gamma,\eps,\delta)$-difference estimator for $F_p$. 

We first utilize the well-known $\countsketch$ algorithm for identifying $\eps\cdot L_2$ heavy-hitters, i.e., any coordinate $i$ such that $x_i\ge\eps\cdot\|\bx\|_2$. 
The algorithm uses a table with $\log\frac{n}{\delta}$ rows, each consisting of $\O{\frac{1}{\eps^2}}$ buckets. 
For each row, each coordinate $i\in[n]$ of the universe is hashed to one of the $\O{\frac{1}{\eps^2}}$ buckets, scaled by a random sign. 
In particular, the signed sum of all items assigned to each bucket across all rows is tracked by the data structure. 
Then to estimate the frequency of each item $i$, the algorithm outputs the median of the values associated with each bucket that $i$ is hashed to, across all rows. 
We recall the following guarantees for $\countsketch$ algorithm:

\begin{theorem}
\thmlab{thm:countsketch}
\cite{CharikarCF04}
Given $\eps\in(0,1)$, there exists a streaming algorithm $\countsketch$ that with probability $1-\frac{1}{\poly(n)}$, outputs an estimate $\widehat{\bx^{(t)}}$ at each time $t$ to a frequency vector $\bx^{(t)}$ such that $|\widehat{\bx^{(t)}}-\bx^{(t)}|_\infty\le\eps\cdot\|\bx^{(t)}\|_2$. 
The algorithm uses $\O{\frac{1}{\eps^2}\log^2 n}$ bits of space. 
\end{theorem}

\begin{figure*}
\begin{mdframed}
\begin{enumerate}
\item
Find a list $\calH$ that includes all $i\in[n]$ with $u_i\ge\frac{\gamma^{1/p}}{16}\|\bu\|_p$.
\item
Using $\countsketch$, obtain an estimate $\widehat{u_i}$ to $u_i$ with additive error $\frac{\eps\gamma^{1/p}}{64\gamma}\|\bu\|_p$ for each $i\in\calH$ and let $\bh\in\mathbb{R}^n$ be the vector such that $h_i=\widehat{u_i}$ if $i\in\calH$ and zero otherwise. 
\item
Perform perfect $L_2$ sampling on $\bu-\bh$ to obtain a set $\calS$ of size $R=\O{\frac{\gamma}{\eps^2}n^{1-2/p}}$. 
\item
Obtain an estimate $\widehat{s_i}$ to $u_i-h_i$ for each $i\in\calS$. 
\item
Let $W$ be a $(1+\eps)$-approximation to $\|\bu-\bh\|_2^2$. 
\item
Output $\sum_{k=1}^{p-1}\binom{p}{k}\left(\sum_{i\in\calH}\widehat{u_i}^{p-k},v_i^k+W\cdot\sum_{i\in\calS}\widehat{s_i}^{p-k-2},v_i^k\right)$. 
\end{enumerate}
\end{mdframed}
\caption{$F_p$ difference estimator for $F_p(\bu+\bv)-F_p(\bu)$ with integer $p>2$.}
\figlab{fig:diff:est:bigp}
\end{figure*}

We first show that for a sampled coordinate $i$, we can acquire a good estimate to $u_i$. 
\begin{lemma}[Moment estimation of sampled items]
\lemlab{lem:coor:exp:var}
\cite{WoodruffZ21}
For integer $p>2$ and failure probability $\delta\in(0,1]$, there exists a one-pass streaming algorithm that outputs an index $i\in[n]$ with probability $\frac{u_i^2}{\|\bu\|_2^2}+\frac{1}{\poly(n)}$, as well as an unbiased estimate to $u_i^p$ with variance $\O{u_i^{2p}}$. 
The algorithm uses $\O{\log^3 n\log\frac{1}{\delta}}$ bits of space.  
\end{lemma}
\begin{proof}
We first describe the construction of the perfect $L_2$ sampler by \cite{JayaramW18}. 
For each coordinate $i\in[n]$, the algorithm $\sampler$ first duplicates $u_i$ a total number of $n^c$ times. 
For each $(i,j)\in[n]\times[n^c]$, $\sampler$ scales $u_i$ by an exponential random variable $e_{i,j}$ to obtain $z_{i,j}=\frac{u_i}{\sqrt{e_{i,j}}}$. 
As a result, this linear transformation to the frequency vector $\bu\in\mathbb{R}^n$ induces a vector $\bz\in\mathbb{R}^{n^{c+1}}$, which can also be viewed as having dimensions $n\times n^c$. 
Afterwards, $\sampler$ only outputs an index $i\in[n]$ if there exists an index $j\in[n^c]$ such that $|z_{i,j}|\ge\Omega(1)\cdot\|\bz\|_2$ for a sufficiently large constant. 
Observe that in this case, if we run an instance of $\countsketch$ with constant factor approximation to obtain an estimate $\widehat{z_{i,j}}$ for the frequency of $z_{i,j}$ and set $\widehat{u_i}=\sqrt{e_{i,j}}\cdot\widehat{z_{i,j}}$, then $\widehat{u_i}$ is an unbiased estimate to $u_i$. 
Formally, each estimate $\widehat{z_{i,j}}$ of $z_{i,j}$ is
\[\sum s_{i,j}\mathbbm{1}[h(a,b)=h(i,j)]\cdot s_{a,b}z_{a,b},\]
where $s_{a,b}\in\{-1,+1\}$ is a random sign and $\mathbbm{1}[h(a,b)=h(i,j)]$ is the indicator random variable for whether $h(a,b)=h(i,j)$, i.e., 
That is, we define $\mathbbm{1}[h(a,b)=h(i,j)]=1$ if $h(a,b)=h(i,j)$ and $\mathbbm{1}[h(a,b)=h(i,j)]=0$ otherwise. 
Therefore, $\widehat{z_{i,j}}$ is an unbiased estimate of $z_{i,j}$, from which it follows that $\widehat{u_i}$ is an unbiased estimate of $u_i$. 
Furthermore, the variance of $\widehat{z_{i,j}}$ is at most $\O{\|\bz\|_2^2}=\O{z_{i,j}^2}$. 
Hence, the variance of $\widehat{u_i}$ is at most $\O{u_i^2}$. 
Therefore, by using $p$ independent instances of $\countsketch$ with estimates $\widehat{u_i}^{(1)},\ldots,\widehat{u_i}^{(p)}$, their product is an unbiased estimate to $u_i^p$ with variance $\O{u_i^{2p}}$, as claimed. 

It remains to analyze the space complexity of this procedure. 
We use $p$ instances of $\countsketch$ with constant factor approximation. 
Each instance uses $\O{\log n\log\frac{n}{\delta}}$ bits of space. 
Moreover, the perfect $L_2$ $\sampler$ uses $\O{\log^3 n\log\frac{1}{\delta}}$ bits of space by \thmref{thm:perfect:sampler}. 
Thus, the total space used is $\O{\log^3 n\log\frac{1}{\delta}}$ bits.
\end{proof}

Unfortunately, perfect $L_2$ sampling coordinates of $\bv$ alone is insufficient because the variance of the resulting procedure is too high to achieve space dependency $\frac{\gamma}{\eps^2}$. 
Hence, we first run a subroutine that removes a set $\calH$ of ``heavy'' coordinates from $\bu$ and we subsequently track the corresponding coordinates of $\bv$. 
Although we have the exact values of $v_i$ for $i\in\calH$, we do not have exact values of $u_i$. 
Nevertheless, we can use estimates $\widehat{u_i}$ for each $u_i$ with $i\in\calH$. 
Let $\bh$ be the sparse vector that contains the estimates $\widehat{u_i}$ for each $i\in\calH$, so that $h$ is intuitively the vector consisting of the heavy-hitters of $\bu$. 
We define $\bw:=\bu-\bh$ and then perform perfect $L_2$ sampling from $\bw$. 

To show the correctness of our difference estimator, observe that we can decompose
\[F_p(\bu+\bv)-F_p(\bu)=\sum_{i\in\calH}\sum_{k=1}^{p}\binom{p}{k}u_i^{p-k}v_i^k+\sum_{i\notin\calH}\sum_{k=1}^{p}\binom{p}{k}u_i^{p-k}v_i^k.\]
Intuitively, the heavy-hitter subroutine allows an accurate estimate to the first summation with sufficiently small variance, while the perfect $L_2$ sampling subroutines allows accurate estimation to the second summation with sufficiently small variance. 
Unfortunately, perfect $L_2$ sampling incurs an additive error of roughly $u_i^2$. 
This is acceptable for the case $k>1$ where each term has $u_i^k$. 
However for $k=1$, we need to perform an alternative procedure. 
Thus, we further consider casework on whether $k=1$ or $k>1$. 

\paragraph{Estimation of $\langle \bu^{p-k},\bv^k\rangle$ for $k>1$.}
We first show that the difference estimator achieves additive error $\eps\cdot F_p(\bu)$ to $\sum_{i\notin\calH}\binom{p}{k}u_i^{p-k}v_i^k$ with $k\ge 2$. 
\begin{lemma}
\lemlab{lem:diff:est:bigp:general:k}
\cite{WoodruffZ21}
For integer $p>2$ and integer $k\ge 2$, there exists an algorithm that uses $\tO{\frac{\gamma}{\eps^2}n^{1-2/p}}$ bits of space and with high probability, outputs an estimate to $\sum_{i\notin\calH}\binom{p}{k}u_i^{p-k}v_i^k$ with additive $\eps\cdot F_p(\bu)$ error. 
\end{lemma}
\begin{proof}
Given an oblivious stream $S$, let $\bu$ be the frequency vector induced by the updates of $S$ from time $t_1$ to $t_2$, and $\bv$ be the frequency vector induced by updates from time $t_2+1$ to $t$, and suppose $F_p(\bu+\bv)-F_p(\bu)\le\gamma\cdot F_p(\bu)$. 
By \thmref{thm:robust:opt:HH}, there exists an algorithm $\heavyhitters$ that uses $\tO{\frac{\gamma^{2-2/p}}{\eps^2}n^{1-2/p}}$ bits of space and outputs a list $\calH\subseteq[n]$ of indices, along with estimates $\widehat{u_i}$ that have additive error $\frac{\eps\gamma^{1/p}}{64\gamma}\|\bu\|_p$. 
Importantly, $\calH$ includes all coordinates $i\in[n]$ such that $u_i\ge\frac{\gamma^{1/p}}{16}\|\bu\|_p$ and no coordinate $j$ such that $u_j\le\frac{\gamma^{1/p}}{32}\|\bu\|_p$. 
Let $\bh$ be the vector consisting of the estimates $\widehat{u_i}$ for each $i\in\calH$ and $0$ in the positions $i\notin\calH$, so that $\bh$ denotes the vector of the estimated values of the heavy-hitters of $\bu$.  
Let $\bw:=\bu-\bh$. 

Consider indices $j_1,\ldots,j_R\in[n]$ acquired from the perfect $L_2$ algorithm $\sampler$, so that each sample is a coordinate $i\in[n]$ with probability $\frac{w_i^2}{\|\bw\|_2^2}+\frac{1}{\poly(n)}$. 
By \lemref{lem:coor:exp:var}, we can obtain unbiased estimates $\widehat{w_{j_1}^{p-k-2}},\ldots,\widehat{w_{j_R}^{p-k-2}}$ to $w_{j_1}^{p-i-2},\ldots,w_{j_R}^{p-k-2}$. 
We can also acquire an unbiased estimate $W$ of $\|\bw\|_2^2$ with variance $\O{\eps^2}\cdot\|\bw\|_2^4$. 
Therefore, for each $b\in[R]$, the product $\widehat{w_{j_b}^{p-k-2}}\cdot W\cdot v_{j_b}^k$ satisfies
\begin{align*}
\mathbb{E}\big[\widehat{w_{j_b}^{p-k-2}}&\cdot W\cdot v_{j_b}^k\big]=\sum_{a\in\calH}\langle\widehat{v_a}^{p-i},u_a^i\rangle\\
&+\sum_{a\in\calH}\left(\frac{w_a^2}{\|\bw\|_2^2}+\frac{1}{\poly(n)}\right)\cdot w_a^{p-k-2}\cdot(1\pm\O{\eps})\|\bw\|_2^2\cdot v_a^k\\
&+\sum_{a\notin\calH}\left(\frac{w_a^2}{\|\bw\|_2^2}+\frac{1}{\poly(n)}\right)\cdot w_a^{p-k-2}\cdot(1\pm\O{\eps})\|\bw\|_2^2\cdot v_a^k.
\end{align*}
Since $\widehat{u_a}$ is a $(1+\eps)$-approximation to $u_a$ for $a\in\calH$, then $|w_a|\le\eps\cdot|u_a|$. 
Hence, the second summation is at most $\O{\eps}\langle \bu^{p-k},\bv^k\rangle$. 
Moreover, $w_a=u_a$ for $a\notin\calH$. 
Therefore, 
\begin{align*}
\Ex{\widehat{w_{j_b}^{p-k-2}}\cdot W\cdot v_{j_b}^k}&\in(1\pm\O{\eps})|\langle \bw^{p-k},\bv^k\rangle|+\frac{1}{\poly(n)}.
\end{align*}
We can also upper bound the variance by 
\begin{align*}
\Var\big(&\widehat{w_{j_b}^{p-k-2}}\cdot W\cdot v_{j_b}^k\big)\\
&\le\sum_{a\in[n]}\left(\frac{w_a^2}{\|\bw\|_2^2}+\frac{1}{\poly(n)}\right)\cdot w_a^{2p-2k-4}\cdot(1\pm\O{\eps})\|\bw\|_2^4\cdot v_a^{2k}\\
&\le\sum_{a\in[n]}(1\pm\O{\eps})w_a^{2p-2k-2}\cdot\|\bw\|_2^2\cdot v_a^{2k}.
\end{align*}
By H\"{o}lder's inequality, the variance is at most
\begin{align*}
(1\pm\O{\eps})\|\bw\|_2^2&\sum_{a\in[n]}w_a^{2p-2k-2}v_a^{2k}\\
&\le(1\pm\O{\eps})\|\bw\|_2^2\left(\sum_{a\in[n]}v_a^p\right)^{2k/p}\left(\sum_{a\in[n]}w_a^{2p}\right)^{1-2k/p}\\
&=(1\pm\O{\eps})\|\bw\|_2^2\cdot\|\bv\|_p^{2k}\cdot\|\bw\|_{2p}^{2p-4k}.
\end{align*}
We have $(1+\eps)\|\bw\|_2^2\le2\|\bu\|_2^2$ for $\eps\le 1$ and $\|\bv\|_p^p\le\gamma\|\bu\|_p^p$. 
Therefore, the variance is at most
\begin{align*}
2\|\bu\|_2^2\cdot\gamma^{2/p}\|\bu\|_p^2\cdot\|\bw\|_{2p}^{2p-4}. 
\end{align*}
Since the vector $w$ is formed by removing from $u$ the coordinates $i\in\calH$, i.e., the coordinates $i\in[n]$ such that $v_i\ge\frac{\gamma^{1/p}}{16}\|\bv\|_p$, then we have $|w_i|\le\frac{\gamma^{1/p}}{16}\|\bv\|_p$ for all $i\in[n]$. 
Given these constraints, it follows that
\[\|\bw\|_{2p}^{2p}\le\frac{16^p}{\gamma}\cdot\frac{\gamma^2}{16^{2p}}\|\bu\|_p^{2p}.\]
Hence,
\[\|\bw\|_{2p}^{2p-4}\le\frac{\gamma^{1-2/p}}{16^{p-2}}\|\bu\|_p^{2p-4}.\]
Thus, the variance is at most 
\begin{align*}
\|\bu\|_2^2\cdot\gamma^{2/p}\|\bu\|_p^2\cdot\gamma^{1-2/p}\|\bu\|_p^{2p-4}\le\frac{\gamma}{n^{1-2/p}}\|\bu\|_p^2\cdot\|\bu\|_p^2\cdot\|\bu\|_p^{2p-4}. 
\end{align*}
Therefore, for $R=\O{\frac{\gamma}{\eps^2}n^{1-2/p}}$, it follows from Chebyshev's inequality that with probability at least $\frac{2}{3}$, the difference estimator achieves an estimate to $F_p(\bu+\bv)-F_p(\bu)$ with additive error $\frac{\eps}{p}\cdot F_p(\bv)$ approximation to $F_p(\bu+\bv)-F_p(\bu)$. 
The success probability of the difference estimator can then be boosted to $1-\frac{\delta}{\poly(n)}$ by repeating $\O{\log\frac{n}{\delta}}$ times. 
We can then take a union bound over all times in the stream of length $\poly(n)$. 
Hence, we have a $(\gamma,\eps,\delta)$-difference estimator for the $F_p$ moment.  

It remains to analyze total space complexity of this component. 
Firstly, observe that we use $\O{\frac{\gamma}{\eps^2}n^{1-2/p}\log n}$ independent instances of an $F_2$ moment estimation algorithm, as well as the subroutines $\sampler$ and $\countsketch$. 
Since we only require constant factor approximation for $\countsketch$, then each instance uses space $\O{\log^2 n}$. 
We also only require constant factor approximation for each $F_2$ moment estimation algorithm. 
Hence, each instance uses $\O{\log^2 n}$ bits of space, by \thmref{thm:strong:F2}. 
By \thmref{thm:perfect:sampler}, each instance of $\sampler$ uses space $\O{\log^3 n}$. 
Hence, the total space is $\tO{\frac{\gamma}{\eps^2}n^{1-2/p}}$ bits of space. 
\end{proof}

\paragraph{Estimation of $\langle \bu,\bv^{p-1}\rangle$.}
To handle the case where $k=1$, we instead perform perfect $L_2$ sampling from $\bv$. 

\begin{figure*}
\begin{mdframed}
\begin{enumerate}
\item 
Use a set of exponential random variables to form a vector $\overline{\bu}$ of duplicated and scaled coordinates of $\bu$. 
\item
Hash the coordinates of $\overline{\bu}$ into a $\countsketch$ data structure with $\O{\log n}$ buckets. 
\item
Perform perfect $L_2$ sampling on $\bv$ to obtain a set $\calS$ of size $R=\O{\frac{\gamma}{\eps^2}n^{1-2/p}}$.
\item
Use $p-3$ independent instances of $\countsketch$ to obtain unbiased estimates $\widehat{v_i^{p-3}}$ to $v_i^{p-3}$ for each $i\in\calS$. 
\item
Let $\widehat{V}$ be an unbiased estimate of $\|\bv\|_2^2$ with second moment $\O{\|\bv\|_2^4}$. 
\item
For each $i\in\calS$, query $\countsketch$ on $\overline{\bu}$ for an unbiased estimate $\widehat{u_i}$ to $u_i$ with variance $\|\bu\|_2^2\cdot\frac{v_i^2}{\|\bv\|_2^2}$.
\item
Output $\sum_{i\in\calS}\widehat{V}\cdot\widehat{u_i}\cdot\widehat{v_i^{p-3}}$. 
\end{enumerate}
\end{mdframed}
\caption{$F_p$ difference estimator for $\langle \bv,\bu^{p-1}\rangle$ with integer $p>2$.}
\figlab{fig:diff:est:bigp:indexone}
\end{figure*}

\begin{lemma}
\lemlab{lem:diff:est:bigp:one}
\cite{WoodruffZ21}
For integer $p>2$, there exists an algorithm that uses $\tO{\frac{\gamma}{\eps^2}n^{1-2/p}}$ bits of space and with high probability, outputs an estimate to $\sum_{i\notin\calH}u_iv_i^{p-1}$ with additive $\eps\cdot F_p(\bu)$ error. 
\end{lemma}
\begin{proof}
Given an oblivious stream $S$, let $\bu$ be the frequency vector induced by the updates of $S$ from time $t_1$ to $t_2$, and $v$ be the frequency vector induced by updates from time $t_2+1$ to $t$, and suppose $F_p(\bu+\bv)-F_p(\bu)\le\gamma\cdot F_p(\bu)$. 
We use the perfect $L_2$ sampling algorithm $\sampler$ on $v$ to acquire indices $j_1,\ldots,j_R$ so that each sample is a coordinate $a\in[n]$ with probability $\frac{v_a^2}{\|\bv\|_2^2}+\frac{1}{\poly(n)}$. 
Let $\calS$ be the set of samples. 
For each $i\in\calS$, we also run $p-3$ independent instances of $\countsketch$ to obtain unbiased estimates $\widehat{v_i^{p-3}}$ to $v_i^{p-3}$, with variance $\O{v_i^{2p-6}}$. 
We also obtain an unbiased estimate $\widehat{V}$ of $\|\bv\|_2^2$. 
Thus, we can compute the expectation of the estimator as
\begin{align*}
\Ex{\widehat{V}\cdot\widehat{u_i}\cdot\widehat{v_i^{p-3}}}&=\|\bv\|_2^2\sum_{a\in[n]}\left(\frac{v_a^2}{\|\bv\|_2^2}+\frac{1}{\poly(n)}\right)\cdot u_a\cdot v_a^{p-3}\\
&=\langle \bu,\bv^{p-1}\rangle+\frac{1}{\poly(n)},
\end{align*}
as desired. 
We can similarly upper bound the variance by
\begin{align*}
\Var\big(&\widehat{V}\cdot\widehat{u_i}\cdot\widehat{v_i^{p-3}}\big)\\
&\le\O{\|\bv\|_2^4}\cdot\sum_{a\in[n]}\left(\frac{v_a^2}{\|\bv\|_2^2}+\frac{1}{\poly(n)}\right)\cdot\|\bu\|_2^2\cdot\frac{v_i^2}{\|\bv\|_2^2}\cdot\O{v_a^{2p-6}}
\\
&\le\sum_{a\in[n]}\|\bu\|_2^2\cdot\O{v_a^{2p-2}}\\
&\le\|\bu\|_2^2\cdot\O{\|\bv\|_p^{2p-2}}\\
&\le n^{1-2/p}\|\bu\|_p^2\cdot\O{\|\bv\|_p^{2p-2}}\\
&\le\O{\gamma}n^{1-2/p}\cdot\|\bu\|_p^{2p},
\end{align*}
where the last inequality results from the fact that $F_p(\bv)\le\gamma F_p(\bu)$ and $p>2$. 
Therefore, by setting $R=\O{\frac{\gamma}{\eps^2}n^{1-2/p}}$ and Chebyshev's inequality, it follows that with probability at least $\frac{2}{3}$, we obtain an estimate to $\langle \bv,\bu^{p-1}\rangle$ with additive error $\frac{\eps}{p}\cdot F_p(\bv)$. 
We can then boost the success probability to $1-\frac{\delta}{\poly(n)}$ by repeating $\O{\log\frac{n}{\delta}}$ times. 
We can then take a union bound over all times in the stream of length $\poly(n)$. 
Therefore, we have a $(\gamma,\eps,\delta)$-difference estimator for the $F_p$ moment.

It remains to analyze total space complexity of this component. 
Observe that we use $\O{\frac{\gamma}{\eps^2}n^{1-2/p}\log n}$ independent instances of an $F_2$ moment estimation algorithm, as well as the subroutines $\sampler$ and $\countsketch$. 
Because we only require constant factor approximation for $\countsketch$, then each instance uses space $\O{\log^2 n}$. 
Similarly, we only require constant factor approximation for each $F_2$ moment estimation algorithm. 
Therefore, each instance uses $\O{\log^2 n}$ bits of space, by \thmref{thm:strong:F2}. 
By \thmref{thm:perfect:sampler}, each instance of $\sampler$ uses $\O{\log^3 n}$ bits of space. 
Hence, the total space is $\tO{\frac{\gamma}{\eps^2}n^{1-2/p}}$ bits of space. 
\end{proof}

\paragraph{Putting it all together.}
We now describe the complete $F_p$ difference estimator for integer $p>2$ using the above subroutines to estimate the difference $F_p(\bu+\bv)-F_p(\bu)=\sum_{k=1}^{p}\binom{p}{k}\langle \bu^{p-k},\bv^k\rangle$. 
\begin{lemma}[$F_p$ difference estimator for integer $p>2$]
\lemlab{lem:diff:est:Fp:bigp}
\cite{WoodruffZ21}
For integer $p>2$, there exists a $(\gamma,\eps,\delta)$-difference estimator for $F_p$ that uses space $\tO{\frac{\gamma}{\eps^2}n^{1-2/p}\log\frac{1}{\delta}}$. 
\end{lemma}
\begin{proof}
Given an oblivious stream $S$, let $\bu$ be the frequency vector induced by the updates of $S$ from time $t_1$ to $t_2$, and $\bv$ be the frequency vector induced by updates from time $t_2+1$ to $t$, and suppose $F_p(\bu+\bv)-F_p(\bu)\le\gamma\cdot F_p(\bu)$. 
Observe that for integer $p$, we can expand 
\[F_p(\bu+\bv)-F_p(\bu)=\sum_{k=1}^{p}\binom{p}{k}\langle \bu^{p-k},\bv^k\rangle,\]
where $\bu^k$ is used to denote the coordinate-wise $k$-th power of $u$. 
Let $\calH$ be the set of coordinates output by $\heavyhitters$. 
Then we can further decompose
\[F_p(\bu+\bv)-F_p(\bu)=\sum_{i\in\calH}\sum_{k=1}^{p}\binom{p}{k}u_i^{p-k}v_i^k+\sum_{i\notin\calH}\sum_{k=1}^{p}\binom{p}{k}u_i^{p-k}v_i^k.\]

By \thmref{thm:robust:opt:HH}, there exists an algorithm $\heavyhitters$ for the $L_2$-heavy hitter algorithm that uses $\tO{\frac{\gamma^{2-2/p}}{\eps^2}n^{1-2/p}}$ space. 
Moreover, the algorithm outputs a list $\calH$ of heavy-hitters of $\bu$, along with estimates $\widehat{u_i}$ for $i\in\calH$ that have additive error $\frac{\eps\gamma^{1/p}}{64\gamma}\|\bu\|_p$. 
Hence, we can use the estimates $\widehat{u_i}$, along with the corresponding coordinates $v_a$, to achieve an additive $\O{\eps}\cdot F_p(\bu)$ approximation to $\sum_{i\in\calH}\sum_{k=1}^{p}\binom{p}{k}u_i^{p-k}v_i^k$ using $\tO{\frac{\gamma\log^2 n}{\eps^2}\,n^{1-2/p}}$ space. 
Moreover, by applying \lemref{lem:diff:est:bigp:one} and \lemref{lem:diff:est:bigp:general:k}, we can also obtain an additive $\O{\eps}\cdot F_p(\bu)$ approximation to $\sum_{i\notin\calH}\sum_{k=1}^{p}\binom{p}{k}u_i^{p-k}v_i^k$ using $\tO{\frac{\gamma}{\eps^2}n^{1-2/p}\log\frac{1}{\delta}}$ space. 
Therefore, we can rescale $\eps$ and obtain an additive $\eps\cdot F_p(\bu)$ approximation to $F_p(\bu+\bv)-F_p(\bu)$ using $\tO{\frac{\gamma}{\eps^2}n^{1-2/p}\log\frac{1}{\delta}}$ bits of space. 
\end{proof}

Finally, we use the difference estimator to obtain a robust algorithm for $F_p$ moment estimation for integer $p>2$. 
\begin{theorem}[Adversarially robust $F_p$ streaming algorithm for integer $p>2$]
\thmlab{thm:robust:opt:Fp:bigp}
Given $\eps\in\left(0,\frac{1}{2}\right)$ and integer $p>2$, there exists an adversarially robust streaming algorithm that outputs a $(1+\eps)$-approximation for the $F_p$ moment that succeeds with probability at least $\frac{2}{3}$ and uses $\tO{\frac{1}{\eps^2}n^{1-2/p}}$ bits of space. 
\end{theorem}
\begin{proof}
For integer $p>2$, the $F_p$ moment is a monotonic function with $(\eps,m)$-flip number $\O{\frac{1}{\eps}\log n}$. 
By \lemref{lem:diff:est:Fp:bigp} and \thmref{thm:strong:Fp:bigp}, there exists a $(\gamma,\eps,\delta)$-difference estimator and a strong tracker for $F_p$ that both use $\tO{\frac{\gamma}{\eps^2}n^{1-2/p}\log\frac{1}{\delta}}$ bits of space. 
By using these subroutines in the framework of \algref{alg:framework} and applying \thmref{thm:framework}, we obtain an adversarially robust streaming algorithm that outputs a $(1+\eps)$-approximation for the $F_p$ moment, which succeeds with probability $\frac{2}{3}$ and uses $\tO{\frac{\gamma}{\eps^2}n^{1-2/p}\log\frac{1}{\delta}}$ bits of space. 
\end{proof}

\subsection{Difference Estimator for \texorpdfstring{$F_0$}{F0} Estimation}
In this section, we use the framework of \algref{alg:framework} to give an adversarially robust streaming algorithm for the distinct elements problem, or equivalently, $F_0$ moment estimation. 
We again require an $F_0$ strong tracker and an $F_0$ difference estimator so that we can apply \thmref{thm:framework}. 
Fortunately, we can use similar sketches for both the $F_0$ strong tracker and the $F_0$ difference estimator.  
We first recall the $F_0$ strong-tracking algorithm on insertion-only streams~\cite{Blasiok20}: 
\thmstrongFzero*
The algorithm corresponding to \thmref{thm:strong:F0} as well as other $F_0$ approximation algorithms use a balls-and-bins argument with various levels of sophistication~\cite{Bar-YossefJKST02, KaneNW10b, Blasiok20}. 
To construct our $F_0$ difference estimator, we use a similar balls-and-bins argument, where each item is subsampled at a level $k$ with probability $\frac{1}{2^k}$. 
By counting the number of items in a level with $\Theta\left(\frac{\gamma}{\eps^2}\right)$ items that survive the subsampling process for the frequency vector $\bu$, it can be shown that the expected number of the distinct items in the frequency vector $\bv$ but not $\bu$ is $\Theta\left(\frac{\gamma^2}{\eps^2}\right)$. 
Therefore, we first run the balls-and-bins experiment on the frequency vector $\bu$ and counting the number of bins that are occupied at some level $k$ with $\Theta\left(\frac{\gamma}{\eps^2}\right)$ survivors. 
Subsequently, we run the same balls-and-bins experiment on $\bv-\bu$ by only counting the additional bins that are occupied at level $k$, i.e., the items in $\bv$ but not $\bu$, and rescaling this number by $2^k$. 
Note that the items in $\bv$ but not $\bu$ correspond exactly to $F_0(\bu+\bv)-F_0(\bu)$. 
Specifically, we achieve a $\left(1+\frac{\eps}{\gamma}\right)$-approximation to $F_0(\bu+\bv)-F_0(\bu)$, which translates to an additive $\eps\cdot F_0(\bu)$ approximation to $F_0(\bu+\bv)-F_0(\bu)$, since $F_0(\bu+\bv)-F_0(\bu)\le\gamma F(u)$. 

\begin{lemma}[$F_0$ difference estimator]
\lemlab{lem:diff:est:F0}
\cite{WoodruffZ21}
There exists a $(\gamma,\eps,\delta)$-difference estimator for the $F_0$ moment, which uses 
\[\O{\frac{\gamma}{\eps^2}\left(\log\frac{1}{\eps}+\log\log n+\log\frac{1}{\delta}\right)+\log n}\] 
bits of space. 
\end{lemma}
\begin{proof}
Suppose we would like to estimate $F_0(\bu+\bv)-F_0(\bu)\le\gamma\cdot F_0(\bu)$ for $\gamma=\Omega(\eps)$ and we subsample each coordinate $i\in[n]$ with probability $\frac{1}{2^k}$, where $k$ is the integer such that $\frac{F_0(\bu)}{2^k}=\Theta\left(\frac{\gamma}{\eps^2}\right)$. 
In expectation, the number of sampled items in $\bv$ that are not in $\bu$ is $\frac{F_0(\bu+\bv)-F_0(\bu)}{2^k}=\Theta\left(\frac{\gamma^2}{\eps^2}\right)$. 
Let $X$ be the number of survivors from $F_0(\bu+\bv)-F_0(\bu)$ at level $k$. 
Then $\Ex{2^k\cdot X}=F_0(\bu+\bv)-F_0(\bu)$. 
Moreover, the variance of $2^k\cdot X$ is at most
\begin{align*}
2^k\cdot(F_0(\bu+\bv)-F_0(\bu))&\le\frac{\O{\eps^2}\cdot F_0(\bu)}{\gamma}\cdot(F_0(\bu+\bv)-F_0(\bu))\\
&\le\O{\eps^2}\cdot (F_0(\bu))^2.
\end{align*}
Therefore by Chebyshev's inequality, with probability at least $\frac{2}{3}$, the arithmetic mean of a constant number of independent instances of $2^k\cdot X$ gives an estimate to $F_0(\bu+\bv)-F_0(\bu)$ with additive error $\eps\cdot F_0(\bu)$. 
We can then boost the probability of success to at least $1-\delta$ by taking the median of $\log\frac{1}{\delta}$ parallel instances. 

It remains to analyze the space complexity of the algorithm. 
The algorithm maintains $\frac{\gamma}{\eps^2}$ sampled items. 
Na\"{i}vely, the total space used is $\O{\frac{\gamma\log n}{\eps^2}\log\frac{1}{\delta}}$. 
However, by hashing to $P=\poly\left(\frac{1}{\eps},\log n,\frac{1}{\delta}\right)$ buckets, we can further improve the space bounds to 
\[\O{\frac{\gamma}{\eps^2}\left(\log\frac{1}{\eps}+\log\log n+\log\frac{1}{\delta}\right)+\log n}.\] 
In particular, we first compose a hash function $h_1:[n]\to[P^3]$ and then a hash function $h_2:[P^3]\to[P]$, along the lines of \cite{KaneNW10b,Blasiok20}.  
Thus the difference estimator produces a good estimate to the difference at a single point in time. 
To obtain the strong tracking property, observe that both the difference estimator and $F_0(\bu+\bv)-F_0(\bu)$ are monotonic. 
Therefore, it suffices to take a union bound over $\log\frac{1}{\eps}$ times when the difference increases by a factor of $(1+\O{\eps})$. 
Hence, the total space used is $\O{\frac{\gamma}{\eps^2}\left(\log\frac{1}{\eps}+\log\log n+\log\frac{1}{\delta}\right)+\log n}$ bits. 
\end{proof}
Finally, observe that the difference estimator in \lemref{lem:diff:est:F0} only requires pairwise independence. 
Hence, we can derandomize the algorithm by using a hash function that can be stored using $\O{\log n}$ bits of space. 
 
\section{Separation for Graph Coloring}
\seclab{sec:graph:color}
So far, we have presented a number of techniques to achieve adversarial robustness for different problems on insertion-only data streams. 
Frameworks such as sketch switching, bounded computation paths, and difference estimators can all be used to design adversarially robust streaming algorithms for a number of central problems such as norm/moment estimation, distinct element estimation, heavy-hitters, and entropy estimation. 
In fact, the difference estimator framework by \cite{WoodruffZ21} showed that up to $\polylog\left(\frac{1}{\eps}\right)$ terms, there is no ``price'' for adversarial robustness on insertion-only streams. 
That is, the space complexity for the aforementioned problems is nearly the same for the non-adaptive setting and the adversarially robust setting. 
A natural question then, is whether adversarial robustness can always be achieved for insertion-only stream with nearly the same space complexity. 
In this section, we present a result by \cite{ChakrabartiGS22}, which shows a significant gap between the non-adaptive setting and the adversarially robust setting for the graph coloring problem. 

\subsection{Graph Coloring in the Streaming Model}
\seclab{sec:separation:graph:coloring}
We first define the graph coloring problem in the streaming model, where we first fix a vertex set $[n]$ prior to the stream. 
The stream consists of a sequence of edges $\{u, v\}$ being added to the graph; no edge deletions occur in the insertion-only model. 
A \emph{semi-streaming algorithm} for graph streams is one that uses $\tO{n}=n\cdot\polylog(n)$ bits of memory.

\begin{definition}[Graph coloring]
In the \emph{$K$-coloring problem}, the input is a graph stream, and the goal is to assign a color from $[K]$ to each vertex such that adjacent vertices receive different colors. 
That is, the output is a vector in $[K]^n$ where no edge $\{u, v\} \in E(G)$ has $\text{color}(u) = \text{color}(v)$. 
\end{definition}
We remark that the value $K$ in the graph coloring problem may depend on a graph parameter such as the maximum degree $\Delta$ of $G$.
Throughout, we assume $\Delta$ is a sublinear function of $n$, for example $\Delta\le n^\alpha$ for some $\alpha\in(0,1)$. 
Since the output requires $\Theta(n \log K)$ bits, we view any semi-streaming algorithm for $K$-coloring that uses $\tO{n}$ space as having \emph{essentially optimal} space complexity.

We first recall the following property upper bounding the maximum degree of a random graph. 
\begin{lemma}
\lemlab{lem:random-graph-max-degree} 
\cite{ChakrabartiGS22}
Consider a graph $G$ with $n$ vertices and $m$ edges, selected uniformly at random. 
Let $\Delta_G$ denote the maximum degree of $G$. 
Then, for all $0 \le \eps \le 1$, we have:
\[\PPr{\Delta_G \ge \frac{2m}{n} (1 + \eps)} \le 2n \exp\left(-\frac{\eps^2}{3} \cdot \frac{2m}{n} \right)\,.\]
\end{lemma}

The one-way communication game $\avoid(t, a, b)$ is defined by \cite{ChakrabartiGS22} as follows:
\begin{itemize}
  \item Alice receives a subset $S \subseteq [t]$ such that $|S| = a$;
  \item Bob must output a subset $T \subseteq [t]$ with $|T| = b$ such that $S \cap T = \emptyset$.
\end{itemize}
Let $\kavoid(t, a, b)$ denote the problem of solving $k$ independent instances of $\avoid(t, a, b)$ simultaneously.
The following result by \cite{ChakrabartiGS22} gives a lower bound on the total amount of communication necessary to solve the $\kavoid$ problem with probability $1-\delta$. 
\begin{lemma}
\lemlab{lem:disjrec-lb}
\cite{ChakrabartiGS22}
Any public-coin one-way communication protocol that solves the $\kavoid(t,a,b)$ problem with probability at least $1-\delta$ must use $\log{(1-\delta)} + k a b / {(t \ln 2)}$ bits of communication. 
\end{lemma}

\subsection{Reducing Multiple Subset Avoidance to Graph Coloring}
\seclab{sec:avoid-to-coloring}
We now show that graph coloring can solve the $\kavoid$ problem. 
This will ultimately show that the graph coloring problem requires a certain amount of space in the adversarially robust setting. 

To describe the reduction from the $\kavoid$ problem to graph coloring, we begin by analyzing a special case. 
Suppose we are given a streaming algorithm $\calA$ that is adversarially robust and maintains a $(\Delta+1)$-coloring of a graph. 
We describe a protocol that uses $\calA$ to solve an instance of $\avoid(t, a, b)$.

Let $t = \binom{n}{2}$ so that the universe corresponds to all possible edges in an $n$-vertex graph. 
Assume Alice’s input set $A$ has size $a \approx\frac{n^2}{8}$. 
Using public randomness, Alice randomly maps her elements to edges, inducing a graph $G$ that, with high probability, has maximum degree approximately $\Delta \approx\frac{n}{4}$, as morally speaking, each of the $\frac{n^2}{8}$ edges has probability roughly $\frac{2}{n}$ of being incident to a fixed vertex. 
Alice streams the edges of $G$ to $\calA$ and then sends the resulting internal state of $\calA$ to Bob.

Bob queries $\calA$ to obtain a $(\Delta+1)$-coloring of $G$ and then pairs vertices of the same color to form a maximal matching. 
Since at most one vertex in each color class can be left unmatched, Bob obtains at least $\frac{n - \Delta - 1}{2}$ disjoint pairs. 
Each such pair corresponds to a non-edge of $G$, i.e., an element that is not in Alice’s set. 
Because the edge-label mapping is determined by public randomness, Bob knows exactly which elements these are.

Bob then constructs a matching from these pairs and adds the corresponding edges to $\calA$. 
Bob then queries the algorithm again to get a new coloring. 
The addition of the matching increases the maximum degree by at most $1$, so the new coloring uses at most $\Delta + 2$ colors. 
Bob can continue this process—forming matchings from monochromatic pairs, inserting the edges, and querying $\calA$—and still expect correct colorings due to the adversarial robustness of the algorithm.
This process continues until the graph reaches maximum degree $n-1$, at which point each vertex may require a unique color and further progress is not possible.

Across all iterations, Bob adds a total of $\Theta((n - \Delta)^2)$ edges corresponding to missing elements. 
When $\Delta \approx\frac{n}{4}$, this total is $\Theta(n^2)$. 
Thus, the protocol solves an instance of $\avoid(t,a,b)$ with $t = \binom{n}{2}$ and $a, b = \Theta(n^2)$, which requires $\Omega\left(\frac{ab}{t}\right) = \Omega(n^2)$ bits of communication. This implies that $\calA$ must use at least $\Omega(n^2) = \Omega(n \Delta)$ bits of space.

With further refinement, this argument can be generalized to work for any value of $\Delta$ in the range $1 \le \Delta \le \frac{n}{2}$, by using the communication complexity of $\kavoid(t,a,b)$ with carefully chosen parameters. 
A more detailed analysis also allows the result to extend to any $f(\Delta)$-coloring algorithm, not just those using $(\Delta+1)$ colors. 

\begin{algorithm}[!ht]
\caption{Protocol for $\avoid\left(\binom{2 K}{2},\left\lfloor\frac{LK}{4}\right\rfloor,\left\lfloor\frac{L}{2}\right\rfloor\left\lceil\frac{K}{2}\right\rceil\right)$~\cite{ChakrabartiGS22}}
\alglab{alg:recovery}
\begin{algorithmic}[1]
\State \textbf{Require:} Algorithm $\calA$ that colors graphs of maximum degree $L$ using at most $K$ colors
\State{$R \gets$ random bits for $\calA$}
\State{$\pi \gets$ uniform random permutation of $\{1, \ldots, \binom{2K}{2}\}$}
\State{$e_1, \ldots, e_{\binom{2K}{2}} \gets$ edges of the complete graph on $2K$ vertices, ordered arbitrarily}
\Statex
\Function{Alice}{$S$}
\State{$Z \gets \calA\texttt{::INIT}(R)$}
\For{$i = 1$ to $\binom{2K}{2}$}
\If{$\pi_i \in S$}
\State{$Z \gets \calA\texttt{::INSERT}(Z, R, e_i)$}
\EndIf
\EndFor
\State{\Return $Z$}
\EndFunction
\Statex
\Function{Bob}{$Z$}
\State{$J \gets [\ ]$} \Comment{initialize empty list}
\For{$i = 1$ to $\left\lfloor \frac{L}{2} \right\rfloor$}
\State{$\coloralg \gets \calA\texttt{::QUERY}(Z, R)$}
\label{step:iter-get-coloring}
\State $M \gets$ maximal pairing of same-colored vertices from $\coloralg$
\For{each $\{u,v\} \in M$}
\State{$Z \gets \calA\texttt{::INSERT}(Z, R, \{u,v\})$} \Comment{insert edges to turn $M$ into matching}\label{step:update-z-with-m}
\EndFor
\State{$J \gets J \cup M$}
\EndFor
\If{$\text{length}(J) \le \left\lfloor \frac{L}{2} \right\rfloor \left\lceil \frac{K}{2} \right\rceil$}
\State{\Return \texttt{Fail}}
\Else
\State{$T \gets \{\pi_i: e_i \in \text{first } \left\lfloor \frac{L}{2} \right\rfloor \left\lceil \frac{K}{2} \right\rceil \text{ edges of } J\}$}
\label{step:graph-to-set}
\State{\Return $T$}
\EndIf
\EndFunction
\end{algorithmic}
\end{algorithm}

\begin{theorem}
\thmlab{thm:lower-bound-core}
\cite{ChakrabartiGS22}
Let $L, n, K$ be integers satisfying $2K \le n$, $L + 1 \le K$, and $L \ge 12 \ln(4n)$. 
Suppose there exists an adversarially robust insertion-only coloring algorithm $\calA$ for graphs on $n$ vertices with maximum degree at most $L$. 
Suppose further that $\calA$ uses at most $K$ colors and outputs a valid coloring with probability at least $1/4$.
Then, any such algorithm $\calA$ must use at least $C$ bits of memory, where
\[C \ge \frac{1}{40 \ln 2} \cdot \frac{n L^2}{K} - 3 \,.\]
\end{theorem}
\begin{proof}
Given such an adversarially robust algorithm $\calA$ as described earlier, we can design a public-coin protocol for solving the communication problem $\avoid^{\left\lfloor \frac{n}{2K} \right\rfloor}\left( \binom{2K}{2}, \left\lfloor \frac{LK}{4} \right\rfloor, \left\lfloor \frac{L}{2} \right\rfloor \left\lceil \frac{K}{2} \right\rceil \right)$ using the same amount of communication as the space used by $\calA$. 
A protocol for the single-instance version $\avoid\left( \binom{2K}{2}, \left\lfloor \frac{LK}{4} \right\rfloor, \left\lfloor \frac{L}{2} \right\rfloor \left\lceil \frac{K}{2} \right\rceil \right)$ is given in \algref{alg:recovery}.

To implement $\calA$ across $s := \left\lfloor \frac{n}{2K} \right\rfloor$ independent instances, select disjoint vertex subsets $V_1, \ldots, V_s \subseteq [n]$, each of size $2K$. 
We simulate the behavior of a streaming algorithm on the vertex set $[2K]$ by mapping it onto each $V_i$ and applying $\calA$. 
Since the sets $V_i$ are disjoint, this simulation can run concurrently across all $s$ instances. 
Moreover, a valid $K$-coloring of the entire $n$-vertex graph naturally induces valid $K$-colorings on each subgraph supported on the $V_i$.
Moreover, to reduce the number of color queries made to the full graph, the protocol in \algref{alg:recovery} can be executed such that matchings from each instance are inserted alternately (in Step~\ref{step:update-z-with-m}), interleaved with global coloring queries (in Step~\ref{step:iter-get-coloring}).

The correctness of the coloring maintained by $\calA$ first relies on the guarantee that the graph stream has maximum degree at most $L$. 
In Bob’s phase, each inserted matching increases the maximum degree of the graph represented by $Z$ by at most $1$. 
Since this is done $\left\lfloor \frac{L}{2} \right\rfloor$ times, the initial graph inserted by Alice must have maximum degree at most $L - \left\lfloor \frac{L}{2} \right\rfloor \le \frac{L}{2}$. 
By \lemref{lem:random-graph-max-degree}, the probability that any graph on a vertex set $V_i$ exceeds this degree is
\[\PPr{\Delta_i \ge \frac{L}{4}(1 + 1)} \le 4K \cdot e^{-L/12} \,.\]
Applying the union bound across all $s$ subsets gives
\[\PPr{\max_{i \in [s]} \Delta_i \ge \frac{L}{2}} \le 4K \left\lfloor \frac{n}{2K} \right\rfloor e^{-L/12} \le 2n \cdot e^{-L/12} \,.\]
To ensure that this event occurs with probability at most $\frac{1}{2}$, it suffices to require
\[L \ge 12 \ln(4n)\,.\]

Assume that every random graph generated by Alice has maximum degree at most $\frac{L}{2}$, and that all $\left\lfloor\frac{L}{2}\right\rfloor$ colorings requested during the protocol are correct. 
Under these conditions, we will show that Bob's phase of the protocol recovers at least $\left\lfloor \frac{L}{2} \right\rfloor \left\lceil \frac{K}{2} \right\rceil$ edges in each instance.
Since the random bits $R$ used by the algorithm $\calA$ and the permutation $\pi$ are drawn independently, the event that the maximum degree is low, i.e., at most $\frac{L}{2}$, and that $\calA$ outputs correct colorings on all such graphs occurs with probability at least $\frac{1}{2}\cdot\frac{1}{4}=\frac{1}{8}$.

The edges Bob inserts during Step~\ref{step:update-z-with-m} are determined entirely by the coloring output of $\calA$ given its internal state $Z$ and the randomness $R$. 
Each such edge links two vertices sharing the same color, and hence none of these edges could have been previously inserted by Alice or Bob. 
Since these edges depend only on the coloring (and not on the full internal state or randomness), the insertion process is independent of $Z$ and $R$ once the coloring is fixed. 
Therefore, the adversarial robustness guarantee of $\calA$ applies, ensuring that the coloring returned in Step~\ref{step:iter-get-coloring} is correct.

Suppose that all queries succeed and Alice's initial graph has maximum degree at most $\frac{L}{2}$. 
Then in each iteration $i \in [\left\lfloor \frac{L}{2} \right\rfloor]$, the coloring returned by $\calA$ uses at most $K$ colors. 
Let $B$ be the set of vertices matched in the pairing $M$, so that the unmatched set is $[2K] \setminus B$. 
Because no two unmatched vertices share a color, we must have $|[2K] \setminus B| \le K$, which implies $|B| \ge K$. 
Since $M$ is a matching, $|M| = \frac{|B|}{2}$ must be an integer, so $|M| \ge \left\lceil \frac{K}{2}\right\rceil$. 
Thus, each iteration adds at least $\left\lceil \frac{K}{2} \right\rceil$ new edges to the list $J$.
Consequently, after all $\left\lfloor \frac{L}{2} \right\rfloor$ iterations, the list $J$ will contain at least $\left\lfloor \frac{L}{2} \right\rfloor \left\lceil \frac{K}{2} \right\rceil$ edges that were not inserted by Alice. 
Finally, Step~\ref{step:graph-to-set} maps the first such edges to corresponding indices in $\{1, \ldots, \binom{2K}{2}\}$ that are disjoint from the set $S$ given to Alice.

It remains to analyze the communication complexity. 
Applying \lemref{lem:disjrec-lb}, the amount of communication $C$ needed to solve $s := \left\lfloor \frac{n}{2K} \right\rfloor$ independent instances of $\avoid\left( \binom{2K}{2}, \left\lfloor \frac{LK}{4} \right\rfloor, \left\lfloor \frac{L}{2} \right\rfloor \left\lceil \frac{K}{2} \right\rceil \right)$ with total failure probability at most $\frac{7}{8}$ satisfies
\[C \ge \log\left(1 - \frac{7}{8}\right) + \left\lfloor\frac{n}{2K}\right\rfloor \cdot \frac{ \left\lfloor \frac{LK}{4} \right\rfloor \cdot \left\lfloor \frac{L}{2} \right\rfloor \left\lceil \frac{K}{2} \right\rceil }{ \binom{2K}{2} \ln 2 } \,.\]
Using the assumption that $K > L \ge 12 \ln(4n) \ge 12 \ln 4$, we can bound the product in the numerator by $\frac{(LK)^2}{20}$, yielding
\[C \ge \frac{n}{4K} \cdot \frac{(LK)^2 / 20}{\frac{1}{2}(2K)^2 \ln 2} - 3= \frac{n L^2}{40 K \ln 2} - 3 \,.\]
\end{proof}

By invoking \thmref{thm:lower-bound-core} with the choice $K = f(L)$, we directly derive the following corollary, which illustrates some particularly insightful parameter regimes.

\begin{corollary}
\corlab{thm:lb-monotonic}
\cite{ChakrabartiGS22}
Let $f$ be a monotonically increasing function and let $L$ be an integer satisfying $L = \Omega(\log n)$ and $f(L) \le\frac{n}{2}$. 
Consider a coloring algorithm $\calA$ for graphs with maximum degree at most $L$ that uses at most $f(\Delta)$ colors at any time, where $\Delta$ is the current maximum degree, and has overall failure probability at most $\frac{3}{4}$ against an adaptive adversary. 
Then, the space complexity $S$ of $\calA$ satisfies the lower bound
\[S=\Omega\left(\frac{n L^2}{f(L)} \right).\]
In particular:
\begin{itemize}
\item
If $f(\Delta) = \Delta + 1$, or more generally $f(\Delta) = \O{\Delta}$, then the required space is $S = \Omega(nL)$.
\item 
To achieve space usage $S=\tO{n}$, we require $f(\Delta)=\widetilde{\Omega}(\Delta^2)$.
\item 
If $f(L) = \Theta(n)$, then the space lower bound becomes $S = \Omega(L^2)$. 
\end{itemize}
\end{corollary}
We remark that the distinction between $L$ and $\Delta$ happens because the graph changes over time, and the goal is to use a number of colors that is a function of the current maximum degree, and not $L$. 

By comparison, it is known that there exists a semi-streaming algorithm, i.e., algorithm that uses $\tO{n}$ bits of space, which gives a $(\Delta+1)$-coloring~\cite{AssadiCK19}. 
\begin{theorem}
\thmlab{thm:graph:color:oblivious}
\cite{AssadiCK19}
There exists an insertion-only semi-streaming algorithm that uses $\tO{n}$ bits of space and gives a $(\Delta+1)$-coloring of an input graph. 
\end{theorem}
\thmref{thm:graph:color:oblivious} follows from an application of the Palette Sparsification Theorem by \cite{AssadiCK19}, which states that for any graph with $n$ vertices and maximum degree $\Delta$, if we independently and uniformly sample $\O{\log n}$ colors for each vertex from $\{1,\ldots,\Delta+1\}$, then with high probability there exists a proper $(\Delta+1)$-coloring in which each vertex receives a color from its sampled palette. 
Unfortunately, such a result is no longer true when the graph edges are not independent of the sampled palette. 
The streaming algorithm then performs a decomposition of the graph into ``dense'' and ``sparse'' vertices, recovering enough edges to color each vertex. 

In summary, \corref{thm:lb-monotonic} states that $\Omega(\Delta^2)$ colors may be necessary to color a graph by an algorithm using $\tO{n}$ space in the adversarially robust setting. 
By comparison, \thmref{thm:graph:color:oblivious} gives an algorithm that uses $\tO{n}$ space to color a graph with $(\Delta+1)$ colors in the oblivious (non-adaptive) setting. 
Thus, \corref{thm:lb-monotonic} and \thmref{thm:graph:color:oblivious} together show a separation for the graph coloring problem between the non-adaptive and the adversarially robust settings. 

\chapter{Differential Privacy and Adaptive Data Analysis}
\chaplab{chap:dp:ada}

\begin{tallchapterbannerbox}
\centering
Differential privacy can be used to hide the randomness of \\
an algorithm from an adversary, improving the number of\\
adversarial queries an algorithm can tolerate. 
\end{tallchapterbannerbox}
\vspace{0.4in}

In this section, we show surprising connections between adversarial robustness and both differential privacy and adaptive data analysis. 
In \secref{sec:dp}, we first describe a framework introduced by \cite{HassidimKMMS22} that utilizes differential privacy to transform non-adaptive streaming algorithms to adversarially robust streaming algorithms; chronologically, these results first appeared in \cite{HassidimKMMS20}. 
In \secref{sec:ada}, we describe a result by \cite{KaplanMNS21}, which showed another separation between the (non-adaptive) insertion-only streaming model and the adversarially robust streaming model. 
This separation is for a more statistical problem, rather than a graph-theoretic problem\footnote{In fact this result by \cite{KaplanMNS21} pre-dated the result by \cite{ChakrabartiGS22} and asked for a separation for a more ``natural'' problem as an open question, which was subsequently resolved by \cite{ChakrabartiGS22}.}

\section{Differential Privacy}
\seclab{sec:dp}
We describe a connection drawn by \cite{HassidimKMMS22} between adversarial robustness in streaming algorithms and the concept of \emph{differential privacy}, a rigorous framework for ensuring privacy when analyzing sensitive data. 
Consider a dataset that includes private information about individuals. 
An algorithm is said to satisfy differential privacy if its output distribution does not significantly change when the data of a single individual is modified.
This means that no individual data has a substantial influence on the algorithmic outcome, thereby limiting the amount of information that can be inferred about any specific person. 
Intuitively, differential privacy ensures that anything learned about an individual could have also been learned had their data been arbitrarily changed or removed altogether. 
Formally, differential privacy provides a probabilistic guarantee that prevents the output from revealing whether any particular individual's data was included in the input:
\defdp*
For preliminaries on differential privacy, see \secref{sec:prelims:dp}

A central conceptual contribution of \cite{HassidimKMMS22} is to demonstrate that differential privacy can be leveraged as a principled tool for designing new adversarially robust streaming algorithms. 
At a high level, the key idea is to apply differential privacy techniques to safeguard the internal state of the algorithm.
Informally, this approach restricts, in a well-defined manner, the extent to which the internal state can depend on the adaptively chosen items in the stream. 
As a result, it becomes possible to reason about the algorithm’s utility guarantees even in adversarial or adaptive settings.

It is important to emphasize that differential privacy is not employed here for the usual purpose of protecting the privacy of the stream data itself. 
Instead, it is used to obscure the internal randomness and state of the algorithm, thereby providing robustness against adaptive inputs.

\paragraph{Framework overview.}
Consider an oblivious streaming algorithm $\calA$ for computing a function $g$. 
\cite{HassidimKMMS22} constructs an adversarially robust version of $\calA$ by running $k$ independent instances of the algorithm in parallel, each with its own independent source of randomness. 
The input stream is fed simultaneously to all $k$ instances.

Upon receiving a query, the framework computes an aggregated response from the $k$ copies in a way that masks their internal randomness using differential privacy. 
This ensures that the aggregation does not leak sensitive information about the individual internal states, thereby achieving robustness.
Furthermore, under the assumption that the stream has a low flip number, then this aggregation only needs to be performed at a small number of time steps. 
To identify the relevant time steps, the framework invokes the sparse vector technique, c.f., \algref{alg:svt}. 
Once such a time step is detected, an approximate median of the $k$ responses is computed using a differentially private selection algorithm.
The framework is given in \algref{alg:dp:robust}. 

\begin{algorithm}[!htb]
\caption{Robust Streaming via Differential Privacy~\cite{HassidimKMMS22}}
\alglab{alg:dp:robust}
\begin{algorithmic}[1]
\Require{Accuracy parameter $\eps\in(0,1)$, an adaptive stream $(a_t,\Delta_t)$ for $t\in[m]$, and an $(\eps,\delta)$-strong tracker for a given function}
\Ensure{Adversarially robust $(1+\eps)$-streaming algorithm}
\State{$\lambda\gets\lambda_{\alpha/10,m}(f)$, $\eps\gets\frac{1}{100}$, $\eps_0\gets\frac{\eps}{16\sqrt{\lambda\ln\frac{1}{\delta}}}$}
\State{$k\gets\Theta\left(\frac{1}{\eps}\sqrt{\lambda\log\frac{1}{\delta}}\log\frac{m}{\alpha\delta}\right)$, $\widehat{f}\gets f(\vec{0})$}
\State{Initialize independent $\left(1+\frac{\alpha}{10}\right)$-approximations for $f$ as $\calA_1,\ldots,\calA_k$, with probability $\frac{9}{10}$}
\For{at most $\lambda$ times}
\label{alg:dp:robust:outer:loop}
\State{$\widehat{t}\gets\frac{k}{2}+\Lap\left(\frac{1}{\eps_0}\right)$}
\While{$\left\lvert\left\{j:\widehat{f}\notin\left(1\pm\frac{\alpha}{2}\right)\cdot y_{i,j}\right\}\right\rvert+\Lap\left(\frac{1}{\eps_0}\right)<\widehat{t}$}
\label{alg:dp:robust:inner:loop}
\State{Update $\calA_1,\ldots,\calA_k$ with next update in the stream and obtain answers $y_{i,1},\ldots,y_{i,k}$}
\State{Output $\widehat{f}$}
\EndWhile
\State{Let $\PrivMed$ be $\eps_0$-DP algorithm for private median}
\Comment{\thmref{thm:dp:median}}
\State{Recompute $\widehat{f}\gets\PrivMed(y_{i,1},\ldots,y_{i,k})$}
\EndFor
\end{algorithmic}
\end{algorithm}

The following statement shows that \algref{alg:dp:robust} preserves the privacy of the internal randomness of the algorithms $\calA_1,\ldots,\calA_k$.
\begin{lemma}
\cite{HassidimKMMS22}
Let $r_1,\ldots,r_k$ be the internal randomness of $\calA_1,\ldots,\calA_k$ in \algref{alg:dp:robust} and let $R=\{r_1,\ldots,r_k\}$. 
Then \algref{alg:dp:robust} is $(\eps,\delta)$-differentially private with respect to $R$.
\end{lemma}
The statement holds from the observation that the inner loop in Line~\ref{alg:dp:robust:inner:loop} is an application of \textsc{AboveThreshold}, c.f., \algref{alg:svt} and \thmref{thm:svt}. 
On the other hand, the outer loop in Line~\ref{alg:dp:robust:outer:loop} applies the private median algorithm $\PrivMed$ with a sufficiently low privacy loss to be satisfactorily amplified by advanced composition, c.f., \thmref{thm:dp:adv:comp}. 

We next show that all outputs of \algref{alg:dp:robust} are sufficiently accurate, while the outer loop has run at most $\lambda$ times. 
\begin{lemma}
\lemlab{lem:robust:dp:correct}
\cite{HassidimKMMS22}
Let $\calA$ be an oblivious strong tracking algorithm for a function $f$ that provides a $\left(1+\frac{\alpha}{10}\right)$-approximation with probability at least $\frac{9}{10}$. 
Then with probability at least $1-\O{\delta}$, \algref{alg:dp:robust} is a $(1+\alpha)$-adversarially robust streaming algorithm for $f$, before the outer loop in Line~\ref{alg:dp:robust:outer:loop} terminates.
\end{lemma}
\begin{proof}
Let $\calE_1$ be the event that all random variables sampled from the Laplace distribution by \algref{alg:dp:robust} are at most $\frac{1}{\eps_0}\log\frac{4m}{\delta}$ in magnitude. 
Observe that for a stream of length $m$, at most $2m$ random variables will be sampled from the Laplace distribution by \algref{alg:dp:robust}. 
Hence, $\calE_1$ holds with probability at least $1-\delta$. 

For $i\in[m]$, let $X_i$ denote the dataset after the first $i$ updates of the stream and let $\calA(r,X_i)$ denote the output of the oblivious streaming algorithm on input $X_i$, using a random string $r$. 
We define the indicator function for an accurate output as 
\[g(r,X_i)=\mathbbm{1}\left\{\calA(r,X_i)\in\left(1+\frac{\alpha}{10}\right)\cdot f(X_i)\right\}.\]
Let $\calE_2$ be the event that for every $i\in[m]$, we have
\[\left\lvert\EEx{r}{g(r,X_i)}-\frac{1}{k}\sum_{j=1}^k g(r_j,X_i)\right\rvert\le10\eps.\]
By the generalization properties of differential privacy, c.f., \thmref{thm:generalization} for $k\ge\frac{1}{\eps^2}\log\frac{2\eps m}{\delta}$, along with a union bound over all $i\in[m]$, then we have that $\calE_2$ holds with probability at least $1-\frac{\delta}{\eps}$. 
Since $\eps=\frac{1}{100}$, then we have $\PPr{\calE_2}\ge 1-\O{\delta}$. 

As each oblivious streaming algorithm $\calA$ is a $\left(1+\frac{\alpha}{10}\right)$-approximation with probability at least $\frac{9}{10}$, we have $\EEx{r}{g(r,X_i)}\ge\frac{9}{10}$. 
Since $\eps\le\frac{1}{100}$, then conditioned on $\calE_2$, at least $\left(\frac{9}{10}-10\eps\right)k\ge\frac{4}{5}k$ independent instances of $\calA$ satisfy $g(r_j,X_i)=1$, so that $y_{i,j}\in\left(1\pm\frac{\alpha}{10}\right)\cdot f(X_i)$. 
We now consider casework on whether the output by \algref{alg:dp:robust} is returned by the inner loop in Line~\ref{alg:dp:robust:inner:loop} or the outer loop in Line~\ref{alg:dp:robust:outer:loop}.

In the first case, the output by \algref{alg:dp:robust} is returned by the inner loop in Line~\ref{alg:dp:robust:inner:loop}. 
Hence, conditioned on $\calE_1$,
\[\left\lvert\left\{j:\widehat{f}\in\left(1\pm\frac{\alpha}{2}\right)\cdot y_{i,j}\right\}\right\rvert\ge\frac{k}{2}-\frac{2}{\eps_0}\log\frac{4m}{\delta}\ge\frac{4k}{10},\]
since $k=\Theta\left(\frac{1}{\eps}\sqrt{\lambda\log\frac{1}{\delta}}\log\frac{m}{\alpha\delta}\right)$. 
Therefore, at least $\frac{4}{5}k$ indices $j\in[k]$ satisfy $y_{i,j}\in\left(1\pm\frac{\alpha}{10}\right)\cdot f(X_i)$ conditioned on $\calE_2$ and at least $\frac{4}{10}k$ indices $j\in[k]$ satisfy $\widehat{f}\in\left(1\pm\frac{\alpha}{2}\right)\cdot y_{i,j}$. 
Thus, there exists $j\in[k]$ that satisfies both conditions, which implies $\widehat{f}\in\left(1\pm\alpha\right)\cdot f(X_i)$, so that the output of the algorithm is a $(1\pm\alpha)$-approximation. 

In the second case, the output by \algref{alg:dp:robust} is returned by the outer loop in Line~\ref{alg:dp:robust:outer:loop}, corresponding to the output of $\PrivMed(y_{i,1},\ldots,y_{i,k})$. 
For $k=\Omega\left(\frac{1}{\eps}\sqrt{\lambda\log\frac{1}{\delta}}\log\frac{m}{\alpha\delta}\right)$ and $m\ge\lambda\log n$, then by \thmref{thm:dp:median}, we have
\[\left\lvert\left\{j:y_{i,j}\ge\widehat{f}\right\}\right\rvert\ge\frac{4k}{10},\qquad\left\lvert\left\{j:y_{i,j}\le\widehat{f}\right\}\right\rvert\ge\frac{4k}{10},\]
with probability at least $1-\frac{\delta}{\lambda}$. 
Conditioned on $\calE_2$, at least $\frac{4}{5}k$ indices $j\in[k]$ satisfy $y_{i,j}\in\left(1\pm\frac{\alpha}{10}\right)\cdot f(X_i)$. 
Therefore, $\widehat{f}\in\left(1\pm\frac{\alpha}{10}\right)\cdot f(X_i)$. 
We then have correctness for all such $\widehat{f}$ output by the outer loop of \algref{alg:dp:robust} by taking a union bound over $\lambda$. 

Putting things together, it follows that with probability at least $1-\O{\delta}$, \algref{alg:dp:robust} outputs a $\left(1\pm\alpha\right)$-approximation until the outer loop terminates. 
\end{proof}

It remains to show that with high probability, the outer loop does not terminate before the stream is completely processed.
\begin{lemma}
\lemlab{lem:robust:dp:complete}
\cite{HassidimKMMS22}
With probability at least $1-\delta$, \algref{alg:dp:robust} does not terminate.
\end{lemma}
\begin{proof}
Similar to the proof of \lemref{lem:robust:dp:correct}, we define $\calE$ to be the event that:
\begin{enumerate}
\item 
All random variables sampled from the Laplace distribution are at most $\frac{1}{\eps_0}\log\frac{2m}{\delta}$ in magnitude. 
\item
All estimates computed by the outer loop are $\left(1\pm\frac{\alpha}{10}\right)$-approximations. 
\item
For each $i\in[m]$, at least $\frac{4}{5}k$ indices $j\in[k]$ satisfy $y_{i,j}\in\left(1\pm\frac{\alpha}{10}\right)\cdot f(X_i)$.
\end{enumerate}
We recall that the event $\calE$ holds with probability $1-\O{\delta}$. 
For each $i\in[m]$, let $\widehat{f_i}$ be the estimate $\widehat{f}$ at time $i$. 
Moreover, let $i_1<i_2$ be fixed times in which the algorithm outputs the same estimate from the inner loop, i.e., $\widehat{f_{i_1}}=\widehat{f_{i_2-1}}$ and $\widehat{f_{i_2-1}}\neq\widehat{f_{i_2}}$. 

Because $\widehat{f_{i_2}}$ is determined from the outer loop, then it follows that
\[\left\lvert\left\{j:\widehat{f_{i_2-1}}\notin\left(1\pm\frac{\alpha}{2}\right)\cdot y_{i_2,j}\right\}\right\rvert\ge\frac{4}{10}k.\]
Since at least $\frac{4}{5}k$ indices $j\in[k]$ satisfy $y_{i_2,j}\in\left(1\pm\frac{\alpha}{10}\right)\cdot f(X_{i_2})$, then there exists an index $j$ such that 
\[\widehat{f_{i_2-1}}\notin\left(1\pm\frac{\alpha}{2}\right)\cdot y_{i_2,j},\qquad y_{i_2,j}\in\left(1\pm\frac{\alpha}{10}\right)\cdot f(X_{i_2}).\]
Therefore, 
\[\widehat{f_{i_2-1}}\notin\left(1\pm\frac{\alpha}{4}\right)\cdot f(X_{i_2}).\] 
Since $\widehat{f_{i_1}}=\widehat{f_{i_2-1}}$, then it follows that $\widehat{f_{i_2-1}}\in\left(1\pm\frac{\alpha}{10}\right)\cdot f(X_{i_1})$, which implies
\[f(X_{i_2})\notin\left(1\pm\frac{\alpha}{10}\right)\cdot f(X_{i_1}).\]
In other words, each time $\widehat{f}$ is computed using the outer loop, the true value of the function $f$ on the data stream must have either increased by a multiplicative factor at least $\left(1+\frac{\alpha}{10}\right)$ or decreased by a multiplicative factor of $\left(1-\frac{\alpha}{10}\right)$. 
Hence, the outer loop is not called more than $\lambda_{\alpha/10,m}(f)$ times with high probability. 
Since $\lambda=\lambda_{\alpha/10,m}(f)$, then it follows that the algorithm does not terminate before the stream is complete, with high probability.  
\end{proof}
Putting together \lemref{lem:robust:dp:correct} and \lemref{lem:robust:dp:complete}, we have the following guarantees for the framework in \algref{alg:dp:robust}. 
\begin{theorem}
\thmlab{thm:frame:dp}
\cite{HassidimKMMS22}
Given an oblivious streaming algorithm $\calA$ for $f$ that uses space $S\left(m,\frac{\alpha}{10},\frac{1}{10}\right)$ for $\left(1\pm\frac{\alpha}{10}\right)$-approximation on a stream of length $m$ with failure probability at most $\frac{1}{10}$, there exists an adversarially robust streaming algorithm for $f$ that produces a $(1+\alpha)$-approximation with probability at least $1-\delta$, using space
\[\O{S\left(m,\frac{\alpha}{10},\frac{1}{10}\right)\cdot\sqrt{\lambda_{\alpha/10,m}(f)\cdot\log\frac{1}{\delta}}\cdot\log\frac{m}{\alpha\delta}}.\]
\end{theorem}

\paragraph{Applications for DP framework.}
We mention that the framework of \algref{alg:dp:robust} can be used to answer a number of adaptive queries in a number of other settings~\cite{BeimelKMNSS22,AttiasCSS23,CherapanamjeriSWZZZ23}, e.g., for graph algorithms in the dynamic setting, for matrix-vector norm queries, linear regression, half-space queries, point queries on turnstile streams, adaptive distance estimation, and adaptive kernel density estimation. 
We provide additional details on the specific primitive in \secref{sec:dp:framework:isolate}. 

We now briefly describe a number of applications of the framework of \algref{alg:dp:robust}. 
First, we observe that by combining the guarantees of the framework in \thmref{thm:frame:dp} with the flip number bounds in \corref{cor:fpflip} and the strong-tracking bounds in either \thmref{thm:strong:F2} or \thmref{thm:strong:Fp:smallp}, we have the following:
\begin{theorem}
\thmlab{thm:fp:smallp:dp}
\cite{HassidimKMMS22}
Given $p\in(0,2]$ and $\eps\in(0,1)$, there exists an adversarially robust algorithm on insertion-only streams of length $m=\poly(n)$ that uses $\tO{\frac{1}{\eps^{2.5}}\log^4 n}$ bits of space and with probability at least $\frac{2}{3}$, outputs a $(1+\eps)$-approximation to the $F_p$ moment at all times.
\end{theorem}
Similarly, by combining the guarantees of the framework in \thmref{thm:frame:dp} with the flip number bounds in \corref{cor:fpflip} and the strong-tracking bounds in \thmref{thm:Fp:bigp}, we have the following:
\begin{theorem}
\thmlab{thm:fp:bigp:dp}
\cite{HassidimKMMS22}
Given $p>2$ and $\eps\in(0,1)$, there exists an adversarially robust algorithm on insertion-only streams of length $m=\poly(n)$ that uses $\tO{\frac{1}{\eps^{2.5}}n^{1-2/p}}$ bits of space and with probability at least $\frac{2}{3}$, outputs a $(1+\eps)$-approximation to the $F_p$ moment at all times.
\end{theorem}

Finally, by combining \thmref{thm:frame:dp} with the flip number bounds in \corref{cor:fpflip} and the strong-tracker of \thmref{thm:strong:F0}, we have the following:

\begin{theorem}
\thmlab{thm:fzero:bigp:dp}
\cite{HassidimKMMS22}
Given $\eps\in(0,1)$, there exists an adversarially robust algorithm on insertion-only streams of length $m=\poly(n)$ that uses $\tO{\frac{1}{\eps^{2.5}}\log^3 n}$ bits of space and with probability at least $\frac{2}{3}$, outputs a $(1+\eps)$-approximation to the number of distinct elements at all times.
\end{theorem}

We summarize the relevant results in \figref{fig:robust:results}. 

\begin{figure*}[!htb]
\begin{center}
\resizebox{\columnwidth}{!}{
{\tabulinesep=1.2mm
\begin{tabu}{|c|c|c|c|}\hline
Problem & \cite{Ben-EliezerJWY22} & \cite{HassidimKMMS22} & \cite{WoodruffZ21} \\\hline\hline
Distinct Elements & $\tO{\frac{\log n}{\eps^3}}$ & $\tO{\frac{\log^4 n}{\eps^{2.5}}}$ & $\tO{\frac{1}{\eps^2}+\frac{\log n}{\eps}}$ \\\hline
$F_p$ Estimation, $p\in(0,2]$ & $\tO{\frac{\log n}{\eps^3}}$ & $\tO{\frac{\log^4 n}{\eps^{2.5}}}$ & $\tO{\frac{\log n}{\eps^2}}$ \\\hline
Shannon Entropy & $\tO{\frac{\log^6 n}{\eps^5}}$ & $\tO{\frac{\log^4 n}{\eps^{3.5}}}$ & $\tO{\frac{\log^3 n}{\eps^2}}$ \\\hline
$L_2$-Heavy Hitters & $\tO{\frac{\log n}{\eps^3}}$ & $\tO{\frac{\log^4 n}{\eps^{2.5}}}$ & $\tO{\frac{\log n}{\eps^2}}$ \\\hline
$F_p$ Estimation, integer $p>2$ & $\tO{\frac{n^{1-2/p}}{\eps^3}}$ & $\tO{\frac{n^{1-2/p}}{\eps^{2.5}}}$ & $\tO{\frac{n^{1-2/p}}{\eps^2}}$ \\\hline
\end{tabu}
}}
\end{center}
\caption{Space bounds for black-box robust streaming algorithms for central problems in the insertion-only model}
\figlab{fig:robust:results}
\end{figure*}

Finally, we emphasize that since the results of \cite{HassidimKMMS22} achieve sublinear dependence on the flip number, they achieve space sublinear in the stream length for the important setting of turnstile streams. 
By comparison, previously discussed techniques such as sketching switching, bounded computation paths, and difference estimators achieve space with a linear dependency in the flip number, and thus not necessarily space sublinear in the stream length for turnstile streams. 

\section{Robust Data Structures}
\seclab{sec:dp:framework:isolate}
We now isolate a specific algorithmic paradigm from the framework in \secref{sec:dp}, for answering a number of adaptive queries. 
In particular, the framework was introduced for the streaming model by~\cite{HassidimKMMS22} and for the dynamic model by~\cite{BeimelKMNSS22,AttiasCSS23}, streamlined in \algref{alg:dp:robust:framework}. 
We remark that this framework can be used to achieve an adversarially robust algorithm for $F_2$ moment estimation on turnstile streams of length $m$, using $\tO{\sqrt{m}}$ space for $\eps=\Omega(1)$. 

\begin{algorithm}[!htb]
\caption{Adversarially Robust Framework}
\alglab{alg:dp:robust:framework}
\begin{algorithmic}[1]
\Require{Oblivious algorithm $\calA$ with success probability $\frac{2}{3}$, number $Q$ of adaptive queries, failure probability $\delta$}
\Ensure{Algorithm robust to $Q$ adaptive queries, with failure probability at most $\delta$}
\State{$\tau\gets\O{\sqrt{Q}\log^2\frac{nQ}{\delta}}$}
\State{Implement $T=\O{\tau}$ independent instances $\calA_1,\ldots,\calA_T$ of $\calA$ on the input}
\For{each query $q_i$, $i\in[Q]$}
\State{Let $Z_{i,j}$ be the output of $\calA_j$ on $q_i$}
\State{Let $\PrivMed$ be $\left(\frac{\log(nQ)}{\tau},0\right)$-DP algorithm for private median}
\Comment{\thmref{thm:dp:median}}
\State{\Return $\PrivMed(\{Z_{i,j}\}_{j\in[T]})$}
\EndFor
\end{algorithmic}
\end{algorithm}

An important ingredient to robust algorithms based on differential privacy is the following property, given by \algref{alg:dp:robust:framework}. 

\begin{restatable}{theorem}{thmrobustdp}
\thmlab{thm:robust:dp}
\cite{HassidimKMMS22,BeimelKMNSS22,AttiasCSS23,CherapanamjeriSWZZZ23}
Given an algorithm $\calA$ that uses $S$ space and answers a query with probability at least $\frac{2}{3}$, there exists a data structure that answers $Q$ adaptive queries, with probability $1-\delta$ using space $\O{S\sqrt{Q}\log^2\frac{nQ}{\delta}}$.
\end{restatable}

To prove \thmref{thm:robust:dp}, we first show \algref{alg:dp:robust:framework} is accurate across all $Q$ rounds of interaction with an adaptive adversary. 
Let $R=\{r_1,\ldots,r_T\}\cup R_0$ be the set of random strings used by the oblivious algorithms $\calA_1,\ldots,\calA_T$ and the additional randomness $R_0$ used by \algref{alg:dp:robust:framework}, e.g., in the private median subroutine $\PrivMed$. 
Let $\Pi(R)=\{\Pi_1,\ldots,\Pi_Q\}$ be a transcript, where for each $i\in[Q]$, $\Pi_i=(q_i,Z_i)$ is the ordered pair consisting of the query $q_i$ and the corresponding answer $Z_i$ by \algref{alg:dp:robust:framework}. 
We first claim the transcript $\Pi(R)$ is differentially private with respect to $R$. 

\begin{lemma}
\lemlab{lem:dp:single:iter}
\cite{HassidimKMMS22}
For each fixed iteration $i\in[Q]$, $\Pi_i$ is $\left(\O{\frac{1}{\sqrt{Q}\log^2\frac{nQ}{\delta}}},0\right)$-differentially private with respect to $R$, conditioned on $\Pi_1,\ldots,\Pi_{i-1}$. 
\end{lemma}
\begin{proof}
The claim follows immediately from the setting of $\PrivMed$ to be $\left(\frac{1}{\tau},0\right)$-differentially private on the outputs of the $T=\O{\sqrt{Q}\log(nQ)}$ algorithms, for $\tau=\O{\sqrt{Q}\log\frac{nQ}{\delta}}$. 
\end{proof}

By advanced composition, it then follows that the entire transcript $\Pi$ is differentially private with respect to the randomness $R$. 
\begin{lemma}
\cite{HassidimKMMS22}
\lemlab{lem:dp:all:iter}
$\Pi$ is $\left(\O{\frac{1}{\log(nQ)}},\frac{1}{\poly(nQ)}\right)$-differentially private with respect to $R$. 
\end{lemma}
\begin{proof}
By \lemref{lem:dp:single:iter}, for each fixed iteration $i\in[Q]$, the transcript $\Pi_i$ is $\left(\O{\frac{1}{\sqrt{Q}\log^2\frac{nQ}{\delta}}},0\right)$-differentially private with respect to $R$. 
Since $\Pi=(\Pi_1,\ldots,\Pi_Q)$, then by the advanced composition of differential privacy, i.e., \thmref{thm:dp:adv:comp}, it follows that the transcript $\Pi$ is $\left(\O{\frac{1}{\log\frac{nQ}{\delta}}},\frac{1}{\poly(nQ)}\right)$-differentially private with respect to $R$. 
\end{proof}

We now prove the correctness of \algref{alg:dp:robust:framework}. 
\thmrobustdp*
\begin{proof}
Consider a sequence of $Q$ adaptive queries and the corresponding transcript $\Pi(R)=\{\Pi_1,\ldots,\Pi_Q\}$ for the randomness $R=\{r_1,\ldots,r_T\}\cup R_0$ used by the oblivious algorithms $\calA_1,\ldots,\calA_T$ and by \algref{alg:dp:robust:framework}. 
For each $i\in[Q]$, let $T_i=(q_i,Z_i)$ be the ordered pair consisting of the query $q_i$ and the corresponding answer $Z_i$ by \algref{alg:dp:robust:framework}. 
By \lemref{lem:dp:all:iter}, the transcript $\Pi$ is $\left(\O{\frac{1}{\log\frac{nQ}{\delta}}},\frac{1}{\poly(nQ)}\right)$-differentially private with respect to $R$. 

For $j\in[T]$, let $Y_j$ be the indicator variable for whether the output $Z_{i,j}$ by algorithm $\calA_j$ is successful on query $q_i$, e.g., in the context of approximation algorithms, within $(1+\alpha)$-approximation of the true answer for query $q_i$. 
From the generalization properties of differential privacy, i.e., \thmref{thm:generalization}, 
\[\PPr{\left\lvert\frac{1}{T}\sum_{j\in[T]}Y_j - \EEx{\calA\sim r}{\mathbbm{1}[\calA(q_i)]}\right\rvert\ge\frac{1}{10}}<\frac{1}{\poly(n,Q)},\]
where the indicator variable $\mathbbm{1}$ denotes correctness of the algorithm $\calA$ on input $q_i$ using randomness $r$. 
Since $\calA$ succeeds with probability $\frac{2}{3}$, then we have
\[\PPr{\frac{1}{T}\sum_{j\in[T]}Y_j\ge\frac{17}{30}}\ge 1-\frac{1}{\poly(n,Q)}.\]
Therefore by the guarantees of $\PrivMed$ with $\left(\frac{\log(nQ)}{\tau},0\right)$-differential privacy, c.f., \thmref{thm:dp:median}, we have that the output $Z_i$ is correct with probability $1-\frac{\delta}{Q}$. 
By a union bound over all $Q$ queries, it follows that all queries are correct with probability at least $1-\delta$.
\end{proof}

We now give a number of applications to \thmref{thm:robust:dp}. 
For additional applications, see \cite{HassidimKMMS22,BeimelKMNSS22,AttiasCSS23,CherapanamjeriSWZZZ23,FengFLSWZ25}. 

\subsection{Applications to Matrix-Vector Norm Approximation}
In the matrix-vector norm query problem, we are given a matrix $\bA \in \mathbb{R}^{n \times d}$ and wish to process $Q$ adaptive queries $\bx^{(1)}, \ldots, \bx^{(Q)}$ for an approximation parameter $\eps > 0$, such that each query $\bx^{(i)} \in \mathbb{R}^d$ is answered with a $(1+\eps)$-approximation to $\|\bA \bx^{(i)}\|_p$. 
Recall that we define the $p$-norm through its $p$-th power as $\|\bv\|_p^p = \sum_{i \in [d]} |v_i|^p$ for any $\bv \in \mathbb{R}^d$. 

Directly computing $\bA\bx^{(i)}$ and then its $p$-norm requires $\O{nd}$ time per query. 
When $n \gg d$, a faster alternative is to compute a subspace embedding: a matrix $\bM \in \mathbb{R}^{m \times d}$ with $m \ll n$ such that for all $\bx \in \mathbb{R}^d$,
\[(1 - \eps)\|\bA\bx\|_p \le \|\bM\bx\|_p \le (1 + \eps)\|\bA\bx\|_p,\]
c.f., \defref{def:lp:subspace}. 
However, since such embeddings must preserve the norm for all possible inputs, the required number of rows is typically $m = \Omega\left(\frac{d}{\eps^2}\right)$, driven by the need to approximate over an $\eps$-net.
Instead, we utilize the following result:
\begin{theorem}
\cite{Indyk06,Li08}
\thmlab{thm:p:stable}
For any $\bA \in \mathbb{R}^{n \times d}$, $p \in (0,2]$, and accuracy parameter $\eps > 0$, there exists a sketching algorithm that constructs a data structure using $\O{\frac{1}{\eps^2}\log n}$ bits and returns a $(1+\eps)$-approximation to $\|\bA \bx\|_p$ for any $\bx \in \mathbb{R}^d$ with high probability, in time $\O{\frac{d}{\eps^2} \log n}$.
\end{theorem}

\thmref{thm:p:stable} works by constructing a random matrix $\bR \in \mathbb{R}^{m \times n}$ with entries drawn from a $p$-stable distribution~\cite{Zolotarev89}, c.f., \defref{def:pstable}, and storing the compressed matrix $\bR\bA$. 
At query time, the procedure computes a fixed function of $\bR\bA\bx$ to approximate $\|\bA\bx\|_p$ up to a $(1+\eps)$ factor. 
The constraint $p \in (0,2]$ arises because $p$-stable distributions are only defined in this range.

Combining \thmref{thm:p:stable} with \thmref{thm:robust:dp}, we obtain the following:

\begin{theorem}
\cite{CherapanamjeriSWZZZ23}
Given a matrix $\bA \in \mathbb{R}^{n \times d}$, $p \in (0,2]$, and accuracy parameter $\eps > 0$, there is an algorithm that constructs a data structure using $\O{\frac{\sqrt{Q}}{\eps^2}\log^2(nQ)}$ bits of space and returns a $(1+\eps)$-approximation to $\|\bA\bx^{(i)}\|_p$ for each of $Q$ adaptive queries $\bx^{(1)}, \ldots, \bx^{(Q)} \in \mathbb{R}^d$, with high probability. 
Each query is answered in time $\tO{\frac{d}{\eps^2} \log^2(nQ) + \log^3(nQ)}$.
\end{theorem}

\subsection{Applications to Linear Regression}
In the linear regression problem, we are given a fixed matrix $\bA \in \mathbb{R}^{n \times d}$ and seek to answer $Q$ adaptive queries $\bb^{(1)}, \ldots, \bb^{(Q)}$ for an accuracy parameter $\eps > 0$. 
For each query vector $\bb^{(i)} \in \mathbb{R}^n$, the goal is to return a $(1+\eps)$-approximation to the regression cost $\min_{\bx \in \mathbb{R}^d} \|\bA\bx - \bb^{(i)}\|_2$.

As with previous problems, one strategy is to compute a subspace embedding $\bM = \bS \bA \in \mathbb{R}^{m \times d}$ using a sketching matrix $\bS$, and then for a given query $\bb^{(i)}$, solve the reduced problem $\min_{\bx \in \mathbb{R}^d} \|\bS \bA \bx - \bS \bb^{(i)}\|_2$~\cite{ClarksonW13}.

\begin{theorem}
\cite{ClarksonW13}
\thmlab{thm:se}
Let $\bA \in \mathbb{R}^{n \times d}$ and $\eps\in(0,1)$ be given. 
Then there is an algorithm that constructs a data structure using $\O{\frac{d^2}{\eps^2} \log^2(nQ)}$ bits of space and, with high probability, returns a $(1+\eps)$-approximation to $\min_{\bx \in \mathbb{R}^d} \|\bA\bx - \bb\|_2$ for any fixed query $\bb\in\mathbb{R}^n$. 
\end{theorem}

However, this approach may not remain reliable under multiple interactions. 
For instance, an adversary could learn the kernel of $\bS$ and then issue a query $\bb^{(i)}$ that lies entirely in the kernel. 
This would cause $\bS\bb^{(i)}$ to be zero, resulting in the algorithm returning the zero vector as an approximate minimizer, which might be a poor approximation to the true solution.

To guard against such failures, a na\"{i}ve solution is to use a separate embedding for each query, leading to total space complexity of $\tO{\frac{Qd^2}{\eps^2}}$. 
In contrast, combining \thmref{thm:se} with \thmref{thm:robust:dp} yields a more efficient solution:

\begin{theorem}
\cite{CherapanamjeriSWZZZ23}
Given $\bA \in \mathbb{R}^{n \times d}$ and accuracy parameter $\eps > 0$, there exists an algorithm that constructs a data structure using $\O{\frac{\sqrt{Q} d^2}{\eps^2} \log^3(nQ)}$ bits of space. With high probability, it provides $(1+\eps)$-approximations to $\min_{\bx \in \mathbb{R}^d} \|\bA\bx - \bb^{(i)}\|_2$ for all $Q$ adaptive queries $\bb^{(1)}, \ldots, \bb^{(Q)}$.
\end{theorem}

Finally, we describe an approach by \cite{FengFLSWZ25} to approximately solve regression for a number of adaptive queries. 
In particular, suppose the design matrix $\bA$ is defined through a number of adaptive, entry-wise perturbations via updates $\bV_t \in \mathbb{R}^{n \times d}$. 
Formally, the guarantees are as follows:
\begin{theorem}
Let $\bA\in \mathbb{R}^{n \times d}$ and $\bb \in \mathbb{R}^n$, and consider a sequence of adaptive updates $\{\bV_1, \ldots, \bV_Q\}$ to $(\bA, \bb)$. 
There exists an adaptive algorithm that pre-processes the initial inputs $(\bA, \bb)$ in time $\tO{\sqrt{Qd} \cdot \left(\nnz(\bA) + \nnz(\bb) + d^3 + \frac{d^2 \kappa^2}{\alpha^2}\right)}$, and uses $\tO{\sqrt{Q} \cdot\frac{d^{2.5} \kappa^2}{\alpha^2}}$ words of space. 
For each query $q \in [Q]$, the algorithm can apply the update $\bV_t$ to transform $(\bA_{t-1}, \bb_{t-1})$ into $(\bA_t, \bb_t)$ in time 
\[\tO{\sqrt{Qd} \cdot \left(\nnz(\bV_t) + d^3 + \frac{d^2 \kappa^2}{\alpha^2}\right)}.\] 
After processing the update, the algorithm outputs a vector $\bx_t$ that, with probability at least $1 - \delta$, is a $(1+\alpha)$-approximate solution to the regression problem defined by $(\bA_t, \bb_t)$, and does so in time $\tO{d^{\omega + 1}\cdot\frac{\kappa^2}{\alpha^2}}$.
\end{theorem}
Here, each update $\bV_t$ is either a matrix in $\mathbb{R}^{n\times d}$ so that $\bA_t=\bA_{t-1}+\bV_t$ (in which case $\bb_t=\bb_{t-1}$) or a vector in $\mathbb{R}^n$ so that $\bb_t=\bb_{t-1}+\bV_t$ (in which case $\bA_t=\bA_{t-1}$). 
Moreover, $\kappa$ is the condition number and $\omega$ is the exponent for matrix multiplication time. 

We now briefly outline the approach; we refer to \cite{FengFLSWZ25} for the full details. 
A natural extension of the private median framework introduced by~\cite{HassidimKMMS22,BeimelKMNSS22} is to produce an approximate solution vector. 
The algorithm follows a generic template: it generates $k$ independent sketching matrices $\bS_1, \ldots, \bS_k$ and pre-processes the data as $(\bS_i\bA, \bS_i\bb)$ for each $i \in [k]$. 
When an update arrives, the corresponding sketches are updated efficiently. 
For each query, the algorithm samples $s=\polylog(n,d,k)$ sketches from the $k$ total sketches. 
The algorithm solves the respective regression problems for each of the sketches, e.g., $\argmin_{\bx}\|\bS_i\bA\bx-\bS_i\bb\|$ for each sampled index $i$, and obtains solution vectors $\bx_{(1)}, \ldots, \bx_{(s)}\in\mathbb{R}^d$. 

These are then aggregated using a coordinate-wise private median mechanism, as follows. 
Namely, to construct the output vector $\bg \in \mathbb{R}^d$, for each coordinate $i \in [d]$, the algorithm takes the private median mechanism with input $((x_{(1)})_i, \ldots, (x_{(s)})_i)$ and sets $g_i$ to be the output. 
Since each entry is computed privately and individually, the advanced composition theorem guarantees overall differential privacy. 
Although this approach requires $\sqrt{Qd}$ sketches—compared to the $\sqrt{Q}$ required in~\cite{BeimelKMNSS22}, it remains efficient when $d \ll Q$.

The main technical challenge lies in proving utility, as the coordinate-wise private median mechanism could destroy the accuracy of the output solution. 
By triangle inequality, the regression error $\|\bA\bg - \bb\|_2$ satisfies
\[\|\bA\bg - \bb\|_2 \leq \|\bA\bx^* - \bb\|_2 + \|\bA(\bx^* - \bg)\|_2,\]
where $\bx^*$ is the optimal solution and the second term is upper bounded by
\[\|\bA(\bx^* - \bg)\|_2 \le \sigma_{\max}(\bA) \cdot \sqrt{d} \cdot \|\bx^* - \bg\|_\infty, \]
where $\sigma_{\max}(\bA)$ is the largest singular value of $\bA$. 
Thus, upper bounding $\|\bx^* - \bg\|_\infty$ is sufficient for upper bounding the total regression error. 

Each vector $\bx_{(i)}$ satisfies $\|\bA \bx_{(i)} - \bb\|_2 \leq (1+\alpha)\|\bA\bx^* - \bb\|_2$, and when $\bA$ is well-conditioned, this implies an upper bound on $\|\bx_{(i)} - \bx^*\|_2$. 
However, in order to effectively run the coordinate-wise median procedure, we need upper bounds on each entry of $\bx_{(i)}-\bx^*$.

While in general $\|\bx_{(i)} - \bx^*\|_\infty$ may be as large as $\|\bx_{(i)} - \bx^*\|_2$, for sketching matrices such as the Subsampled Randomized Hadamard Transform (SRHT), it has been shown~\cite{PriceSW17} that 
\[\|\bx_{(i)} - \bx^*\|_\infty \leq \frac{\alpha}{\sqrt{d}} \cdot \frac{\|\bA\bx^* - \bb\|_2}{\sigma_{\min}(\bA)}.\]
Therefore, we use the SRHT for the sketching matrix $\bS$, after which it follows that the regression error can be upper bounded by
\[\|\bA(\bx^* - \bg)\|_2 \leq \alpha \kappa(\bA) \cdot \|\bA\bx^* - \bb\|_2,\]
where $\kappa(\bA)$ is the condition number. 
By scaling $\alpha$ down by a factor of $\kappa(\bA)$, we achieve the originally desired utility guarantee. 

This highlights the private median as a novel use of $\ell_\infty$ guarantees in sketching, an aspect that has received limited attention in the literature.
To further accelerate both pre-processing and update time, the SRHT is composed with CountSketch~\cite{CharikarCF04}. 

\subsection{Applications to Half-Space Queries}
Consider a set $P$ of $n$ points in $\mathbb{R}^d$. 
The range search problem asks us to pre-process $P$ so that, given a query region $R$ from some fixed family, we can efficiently determine or count the points in the intersection $P \cap R$. 
This is a foundational problem in computational geometry~\cite{toth2017handbook}. 
A particularly important case is when the query regions are half-spaces, i.e., one side of a hyperplane, since many algebraic constraints can be represented as hyperplanes in a lifted space.

However, exact solutions to this problem are notoriously difficult in high dimensions, as the query time typically scales exponentially with $d$—a manifestation of the curse of dimensionality~\cite{bronnimann1993hard,chazelle2000discrepancy}. 

To address this, \cite{chazelle2008approximate} proposed an approximate data structure that can answer half-space queries in polynomial time. 
Their notion of approximation works as follows: for a point set $P$ contained in the unit $L_2$ ball, a hyperplane $R$, and accuracy parameter $\eps > 0$, the data structure returns an estimate of the number of points on one side of $R$, with an additive error at most equal to the number of points within distance $\eps$ of the hyperplane boundary. 
This is called an $\eps$-approximate half-space query.

\begin{theorem}
\cite{chazelle2008approximate}
Let $P$ be a set of $n$ points in the unit $L_2$ ball. There exists a data structure using $\tO{dn^{\O{\eps^{-2}}}}$ space that answers $\eps$-approximate hyperplane queries with high probability. The query time is $\tO{d/\eps^2}$.
\end{theorem}

The construction in~\cite{chazelle2008approximate} is randomized, relying on techniques such as random projection for dimensionality reduction. 
Because of this randomness, the data structure may not remain accurate under multiple adaptive queries. 
To address this, by applying the general framework from \secref{sec:dp:framework:isolate} and \thmref{thm:robust:dp}, we obtain a more robust guarantee:

\begin{theorem}
\cite{CherapanamjeriSWZZZ23}
Let $P$ be a set of $n$ points contained in the unit $L_2$ ball. 
Then there exists a data structure using $\tO{\sqrt{Q}dn^{\O{\eps^{-2}}}}$ space that can answer $Q$ adaptive $\eps$-approximate hyperplane queries with high probability. 
Each query is processed in $\tO{\frac{d}{\eps^2}}$ time.
\end{theorem}

\subsection{Applications to Point Queries in Turnstile Streams}
In the turnstile streaming model, we observe a sequence of $m$ updates to a frequency vector $\bx\in \mathbb{R}^n$. 
Each update affects a coordinate $i \in [n]$, modifying $f_i$ by some value $\Delta_i \in [-\Delta, \Delta]$, where $\Delta$ is a polynomial in $n$. 
For any time step $t \in [m]$, let $\bx^{(t)}$ denote the frequency vector after the first $t$ updates.

The point query problem is to return the value of $\bx^{(t)}_i$ for various choices of $i$ and $t$, up to an additive error of $\eps\|\bx^{(t)}\|_1$, for a fixed constant $\eps > 0$.

\begin{theorem}
\cite{AlmanY20}
\thmlab{thm:turnstile:point}
There exists an algorithm that, with high probability and for $\eps = 0.1$, uses $\O{\log^2 n}$ bits of space, has a worst-case update time of $\O{\log^{0.582}n}$, and supports point queries in $\O{\log^{1.582}n}$ time.
\end{theorem}

A key feature of \thmref{thm:turnstile:point} is its improved update time compared to earlier data structures such as~\cite{CharikarCF04}, though this comes with increased query time. 
By leveraging the general adaptive framework of \thmref{thm:robust:dp}, we can extend this algorithm to handle adaptive queries more robustly, without further increasing the query time:

\begin{theorem}
There exists an algorithm that supports $Q$ adaptive point queries with $\eps = 0.1$ and high probability, using $\O{\sqrt{Q} \log^3(nQ)}$ bits of space, worst-case update time $\O{\sqrt{Q} \log^{1.582}(nQ)}$, and query time $\tO{\log^3(nQ)}$.
\end{theorem}

\subsection{Applications to Nearest Neighbor Search}
In the approximate nearest neighbor (ANN) problem, we are given a dataset of $n$ points in $\mathbb{R}^d$, represented as a matrix $\bU \in \mathbb{R}^{n \times d}$, which may be preprocessed in advance. 
The task is to construct a data structure that, given a sequence of possibly adaptive queries $\bv_1, \ldots, \bv_Q$, answers each query efficiently and with high probability of correctness. 
Specifically, for each query $\bv_q$, where $q \in [Q]$, the data structure should return a $c$-approximate near neighbor from $\bU$, meaning that if the true nearest neighbor to $\bv_q$ lies at distance $r$, the output should be a point in $\bU$ at distance at most $cr$, where $c > 1$.

We first recall prior work on ANN search data structures. 
A natural starting point is deterministic data structures, which have been studied extensively since the 1970s~\cite{mark2008computational,cormen2022introduction} and are adversarially robust by design. 
However, in high-dimensional settings, and without making any assumptions about the structure of the query distribution or dataset, these methods encounter the curse of dimensionality: both pre-processing time and space requirements grow exponentially with the dimension $d$, rendering them impractical in moderate dimensions. 
More efficient deterministic structures can be constructed under strong geometric assumptions, such as bounded growth conditions, e.g., when the number of points within a ball of radius $2r$ centered at a query point is only a constant factor larger than that within radius $r$~\cite{Clarkson97,KargerR02,KrauthgamerL04,BeygelzimerKL06}. 
These assumptions permit data structures with polynomial pre-processing time and space complexity and logarithmic query time. 
However, such assumptions are quite restrictive and do not apply to many real-world datasets.

Instead, \cite{FengFLSWZ25} considered solutions based on differential privacy (DP), which motivated them to look at the private selection problem.
Suppose we are given $n$ categories and $s$ binary vectors $\bb_{(1)}, \ldots, \bb_{(s)} \in \{0,1\}^n$. 
The task is to identify the category $j^* \in [n]$ that appears most frequently across these vectors, i.e., $j^* = \argmax_j \sum_{i=1}^s b_{(i),j}$. 
A standard approach is to first compute the total count vector $\bB = \sum_{i=1}^s \bb_{(i)}$, then add independent Laplace noise $\mathrm{Lap}\left(\frac{1}{\eps}\right)$ to each coordinate of $\bB$, and finally report the index with the highest noisy value. 
Importantly, the privacy parameter $\eps$ remains independent of both $n$ and $s$, since only a single index is released.

\cite{FengFLSWZ25} now connects approximate nearest neighbor (ANN) search to this selection task. 
By assigning a category to each point $\bu\in\bU$, and instructing each ANN data structure to return all nearby neighbors (instead of just one), we obtain a binary indicator vector for each structure, where the $i$-th entry is 1 if $u_i$ is found. 
Collecting these indicator vectors across multiple data structures, we can apply the DP selection mechanism described above and return the point with the highest noisy count. 
This construction ensures differential privacy with respect to the internal randomness of the data structures: for fixed dataset $U$ and adaptive query $\bv$, the indicator vector depends only on the randomness within each structure. 
Consequently, the mechanism satisfies $(\eps, 0)$-DP, allowing us to invoke the advanced composition theorem and reduce the number of independent data structures from $Q$ to $\tO{\sqrt{Q}}$.

To show utility, \cite{FengFLSWZ25} first proves that the selection mechanism is differentially private, and then applies generalization guarantees from DP, similar to the argument in \secref{sec:dp:framework:isolate}, to conclude that a constant fraction of the data structures must succeed. 
There are two technical issues however. 
First, assuming only one near neighbor is overly restrictive. 
Fortunately, the approach generalizes: if there are at most $s$ approximate near neighbors, then querying $\omega(s \log n)$ data structures suffices. 
By the pigeonhole principle, some point will have count significantly above the noise. 
Since LSH-based ANN methods return at most $n^\rho$ candidates, \cite{FengFLSWZ25} notes that one can select $s = \O{n^\rho}$. 
Equivalently, a structural property can be imposed: that the ball of radius $cr$ around any point intersects at most $s$ other such balls. 
This condition is satisfied, for example, in the Hamming setting and in the model of~\cite{KapralovMS25}.

Second, a na\"{i}ve implementation of the DP selection process requires adding noise to each of the $n$ entries in the count vector, resulting in $\Omega(n)$ time per query. 
On the other hand, the vector $\bB$ is $s$-sparse and its nonzero entries typically exceed the noise level. 
Prior work on private sparse vector release~\cite{CormodePST12} has considered this issue, but existing methods either weaken the privacy guarantee or convert the algorithm to Las Vegas form with only expected runtime bounds. 
To address this, \cite{FengFLSWZ25} introduces a new \emph{sparse argmax mechanism}, which efficiently simulates exponential noise addition over sparse vectors. 
The algorithm works by (1) adding $s$ exponential $\mathrm{Exp}\left(\frac{1}{\eps}\right)$ samples to the nonzero entries, (2) drawing a threshold $X$ from the $n$-th order statistic of $\mathrm{Exp}\left(\frac{1}{\eps}\right)$, and (3) flipping a biased coin (heads with probability $\frac{s}{n}$). 
If heads, it generates $s-1$ additional exponential samples (none exceeding $X$) and adds them to the support, including $X$, then returns the maximum index. 
If tails, it generates $s$ such values, includes $X$ as a random index from the complement support, and returns the maximum index overall. 
\cite{FengFLSWZ25} shows that with high probability, this procedure runs in $\O{s \log n}$ time and faithfully reproduces the output distribution of full-size exponential sampling.
Putting things together, the final guarantees for approximate nearest neighbor search is as follows:
\begin{theorem}
\thmlab{thm:adaptive:ann}
\cite{FengFLSWZ25}
Let $\bU \subseteq \mathbb{R}^d$ be an $n$-point dataset, and suppose $f_{\bv}: (\mathbb{R}^d)^n \to (\mathbb{R}^d)^s$ is a predicate function such that for any query $\bv$, the output $f_{\bv}(\bU)$ contains at most $s$ candidates. 
Let $\calA$ be an oblivious algorithm which, for any fixed query $\bv \in \mathbb{R}^d$, returns a point (or subset) in $f_{\bv}(\bU)$ with probability at least $1 - \delta$ whenever $f_{\bv}(\bU)$ is non-empty. 
Suppose $\calA$ has pre-processing time $\calT_{\mathrm{prep}}$, space complexity $\calS_{\mathrm{space}}$, and query time $\calT_{\mathrm{query}}$. 
Then there exists an adaptive algorithm $\widetilde{\calA}$ that:
\begin{enumerate}
\item
pre-processes $\bU$ in time $\tO{\sqrt{Q} \cdot s} \cdot \calT_{\mathrm{prep}}$;
\item
uses space $\tO{\sqrt{Q} \cdot s} \cdot \calS_{\mathrm{space}}$;
\item 
for any query $\bv_q$ with $q \in [Q]$, it returns a point (or subset) in $f_{\bv_q}(\bU)$ with probability at least $1 - \delta$, using query time $\tO{s}\cdot\calT_{\mathrm{query}}$.
\end{enumerate}
In particular, the amortized cost per query is $\tO{\frac{s}{\sqrt{T}}}\cdot\calT_{\mathrm{prep}} + \tO{s}\cdot\calT_{\mathrm{query}}$.
\end{theorem}

\section{Adaptive Data Analysis}
\seclab{sec:ada}
Adaptive data analysis refers to the common scenario where an analyst makes a sequence of queries on a dataset, with each query potentially depending on the outcomes of previous ones. 
This adaptivity arises in many real-world workflows, such as scientific discovery, online experimentation, and iterative model development. 
Although the dataset itself may be static, classical statistical methods can still break down in adaptive settings, as they typically assume queries are fixed beforehand. 
Hence, the setting is inherently related to adversarial robustness and there has been a large body of work exploring attacks and robust algorithms~\cite{HardtU14,DworkFHPRR15,DworkFHPRR15b,SteinkeU15,BassilyNSSSU16,FeldmanS18,CherapanamjeriN20,CohenLNSSS22,KontorovichSS22,CherapanamjeriSWZZZ23,CohenNSS23,DinurSWZ23,NissimST23,CohenNSSS24,FengFLSWZ25}. 

Let $\calX$ be a data domain and let $S \in \calX^n$ be a dataset drawn i.i.d. from an unknown distribution $\calD$.
In adaptive data analysis, an analyst interacts with an algorithm over multiple rounds. 
In each round $t$, the analyst selects a query $q_t : \calX \to [0,1]$, where the choice of $q_t$ may depend arbitrarily on all previous queries and answers.
The algorithm returns an answer $a_t$, possibly randomized, based on $S$ and the interaction history.
The goal is to ensure that, with high probability, the answers $a_t$ remain close to the corresponding population values $\EEx{x \sim \calD}{q_t(x)}$ for all adaptively chosen queries.

\subsection{Simple Attacks on Empirical Means}
To motivate the formal lower bounds for adaptive data analysis, we first describe two fundamental attacks. 
These examples demonstrate how an analyst, by choosing queries based on previous results, can force a simple algorithm, namely, the algorithm that merely returns empirical means, to output answers that fail to generalize to the underlying distribution $\calD$.

\paragraph{The Decoding Attack (High Precision).} 
When an algorithm provides the empirical mean of linear queries with arbitrarily high precision, an analyst can effectively ``decode'' the entire sample $S = \{x_1, \ldots, x_n\}$ in a very small number of steps. 
For instance, if the data points $x_i$ are bits, the analyst can query the empirical mean of $q(x_i) = 2^i$. 
The resulting answer $a = \frac{1}{n} \sum_{i=1}^n 2^i x_i$ is a value whose binary representation reveals every $x_i$ in the sample. 

Once the sample is decoded, the analyst can define a ``targeted'' counting query $q_{target}(x) = \mathbb{I}[x \in S]$. 
The empirical mean on the sample is exactly $1$, yet for any sufficiently large domain, the true expectation $\EEx{x \sim\calD}{q_{target}(x)}$ will be nearly $0$. 
This shows that even a single high-precision answer allows an adaptive analyst to construct a query with maximal generalization error.

\paragraph{The Correlation Attack (Linear Number of Queries).} 
If the analyst is restricted to counting queries, i.e., $q(x) \in \{0, 1\}$, and the algorithm returns empirical means, a similar attack is possible using a number of queries linear in the sample size $n$. 
In this attack, the analyst first issues $k = \O{n}$ random counting queries $q_1, \ldots, q_k$ and receives their empirical means $a_1, \ldots, a_k$. 

By comparing each answer $a_j$ to its known population mean $\mu_j$, the analyst can identify which queries happen to have a positive or negative bias on the specific sample $S$. 
The analyst can then construct a new query $Q(x) = \sum_{j=1}^k w_j q_j(x)$, where the weights $w_j$ are chosen to correlate with the sampling error, e.g., $w_j = \text{sign}(a_j - \mu_j)$. 
This constructed query $Q$ will, by design, have an empirical mean significantly higher than its true expectation. 

These simple attacks illustrate that adaptivity allows an analyst to ``search'' for the sampling noise inherent in any finite dataset. 
While the fingerprinting attack discussed in the following sections provides a more general $\O{n^2}$ lower bound for any mechanism, these basic examples highlight why standard empirical estimation is fundamentally insecure over adaptive interactions. 

\subsection{The Streaming Adaptive Data Analysis (SADA) Problem}
\seclab{sec:SADA}
In this section, we present a result by \cite{KaplanMNS21} that uses adaptive data analysis to demonstrate a separation between non-adaptive insertion-only data streams and adversarial insertion-only data streams. 
Namely, they constructed a streaming problem that admits an efficient oblivious algorithm, but any robust algorithm must use polynomially larger space. 
Throughout this section, we follow the notation and exposition of \cite{KaplanMNS21}.
We first present a number of necessary preliminaries for adaptive data analysis. 
Recall the definition of the accuracy game $\accgame$, as defined by \cite{KaplanMNS21}. 

\begin{algorithm}[H]
\caption{Accuracy game $\accgame_{n,\ell,\calM,\mathbb{A}}$}
\alglab{alg:acc:game}
\begin{algorithmic}[1]
\State{\textbf{Input:} A dataset $S \in X^n$.}
\For{$i = 1$ to $\ell$}
\State{The adversary $\mathbb{A}$ adaptively selects a statistical query $q_i$.}
\State{The mechanism $\calM$ uses $S$ to compute an answer $z_i$.}
\State{The answer $z_i$ is disclosed to the adversary $\mathbb{A}$.}
\EndFor
\State{\Return the transcript $(q_1, z_1, \ldots, q_\ell, z_\ell)$.}
\end{algorithmic}
\end{algorithm}

\paragraph{Pseudorandom generators.}
We next recall a number of results regarding information-theoretic security of pseudorandom generators in the \emph{bounded storage model}, i.e., against adversaries with bounded space. 
Per the exposition of \cite{KaplanMNS21}, we give the formulation presented by Vadhan~\cite{Vadhan04}.

In the bounded storage model, a short secret seed $K \in \{0,1\}^b$ (hidden from the adversary) is used alongside a long sequence of public random bits $X_1, X_2, \ldots$ (accessible to all parties). 
A {\em bounded storage model (BSM) pseudorandom generator} is a function
\[\PRG: \{0,1\}^a \times \{0,1\}^b \rightarrow \{0,1\}^c,\]
where typically $b, c \ll a$. 
Two honest parties begin with a shared seed $K \in \{0,1\}^b$ hidden from the adversary. 
At each time step $t \in [T]$, the next $a$ bits from the public stream, $(X_{(t-1)a}, \ldots, X_{ta})$, are revealed. 
Although the adversary observes this stream, it is limited in memory and cannot retain the entire sequence. 
The honest parties apply the pseudorandom generator to the observed bits and the seed $K$ to produce an output $Y_t \in \{0,1\}^c$ of pseudorandom bits via $\PRG(\cdot, K)$.

To define security in this setting, suppose the adversary has storage capacity limited to $\beta a$ bits. 
Let $S_t \in \{0,1\}^{\beta a}$ denote the adversary's internal state at time $t$. 
Informally, security is defined by comparing two scenarios: the {\em real} setting, where $\PRG$ is applied, and an {\em ideal} setting, where the output is truly random. 
The adversary $\calA$ updates its state and, based on its stored information, attempts to distinguish between these two cases at the end of the process.

\begin{figure*}[!htb]
\begin{mdframed}
Real Experiment:
\begin{itemize}
\item 
Let $X = (X_1, X_2, \ldots, X_{Ta})$ be a sequence of uniformly random bits, let the secret key be $K \leftarrow \{0,1\}^b$, and initialize the adversary’s memory state as $S_0 = 0^{\beta a}$.
\item 
For each round $t = 1, \ldots, T$:
\begin{itemize}
\item 
Compute the pseudorandom output $Y_t = \PRG\left( X_{(t-1)a+1}, \ldots, X_{ta}, K \right) \in \{0,1\}^c$.
\item 
Update the adversary’s state as $S_t = \calA\left( Y_1, \ldots, Y_{t-1}, S_{t-1}, X_{(t-1)a+1}, \ldots, X_{ta} \right) \in \{0,1\}^{\beta a}$.
\end{itemize}
\item 
Finally, the adversary outputs $\calA\left(Y_1, \ldots,Y_T,S_T,K\right)\in\{0,1\}$.
\end{itemize}    
\end{mdframed}

\begin{mdframed}
Ideal Experiment:
\begin{itemize}
\item 
Let $X = (X_1, X_2, \ldots, X_{Ta})$ be a sequence of uniformly random bits, let the secret key be $K \leftarrow \{0,1\}^b$, and initialize the adversary’s memory state as $S_0 = 0^{\beta a}$.
\item 
For each round $t = 1, \ldots, T$:
\begin{itemize}
\item 
Acquire uniformly random bits $Y_t\in\{0,1\}^c$. 
\item 
Update the adversary’s state as $S_t = \calA\left( Y_1, \ldots, Y_{t-1}, S_{t-1}, X_{(t-1)a+1}, \ldots, X_{ta} \right) \in \{0,1\}^{\beta a}$.
\end{itemize}
\item 
Finally, the adversary outputs $\calA\left(Y_1, \ldots,Y_T,S_T,K\right)\in\{0,1\}$.
\end{itemize}  
\end{mdframed}
\end{figure*}

Observe that at every time step, the adversary is granted access to all previously generated values $Y_i$ without this access counting toward its storage limit. 
Additionally, at the final step, the adversary is also given the key $K$.

\begin{definition}\cite{Vadhan04}
Given $\beta\in[0,1]$, we say that $\PRG:\{0,1\}^a\times\{0,1\}^b\rightarrow\{0,1\}^c$ is an {\em $\eps$-secure BSM pseudorandom generator} against storage rate $\beta$ if, for any adversary $\calA$ with at most $\beta a$ bits of memory and for every $T\in\mathbb{N}$, the distinguishing advantage between the real and ideal experiments is at most $T\eps$. 
That is,
\[\left|\mathbf{\Pr}_{\rm r}\left[ \calA\left( Y_1,\ldots,Y_T,  S_T,  K  \right)=1 \right]-\mathbf{\Pr}_{\rm i}\left[ \calA\left( Y_1,\ldots,Y_T,  S_T,  K  \right)=1 \right]
\right| \le T\cdot\eps,\]
where $\mathbf{\Pr}_{\rm r}$ denotes the probability over the real experiment and $\mathbf{\Pr}_{\rm i}$ denotes the probability over the ideal experiment. 
\end{definition}

We recall a result of \cite{Vadhan04}, which gives a construction of a BSM PRG with the following guarantees:

\begin{theorem}
\cite{Vadhan04}
\thmlab{thm:prg:vadhan}
For every $\eps>\exp\left( -a/2^{\O{\log^{*}a}} \right)$, every $a\in\mathbb{N}$, and every $c\leq a/4$, there exists a BSM pseudorandom generator $\PRG:\{0,1\}^a \times\{0,1\}^b\rightarrow\{0,1\}^c$ satisfying the following properties:
\begin{enumerate}
\item 
The generator $\PRG$ is $\eps$-secure against adversaries with storage rate $\beta\le\frac{1}{2}$.
\item 
The key length is $b=\O{\log\frac{a}{\eps}}$.
\item 
For any fixed key $K$, the function $\PRG(\cdot,K)$ reads at most $t=\O{c+\log\frac{1}{\eps}}$ bits from the public stream, chosen nonadaptively.
\item 
The generator can be evaluated in time $\poly(t,b)$ and requires only $\poly(\log t, \log b)$ bits of workspace beyond the $t$ bits from the public stream and the key of length $b$.
\end{enumerate}
\end{theorem}

We now define the \emph{Streaming Adaptive Data Analysis (SADA)} problem introduced by \cite{KaplanMNS21}, who showed a strong positive result for this problem in the oblivious setting and a strong negative result in the adversarial setting.
Again, we follow the exposition of \cite{KaplanMNS21} throughout the section. 

Let $X=\{0,1\}^d\times\{0,1\}^b$ denote the data domain, let $\gamma\ge 0$ be a fixed constant, and let $\PRG:\{0,1\}^a\times\{0,1\}^b\rightarrow\{0,1\}^c$ be a BSM pseudorandom generator with output length $c=1$. 
At each time step $i\in[m]$, the stream provides an update $x_i=(p_i,k_i)\in X$.
The first $n$ updates, $x_1,\ldots,x_n$, are interpreted as pairs of ``data points'' and their associated ``keys''. 
We let $S$ denote the multiset formed from these initial $n$ updates. 
Additionally, for technical purposes, $S$ is augmented with $\frac{\gamma n}{1-\gamma}$ copies of a special placeholder element $\bot$. 
This multiset $S$ remains fixed after time step $n$.

Starting at time step $j=n+1$, each block of $(a+1)\cdot 2^d$ updates specifies a new ``function'' (or ``query'' in the context of the adaptive data analysis problem) that the streaming algorithm must evaluate over the fixed multiset $S$. 
Specifically:

\begin{itemize}
\item 
For each $p\in\{0,1\}^d$ (processed in lexicographic order), perform the following steps:
\begin{enumerate}
\item 
Let $u^1,\ldots,u^a\in\{0,1\}$ be the next $a$ stream updates, and define $\Gamma_p\in\{0,1\}^a=u^1\circ\ldots\circ u^a$ to be the corresponding bitstring. 
\item 
Let $\sigma$ be the following stream update.
\item 
For every $k\in\{0,1\}^b$, compute $Y_k = \PRG(\Gamma_p, k)$ and define the function value $f(p,k) = \sigma \oplus Y_k$.
\end{enumerate}
\item
In addition, define $f(\bot) = 1$.
\end{itemize}
Observe that these updates implicitly define a truth table for the function $f:\{0,1\}^d \times \{0,1\}^b\rightarrow\{0,1\}$. 
Using these truth tables, the SADA problem is defined as follows:
\begin{definition}[The $(a,b,d,m,n,\gamma)$-SADA Problem]
At the end of each block that defines a function $f$, the streaming algorithm must compute or approximate the average of $f$ on the multiset $S$, i.e., $\frac{1}{|S|}\sum_{x_i=(p_i,k_i)\in S}f(p_i,k_i)$. 
For the other time steps, the algorithm is permitted arbitrary output. 
Here, $m$ is the total number of updates, i.e., the length of the stream, $n$ is the total number of data points in $S$, $\gamma$ is a small constant, and $a$, $b$, and $d$ are the parameters defining the domain and the $\PRG$.
\end{definition}
The main point of the design is to link the query's output to a pseudorandom bit derived from a hidden key, making it difficult for an algorithm to learn about the underlying data without knowing the key.
 
\paragraph{An oblivious algorithm for the SADA problem.}
We briefly describe a streaming algorithm for the SADA problem in the oblivious setting. 
The algorithm performs as follows. 
For $s=\O{\frac{1}{\alpha^2\gamma}\log\frac{m}{\beta}}$, we independently sample and store $s$ items from $S=\{x_1,\ldots,x_n\}$ uniformly at random, e.g., by reservoir sampling~\cite{Vitter85}. 
For each query $f$, we then output the empirical average of $f$ on the stored samples. 

\begin{theorem}
\thmlab{thm:sada:oblivious}
\cite{KaplanMNS21}
For constant $\alpha,\beta,\gamma$, there exists an algorithm that uses space $\O{(b+d)\log m}$ and is $(\alpha,\beta)$-accurate for the SADA problem on a non-adaptive data stream. 
\end{theorem}
\begin{proof}
For $Q\le m$, let $f_1,\ldots,f_Q$ be the non-adaptive queries fixed by the data stream. 
Let $D$ be the set of $s$ samples maintained by our algorithm. 
By a standard Chernoff bound argument, observe that if $s\succeq\frac{C}{\alpha^2\gamma}\log\frac{m}{\beta}$ for a sufficiently large constant $C>0$, then for each fixed $q\in[Q]$, we have
\[\PPr{\left\lvert\frac{1}{|D|}\sum_{x_i=(p_i,k_i)\in D}f_q(p_i,k_i)-\frac{1}{|S|}\sum_{x_i=(p_i,k_i)\in S}f_q(p_i,k_i)\right\rvert\le\alpha}\ge1-\frac{\beta}{m}.\]
Then by a union bound, we have that with probability at least $1-\beta$, the empirical average $\frac{1}{|D|}\sum_{x_i=(p_i,k_i)\in D}f_q(p_i,k_i)$ is within additive $\alpha$ of the true value $\frac{1}{|S|}\sum_{x_i=(p_i,k_i)\in S}f_q(p_i,k_i)$ for all queries  $q\in[Q]$ in the data stream. 
Hence, the algorithm is $(\alpha,\beta)$-correct. 

It remains to analyze the space complexity. 
Since $s=\O{\frac{1}{\alpha^2\gamma}\log\frac{m}{\beta}}=\O{\log m}$ for constant $\alpha$ and $\beta$, then the algorithm maintains $\O{\log m}$ samples in $S$. 
Each element $x_i=(p_i,d_i)$ can be stored using $b+d$ bits, since $p_i\in\{0,1\}^d$ and $d_i\in\{0,1\}^b$. 
Finally, we remark that if the SADA problem is instantiated using the PRG of \thmref{thm:prg:vadhan}, then the PRG itself uses $\O{b+d}$ bits to store. 
Therefore for constant $\alpha$ and $\beta$, the total memory required is $\O{(b+d)\log m}$ bits. 
\end{proof}

\subsection{An Impossibility Result for Adaptive Streaming}
\seclab{sec:negative}
In this section, we show that an adversarially-robust streaming algorithm for the SADA problem requires significantly more space than a streaming algorithm for the SADA problem in the non-adaptive setting. 
The underlying intuition is as follows. 
Any space-bounded streaming algorithm can only retain a small subset of the elements it processes and must effectively forget the rest of the stream. 
An adaptive adversary can then define queries over the entire dataset. 
Because the algorithmic outputs are based only on the limited subset it remembers, it quickly overfits to those elements and fails to answer the adaptive queries correctly. 
In contrast, in the non-adaptive setting an oblivious algorithm is easy to construct: standard concentration bounds together with a union bound imply that storing only a logarithmic number of elements already suffices.
We now formalize this notion. 

Suppose there exists a streaming algorithm $\calA$ that is robust against adversarial inputs for the SADA problem. 
We describe in Algorithm \texttt{AnswerQueries}, c.f., \algref{alg:AnswerQueries}, how to use $\calA$ to build an algorithm that takes as input a dataset $P$ of $n$ elements from $\{0,1\}^d$ and is able to respond to a sequence of adaptively chosen queries of the form $q: \{0,1\}^d \to \{0,1\}$. 

\begin{algorithm}[t!]
\caption{\texttt{AnswerQueries}, c.f., \cite{KaplanMNS21}}
\alglab{alg:AnswerQueries}
\begin{algorithmic}[1]
\State{\textbf{Input:} A database $P \in (\{0,1\}^d)^n$ consisting of $n$ elements from $\{0,1\}^d$.}
\State{\textbf{Setting:} A stream of queries $q: \{0,1\}^d \to \{0,1\}$ arrives one per time step.}
\State{\textbf{Algorithm:} An adversarially robust streaming algorithm $\mathcal{A}$ for the SADA problem with $(\alpha,\beta)$-accuracy for streams of length $m$, using two random bitstrings $r_1$ and $r_2$ for randomness.}
\State{\textbf{Pseudorandom Generator:} A BSM pseudorandom generator $\PRG: \{0,1\}^a \times \{0,1\}^b \to \{0,1\}$.}
\Statex
\State{For each $p \in \{0,1\}^d$, sample a key $k_p \in \{0,1\}^b$ uniformly at random.}
\label{step:loopp}
\State{Sample a random string $r_1 \in \{0,1\}^\nu$ and initialize $\mathcal{A}$ to use $r_1$ for its randomness.}
\State{Feed each item $(p, k_p)$ from the dataset $P$ to $\mathcal{A}$ as an update.}
\State{Sample a new random string $r_2 \in \{0,1\}^\nu$ and switch $\mathcal{A}$ to use $r_2$ for further coin tosses.}
\Statex
\State{Repeat $\ell \gets \frac{m-n}{(a+1)\cdot 2^d}$ times:}
\State{\quad Receive the next query $q: \{0,1\}^d \rightarrow \{0,1\}$.}
\State{\quad For each $p \in \{0,1\}^d$:}
\State{\quad\quad Sample $\Gamma \in \{0,1\}^a$ uniformly at random.}
\State{\quad\quad Feed $a$ updates into $\mathcal{A}$ whose first bits concatenate to $\Gamma$.}
\State{\quad\quad For simplicity, do the following:}
\State{\quad\quad\quad \textcolor{blue}{Compute $Y = \PRG(\Gamma, k_p)$.}}
\State{\quad\quad For a \emph{natural} algorithm, instead do the following:}
\State{\quad\quad\quad \textcolor{red}{If $p\in P$ then let $Y=\PRG(\Gamma,k_p)$. Otherwise sample $Y\in\{0,1\}$ uniformly.}}
\State{\quad\quad Feed an additional update to $\mathcal{A}$ with first bit set to $Y \oplus q(p)$.}
\State{\quad Get the output $z$ from $\mathcal{A}$.}
\State{\quad Output $z$.}
\end{algorithmic}
\end{algorithm}

We now show that if $\calA$ is accurate for the SADA problem, then \texttt{AnswerQueries} (with the blue text) outputs a good estimate of the average of the query $q$ on the input dataset $P$. 
This fact shall ultimately be used to show a contradiction to the adaptive data analysis problem. 

\begin{lemma}
\lemlab{lem:empiricalAccuracy}
\cite{KaplanMNS21}
Suppose $\calA$ is $(\alpha,\beta)$-accurate for the SADA problem. 
Then the simplified instance of \texttt{AnswerQueries} is $\left(\frac{\alpha}{1 - \gamma}, \beta\right)$-accurate over $\frac{m-n}{(a+1) \cdot 2^d}$ adaptive queries, with respect to an input dataset of size $n$.
\end{lemma}
\begin{proof}
Let $q$ be the query for a fixed iteration, corresponding to a function $f$ whose truth table is given to $\calA$. 
Observe that for every pair $(p,k)$, we have $f(p, k) = q(p)$ by construction. 
In particular, the answers produced by $\calA$ are $\alpha$-accurate with respect to the extended dataset $P\cup\{\bot, \ldots, \bot\}$ with high probability. 
Since at most $\gamma$ fraction of this extended dataset consists of $\bot$, then the algorithm is $\frac{\alpha}{1-\gamma}$-accurate with respect to the original dataset $P$.
\end{proof}

\paragraph{Transcript compressibility.}
A key concept used to analyze the utility of algorithms that respond to adaptively chosen queries is transcript compressibility, which we define next. 
\begin{definition}
\cite{DworkFHPRR15b}
A mechanism $\calM$ that supports $\ell$ queries on a dataset of size $n$ is \emph{transcript compressible} to $b(n,\ell)$ bits if, for every deterministic adversary $\mathbb{A}$, there exists a set of transcripts $H_{\mathbb{A}}$ with size at most $2^{b(n,\ell)}$ such that, for every dataset $S \in X^n$, 
\[\PPr{\accgame_{n, \ell,\calM, \mathbb{A}}(S)\in H_{\mathbb{A}}}=1,\]
where we recall the definition of $\accgame$ in \algref{alg:acc:game}. 
\end{definition}

Notably, the empirical average of every query issued during an interaction with a transcript-compressible mechanism is close to its expected value (with high probability). 

\begin{theorem}
\cite{DworkFHPRR15b}
\thmlab{thm:transcriptCompression}
Suppose that $\calM$ is transcript compressible to $b(n,\ell)$ bits, and let $\beta > 0$. 
Then, for every adversary $\mathbb{A}$ and every distribution $\calD$, it holds that
\[\PPPr{\substack{S\sim\calD^n\\ \accgame_{n, \ell,\calM, \mathbb{A}}(S)}}{\exists i \text{ such that } \left|q_i(S)-q_i(\calD)\right|>\alpha}\leq\beta,\]
where
\[\alpha=\O{\sqrt{\frac{b(n,\ell)+\ln(\ell/\beta)}{n}}}.\]
\end{theorem}

We now show the simplified version of \texttt{AnswerQueries} is transcript-compressible. 
For any fixed choices of $\vec{\Gamma}$, $\vec{k}$, and the random strings $r_1$ and $r_2$ used during the execution, we use $\texttt{AnswerQueries}_{\vec{\Gamma}, \vec{k}, r_1, r_2}$ to denote the instantiation of the \texttt{AnswerQueries} algorithm with these components fixed.

\begin{lemma}
\lemlab{lem:compress}
\cite{KaplanMNS21}
If the algorithm $\calA$ operates using at most $w$ bits of memory, then for any fixed choice of $\vec{\Gamma}$, $\vec{k}$, $r_1$, and $r_2$, the algorithm $\texttt{AnswerQueries}_{\vec{\Gamma}, \vec{k}, r_1, r_2}$ is transcript-compressible into $w$ bits.
\end{lemma}
\begin{proof}
Note that we can assume the adversary generating the queries $q$ is deterministic without loss of generality. 
Then it follows that the full transcript of the interaction is entirely determined by the internal state of algorithm $\calA$ at the conclusion of Step 3.
\end{proof}

\cite{KaplanMNS21} emphasizes that the ``switch'' from $r_1$ to $r_2$ is convenient in the proof of \lemref{lem:compress} as otherwise, the transcript would need to store additional information to describe the state of the algorithm after Step 3. 
In particular, both the internal state of $\calA$ and the position for the next coin from $r_1$ would need to be maintained by the transcript. 

Putting together \lemref{lem:empiricalAccuracy} and \lemref{lem:compress}, \cite{KaplanMNS21} then shows that \texttt{AnswerQueries} is accurate, as transcript-compressible mechanisms have their empirical averages close to the expectation. 

\begin{lemma}
\lemlab{lem:streamingAccuracy}
\cite{KaplanMNS21}
Suppose that $\calA$ uses $w$ memory and is $(\alpha,\beta)$-accurate for the SADA problem on streams of length $m$. 
Then for any $\beta' > 0$, the algorithm \texttt{AnswerQueries} achieves $\left( \frac{\alpha}{1-\gamma} + \alpha',\, \beta + \beta' \right)$-accuracy over $\ell = \frac{m - n}{(a + 1)\cdot 2^d}$ queries, where
\[\alpha' = \O{\sqrt{ \frac{w + \ln(\frac{\ell}{\beta'})}{n}}}.\]
\end{lemma}
\begin{proof}
Let $\calD$ be a fixed distribution over $\{0,1\}^d$ and let $\calA$ be a fixed adversary that generates the queries $q_i$. 
We analyze the accuracy game $\accgame$. 
By \lemref{lem:empiricalAccuracy}, we have:
\[\mathbf{\Pr}_{\substack{S \sim \calD^n\\ \accgame_{n, \ell,\texttt{AnswerQueries}, \mathbb{A}}(S)}}\left[ \exists i \text{ such that } \left| q_i(S) - z_i \right| > \frac{\alpha}{1 - \gamma} \right] \leq \beta,\]
where $z_i$ denotes the response returned by the algorithm for query $q_i$. 
Moreover, by \lemref{lem:compress} and \thmref{thm:transcriptCompression}, for any fixed choice of $\vec{\Gamma}, \vec{k}, r_1, r_2$, it holds that:
\[\mathbf{\Pr}_{\substack{S \sim \calD^n\\ \accgame_{n, \ell,\texttt{AnswerQueries}, \mathbb{A}}(S)}}\left[ \exists i \text{ such that } \left| q_i(S) - q_i(\calD) \right| > \alpha' \,\Big|\, \vec{\Gamma}, \vec{k}, r_1, r_2 \right] \leq \beta',\]
where $\alpha' = \O{\sqrt{\frac{w + \ln(\ell/\beta')}{n}}}$. 
Since the above inequality holds for every fixed choice of $\vec{\Gamma}, \vec{k}, r_1, r_2$, then it also holds when these values are chosen randomly. 
Thus, applying the triangle inequality and a union bound, we conclude:
\[\mathbf{\Pr}_{\substack{S \sim \calD^n\\ \accgame_{n, \ell,\texttt{AnswerQueries}, \mathbb{A}}(S)}}\left[ \exists i \text{ such that } \left| z_i - q_i(\calD) \right| > \frac{\alpha}{1 - \gamma} + \alpha' \right]\leq \beta + \beta'.\]
\end{proof}

We now recall the following impossibility result for the adaptive data analysis (ADA) problem. 
Consider a mechanism $\calM$ designed for the ADA setting, which receives an input sample $P = (p_1, \ldots, p_n)$ and is required to answer a sequence of adaptive queries.
It is known that if $\calM$ produces its responses based solely on the evaluations of each query $q$ on the sample points, i.e., as a function of the values $q(p_1), \ldots, q(p_n)$, then, in general, it cannot reliably answer more than $n^2$ adaptively chosen queries. 
We call such a mechanism a {\em natural} mechanism. 

\begin{definition}
\cite{HardtU14}
An algorithm that receives a sample $P$ and responds to queries $q$ is called \emph{natural} if the following holds: 
For any sample $P$ and any pair of queries $q$ and $q'$ such that $q(p) = q'(p)$ for all $p \in P$, the algorithm produces the same output on $q$ and $q'$. 
That is, the outputs $z$ and $z'$ are equal if the algorithm is deterministic, and identically distributed if the algorithm is randomized. 
If the algorithm maintains internal state, then this requirement must hold regardless of its current state.
\end{definition}

We recall the following impossibility result for adaptive data analysis by Steinke and Ullman~\cite{SteinkeU15} though we observe that less optimal impossibility results~\cite{HardtU14, UllmanSNSS18}. 
These attacks thus show that storing a small set of elements cannot answer a large number of adaptive queries about the whole universe. 

\begin{theorem}
\cite{SteinkeU15}
\thmlab{thm:adaNegative}
There exists an absolute constant $c\in(0,1)$ such that no natural algorithms can be $(c,c)$-statistically-accurate for $\O{n^2}$ adaptive queries, given $n$ samples over a universe of size $\Omega(n)$.
\end{theorem}

As shown earlier in \lemref{lem:streamingAccuracy}, the algorithm \texttt{AnswerQueries} is statistically accurate for $\ell = \frac{m - n}{(a+1)\cdot 2^d}$ adaptively chosen queries, where $\ell$ can be made larger than $n^2$ by choosing $m$ sufficiently large. 
To show, we aim to apply \thmref{thm:adaNegative}. 
However, the algorithm \texttt{AnswerQueries} is not strictly a natural algorithm since the output of the streaming algorithm $\calA$ might depend on the values of the query function outside the sample $P$. 
To address this, we modify \texttt{AnswerQueries} to ensure it becomes a natural algorithm, c.f., \algref{alg:AnswerQueries} using the red text instead. 

\begin{lemma}
\lemlab{lem:natural}
\cite{KaplanMNS21}
Algorithm \texttt{AnswerQueries} (using the red text) is natural.
\end{lemma}
\begin{proof}[Proof sketch]
The claim follows from the fact that the values of the queries outside the input sample $P$ are entirely concealed from algorithm $\calA$, effectively using the classical one-time pad encryption scheme. 
Additionally, note that the output $z$ produced by \texttt{AnswerQueries} using the red text for a query $q$ is fully determined by the internal state of algorithm $\calA$ following the completion of the corresponding iteration of Step~5. 
\end{proof}

We now show that the modification to \texttt{AnswerQueries} in the red text has a negligible effect on the correctness of the execution and thus is both natural and accurate, ultimately leading to a contradiction to known adaptive data analysis impossibility results. 

\begin{lemma}
\lemlab{lem:TVdistance}
\cite{KaplanMNS21}
Suppose that $\calA$ uses at most space $w$, and let $\ell = \frac{m - n}{(a + 1)\cdot 2^d}$. 
If $\PRG$ is an $\eps$-secure BSM pseudorandom generator against adversaries with storage bounded by $\O{w + \ell + b \cdot 2^d}$, then for any input dataset $P$ and any adversary $\mathbb{A}$, the output distributions of the simplified and natural versions of $\accgame_{n, \ell,\texttt{AnswerQueries}, \mathbb{A}}(P)$ differ in total variation distance by at most $2^d m \eps$.
\end{lemma}
\begin{proof}
For the clarity of discussion, we use \texttt{AnswerQueriesSimp} to refer to the simplified version and \texttt{AnswerQueriesNat} to refer to the natural version. 
Recall that the output of $\accgame_{n, \ell,\texttt{AnswerQueries}, \mathbb{A}}(P)$ is the transcript $(q_1, z_1, \ldots, q_\ell, z_\ell)$, where $q_i$ are the queries issued by $\mathbb{A}$ and $z_i$ are the corresponding answers returned by $\texttt{AnswerQueries}$. Our goal is to show that the distributions over these transcripts, when using $\texttt{AnswerQueriesSimp}$ and $\texttt{AnswerQueriesNat}$, are close. 
Without loss of generality, we may assume that $\mathbb{A}$ is deterministic, since if the claim holds for all deterministic adversaries, it also holds for randomized ones. 
Under this assumption, the full transcript $(q_1, z_1, \ldots, q_\ell, z_\ell)$ is completely determined by the sequence of answers $(z_1, \ldots, z_\ell)$. 
Therefore, it suffices to show that the distributions of the answer sequences are similar in both executions.
Furthermore, because we are interested in constant-accuracy guarantees, we may assume that each answer $z_i$ is encoded using a constant number of bits. 
If not, we can slightly modify algorithm $\calA$ to ensure this property while preserving its correctness.

Now, for each $g \in \{0, 1, \ldots, 2^d\}$, define $\texttt{AnswerQueries}_g$ to be a variant of $\texttt{AnswerQueriesNat}$ where we instead set $Y = \PRG(\Gamma, k_p)$ if $p \in P$ or $p \geq g$, and otherwise we set $Y$ to a uniformly random bit. 
Observe that $\texttt{AnswerQueries}_0$ is equivalent to $\texttt{AnswerQueriesSimp}$, and $\texttt{AnswerQueries}_{2^d}$ is equivalent to $\texttt{AnswerQueriesOTPNat}$. 
We will now prove that for each $g$, the statistical distance between the outcomes of $\accgame_{n, \ell,\texttt{AnswerQueries}_g, \mathbb{A}}(P)$ and $\accgame_{n, \ell,\texttt{AnswerQueries}_{g+1}, \mathbb{A}}(P)$ is at most $\eps m$. 
This suffices to prove the desired claim via the triangle inequality.

To that end, we fix an index $g \in \{0,1,\ldots,2^d - 1\}$, and consider a hybrid algorithm $\accgame_g^*$ that simulates the interaction between $\mathbb{A}$ and $\texttt{AnswerQueries}_g$ on database $P$. 
During any iteration in Step~\ref{step:loopp} with $p = g$, the algorithm receives as input a string $\Gamma$ and a value $Y$, where $\Gamma$ is uniformly sampled and $Y$ is either a uniformly random bit or the output of $\PRG(\Gamma, k)$ for a randomly sampled key $k \in \{0,1\}^b$ (which remains hidden from the simulated algorithm). 
These two cases correspond to the behaviors of 
$\texttt{AnswerQueries}_{g+1}$ and $\texttt{AnswerQueries}_g$, respectively.

We note that $\accgame_g^*$ can be implemented using at most 
\[\hat{W} = \O{w + \ell + b \cdot 2^d}\]
bits of memory, which suffices to store:
\begin{itemize}
\item
the internal state of $\calA$ (using $w$ bits),
\item
the past answers $z_1, \ldots, z_i$ (using $\O{\ell}$ bits),
\item
the secret keys $k_p$ for all $p \neq g$ (using $b \cdot 2^d$ bits).
\end{itemize}
Since $\mathbb{A}$ is deterministic, each next query can be derived from the previous answers.

Now consider the behavior of $\accgame_g^*$ in the two scenarios: when $Y$ is uniformly random and when $Y = \PRG(\Gamma, k)$. 
These match the ideal and real experiments for the pseudorandom generator $\PRG$, respectively. 
By \thmref{thm:prg:vadhan}, if $\PRG$ is $\eps$-secure against adversaries with memory at most $\hat{W}$, then the total variation distance between these two distributions over the internal memory states of $\accgame_g^*$ is at most $\eps m$. 
Since the sequence of answers $(z_1, \ldots, z_\ell)$ is included in the state of $\accgame_g^*$, this bounds the distance between the output distributions as well, completing the proof.
\end{proof}

By combining \lemref{lem:streamingAccuracy}, which establishes the accuracy of the procedure \texttt{AnswerQueries}, with \lemref{lem:TVdistance}, which shows that the procedure \texttt{AnswerQueries} does not change much in total variation distance between the simplified and natural versions, it follows that \texttt{AnswerQueriesOTP} is also statistically accurate. 

\begin{lemma}
\lemlab{lem:OTPAccuracy}
\cite{KaplanMNS21}
Suppose that $\calA$ is an $(\alpha,\beta)$-accurate algorithm for the SADA problem over streams of length $m$ with memory usage $w$, and that $\PRG$ is an $\eps$-secure BSM pseudorandom generator against adversaries with storage bounded by $\O{w + \ell + b \cdot 2^d}$, where $\ell = \frac{m - n}{(a+1) \cdot 2^d}$. 
Then, for any $\beta', \eps > 0$, the natural algorithm \texttt{AnswerQueries} achieves $(\frac{\alpha}{1 - \gamma} + \alpha', \beta + \beta' + 2^d m \eps)$ statistical accuracy for $\ell$ queries, where
\[\alpha' = \O{\sqrt{\frac{w + \ln(\frac{\ell}{\beta'})}{n}}}.\]
\end{lemma}

Thus, \lemref{lem:natural} and~\lemref{lem:OTPAccuracy} collectively show that the procedure \texttt{AnswerQueries} can be both natural and accurate. 
As a result, we have a contradiction to \thmref{thm:adaNegative} by implementing the algorithm using the PRG from \thmref{thm:prg:vadhan}:

\begin{theorem}
\thmlab{thm:sada:separation}
\cite{KaplanMNS21}
For any given $w$, there exists a streaming problem defined over a universe of size $\poly(w)$ and a stream of length $\O{w^5}$ that can be approximated within a sufficiently small constant additive error using space $\O{\log^2(w)}$ in the oblivious setting, but requires space at least $w$ to be solved in the adversarial setting.  
\end{theorem}
\begin{proof}
To achieve a contradiction to \thmref{thm:adaNegative}, we must show that the natural algorithm \texttt{AnswerQueries} can answer more than $n^2$ adaptive queries given $n$ samples over a domain of size $\Omega(n)$. 
To do this, we set 
\[\ell = \frac{m - n}{(a + 1) \cdot 2^d} = \Omega(n^2) \quad \text{and} \quad d = \O{1} + \log n.\]
With these settings, we get $m = \Theta(n^3 \cdot a)$.

By \lemref{lem:OTPAccuracy}, we set $n=\Theta(w+\log n)$ to guarantee that the procedure \texttt{AnswerQueries} is accurate within a small constant additive error. 
Assuming without loss of generality that $w \geq \log m$ since we can always increase the space usage of $\calA$, this simplifies to $n = \Theta(w)$ and thus $m = \Theta(w^3 \cdot a)$.

Next, to apply \lemref{lem:OTPAccuracy}, we need the pseudorandom generator $\PRG$ to be sufficiently secure. 
We use the construction from \thmref{thm:prg:vadhan} with error 
\[\eps = \frac{\tau}{m \cdot 2^d} = \O{\frac{1}{m n}} = \O{\frac{1}{m w}},\]
for some small constant $\tau > 0$. 
To ensure security against adversaries with storage 
\[\O{w + \ell + b \cdot 2^d} = \O{w^2 + b w},\]
we further need to choose $a = \Omega(w^2 + b w)$ and $b = \Omega\left(\log\frac{a}{\eps}\right) = \Theta(\log(am))$. 
Hence, it suffices to set $a = \Theta(w^2)$ and $b = \Theta(\log(w m)) = \Theta(\log w)$.

Putting everything together, with these parameters, \lemref{lem:OTPAccuracy} guarantees that \texttt{AnswerQueries} answers $\ell = \Omega(n^2)$ adaptive queries over a domain of size $\Omega(n)$, contradicting \thmref{thm:adaNegative}. 
Therefore, no algorithm with space complexity $w$ can solve the $(a, b, d, m, n, \gamma)$-SADA problem to within small constant additive error when:
\[a = \Theta(w^2), \quad b = d = \O{\log w}, \quad m = \Theta(w^5), \quad n = \Theta(w).\]

In contrast, by \thmref{thm:sada:oblivious}, for constant values of $\alpha, \beta, \gamma$, there exists an oblivious algorithm that uses only $\O{(b+d)\log m}=\O{\log^2 w}$ bits of space in this setting.
\end{proof}

\chapter{Turnstile Streams}
\chaplab{chap:turnstile}

\begin{chapterbannerbox}
\centering
For many fundamental problems, any linear sketch can be broken with a polynomial length stream with insertions and deletions.
\end{chapterbannerbox}
\vspace{0.4in}

In this chapter, we continue to study the black-box adversarial model. 
However, we depart from the insertion-only setting discussed in the previous chapters. 
Instead, we focus on turnstile streams, also called insertion-deletion streams, where updates may either increase or decrease coordinates of an underlying frequency vector. 
Crucially, many of the previous frameworks, such as sketch switching, bounded computation paths, difference estimators, or differential privacy, have good performance on insertion-only streams because they have explicit dependencies on the flip number of the stream, c.f., \defref{def:flipno}, and \lemref{lem:monfpflip} shows that the flip number must be small for insertion-only streams. 
In particular, for a stream of length $m=\poly(n)$, the $(\eps,m)$-flip number of many interesting problems such as norm/moment estimation or distinct element estimation is at most $\O{\frac{1}{\eps}\log n}$. 
This is not necessarily the case for turnstile streams, where the $(\eps,m)$-flip number can be as large as $m$. 
Thus although the results in the previous sections can still be used for turnstile streams, their resulting bounds may be polynomial in $m$, which is undesirable. 
In this section, we study whether better adversarially robust algorithms can be designed for the turnstile setting. 

\section{Attack on Specific Sketches}
\seclab{sec:attack:specific:turnstile}
In this section, we analyze adversarial attacks on standard linear sketches within the streaming model. 
In \secref{sec:robust:gauss}, we demonstrate that the Gaussian sketch for $L_2$ estimation with $r$ rows remains robust against $\O{r}$ adaptive queries, meaning it continues to return a constant-factor approximation under such adaptivity. 
This implies that any successful attack on a linear sketch must make at least $\Omega(r)$ adaptive queries. 
Subsequently, in \secref{sec:attack:norm}, we construct a general attack against norm-based estimators using only $\O{r}$ adaptive queries, thereby establishing that $\Theta(r)$ queries are both necessary and sufficient in this setting.

\subsection{Robustness of Gaussian Sketches}
\seclab{sec:robust:gauss}

To establish the existence of a linear sketch that remains accurate under $\O{r}$ adaptive queries, we first recall a well-known result establishing that a scaled Gaussian matrix $\bA \in \mathbb{R}^{m \times n}$ serves as an $L_2$ subspace embedding for sufficiently large $m = \O{n}$, guaranteeing that $\|\bA\bx\|_2 \approx \|\bx\|_2$ holds simultaneously for all $\bx \in \mathbb{R}^n$, as stated for instance in Theorem 2.3 of \cite{Woodruff14}.

\begin{theorem}
\thmlab{thm:gaussian:se}
Let $\bM\in\mathbb{R}^{n \times d}$ be a matrix and let $\eps\in (0,1)$ be a desired approximation parameter. 
Suppose $\bA \in \mathbb{R}^{m \times n}$ is a random matrix whose entries are drawn independently from the normal distribution $\calN\left(0, \frac{1}{m}\right)$. 
If $m = \O{\frac{d + \log(1/\delta)}{\eps^2}}$, then with probability at least $1 - \delta$, the matrix $\bA\bM$ is a $(1+\eps)$ subspace embedding for $\bM$.
\end{theorem}
Moreover, when $m = \O{\frac{1}{\eps^2} \log \frac{1}{\delta}}$, it holds for any fixed $\bx \in \mathbb{R}^d$ that
\[\PPr{(1-\eps)\|\bx\|_2^2 \leq \|\bA\bx\|_2^2 \leq (1+\eps)\|\bx\|_2^2} \geq 1 - \delta.\]
The classical Gaussian sketch sets $\delta$ small enough to ensure correctness over many queries using a standard union bound.

We prove that the Gaussian sketch remains accurate against $\O{r}$ adaptive queries, even when the adversary is given full access to the sketch responses $\bA\bx \in \mathbb{R}^m$, not just the norm $\|\bA\bx\|_2$. 
After $\O{r}$ queries, the adversary learns only an $\O{r} \times \O{r}$ submatrix $\bM$ of $\bA$, since any adaptive query sequence can be simulated by queries on standard basis vectors. 
Importantly, the submatrix $\bM$ is itself Gaussian and thus forms a valid subspace embedding for vectors supported on its rows. 
Because subspace embeddings guarantee norm preservation for \emph{all} vectors in the relevant subspace, we have $\|\bM\bz\|_2 \approx \|\bz\|_2$ for all $\bz$ supported on the corresponding coordinates. 
The adversary gains no information about the remaining rows of $\bA$, and so cannot exploit them. 
Thus, the sketch continues to yield a constant-factor approximation of the $L_2$ norm for any adaptive query $\bx$.

\begin{restatable}{theorem}{thmgaussianrobust}
\thmlab{thm:gaussian:robust}
Any attack that finds $\bx$ such that $\|\bA\bx\|_2\notin\left(\frac{\|\bx\|_2}{2},2\|\bx\|_2\right)$ with probability at least $\frac{9}{10}$ requires $\Omega(r)$ adaptive queries.
\end{restatable}
\begin{proof}
Let $\bA \in \mathbb{R}^{r \times n}$ be a random Gaussian matrix. 
Suppose an adversary is allowed to adaptively query the sketch and receives both $\bA\bx^{(i)}$ and $\|\bA\bx^{(i)}\|_2$ for each query $\bx^{(i)}$, for $i \in [t]$, where $t = cr$ for some sufficiently small constant $c < 1$.

We can assume without loss of generality that the queries $\bx^{(1)}, \ldots, \bx^{(t)}$ form an orthonormal sequence. 
Indeed, any query can be decomposed into a component orthogonal to the previous ones and a component lying in their span. 
The response to the latter can be inferred from previous outputs, and the orthogonal component can be normalized and used as the new query.

Further, since the adversary’s queries can be spanned by elementary vectors, we may assume $\bx^{(i)} = \be_i$ for all $i \in [t]$. 
Letting $\bX = \bx^{(1)} \circ \ldots \circ \bx^{(t)}$, the matrix $\bA\bX \in \mathbb{R}^{r \times t}$ is a submatrix of $\bA$, and due to the rotational invariance of Gaussian distributions, it remains a random Gaussian matrix.

According to \thmref{thm:gaussian:se}, with high probability (at least $0.99$), this submatrix is a constant-factor subspace embedding for $\mathbb{R}^t$, assuming $r = \O{t}$. 
Therefore, for all $\by \in \mathbb{R}^t$, we have:
\[\frac{99}{100} \|\by\|_2^2 \leq \|\bA\bX\by\|_2^2 \leq \frac{101}{100} \|\by\|_2^2.\]
This implies that for any vector $\bw \in \mathbb{R}^n$ that lies within the span of $\{\be_1, \ldots, \be_t\}$, we also have:
\[\frac{99}{100} \|\bw\|_2^2 \leq \|\bA\bw\|_2^2 \leq \frac{101}{100} \|\bw\|_2^2.\]

Now consider any vector $\bz \in \mathbb{R}^n$ orthogonal to the span of the previous queries. 
Since the adversary lacks any information about the corresponding rows of $\bA$, standard concentration bounds imply that with probability at least $0.9$:
\[\frac{99}{100} \|\bz\|_2^2 \leq \|\bA\bz\|_2^2 \leq \frac{101}{100} \|\bz\|_2^2.\]
Furthermore, due to this independence, the cross term $2\langle\bA\by, \bA\bz\rangle$ concentrates around zero, and for sufficiently large $r$, satisfies $|2\langle\bA\by, \bA\bz\rangle| \leq \frac{1}{100}(\|\by\|_2^2 + \|\bz\|_2^2)$ with high probability.

Hence, for the final adaptive query $\bx = \by + \bz$, composed of known and unknown components, we can expand $\|\bA\bx\|_2^2 = \|\bA\by\|_2^2 + \|\bA\bz\|_2^2 + 2\langle\bA\by, \bA\bz\rangle$ to obtain:
\[\PPr{\frac{98}{100} \|\bx\|_2 \leq \|\bA\bx\|_2 \leq \frac{102}{100} \|\bx\|_2} \geq \frac{9}{10}.\]
This demonstrates that the sketch still yields an accurate approximation to $\|\bx\|_2$ with high probability, contradicting the possibility of a successful attack within $t = \O{r}$ queries.
\end{proof}

\subsection{Attack on Norm-Based Estimators}
\seclab{sec:attack:norm}
In this section, we give an attack on norm-based estimators, which use a sketch matrix $\bA\in\mathbb{R}^{r\times n}$ to maintain $\bA\bx$ for the underlying frequency vector $\bx\in\mathbb{R}^n$, and then returns the fixed function $\|\bA\bx\|_2^2$. 
Given a linear sketch with $r$ rows, our attack focuses only on the first $Cr$ coordinates. 
Therefore, we may assume without loss of generality that $\bA \in \mathbb{R}^{r \times Cr}$. 
The attack proceeds as follows. 
We sample a random subset $S \subseteq [Cr]$ and construct a vector $\bx \in \mathbb{R}^{Cr}$ such that $x_j = 0$ for $j \notin S$, and $x_j$ takes a uniformly random sign for $j \in S$. 
Next, we construct a vector $\by \in \mathbb{R}^{Cr}$ satisfying $y_j = 0$ for $j \in S$, and for each $j \notin S$, we set the sign of $y_j$ to align with the sign of the bias in the inner product $\langle \bA \bx, \bA \be_j \rangle$. 
The complete algorithm is presented in \algref{alg:norm:attack}.

\begin{algorithm}[!htb]
\caption{Adversarial Attack on Norm-Based Estimators}
\alglab{alg:norm:attack}
\begin{algorithmic}[1]
\Require{Black-box access to a norm-based estimator $\calA$ over inputs in $\mathbb{R}^{Cr}$}
\Ensure{A vector $\bz$ on which $\calA$ fails to approximate $\|\bz\|_2^2$ within a $1 + \Omega(1)$ factor}
\State{Sample a random sign vector $\bx \in \mathbb{R}^{Cr}$}
\State{Query $\calA(\bx)$}
\For{each $i \in [Cr]$}
  \State{Query $\calA(\be_i)$ and $\calA(\bx + \be_i)$}
  \State{Set $y_i \gets \frac{1}{2} \left( \calA(\bx + \be_i) - \calA(\be_i) - \calA(\bx) \right)$}
  \Comment{$y_i=\langle\bA\bx,\bA\be_i\rangle$}
\EndFor
\State{Query $\calA(\by)$}
\If{any query has failed}
  \State{Set $\bz$ to be the corresponding input vector}
\Else
  \State{Set $\bz \gets \bx + \by$}
\EndIf
\State{\Return $\bz$}
\end{algorithmic}
\end{algorithm}

Before presenting our attack on general norm-based estimators, we establish a structural lemma that guarantees sufficient spectral mass is captured by a random subset of columns from the Gram matrix $\bA^\top\bA$, assuming that the columns of $\bA$ have approximately unit norm.
Here we use $\|\bA^\top\bA\|_{1,2}$ to denote the $L_{1,2}$ norm of $\bA^\top\bA$, which is the sum of the $L_2$ norms of the rows of $\bA^\top\bA$, which is equivalent to the sum of the $L_2$ norms of the columns of $\bA^\top\bA$, since the matrix is symmetric. 
\begin{lemma}
\lemlab{lem:attack:large:coords}
Let $\eps \in \left(0, \frac{1}{2}\right)$, and let $C$ be a sufficiently large constant. 
Suppose $\bA \in \mathbb{R}^{r \times Cr}$ is a matrix whose column $\ell_2$ norms lie in the interval $[(1 - \eps), (1 + \eps)]$. 
Then,
\[\frac{\sqrt{C}}{4} \cdot \|\bA^\top \bA\|_{1,2} \le \|\bA^\top \bA\|_F^2.\]
\end{lemma}
\begin{proof}
We begin by observing that the Frobenius norm of $\bA$ satisfies:
\[\|\bA\|_F \ge \sqrt{Cr(1 - \eps)}.\]
This implies that the sum of the squared singular values of $\bA$ satisfies:
\[\sigma_1(\bA)^2 + \ldots + \sigma_r(\bA)^2 \ge Cr(1 - \eps),\]
and therefore,
\[\|\bA^\top \bA\|_F^2 = \sum_{i=1}^r \sigma_i(\bA)^4 \ge C^2r(1 - \eps)^2 \ge \frac{C^2 r}{4},\]
where the final inequality uses $\eps < \frac{1}{2}$.

Now let $\bw^{(i)}$ denote the $i$-th column of $\bA^\top \bA$ for $i \in [Cr]$. Then we have:
\begin{align*}
\|\bA^\top \bA\|_F^2 &= \sum_{i=1}^{Cr} \|\bw^{(i)}\|_2^2,\\
\|\bA^\top \bA\|_{1,2} &= \sum_{i=1}^{Cr} \|\bw^{(i)}\|_2.
\end{align*}
By the RMS-AM inequality, we get:
\[\sqrt{\frac{\|\bA^\top \bA\|_F^2}{Cr}} \ge \frac{\|\bA^\top \bA\|_{1,2}}{Cr},\]
which rearranges to:
\[\|\bA^\top \bA\|_{1,2} \le \sqrt{Cr \cdot \|\bA^\top \bA\|_F^2}.\]
Combining this with the lower bound $\|\bA^\top \bA\|_F^2 \ge \frac{C^2 r}{4}$, we obtain:
\[\frac{\sqrt{C}}{4} \cdot \|\bA^\top \bA\|_{1,2} \le \|\bA^\top \bA\|_F^2,\]
as claimed.
\end{proof}
We also recall the following formulation of the Paley-Zygmund Inequality. 
\begin{lemma}[Paley-Zygmund Inequality]
Let $X$ be a nonnegative random variable with $\Ex{X^2}<\infty$.
Then for any $\theta \in (0,1)$,
\[\PPr{X\ge\theta\cdot\Ex{X}}\ge(1-\theta)^2\cdot \frac{\Ex{X}^2}{\Ex{X^2}}.\]
\end{lemma}
\begin{proof}
Let $\mu = \Ex{X}$.
By the Cauchy-Schwarz inequality,
\[\Ex{X\cdot\mathbf{1}[X\ge\theta \mu]}\le\sqrt{\Ex{X^2}\cdot\PPr{X\ge\theta\mu}}.\]
On the other hand,
\[\Ex{X\cdot\mathbf{1}[X\ge\theta\mu]}=\Ex{X}-\Ex{X\cdot \mathbf{1}[X<\theta \mu]}\ge\mu-\theta \mu = (1-\theta)\mu.\]
Combining the inequalities and squaring both sides yields
\[(1-\theta)^2 \mu^2\le\Ex{X^2} \cdot \PPr{X \ge \theta \mu},\]
which implies the desired claim.
\end{proof}

Now, we give the full guarantees of the attack in \algref{alg:norm:attack}. 
\begin{restatable}{theorem}{thmnormattack}
\thmlab{thm:norm:attack}
Suppose $\bA \in \mathbb{R}^{r \times Cr}$ for some sufficiently large constant $C > 1$. 
Then there exists an attack that issues $\O{r}$ non-adaptive queries, followed by just two adaptive queries, which causes any norm-based $L_2$ estimator that relies on $\bA$ as a linear sketch to fail to provide a $\Omega(1)$-approximation.
\end{restatable}
\begin{proof}
Let $\bA \in \mathbb{R}^{r \times Cr}$ for some sufficiently large constant $C > 1$. 
Suppose that any column of $\bA$ has $\ell_2$ norm outside the range $(1 - \eps)$ to $(1 + \eps)$; then querying $\calA$ on the corresponding standard basis vector $\be_i$ would reveal this discrepancy and cause the sketch to fail. 
Therefore, without loss of generality, we assume all columns of $\bA$ have norms within this range, satisfying the conditions of \lemref{lem:attack:large:coords}.

Let $\bx \in \mathbb{R}^{Cr}$ be a random vector with i.i.d. Rademacher entries, i.e., $\PPr{x_i = \pm 1} = \frac{1}{2}$ for all $i \in [Cr]$. 
Define $y_i = \langle \bA \bx, \bA \be_i \rangle$, and let $\by$ be the vector with these entries. 
Then
\[\langle \bA \bx, y_i \cdot \bA \be_i \rangle = \langle \bA \bx, \bA \be_i \rangle^2,\]
and summing over $i$ gives
\[\langle \bA \bx, \bA \by \rangle = \sum_{i=1}^{Cr} \langle \bA \bx, \bA \be_i \rangle^2.\]

Let $\bw^{(i)}$ denote the $i$-th column of the Gram matrix $\bA^\top \bA$. 
Since $\bx$ has random signs, by Khintchine's inequality
\begin{align*}
\Ex{\langle \bA \bx, \bA \be_i \rangle^2} &= \Theta(\|\bw^{(i)}\|_2^2),\\
\Ex{\langle \bA \bx, \bA \be_i \rangle^4} &= \Theta(\|\bw^{(i)}\|_2^4).
\end{align*}
Also, by Cauchy-Schwarz,
\[\Ex{\langle \bA \bx, \bA \be_i \rangle^2 \langle \bA \bx, \bA \be_j \rangle^2}\le \sqrt{\Ex{\langle \bA \bx, \bA \be_i \rangle^4} \cdot \Ex{\langle \bA \bx, \bA \be_j \rangle^4}}.\]
Thus, it follows from the earlier observation of $\langle \bA \bx, \bA \by \rangle = \sum_{i=1}^{Cr} \langle \bA \bx, \bA \be_i \rangle^2$ that
\[\Ex{\langle \bA \bx, \bA \by \rangle} = \sum_{i=1}^{Cr} \Theta(\|\bw^{(i)}\|_2^2) = \Theta(\|\bA^\top \bA\|_F^2),\]
and similarly,
\[\Ex{\langle \bA \bx, \bA \by \rangle^2} = \sum_{i,j=1}^{Cr} \Theta(\|\bw^{(i)}\|_2^2 \|\bw^{(j)}\|_2^2) = \Theta(\|\bA^\top \bA\|_F^4).\]

By the Paley-Zygmund inequality, this implies
\[\PPr{\langle \bA \bx, \bA \by \rangle \ge \xi_1 \cdot \|\bA^\top \bA\|_F^2} \ge p,\]
for some fixed constant $\xi_1$ and $p = \Omega(1)$ is a constant independent of $C$. 

Next, observe that since $y_i = \langle \bA \bx, \bA \be_i \rangle$, Khintchine's inequality gives
\[\Ex{y_i^2} = \Theta(\|\bw^{(i)}\|_2^2),\]
and hence
\[\Ex{\|\by\|_2^2} = \sum_{i=1}^{Cr} \Theta(\|\bw^{(i)}\|_2^2) = \Theta(\|\bA^\top \bA\|_F^2).\]
Applying Markov's inequality, there exists a constant $\xi_2$ such that:
\[\PPr{\|\by\|_2^2 \le \xi_2 \cdot \|\bA^\top \bA\|_F^2} \ge 1 - \frac{p}{3}.\]

Moreover, by Khintchine and the triangle inequality
\[\Ex{\langle \bx, \by \rangle} \le \sum_{i=1}^{Cr} \Ex{|\langle \bA \bx, \bA \be_i \rangle|} \le \sum_{i=1}^{Cr} \|\bw^{(i)}\|_2 = \|\bA^\top \bA\|_{1,2}.\]
Another application of Markov yields
\[\PPr{\langle \bx, \by \rangle \le \tfrac{3}{p} \cdot \|\bA^\top \bA\|_{1,2}} \ge 1 - \frac{p}{3}.\]

Define $\calE$ as the event that all of the following occur:
\begin{enumerate}
\item
$\langle \bx, \by \rangle \le \frac{3}{p} \cdot \|\bA^\top \bA\|_{1,2}$
\item
$\langle \bA \bx, \bA \by \rangle \ge \xi_1 \cdot \|\bA^\top \bA\|_F^2$
\item 
$\|\by\|_2^2 \le \xi_2 \cdot \|\bA^\top \bA\|_F^2$.
\end{enumerate}
Then $\PPr{\calE} \ge \frac{p}{3}$ by the above argument. 

The attack now queries $\bx$, $\by$, and $\bx + \by$. 
If the estimator $\calA$ fails on either $\bx$ or $\by$, it already fails. 
So assume it succeeds on both with some fixed error $\eps = \Theta(1)$. 
Observe that
\begin{align*}
\|\bx + \by\|_2^2 &= \|\bx\|_2^2 + 2\langle \bx, \by \rangle + \|\by\|_2^2,\\
\|\bA(\bx + \by)\|_2^2 &= \|\bA \bx\|_2^2 + 2\langle \bA \bx, \bA \by \rangle + \|\bA \by\|_2^2.
\end{align*}
Conditioned on $\calE$, we have $\|\bx\|_2^2 + \|\by\|_2^2 \le \|\bx\|_2^2 + \|\bA^\top \bA\|_F^2$
and $\langle \bx, \by \rangle \le \frac{3}{p} \cdot \|\bA^\top \bA\|_{1,2}$, but
\[\|\bA(\bx + \by)\|_2^2 \ge (1 - \eps)\|\bx\|_2^2 + 2\xi_1 \cdot \|\bA^\top \bA\|_F^2 + (1 - \eps)\|\by\|_2^2.\]

By \lemref{lem:attack:large:coords}, we know $\|\bA^\top \bA\|_F^2 \ge \frac{\sqrt{C}}{4} \cdot \|\bA^\top \bA\|_{1,2}$. 
So for large enough $C$ and small enough $\eps = \O{1}$, the estimate $\calA(\bx + \by)$ will violate the $(1 + \eps)$-approximation condition.
Since the attack succeeds with constant probability $q = \Omega(1)$, repeating it $\O{\frac{1}{q}}$ times suffices to ensure success with high probability, e.g., $0.99$.

Finally, observe that the attack uses $\O{r}$ non-adaptive queries, for $\calA(\be_i)$ and $\calA(\bx + \be_i)$, followed by two adaptive queries: one for $\calA(\by)$ and one for the final attack vector $\bz = \bx + \by$.
\end{proof}

\section{Lower Bound for \texorpdfstring{$F_0$}{F0} Estimation with Real-Valued Linear Sketches}
In this section, we study the setting where the sketching matrix $\bA \in \mathbb{R}^{r \times n}$ has all nonzero subdeterminants lower bounded by $\frac{1}{\poly(r)}$, which is a property satisfied by all known sketching algorithms. 
We emphasize that we focus on linear sketches in this section, as opposed to integer sketches. 
Thus, the results in this section are analogous to those of \secref{sec:fp:lb:real}
We describe the following result by \cite{GribelyukLWYZ24}. 

\begin{restatable}{theorem}{thmreal}
\thmlab{thm:real}
\cite{GribelyukLWYZ24}
Let $\mathcal{A}$ be a linear sketching algorithm with sketching matrix $\bA \in \mathbb{R}^{r \times n}$, and suppose all nonzero subdeterminants of $\bA$ are at least $\frac{1}{\poly(r)}$. 
Suppose $\mathcal{A}$ uses an estimator $f: \mathbb{R}^{r\times n}\times\mathbb{R}^r \rightarrow \{-1, +1\}$ to solve the $(\alpha + c, \beta - c)$-gap $F_0$ problem for constants $\alpha, \beta, c$, and returns $f(\bA, \bA\bx)$ on each input $\bx$.

Then, there exists a randomized algorithm that, with high constant probability and using at most $\poly(r)$ adaptive queries to $\mathcal{A}$, finds a distribution $D$ over $\mathbb{R}^n$ such that $\mathcal{A}$ fails on $D$ with constant probability. 
Moreover, this algorithm runs in $\poly(r)$ time.
\end{restatable}

The analysis of \cite{GribelyukLWYZ24} follows a similar attack strategy as in the integer-valued case $\bx \in \mathbb{Z}^n$, aiming to identify the significant columns of $\bA$ through carefully chosen queries. 
However, since $\bA$ is now real-valued, we need to redefine the notion of column significance and design new hard input distributions for the insignificant coordinates. 
In particular, we say column $i$ is significant if there exists $\by^\top \in \mathbb{R}^r$ such that
\[(\by^\top \bA)_i^2 \ge \frac{1}{s} \cdot \|\by^\top \bA\|_2^2.\]
To handle the real-valued case, we show how to iteratively eliminate a small number of columns so that the resulting matrix $\bA' \in \mathbb{R}^{r \times n'}$ has all leverage scores at most $\frac{1}{s}$, c.f., \defref{def:leverage:score}. 
Due to real-valued entries in the matrix $\bA$, the earlier information-theoretic arguments from \secref{sec:fzero:preprocess} no longer apply. 
Instead, we rely on a volume-based argument. 
To that end, observe the total leverage score is at most $r$, so one might hope to remove at most $rs$ columns since each removed column will have leverage score at least $\frac{1}{s}$. 
However, this is complicated by the fact that leverage scores can increase when other columns are zeroed out. 
To manage this, \cite{GribelyukLWYZ24} uses a more refined analysis based on volume change, previously applied to online leverage score bounds~\cite{CohenMP20,BravermanDMMUWZ20}.

\begin{lemma}[Matrix determinant lemma]
\lemlab{lem:matrix:det}
For any vector $\bu \in \mathbb{R}^d$ and matrix $\bM \in \mathbb{R}^{d \times d}$, we have
\[\det(\bM + \bu\bu^\top) = \det(\bM) \cdot (1 + \bu^\top \bM^{-1} \bu).\]
\end{lemma}
\cite{GribelyukLWYZ24} shows that for matrices with both bounded entries and bounded subdeterminants, only a bounded number of high-leverage columns need to be eliminated before the remaining ones have low leverage scores. 
This includes the important case of integer matrices with bounded entries. 
We also note that for general matrices with entries represented using $b$ bits, rescaling can convert them into integer matrices with entries of magnitude at most $2^b$.

We further remark that while the lemma below is stated for matrices whose subdeterminants are at least $\frac{1}{\poly(r)}$, the result naturally generalizes to matrices with subdeterminants lower bounded by any constant $\kappa > 0$. 
In such cases, the number of columns removed increases to $\O{r^2s \log(\kappa nr)}$. 
For example, if the subdeterminants are at least $\frac{1}{n^{\poly(r)}}$, this results in removing $s\cdot\poly(r)\cdot \log n$ columns.

\begin{lemma}
\cite{GribelyukLWYZ24}
Let $\bA \in \mathbb{R}^{r \times n}$ be a matrix with entries bounded by $\poly(r)$ and all subdeterminants either zero or at least $\frac{1}{\poly(r)}$. 
For any parameter $s \ge 1$, there exists a preprocessing algorithm that outputs a matrix $\bA' \in \mathbb{R}^{r \times n}$ by zeroing out at most $\O{r^2 s \log(nr)}$ columns of $\bA$, such that all remaining columns of $\bA'$ have leverage score at most $\frac{1}{s}$.
\end{lemma}
\begin{proof}
Let $\bS = \bA \bA^\top \in \mathbb{R}^{r \times r}$. 
Using the matrix determinant lemma, c.f., \lemref{lem:matrix:det}, we have for any $\bu \in \mathbb{R}^r$:
\[\det(\bS + \bu \bu^\top) = \det(\bS) \cdot (1 + \bu^\top \bS^{-1} \bu).\]
Now suppose we remove a column $\bA_i$ from $\bA$, which corresponds to updating $\bS$ to $\bS - \bA_i \bA_i^\top$. 
Then,
\[\det(\bS - \bA_i \bA_i^\top) = \det(\bS) \cdot (1 - \ell_i),\]
where $\ell_i = \bA_i^\top \bS^{-1} \bA_i$ is the leverage score of column $i$.

If $\ell_i = 1$, then removing $\bA_i$ reduces the rank of $\bS$, and we can restart the argument on the resulting linearly independent submatrix. 
This can happen at most $r$ times. 
For the remainder, we assume all $\ell_i < 1$ and focus on removing columns with leverage score $\ell_i > \frac{1}{s}$.

In this case,
\[|\det(\bS - \bA_i \bA_i^\top)| \le |\det(\bS)| \cdot \left(1 - \frac{1}{s} \right).\]
Since the entries of $\bA$ are bounded by $\poly(r)$, we have
\[|\det(\bS)| \le \|\bS\|_F^r \le (n \cdot \poly(r))^r.\]
Thus, after $\O{rs \log(nr)}$ such removals, we would obtain
\[|\det(\bS - \bA_i \bA_i^\top)| < \frac{1}{\poly(r)},\]
which contradicts the assumption that every nonzero subdeterminant is at least $\frac{1}{\poly(r)}$. 
Since rank can decrease at most $r$ times, we conclude that no more than $\O{r^2s \log(nr)}$ columns are removed in total before all remaining leverage scores are at most $\frac{1}{s}$.
\end{proof}

We now turn to the construction of the hard distribution for the insignificant coordinates. 
Let $D = \calN(0,1)$ denote the standard normal distribution. 
Define a parameterized distribution $D_p$ for some constant $p \in (0,1)$ as follows: sample $x \sim D_p$ by letting
\[x = \frac{1}{\sqrt{p}} \cdot \Bern(p) \cdot \calN(0,1),\]
where $\Bern(p)$ is a Bernoulli random variable taking value $1$ with probability $p$ and $0$ otherwise. 
Equivalently, with probability $1 - p$, we have $x = 0$ and otherwise with probability $p$, we have $x \sim \calN\left(0,\frac{1}{p}\right)$.
This construction ensures the following properties:
\[\EEx{x \sim D_1}{x} = \EEx{x \sim D_2}{x} = 0, \quad
\EEx{x \sim D_1}{x^2} = \EEx{x \sim D_2}{x^2} = 1.\]

We next upper bound the total variation distance $\TVD(\bA \bx^{(1)}, \bA \bx^{(2)})$ for $\bx^{(1)} \sim D_p$ and $\bx^{(2)} \sim D_q$, where $p$ and $q$ are drawn uniformly at random from an interval $(\alpha, 1)$ for some small constant $\alpha\in(0,1)$.
To support this analysis, we recall a concentration bound in the form of Azuma's inequality.

\begin{theorem}[Azuma's inequality]
\thmlab{thm:azuma:ineq}
Let $Z_1, \ldots, Z_n$ be mean-zero random variables, and suppose that $|Z_i| \le \beta_i$ for each $i \in [n]$. 
Then for any $t > 0$,
\[\PPr{\left| \sum_{i=1}^n Z_i \right| > t} \le \exp\left( -\frac{t^2}{2 \sum_{i=1}^n \beta_i^2} \right).\]
\end{theorem}
We now show that randomly sampling and rescaling columns from a matrix $\bA$ yields a good subspace embedding with high probability.
\begin{lemma}
\lemlab{lem:spectral:preserve}
\cite{GribelyukLWYZ24}
Let $\gamma \ge 1$ be a fixed constant and let $p\in(0,1)$ be a fixed probability. 
Let $\bA \in \mathbb{R}^{r \times n}$ be a matrix such that no column has leverage score larger than $\frac{1}{s}$, where $s = \Theta\left(\frac{\gamma^2}{p^2}\cdot r^4 \log r\right)$. 
Construct a random matrix $\bB \in \mathbb{R}^{r \times n}$ by sampling each column of $\bA$ independently with probability $p$, and scaling selected columns by $\frac{1}{\sqrt{p}}$; otherwise, set the column to zero. 
Then with high probability, for all $\bx \in \mathbb{R}^r$,
\[\left(1 - \frac{1}{\gamma r}\right) \cdot \|\bx^\top \bB\|_2^2 \le \|\bx^\top \bA\|_2^2 \le \left(1 + \frac{1}{\gamma r}\right) \cdot \|\bx^\top \bB\|_2^2.\]
\end{lemma}
\begin{proof}
Fix any vector $\bx \in \mathbb{R}^r$ and define $\by = \bA^\top\bx\in\mathbb{R}^n$. 
If $\by = \mathbf{0}^n$, then $\bx^\top \bA = \bx^\top \bB = \mathbf{0}^n$ and the claim trivially holds. 
Otherwise, we may assume without loss of generality that $\|\by\|_2 = 1$.

For each $i \in [n]$, define the random variable
\[Z_i = \frac{1}{p} \cdot y_i^2 \cdot X_i - y_i^2,\]
where $X_i \sim \mathrm{Bern}(p)$. 
Then $\mathbb{E}[Z_i] = 0$. 
Since $y_i^2 \le \ell_i$, where $\ell_i$ is the leverage score of column $i$, we have
\[|Z_i| \le \left( \frac{1}{p} - 1 \right) y_i^2 \le \frac{\ell_i}{p} \le \frac{1}{ps},\]
using the assumption that $\ell_i \le \frac{1}{s}$. 
Define $\beta_i = \frac{\ell_i}{p}$, so that
\[\sum_{i=1}^n \beta_i \le \frac{r}{p}.\]

Applying Azuma's inequality (\thmref{thm:azuma:ineq}) gives
\[\PPr{\left| \sum_{i=1}^n Z_i \right| > \frac{1}{\gamma r}}\le \exp\left( -\frac{p^2}{2 \cdot \gamma^2 r^2 \cdot \frac{r}{s}} \right) = \exp\left( -\Theta(\gamma^2 r \log r) \right),\]
where we used the assumption $s = \Theta\left( \frac{\gamma^2}{p^2} \cdot r^4 \log r \right)$.

To ensure this holds uniformly over all directions $\bx$, consider a $\frac{1}{\gamma r}$-net $\mathcal{N}$ over the unit sphere in the row span of $\bA$. 
This net has size at most $|\mathcal{N}| \le (\gamma r)^{\mathcal{O}(r)}$, so a union bound over $\mathcal{N}$ implies that the above bound holds for all $\by' \in \mathcal{N}$ simultaneously with high probability.

Now consider any unit vector $\by$ in the row span of $\bA$, and let $\by' \in \mathcal{N}$ be the closest net point. 
Define the diagonal sampling matrix $\bD \in \mathbb{R}^{n \times n}$ with $\sqrt{\frac{1}{p}}$ on the diagonal for sampled columns and $0$ elsewhere. Then
\[\|\bD \by\|_2 \le \|\bD \by'\|_2 + \|\bD(\by - \by')\|_2 \le 1 + \frac{1}{\gamma r} + \sqrt{2} \cdot \|\by - \by'\|_2 \le 1 + \O{\frac{1}{\gamma r}},\]
and similarly,
\[\|\bD \by\|_2 \ge \|\bD \by'\|_2 - \|\bD(\by - \by')\|_2 \ge 1 - \frac{1}{\gamma r} - \sqrt{2} \cdot \|\by - \by'\|_2 \ge 1 - \O{\frac{1}{\gamma r}}.\]
Since $\bx^\top \bB = \bD \by$, then the lemma follows.
\end{proof}

We begin by recalling the definition of Kullback-Leibler (KL) divergence.

\begin{definition}[Kullback-Leibler Divergence]
Let $P$ and $Q$ be continuous probability distributions over a random variable, with corresponding densities $p$ and $q$ supported on a domain $\Omega$. 
The KL divergence between $P$ and $Q$ is defined as
\[\KLD(P \| Q) = \int_{x \in \Omega} p(x) \log \frac{p(x)}{q(x)} \, dx.\]
\end{definition}

A standard result gives a closed-form expression for the KL divergence between two multivariate Gaussian distributions; see, for example, \cite{duchi2020derivations}.

\begin{lemma}
\lemlab{lem:kl:multivariates}
Let $P = \calN(\mu_1, \Sigma_1)$ and $Q = \calN(\mu_2, \Sigma_2)$. 
Then
\[\KLD(P \| Q) = \frac{1}{2} \left( \log \frac{\det(\Sigma_2)}{\det(\Sigma_1)} - r + \Trace(\Sigma_2^{-1} \Sigma_1) + (\mu_2 - \mu_1)^\top \Sigma_2^{-1} (\mu_2 - \mu_1) \right).\]
\end{lemma}

We also recall the following useful inequality relating KL divergence to total variation distance.

\begin{theorem}[Pinsker's Inequality]
\thmlab{thm:pinsker}
For any distributions $P$ and $Q$, the total variation distance satisfies
\[\TVD(P, Q) \le \sqrt{ \frac{1}{2} \KLD(P \| Q) }.\]
\end{theorem}

We now bound the total variation distance between the distributions obtained by applying a fixed matrix $\bA$ to two different random vectors: one standard Gaussian and one sparse scaled Gaussian.

\begin{lemma}
\lemlab{lem:tvd:p}
\cite{GribelyukLWYZ24}
Let $\gamma \ge 1$ be a fixed constant, and set $s = \Theta\left( \frac{\gamma^2}{p^2} \cdot r^4 \log r \right)$. 
Let $D = \calN(0, 1)$ and $D_p = \Bern(p) \cdot \calN\left(0, \frac{1}{p}\right)$ for some constant $p \in (0, 1)$. 
Let $\bx^{(1)} \sim D^n$ and $\bx^{(2)} \sim D_p^n$. 
Let $\bA \in \mathbb{R}^{r \times n}$ be a matrix whose columns all have leverage score at most $\frac{1}{s}$. Then
\[\TVD(\bA \bx^{(1)}, \bA \bx^{(2)}) \le \O{\frac{1}{\gamma}}.\]
\end{lemma}
\begin{proof}
The vector $\bx^{(1)}$ is a standard Gaussian, so $\bA \bx^{(1)}$ is a multivariate Gaussian with mean $\mathbf{0}^r$ and covariance $\bA \bA^\top$.
For $\bx^{(2)} \sim D_p^n$, define $S$ to be its support. 
Each nonzero entry of $\bx^{(2)}$ is distributed as $\calN\left(0, \frac{1}{p}\right)$. 
Thus, $\bA \bx^{(2)}$ is also Gaussian with mean $\mathbf{0}^r$ and some covariance matrix $\bB \bB^\top$ for an appropriate matrix $\bB$. 
Applying \lemref{lem:kl:multivariates}, we obtain:
\[\KLD(\bA \bx^{(1)} \| \bA \bx^{(2)}) = \frac{1}{2} \left( \log \frac{\det(\bB\bB^\top)}{\det(\bA\bA^\top)} - r + \Trace\left((\bB\bB^\top)^{-1} \bA\bA^\top\right) \right).\]
Let $\calE$ be the event that
\[\left(1 - \frac{1}{\gamma r} \right)^2 \bB\bB^\top \preceq \bA\bA^\top \preceq \left(1 + \frac{1}{\gamma r} \right)^2 \bB\bB^\top.\]
From \lemref{lem:spectral:preserve}, we know that $\PPr{\calE} \ge 1 - \frac{1}{\poly(r)}$. 

Conditioned on $\calE$, let $\lambda_1, \ldots, \lambda_r$ be the eigenvalues of $(\bB \bB^\top)^{-1} \bA \bA^\top$. 
The event $\calE$ implies that for all $i \in [r]$,
\[\left(1 - \frac{1}{\gamma r} \right)^2 \le \lambda_i \le \left(1 + \frac{1}{\gamma r} \right)^2.\]
By expanding the squares, we can write $\lambda_i = 1 + \delta_i$ for some $\delta_i$ satisfying $|\delta_i| = \O{\frac{1}{\gamma r}}$. 
Since the trace of a matrix is the sum of its eigenvalues and the determinant is their product, we have $\Trace\left((\bB \bB^\top)^{-1} \bA \bA^\top\right) = \sum_{i=1}^r \lambda_i$ and $\log \frac{\det(\bB \bB^\top)}{\det(\bA \bA^\top)} = -\log \det\left((\bB \bB^\top)^{-1} \bA \bA^\top\right) = -\sum_{i=1}^r \log \lambda_i$. 
Therefore, the KL divergence can be expressed directly in terms of these eigenvalues as:
\begin{align*}
\KLD(\bA \bx^{(1)} \| \bA \bx^{(2)} \mid \calE) &= \frac{1}{2} \sum_{i=1}^r (\lambda_i - 1 - \log \lambda_i) \\
&= \frac{1}{2} \sum_{i=1}^r (\delta_i - \log(1+\delta_i)).
\end{align*}
Using the second-order Taylor expansion $x - \log(1+x) = \O{x^2}$ for small $x$, we have $\delta_i - \log(1+\delta_i) = \O{\delta_i^2}$. 
Thus, the KL divergence is upper bounded by
\begin{align*}
\KLD(\bA \bx^{(1)} \| \bA \bx^{(2)} \mid \calE) &\le \frac{1}{2} \sum_{i=1}^r \O{\delta_i^2} \\
&= \O{r \cdot \frac{1}{\gamma^2 r^2}} \\
&= \O{\frac{1}{\gamma^2 r}} \le \O{\frac{1}{\gamma^2}}.
\end{align*}
Applying \thmref{thm:pinsker} gives
\[\TVD(\bA \bx^{(1)}, \bA \bx^{(2)} \mid \calE) \le \sqrt{\frac{1}{2} \KLD(\bA \bx^{(1)} \| \bA \bx^{(2)} \mid \calE)} \le \O{\frac{1}{\gamma \sqrt{r}}} \le \O{\frac{1}{\gamma}}.\] 
Finally, since $\PPr{\calE} \ge 1 - \frac{1}{\poly(r)}$, we conclude that
\[\TVD(\bA \bx^{(1)}, \bA \bx^{(2)}) \le \O{\frac{1}{\gamma}}.\]
\end{proof}

By combining \lemref{lem:tvd:p} with the triangle inequality, we obtain the following result directly.
\begin{lemma}
\lemlab{lem:tvd:pq}
\cite{GribelyukLWYZ24}
Let $\gamma \ge 1$ be a fixed constant, and define $s = \Theta(\gamma^2 r^4 \log r)$. 
Consider the distributions $D_p = \Bern(p) \cdot \calN\left(0, \frac{1}{p}\right)$ and $D_q = \Bern(q) \cdot \calN\left(0, \frac{1}{q}\right)$ for constants $p, q \in (0,1)$. 
Let $\bx^{(1)} \sim D_p^n$ and $\bx^{(2)} \sim D_q^n$. 
Suppose $\bA \in \mathbb{R}^{r \times n}$ has leverage scores at most $\frac{1}{s}$. 
Then
\[\TVD(\bA \bx^{(1)}, \bA \bx^{(2)}) \le \O{\frac{1}{\gamma}}.\]
\end{lemma}

We now describe our attack algorithm over real-valued inputs, presented in \figref{fig:fzero:real:attack}. 
This construction mirrors the attack used for inputs in $\mathbb{Z}^{r \times n}$, with the key differences being the choice of input distribution and the tuning of the parameters $h$ and $\sigma$.

\begin{figure}[!htb]
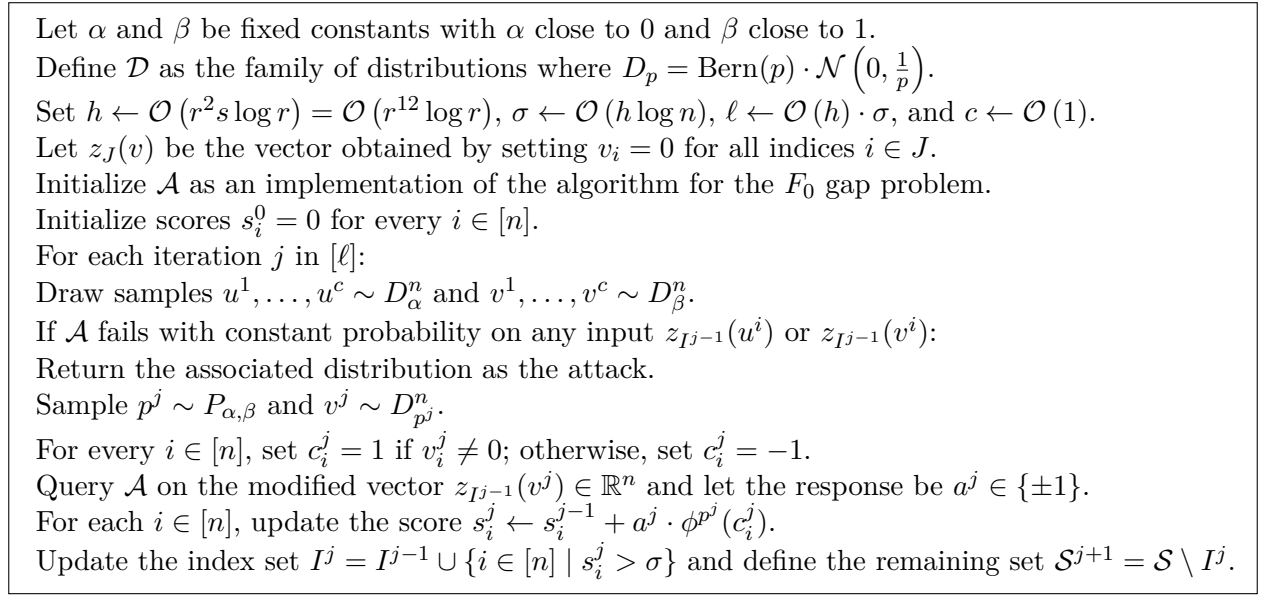

\begin{mdframed}
Let $\alpha$ and $\beta$ be fixed constants with $\alpha$ close to $0$ and $\beta$ close to $1$.
\newline
Define $\mathcal{D}$ as the family of distributions where $D_p = \Bern(p) \cdot \calN\left(0,\frac{1}{p}\right)$.
\newline\noindent
Set $h \gets \O{r^2 s \log r} = \O{r^{12} \log r}$, $\sigma \gets \O{h \log n}$, $\ell \gets \O{h} \cdot \sigma$, and $c \gets \O{1}$.
\newline\noindent
Let $z_J(v)$ be the vector obtained by setting $v_i = 0$ for all indices $i \in J$.
\newline
Initialize $\mathcal{A}$ as an implementation of the algorithm for the $F_0$ gap problem.
\newline\noindent
Initialize scores $s_i^0 = 0$ for every $i \in [n]$.
\newline\noindent
For each iteration $j$ in $[\ell]$:
\newline\indent
Draw samples $u^1, \ldots, u^c \sim D_\alpha^n$ and $v^1, \ldots, v^c \sim D_\beta^n$.
\newline\indent
If $\mathcal{A}$ fails with constant probability on any input $z_{I^{j-1}}(u^i)$ or $z_{I^{j-1}}(v^i)$:
\newline\indent\indent
Return the associated distribution as the attack.
\newline\indent
Sample $p^j \sim P_{\alpha,\beta}$ and $v^j \sim D_{p^j}^n$.
\newline\indent
For every $i \in [n]$, set $c_i^j = 1$ if $v_i^j \ne 0$; otherwise, set $c_i^j = -1$.
\newline\indent
Query $\mathcal{A}$ on the modified vector $z_{I^{j-1}}(v^j) \in \mathbb{R}^n$ and let the response be $a^j \in \{\pm 1\}$.
\newline\indent
For each $i \in [n]$, update the score $s_i^j \gets s_i^{j-1} + a^j \cdot \phi^{p^j}(c_i^j)$.
\newline\indent
Update the index set $I^j = I^{j-1} \cup \{i \in [n] \mid s_i^j > \sigma\}$ and define the remaining set $\mathcal{S}^{j+1} = \mathcal{S} \setminus I^j$.
\end{mdframed}
\caption{Attack on Real-Valued Sketches for $F_0$ Estimation}
\figlab{fig:fzero:real:attack}
\end{figure}

We now prove \thmref{thm:real}.
\thmreal*
\begin{proof}
The proof follows a similar structure to that in \secref{sec:fzero:fingerprinting}. 
Recall from \lemref{lem:tvd:pq} that we chose $\gamma = r^3$, which implies $s = \O{r^{10} \log r}$ and $r^2 s = \O{r^{12} \log r}$. 
Therefore, we may assume the sketching matrix $\bA$ has the form
\[\bA = \begin{bmatrix}
        \bD \\
        \bS
    \end{bmatrix} \;,\]
where $\bS$ has at most $r^2 s$ non-zero columns and $\bD$ satisfies the property:
\begin{equation}
\label{eq:real}
\forall \by^\top \in \mathbb{R}^r, \quad (\by^\top \bD)_i^2 \le \frac{1}{s} \cdot \|\by^\top \bD\|_2^2 \;.
\end{equation}
Let $\mathcal{S}$ denote the indices corresponding to the non-zero columns in $\bS$.

\paragraph{Soundness.}
Consider indices $i \in I \setminus \mathcal{S}$. 
Given our parameter settings, the total variation distance between distributions $\bD \bx^D$ for different $p \in [\alpha, \beta]$ (with $\bx \sim D_p$) is at most $\O{\frac{1}{r^3}}$. 
Consequently, only $\O{r^9}$ coordinates in $I \setminus \mathcal{S}$ can have expected increments $\Ex{s_i^t - s_i^{t-1}} = \Omega\left(\frac{1}{r^{12}}\right)$ conditioned on $\bD \bx^D$. 
Since the total number of queries is $\tO{r^2 s^4}$, with high probability, at most $\tO{r^9} = o(s)$ coordinates outside $\mathcal{S}$ will be falsely flagged.

Suppose that $\bD$ satisfies the condition in \eqref{eq:real}, and let $\bD'$ be the matrix obtained by zeroing out $o(s)$ columns of $\bD$. 
Since for every remaining index $i$ and any $\by^\top \in \mathbb{R}^r$ we have $(\by^\top \bD)_i^2 \le \frac{1}{s} \cdot \|\by^\top \bD\|_2^2$, then it follows that
\[\forall \by^\top \in \mathbb{R}^r, \quad (\by^\top \bD')_i^2 \le \frac{1.1}{s} \cdot \|\by^\top \bD'\|_2^2 \;.\]
Thus, the key property of $\bD$ remains essentially intact in $\bD'$, up to a small multiplicative factor.

\FloatBarrier

\paragraph{Completeness.}
Suppose that the target algorithm $\mathcal{A}$ uses an estimator $f$. 
Consider a modified algorithm $\mathcal{A}'$ that also uses $f$ but, instead of directly taking the sketch $\bA \bx$, it takes
\[\begin{bmatrix}
\bD' \bx' \\
\bS \bx_{\mathcal{S}}
\end{bmatrix},\]
where $\bx' \sim D_\gamma^{|D|}$ for some fixed $\gamma \in [\alpha, \beta]$. 
This sample $\bx'$ is generated independently of the actual input by the algorithm.
By \lemref{lem:distribution:tvd}, the total variation distance between the actual sketch 
$\begin{bmatrix}
        \bD' \bx^{(t)}_{D} \\
        \bS \bx^{(t)}_{\mathcal{S}}
\end{bmatrix}$
and 
$\begin{bmatrix}
        \bD' \bx' \\
        \bS \bx^{(t)}_{\mathcal{S}}
\end{bmatrix}$
is at most $\O{\frac{1}{r^3}}$ in each round, for a sufficiently small constant $\gamma$.  
Thus, if $\mathcal{A}$ succeeds with probability at least $1 - \delta$ under some input distribution, then $\mathcal{A}'$ succeeds with probability at least $1 - \delta - \O{\frac{1}{r^3}}$.

Now consider attacking $\mathcal{A}'$. 
Since its only dependence on $\bx$ is through $\bx_{\mathcal{S}}$, \lemref{lem:soundness} guarantees that, with probability at least $1 - \frac{1}{n}$, no index outside $\mathcal{S}$ is incorrectly flagged. 
Moreover, \lemref{lem:completeness} ensures that, with high probability, our attack correctly identifies (some or all) of the relevant coordinates in $\mathcal{S}$ and outputs a distribution on which $\mathcal{A}'$ fails. 
The increase in error probability due to the variation distance is only $\O{\frac{1}{r^3}}$, which minimally impacts the function gap $g(\beta) - g(\alpha)$ in \lemref{lem:function:gap}, preserving its $\Omega(1)$ gap.

It follows that our attack identifies a distribution $\bq$ where $\mathcal{A}'$ fails with constant probability. 
Since the total variation distance between the output distributions of $\mathcal{A}$ and $\mathcal{A}'$ is $\O{\frac{1}{r^3}}$, and this randomness is only over the input $\bx$, the outputs of both algorithms are nearly indistinguishable. 
Consequently, $\mathcal{A}$ also fails on this distribution with constant probability, completing the proof.
\end{proof}

\section{Lower Bound for \texorpdfstring{$F_p$}{Fp} Estimation with Real-Valued Linear Sketches}
\seclab{sec:fp:lb:real}
Although the results in \secref{sec:dp} show that there exist non-trivial algorithms that achieve robustness on turnstile streams of length $m$ using $\tO{\sqrt{m}}$ space, this is still quite far from non-adaptive turnstile streams that solve the corresponding problems using $\polylog(m)$ space. 
Thus it is natural to ask whether there is an inherent limitation for adversarial inputs on turnstile streams. 

A common technique for algorithmic design for non-adaptive turnstile streams is the use of \emph{linear sketches}. 
The general approach involves defining a distribution $\Pi$ over linear mappings $\bA: \mathbb{R}^n \to \mathbb{R}^r$, where $r \ll n$. 
A matrix $\bA$ is sampled from $\Pi$, and during the online phase, when a vector $\bx\in\mathbb{R}^n$ is updated, the algorithm maintains the sketch $\bA\bx$.
This yields a compact representation of $\bx$, which enables approximate answers to a variety of queries about $\bx$, given various choices of post-processing on the sketch vector $\bA\bx$. 
Indeed, linear sketches have been used for distance estimation~\cite{Johnson84}, distinct element estimation~\cite{KaneNW10b}, norm/moment estimation~\cite{AlonMS99,IndykW05,Indyk06,KaneNW10a,AndoniKO11,GangulyW18}, heavy-hitters~\cite{CharikarCF04,CormodeM05}, compressed sensing~\cite{PriceW11,PriceW13}, subspace embeddings~\cite{ClarksonW13,Woodruff14}, eigenvalue estimation and PSD testing~\cite{NeedellSW22,SwartworthW23}. 
In this section, we present a result by \cite{HardtW13} that shows linear sketches are not robust to adaptive queries:

\begin{theorem}[Informal version of \thmref{thm:attack}]
\cite{HardtW13}
Given a parameter $B>2$ and oracle access to a linear sketch consisting of at most $r = n - \O{\log(nB)}$ rows, there exists a randomized algorithm that with high probability, constructs a distribution over inputs on which the sketch fails to provide a $B$-approximation to $\|\bv\|_2^2$, where $v$ is the frequency vector defined by the (adaptive) turnstile stream. 

The algorithm performs at most $\poly(rB)$ queries to the oracle, with each query chosen adaptively, and the total runtime is bounded by $\poly(rB)$. 
Furthermore, the adaptivity is structured: the full set of queries can be divided into at most $r$ batches, each consisting of non-adaptive queries. 
Thus, the algorithm operates in at most $r$ rounds of adaptivity.
\end{theorem}

\subsection{Overview of Attack on Real-Valued Linear Sketches}
The attack is established by analyzing a communication protocol between two entities, referred to as Alice and Bob. 
Alice selects a matrix $\bA \in \mathbb{R}^{r \times n}$ drawn from a distribution $\pi$. 
Bob submits a sequence of queries $\bx^1, \ldots, \bx^s \in \mathbb{R}^n$, and for each query $\bx^i$, Alice observes only the sketch $\bA\bx^i$ and returns a value $f(\bA\bx^i)$, where $f$ is an arbitrary deterministic function (this assumption is removed in \secref{sec:net}). 
Bob's objective is to compute the row space $R(\bA)$ of the matrix $\bA$, which is an $r$-dimensional subspace of $\mathbb{R}^n$.

If $R(\bA)$ were known, Bob could alternate between submitting the zero vector and a vector from the kernel of $\bA$, thereby creating two indistinguishable cases for Alice that lead to significantly different outputs. 
This would force any approximation algorithm to fail under a relative error metric. 
The core result presents an efficient algorithm that enables Bob to recover $r - \O{1}$ orthonormal vectors that are nearly contained in $R(\bA)$, thereby constraining the effective dimension of Alice's sketch and increasing the likelihood of error on future inputs.

\paragraph{The Conditional Expectation Lemma.}
To begin identifying $R(\bA)$, Bob samples a query from the multivariate Gaussian distribution $N(0, \tau \mathbb{I}_n)$, where $\tau > 0$ is a scalar and $\mathbb{I}_n$ is the identity matrix. 
This ensures that Alice's observation, i.e., the projection $P_{\bA}\bx$ of $\bx$ onto $R(\bA)$, is spherically symmetric and determined solely by the norm $\|\bP_{\bA} \bx\|_2$. 
Alice must respond based on this projection, attempting to infer the norm of $\bx$ and outputting a binary decision.

The key insight is that when $\bx$ has a non-trivial component in $R(\bA)$, the observed norm $\|\bP_{\bA}\bx\|_2$ may be slightly larger than its expected value, potentially leading Alice to overestimate the norm of $\bx$. 
\cite{HardtW13} proves a conditional expectation lemma, which guarantees the existence of a choice of $\tau$ such that
\[\EEx{\bx \sim N(0, \tau \mathbb{I}_n)}{\|\bP_{\bA} \bx\|_2^2 \mid f(\bA\bx) = 1} - \EEx{\bx \sim N(0, \tau \mathbb{I}_n)}{\|\bP_{\bA} \bx\|_2^2}\]
is non-negligible. 
This result is derived by considering the aggregate contribution over a range $\tau \in [1, B]$, where $B$ denotes the desired approximation factor. 
For each value $v = \|\bP_{\bA}\bx\|_2^2$, the probability $q(v)$ of outputting $1$ affects the conditional expectation difference. 
By analyzing the weight of $v$ under the corresponding $\chi^2$ distribution, \cite{HardtU14} shows that the overall sum is significantly positive, indicating that some values of $\tau$ produce meaningful gaps in the expectations. 
Moreover, correctness assumptions imply that $q(v)$ is small for small $v$ and large for large $v$, further amplifying the net positive contribution.

\paragraph{Boosting correlation.}
After identifying a collection of queries $\bx^1, \ldots, \bx^m$ that exhibit slightly higher correlations with $R(\bA)$, these vectors are aggregated into a matrix $\bG \in \mathbb{R}^{m \times n}$. 
The top right singular vector $\bv^*$ of $\bG$ is then computed. 
This step, which is computationally efficient, yields a unit vector that is nearly contained in $R(\bA)$; formally, $\|\bP_{\bA} \bv^*\|_2 \ge 1 - \frac{1}{\poly(n)}$. 
This allows the dimension of Alice's effective sketch space to be reduced by one, laying the groundwork for iterative refinement.

\paragraph{Iterative refinement.}
Identifying a single direction in $R(\bA)$ is not sufficient, as Alice may dynamically adjust the component of the sketching matrix used in the post-processing function over the course of the data stream. 
Therefore, the procedure is repeated: Bob generates new queries drawn from a Gaussian distribution restricted to the orthogonal complement of the subspace spanned by the previously recovered vectors. 
Each iteration reduces the dimension of $R(\bA)$ by one.

This iterative attack introduces additional challenges. 
The recovered vectors are only approximately aligned with $R(\bA)$, and such approximation errors could, in theory, be exploited by Alice. 
To mitigate this, global Gaussian noise is added to each query to ensure that the query distribution remains statistically indistinguishable from one generated using vectors fully contained in $R(\bA)$. 
A generalized version of the conditional expectation lemma is then invoked to guarantee the robustness of this approach in the presence of noise.

\subsection{Certain Averages of \texorpdfstring{$\chi^2$}{Chi-Squared} Distributions}
This section introduces the foundational components required to establish the conditional expectation lemma. 
The analysis is conducted in $\mathbb{R}^d$ and involves examining weighted averages of the $\chi^2$-distribution over specific intervals.

Let $\nu(s)$ denote the density function of the squared Euclidean norm of a standard $d$-dimensional Gaussian random variable. This density is given by:
\[\nu(s) = \frac{s^{d/2 - 1} e^{-s/2}}{2^{d/2} \Gamma(d/2)}.\]
Define $\nu_{\tau,d}: [0, \infty) \to [0,1]$ as the density function of the $\chi^2$-distribution with $d$ degrees of freedom and expected value $\tau$. 
This coincides with the distribution of the squared norm of a $d$-dimensional Gaussian vector sampled from $N(0, \tau/d)^d$. 
The corresponding density function is:
\[\nu_{\tau,d}(s) = \frac{d \left(\frac{sd}{\tau}\right)^{d/2 - 1} e^{- \frac{sd}{2\tau}}}{\tau 2^{d/2} \Gamma(d/2)}.\]
This expression follows from the identity $\nu_{\tau,d}(s) = \frac{d}{\tau} \nu\left(\frac{sd}{\tau}\right)$. 
For notational simplicity, the subscript $d$ is omitted when unambiguous. 
The following function shows that if a weighting function $h(s)$ is ``small'' at ``small'' values and ``large'' and ``large'' values, then over all scales $\tau$, the average difference between the observed squared norm $s$ and its expected value $\tau$ is noticeably positive.
\begin{lemma}
\lemlab{lem:s-tau}
\cite{HardtW13}
Let $B>4$ and $d_0$ be a sufficiently large constant. 
Suppose $d\ge d_0$ and let $h:[0,\infty)\to[0,1]$ be any function satisfying the properties: 
\begin{enumerate}
\item
$\int_{Bd/2}^{2Bd} (1-h(s))\,ds\le\frac{1}{Bd}$,
\item
$\int_0^{2d}h(s)\,ds\le\frac{1}{d}$.
\end{enumerate}
Then, 
\[\int_{s=0}^\infty\int_{\tau=d}^{Bd} (s-\tau)\nu_\tau(s)h(s)\,d\tau\,ds \ge\frac {d}{4}.\]
\end{lemma}
This lemma is useful because it guarantees that some values of $s$ will consistently look bigger than expected, which is exactly the kind of statistical signal we need to detect structure, e.g., whether a query is correlated with a hidden subspace, such as $R(\bA)$ from earlier sections. 

\subsection{Conditional Expectation Lemma}
\seclab{sec:real:cond:exp:lemma}
The key analytic tool for the attack on real-valued linear sketches by \cite{HardtW13} is the so-called \emph{conditional expectation lemma}. 
Intuitively, it shows that the attack can find a distribution over inputs that have a non-trivially large correlation with the unknown subspace used by the linear sketch. 
Formally, let $\bU\subseteq\mathbb{R}^n$ be a fixed $d$-dimensional linear subspace, where $d$ is at least some sufficiently large constant. 
We suppose the linear sketch uses a post-processing function $f:\mathbb{R}^n\to\{0,1\}$ which satisfies the identity $f(\bx) = f(P_{\bU}\bx)$ for all $\bx\in\mathbb{R}^n$. 

\begin{definition}[Subspace Gaussian]
Let $\bU\subseteq\mathbb{R}^n$ be a linear subspace of $\mathbb{R}^n$. 
We call a family of distributions $\calG(\bU)=\{\bg_\tau\}_{\tau\in(0,\infty)}$  is a \emph{subspace Gaussian} family if:
\begin{enumerate}
\item 
$P_{\bU}\bg_\tau$ is distributed like a standard Gaussian variable inside $\bU$, with $\Ex{\|\bP_{\bU}\bg_\tau\|^2}=\tau$. 
\item 
$P_{\bU^\bot}\bg_\tau$ is a spherical Gaussian distribution that does not depend on $\tau$. 
\item 
$P_{\bU^\bot}\bg_\tau$ is statistically independent of $P_{\bU}\bg_\tau$.
\end{enumerate}
\end{definition}
The following claim shows that the norm is a sufficient statistic to characterize a subspace Gaussian family. 
\begin{lemma}
\lemlab{lem:suffstat}
\cite{HardtW13}
For a subspace Gaussian family $\calG(\bU)=\{\bg_\tau\}_\tau$, the norm $\|\bP_{\bU}\bg_\tau\|^2$ is a sufficient statistic. 
Specifically, for every $s>0$, the distribution of $\bg_\tau$ is independent of $\tau$, conditioned on $s=\|\bP_{\bU}\bg_\tau\|^2$. 
\end{lemma}
\begin{proof}
First, observe that we can decompose $\bg_\tau=\bg_1+\bg_2$, where $\bg_1$ is a Gaussian distribution supported on $U$ and $\bg_2$ is some Gaussian distribution independent of $\tau$ supported on $U^\bot$. 
By rotational invariance of both $\bg_1$ and $\bg_2$, we may assume without loss of generality that $\bU$ is a coordinate subspace. 
In particular, we may assume $\bU$ is the first $d=\dim(\bU)$ coordinates of the standard basis. 
Because $\bg_2$ is independent of $\bg_\tau$ and supported on a disjoint set of coordinates, it suffices to verify the claim for $\bg_1$. 
By the factorization theorem for sufficient statistics \cite{ChangP97}, a statistic $T(\bx)$ is sufficient for a parameter $\tau$ if and only if the density $p_\tau(\bx)$ can be written as 
\[p_\tau(\bx) = f(\bx)\, h_\tau(T(\bx)),\] 
where $f$ does not depend on $\tau$.  
In our case, the Gaussian density depends on $\bx$ only through $\|\bx\|^2$, so taking $T(\bx)=\|\bx\|^2$ gives the desired factorization.
Note that this follows directly from the fact that the Gaussian density at a point $\bx$ depends only on $\|\bx\|^2$, and so the desired claim holds. 
\end{proof}

The following soundness definition captures the notion that the output of the linear sketch should evaluate to $1$ on inputs that have large norm and $0$ on inputs that have small norm. 

\begin{restatable}[Soundness]{definition}{defsoundness}
\deflab{def:soundness}
We say that a function $f:\bU\to\{0,1\}$, where $\dim(\bU)=d$, is \emph{$B$-sound} for a subspace Gaussian family $\calG(\bU)$ if:
\begin{enumerate}
\item 
$\int_{Bd/2}^{2Bd} \Ex{f(\bg_\tau)\mid \|\bP_{\bU}\bg_\tau\|^2=s}\,ds\ge\frac{1}{Bd}$.
\item 
$\int_0^{2d}\Ex{f(\bg_\tau)\mid \|\bP_{\bU}\bg_\tau\|^2= s}\,ds\le\frac{1}{d}$
\end{enumerate}
\end{restatable}

We now state the first version of the Conditional Expectation Lemma.

\begin{restatable}[Conditional Expectation Lemma]{lemma}{lemconditional}
\lemlab{lem:conditional}
\cite{HardtW13}
Let $B\ge 4$ and $d_0$ be some sufficiently large constant. 
Let $\calG(\bU)$ be a subspace Gaussian family where $\bU$ has dimension $d\ge d_0$ and suppose $f:\bU\to\{0,1\}$ is $B$-sound for $\calG(\bU)$.  
Then there exists $\tau\in[d,Bd]$ such that 
\begin{enumerate}
\item 
$\Ex{\|\bP_{\bU}\bg_\tau\|^2 \, \Big|\, f(\bg_\tau)=1}\ge \Ex{\|\bP_{\bU}\bg_\tau\|^2} + \frac{1}{4B}$.
\item 
$\PPr{f(\bg_\tau)=1}\ge \frac{1}{40B^2d}$.
\end{enumerate}
\end{restatable}
\begin{proof}
We first define the function $h:(0,\infty)\to\mathbb{R}$ by
\[h(s) = \Ex{f(\bg_\tau)\mid \|\bP_{\bU}\bg_\tau\|^2=s}.\]
By \lemref{lem:suffstat}, this definition for the function $h$ is well-defined. 
Hence, we proceed by expanding the conditional expectation as follows.
\begin{align*}
\mathbb{E}\big[&\|\bP_{\bU}\bg_\tau\|^2 \, \Big|\, f(\bg_\tau)=1\big] = \int_0^\infty s\PPr{\|\bP_{\bU}\bg_\tau\|^2=s \mid f(\bg_\tau)=1}\,ds\\
& = \int_0^\infty s\PPr{f(\bg_\tau)=1 \mid \|\bP_{\bU}\bg_\tau\|^2=s}\cdot \frac{\nu_\tau(s)}{\PPr{ f(\bg_\tau)=1}}\,ds,
\end{align*}
by Bayes' rule, where $\nu_\tau=\nu_{\tau,d}$ denotes the $\chi^2$-distribution that has $d$ degrees of freedom, corresponding to the dimension of $U$. 
Thus, 
\begin{align*}
\Ex{\|\bP_{\bU}\bg_\tau\|^2 \, \Big|\, f(\bg_\tau)=1}
& = \int_0^\infty  \frac{sh(s)\nu_\tau(s)}{\PPr{ f(\bg_\tau)=1}}\,ds.
\end{align*}
We first claim that the proof follows from the inequality
\[\int_l^u\int_0^\infty(s-\tau)\nu_\tau(s)h(s)\,ds\,d\tau
\ge \frac{d}{4}.\]
Indeed, assuming the previous inequality, there must exist $\tau\in[d,Bd]$ such that 
\[\int_0^\infty s\nu_\tau(s)h(s)\,ds\ge 
\tau \int_0^\infty \nu_\tau(s)h(s)\,ds+ \frac{d}{4Bd}=\tau\PPr{f(\bg_\tau)=1} + \frac {1}{4B}.\]
Therefore,
\begin{align*}
\Ex{\|\bP_{\bU}\bg_\tau\|^2 \, \Big|\, f(\bg_\tau)=1}&=\int_0^\infty  \frac{sh(s)\nu_\tau(s)}{\PPr{ f(\bg_\tau)=1}}\,ds \\
&\ge \frac{\tau\PPr{f(\bg_\tau)=1}+1/4B}{\PPr{ f(\bg_\tau)=1}}\\
&\ge \tau + \frac{1/4B}{\PPr{f(\bg_\tau)=1}}.
\end{align*}
Since $\PPr{f(\bg_\tau)=1}\le1$, then the first claim of the lemma holds. 
Thus, it remains to lower bound $\PPr{f(\bg_\tau)=1}$. 
To that end, recall that $\int_{5Bd}^\infty
s\nu_\tau(s)h(s)\,ds\le \frac{1}{2}\int_0^\infty s\nu_\tau(s)h(s)\,ds,$ by standard concentration properties of $\nu_\tau$. 
Therefore,
\begin{align*}
\PPr{f(\bg_\tau)=1} &= \int_0^\infty \nu_\tau(s)h(s)\,ds\\
&\ge \frac{1}{10Bd}\int_0^\infty s\nu_\tau(s)h(s)\,ds\\
&\ge \frac{1/4B}{10Bd} = \frac{1}{40B^2d},
\end{align*}
which gives the second part of the claim. 

It remains to prove the inequality 
\[\int_l^u\int_0^\infty(s-\tau)\nu_\tau(s)h(s)\,ds\,d\tau
\ge \frac{d}{4}.\]
To that end, we will apply \lemref{lem:s-tau}, which in fact directly implies the claim, provided that the function $h$ satisfies the necessary conditions required in \lemref{lem:s-tau}. 
It is straightforward to verify that these properties coincide with the soundness assumption on $f$. 
Hence, the desired inequality holds, which implies the two claimed conditions by the above argument. 
\end{proof}

We have the following corollary of the Conditional Expectation Lemma in \lemref{lem:conditional}, which states that there exists some direction in the subspace that has increased variance.

\begin{corollary}
\corlab{cor:conditional}
Suppose $\calG(\bU)$ satisfies the assumptions of \lemref{lem:conditional}. 
Then there exists $\tau\in[d,Bd]$ and a vector $\bu\in \bU$ such that
\begin{enumerate}
\item 
$\Ex{\langle \bu,\bg_\tau\rangle^2 \, \Big|\, f(\bg_\tau)=1}\ge \Ex{\langle \bu,\bg_\tau\rangle^2} + \frac{1}{4Bd}$.
\item 
$\PPr{f(\bg_\tau)=1}\ge \frac{1}{40B^2d}$.
\end{enumerate}
\end{corollary}
\begin{proof}
Let $\bu_1,\ldots,\bu_d$ be an arbitrary orthonormal basis of $\bU$. 
Since $\|\bP_{\bU}\bg_\tau\|^2=\sum_{i=1}^d\langle \bu_i,\bg_\tau\rangle^2$, then by an averaging argument, at least one of the basis vectors must satisfy the desired claim. 
\end{proof}

\subsubsection{Noisy Orthogonal Complements}
In this section, we extend the Conditional Expectation Lemma to a family of distributions that is subspace Gaussian on a subspace of $\bA$, rather than subspace Gaussian on the entirety of $\bA$. 
Specifically, we consider the family of
distributions that is subspace Gaussian on $\bA\cap \bV^\bot,$ where $\bV\subseteq \bA$ is a linear subspace of $A$. 

Each distribution in this family is parameterized by a subspace $\bV$ and a variance $\sigma^2$, intuitively corresponding to a Gaussian distribution on the subspace $\bV^\bot$ with variance $\sigma^2$, plus a small Gaussian supported on all of $\mathbb{R}^n$ of constant variance, independent of $\sigma^2$. 
Formally, we define:
\begin{restatable}{definition}{defcomplementgaussian}
\deflab{def:complement-gaussian}
Let $\sigma>0$ be a parameter, $\bV\subseteq \bA$ be a subspace of $\bA$ of dimension $t\le r-1$. 
Let $G(\bV^\bot,\sigma^2)$ be the distribution obtained from sampling $g_1\sim N(0,\sigma^2)^n, g_2\sim N(0,1/4)^n$ independently and outputting $g=P_{\bV^\bot}g_1 + g_2$.  

Furthermore, for $d=r-t$, we define the family of distributions $\calG(\bA\cap \bV^\bot)=\{\bg_\tau\}$ by setting $\bg_\tau=P_{\bA} \bg$, for $\bg\sim G(\bV^\bot,\tau/d - 1/4)$ if $\frac{\tau}{d}>\frac{1}{4}$ and otherwise setting $\bg_\tau = P_{\bA} \bg$ where $\bg\sim G(\bV^\bot,\tau/d)$.
\end{restatable}

We first show that $\calG(\bA\cap \bV^\bot)$ is subspace Gaussian.
\begin{lemma}
\lemlab{subspace-gaussian}
\cite{HardtW13}
$\calG(\bA\cap \bV^\bot)$ is a subspace Gaussian family. 
\end{lemma}
\begin{proof}
Let $\bU=\bA\cap \bV^\bot$. 
For $\frac{\tau}{d}\le\frac{1}{4}$, we have that $\Ex{\|\bP_{\bU}g_\tau\|^2}=d\cdot\frac{\tau}{d}=\tau$. 
Moreover, since $g_\tau\sim G(\bV^\bot,\tau/d)$ and $\bU\subseteq\bV^\bot$, then $g_\tau$ is orthogonal to $\bU^\bot$, so that $\|\bP_{\bU^\bot}g_\tau\|=0$, as required. 
For $\frac{\tau}{d}>\frac{1}{4}$, we have $g=P_{\bV^\bot}g_1+g_2$, so that $P_{\bU}g$ is distributed like a spherical Gaussian with variance $\frac{\tau}{d}$ in each direction. 
Hence,
\[\Ex{\|\bP_{\bU} g_\tau\|^2} = \Ex{\|\bP_{\bU} (g_1 + g_2)\|^2} = d\cdot\frac{\tau}{d}=\tau.\] 
On the other hand, $P_{\bU^\bot}g$ only depends on $g_2$ and is thus independent of $\tau$. 
Therefore, $\calG(\bA\cap \bV^\bot)$ is a subspace Gaussian family. 
\end{proof}

We now formalize the notion of correctness for our linear sketching algorithm $f$ for inputs drawn from the distribution $G(\bV^\bot,\sigma^2)$.

\begin{restatable}[Correctness]{definition}{defcorrectness}
\deflab{def:correctness}
Let $\bA$ be a subspace and $\bV\subseteq \bA$ be a subspace with $d=\dim(\bV^\bot\cap \bA)$. 
We call a function $f:\bA\to\{0,1\}$ to be \emph{$(\eps,B)$-correct} on $\bV^\bot$:
\begin{enumerate}
\item 
For all $\sigma^2\in[B/2,2B]$ and $g\sim G(\bV^\bot,\sigma^2)$, we have 
\[\PPr{f(g)=1}\ge 1-\eps.\]
\item 
For all $\sigma^2\in[0,2]$ and $g\sim G(\bV^\bot,\sigma^2)$, we have 
\[\PPr{f(g)=1}\le\eps.\] 
\end{enumerate}
We simply call $f$ to be \emph{$B$-correct} on $\bV^\bot$ if it is $(\eps,B)$-correct for some $\eps\le\frac{1}{10(Bd)^2}$. 
\end{restatable}

We now relate the correctness and soundness definitions.

\begin{lemma}
\lemlab{lem:correct2sound}
\cite{HardtW13}
Suppose $f$ is $B$-correct on $\bV^\bot$. 
Then $f$ is $B$-sound for $\calG(\bA\cap \bV^\bot)$. 
\end{lemma}
\begin{proof}
We prove the contrapositive of the claim. 
To that end, suppose that $f$ is not $B$-sound for $\calG(\bA\cap \bV^\bot)$. 
In other words, at least one of the two
requirements in \defref{def:soundness} is not satisfied. 
Suppose the first requirement is not satisfied, so that for $I=[Bd/2,2Bd]$ and $h(s)=\PPr{f(\bg_\tau)=1\mid \|\bg_\tau\|^2=s}$, we have $\EEx{s\in I}{(1-h(s))}<\frac{1}{2(Bd)^2}$. 
Suppose we choose $\sigma^2$ uniformly at random from $\left\{\frac{B}{2},2B\right\}$ and then sample $\bg\sim G(\bV^\bot,\sigma^2)$. 
It can be shown that the resulting distribution of $\|\bg\|^2$ is point-wise within a factor $5$ of the uniform distribution inside the interval $\left[\frac{B}{2},2B\right]$. 
Thus, $\Ex{h(\|\bg\|^2)}<\frac{1}{10(Bd)^2}$, which violates the first condition of correctness. 
The case where the second requirement is not satisfied follows from an analogous argument.
\end{proof}

We now prove a variant of the Conditional Expectation Lemma for inputs drawn from distributions of the form $G(\bV^\bot,\sigma^2)$. 
Moreover, the result applies to any $t\le \dim(\bA)$, rather than $t\le r-d_0$. 

\begin{lemma}
\lemlab{lem:conditional-main}
\cite{HardtW13}
Let $d_0$ be a sufficiently large constant. 
Let $\bA\subseteq\mathbb{R}^n$ be a subspace with dimension $\dim(\bA)=r\le n-d_0$ and let $\bV\subseteq \bA$ be a subspace of $A$ of dimension $t\le r$.  
Suppose $f:\bA\to\{0,1\}$ is $\left(\frac{1}{10(d_0B)^2},B\right)$-correct on $\bV^\bot$ and let $d=\max\{r-t,d_0\}$. 
Then there exists a scalar $\sigma^2\in\left[\frac{3}{4},B\right]$ and a vector $\bu\in \bA\cap \bV^\bot$, so that for $\bg\sim G(\bV^\bot,\sigma^2)$:
\begin{enumerate}
\item 
$\Ex{\langle \bu,\bg\rangle^2 \, \Big|\, f(\bg)=1} \ge \Ex{\langle \bu,\bg\rangle^2} + \frac{1}{4Bd}$
\item 
$\PPr{f(\bg)=1}\ge \frac{1}{40B^2d}$
\end{enumerate}
\end{lemma}
\begin{proof}
Without loss of generality, we can consider a subspace $\bA'\supseteq \bA$ of dimension $r+d_0$ by extending $\bA$ arbitrarily to $r+d_0$ dimensions. 
In particular, this is possible because $n\ge r+d_0$.
Hence, we can assume $\dim(\bA\cap \bV^\bot)\ge d_0$.  

Consider the function $f'(\bx)=f(P_{\bA}\bx)$ and observe that $f'(\bx)=f(\bx)$ for all $\bx\in\mathbb{R}^n$. 
Therefore, $f'$ is still $\left(\frac{1}{10(d_0B)^2},B\right)$-correct on $\bV^\bot$. 
Moreover, we have $\dim(\bV^\bot\cap \bA)=d_0$. 
Thus by \lemref{lem:correct2sound}, $f$ is sound for the subspace Gaussian family $\calG(\bA'\cap \bV^\bot)$. 
Let $\bU=\bA'\cap \bV^\bot$. 
By \corref{cor:conditional} on $\calG(\bU)$, there exists $\tau\in[d,Bd]$ and $\bu\in \bU$ such that
\[\Ex{\langle \bu,\bg_\tau\rangle^2 \, \Big|\, f'(\bg_\tau)=1}\ge\Ex{\langle \bu,\bg_\tau\rangle^2} + \frac1{4Bd}\]
and $\PPr{f'(\bg_\tau)=1}\ge \frac{1}{40B^2d}$. 
By the construction of $f'$, the condition $f'(\bg_\tau)=1$ is equivalent to $f(\bg_\tau)=1$. 
Moreover, the condition $f(\bg_\tau)=1$ does not affect any vector that is orthogonal to $\bA$. 
Hence, we can assume $\bu\in \bA\cap \bV^\bot$ without loss of generality. 
Furthermore, observe that $\bg_\tau=P_{\bA'}\bg$ for some $\bg\sim G(\bV^\bot,\sigma^2)$ with $\sigma^2\in\left[\frac{3}{4},B\right]$. 
Since $f(\bg)=f'(\bg_\tau)$ and $\langle \bu,\bg_\tau\rangle = \langle \bu,\bg\rangle$, due to  $\bu\in \bU$, then we have
\[\Ex{\langle \bu,\bg\rangle^2 \, \Big|\, f(\bg)=1}\ge \Ex{\langle \bu,\bg\rangle^2} + \frac{1}{4Bd},\]
with $\PPr{f(\bg)=1}\ge \frac{1}{40B^2d}$, as desired.
\end{proof}

\subsubsection{Distance Between Subspaces}
In this section, we relate distributions of the form $G(\bV^\bot,\sigma^2)$ to distributions of the form $G(\bW^\bot,\sigma^2)$, for subspaces $\bV$ and $\bW$.  
To this end, for subspaces $\bV,\bW\subseteq\mathbb{R}^n$, we consider the distance function as follows.

\begin{restatable}[Distance between subspaces]{definition}{defsubspacedist}
\deflab{def:subspace:dist}
We define the distance between subspaces $\bV,\bW\subseteq\mathbb{R}^n$ by
\[d(\bV,\bW)=\|\bP_{\bV}-P_{\bW}\|_2:=\sup_{\bv\in\mathbb{R}^n}\frac{\|\bP_{\bV}\bv-P_{\bW}\bv\|_2}{\|\bv\|_2}.\]
\end{restatable}

Intuitively, if $\bV$ and $\bW$ are close in this distance function, then the two distributions $G(\bV^\bot,\sigma^2)$ and $G(\bW^\bot,\sigma^2)$ are also statistically close. 
We require the following relationship upper bounding the total variation distance between two normal distributions.
\begin{fact}
\factlab{fact:shifted-gaussians}
Let $\bv\in\mathbb{R}^n$. Then
\[\TVD(N(0,\sigma^2)^n,N(\bv,\sigma^2)^n)\le \frac{\|\bv\|_2}{\sigma}.\]
\end{fact}
Using this fact, we can bound the total variation distance between $G(\bV^\bot,\sigma^2)$ and $G(\bW^\bot,\sigma^2)$ in terms of the distance $d(\bV,\bW)$, as follows:
\begin{lemma}
\lemlab{lem:stat-dist}
\cite{HardtW13}
For every $\sigma^2\in(0,B]$, 
\[\TVD(G(\bV^\bot,\sigma^2),G(\bW^\bot,\sigma^2))\le 20\sqrt{Bn\log(Bn)}\cdot d(\bV,\bW)+\frac{1}{(Bn)^5}.\]
\end{lemma}
\begin{proof}
Let $\bg_1\sim N(0,\sigma^2)^n$ and $\bg_2,\bg_2'\sim N(0,1/4)^n$. 
Let $\bx = P_{\bV^\bot}\bg_1 + \bg_2$ and $\by=P_{\bW^\bot}\bg_1 + \bg_2'$. 
Observe that $\bx$ is distributed like a random draw from $G(\bV^\bot,\sigma^2)$, while $\by$ is distributed like a random draw from $G(\bW^\bot,\sigma^2)$. 
Hence, it suffices to bound the total variation distance of these distributions, analyzing the dependency due to the coupled variable $\bg_1$. 
We have
\[\| P_{\bV^\bot}\bg_1 - P_{\bW^\bot} \bg_1 \|_2= \| P_{\bV}\bg_1 - P_{\bW} \bg_1 \|_2\le \|\bg_1\|\cdot d(\bV,\bW).\]
By standard Gaussian concentration bounds,
\[\PPr{\|\bg_1\|_2>10\sqrt{Bn\log(Bn)}} \le \frac {1}{(Bn)^5}.\]
Thus if $\calE$ is the event that $\|\bg_1\|\le 10\sqrt{Bn\log(Bn)}$, then we have $\PPr{\calE}\ge1-\frac{1}{(Bn)^5}$. 
Conditioned on $\calE$, then for every possible value $\bu=P_{\bV^\bot}\bg_1-P_{\bW^\bot}\bg_1,$, we have by \factref{fact:shifted-gaussians}
\begin{align*}
\TVD\left(\calN\left(\bu,\frac{1}{4}\cdot\mathbb{I}\right), \calN\left(0,\frac{1}{4}\right)^n\right) & \le 2\|\bu\|_2 \\
& \le 2\|\bg_1\|_2\cdot d(\bV,\bW) \\
& \le 20\sqrt{Bn\log(Bn)}\cdot d(\bV,\bW).
\end{align*}
Since $\bu + \calN\left(0,\frac{1}{4}\right)^n = \calN\left(\bu,\frac{1}{4}\right)^n$ and $\TVD(p,q)=\frac{1}{2}\cdot\|p-q\|_1$ for probability distributions $p$ and $q$, then
\begin{align*}
\TVD\Bigg(P_{\bV^{\bot}}\bg_1 + \calN\left(0,\frac{1}{4}\right)^n, P_{\bW^{\bot}}\bg_1 &+  \calN\left(0,\frac{1}{4}\right)^n\Bigg) \\
&= \TVD\left(\calN\left(\bu,\frac{1}{4}\cdot\mathbb{I}\right), \calN\left(0,\frac{1}{4}\right)^n\right) \\
&\le 20\sqrt{Bn\log(Bn)}\cdot d(\bV,\bW).
\end{align*}
Finally, since the event $\calE$ has probability at least $1-\frac{1}{(Bn)^5}$, then failure of the event can only increase the statistical distance of the two variables by an additive $\frac{1}{(Bn)^5}$.  
\end{proof}

\subsection{An Adaptive Reconstruction Attack}
In this section, we show that no linear sketching post-processing function $f:\mathbb{R}^n\to\{0,1\}$ that depends only on a lower dimensional subspace can correctly predict the $F_2$-moment up to a multiplicative factor $B$ on a polynomial number of adaptive queries. 
We emphasize that $B$ can be any parameter, but the complexity of the attack depends on $B$, as well as the dimension of the subspace. 
Formally, the following definition characterizes the ability of the attack to induce failure from the linear sketch.

\begin{definition}[Failure certificate]
\deflab{def:failure:real}
Let $B\ge 8$ and let $f:\mathbb{R}^n\to\{0,1\}$. 
We call a pair $(\bV,\sigma^2)$ a \emph{$d$-dimensional failure certificate for $f$} if $\bV\subseteq\mathbb{R}^n$ is $d$-dimensional subspace and $\sigma^2\in[0,2B]$ such that for some constant $C>0$, we have: 
\begin{itemize}
\item 
$n\ge d+10C\log(Bn)$
\item 
Either $\sigma^2\in\left[\frac{B}{2},50B\right]$ and 
$\PPPr{\bg\sim G(\bV^\bot,\sigma^2)}{f(\bg)=1}\le 1- \frac{1}{(Bn)^C}$
\item or $\sigma^2\le 2$ and
$\PPPr{\bg\sim G(\bV^\bot,\sigma^2)}{f(\bg)=1}\ge \frac{1}{n^C}$.
\end{itemize}
\end{definition}
We now formally define the following formulation of the $\GapNorm$ problem.
\begin{restatable}[$\GapNorm(B, \alpha)$ promise problem]{definition}{defgapnorm}
Given any fixed $B \ge 8$ and $\alpha = \poly(n)$, we say that an algorithm $\mathcal{A}$ solves the problem $\GapNorm(B, \alpha)$ if it satisfies the following condition: for every input vector $\bx \in \mathbb{Z}^n$, the algorithm outputs $1$ whenever $\|\bx\|_2^2 \ge \alpha \cdot B$, and outputs $0$ whenever $\|\bx\|_2^2 \le \alpha$. 
If $\|\bx\|_2^2$ falls within the interval $(\alpha, \alpha \cdot B)$, then $\mathcal{A}$ is permitted to return either $0$ or $1$.
\end{restatable}

The intuition for the definition of a failure certificate is given by the next simple fact, which shows that a failure certificate corresponds to a distribution on which the linear sketching algorithm $f$ does not decide the following formulation of the $\GapNorm$ problem up to a multiplicative factor $\Omega(B)$ on a polynomial number of queries. 
\begin{fact}
\cite{HardtW13}
\factlab{fact:real:failure:cert:attack}
Given a $d$-dimensional failure certificate for a linear sketching algorithm $f$, we can use $\poly(Bn)$ non-adaptive queries and with probability at least $2/3$, find an input $\bx$ such that either (1) $\|\bx\|_2^2\ge B(n-d)/3$ and $f(\bx)=0$ or (2) $\|\bx\|_2^2\le 3(n-d)$ and $f(\bx)=1$. 
\end{fact}
\begin{proof}
Consider a set of $\O{(Bn)^C}$ samples from a $d$-dimensional failure certificate $G(\bV^\bot,\sigma^2)$. 
We first consider the case $\sigma^2\le 2$. 
Since $n-d$ is sufficiently large compared to $d$, then by a union bound and Gaussian concentration, we have that $\|\bx\|^2\le 3(n-d)$ simultaneously for all queries $\bx$, with high probability. 
On the other hand, $f$ outputs $1$ on at least one of the queries, with high probability, which produces the desired claim. 
The case where $\sigma^2\ge B/2$ follows using a similar argument.
\end{proof}

\begin{figure*}[!htb]
\begin{mdframed}
\textbf{Input}: Oracle $\calA$ providing access to a function $f:\mathbb{R}^n\to\{0,1\}$, parameter $B\ge 4$.
\newline\noindent
\textbf{Attack}: Let $\bV_1=\emptyset$, $m=\O{B^{13}n^{11}\log^{15}(n)}$, $S=\left[\frac{3}{4},\alpha \cdot B\right]\cap\eps\mathbb{Z}$ where $\eps=\frac{1}{20(Bn)^2\log(Bn)}$. 
\newline\noindent
\textbf{For $t\in[r+1]$:}
\begin{enumerate}
\item
For each $\sigma^2\in S$:
\begin{enumerate}
\item
Sample $\bg_1,\ldots,\bg_m\sim G(\bV^\perp,\sigma^2)$. 
Query $\calA$ on each $\bg_i$ and let $a_i=\calA(\bg_i)$.
\item
Let $s(t,\sigma^2)=\frac{1}{m}\sum_{i=1}^m a_i$ denote the fraction of samples that are positively labeled.
\begin{enumerate}
\item
If either (1) $\sigma^2\ge B/2$ and $s(t,\sigma^2)\le 1-\eps$ or (2) $\sigma^2\le 2$ and $s(t,\sigma^2)\ge\eps$, then terminate and \textbf{return} $(\bV_t^\perp,\sigma^2)$.
\item
Else let $\bg'_1,\ldots,\bg'_{m'}$ be the vectors such that $\calA(\bg'_i)=1$ for all $i\in[m']$. 
\end{enumerate}
\item
If $m'<\frac{m}{100B^2n}$, increment $\sigma^2$. 
Else, compute $\bv_{\sigma}=\argmax_{\bv\in\mathbb{R}^n}z(\bv)$ for $z(\bv):=\frac{1}{m'}\sum_{i=1}^{m'}\langle \bv,\bg'_i\rangle^2$
\end{enumerate}
\item
Let $\bv'$ represent the first vector $\bv_\sigma$ with $z(\bv_\sigma)\ge\sigma^2+\frac{\sigma^2}{4}+\frac{1}{14Br}$.
\begin{enumerate}
\item
If no such $\bv_\sigma$ was found, set $\bV_{t+1}=\bV_t$ and proceed to the next round.
\item 
Otherwise, let $\bv^* = \bv'$. 
Compute $\bv_t=\bv^*-\frac{\sum_{\bv\in \bV_t}\bv\langle \bv,\bv^*\rangle}{\left\|\sum_{\bv\in \bV_t}\bv\langle \bv,\bv^*\rangle\right\|_2}$ and set $\bV_{t+1}=\bV_t\cup\{\bv_t\}$. 
\end{enumerate}
\end{enumerate}
\end{mdframed}
\caption{Reconstruction attack on linear sketches by \cite{HardtW13}. 
The algorithm iteratively builds a subspace $\bV_t$ that is approximately contained in the unknown subspace $\bA$. 
In each round, the algorithm queries $\calA$ on a sequence of queries that are mostly contained within the orthogonal complement of $\bV_t$. 
As the dimension of $\bV_t$ increases, the oracle must eventually make a mistake.}
\figlab{fig:hw13:attack}
\end{figure*}

\subsubsection{Proof of Main Theorem}
The main purpose of the next few sections is to establish that the attack can always find a failure certificate, using a polynomial number of queries:
\begin{restatable}{theorem}{thmattack}
\thmlab{thm:attack}
\cite{HardtW13}
Let $B\ge 8$ and suppose that $B\le\poly(n)$. 
Let $\bA\subseteq\mathbb{R}^n$ be a $r$-dimensional subspace of $\mathbb{R}^n$, such that $n\ge r+90\log(Br)$.
Let $f:\mathbb{R}^n\to\{0,1\}$ such that $f(\bx)=f(P_{\bA}\bx)$ for all $\bx\in\mathbb{R}^n$. 
Then there exists an algorithm that uses only  oracle access to $f$ and finds a failure certificate for $f$, with probability $\frac{9}{10}$. 
Moreover, the time and query complexity of the algorithm is bounded by $\poly(B,r)$ and all queries of the algorithm are sampled from
$G(\bV^\bot,\sigma^2)$, for some $\bV\subseteq\mathbb{R}^n$ and $\sigma^2\in(0,B]$. 
\end{restatable}
\begin{proof}
Note that we can assume $n=r+90\log(Br)$ without loss of generality, by only working with the first $r+90\log(Br)$ coordinates of $\mathbb{R}^n$. 
Consequently, a polynomial dependence on $n$ in the attack translates to a polynomial dependence on $r$. 

For each $1\le t\le r+1$, let $\bW_t\subseteq\bA$ be the closest $(t-1)$-dimensional subspace to $\bV_t$ that is contained in $\bA$. 
Formally, we define $\bW_t$ by: 
\[d(\bV_t,\bW_t) = \min\{d(\bV_t,\bW)\,\mid\,\dim(\bW)=t-1, \bW\subseteq \bA\},\]
where we use $\bV_t$ to also denote the subspace that is spanned by the vectors contained in $\bV_t$. 
Our goal is to maintain the following invariant throughout the course of the attack, with high probability:
\begin{invariant}[Invariant at step $t$:] 
\invlab{inv:invariant}
\[\dim(\bV_t) = t-1\qquad\text{and}\qquad
d(\bV_t,\bW_t)\le 
\frac{t}{20(Bn)^{3.5}\log(Bn)^{2.5}}.\]
\end{invariant}
Observe that the invariant holds vacuously at the first step, since $\bV_1=\{0\}\subseteq \bA$. 
Intuitively, our goal is to show that either the algorithm terminates with a failure certificate or \invref{inv:invariant} continues to hold.
Specifically, if \invref{inv:invariant} holds in a step $t$, then 
\[d(\bV_t,\bW_t)\le \frac{1}{20B^{3.5}n^{2.5}\log(Bn)^{2.5}}.\]
Thus, \lemref{lem:stat-dist} implies that for every $\sigma^2\in(0,B]$,
\begin{align}
\eqnlab{eqn:VtWt}
\TVD(G(\bV_t^\bot,\sigma^2),G(\bW_t^\bot,\sigma^2)) &\le 20\sqrt{B n\log(Bn)}\cdot d(\bV_t,\bW_t)+\frac{1}{(Bn)^5}\notag\\
&\le \frac{1}{B^3n^2\log(Bn)^2}.
\end{align}
We can then show the following useful lemma.
\begin{lemma}
\lemlab{lem:correct}
\cite{HardtW13}
Suppose the invariant holds at step $t\in[r+1]$. 
If $f$ is $(\alpha,B)$-correct on $\bV_t^\bot$, then $f$ is $(\alpha+\eps,B)$-correct on $\bW_t^\bot$. 
\end{lemma}
\begin{proof}
Equation \eqnref{eqn:VtWt} implies that for every $\sigma^2\in(0,B]$, the total variation distance between $G(\bV_t^\bot,\sigma^2)$ and $G(\bW_t^\bot,\sigma^2)$ is at most $\eps$. 
Therefore, the correctness conditions from \defref{def:correctness} hold up to an additive $\eps$-loss in the probabilities.
\end{proof}
Let $\calE$ be the event that the empirical estimate $s(t,\sigma^2)$ is accurate at all steps of the algorithm, i.e., for all $t\in[r+1]$ and all $\sigma^2\in S$:
\[\left\lvert s(t,\sigma^2) - \PPPr{G(\bV_t^\bot,\sigma^2)}{f(\bg)=1}\right\rvert\le\eps.\]
\begin{lemma}
\lemlab{lem:real:good:estimate}
\cite{HardtW13}
$\PPr{\calE}\ge 1-\exp(-n)$.
\end{lemma}
\begin{proof}
Note that the claim follows from standard Chernoff bounds, since the number of samples is $m\gg\left(\frac{Bn}{\eps}\right)^2$.
\end{proof}
\begin{lemma}
\lemlab{lem:empirical}
\cite{HardtW13}
Conditioned on $\calE$, if the algorithm terminates in round $t$ and outputs $G(\bV_t^\bot,\sigma^2)$, then $G(\bV_t^\bot,\sigma^2)$ is a failure certificate for $f$. 
Furthermore, if \invref{inv:invariant} holds in round $t$ but the algorithm does not terminate in round $t$, then $f$ is $B$-correct on $\bW^\bot$.
\end{lemma}
\begin{proof}
Conditioned on $\calE$, the empirical error given by $s(t,\sigma^2)$ is $\eps$-close to the actual error. 
Thus, the first claim follows directly from the definition of a failure certificate, given the setting of $\eps=\frac{1}{20(Bn)^2\log(Bn)}$.

We now prove the second claim. 
Conditioned on $\calE$, we must have that $f$ is $(2\eps,B)$-correct on $\bV_t^\bot$, since the algorithm did not terminate. 
By \lemref{lem:correct}, it follows that $f$ is $(3\eps,B)$-correct on $\bW_t^\bot$. 
Since $3\eps\le\frac{1}{10(Bn)^2}$, it follows that $f$ is $B$-correct on $\bW_t^\bot$, which implies the second claim. 
\end{proof}
We next prove the Progress Lemma, which shows that the invariant continues to hold with high probability, provided that $f$ continues to be correct. 
\begin{restatable}[Progress]{lemma}{lemprogress}
\lemlab{lem:progress}
\cite{HardtW13}
Let $t\in[r]$ and suppose that \invref{inv:invariant} holds in round $t$ and that $f$ is $B$-correct on $\bW_t^\bot$. 
Then \invref{inv:invariant} holds in round $t+1$, with probability at least $1-\frac{1}{n^2}$.  
\end{restatable}
We shall prove \lemref{lem:progress} in \secref{sec:real:progress}. 
Here, we continue the proof of \thmref{thm:attack}, assuming the Progress Lemma, i.e., \lemref{lem:progress}. 
We claim that if the attack progresses to the final round and \invref{inv:invariant} still holds, then we have effectively reconstructed the entirety of $\bA$. 
Thus, $f$ cannot be correct on $\bW_t^\bot$, as follows:
\begin{lemma}
\lemlab{lem:final}
\cite{HardtW13}
If \invref{inv:invariant} holds for $t=r+1$, then $f$ is not $B$-correct on $\bW_t$. 
\end{lemma}
\begin{proof}
Since $t=r+1$ and \invref{inv:invariant} holds, then $\dim(\bV_t)=\dim(\bW_t)=r$. 
Moreover, $\bW_t\subseteq \bA$ and $\dim(\bA)=r$, which implies $\bW_t=\bA$. 
Thus, the function $f$ cannot distinguish between samples from $G(\bW_t^\bot,2)$ and samples from $G(\bW_t^\bot,B)$, and so, $f$ must make a mistake with constant probability on one of the distributions. 
Therefore, $f$ is not $B$-correct on $\bW_t$. 
\end{proof}
Note that since $\calE$ fails with probability $1-\exp(-n)$, conditioning on $\calE$ only affects the success probability of our algorithm by a negligible amount. 
Now, conditioning on $\calE$, if the algorithm terminates in a round $t$ with $t\le r$, then the algorithm outputs a failure certificate for $f$, by \lemref{lem:empirical}. 
On the other hand, if the attack does not terminate in any of the rounds $t\le r$ and assuming \invref{inv:invariant} holds, then by the second part of \lemref{lem:empirical}, $f$ must be $B$-correct on $\bW_t^\bot$ for all rounds $t\in[r]$. 
In this case, \lemref{lem:progress} implies that \invref{inv:invariant} continues to hold in round $t+1$ with high probability. 
Specifically, if the algorithm does not terminate before round $r$, then with probability $\left(1-\frac{1}{n^2}\right)^r\ge 1-\frac{1}{n}$, \invref{inv:invariant} continues to hold at step $r+1$. 
However, by \lemref{lem:final}, $\bW_{r+1}$ is not correct for $f$. 
Thus by \lemref{lem:empirical}, the attack outputs a failure certificate with probability $1-\exp(-n)$. 
Putting the two cases together, it follows that the attack successfully finds a failure certificate for $f$ with probability at least $1-\frac{2}{n}$, as claimed by the statement of \thmref{thm:attack}.

It remains to analyze the query complexity and runtime of the attack. 
The query complexity is polynomially bounded in $n$. 
However, since we may assume without loss of generality that $n\le\O{r}$ by only working with the first $r+90\log(Br)$ coordinates of $\mathbb{R}^n$, then the query complexity is also polynomially bounded in $r$. 
Computationally, the only non-trivial step is finding the vector $\bv_\sigma$, which maximizes $z(\bv)=\frac{1}{m'}\sum_{i=1}^{m'}\langle \bv_\sigma,\bg_i\rangle^2$. 
We note that this vector can be efficiently computed using singular vector computation. 
In particular, let $\bG$ be the $m'\times n$ matrix with the rows $\bg_1,\ldots,\bg'_m$. 
By definition, the top singular vector $\bv$ of $\bG$ maximizes $\|\bG\bv\|_2^2= \sum_{i=1}^{m'}\langle \bg_i,\bv\rangle^2$. 
Thus, the top singular vector of $\bG$ must also maximize the quantity $z(\bv)$. 
Since the top singular vector can be computed in polynomial time, it follows that the overall attack can be implemented in time polynomial in $r$, concluding the proof of \thmref{thm:attack}.
\end{proof}

\subsubsection{Proof of the Progress Lemma \texorpdfstring{(\lemref{lem:progress})}{}}
\seclab{sec:real:progress}
Let $t\le r$ and suppose \invref{inv:invariant} holds up to round $t$. 
Moreover, suppose $f$ is $B$-correct on $\bW_t^\bot$.  
Recall that under these assumptions, by \eqnref{eqn:VtWt}, we have that for every $\sigma^2\in(0,B]$,
\begin{equation}
\eqnlab{eqn:VtWt:two}
\delta:=\TVD(G(\bV_t^\bot,\sigma^2),G(\bW_t^\bot,\sigma^2))\le \frac{1}{B^3n^2\log(Bn)^2}.    
\end{equation}
We shall show that with probability at least $1-\frac{1}{n^2}$, \invref{inv:invariant} also holds in round $t+1$. 
To this end, we require the following formulation of the Conditional Expectation Lemma, i.e., \lemref{lem:conditional-main}. 

\begin{lemma}
\lemlab{lem:sigmatilde}
\cite{HardtW13}
Suppose $f$ is correct on $\bW_t^\bot$. 
Then there exist $\tilde\sigma^2\in S$, $\Delta\ge\frac{1}{7Br}$, and $\bu\in \bV_t^\bot\cap \bA$, such that for $\bg\sim G(\bV_t^\bot,\tilde\sigma^2)$, we have $\PPr{f(\bg)=1}\ge\frac{1}{60B^2r}$ and 
\[\Ex{\langle \bu,\bg\rangle^2 \, \Big|\, f(\bg)=1}\ge \Ex{\langle \bu,\bg\rangle^2} + \Delta.\]
\end{lemma}
\begin{proof}
By the assumption that $f$ is correct on $\bW_t^\bot$, then the Conditional Expectation Lemma, i.e., \lemref{lem:conditional-main}) implies there exists $\bu\in\bU=\bW_t^\bot\cap\bA$ and $\sigma\in\left[\frac{3}{4},B\right]$ such that 
\[\EEx{G(\bW_t^\bot,\sigma^2)}{\langle \bu,\bg\rangle^2 \, \mid|\, f(\bg)=1}\ge \Ex{\langle \bu,\bg\rangle^2} + \frac{1}{4Br}\]
and $\PPr{f(\bg)=1}\ge\frac{1}{40B^2r}$.  
On the other hand, by \eqnref{eqn:VtWt:two}, we have that $G(\bW_t^\bot,\sigma^2)$ is $\delta$-statistically close to $G(\bV_t^\bot,\sigma^2)$ for $\delta=o\left(\frac{1}{B^2n}\right)$. 
We claim that as a result,
\begin{equation}
\eqnlab{eqn:condex}
\EEx{G(\bV_t^\bot\cap\bA,\sigma^2)}{\langle \bu,\bg\rangle^2 \, \Big|\, f(\bg)=1}\ge \Ex{\langle \bu,\bg\rangle^2} + \frac{1}{6Br}
\end{equation}
and $\PPr{f(\bg)=1}\ge\frac{1}{50B^2r}$. 
Since $f(\bg)\in\{0,1\}$, and so $\PPr{f(\bg)=1}$ differs by at most $\delta = o\left(\frac{1}{B^3r\log(rB)}\right)$ between the two distributions, then it follows that $\PPr{f(\bg)=1}\ge\frac{1}{50B^2r}$. 
This also implies that the total variation distance between the two distributions conditioned on $f(\bg)=1$ can only increase by a factor of $50B^2r$. 
In other words, for any distinguishing function $R:\mathbb{R}^n\to[0,D]$,
\begin{equation}
\eqnlab{eqn:expectation-diff}
\left\lvert\EEx{G(\bW_t^\bot,\sigma^2)}{R(\bg) \, \Big|\, f(\bg)=1}-\EEx{G(\bV_t^\bot\cap \bA,\sigma^2)}{R(\bg) \, \Big|\, f(\bg)=1}\right\rvert\le 50B^2r D\delta.
\end{equation}
By standard Gaussian concentration, we have $\PPr{\langle \bu,\bg\rangle^2>10\ell B}\le \exp\left(-\ell\right)$. 
Thus, we can truncate $\langle \bu,\bg\rangle^2$ at $D=10B\log(rB)$ without changing either expectation by more than $o\left(\frac{1}{Br}\right)$.  
Since $\O{B^3r\log(rB)}\cdot\delta=o\left(\frac{1}{Br}\right)$, then by \eqnref{eqn:expectation-diff}
\begin{equation}
\eqnlab{eqn:expectation-diff2}
\left\lvert\EEx{G(\bW_t^\bot,\sigma^2)}{\langle \bu,\bg\rangle^2 \, \Big|\, f(g)=1}
-\EEx{G(\bV_t^\bot\cap \bA,\sigma^2)}{\langle \bu,\bg\rangle^2 \, \Big|\, f(\bg)=1}
\right\rvert\le o\left(\frac{1}{Br}\right).
\end{equation}
The desired inequality then holds for $\sigma^2$ and $\bu\in \bW_t\cap \bA$ by a triangle inequality. 
By a similar argument, it follows that changing $\sigma^2$ by only $o\left(\frac{1}{B^2n^2}\right)$ additively, \eqnref{eqn:expectation-diff2} continues to hold up to a lower-order term in the parameters. 
Thus, there exists $\tilde\sigma^2$ in the discretization defined by $S$, for which the claim is true. 
Finally, since $\bu\in \bW_t\cap \bA$ and \invref{inv:invariant} holds for $(\bV_t,\bW_t)$, then it follows that $\|\bP_{\bV_t}\bu\|_2\ge 1-\frac{1}{B^2n^2}$. 
Therefore, the conclusion of the lemma also holds for some $\bu\in \bV_t\cap \bA$ up to an additive $o\left(\frac{1}{Bn}\right)$ loss in the expectation.
\end{proof}

Intuitively, \lemref{lem:sigmatilde} proves that there exists $\tilde\sigma\in S$ so that the vector $\bv_{\tilde\sigma}$ has high objective value $z(\bv_{\tilde\sigma})$. 
On the other hand, it could be that many vectors have high objective value. 
Hence, we next show that \emph{any} vector that has high objective value must be very close to subspace $\bV_t^\bot\cap \bA$:

\begin{restatable}{lemma}{lembiasednet}
\lemlab{lem:biased:net}
\cite{HardtW13}
Let $\tau\ge0$ and $\bV$ be a subspace of $\mathbb{R}^n$. 
Let $G$ be distribution over $\mathbb{R}^n$ such that for $\bg\sim G$:
\begin{enumerate}
\item 
For every unit vector $\bw\in \bV^\bot$, we have $\Ex{\langle \bw,\bg\rangle^2} \le \tau$.
\item 
For every two unit vectors $\bv\in V$ and $\bw\in \bV^\bot$, we have 
\[\Ex{\langle \bv,\bg\rangle\langle \bw,\bg\rangle} = 0.\] 
\item 
The maximum of $\Ex{\langle \bv,\bg\rangle^2}$ over all unit vectors $\bv\in \bV$ is at least $\tau+\Delta$, for some $\Delta>\frac{1}{\poly(n)}$. 
\item 
For every unit vector $\bu\in\mathbb{R}^n$, we have $\|\bg\|_2^2\le\xi$ almost surely, as well as $\Var\left(\langle \bu,\bg\rangle^2\right)\le \xi^2$. 
\end{enumerate}
Let $\gamma>1/\poly(n)$ and suppose we draw $m=\O{\frac{n\log^2(n)\xi^2}{\gamma^2\Delta^2}}$ i.i.d. samples $\bg_1,\ldots,\bg_m\sim G$ from $G$. 
Let 
\[\bu^* = \arg\max_{\|\bu\|_2=1}\sum_{i=1}^m \langle \bg_i,\bu\rangle^2.\]
Then, with probability $1-\exp(-n\log^2 n)$, we have $\|\bP_{\bV}\bu^*\|^2_2\ge1-\gamma$ and
\[\frac{1}{m}\sum_{i=1}^m \langle \bg_i,\bu^*\rangle^2\ge\tau + \frac{\Delta}{2}.\]
\end{restatable}

The property follows from analyzing the top singular vector of the biased Gaussian matrix consisting of the positively labeled examples. 
The analysis uses standard techniques and thus we defer the proof of \lemref{lem:biased:net} to \secref{sec:net}. 
Nevertheless we shall require the property in the proof of the following statement:
\begin{lemma}
\cite{HardtW13}
With probability $1-\exp(-n)$, the vector $\bv^*$ computed by the algorithm in round $t$ satisfies
\begin{equation}
\eqnlab{eqn:PUvt}
\|\bP_{\bV_t^\bot\cap \bA}\bv^*\|_2^2\ge 1-\frac{1}{20(Bn)^{3.5}\log^4(Bn)}.
\end{equation}
\end{lemma}
\begin{proof}
Let $t\in[r]$ be a fixed round by the attack. 
Let $z^*=\max_{\sigma^2\in S}z(\bv_\sigma)$ denote the maximum objective value achieved across all values of $\sigma$ in the discretization $S$ in round $t$. 
We first lower bound $z^*$ by applying  \lemref{lem:biased:net} to the conditional distribution of $\bg\sim G(\bV_t^\bot,\tilde\sigma^2)$ conditioned on $f(\bg)=1$. 
Specifically, $\bg_1',\ldots,\bg'_{m'}$ are uniformly sampled from this conditional distribution. 
Moreover, the probability that $\calA$ outputs $1$ on any sample is at least $p\ge\Omega\left(\frac{1}{B^2n}\right)$. 
Hence by a standard Chernoff bound argument, we have that with probability $1-\exp(-n)$,
\[m'\ge \frac{pm}{10} \ge \Omega\left(B^{11}n^{10}\log^{15}(n)\right),\]
which implies that we have a large number of samples from the conditional distribution.
We shall apply \lemref{lem:biased:net} with
\[\gamma = \frac{1}{(Bn)^{3.5}\log^4(Bn)}.\]
Thus, we need to show that the prerequisites of \lemref{lem:biased:net} hold. 
To that end, let $\bV=\bV_t^\bot\cap \bA$ and $\bW=\bV^\bot = \bV_t + \bA^\bot$. 
Moreover, let $\tau=\tilde\sigma^2+\frac{1}{4}$ and $\Delta$ be the parameter from \lemref{lem:sigmatilde}. 
\begin{enumerate}
\item 
Note that any unit vector $\bw\in \bW$ can be written as $\alpha \bv + \beta \bw'$, where $\bv,\bw'$ are unit vectors with $\bv\in \bV_t$ and $\bw'\in \bA^\bot$ is orthogonal to $\bv$, and the coefficients satisfy $\alpha^2+\beta^2=1$. 
Since $\bg\sim G(\bV_t^\bot,\tilde\sigma^2)$, then $\Ex{\langle \bv,\bg\rangle^2}\le\frac{1}{4}$.  
Further, the condition $f(\bg)=1$ does not bias the distribution along directions inside $\bA^\bot$, since $f$ only considers directions inside $\bA$. 
Thus, $\Ex{\langle \bw',\bg\rangle^2}\le\tau$. 
Since $\Ex{\langle \bw',\bg\rangle\langle
\bv,\bg\rangle}=0$, as $\bv$ and $\bw'$ are orthogonal and $\bg$ can be written as the sum of two independent spherical Gaussians, then we also have $\Ex{\langle \bw,\bg\rangle^2}\le\tau$. 
\item 
For every $\bv\in \bV$ and $\bw\in \bW$, $\bv$ and $\bw$ are orthogonal and so, $\bg\sim G(\bV_t^\bot,\tilde\sigma^2)$ can be written as the sum of two independent spherical Gaussians. 
Thus, $\Ex{\langle \bv,\bg\rangle\langle \bw,\bg\rangle}=0$. 
\item 
Let $\Delta\ge\frac{1}{7Br}$. 
Then by \lemref{lem:sigmatilde}, there exists $\bv\in \bV$ such that $\Ex{\langle \bv,\bg\rangle^2}\ge \tau + \Delta$.  
\item 
For every $\bu\in\mathbb{R}^n$, observe that $\xi^2:=\Var\left(\langle \bu,\bg\rangle^2\right)\le \O{B^2\log^2 n}$ corresponds to the fourth moment of a Gaussian with variance at most $B$, conditioned on an event of probability $\frac{1}{\poly(n)}$. 
Thus the fourth moment of $\O{B^2}$ is at most a multiplicative $\O{\log^2 n}$ factor larger. 
\end{enumerate}
Now, to apply \lemref{lem:biased:net} for the specified values of $\gamma$, $\Delta$, and $\xi^2$, we require 
\[\Theta\left(\frac{n\log^2(n)\xi^2}{\gamma^2\Delta^2}\right)\le\Theta\left(B^{11}n^{10}\log^{14}(n)\right).\]
samples from the conditional distribution. 
Note that this quantity is $o(m')$ and thus all the prerequisites of \lemref{lem:biased:net} hold. 
Therefore, with probability $1-\exp(-n\log n)$, we have that
\[z^* \ge z(\tilde\sigma) \ge 1 + \frac{\Delta}{14}.\]
Now, we call a discretized variance $\sigma^2\in S$ \emph{bad} if, for every unit vector $\bu\in \bV_t^\bot\cap A$, we have for $\bg\sim G(\bV_t^\bot,\sigma^2)$,
\[\Ex{\langle \bu,\bg\rangle^2 \, \Big|\, f(\bg)=1}\le \Ex{\langle \bu,\bg\rangle^2} + \frac{\Delta}{20}.\]
We claim that with probability $1-\exp(-n)$, every bad $\sigma^2$ will achieve strictly smaller objective value, i.e., $z(\sigma)\le 1+\frac{\Delta}{18}$. 
This follows from applying a standard concentration similar to the analysis of \lemref{lem:biased:net}, though we now use an upper bound on $\Ex{\langle \bu,\bg\rangle^2}$ also for $\bu\in \bV_t^\bot\cap A$.  

Therefore, the maximizer of the objective function corresponds to a $\sigma^*$ that satisfies the assumptions of \lemref{lem:biased:net}. 
Thus by \lemref{lem:biased:net}, with probability $1-\exp(-n)$, $\bv^*$ satisfies 
\[\|\bP_{\bV_t^\bot\cap \bA}\bv^*\|_2^2\ge 1-\gamma.\]
\end{proof}
Finally, we show that the invariant continues to hold with high probability assuming that $f$ continues to be correct. 
\begin{lemma}
\lemlab{lem:progress2}
\cite{HardtW13}
Suppose $(\bV_t,\bW_t)$ satisfies \invref{inv:invariant} and suppose the vector $\bv^*$ computed in round $t$ satisfies \eqnref{eqn:PUvt}. 
Then, \invref{inv:invariant} holds for $(\bV_{t+1},\bW_{t+1})$. 
\end{lemma}
\begin{proof}
Let $\gamma=\frac{1}{(Bn)^{3.5}\log^4(Bn)}$ and let $\bU=\bV_t^\bot\cap \bA$. 
Then by \eqnref{eqn:PUvt}, $\|\bP_{\bU}\bv^*\|_2^2\ge 1-\gamma$. 
In particular, $\|\bP_{\bA}\bv^*\|_2^2\ge 1-\gamma$ and $\|\bP_{\bV_t}\bv^*\|_2^2\le\gamma$. 
Hence,
\[\|\bP_{\bA} \bv_t\|_2 \ge 1- \O{\gamma}.\]
On the other hand, observe that $\bP_{\bV_{t+1}}=\bP_{V_t}+\bv_t\bv_t^\top$. 
Further, $\bP_{W_{t+1}}=\bP_{W_t} + \bw\bw^\top$ where $\bw$ is a unit vector orthogonal to $\bW_t$ such that $\|\bv_t-\bw\|\le\O{\gamma}$. 
Thus,
\begin{align*}
d(\bV_{t+1},\bW_{t+1}) &= \|\bP_{\bV_{t+1}}- \bP_{\bW_{t+1}}\|_2 \\
&\le \|\bP_{\bV_{t+1}}- \bP_{\bW_{t+1}}\|_2 + \|\bv_t\bv_t^\top - \bw\bw^\top\|_2 \\
&\le d(\bV_t,\bW_t)+\O{\gamma}.
\end{align*}
Therefore, we have
\begin{align*}
d(\bV_{t+1},\bW_{t+1})&\le d(\bV_t,\bW_t)+\O{\gamma}\\
&\le \frac t{(Bn)^{3.5}\log^3(Bn)}+\O{\frac1{(Bn)^{3.5}\log^4(Bn)}}\\
&\le\frac
{t+1}{(Bn)^{3.5}\log(Bn)^3},
\end{align*}
for sufficiently large $n$. 
\end{proof}
In summary, we have shown that with probability $1-\O{\exp(-n)}$, the prerequisites of \lemref{lem:progress2} hold, which implies that \invref{inv:invariant} continues to hold for round $t+1$, which in turn concludes the proof of \lemref{lem:progress}.

\subsubsection{Top Singular Vector of Biased Gaussian Matrices}
\seclab{sec:net}
Recall that to prove \lemref{lem:progress2}, we needed to show that \emph{any} vector that has high objective value must be very close to subspace $\bV_t^\bot\cap \bA$. 
In this section, we will prove the necessary statements to understand the top singular vector of biased Gaussian matrices. 
We first recall the following standard discretization of the unit sphere.
\begin{lemma}[$\eps$-net for the sphere]
\lemlab{lem:sphere:eps-net}
\cite{HardtW13}
For every $c>0$, there exists a set $N\subseteq\mathbb{S}^{n-1}$ of size $|N|\le \exp(\O{n\log (1/c)})$, such that for every unit vector $u\in\mathbb{R}^n$, there exists a corresponding unit vector $v\in N$ such that $\langle u,v\rangle^2\le c$.  
\end{lemma}

We also utilize the following formulation of the Chernoff-Hoeffding bound.
\begin{theorem}[Chernoff-Hoeffding]
\thmlab{thm:chernoff:hoeffding}
Let $X_1,\ldots,X_m$ be independent random variables.
Let $X = \sum_{i=1}^m X_i$ and let $\xi^2=\Var(X)$. 
Then for any $t>0$,
\[\PPr{\left|X-\Ex{X}\right| > t} \le\exp\left(-\frac{t^2}{4\xi^2}\right).\]
\end{theorem}

We now prove the desired statement about the top singular vector for biased Gaussian matrices. 
\lembiasednet*
\begin{proof}
Suppose we draw $m$ i.i.d. samples $\bg_1,\ldots,\bg_m\sim G$ from $G$. 
Let $L = \max_{\bv\in\bV, \|\bv\|_2=1} \Ex{\langle \bv,\bg\rangle^2}$. 
By the third assumption, $L \ge \tau + \Delta$. 
Furthermore, since we have $\|\bg\|_2^2 \le \xi$ almost surely, then it follows that $L \le \xi$. 

Let $\bv^*\in \bV$ be a unit vector achieving this maximum, and let $X = \sum_{i=1}^m \langle \bv^*,\bg_i\rangle^2$. 
Then $\Ex{X} = Lm$. 
Because each term $\langle \bv^*,\bg_i\rangle^2$ is contained in $[0, \xi]$ almost surely, we can apply Hoeffding's inequality (\thmref{thm:chernoff:hoeffding}) to obtain:
\[\PPr{X\le Lm - \frac{\gamma\Delta m}{4}} \le \exp\left(-\frac{2(\gamma\Delta m/4)^2}{m\xi^2}\right) = \exp\left(-\frac{\gamma^2\Delta^2 m}{8\xi^2}\right) \le \exp(-\Omega(n\log^2 n)),\]
where the last inequality follows from our choice of sample complexity $m = \Omega\left(\frac{n\log^2(n)\xi^2}{\gamma^2\Delta^2}\right)$.

Let $M = \{\bu \in \mathbb{S}^{n-1} \mid \|\bP_{\bV}\bu\|_2^2 \le 1-\gamma\}$. 
For any $\bu \in M$, we decompose it as $\bu=\alpha \bv +\beta \bw$ for unit vectors $\bv \in \bV, \bw \in \bV^\bot$, with $\alpha^2+\beta^2=1$ and $\alpha^2\le 1-\gamma$. 
By linearity of expectation,
\begin{align*}
\Ex{\langle \bu,\bg\rangle^2} &= \alpha^2\Ex{\langle \bv,\bg\rangle^2} + \beta^2\Ex{\langle \bw,\bg\rangle^2} + 2\alpha\beta\Ex{\langle \bv,\bg\rangle\langle \bw,\bg\rangle} \\
&= \alpha^2\Ex{\langle \bv,\bg\rangle^2} + \beta^2\Ex{\langle \bw,\bg\rangle^2},
\end{align*}
using the second assumption that $\Ex{\langle \bv,\bg\rangle\langle \bw,\bg\rangle}=0$. 
Since $\bv\in\bV$, we have $\Ex{\langle \bv,\bg\rangle^2} \le L$. 
By the first assumption, $\Ex{\langle \bw,\bg\rangle^2} \le \tau$. 
Thus, 
\begin{align*}
\Ex{\langle \bu,\bg\rangle^2} &\le \alpha^2 L + (1-\alpha^2)\tau = \tau + \alpha^2(L-\tau).
\end{align*}
Because $L \ge \tau+\Delta > \tau$, this expression is increasing with respect to $\alpha^2$. 
Using the constraint $\alpha^2 \le 1-\gamma$, we have an upper bound on the expectation of any vector in $M$:
\begin{align*}
\Ex{\langle \bu,\bg\rangle^2} &\le \tau + (1-\gamma)(L-\tau) = L - \gamma(L-\tau) \le L - \gamma\Delta.
\end{align*}
Now, let $N$ be a $c$-net of the unit sphere $\mathbb{S}^{n-1}$ given by \lemref{lem:sphere:eps-net} with granularity $c=\frac{\gamma\Delta}{32\xi}$. 
Because $\xi \le \poly(n)$ and $\gamma,\Delta \ge \frac{1}{\poly(n)}$, we have $c \ge\frac{1}{\poly(n)}$, which implies $|N| = \exp(\O{n\log(1/c)}) = \exp(\O{n\log n})$. 

For any fixed $\tilde{\bu} \in N$, let $Y(\tilde{\bu}) = \sum_{i=1}^m \langle \tilde{\bu},\bg_i\rangle^2$. 
Because $\langle \tilde{\bu},\bg_i\rangle^2 \in [0, B]$ almost surely, Hoeffding's inequality implies
\[ \PPr{Y(\tilde{\bu}) - m\Ex{\langle \tilde{\bu},\bg\rangle^2} \ge \frac{\gamma\Delta m}{16}} \le \exp\left(-\frac{2(\gamma\Delta m/16)^2}{m \xi^2}\right) \le \exp(-\Omega(n\log^2 n)). \]
Taking a union bound over all points in $N$, this bound holds simultaneously for all $\tilde{\bu} \in N$ with probability at least $1 - \exp(-\Omega(n\log^2 n))$.
We assume this high-probability event holds. 

For any vector $\bu \in M$, there exists a nearest point $\tilde{\bu} \in N$ such that $\|\bu - \tilde{\bu}\|_2 \le c$. 
Then since $\|\bg_i\|_2^2 \le \xi$, we have
\begin{align*}
\left|\sum_{i=1}^m \langle \bu,\bg_i\rangle^2 - \sum_{i=1}^m \langle \tilde{\bu},\bg_i\rangle^2\right| &\le \sum_{i=1}^m |\langle \bu - \tilde{\bu}, \bg_i \rangle \langle \bu + \tilde{\bu}, \bg_i \rangle| \\
&\le \sum_{i=1}^m \|\bu - \tilde{\bu}\|_2 \|\bu + \tilde{\bu}\|_2 \|\bg_i\|_2^2 \\
&\le \sum_{i=1}^m c \cdot 2 \cdot B = 2m\xi c = \frac{\gamma\Delta m}{16}.
\end{align*}
By taking expectations on both sides, we thus have
\[ m\left|\Ex{\langle \bu,\bg\rangle^2} - \Ex{\langle \tilde{\bu},\bg\rangle^2}\right| \le 2m\xi c = \frac{\gamma\Delta m}{16}.\]
Therefore, for any $\bu \in M$, we can bound its empirical sum by passing through its net point $\tilde{\bu}$:
\begin{align*}
\sum_{i=1}^m \langle \bu,\bg_i\rangle^2 &\le Y(\tilde{\bu}) + \frac{\gamma\Delta m}{16} \\
&\le m\Ex{\langle \tilde{\bu},\bg\rangle^2} + \frac{2\gamma\Delta m}{16} \\
&\le m\Ex{\langle \bu,\bg\rangle^2} + \frac{3\gamma\Delta m}{16}.
\end{align*}
Substituting our earlier upper bound $\Ex{\langle \bu,\bg\rangle^2} \le L - \gamma\Delta$, we have
\[\sum_{i=1}^m \langle \bu,\bg_i\rangle^2 \le m(L - \gamma\Delta) + \frac{3\gamma\Delta m}{16} = Lm - \frac{13\gamma\Delta m}{16}.\]
Crucially, there is a margin of $\frac{9\gamma\Delta m}{16}$ between the lower bound at $\bv^*$ (which achieved $X \ge Lm - \frac{4\gamma\Delta m}{16}$) and the upper bound for any vector in $M$. 
Thus, any vector $\bu$ with $\|\bP_{\bV}\bu\|_2^2 \le 1-\gamma$ results in an objective value smaller than $X$, implying the global empirical maximizer $\bu^*$ must be outside $M$, satisfying $\|\bP_{\bV}\bu^*\|_2^2 > 1-\gamma$.

Finally, the objective value evaluated at $\bu^*$ must be at least $\frac{X}{m}$. 
Since $\gamma \le 1$, we have
\[\frac{1}{m}\sum_{i=1}^m \langle \bg_i,\bu^*\rangle^2 \ge \frac{X}{m} \ge L - \frac{\gamma\Delta}{4} \ge \tau + \Delta - \frac{\Delta}{4} > \tau + \frac{\Delta}{2},\]
completing the proof.
\end{proof}

With \lemref{lem:biased:net} in place, this concludes the proof of \thmref{thm:attack}, summarized as follows:
\thmattack*

\paragraph{On randomized algorithms.}
\cite{HardtW13} notes that although the proof is stated for algorithms that return a deterministic function $f$ of the sketch $\bA\bx$, the conclusions remain valid for algorithms that incorporate additional randomness and output a randomized function $f$ of $\bA\bx$. 
The key observation is that the attack, with probability $1$, never repeats a query. 
Therefore, for any fixed setting of the randomness used by $f$ across all possible inputs, we obtain a deterministic function to which \thmref{thm:attack} can be directly applied.

Now consider the attack described in \figref{fig:hw13:attack}. 
In each round $t$, we allow the algorithm to apply a possibly new function $f_t: \mathbb{R}^n \to \{0,1\}$, provided that $f_t(\bx)$ depends only on the projection $\bP_{\bA}\bx$. 
Under this condition, the proof of \thmref{thm:attack} goes through without change, except that $f$ is replaced by $f_t$ in round $t$, i.e., $f(\bx)=f_t(\bA\bx)$ at each time $t$. 
Note that since $f$ is a norm-based estimator, it can use different randomness at time $t$, but it cannot be a function of other parameters such as the history of previous queries. 
The key requirement for the attack to succeed is simply that the subspace $\bA$ remains fixed.

\section{Lower Bound for \texorpdfstring{$F_p$}{Fp} Estimation with Integer Sketches}
Although the attack by \cite{HardtW13} in \secref{sec:fp:lb:real} describes a real-valued adaptive attack on linear sketches for $L_p$ norms, streaming algorithms realistically require discrete-valued input data streams. 
In this section, we demonstrate a similar attack by \cite{GribelyukLWYZ25} within the streaming model, in particular on the following notion of integer sketches:
\begin{definition}[Integer sketch]
Given a data stream of length at most $m = \poly(n)$ that implicitly defines a vector in $\mathbb{R}^n$ with integer entries bounded in magnitude by $\poly(n)$, an \emph{integer sketch} is a linear sketching algorithm whose sketching matrix consists entirely of integers with magnitude at most $M = \poly(n)$. 
\end{definition}
More generally, the lifting technique of \cite{GribelyukLWYZ25} is able to extend linear sketch lower bounds from the continuous setting to the discrete setting, thereby achieving a number of additional lower bounds for streaming algorithms. 

A standard method for generating hard input distributions in $\mathbb{R}^n$ for real-valued linear sketches $\bA \in \mathbb{R}^{r \times n}$ involves drawing inputs $\bx$ from a continuous Gaussian distribution $\calN(0, \sigma^2 \cdot \mathbb{I}_n)$. 
Due to the rotational invariance of Gaussian distributions, the sketch output $\bA\bx$ is itself Gaussian, specifically distributed as $\calN(0, \sigma^2 \bA \bA^\top)$. 
This property allows lower bounds for sketching to be established by analyzing the total variation distance between low-dimensional Gaussian distributions.

One might naturally attempt to extend such lower bounds to discrete input settings by applying discretization techniques, such as rounding Gaussian samples to the nearest multiple of $\frac{1}{\poly(n)}$. 
However, this approach breaks the rotational invariance of the distribution, making the distribution of $\bA\bx$ harder to analyze. 
The truncation may inadvertently leak information about $\bx$, complicating the analysis of sketch outputs.

Alternatively, one could consider using discrete Gaussian distributions directly. 
However, if $\bx \sim \calD(0, \sigma^2 \mathbb{I}_n)$ is drawn from a discrete Gaussian, it does not follow that $\bA\bx$ is distributed as a discrete Gaussian over the lattice $\bA\mathbb{Z}^n$ with covariance $\sigma^2 \bA \bA^\top$. 
In fact, this discrepancy motivates a line of work in lattice-based analysis, which studies the behavior of discrete Gaussians through the geometry of lattices—particularly via bounds on the successive minima of the orthogonal lattice to the row span of $\bA$. 
Indeed, at a high level, \cite{GribelyukLWYZ24} uses lattice theory to prove a ``cell'' lemma, which allow the simulation of a continuous Gaussian distribution using a discrete Gaussian distribution plus additional uniform continuous noise. 
The integer sketch lower bounds then follow by simulating the previously known attacks over the reals, which originally required continuous Gaussian queries. 
In particular, the attack on real-valued linear sketches by \cite{HardtW13} from \secref{sec:fp:lb:real} uses continuous Gaussian queries and thus can be lifted by the techniques of \cite{GribelyukSWY25} to obtain an attack on integer-valued sketches. 

\subsection{Lattice Theory}
We review several fundamental concepts from lattice theory, such as the definitions of a lattice, its dual lattice, the orthogonal lattice, the smoothing parameter, and the successive minima.
\begin{definition}[Lattice]
A lattice is a discrete additive subgroup of $\mathbb{R}^m$. 
Given a basis of linearly independent vectors $\bB = \{\bb_1, \ldots, \bb_n\} \subset \mathbb{R}^m$ with $m \ge n \ge 1$, the lattice generated by $\bB$ is defined as
\[\mathcal{L} = \calL(\bB) = \left\{ \bB \bz = \sum_{i=1}^n z_i \bb_i \mid \bz \in \mathbb{Z}^n \right\}.\]
\end{definition}

\begin{definition}[Successive minima]
For each $i \in [n]$, the $i$-th successive minimum $\lambda_i(\calL)$ of a lattice $\calL$ is defined as the minimal radius $r$ such that the ball of radius $r$ centered at the origin contains at least $i$ linearly independent vectors from $\calL$.
\end{definition}

\begin{definition}[Orthogonal lattice]
For a lattice $\calL \subset \mathbb{Z}^n$, the orthogonal lattice $\calL^{\perp}$ is defined as
\[\calL^{\perp} = \{ \by \in \mathbb{Z}^n \mid \langle \by, \bx \rangle = 0 \text{ for all } \bx \in \calL \}.\]
\end{definition}
Recall from \defref{def:continuous:gaussian} that we define for any $s>0$, the probability density function of the spherical Gaussian function with center $\mu \in \mathbb{R}^n$ by
\[\rho_s(\bx) = \exp\left(-\frac{\|\bx - \mu\|_2^2}{2s^2}\right),\]
for all $\bx \in \mathbb{R}^n$, and $\rho_s(A) = \sum_{\bx \in A} \rho_s(\bx)$ for any subset $A \subseteq \mathbb{R}^n$. 
\begin{definition}[Dual lattice and smoothing parameter]
For any lattice $\calL$, its dual lattice $\calL^*$ is defined by
\[\calL^* = \{ \bx \in \mathbb{R}^n \mid \langle \bx, \by \rangle \in \mathbb{Z} \text{ for all } \by \in \calL \}.\]
Furthermore, given a lattice $\calL$ and $\eps > 0$, the smoothing parameter $\eta_{\eps}(\calL)$ is the smallest positive real number $s$ such that
\[\rho_{1/s}(\calL^* \setminus \{0\}) \le \eps.\]
\end{definition}

Let $\bA \in \mathbb{Z}^{m \times n}$ and $\bM \in \mathbb{R}^{m \times n}$ be full row-rank matrices, and let $\bv \in \mathbb{R}^m$. We define
\[\calE_{\bA, \bM, \bv} = \{\bA \bx: \bx \sim \calD_{\mathbb{Z}^n + \bv, \bM} \}.\]
Here, $\calE_{\bA, \bM, \bv}$ is constructed by first sampling a discrete Gaussian vector $\bx$ and then applying $\bA$ to it, so that the sample is $\bA\bx$. 
This differs from directly sampling a discrete Gaussian supported on $\bA \mathbb{Z}^n$ with the corresponding covariance.  
Nevertheless, the following theorem shows that under certain conditions on the successive minima of $\bA$, these two sampling methods yield distributions that are close point-wise.
\begin{theorem}[Lemma 4 in \cite{AgrawalGHS13}]
\lemlab{discrete_gaussian_tvd}
For any rank-$n$ lattice $\calL$, parameter $\eps \in (0,1)$, vector $\bv \in \mathbb{R}^n$, and full-rank matrix $\bS \in \mathbb{R}^{n \times n}$ satisfying $\sigma_n(\bS) \ge \eta_\eps(\calL)$, we have
\[\rho_{\bS}(\calL+\bv)\in \left[\frac{1-\eps}{1+\eps},\, 1 \right] \cdot\rho_{\bS}(\calL).\]
\end{theorem}
We state the following result contained within the proof of Lemma 3.3 in \cite{AggarwalR16}, which upper bounded the total variation distance between $\calE_{\bA,\bS,\bv}$ and $\calD_{\bA\mathbb{Z}^n+\bA\bv,\bS\bA^\top}$ using the following stronger point-wise bound. 
\begin{theorem}[Lemma 3.3 in \cite{AggarwalR16}]
\thmlab{thm:disc:tvd}
Let $\bA \in \mathbb{R}^{r \times n}$ be a full-rank matrix with $r < 0.75 n$, and let $\calL^\perp(\bA)$ denote the orthogonal lattice of $\bA$.  
Consider any vector $\bv \in \mathbb{R}^n$ and a full-rank matrix $\bS \in \mathbb{R}^{n \times n}$ whose smallest singular value $\sigma_n(\bS)$ satisfies
\[\sigma_n(\bS) > \lambda_{n - r}(\calL^\perp(\bA)) \cdot \sqrt{\frac{\ln\left(2n(1 + 1/\eps)\right)}{\pi}}.\]
For any vector $\bx \in \bA \mathbb{Z}^n + \bA \bv$, let $\rho_1(\bx)$ be the probability mass function of $\calE_{\bA, \bS, \bv}$ and $\rho_2(\bx)$ be that of $\calD_{\bA \mathbb{Z}^n + \bA \bv, \bS \bA^\top}$. 
Then
\[\frac{1-\eps}{1+\eps}\le\frac{\rho_1(\bx)}{\rho_2(\bx)}\le\frac{1+\eps}{1-\eps}.\]
\end{theorem}
We also restate Siegel's lemma, which provides an upper bound on the size of the entries of integer vectors that lie in the kernel of a matrix $\bA$.
\begin{lemma}[Siegel's lemma]
\lemlab{lem:siegel}
\cite{silverman2000diophantine}
Let $\bA \in \mathbb{Z}^{r \times n}$ be a nonzero integer matrix with $r < n$, and suppose all entries of $\bA$ are bounded in magnitude by $M$.  
Then there exists a nonzero integer vector $\bx \in \mathbb{Z}^n$ satisfying $\bA \bx = \boldsymbol{0}^r$, whose entries are bounded in magnitude by $(nM)^{r/(n - r)}$.
\end{lemma}

\subsection{Lifting Framework}
In this section, we introduce the lifting framework of \cite{GribelyukLWYZ25}, to prove a lower bound against adversarially robust algorithms based on integer-valued sketches on turnstile streams.  
We first give a technical overview of the approach, following the exposition of \cite{GribelyukLWYZ25}. 

\paragraph{Smoothing parameter and successive minima.}
Let $\bA \in \mathbb{Z}^{r \times n}$ have columns $\bA_i$, and define the lattice $\calL(\bA) = \{ \sum_{i=1}^n z_i \bA_i: \bz \in \mathbb{Z}^n \}$. 
The \emph{successive minima} $\lambda_i(\calL)$ describe the smallest radius needed to capture $i$ linearly independent lattice vectors centered at the origin. 
For convenience, we use $\lambda_{\max}(\calL(\bA))$ to denote $\lambda_{\rank(\bA)}(\calL(\bA))$.

To argue that the total variation distance between $\bA \bx$ and $\by$ is small for $\bx \sim \calD_{\mathbb{Z}^n, \bS}$ and $\by \sim \calD_{\bA\mathbb{Z}^n, \bS\bA^\top}$, we rely on the previously introduced tools from lattice theory~\cite{AggarwalR16, AgrawalGHS13}. 
These works often reduce the task of upper bounding the total variation distance between $\bA\bx$ and $\by$ to the task of upper bounding the \emph{smoothing parameter}, which in turn depends on the successive minima of the \emph{orthogonal lattice} $\calL^\perp(\bA)$, i.e., the set of integer vectors orthogonal to all rows of $\bA$. 
The main goal is to find a short basis for $\calL^\perp(\bA)$, i.e., with $\lambda_{n-r}(\calL^\perp(\bA)) \leq \poly(n)$. 
However, arbitrary sketching matrices $\bA$ may not admit such a basis directly, so \cite{GribelyukLWYZ25} designs a pre-processing step to ensure this property.

\paragraph{Pre-processing to upper bound the successive minima.}
Given an integer matrix $\bA \in \mathbb{Z}^{r \times n}$ with $r < n$ and entries bounded by $M \leq \poly(n)$, the goal is to augment $\bA$ such that the resulting matrix $\bA'$ satisfies $\lambda_{\max}(\calL^\perp(\bA')) \leq \poly(n)$, has at most $m \leq n - \Omega(n)$ rows, and can be assumed without loss of generality in the lower bounds. 
The approach involves iteratively appending kernel vectors to $\bA$, using Siegel’s lemma to guarantee short integer solutions. 
After $T = 0.49n - r$ such steps, we obtain a set of short kernel vectors, each bounded in norm by $(nM)^2$. 
\cite{GribelyukLWYZ24} then generates an additional $0.51n$ vectors using Siegel’s lemma and adds them to $\bA$ to form $\bA'$, yielding an orthogonal lattice with polynomially bounded maximum successive minimum.
Since these rows only strengthen the sketch, we can assume without loss of generality that the final sketching matrix is $\bA'$.

\paragraph{A better pre-processing.}
The previous method adds $\Theta(n)$ rows, which is too costly for sublinear lower bounds. 
To reduce this overhead, \cite{GribelyukLWYZ25} constructs $n - 4r$ short, linearly independent vectors in $\calL^\perp(\bA)$ by a probabilistic method: sampling $\Theta(M^{1.5r})$ integer vectors with entries in $\{0, 1, \ldots, M-1\}$ and exploiting the birthday paradox to find a pair with matching images under $\bA$ but linearly independent differences. 
This is repeated until $n - 4r$ such vectors are obtained. 
Then, Siegel’s lemma can again be used to generate $3r$ additional vectors orthogonal to both the row span of $\bA$ and the constructed vectors, not necessarily with polynomially bounded entries. 
This gives a final sketch matrix $\bA' \in \mathbb{Z}^{4r \times n}$ for which $\lambda_{\max}(\calL^\perp(\bA')) \leq \sqrt{nM^2} = \poly(n)$. 
Although $\bA'$ may contain entries larger than $\poly(n)$ due to the additional $3r$ vectors, this is not an issue, as the subsequent argument only requires bounded successive minima.

\paragraph{Cell lemma.}
\seclab{cell-lemma}
Let $\bA \in \mathbb{Z}^{r \times n}$ be the sketching matrix obtained after applying the pre-processing described above, so that the maximum successive minimum $\lambda_{\max}(\calL^\perp(\bA))$ is at most polynomial in $n$. 
Define the fundamental parallelepiped $\calF$ of the lattice $\calL(\bA)$ as 
\[\calF = \left\{ \bx \in \mathbb{R}^r \ \middle|\ x_i \in [0, \ell_i) \right\},\]
where $\ell_i$ is the smallest non-zero value of $|(\bA \bx)_i|$ over all $\bx \in \mathbb{Z}^n$. 
This region tiles $\mathbb{R}^r$ into disjoint ``cells'', each centered at a lattice point $\bA \bz$ for $\bz \in \mathbb{Z}^n$.
We may also assume without loss of generality that $\bA$ has orthonormal rows. 
This can be achieved via a change of basis in a post-processing step. 
Crucially, this does not alter the structure of the fundamental parallelepiped: let $\bU^\top$ be a matrix with orthonormal rows spanning the same row space as $\bA$. 
Then the projection onto this row space can be written as $\bA^\top (\bA \bA^\top)^{-1} \bA = \bU \bU^\top$. 
Because the rows of $\bA$ span a rational subspace, the projection matrix $\bU \bU^\top$ has rational entries, and the projection $\bU \bU^\top \bx$ is lower bounded for any non-zero $\bx \in \mathbb{Z}^n$. 
Hence, the fundamental parallelepiped defined by $\bU^\top$ remains well-formed.

This setup allows us to compare the probability mass functions of $\bA \bx$ and $\by$, where $\bx \sim \calD_{\mathbb{Z}^n, \bS}$ is a discrete Gaussian and $\by \sim \calD_{\bA \mathbb{Z}^n, \bS \bA^\top}$ is a projected discrete Gaussian. 
Let $\rho_1$ denote the probability mass function of $\bA \bx$, and $\rho_2$ that of $\by$. 
Suppose the covariance matrix $\Sigma = \bS^\top \bS$ satisfies $\sigma_n(\Sigma) \geq \alpha$ for 
\[\alpha = \poly(n) > \lambda_{n-m}(\calL^\perp(\bA))^2 \cdot \frac{\ln(2n(1+1/\eps))}{\pi},\]
which is guaranteed by our pre-processing. 
Then, applying the lattice theory result \lemref{discrete_gaussian_tvd} from \cite{AgrawalGHS13}, point-wise closeness between the distributions can be obtained:
\[1 - \frac{1}{\poly(n)} \leq \frac{\rho_1(\bx)}{\rho_2(\bx)} \leq 1 + \frac{1}{\poly(n)}.\]
While the discrete and continuous Gaussians are close on lattice points, their total variation distance over each unit cell of $\calL(\bA)$ is large. 
Specifically, the discrete distribution has no support in the interior of these cells. 
This poses a challenge when trying to lift lower bounds that rely on continuous Gaussian inputs.

To overcome this, \cite{GribelyukLWYZ25} notes that for each integer input $\bx \in \mathbb{Z}^n$, the sketching algorithm can be assumed (without loss of generality) to observe $\bA \bx + \bfEta$, where $\bfEta$ is sampled uniformly from the fundamental cell $\calF$. 
This is valid because the algorithm can always recover $\bA \bx$ by rounding $\bA \bx + \bfEta$ to the nearest lattice point in $\calL(\bA)$. 
Define $p$ as the probability density function of $\bA \bx + \bfEta$, where $\bx \sim \calD_{\mathbb{Z}^n, \bS}$ and $\bfEta$ is drawn uniformly from $\calF$. 
Let $q$ be the probability density function of the continuous Gaussian $\calN(0, \bA \Sigma \bA^\top)$. 
Then, with high probability over the randomness of $\bfEta$, 
\[1 - \frac{1}{\poly(n)} \leq \frac{p(\bx)}{q(\bx)} \leq 1 + \frac{1}{\poly(n)}.\]
This lemma shows that the distribution of $\bA \bx + \bfEta$ closely approximates a continuous Gaussian, allowing us to simulate Gaussian inputs even when restricted to integer-valued queries. 
As a result, we can simulate lower bounds based on real-valued Gaussian inputs to the integer setting using the lifting framework in \thmref{thm:lifting}.

The above argument establishes point-wise multiplicative closeness of probability densities \emph{after smoothing}, namely, for a fixed realization of the random shift $\bfEta$, the density of $\bA\bx + \bfEta$ is close to that of a continuous Gaussian.
However, this form of closeness is \emph{not preserved under rounding} or other nonlinear post-processing, and therefore does not imply small total variation distance between the corresponding discrete distributions.

In particular, although $\bA\bx + \bfEta$ is close in density to a Gaussian, it does not follow that rounding a Gaussian sample to the nearest lattice point yields a distribution close (in total variation distance) to that obtained by first rounding an independent Gaussian sample and then applying $\bA$.
That is, sampling a Gaussian $\bg$, rounding it to obtain $\overline{\bg} \in \mathbb{Z}^n$, and outputting $\bA\overline{\bg}$ need not produce a distribution close to rounding $\bA \bg'$ for an independent Gaussian $\bg'$.

As a concrete example, let $\bA=[1,N]$ and consider $\bA\bx$ for $\bx\sim\calD_{\mathbb{Z}^2, \mathbb{I}_2}$ and $\by\sim\calD_{\bA\mathbb{Z}^2,\bA^\top}$. 
Even though the support of $\bA\bx$ and $\by$ are both $\bA\mathbb{Z}^2$, the probability mass of $\by$ concentrates around $0$, while the probability mass of $\bA\bx$ concentrates around multiples of $N$. 
Hence, the point-wise values of the probability mass functions are quite different, c.f., \figref{fig:disc:gauss:pmf}. 

\begin{figure}[!htb]
    \centering
    \begin{subfigure}[b]{0.45\textwidth}
        \includegraphics[scale=0.95]{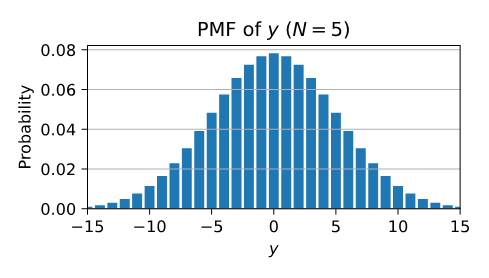}
    \end{subfigure}
    \begin{subfigure}[b]{0.45\textwidth}
        \includegraphics[scale=0.95]{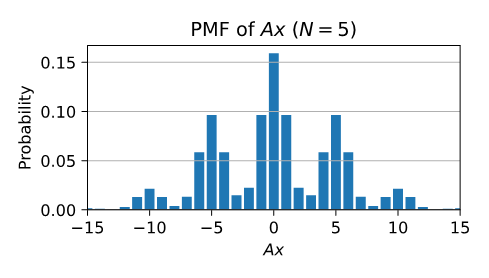}
    \end{subfigure}
    \caption{Differences in probability mass functions}
    \figlab{fig:disc:gauss:pmf}
\end{figure}

\paragraph{Formal analysis.}
We begin with the following matrix pre-processing lemma, which proves that given a matrix $\bA \in \mathbb{Z}^{r \times n}$, it is possible to append at most $3r$ integer vectors to its rows to obtain a new matrix $\bA'$ such that $\lambda_{n - r}(\calL^\perp(\bA')) \le \sqrt{n M^2}$.

\begin{lemma}[Matrix pre-processing lemma]
\lemlab{lem:preprocessing}
\cite{GribelyukLWYZ25}
Let $\bA\in\mathbb{Z}^{r \times n}$ be a full-rank integer matrix with $r \le \frac{n}{4}$ and entries bounded in magnitude by $M$, for some $M \ge n$.  
Then there exists a pre-processing procedure that extends $\bA$ by adding additional rows to obtain a matrix $\bA' \in \mathbb{Z}^{m \times n}$ with $m \le 4r$, such that
\[\lambda_{n - r}(\calL^\perp(\bA')) \le \sqrt{nM^2}.\]
\end{lemma}
\begin{proof}
We first show that it is possible to find $n-4r$ linearly independent integer vectors with entries bounded in magnitude by $M$ in the kernel of $\bA$. 
To do this, we use a probabilistic construction to build these vectors iteratively. 
Suppose we have already selected $t\le n-4r$ linearly independent vectors in the kernel of $\bA$. 
Let $\bB \in \mathbb{R}^{(n-t) \times n}$ be a matrix whose rows form a basis for the orthogonal complement of the span of these $t$ vectors. 
Without loss of generality, we can choose $\bB$ to contain $n-t$ columns that form a diagonal matrix with non-zero diagonal entries, as the row space of $\bB$ can be selected arbitrarily.

Now, consider the following sampling process: draw $\Theta(M^{1.5r})$ random integer vectors $\bv^1, \bv^2, \ldots, \bv^s$ with entries in $\{0,1,\ldots,M-1\}$. 
Let $\calE_1$ denote the event that there exist indices $i$ and $j$ such that $1 \le i < j \le s$ and $\bA \bv^i = \bA \bv^j$. 
Each coordinate of $\bA\bv^i$ lies in $\{-nM^2, \ldots, nM^2\}$ and $M\ge n$, so there are at most $M^{3r}$ possible values of $\bA\bv^i$. 
Therefore, by the birthday paradox, we have $\PPr{\calE_1}\ge 0.9$.

Next, consider the difference $\bx = \bv^i - \bv^j$, which is an integer vector with entries in $\{-(M - 1), \ldots, M - 1\}$. 
Since $\bB$ is a diagonal matrix with $n-t$ non-zero entries, and the remaining $t$ coordinates of $\bx$ are fixed, there is at most one assignment to the remaining coordinates that satisfies $\bB\bx = \mathbf{0}^{n-t}$. 
Hence, we have $\PPr{\bB\bx = \mathbf{0}^{n-t}}\le\left(\frac{1}{M}\right)^{n - t}$. 
Let $\calE_2$ denote the event that $\bB \bv^i \neq \bB \bv^j$ for all $1\le i<j\le s$. 
Since $n-t \ge 4r$, a union bound gives $\PPr{\calE_2}\ge 0.9$.

Combining the two events, we conclude that $\PPr{\calE_1 \cap \calE_2} \ge 0.8$. 
This means there exists a vector $\by = \bv^i - \bv^j$ such that (1) $\bA \by = \mathbf{0}^r$ and (2) $\bB \by \neq \mathbf{0}^{n-t}$. 
Since the rows of $\bB$ span the orthogonal complement of the span of the first $t$ vectors, this implies that $\by$ is linearly independent of them. 
Thus, $\by$ can be selected as the $(t+1)$-th vector in the kernel.

Repeating this process, we obtain $n-4r$ linearly independent integer vectors with entries in $\{-(M - 1), \ldots, M - 1\}$, so that each vector has length at most $\sqrt{nM^2}$. 
Let $V$ be the subspace they span. 
Then the dimension of $\ker(\bA) \setminus V$ is $3r$. 
We can now iteratively apply Siegel’s lemma to construct a basis of $3r$ integer vectors for $\ker(\bA) \setminus V$, and add them as rows to $\bA$, forming a matrix $\bA' \in \mathbb{Z}^{4r \times n}$. 
Since these new rows are orthogonal to the original ones, we have
\[\lambda_{\max}(\calL^\perp(\bA')) \le \sqrt{nM^2}.\]
\end{proof}

\begin{lemma}
\lemlab{lem:main:disc:to:cont}
\cite{GribelyukLWYZ25}
Let $C > 0$ be any fixed constant, and let $\bS \in \mathbb{R}^{n \times n}$ be a full-rank matrix. 
Let $\bA \in \mathbb{R}^{r \times n}$ be an orthonormal matrix with $r \le n$, satisfying the following conditions:
\begin{enumerate}
\item 
There exists a matrix $\bM \in \mathbb{R}^{r \times n}$ with rational entries such that $\bM$ and $\bA$ share the same row space.
\item 
The largest successive minimum of the orthogonal lattice to $\bA$ satisfies $\lambda_{\max}(\calL^\perp(\bA)) \le \frac{\sigma_n(\bS)}{10C \log n}$.
\item 
The singular values of $\bS$ satisfy the bounds $n^{6C} \ge \sigma_1(\bS) \ge \sigma_n(\bS) \ge n^{5C + 3}$.
\end{enumerate}
Let $\bx \sim \calD(0, \bS^\top \bS)$, and let $\bfEta$ be sampled uniformly from the fundamental parallelepiped of $\calL(\bA)$. 
Define $p$ to be the probability density function of the random variable $\bA\bx + \bfEta$, and let $q$ denote the density function of the Gaussian distribution $\calN(0, \bA\bS^\top \bS \bA^\top)$. 
Then, with probability at least $1 - \frac{1}{\poly(n)}$ over $\bv = \bA\bx \in \mathbb{R}^r$ for $\bx \sim \calD(0, \bS^\top \bS)$, we have
\[\frac{p(\bv)}{q(\bv)}\in\left[1-\frac{1}{n^C}, 1+\frac{1}{n^C}\right].\]
\end{lemma}
\begin{proof}
Let $p'$ denote the probability density function of the random variable $\by + \eta$, where $\by \sim \calD_{\bA\mathbb{Z}^n, \bS\bA^\top}$. 
Assuming that $\lambda_{\max}(\calL^\perp(\bA)) \le \frac{\sigma_n(\bS)}{10C\log n}$, we invoke \thmref{thm:disc:tvd} to obtain
\[\frac{p(\bv)}{p'(\bv)}\in\left[1-\frac{1}{n^{2C}}, 1+\frac{1}{n^{2C}}\right]\]
for every vector $\bv \in \mathbb{R}^r$.

We begin by verifying that the unit cell associated with $\bA$ is well-defined, i.e., a fundamental parallelepiped of the corresponding lattice exists. 
Suppose there exists a matrix $\bM \in \mathbb{R}^{r \times n}$ with rational entries sharing the same row space as $\bA$. 
Then, the fundamental parallelepiped of $\bM$ is well-defined and non-degenerate. 
The projection matrix onto the row space of $\bM$ is given by $\bM^\top(\bM\bM^\top)^{-1}\bM$, which retains the same row span. 
Since $\bM$ has rational entries, the projection matrix also consists entirely of rational values. 
Letting $\bU\bU^\top = \bM^\top(\bM\bM^\top)^{-1}\bM$ with $\bU$ orthonormal and spanning the same subspace, and noting that $\bx$ is an integer vector, it follows that the non-zero coordinates of $\bU\bU^\top\bx$ are bounded below by some rational constant. 
Hence, $\bU$ defines a valid fundamental parallelepiped. 
As $\bA$ is merely a rotation of $\bU$, the same holds for $\bA$, and its unit cell is well-defined.

Now consider two vectors $\bv$ and $\bv'$ lying within the same unit cell of $\bA$. 
Let $A^{(i_1)}, \ldots, A^{(i_r)}$ denote $r$ linearly independent columns of $\bA$, and define $\calC$ as the unit cell generated by these columns. 
Any additional columns in $\bA$ can only reduce the cell volume. 
The length of the diagonal of $\calC$ therefore upper bounds the maximum distance between any two points $\bv$ and $\bv'$ in the unit cell, and this length is at most $\sum_{j=1}^r \|A^{(i_j)}\|_2 \le r\sqrt{r}$, since $\bA$ has orthonormal rows.

Thus, for all $\bv$ and $\bv'$ in a unit cell, we may express $\bv' = \bv + \bw$ for some $\bw$ with $\|\bw\|_2 \le r\sqrt{r}$. 
Then,
\begin{align*}
\bv'&(\bA\bS^\top\bS\bA^\top)^{-1}(\bv')^\top = \bv'\bA(\bS^\top\bS)^{-1}\bA^\top(\bv')^\top \\
&= (\bv + \bw)\bA(\bS^\top\bS)^{-1}\bA^\top(\bv + \bw)^\top \\
&= \bv\bA(\bS^\top\bS)^{-1}\bA^\top\bv^\top + 2\bv\bA(\bS^\top\bS)^{-1}\bA^\top\bw^\top + \bw\bA(\bS^\top\bS)^{-1}\bA^\top\bw^\top.
\end{align*}
For $\bx \sim \calD(0, \bS^\top\bS)$, we have $\|\bx\|_2 = \O{n^{6C+1}}$ with high probability, since $\sigma_1(\bS) \le n^{6C}$. 
As $\bv = \bA\bx$ and $\bA$ has orthonormal rows, it follows that $\|\bv\|_2 \le 2n^{6C+1}$ with high probability. 
We also have $\|\bw\|_2 \le r\sqrt{r}$ and $\sigma_n(\bS) \ge n^{5C+3}$. 
Putting this together, we obtain:
\begin{align*}
\left|\bv'(\bA\bS^\top\bS\bA^\top)^{-1}(\bv')^\top - \bv\bA(\bS^\top\bS)^{-1}\bA^\top\bv^\top\right|&\le 6r\sqrt{r}n^{6C+1} \cdot \frac{1}{n^{10C+6}}\\
&\le \frac{1}{n^{4C}},
\end{align*}
for sufficiently large $n$. 
Therefore,
\[\frac{\rho_{\bA^\top}(\bv)}{\rho_{\bA^\top}(\bv')}\in\left[1-\frac{1}{n^{2C}}, 1+\frac{1}{n^{2C}}\right].\]
On the other hand, all points $\bv$ and $\bv'$ in a unit cell have identical density under $p'$. 
Hence, we conclude that
\[\frac{p'(\bv)}{q(\bv)} \in \left[1 - \frac{1}{n^{2C}}, 1 + \frac{1}{n^{2C}}\right]\]
for all $\bv \in \mathbb{R}^r$, and thus,
\[\frac{p(\bv)}{q(\bv)} \in \left[1 - \frac{1}{n^C}, 1 + \frac{1}{n^C}\right].\]
\end{proof}
We introduce the following notion of smoothness to formally characterize functions that, informally speaking, exhibit stability under small perturbations.
\begin{definition}[Smoothness]
Let $\calI$ be any distribution over $\mathbb{Z}^n$. 
A function $f:\mathbb{R}^n \to S$, where $S$ is a finite subset of $\mathbb{Z}$, is said to be $\delta$-smooth with respect to the distribution $\calD(\mu,\bSigma) + \calI$ if
\[\mathbf{\Pr}_{\substack{X \sim \calD(\mu, \bSigma) + \calI,\, Y \sim \calN(\mu, \bSigma) + \calI}} \left[f(X) \ne f(Y)\right] \le \delta.\]
\end{definition}
For instance, consider a function that behaves very differently on integers compared to real numbers, such as $f(x) = 1$ if $x \in \mathbb{Z}$ and $f(x) = 0$ if $x \in \mathbb{R} \setminus \mathbb{Z}$. 
Such a function is not smooth, as even a small perturbation can cause a drastic change in the output. 
These types of functions are not well-suited to our framework, since minor changes within the fundamental parallelepiped of a lattice (defined by the sketch matrix) can yield entirely different outcomes after post-processing.

We are now ready to prove the main result of this section.
\begin{restatable}{theorem}{thmlifting}
\thmlab{thm:lifting}
\cite{GribelyukLWYZ25}
Let $\calI$ be an arbitrary distribution over $\mathbb{Z}^n$, and let $\delta \ge \frac{1}{\poly(n)}$ denote a failure probability. 
Suppose $f$ is a $\frac{\delta}{3}$-smooth function with respect to the distribution $\calD(0, \bS^\top \bS) + \calI$, where $\bS^\top \bS \in \mathbb{R}^{n \times n}$ is a full-rank covariance matrix. 
Assume further that $f$ can be computed via an orthonormal sketch $\bA \in \mathbb{R}^{r \times n}$ for some $r \le n$, followed by an arbitrary post-processing function $g$, such that:
\begin{enumerate}
\item 
There exists a matrix $\bM \in \mathbb{R}^{r \times n}$ with rational entries that spans the same row space as $\bA$.
\item
The post-processed output satisfies $g(\bA \bx) = f(\bx)$ with probability at least $1 - \frac{\delta}{3}$ over $\bx \sim \calD(0, \bS^\top \bS) + \calI$.
\item
The largest successive minimum of the dual lattice satisfies 
\[\lambda_{\max}(\calL^\bot(\bA)) \le \frac{\sigma_n(\bS)}{10C \log n}.\]
\item
The spectrum of $\bS$ is bounded such that $n^{6C} \ge \sigma_1(\bS) \ge \sigma_n(\bS) \ge n^{5C+3}$.
\end{enumerate}
Then, there exists a new sketching matrix $\bA' \in \mathbb{R}^{4r \times n}$ and a post-processing function $h$ such that, with probability at least $1 - \delta$ over $\bx \sim \calN(0, \bSigma) + \calI$, the reconstruction satisfies $h(\bA' \bx) = f(\bx)$.
\end{restatable}
\begin{proof}
Let $\bSigma = \bS^\top \bS$ denote the covariance matrix. 
Define a mapping $\phi: \mathbb{R}^r \to \mathbb{R}^r$ that rounds each vector $\bv \in \mathbb{R}^r$ to the lattice point in the unit cell of $\bv$ induced by the lattice $\calL(\bA)$. 
Given $\bx \sim \calN(0, \bSigma) + \calI$, consider the post-processing function defined by $h(\bA \bx) = g(\phi(\bA \bx))$. 
We aim to show that $h$ correctly recovers $f(\bx)$ with probability at least $1 - \delta$ over the randomness of $\bx \sim \calN(0, \bSigma) + \calI$. 

The final two conditions of the theorem ensure that the assumptions of \lemref{lem:main:disc:to:cont} are satisfied. 
Consequently, for $\bx \sim \calN(0, \bSigma)$, the distribution of $\bA\bx$ is within total variation distance $\frac{1}{\poly(n)}$ of the distribution obtained by drawing $\bx \sim \calD(0, \bSigma)$ and then outputting $\bA \bx + \bfEta$, where $\bfEta$ is sampled uniformly from the fundamental parallelepiped of $\calL(\bA)$. 

Since $\calI$ is supported only on the integer lattice $\mathbb{Z}^n$, the same conclusion holds when sampling $\bx \sim \calN(0, \bSigma) + \calI$: namely, the distribution of $\bA \bx$ is $\frac{1}{\poly(n)}$-close in total variation to the distribution $\overline{\calD}$ obtained by first sampling $\bx \sim \calD(0, \bSigma) + \calI$, then adding a random offset $\bfEta$ from the fundamental parallelepiped of $\calL(\bA)$. 

Now observe that the distribution of $h(\bA\bx)$ for $\bx\sim\overline{\calD}$ is the same as the distribution of $g(\bA\bx)$ under the discrete Gaussian distribution $\bx\sim\calD(0, \bSigma) + \calI$.  
By assumption, $g(\bA\bx)$ correctly outputs $f(\bx)$ with probability at least $1 - \frac{\delta}{3}$ over $\bx \sim \calD(0, \bSigma) + \calI$. 

Furthermore, since $f$ is assumed to be $\frac{\delta}{3}$-smooth with respect to $\calD(0, \bSigma) + \calI$, replacing the discrete Gaussian with the continuous Gaussian (i.e., sampling $\bx \sim \calN(0, \bSigma) + \calI$) changes the output with probability at most $\frac{\delta}{3}$. 

Combining these facts, and accounting for the total variation error of at most $\frac{1}{\poly(n)}$, it follows that $h(\bA\bx) = f(\bx)$ with probability at least
\[1 - \frac{\delta}{3} - \frac{\delta}{3} - \frac{1}{\poly(n)} \ge 1 - \delta,\]
as claimed.
\end{proof}
We also present the following useful statement, which informally states that when the associated lattice has sufficiently small smoothing parameter, the image of a rounded continuous Gaussian is close in distribution to the image of a discrete Gaussian. 
Note that this is precisely why the examples in \figref{fig:disc:gauss:pmf} have large total variation distance. 
\begin{restatable}{lemma}{lemdistroundclose}
\lemlab{thm:dist:round:close}
\cite{GribelyukLWYZ25}
Let $\bS^\top\bS \in \mathbb{R}^{n \times n}$ be a full-rank covariance matrix, and let $\bA \in \mathbb{R}^{r \times n}$ with $r \le n$ satisfy the following:
\begin{enumerate}
\item 
There exists a matrix $\bM \in \mathbb{R}^{r \times n}$ with rational entries that has the same row space as $\bA$.
\item
$\lambda_{\max}(\calL^\bot(\bA)) \le \frac{\sigma_n(\bS)}{10C \log n}$.
\item
$n^{6C} \ge \sigma_1(\bS) \ge \sigma_n(\bS) \ge n^{5C+3}$.
\end{enumerate}
Let $\calD_1$ denote the distribution of $\bA\bx$ where $\bx \sim \calD(0, \bS^\top\bS)$, and let $\calD_2$ denote the distribution of $\by$, where $\by$ is obtained by rounding $\bA\bx$ for $\bx \sim \calN(0, \bS^\top\bS)$ to the closest lattice point within the unit cell of $\bA \cdot \mathbb{Z}^n$.  
Then the total variation distance between $\calD_1$ and $\calD_2$ satisfies
\[\TVD(\calD_1, \calD_2) \le \frac{1}{n^C}.\]
\end{restatable}
\begin{proof}
First, we observe that by the same reasoning used in the proof of \lemref{lem:main:disc:to:cont}, the unit cell of $\bA$ is well-defined due to the first assumption. 
Without loss of generality, we may also assume that $\bA$ has orthonormal rows.

For any vector $\bv$ in the integer row space of $\bA$, let $p(\bv)$ denote the probability mass function of $\calD_1$, and let $q(\bv)$ denote the probability mass function of $\calD_2$. 
Define $\bSigma = \bS^\top \bS$ as the covariance matrix, and let $\phi: \mathbb{R}^r \to \mathbb{R}^r$ be the map sending any vector $\bv$ to the lattice point in the unit cell of $\bv$ induced by $\calL(\bA)$. 

Let $p'$ be the probability density function of $\by$, where $\by \sim \calD_{\bA \mathbb{Z}^n, \bS \bA^\top}$. 
Assuming that 
\[\lambda_{\max}(\calL^\bot(\bA)) \le \frac{\sigma_n(\bS)}{10C \log n},\]
we apply \thmref{thm:disc:tvd} to obtain
\[\frac{p(\bv)}{p'(\bv)} \in \left[1 - \frac{1}{n^{2C}}, 1 + \frac{1}{n^{2C}}\right]\]
for all $\bv \in \mathbb{R}^r$.

Using the rotational invariance property of Gaussian distributions, we have $\bA \bx \sim \calN(0, \bA \bA^\top)$. 
By the same argument as in the proof of \lemref{lem:main:disc:to:cont}, for $\bv = \bA \bx$, it holds with high probability over $\bx \sim \calD(0, \bSigma)$ that
\[\frac{\rho_{\bA^\top}(\bv)}{\rho_{\bA^\top}(\bv')} \in \left[1 - \frac{1}{n^{2C}}, 1 + \frac{1}{n^{2C}}\right]\]
for any two vectors $\bv, \bv'$ in the same unit cell of $\bA$. 

Hence, it follows that
\[\frac{p'(\bv)}{q(\bv)} \in \left[1 - \frac{1}{n^{2C}}, 1 + \frac{1}{n^{2C}}\right]\]
and therefore,
\[\TVD(\calD_1, \calD_2) \le \frac{1}{n^C}.\]
\end{proof}

\subsection{Preliminaries for Adversarial Attack}
Recall that a linear sketch consists of a distribution $\mathcal{M}$ over $r \times n$ matrices, along with an estimator $F: \mathbb{R}^{r \times n} \times \mathbb{R}^n \rightarrow \{0,1\}$, which can be arbitrary. 
A streaming algorithm $\mathcal{A}$ proceeds by first sampling a matrix $\bA \sim \mathcal{M}$, and then for any query vector $\bx \in \mathbb{Z}^n$, it returns $F(\bA, \bA \bx)$ though often $F(\bA\bx)$ suffices. 
We may assume without loss of generality that $\bA$ has full row rank, and we denote the row space of $\bA$ by $R(\bA)$. 
For simplicity, we will identify the matrix $\bA$ with its row space, which forms an $r$-dimensional subspace of $\mathbb{R}^n$. 
As a result, we can represent the sketch as a function $f: \mathbb{Z}^n \rightarrow \{0,1\}$ such that $f(\bx) = f(\bP_{\bA} \bx)$, where $\bP_{\bU}$ is the orthogonal projection operator onto the subspace $\bU$. 
We now describe our attack in the context of the $\GapNorm$ promise problem, which we restate below:

\defgapnorm*

For convenience in the analysis of our discrete attack, we recall the family of distributions $G(\bV^\perp, \sigma^2)$ that was used for the real-valued attack by \cite{HardtW13}.

\defcomplementgaussian*

Now, we define the analogous family of discrete Gaussian distributions from which queries are drawn in our attack for integer sketches. 

\begin{definition}[Distribution $D(\bV^{\perp}, \sigma^2)$]
Let $\bV \subseteq \bA$ be a subspace of dimension $t \le r - 1$, and define $d = r - t$. 
We define the distribution $D(\bV^\perp, \sigma^2)$ as $\calD_{\mathbb{Z}^n}(0, \Sigma_{\sigma^2})$, where $\Sigma_{\sigma^2}$ is the covariance matrix associated with the Gaussian distribution $G(\bV^\perp, \sigma^2)$. 
Specifically, the covariance is given by
\[\Sigma_{\sigma^2} = \frac{3\sigma^2}{4} \bP_{\bV^\perp}^\top \bP_{\bV^\perp} + \frac{\sigma^2}{4} \cdot\mathbb{I}_n.\]
\end{definition}

\subsection{Discrete Conditional Expectation Lemma}
In this section, we establish a discrete analogue of the Conditional Expectation Lemma from \cite{HardtW13}, c.f., \lemref{lem:conditional}, that we showed in \secref{sec:real:cond:exp:lemma}. 
To begin, we recall the definition of correctness to intuitively characterize the expected behavior of an estimator $f$ on inputs sampled from the distribution $D(\bV^\perp, \sigma^2)$ or more generally any distribution $\calD$.

\defcorrectness*

Recall the following definition of soundness:

\defsoundness*

We now restate the Conditional Expectation Lemma from \cite{HardtW13}, which observes that there exist specific directions, say $\bu$, that are more closely aligned with the sketch matrix $\bA$. 
This alignment leads to a slight deviation in the expected value of $\langle \bu, \bg \rangle^2$ for a Gaussian random vector $\bg$, conditioned on the event that $f(\bg) = 1$.

\lemconditional*

Let $\Sigma$ represent the covariance matrix of the distribution family $D(\bV^{\perp}, \sigma^2)$. 
Note that $\Sigma$ implicitly depends on the parameter $\sigma^2$. 
We prove a discrete counterpart to the Conditional Expectation Lemma stated in \lemref{lem:conditional}, where the input $\bx$ is sampled from a discrete Gaussian distribution. 
We highlight that establishing such a result was previously considered quite challenging, and this emphasizes the strength of \lemref{thm:dist:round:close}, as the continuous version in \cite{HardtW13} relied heavily on the rotational symmetry property of continuous Gaussians.

\begin{lemma}[Discrete Conditional Expectation Lemma]
\lemlab{lem:disc_cond:exp}
\cite{GribelyukLWYZ25}
Let $\bA$ be a subspace with dimension $\dim(\bA) = r \le n - d_0$, where $d_0$ is a sufficiently large constant. 
Let $\bV \subseteq \bA$ be a subspace of dimension $t \le r$. 
Suppose that the smallest singular value of $\Sigma$ satisfies 
\[\sigma_n(\Sigma) \ge \lambda_{\max}\left( \calL^{\perp}(\bA)^2 \cdot \frac{\ln(2n(1 + 1/\eps))}{\pi} \right).\]
Suppose $f: \bA \to \{0,1\}$ is a $\left( \frac{1}{10 B^2 d_0^2} + \frac{1}{\poly(n)}, \alpha, B \right)$-correct function on $\bV^\perp$. 

Then there exists a value $\sigma^2 \in [\alpha, \alpha B]$ and a vector $\bu \in \bA \cap \bV^\perp$ such that for $\bx \sim D(\bV^\perp, \sigma^2)$, the following two properties hold, for $d = \max(r - d_0, d_0)$:
\begin{enumerate}
\item 
$\Ex{ \langle \bu, \bx \rangle^2 \,\mid\, f(\bx) = 1 } \ge \Ex{ \langle \bu, \bx \rangle^2 } + \frac{\alpha}{4 B d}$
\item 
$\PPr{ f(\bx) = 1 } \ge \frac{1}{40 B^2 d}$
\end{enumerate}
\end{lemma}
\begin{proof}
We will prove the discrete version of the Conditional Expectation Lemma by contradiction. 
Suppose, by way of contradiction, that there exists a discrete linear sketch $f: \bA \rightarrow \{0,1\}$ such that for all $\sigma^2 \in [\alpha, \alpha B]$, no vector $\bu \in \bA \cap \bV^\perp$ satisfies both properties (1) and (2) when $\bx \sim D(\bV^\perp, \sigma^2)$.

Before proceeding, we define a partition of $\calL(\bA)$ as follows. 
Fix a fundamental parallelepiped $\calF$ within $\calL(\bA) \subseteq \mathbb{R}^r$, having diameter $\poly(r)$. 
This parallelepiped partitions $\mathbb{R}^r$ into cells such that each lattice point $\bz = \bA \by$ for some $\by \in \mathbb{Z}^n$ belongs to one cell. 
Let $\mathcal{P}(\calL(\bA))$ denote this partition of $\calL(\bA)$ into fundamental parallelepipeds.

Define the rounding map $\phi: \mathbb{R}^r \rightarrow \mathbb{R}^r$ which sends any vector $\bv$ to the canonical representative of its cell in the partition. 
Let $p$ be the probability mass function of $\phi(\bA \bg)$ for $\bg \sim G(\bV^\perp, \sigma^2)$, and let $q$ be the probability mass function of $\bA \by$ for $\by \sim D(\bV^\perp, \sigma^2)$. 

Using the assumption 
\[\sigma_n(\Sigma) \ge \lambda_{\max} \left( \calL^\perp(\bA)^2 \cdot \frac{\ln(2n(1+1/\eps))}{\pi} \right),\]
\lemref{thm:dist:round:close} implies that the total variation distance between $\phi(\bA \bg)$ and $\bA \by$ is at most $\frac{1}{\poly(n)}$.

Suppose $f: \bA \to \{0,1\}$ is a $\left( \frac{1}{10(d_0B)^2}, \alpha, B \right)$-correct discrete sketch. 
We define a continuous sketch $f': \bA \to \{0,1\}$ as follows: given a query $\bg \sim G(\bV^\perp, \sigma^2)$, the sketch observes $\bA\bg$, applies $\phi$ to obtain the corresponding lattice point $\by$, and returns $f'(\bg) = f(\by)$. 

Because the distribution $\bA \bg$ for $\bg \sim G(\bV^\perp, \sigma^2)$ is within $\frac{1}{\poly(n)}$ total variation distance of the distribution $\bA \by + \bfEta$ for $\by \sim D(\bV^\perp, \sigma^2)$, and any sketch can round $\bA \by + \bfEta$ to recover $\bA \by$, it follows that $f'$ is a $\left( \frac{2}{10(d_0B)^2} + \frac{1}{\poly(n)}, \alpha, B \right)$-correct sketch over $G(\bV^\perp, \sigma^2)$.

By applying \lemref{lem:conditional}, there exists some direction $\bu \in \bA \cap \bV^\perp$ such that the Conditional Expectation Lemma applies to $\bg \sim G(\bV^\perp, \sigma^2)$ for some $\sigma^2 \in [\alpha, \alpha B]$.

Note that conditioning on $\calE_1 = \{ f'(\bg) = 1 \}$ or $\calE_2 = \{ f(\by) = 1 \}$ affects the distribution of $\langle \bu, \bg \rangle^2$ by at most an additive $\frac{1}{\poly(n)}$. 
Further, since $\mathbb{E}[\langle \bu, \bg \rangle^2] \le \sigma^2$ and $\mathbb{E}[\langle \bu, \by \rangle^2] \le \sigma^2$, conditioning on $\| \bg \|_2^2 \le \sigma^2 \log(n)$ ensures that the shift in the expectation between $\langle \bu, \by \rangle^2$ and $\langle \bu, \bg \rangle^2$ under conditioning is bounded by $\frac{\sigma^2}{\poly(n)}$.

Hence, we conclude:
\[\mathbb{E}\left[ \langle \bu, \bg \rangle^2 \mid f'(\bg) = 1 \right] - \mathbb{E}\left[ \langle \bu, \bg \rangle^2 \right] < \frac{\alpha}{10Bd} + \frac{\sigma^2}{\poly(n)} + \frac{1}{\poly(n)} \le \frac{\alpha}{4Bd}.\]
However, since $f'$ is a $\left( \frac{2}{10(d_0B)^2} + \frac{1}{\poly(n)}, \alpha, B \right)$-accurate sketch for $G(\bV^\perp, \sigma^2)$, this contradicts the Conditional Expectation Lemma from \cite{HardtW13} in \lemref{lem:conditional}. 
\end{proof}

\subsection{Algorithm for Adversarial Attack}
In this section, we present an adversarial attack for the $\GapNorm$ promise problem, succinctly described in \figref{fig:attack}. 
\begin{figure*}[!htb]
\begin{mdframed}
\textbf{Input:} Access to an oracle $\calA$ computing $f:\mathbb{R}^n \to \{0,1\}$, parameter $B \ge 4$, and a sufficiently large $\alpha = \poly(n)$ such that $\alpha \ge \ell_{\bA}^2 \cdot \frac{\ln(2n(1+1/\eps))}{\pi}$ (ensured via preprocessing for all integer matrices $\bA$ with $\poly(n)$-bounded entries).

\textbf{Attack Initialization:}  
Set $\bV_0 = \emptyset$, let $m = \widetilde{O}(B^{13}n^{11}\log^{15} n)$, and define a discretized set of variances $S = [\alpha, \alpha B] \cap \zeta \mathbb{Z}$ where $\zeta = \frac{1}{20(Bn)^2 \log(Bn)}$.

\textbf{Iterative Procedure (for $t = 1$ to $r+1$):}
\begin{enumerate}
\item
For each $\sigma^2 \in S$:
\begin{enumerate}
\item 
Sample $m$ points $\bx_1,\ldots,\bx_m \sim D(\bV^\perp, \sigma^2)$ and query the oracle to obtain labels $a_i = \calA(\bx_i)$.
\item 
Let $s(t, \sigma^2) = \frac{1}{m} \sum_{i=1}^m a_i$ be the fraction of samples that are positively labeled.
\begin{enumerate}
\item 
If either (1) $\sigma^2 \ge \alpha B / 2$ and $s(t, \sigma^2) \le 1 - \zeta$, or (2) $\sigma^2 \le 2\alpha$ and $s(t, \sigma^2) \ge \zeta$, then \textbf{terminate} and \textbf{return} $(\bV_t^\perp, \sigma^2)$.
\item 
Otherwise, collect the positively labeled points $\bx'_1, \ldots, \bx'_{m'}$.
\end{enumerate}
\item 
If $m' < \frac{m}{100B^2n}$, increment $\sigma^2$; else, compute the direction
\[\bv_\sigma = \arg\max_{\bv \in \mathbb{R}^n} \left( \frac{1}{m'} \sum_{i=1}^{m'} \langle \bv, \bx'_i \rangle^2 \right).\]
\end{enumerate}
\item 
Let $\bv'$ be the first $\bv_\sigma$ such that 
\[\frac{1}{m'} \sum_{i=1}^{m'} \langle \bv_\sigma, \bx'_i \rangle^2 \ge \sigma^2 + \frac{\sigma^2}{4} + \frac{1}{14Br}.\]
\begin{enumerate}
\item 
If no such $\bv'$ is found, set $\bV_{t+1} = \bV_t$ and continue.
\item 
Else, let $\bv^* = \bv'$ and define $\bv_t$ as the projection of $\bv^*$ orthogonal to $\bV_t$. 
Set $\bV_{t+1} = \bV_t \cup \{ \bv_t \}$.
\end{enumerate}
\end{enumerate}
\end{mdframed}
\caption{Algorithm that generates an adaptive attack via a turnstile data stream.}
\figlab{fig:attack}
\end{figure*}

Ultimately, the goal of the attack is to find the following notion of a failure certificate, which slightly differs from \defref{def:failure:real} due to the input having support over a discrete distribution rather than a continuous distribution. 

\begin{definition}[Failure certificate]
Let $B \ge 8$, $\sigma^2 \in [\alpha, 2\alpha B]$, and consider a function $f: \mathbb{R}^n \to \{0,1\}$.  
We define a pair $(\bV, \sigma^2)$ to be a \emph{$d$-dimensional failure certificate} for $f$ if $\bV \subseteq \mathbb{R}^n$ is a subspace of dimension $d$, and there exists a constant $C > 0$ such that $n \ge d + 10C \log(Bn)$ and one of the following holds:
\begin{itemize}
\item 
If $\sigma^2 \in [\alpha B/2, 50\alpha B]$, then:
\[\mathbf{\Pr}_{\bg \sim D(\bV^\perp, \sigma^2)}[f(\bg) = 1] \le 1 - (Bn)^{-C},\]
\item 
Or, if $\sigma^2 \in [\alpha, 2\alpha]$, then:
\[\mathbf{\Pr}_{\bg \sim D(\bV^\perp, \sigma^2)}[f(\bg) = 1] \ge n^{-C}.\]
\end{itemize}
\end{definition}
For the purposes of our attack, we will consider subspaces $\bV$ spanned by vectors with bounded precision, i.e., appropriately scaled integer vectors. 
To that end, we begin by demonstrating that a failure certificate for $f$ can be used to construct a discrete input vector $\bx$ on which $f$ makes an error. 
This mirrors Fact 5.2 from \cite{HardtW13} in \factref{fact:real:failure:cert:attack}, which establishes a similar result but only for real-valued vectors $\bx$. 
\begin{theorem}
\cite{GribelyukLWYZ25}
Let $\eta = \frac{1}{\poly(n)}$ be a fixed small parameter, and suppose $\alpha = \poly(n)$ is sufficiently large. 
Then, given a $d$-dimensional failure certificate for $f$, there exists an algorithm that makes $\poly(Bn)$ non-adaptive queries and, with probability at least $\frac{2}{3}$, returns a vector $\bx \in (\eta \cdot \mathbb{Z})^n$ such that either $\|\bx\|_2 > \frac{\alpha B(n - d)}{3}$ and $f(\bx) = 0$, or $\|\bx\|_2 < 3\alpha(n - d)$ and $f(\bx) = 1$.
\end{theorem}
\begin{proof}
Suppose we draw a set $X$ of $\O{(\alpha Bn)^C}$ independent samples from the distribution $D(\bV^\perp, \sigma^2)$ for some $\sigma^2 \in [\alpha, 2\alpha]$. 
Since $n - d \ge d$, standard sub-Gaussian tail bounds combined with a union bound imply that, with high probability, all vectors $\bx \in X$ satisfy $\|\bx\|_2^2 < 3\alpha(n - d)$. 
Moreover, since $D(\bV^\perp, \sigma^2)$ is a failure certificate, with high probability there exists a sample $\bx \in X$ such that $f(\bA \bx) = 1$.

The case when $\sigma^2 \ge \frac{B\alpha}{2}$ is analogous. 
Suppose again we sample a set $X$ of $\O{(\alpha Bn)^C}$ points from $D(\bV^\perp, \sigma^2)$ for $\sigma^2 \ge \frac{B\alpha}{2}$. 
Since $n - d \ge d$, a similar concentration argument shows that with high probability, every $\bx \in X$ satisfies $\|\bx\|_2^2 > \frac{\alpha B(n - d)}{3}$. 
Furthermore, since $D(\bV^\perp, \sigma^2)$ is a failure certificate, with high probability there exists $\bx \in X$ such that $f(\bA \bx) = 0$.
\end{proof}

\FloatBarrier 

We now state the guarantees of the adaptive attack described in \figref{fig:attack}, demonstrating that it successfully identifies a failure certificate. 
Since the proof relies on several structural lemmas that will be established in the following sections, we postpone the proof of \thmref{thm:adaptive-attack} until \secref{sec:attack:final}.

\begin{restatable}[Adaptive attack against integer linear sketches]{theorem}{thmadaptiveattack}
\thmlab{thm:adaptive-attack}
\cite{GribelyukLWYZ25}
Suppose that $8 \le B \le \poly(n)$ and $\alpha \ge 4\ell_{\bA}$, where $\lambda_{\max}(\calL^{\perp}(\bA)) \le \ell_{\bA} = \sqrt{nM^2}$. 
Let $\bA \subseteq \mathbb{R}^n$ be an $r$-dimensional subspace with $n \ge \O{r} + 90\log(Bn)$, and suppose $f: \bA \to \{0,1\}$ is a linear sketch defined over all $\bx \in \mathbb{Z}^n$. 
Then there exists an attack algorithm that, using only oracle access to $f$, identifies a failure certificate for $f$ with probability at least $\frac{9}{10}$. 
The algorithm's runtime and number of queries are bounded by $\poly(r \log n)$. 
Moreover, each query made by the algorithm is drawn from a distribution $D(\bV^\perp, \sigma^2)$ for some subspace $\bV \subseteq \mathbb{R}^n$ and variance $\sigma^2 \in [\alpha, B\alpha]$.
\end{restatable}

\subsection{Distance Between Subspaces}
In this section, we provide a brief overview of subspace distance, which will be used to quantify how close the subspace generated by the attack's collected vectors is to the nearest subspace contained in the sketch matrix $\bA$, as discussed in \secref{sec:int:progress}. 
This notion is key to establishing an invariant that remains valid throughout the execution of the attack. 
We recall the formal definition of the distance between subspaces:

\defsubspacedist*

Next, we present a structural result that upper bounds the total variation distance between two discrete subspace Gaussian distributions in terms of the distance between their corresponding orthogonal subspaces. 
This result serves as a discrete analogue of Lemma 4.14 from~\cite{HardtW13} in \factref{fact:shifted-gaussians}.

\begin{lemma}
\lemlab{lem:tvd:subspaces}
\cite{GribelyukLWYZ25}
For every $\sigma^2\in(\alpha,B \alpha]$, we have
\[\TVD(D(\bV^\perp,\sigma^2),D(\bW^\perp,\sigma^2))\le40\sqrt{ Bn\log (Bn)}\cdot d(\bV,\bW)+\frac{1}{(Bn)^4}.\]
\end{lemma}
\begin{proof}
Let $\bx_1 \sim D(\bV^\perp, \sigma^2)$ and $\bx_2 \sim D(\bW^{\perp}, \sigma^2)$ be independently sampled. 
By choosing $\bA$ as the identity in \lemref{thm:dist:round:close}, we establish that $D(\bV^\perp, \sigma^2)$ is point-wise close to the rounded version of $G(\bV^\perp, \sigma^2)$, differing by at most a multiplicative factor of $1 \pm \frac{1}{\poly(n)}$, assuming sufficiently large $\alpha$. 
Applying the same reasoning to $D(\bW^\perp, \sigma^2)$ and $G(\bW^\perp, \sigma^2)$, we obtain:
\[\TVD(D(\bV^\perp, \sigma^2), D(\bW^\perp, \sigma^2)) \le \TVD(G(\bV^\perp, \sigma^2), G(\bW^\perp, \sigma^2)) + \frac{1}{\poly(n)}.\]
From Lemma 4.14 in~\cite{HardtW13}, c.f., \factref{fact:shifted-gaussians}, for $\sigma^2$ in the interval $[\alpha, \alpha B]$, we have:
\[\TVD(G(\bV^\perp, \sigma^2), G(\bW^\perp, \sigma^2)) \le 20 \sqrt{Bn \log(Bn)} \cdot d(\bV,\bW) + \frac{1}{(Bn)^5}.\]
Combining these using the triangle inequality, the claimed bound follows, up to an additive loss of $\frac{1}{\poly(n)}$ in total variation distance.
\end{proof}

\subsection{Progress Lemma}
\seclab{sec:int:progress}
We now demonstrate that each iteration of the adaptive attack advances toward finding a subspace that closely approximates the row space of the sketch matrix $\bA$. 
For each $1 \le t \le r$, define $\bW_t \subseteq \bA$ as the $(t-1)$-dimensional subspace of $\bA$ that is nearest to $V_t$. 
Formally, we implicitly define $W_t$, so that:
\begin{align*}
d(\bV_t, \bW_t) = \min\{d(\bV_t, \bW): \dim(\bW) = t-1,\, \bW \subseteq \bA \},
\end{align*}
where we slightly abuse notation so that $\bV_t$ refers to the subspace spanned by the vectors in $\bV_t$, i.e., $\Span(\bV_t)$. 
To establish the correctness of the attack, we show that the algorithm preserves the following invariant with high probability throughout its execution:

\begin{invariant}
\invarlab{invar:step}
For each step $t\in[r+1]$, we have $\dim(\bV_t)=t-1$ and $d(\bV_t,\bW_t)\le\frac{t}{40(Bn)^{3.5}\log^{2.5}(Bn)}$.
\end{invariant}

Observe that if \invarref{invar:step} holds at step $t$, then we can upper bound the total variation distance between the two subspace Gaussians as follows: 
\begin{lemma}
\lemlab{lem:invar:tvd}
\cite{GribelyukLWYZ25}
If \invarref{invar:step} holds at step $t$, then we have 
\[\TVD(D(\bV_t^\perp,\sigma^2),D(\bW_t^\perp,\sigma^2))\le\frac{1}{B^3n^2\log^2(Bn)}.\]
\end{lemma}
\begin{proof}
Suppose \invarref{invar:step} holds at step $t$. 
Then it follows that
\[d(\bV_t,\bW_t)\le\frac{t}{40(Bn)^{3.5}\log^{2.5}(\alpha Bn)}\le\frac{1}{40 B^{3.5}n^{2.5}\log^{2.5}(Bn)}.\]
Therefore, by \lemref{lem:tvd:subspaces},  we have:
\[\TVD(D(\bV_t^\perp,\sigma^2),D(\bW_t^\perp,\sigma^2))\le\frac{1}{B^3n^2\log^2(Bn)}.\]
\end{proof}
Next, we show that given \invarref{invar:step}, then correctness on the subspace $\bV_t^\perp$ implies correctness on the subspace $\bW_t^\perp$ with some small loss in the probability, similar to Lemma 5.4 in \cite{HardtW13}, c.f., \lemref{lem:correct}, which gave a similar statement for inputs that are generated from a continuous Gaussian distribution. 
\begin{lemma}
\lemlab{lem:correct:before:after}
\cite{GribelyukLWYZ25}
Suppose \invarref{invar:step} holds at step $t$ and let $\zeta=\frac{1}{20(Bn)^2\log(Bn)}$. 
Suppose $f$ is $(\alpha,B)$-correct on $\bV_t^\perp$. 
Then $f$ is $(\alpha+\zeta,B)$-correct on $\bW_t^\perp$. 
\end{lemma}
\begin{proof}
Assuming that the invariant in \invarref{invar:step} holds at step $t$, \lemref{lem:invar:tvd} implies that 
\[\TVD(D(\bV_t^\perp, \sigma^2), D(\bW_t^\perp, \sigma^2)) \le \frac{1}{B^3 n^2 \log^2(Bn)}.\]
Given that $\zeta = \frac{1}{20(Bn)^2 \log(Bn)}$, this bound ensures that the total variation distance is at most $\zeta$. 
Consequently, if $f$ is $(\alpha, B)$-correct on $\bV_t^\perp$, then it is also $(\alpha + \zeta, B)$-correct on $\bW_t^\perp$, since the difference in behavior between the two distributions contributes an additive error of at most $\zeta$ to the failure probability.
\end{proof}

We now show that in each iteration of the attack, one of two outcomes occurs: (1) either the algorithm halts and returns a failure certificate, or (2) assuming the invariant continues to hold in the following round, then the function $f$ remains correct on the corresponding orthogonal subspace. 
This result mirrors Lemma 5.6 in~\cite{HardtW13} in \lemref{lem:empirical}. 

Note that we have not yet established that the invariant holds in the next round; this will be addressed later in the ``Progress Lemma'' (see \lemref{lem:progress:lem}).

\begin{lemma}
\lemlab{lem:end:or:invar}
\cite{GribelyukLWYZ25}
Assuming the event $\calE$ from \lemref{lem:event:whp} holds, then one of the following must be true for any round $t$:
\begin{itemize}
\item 
The algorithm halts during round $t$ and returns a failure certificate $D(\bV_t^\perp,\sigma^2)$ for the function $f$, or
\item 
If the algorithm continues past round $t$ and the invariant holds at that step, then $f$ is $(\alpha, B)$-correct on the subspace $\bW_t^\perp$.
\end{itemize}
\end{lemma}
\begin{proof}
The first part follows immediately from the definition of a failure certificate and the assumption that the empirical estimate $s(t,p)$ is within $\zeta$ of the true error rate.

On the other hand, if the algorithm does not halt in round $t$, it must be because $f$ is $(2\zeta, B)$-correct on $\bV_t^\perp$. 
Then, applying \lemref{lem:correct:before:after} and assuming event $\calE$ holds, we conclude that $f$ is $(3\zeta, B)$-correct on $\bW_t^\perp$. 
Given that $\zeta = \frac{1}{20(Bn)^2 \log(Bn)}$, we have $3\zeta \le \frac{1}{10(Bn)^2}$ for sufficiently large $n$, which ensures that $f$ satisfies $(\alpha, B)$-correctness on $W_t^\perp$, establishing the second part of the claim.
\end{proof}
Now, we prove that if the attack continues until round $r+1$, then $f$ cannot be correct on the orthogonal subspace afterwards, similar to Lemma 5.8 in \cite{HardtW13}, c.f., \lemref{lem:final}. 
\begin{lemma}
\lemlab{lem:invar:end:wrong}
\cite{GribelyukLWYZ25}
Suppose \invarref{invar:step} holds at step $t=r+1$. 
Then $f$ is not $(\alpha, B)$-correct on $\bW_{r+1}$.
\end{lemma}
\begin{proof}
Observe that both $\bV_{r+1}$ and $\bW_{r+1}$ have dimension $r$. 
Since $\bW_{r+1}$ is a subspace of $\bA$ and $\bA$ itself has dimension $r$, it follows that $\bW_{r+1} = \bA$. 
Furthermore, $f$ cannot distinguish between samples $\bg$ drawn from $D(\bW_{r+1}^\perp, 2\alpha)$ and those from $D(\bW_{r+1}^\perp, B\alpha)$. 
As a result, $f$ must be incorrect with constant probability on inputs $\bg$ drawn from at least one of these distributions.
\end{proof}

Next, we show that the Conditional Expectation Lemma still holds for a discretization of the variance in the set $S$ of the attack, analogous to Lemma 5.9 in \cite{HardtW13}, c.f., \lemref{lem:sigmatilde}. 
We require such a statement to ultimately prove the progress lemma in \lemref{lem:progress:lem},
\begin{lemma}
\lemlab{lem:exist:var}
\cite{GribelyukLWYZ25}
Suppose that $f$ is $(\alpha, B)$-correct on the subspace $\bW_t^\perp$.  
Then there exists a variance parameter $\widetilde{\sigma}^2 \in S$, a gap $\Delta \ge \frac{\alpha}{7Br}$, and a vector $\bu \in \bV_t^\perp \cap \bA$ such that, for a sample $\bg \sim D(\bV_t^\perp, \widetilde{\sigma}^2)$, the function satisfies $\PPr{f(\bg) = 1} \ge \frac{1}{60B^2 r}$ and
\[\Ex{\langle \bu, \bg \rangle^2 \mid f(\bg) = 1} \ge \Ex{\langle \bu, \bg \rangle^2} + \Delta.\]
\end{lemma}
\begin{proof}
Since $f$ is correct on $\bW_t^\perp$, by the (continuous) Conditional Expectation Lemma in \lemref{lem:conditional}, there exists $\bu \in \bU = \bW_t^\perp \cap \bA$ and $\sigma \in [\alpha, \alpha B]$ such that
\[\Ex{\langle \bu, \bg \rangle^2 \mid f(\bg) = 1} \ge \Ex{\langle \bu, \bg \rangle^2} + \frac{\alpha}{4Br}\]
and
\[\PPr{f(\bg) = 1} \ge \frac{1}{40 B^2 r}.\]
Since \invarref{invar:step} holds at step $t$, by \lemref{lem:invar:tvd} we have
\[\TVD(D(\bV_t^\perp, \sigma^2), D(\bW_t^\perp, \sigma^2)) \le \frac{1}{B^3 n^2 \log^2(Bn)}.\]
Hence,
\[\PPr{f(\bg) = 1} \ge \frac{1}{50 B^2 r}.\]
Moreover, conditioning on $f(\bg) = 1$ multiplicatively increases the total variation distance by at most a factor of $50B^2 r$. 
For any $F: \mathbb{R}^n \to [0,M]$, it follows that
\begin{align*}
\big| \EEx{\bg \sim D(\bW_t^\perp, \sigma^2)}{F(\bg) \mid f(\bg) = 1} &- \EEx{\bg \sim D(\bV_t^\perp, \sigma^2)}{F(\bg) \mid f(\bg) = 1} \big| \\
&\le \frac{M (50 B^2 r)}{B^3 n^2 \log^2(Bn)}.
\end{align*}
By sub-Gaussian concentration, for any $\beta > 0$,
\[\PPr{\langle \bu, \bg \rangle^2 > 10 \beta C} \le \exp(-\beta),\]
so truncating $\langle \bu, \bg \rangle^2$ at $M = 10 B \log(rB)$ only changes the expectations by $o\left(\frac{\alpha}{Br}\right)$. 
Thus,
\[\left| \EEx{\bg \sim D(\bW_t^\perp, \sigma^2)}{F(\bg) \mid f(\bg) = 1} - \EEx{\bg \sim D(\bV_t^\perp, \sigma^2)}{F(\bg) \mid f(\bg) = 1} \right| \le o\left(\frac{1}{Br}\right).\]
This also holds when rounding $\sigma^2$ to some $(\widetilde{\sigma})^2 \in S = [\alpha, B\alpha] \cap \zeta \mathbb{Z}$ with $\zeta = \frac{1}{20 (Bn)^2 \log(Bn)}$.

Finally, since $\bu \in \bW_t^\perp \cap \bA$ and the invariant holds for $(\bV_t,\bW_t)$, then $\|\bP_{\bV_t} \bu\|_2 \le \frac{1}{B^2 n^2}$. 
Hence, there exists $\bu \in \bV_t^\perp \cap \bA$ such that
\begin{align*}
\Ex{\langle \bu, \bg \rangle^2 \mid f(\bg) = 1} 
&\ge \Ex{\langle \bu, \bg \rangle^2} + \frac{\alpha}{4Br} - o\left(\frac{1}{Br}\right) - \frac{1}{B^2 n^2} \\
&\ge \Ex{\langle \bu, \bg \rangle^2} + \Delta,
\end{align*}
for any $\Delta \ge \frac{\alpha}{7Br}$, as claimed.
\end{proof}

We now provide a lower bound on the norm of the output vector $\bv^*$ after projecting onto $\bV_t^\perp \cap \bA$, adapting Lemma 5.10 from \cite{HardtW13} to accommodate discrete Gaussians in the computation of the maximizer $\bv^*$.  
In the proof of \lemref{lem:found:vec}, we rely on the supporting result \lemref{lem:top:vector}, which establishes that the vector identified in each round has a substantial projection onto the portion of the rowspan of $\bA$ orthogonal to previously found vectors.  
The proof of this supporting lemma is deferred to \secref{sec:attack:top:singular}.
\begin{lemma}
\lemlab{lem:found:vec}
\cite{GribelyukLWYZ25}
Let $N>0$ be sufficiently large and assume that 
\[\sigma_n(\Sigma_t) > \lambda_{\max}(\calL^{\perp}(\bA))^2 \cdot \frac{\ln(2n(1+1/\eps))}{\pi}.\]
Then, with probability at least $1 - \frac{1}{\poly(n)}$, the vector $\bv^*$ identified at step $t$ satisfies
\[\|\bP_{\bV_t^\perp \cap \bA} \bv^*\|_2^2 \ge 1 - \frac{1}{200 (Bn)^{3.5} \log^4(Bn)}.\]
\end{lemma}
\begin{proof}
Consider the variance parameter $\widetilde{\sigma}^2$ in \lemref{lem:exist:var}.  
For a random vector $\bg$ drawn from the distribution $D(\bV_t^\perp, \widetilde{\sigma}^2)$, \lemref{lem:exist:var} guarantees that 
\[\PPr{f(\bg) = 1} \geq \frac{1}{60 B^2 r}.\]
Now, focus on the distribution of $\bg \sim D(\bV_t^\perp, \widetilde{\sigma}^2)$ conditioned on the event $f(\bg) = 1$.  
Note that the samples $\bg'_1,\ldots,\bg'_{m'}$ are independently drawn from this conditional distribution.  

Since the algorithm $\calA$ outputs $1$ on each sample with probability at least $\frac{1}{70 B^2 r}$, and given parameters $\Delta = \O{\frac{1}{B r}}$, $\xi^2 = \O{B^2 \log^2 n}$,  and $\gamma = \O{\frac{1}{(B r)^{3.5} \log^4(B n)}}$, we have that the number of samples satisfies
\[m' \geq \frac{m}{700 B^2 r} = \Omega\left(B^{11} n^{10} \log^{15}(r)\right),\]
where these samples come from $D(\bV_t^\perp, \widetilde{\sigma}^2)$ conditioned on $f(\trunc_\eta(\bg)) = 1$.

Define the vector $\bv^*$ as the first vector $\bv_\sigma = \arg\max_{\bv \in \mathbb{R}^n} z(\bv)$ such that 
\[z(\bv) \geq \sigma^2 + \frac{\sigma^2}{4} + \frac{1}{14 B r},\]
where 
\[z(\bv) = \frac{1}{m'} \sum_{i=1}^{m'} \langle \bv, \bg'_i \rangle^2.\]
We show that the conditions required in \lemref{lem:top:vector} hold, allowing us to conclude a large portion of the mass of the projection of $\bv^*$ onto $\bV_t^\perp$, using the choice 
\[\gamma = \frac{1}{(B n)^{3.5} \log^4(B n)}.\]
Specifically, we set 
\[\bV = \bV_t^\perp \cap \bA, \quad \bW = \bV_t + \bA^\perp, \quad \tau = \widetilde{\sigma}^2 + \frac{\widetilde{\sigma}^2}{4} + \frac{1}{\poly(n)},\]
and let $\Delta$ be the parameter from \lemref{lem:exist:var}.
\begin{itemize}
\item 
We begin by verifying the second condition.  
Take any unit vector $\bw \in W$ and decompose it as $\bw = \alpha \bw_1 + \beta \bw_2$, where $\bw_1 \in V_t$ and $\bw_2 \in \bA^\perp$ are orthogonal unit vectors.  
Since $\sigma_n(\Sigma_t) > \lambda_{\max}(\calL^\perp(\bA)) \cdot \sqrt{\frac{\ln(2n(1+1/\eps))}{\pi}}$, the distribution of $\bg$ can be approximated (up to an additive $\frac{1}{\poly(n)}$ variation distance) by a product of independent continuous Gaussians $\bg_1$ in $\bA$ and $\bg_2$ in $\bA^\perp$, followed by rounding, as argued similarly in \lemref{thm:dist:round:close}.  
Therefore, we get 
\[\Ex{\langle \bw_1, \bg \rangle \langle \bw_2, \bg \rangle} \leq \frac{1}{\poly(n)}.\]
Also, with high probability, 
\[|\langle \bw_1, \bg \rangle| \leq \O{\sigma \log n}, \quad \text{and} \quad |\langle \bw_2, \bg \rangle| \leq \O{\log n}.\]
Hence, conditioning on $f(\bg) = 1$, we have
\[\EEx{\bg \sim D(\bV_t^\perp, \widetilde{\sigma}^2)}{\langle \bw_1, \bg \rangle \langle \bw_2, \bg \rangle \mid f(\bg) = 1} \leq \frac{1}{\poly(n)} + \O{\sigma \log^2 n}.\]

\item 
Next, we verify the first condition.  
For any fixed unit vector $\bw \in \bW$, again write $\bw = \alpha \bw_1 + \beta \bw_2$ with $\bw_1 \in \bV_t$ and $\bw_2 \in \bA^\perp$ orthogonal unit vectors.  
Our goal is to upper bound 
\[\EEx{\bg}{\langle \bw_2, \bg \rangle^2 \mid P_{\bA} \bg = \bz}\]
for a fixed vector $\bz$ in the range of $\bA$. 
Consider another vector $\bz'$ in the range of $\bA$ such that 
\[|\|\bz\|_2^2 - \|\bz'\|_2^2| \leq \mathcal{O}(\sigma \log n).\]
The ratio of the probability densities at $\bz$ and $\bz'$ is 
\[\exp\big((\bz + \bq)^\top \bSigma^{-1} (\bz + \bq) - (\bz' + \bq)^\top \bSigma^{-1} (\bz' + \bq)\big)\]
for all $\bq$ orthogonal to $\bA$.  
Since 
\[|\|\bz\|_2^2 - \|\bz'\|_2^2| \leq \O{\sigma \log n}\]
and $\sigma_{\min}(\bSigma^{-1}) \geq \frac{\sigma^2}{\poly(n)} \gg \sigma \log n$, the above ratio is at most $1 + \frac{1}{\poly(n)}$.  
Furthermore,
\[\langle \bw, \bq + \bz \rangle^2 - \langle \bw, \bq + \bz' \rangle^2 \leq \O{\sigma \log n}\]
since $\bq$ is orthogonal to $\bw$ and has variance bounded by $\O{1}$.  
Using sub-Gaussian tail bounds, it follows that
\[\EEx{\bg}{\langle \bw_2, \bg \rangle^2 \mid P_{\bA} \bg = \bz} - \EEx{\bg}{\langle \bw_2, \bg \rangle^2 \mid P_{\bA} \bg = \bz'} \leq \O{\sigma \log n}.\]
Therefore, by averaging over $\bz$ with $f(\bz) = 1$, we get
\begin{align*}
\underset{\bg}{\mathbb{E}}[\langle \bw_2, &\bg \rangle^2 \mid f(\bg) = 1] \\
&= \sum_{\bz: f(\bz) = 1} \EEx{\bg}{\langle \bw_2, \bg \rangle^2 \mid P_{\bA} \bg = \bz} \frac{\PPr{P_{\bA} \bg = \bz}}{\PPr{f(\bz) = 1}} \\
&\leq \O{\sigma \log n} + \sum_{\bz: f(\bz) = 1} \EEx{\bg}{\langle \bw_2, \bg \rangle^2} \frac{\PPr{P_{\bA} \bg = \bz}}{\PPr{f(\bz) = 1}} \\
&\leq \O{\sigma \log n} + \EEx{\bg}{\langle \bw_2, \bg \rangle^2}.
\end{align*}
On the other hand, we have $\EEx{\bg \sim G(\bV_t^\perp, \sigma^2)}{\langle \bw_2, \bg \rangle^2} \leq \sigma^2$. 
Moreover, $\langle \bw_2, \bg \rangle^2$ is a continuous function and the difference is small between the probability density function and probability mass function for continuous and our discrete Gaussian distribution, particularly for the granularity $\eta=\frac{1}{\poly(n)}$ of the support of the latter. 
Thus, we have \[\EEx{\bg}{\langle \bw_2, \bg \rangle^2} \leq n + \sigma^2,\]
Similarly, 
\[\EEx{\bg}{\langle \bw_1, \bg \rangle^2} \ll \mathcal{O}(n),\]
as $\bw_1$ lies in $\bV_t$. 
Combining this with the earlier bound, we have
\[\Ex{\langle \bw_1, \bg \rangle \langle \bw_2, \bg \rangle \mid f(\bg) = 1} \leq \frac{1}{\poly(n)} + \O{\sigma \log^2 n},\]
and we conclude
\[\EEx{\bg \sim D(\bV_t^\perp, \widetilde{\sigma}^2), f(\bg)=1}{\langle \bw, \bg \rangle^2} \leq \tau,\]
since $\tau = \widetilde{\sigma}^2 + \frac{\widetilde{\sigma}^2}{4} + \frac{1}{\poly(n)}$ and $\widetilde{\sigma}^2\ge\alpha$ and $\alpha$ is a sufficiently large polynomial in $n$. 

\item 
For the third condition, \lemref{lem:exist:var} guarantees the existence of a vector $\bv \in V$ satisfying
\[\EEx{\bg \sim D(\bV_t^\perp, \widetilde{\sigma}^2), f(\bg)=1}{\langle \bv, \bg \rangle^2} \geq \tau + \frac{\Delta}{2},\]
where $\Delta \geq \frac{1}{7 B r}$.

\item 
To verify the fourth condition, note that for every unit vector $\bu \in \mathbb{R}^n$, the conditional variance satisfies
\[\xi^2 = \Var(\langle \bu, \bg \rangle^2) \leq \O{B^2 \log^2 n},\]
which follows from standard sub-Gaussian concentration inequalities.

\item 
Finally, we confirm that the number of samples $m'$ is sufficiently large.  
Recall that
\[m' = \Omega\left(B^{11} n^{10} \log^{15}(r)\right) = \Omega\left(\frac{n \log^2 n \, \xi^2}{\gamma^2 \Delta^2}\right),\]
for parameters 
\[\Delta = \Omega\left(\frac{1}{B r}\right), \quad \gamma = \Omega\left(\frac{1}{(B r)^{3.5} \log^4(B n)}\right), \quad \xi^2 = \O{B^2 \log^2 n}.\]
\end{itemize}
Hence, by applying \lemref{lem:top:vector}, with probability at least $1-\exp(-n)$, we have both 
\[z(\bv) \geq \sigma^2 + \frac{\sigma^2}{14 B r}\]
and 
\[\| \bP_{\bV_t^\perp \cap \bA} \bv^* \|_2^2 \geq 1 - \frac{1}{200 (B n)^{3.5} \log^4(B n)},\]
where $\widetilde{\sigma}^2$ is the variance parameter and $\bv^*$ is the vector that maximizes the quantity $z(\bv)$. 

Furthermore, we define a variance $\sigma^2 \in S$ to be \emph{bad} if for every unit vector $\bv \in \bV_t^\perp \cap \bA$ and for $\bg \sim D(\bV_t^\perp, \sigma^2)$,
\[\Ex{\langle \bu, \bg \rangle^2 \mid f(\bg) = 1} \leq \Ex{\langle \bu, \bg \rangle^2} + \frac{\Delta}{20}.\]
Using a standard concentration inequality combined with a union bound over a suitable net (similar to the approach in \lemref{lem:top:vector}), we conclude that with probability at least $1 - \exp(-n)$, for all such bad variances $\sigma^2$ and all vectors $\bv$,
\[z(\bv) < \sigma^2 + \frac{\sigma^2}{18 B r}.\]
Therefore, the vector $\bv^*$ corresponding to $\widetilde{\sigma}^2$ will be the first vector to reach this sufficiently high objective value and will be selected at step $t$.
\end{proof}

We now argue that if \invarref{invar:step} holds at round $t$, then it continues to hold at round $t+1$, assuming the algorithm has not yet terminated. 
This parallels the argument of Lemma 5.11 from \cite{HardtW13}, and completes the proof of the progress lemma.

\begin{lemma}
\lemlab{lem:invar:next:step}
\cite{GribelyukLWYZ25}
Let the granularity $\eta = \frac{1}{\poly(n)}$ for the discrete Gaussian be a fixed parameter. 
Suppose that $\sigma_n(\Sigma_t) > \lambda_{\max}(\calL^{\perp}(\bA))^2 \cdot \frac{\ln(2n(1+1/\eps))}{\pi}$. 
Suppose the pair $(V_t, W_t)$ satisfies \invarref{invar:step} and that the vector $\bv^*$ computed in round $t$ meets the condition
\[\|\bP_{\bV^\perp_t \cap \bA} \bv^*\|_2^2 \ge 1 - \frac{1}{200(Bn)^{3.5} \log^4(Bn)}.\]
Then \invarref{invar:step} continues to hold for $(V_{t+1}, W_{t+1})$.
\end{lemma}
\begin{proof}
Let $\gamma = \frac{1}{20(Bn)^{3.5}\log^4(Bn)}$.  
From \lemref{lem:found:vec}, we know that  
\[\|\bP_{V_t^\perp \cap \bA} \bv^*\|_2^2 \ge 1 - \frac{\gamma}{10}.\] 
This implies  
\[\|\bP_{\bA} \bv^*\|_2^2 \ge 1 - \frac{\gamma}{10}, \qquad \|\bP_{V_t} \bv^*\|_2^2 \le \frac{\gamma}{10}.\]  
Given that 
\[\bv_t = \trunc_{\eta}\left(\bv^* - \frac{\sum_{\bv \in V_t} \bv \langle \bv, \bv^* \rangle}{\left\|\sum_{\bv \in V_t} \bv \langle \bv, \bv^* \rangle \right\|_2}\right),\] 
it follows from the Pythagorean theorem, assuming the granularity $\eta=\frac{1}{\poly(n)}$ is sufficiently small, that  
\[\|\bP_{\bA} \bv_t\|_2^2 \ge 1 - \frac{\gamma}{4}.\]  
Hence, we can write $P_{W_{t+1}} = P_{W_t} + \bw_t \bw_t^\top$ for some unit vector $\bw_t$ orthogonal to $W_t$ such that $\|\bv_t - \bw_t\|_2 \le \frac{\gamma}{4}$.  

Note that $P_{V_{t+1}} = P_{V_t} + \bv_t \bv_t^\top$.  
Applying the triangle inequality, we have:  
\begin{align*}
d(V_{t+1}, W_{t+1}) &= \|\bP_{V_{t+1}} - P_{W_{t+1}}\|_2 \\
&\le \|\bP_{V_t} - P_{W_t}\|_2 + \|\bv_t \bv_t^\top - \bw_t \bw_t^\top\|_2 \\
&\le d(V_t, W_t) + \|\bv_t \bv_t^\top - \bv_t \bw_t^\top\|_2 + \|\bv_t \bw_t^\top - \bw_t \bw_t^\top\|_2.
\end{align*}  
Using the submultiplicativity of the spectral norm and the fact that both $\bv_t$ and $\bw_t$ are unit vectors, we obtain:  
\begin{align*}
d(V_{t+1}, W_{t+1}) &\le d(V_t, W_t) + \|\bv_t\|_2 \cdot \|\bv_t^\top - \bw_t^\top\|_2 + \|\bw_t^\top\|_2 \cdot \|\bv_t - \bw_t\|_2 \\
&\le d(V_t, W_t) + \frac{\gamma}{2}.
\end{align*}  
From the assumption in \invarref{invar:step}, we have  
\[d(V_t, W_t) \le \frac{t}{20(Bn)^{3.5} \log^{2.5}(Bn)}.\]  
Combining the bounds gives:  
\begin{align*}
d(V_{t+1}, W_{t+1}) &\le \frac{t}{20(Bn)^{3.5} \log^{2.5}(Bn)} + \frac{1}{40(Bn)^{3.5} \log^4(Bn)} \\
&\le \frac{t+1}{20(Bn)^{3.5} \log^{2.5}(Bn)}.
\end{align*}
Therefore, the invariant continues to hold for $(V_{t+1}, W_{t+1})$.
\end{proof}

We now combine the previous statements to establish the progress lemma, analogous to Lemma 5.7 in \cite{HardtW13}, but enabling the analysis to handle truncated vectors rather than real-valued vectors. 
\begin{lemma}[Progress lemma]
\lemlab{lem:progress:lem}
\cite{GribelyukLWYZ25}
Let the granularity $\eta = \frac{1}{\poly(n)}$ be a sufficiently small fixed parameter. 
Suppose $\sigma_n(\Sigma_t) > \lambda_{\max}(\calL^{\perp}(\bA))^2 \cdot \frac{\ln(2n(1+1/\eps))}{\pi}$. 
Let $t \in [r]$, and suppose the invariant \invarref{invar:step} is satisfied at round $t$, and that the function $f$ is $B$-correct on $W_t^\perp$. 
Then, with probability at least $1 - \frac{1}{n^2}$, the invariant \invarref{invar:step} also holds at round $t+1$.
\end{lemma}
\begin{proof}
The claim follows immediately from \lemref{lem:exist:var}, \lemref{lem:found:vec}, and \lemref{lem:invar:next:step}. 
\end{proof}

\subsection{Top Right Singular Vector of Biased Discrete Gaussian Matrices}
\seclab{sec:attack:top:singular}
In this section, we prove \lemref{lem:top:vector}, which asserts that if we form a matrix whose rows are vectors that exhibit a small correlation with the row space of $\bA$, then the top right singular vector of this matrix has a significant correlation with the row space of $\bA$. 
Recall that \lemref{lem:top:vector} played a key role in the proof of \lemref{lem:found:vec}, which in turn was essential for establishing the Progress Lemma in \lemref{lem:progress:lem}.

We establish the following result concerning the top right singular vector of discrete Gaussian matrices with a bias. 
In the setting of \lemref{lem:found:vec}, the subspace $V$ represents the portion of the row space of $\bA$ that remains undiscovered by the adversary, the parameter $\tau$ corresponds to the expected value $\EEx{\bg\sim D}{\langle\bw,\bg\rangle^2}$, and $\Delta$ reflects the gap described in the Conditional Expectation Lemma.
\begin{lemma}
\lemlab{lem:top:vector}
\cite{GribelyukLWYZ25}
Let $\tau \ge 0$ be a fixed threshold, $\eta=\frac{1}{\poly(n)}$ be a sufficiently small term, and let $V$ be a subspace of $\mathbb{R}^n$. 
Consider a distribution $D$ over $\left(\eta \cdot \mathbb{Z}\right)^n$ such that for a random vector $\bg \sim D$, the following conditions hold:
\begin{enumerate}
\item 
For every unit vector $\bw \in \bV^\perp$, the expected squared projection satisfies $\mathbb{E}_{\bg \sim D}[\langle \bw, \bg \rangle^2] \le \tau$.
\item 
For every pair of unit vectors $\bv \in V$ and $\bw \in \bV^\perp$, the expected cross term is small: $\left|\mathbb{E}_{\bg \sim D}[\langle \bv, \bg \rangle \cdot \langle \bw, \bg \rangle]\right| \le \eta$.
\item 
There exists a unit vector $\bv \in V \cap \left(\eta \cdot \mathbb{Z}\right)^n$ such that $\mathbb{E}_{\bg \sim D}[\langle \bv, \bg \rangle^2] \ge \tau + \Delta$, where $\Delta > \frac{1}{\poly(n)}$.
\item 
For every unit vector $\bu \in \mathbb{R}^n$, the variance of the squared projection is bounded: $\mathbb{V}_{\bg \sim D}[\langle \bu, \bg \rangle^2] \le \xi^2$.
\end{enumerate}
Let $\gamma \in \left(\frac{1}{\poly(n)}, \frac{1}{2n} \right)$ and let $m\succeq \frac{Cn \log^2 n \cdot \xi^2}{\gamma^2 \Delta^2}$ for some fixed constant $C>0$. 
Let $\bg_1, \ldots, \bg_m \sim D$ be i.i.d.\ samples, and define
\[\bu^* = \arg\max_{\|\bu\|_2 \le 1 + n\eta} \sum_{i=1}^m \langle \bg_i, \bu \rangle^2.\]
Then, with probability at least $1 - \exp(-n \log^2 n)$, we have
\[\|\bP_V \bu^*\|_2^2 \ge 1 - \gamma, \qquad \text{and} \qquad \frac{1}{m} \sum_{i=1}^m \langle \bg_i, \bu^* \rangle^2 \ge \tau + \frac{\Delta}{2}.\]
\end{lemma}
\begin{proof}
Consider the vector $\bv \in V \cap (\eta \cdot \mathbb{Z})^n$ satisfying $|\|\bv\|_2^2 - 1| \le n\eta$ and $\mathbb{E}_{\bg \sim D}[\langle \bv, \bg \rangle^2] \ge \tau - n^2 \eta + \Delta$ for some $\Delta > \frac{1}{\poly(n)}$, as ensured by our assumptions.  
Define $X = \sum_{i \in [m]} \langle \bv, \bg_i \rangle^2$, so that $\mathbb{E}[X] \ge \tau m + n^2 \eta m + \Delta m$ and $\mathbb{V}[X] \le m \xi^2$.  
Applying \thmref{thm:chernoff:hoeffding} and noting that $\eta \ll \frac{\tau}{n^2}$, we have:
\begin{align*}
\PPr{X \le (\tau - n^2 \eta + \Delta)m - \frac{\gamma m \Delta}{4}} &\le \exp\left( - \frac{\Delta^2 \gamma^2 m^2}{\O{\xi^2 m}} \right)\\
&\le \exp(-\Omega(n \log^2 n)).
\end{align*}
This implies that with high probability, there exists a nearly unit-length vector well-aligned with $V$ whose squared projections onto the samples $\bg_1, \ldots, \bg_m$ are significantly large.

Next, we argue that vectors poorly aligned with $V$ are unlikely to achieve such large squared projections.  
Let $\bu = \alpha \bv + \beta \bw$ be a truncated unit vector satisfying $\alpha^2 + \beta^2 \le 1 + n\eta$, where $\bv \in V \cap (\eta \cdot \mathbb{Z})^n$, $\bw \in \bV^\perp \cap (\eta \cdot \mathbb{Z})^n$, and both $\|\bv\|_2^2-1|\le n\eta$ and $|\|\bw\|_2^2-1|\le n\eta$. 
Suppose $\alpha^2 < 1 - \gamma$ and define $Y = \sum_{i=1}^m \langle \bu, \bg_i \rangle^2$.  
Then:
\[\mathbb{E}_{\bg \sim D}[\langle \bu, \bg \rangle^2] = 
\alpha^2 \cdot \mathbb{E}[\langle \bv, \bg \rangle^2] 
+ \beta^2 \cdot \mathbb{E}[\langle \bw, \bg \rangle^2] 
+ 2\alpha\beta \cdot \mathbb{E}[\langle \bv, \bg \rangle \cdot \langle \bw, \bg \rangle].\]
Since $|\mathbb{E}[\langle \bv, \bg \rangle \cdot \langle \bw, \bg \rangle]| \le n^2 \eta$ by assumption, it follows that:
\begin{align*}
\mathbb{E}[\langle \bu, \bg \rangle^2] 
&\le (1 - \gamma)(\tau + n^2 \eta + \Delta) + \frac{\tau + n^2 \eta}{2n} + 4n^2 \eta \\
&\le \tau + (1 - \gamma) \Delta.
\end{align*}
Hence, $\mathbb{E}[Y] \le (\tau + (1 - \gamma)\Delta) m \le (\tau + \Delta)m$, and by \thmref{thm:chernoff:hoeffding},
\[\PPr{Y \ge (\tau + \Delta)m - \frac{3\gamma m \Delta}{4}} \le \exp\left( - \frac{\Delta^2 \gamma^2 m^2}{\O{\xi^2 m}} \right)\le \exp(-\Omega(n \log^2 n)).\]
Therefore, with high probability, there is a gap of at least $\frac{\gamma m \Delta}{4}$ between the lower bound on $X$ and the upper bound on $Y$, assuming $\eta n^2 \ll \gamma m \Delta$.

Let $M = \{ \bu: \|\bP_V \bu\|_2^2 \ge 1 - \gamma \}$ and construct a $\frac{\gamma \Delta}{8}$-net $N$ over $M$ such that $|N| \le \exp(\O{n \log n})$.  
Then, applying the union bound:
\begin{align*}
\PPr{\max_{\bu \in N} \sum_{i=1}^m \langle \bu, \bg_i \rangle^2 \ge (\tau + \Delta)m - \frac{3\gamma m \Delta}{4}}&\le |N| \cdot \exp(-\Omega(n \log^2 n))\\
&\le \exp(-n \log^2 n),
\end{align*}
for sufficiently large constant in the choice of $m$.

Moreover, since each term in the sum differs by at most $\frac{\gamma \Delta m}{8}$ from its closest net point, we have:
\[\max_{\bu \in M} \sum_{i=1}^m \langle \bu, \bg_i \rangle^2 
\le \max_{\bu \in N} \sum_{i=1}^m \langle \bu, \bg_i \rangle^2 + \frac{\gamma \Delta m}{8}.\]
Putting everything together, it follows that with probability at least $1 - \exp(-n \log^2 n)$, we have:
\[\|\bP_V \bu^*\|_2^2 \ge 1 - \gamma, \qquad \text{and} \qquad \frac{1}{m} \sum_{i=1}^m \langle \bg_i, \bu^* \rangle^2 \ge \tau + \frac{\Delta}{2}.\]
\end{proof}
 
\subsection{Putting It All Together}
\seclab{sec:attack:final}
We now show that the empirical value $s(t, \sigma^2)$ is a good estimate of the actual probability that $f(\bg) = 1$ when $\bg$ is drawn from a discrete Gaussian distribution. 
This mirrors Lemma 5.5 in~\cite{HardtW13}, c.f., \lemref{lem:real:good:estimate}, which establishes the accuracy of such empirical estimates in approximating the probability that $f(\bg) = 1$ for a continuous Gaussian sample.
\begin{lemma}
\lemlab{lem:event:whp}
\cite{GribelyukLWYZ25}
Let the granularity $\eta=\frac{1}{\poly(n)}$ be sufficiently small and define $\calE$ as the event that, for every $t\in[r+1]$ and each $\sigma^2\in S$, the following holds:
\[\left|s(t, \sigma^2) - \PPPr{\bg \sim D(\bV_t^\perp, \sigma^2)}{f(\bg) = 1}\right| \le \frac{1}{20(Bn)^2 \log(Bn)}.\]
Then $\PPr{\calE}\ge1 - \exp(-n)$.
\end{lemma}
\begin{proof}
Let $\zeta=\frac{1}{20(Bn)^2\log(Bn)}$. 
Because the number of samples satisfies $m\gg\left(\frac{\alpha Bn}{\zeta}\right)^2$, then the desired claim follows from standard Chernoff bounds. 
\end{proof}

We conclude the proof of \thmref{thm:adaptive-attack}, which provides the final guarantees of our adaptive attack. 
\thmadaptiveattack*
\begin{proof}
First, if $r \le n^{o(1)}$, then for each row $\bA_k$ of the pre-processed matrix $\bA$, we represent each entry in binary using $\O{\log n}$ bits. 
Then, for each bit index $i \in [\O{\log n}]$, we define a new row $B_i = [b_{1,i}, \ldots, b_{n,i}]$. 
This reformulation is without loss of generality, as we can reconstruct the original row via $\sum_{i=0}^{\O{\log n}} 2^i \cdot B_i = \bA_k$.

Next, we apply the pre-processing step from \lemref{lem:preprocessing} to the resulting matrix $\bA$ without loss of generality, as adding rows can only strengthen the sketch. 
At this point, the matrix $\bA$ has at most $m \le 4r \log n$ rows and satisfies $\lambda_{\max}(\calL^{\perp}(\bA)) \le \sqrt{n}$.

In the case where $r \ge n^{\O{1}}$, we directly apply the pre-processing from \lemref{lem:preprocessing} to the original matrix $\bA$, yielding a sketch with $m \le 4r$ rows. 
Without loss of generality, we assume $n = 4m + 90\log(Bm)$ by restricting attention to the first $4m + 90\log(Bm)$ coordinates in $\mathbb{R}^n$, ensuring that any polynomial dependence on $n$ is also polynomial in $r\log n$.

We now verify the invariant
\[\sigma_n(\Sigma_{\sigma^2}^t) \ge \lambda_{\max}(\calL^{\perp}(\bA))^2 \cdot \frac{\ln(2n(1+1/\eps))}{\pi}\]
holds at each step of the attack. 
Define the covariance matrix at round $t$ of the attack as
\[\Sigma_{\sigma^2}^t = \frac{3\sigma^2}{4} \bP_{\bV^\perp}^\top \bP_{\bV^\perp} + \frac{\sigma^2}{4} \mathbb{I}_n.\]
Since $\sigma^2 \in [\alpha, B\alpha]$ throughout the attack, we obtain
\[\sigma_n(\Sigma_{\sigma^2}^t) \ge \frac{\alpha}{4} > \lambda_{\max}(\calL^{\perp}(\bA))^2 \cdot \frac{\ln(2n(1+1/\eps))}{\pi},\]
which ensures the invariant holds by construction.

For each $t \in [r]$, let $\bW_t \subseteq \bA$ denote the closest $(t-1)$-dimensional subspace to $\bV_t$ contained in the row span of $\bA$, defined by
\[W_t = \arg\min \left\{ d(\bV_t, \bW) \mid \dim(\bW) = t - 1,\; W \subseteq \bA \right\}.\]
We claim that \invarref{invar:step} holds throughout the attack. 
At $t=1$, this holds trivially since $\bV_0 = \{0\} \subseteq \bA$.

Let $\calE$ be the event that the empirical estimate $s(t, \sigma^2)$ is accurate for all rounds. 
By \lemref{lem:event:whp}, this event occurs with probability at least $1 - \exp(-n)$.

Assume inductively that \invarref{invar:step} holds up to round $t-1$. 
Then, by \lemref{lem:end:or:invar}, if the algorithm halts in round $t$, it outputs a failure certificate $D(\bV_t^\perp, \sigma^2)$ for $f$. 
Otherwise, assuming the invariant holds at round $t$, it follows that $f$ is $B$-correct on $W_t^\perp$. 
Moreover, by the progress lemma, the invariant continues to hold at round $t+1$ with probability at least $1 - \frac{1}{n^2}$.

By a union bound over all rounds, with probability at least $1 - \frac{1}{n}$, either the algorithm halts and returns a failure certificate at some step $t$, or \invarref{invar:step} holds through round $r+1$. 
In the latter case, since $\bW_{r+1}$ is not correct for $f$ by \lemref{lem:invar:end:wrong}, it follows from \lemref{lem:end:or:invar} that the algorithm must halt in round $r+1$ and output a failure certificate. 
Hence, the algorithm returns a failure certificate with probability at least $1 - \frac{2}{n}$.

Finally, observe that the query complexity is polynomial in $\alpha$, $B$, and $m$, where $\alpha \ge \max_{\bA \in \mathbb{Z}^{r \times n}}(\ell_{\bA})$, and $\ell_{\bA} = \poly(n)$ is an upper bound on $\lambda_{n - r}(\calL^{\perp}(\bA))$ after pre-processing. 
Therefore, the total number of queries is bounded by $\poly(r \log n)$. 
The runtime is also polynomial in $r\log n$, since optimizing $z(\bv)$ can be done via singular vector computation, which can be done in $\poly(r\log n)$ time. 
\end{proof}

\section{Lower Bound for \texorpdfstring{$F_0$}{F0} Estimation with Integer Sketches}
In this section, we show lower bounds for integer and linear sketches for the $F_0$ estimation problem, where the goal is to estimate the number of non-zero entries $\|\bx\|_0 = |\{ i: x_i \neq 0 \}|$ in the vector $\bx$ defined by the data stream, c.f., \defref{def:distinct:elements}. 
We first define the gap version of the $F_0$ estimation problem. 
Formally, the problem is defined as follows:
\begin{definition}[$F_0$ gap problem]
Let $0 \le \alpha < \beta \le 1$. 
An algorithm $\calA$ solves the $(\alpha, \beta)$-$F_0$ gap problem if for any input vector $\bx\in\mathbb{Z}^n$, $\calA$ outputs $0$ if $\|\bx\|_0 \le \alpha n$ and outputs $1$ if $\|\bx\|_0 \ge \beta n$. 
If $\|\bx\|_0\in(\alpha n,\beta n)$, then $\calA$ may return either $0$ or $1$.
\end{definition}
We shall show that for any integer sketching algorithm $\calA$ that uses an integer sketching matrix $\bA\in \mathbb{Z}^{r \times n}$, where $n$ is the size of the universe, there exists an attack by \cite{GribelyukLWYZ24} that makes $\tilde{\mathcal{O}}(r^8)$ queries and succeeds with high constant probability in breaking the sketch, in the sense that the algorithm $\calA$ will fail to solve the $F_0$ gap problem. 
We also describe an adaptive attack by \cite{GribelyukLWYZ24} over $\mathbb{R}^n$ against linear sketches $\bA \in \mathbb{R}^{r \times n}$ for $F_0$-estimation, in the setting where $\bA$ has all nonzero subdeterminants at least $\frac{1}{\poly(r)}$, i.e., every $k\times k$ submatrix obtained by selecting any $k$ rows and $k$ columns of $\bA$ has determinant either zero or at least $\frac{1}{\poly(r)}$. 

\subsection{Overview of Attack on \texorpdfstring{$F_0$}{F0} Estimation}
In this section, we describe the adaptive attack of \cite{GribelyukLWYZ24} against integer sketching algorithms designed for the $F_0$ gap problem. 
The strategy of the adaptive adversary is to incrementally learn the structure of the sketching matrix $\bA$, and then using this knowledge to construct increasingly challenging query vectors. 
A sketching matrix $\bA$ may encode a substantial amount of information about certain coordinates $\bx_i$ in the sketch $\bA\bx$. 
For instance, when a row in $\bA$ has nonzero entries in only one column, it can directly recover that coordinate. 
On the other hand, its dependence on other coordinates may be minimal, such as when each coordinate is always involved in linear combinations with many other coordinates. 
The coordinates that the sketch maintains substantial information about, referred to as \emph{significant coordinates}, are particularly useful for $F_0$-estimation in a non-adaptive setting. 
For example, one might design $\bA$ so that it captures information about a random subset of $\O{1}$ coordinates. 
By examining whether these sampled coordinates are zero or nonzero, one could estimate the total number of nonzeros in $\bx$ within additive error $0.1n$, solving the $F_0$ gap problem. 

Therefore, the attack of \cite{GribelyukLWYZ24} iteratively identifies significant coordinates and ensures they are set to zero in all subsequent queries. 
Doing so makes future queries harder for $\bA$ to interpret: once a coordinate is always zero, $\bA$ may still waste its capacity storing information about it, thus effectively reducing its dimensionality on the remaining coordinates. 
When $r \ll n$, the sketch cannot preserve significant information about too many coordinates. 
Eventually, the sketching process is forced to rely solely on \emph{insignificant coordinates}, which the sketch can only weakly reflect.

The attack hinges on solving the three following key subproblems:
\begin{itemize}
\item 
Formally define the notion of a ``significant'' coordinate, and prove that when $r \ll n$, only a small number of coordinates can be significant.
\item 
Design an algorithm that reliably identifies these significant coordinates using a polynomial number of queries.
\item 
Demonstrate that any algorithm must fail to estimate the $F_0$ value accurately when the input is supported only on the insignificant coordinates. 
To do this, we construct distributions over $\bx$ with drastically different $F_0$ values whose corresponding sketches $\bA\bx$ are nearly indistinguishable.
\end{itemize}
Informally, a notion of ``significant'' coordinates will be defined conceptually, and then a pre-processing argument will upper bound the number of significant coordinates. 
A fingerprinting attack can be applied to identify the significant coordinates. 
Finally, a moment-matching argument will be used to construct ``hard'' distributions for sketches with support on the insignificant coordinates. 
We now further outline how \cite{GribelyukLWYZ24} addresses each of these components.

\paragraph{Fingerprinting codes.} 
To learn significant coordinates, first consider a special case: when $\bA\bx$ simply returns a subset of $r$ coordinates of $\bx$, i.e., each row of $\bA$ is a unit vector. 
Here, these $r$ coordinates are extremely significant, while the rest are entirely insignificant.

This problem aligns with the setting addressed by interactive fingerprinting codes~\cite{SteinkeU15}. 
In that setting, an adversary $\mathcal{P}$ selects a hidden subset $\mathcal{S} \subset [n]$ of size $k$, and a fingerprinting code $\mathcal{F}$ must identify $\mathcal{S}$ through adaptive queries $c^t \in \{\pm 1\}^n$. 
Each response $a^t$ must be consistent with some coordinate $c^t_i$ for $i \in \mathcal{S}$. 
Over repeated interactions, $\mathcal{F}$ tracks correlations between queried indices and the responses, accumulating to identify the hidden subset $\mathcal{S}$ with high probability in $\O{k^2}$ queries.
For intuition, we remark that in the special case where each row of $\bA$ equals some elementary vector $\be_i$, the sketch provides exact values of individual $\bx_i$. 
Moreover, solving the $F_0$-gap problem demands distinguishing between $\|\bx\|_0 \leq \alpha n$ and $\|\bx\|_0 \geq \beta n$ and is strictly harder than the fingerprinting code task, so the same strategy suffices for the attack.

\paragraph{Significant coordinates.}
The analysis is then extended to general matrices $\bA$ by considering what constitutes significant information  for a coordinate. 
Beyond the elementary vector case, if $\bA$ is linear, then $\bA\bx$ lets us compute any $\bw^\top \bx$ where $\bw$ lies in the row span of $\bA$. 
This motivates the following definition: coordinate $i$ is $s$-significant if there exists $\by^\top \in \mathbb{R}^r$ such that
\[(\by^\top \bA)_i^2 \geq \frac{1}{s} \cdot \|\by^\top \bA\|_2^2.\]
This definition is equivalent to saying that the leverage score of column $i$ is at least $\frac{1}{s}$, which accurately captures significance if real-valued queries are allowed.

However, when query vectors must have integer coordinates bounded by $\poly(n)$, leverage scores are insufficient. 
For example, consider a row vector $(C, \ldots, C, 1)$ in $\bA$. 
Most leverage scores are low, but $\bA\bx$ still leaks $\bx_n \mod C$. 
This is because although the vector $\bw^\top=\left(1,1,1,\ldots,1,\frac{1}{C}\right)$ is in the row span of $\bA$, the first $n-1$ coordinates never contribute to the \emph{fractional part} of the inner product $\bw^\top \bx$ regardless of the integer vector $\bx$. 
That is, the last coordinate of $\left(0,0,0,\ldots,0,\frac{1}{C}\right)$ is heavy in the \emph{fractional part} of $\bw^{\top}$. 
This motivates the final definition of significance, based on the fractional part:
\begin{equation}
\label{eq:frac}
\exists \by^\top\in\mathbb{R}^r, \quad |\fracpart((\by^\top\bA)_i)|^2 \geq \frac{1}{s}\cdot\|\fracpart(\by^\top\bA)\|_2^2,
\end{equation}
where $\fracpart(\cdot)$ is taken coordinate-wise. 

\paragraph{Matrix pre-processing.}
To aid the analysis, \cite{GribelyukSWY24} transforms $\bA$ into a matrix $\bA'$ that separates significant and insignificant columns while preserving the information available to the sketch. 
Consider the following iterative procedure: while there exists a column satisfying~\eqref{eq:frac}, zero it out and record its index in a significant set $\mathcal{S}$. 
After no such column remains, add a new row $\be_i$ for each $i \in \mathcal{S}$. 
This results in a sketching matrix:
\[\bA' = \begin{bmatrix}
\bD \\
\mathbf{S}\end{bmatrix},\]
where $\bD$ is a dense matrix that contains only insignificant columns, and $\mathbf{S}$ is a sparse matrix that has at most one non-zero entry $1$ in each row and column, with the set of non-zero columns corresponding exactly with the set of significant coordinates $\mathcal{S}$. 
Moreover, the columns of $\bD$ and $\bS$ are disjoint.

Observe that the sparse component corresponds precisely to the extreme scenario discussed earlier, and the set $\calS$ can be recovered using the fingerprinting code in the absence of a dense component. 
Additionally, \cite{GribelyukLWYZ24} establishes that the notion of significant coordinates, along with the pre-processing step, ensures that the sparse set remains small, i.e., $|\calS| \ll n$. 
This guarantees that once $\calS$ is identified and its coordinates are zeroed out in the query, the $F_0$ estimation problem on the remaining coordinates remains non-trivial.

Informally, the existence of such a pre-processing step is shown by arguing that under the uniform distribution over $\bx \in \{-1, 0, 1\}^n$, any column $i$ satisfying~\eqref{eq:frac} must induce a non-negligible mutual information between $\bA \bx$ and $\bx_i$, namely, $I(\bA \bx; \bx_i) \geq \Omega\left(\frac{1}{s}\right)$. 
Therefore, if the pre-processing procedure iteratively removes $T$ such columns, the chain rule for mutual information implies that the total information between $\bA \bx$ and these $T$ coordinates is at least $\Omega\left(\frac{T}{s}\right)$.
However, since $\bA \bx$ can be described using at most $\O{r \log n}$ bits, the overall mutual information cannot exceed $\O{r \log n}$. 
As a result, we conclude that the number of columns added to the sparse part must satisfy $T = \O{r s \log n}$.

\paragraph{Attack description.}
The final step is to demonstrate that the dense component, i.e., the insignificant coordinates, does not contribute meaningful information to the algorithm. 
This is achieved by carefully constructing an appropriate query distribution.

Specifically, we define a family of distributions $\calD$ supported on the set $\{-R, -(R-1), \ldots, R\}$, where $R = \poly(n)$ is bounded by a small polynomial in $n$. 
The distributions in $\calD$ are designed to satisfy the following properties:
\begin{enumerate}
\item 
For any $D_p \in \calD$ with parameter $p \in [\alpha, \beta]$ for some constants $0 < \alpha < \beta < 1$, it holds that $\PPPr{X \sim D_p}{X = 0} = p$;
\item 
For any two parameters $p, q \in [\alpha, \beta]$, and corresponding product distributions $\bx_p \sim D_p^n$ and $\bx_q \sim D_q^n$, the total variation distance between $\bD \bx_p$ and $\bD \bx_q$ satisfies $d_{\mathrm{tv}}(\bD \bx_p, \bD \bx_q) \le \frac{1}{\poly(n)}$.
\end{enumerate}
We defer the construction of such a distribution family to the subsequent paragraph.
Using this setup, we consider the sketch
\[\bA' \bx = \begin{bmatrix}
    \bD \bx \\
    \bS \bx
\end{bmatrix},\]
where the query vectors $\bx$ are drawn from $D_p^n$ for different values of $p$. 
Due to the properties of $\calD$, the marginal distribution of $\bD \bx$ remains nearly invariant with respect to $p$. 
Additionally, since the nonzero columns of $\bD$ and $\bS$ are disjoint, the outputs $\bD \bx$ and $\bS \bx$ are conditionally independent given $p$.
Consequently, if queries are sampled from this distribution family, the algorithm cannot utilize the dense part $\bD \bx$ to distinguish different values of $p$, and must instead rely entirely on the sparse part $\bS \bx$ to approximate $\|\bx\|_0$.
Importantly, the distributions $D_p$ can be incorporated into the fingerprinting code construction, ensuring that the dense component remains uninformative during the attack on the sparse part. 
This enables gradual identification of the significant coordinate set $\calS$ and once all such coordinates have been identified and zeroed out, we issue a final query in which all entries in $\calS$ are set to zero. 
At this point, the algorithm observes only $\bD \bx$, which, by design, is nearly independent of $p$, and hence must fail to produce a correct output with high probability.

\paragraph{Hard distribution for the insignificant coordinates.}
It remains to construct a distribution family such that the total variation distance between $\bD \bx_p$ and $\bD \bx_q$ remains small for vectors $\bx_p \sim D_p^n$ and $\bx_q \sim D_q^n$, where $\bD$ is the dense part of the sketch matrix $\bA$. 
The construction of \cite{GribelyukLWYZ24} crucially aims to ensure that the first $K = \O{r \log n}$ moments of each pair $D_p, D_q \in \calD$ are identical:
\[\EEx{X \sim D_p}{X^k} = \EEx{X \sim D_q}{X^k} \qquad \text{for all } k \in [K].\]
Additionally, symmetry is enforced for each $D_p \in \calD$, meaning $D_p(t) = D_p(-t)$ for all $t$. 
This symmetry ensures that all odd moments vanish, so it suffices to match only the even moments. 
For these, the moment-matching condition becomes
\[\sum_{i=0}^R i^k \cdot (D_p(i) - D_q(i)) = 0 \qquad \text{for } k \le K.\]
From polynomial theory, it is known, c.f., \cite{LarsenWY20}, that there exists a polynomial $Q$ of degree at most $R - \Omega(\sqrt{R})$ satisfying
\[|Q(0)| = \Omega(1), \qquad \sum_{i=0}^R \left\lvert \binom{R}{i} \cdot Q(i) \right\rvert = \O{1}.\]
Since $\deg(Q(i) \cdot i^t) < R$ for $t < R - \deg(Q)$, and the alternating binomial sum of any polynomial of degree less than $R$ vanishes, it follows that
\[\sum_{i=0}^R (-1)^i \binom{R}{i} Q(i) i^t = 0 \qquad \text{for all } t < R - \deg(Q).\]
For $R = \Theta(K^2)$ with a sufficiently large constant, define the family $\calD = \{D_p\}$ by perturbing a base distribution $D$ as follows:
\[D_p(i) = D(i) + c_p \cdot (-1)^i \binom{R}{i} Q(i),\]
for some constants $c_p$.
The difference between any two such distributions is then
\[D_p(i) - D_q(i) = (c_p - c_q) \cdot (-1)^i \binom{R}{i} Q(i).\]
This guarantees moment matching:
\[\sum_{i=0}^R i^k \cdot (D_p(i) - D_q(i)) = (c_p - c_q) \cdot \sum_{i=0}^R i^k (-1)^i \binom{R}{i} Q(i) = 0\]
for all $k \le K \le \O{\sqrt{R}}$.
Finally, we remark that the bounds on $\sum_{i=0}^R \left\lvert \binom{R}{i} Q(i) \right\rvert$ and $|Q(0)|$ can be used to show that the gap $\beta - \alpha$ between the largest and smallest probability mass at $0$ over the family $\calD$ can be made $\Omega(1)$ by choosing the base distribution $D$ appropriately, where $\alpha$ and $\beta$ denote the minimum and maximum mass at $0$, respectively. 
This results in a gap in the $F_0$ values of the resulting query vectors. 

\paragraph{Bounding the total variation distance.}
Let $P = D_p$ and $Q = D_q$ be two distributions from the family $\calD$, with matching moments up to order $K$, for some $p, q \in [\alpha, \beta]$. 
Consider the product distributions $P^n$ and $Q^n$ over $n$ independent samples from $P$ and $Q$, respectively. 
Let $\bD$ be a dense matrix where no column satisfies condition~\eqref{eq:frac} with parameter $s$. 
For $\bx \sim P^n$ and $\bx' \sim Q^n$, let $P_{\bD}$ and $Q_{\bD}$ denote the distributions of $\bD \bx$ and $\bD \bx'$, respectively.

We aim to show that the total variation distance between $P_{\bD}$ and $Q_{\bD}$ satisfies $\TVD(P_{\bD}, Q_{\bD}) \le \frac{1}{\poly(n)}$. 
This follows from the following Fourier-analytic bound:
\begin{align*}
|P_{\bD}(x) - Q_{\bD}(x)| &= \left\lvert \frac{1}{(2\pi)^r} \int_{[-\pi, \pi)^r} e^{i \langle \bu, x \rangle} \left(\widehat{P_{\bD}}(\bu) - \widehat{Q_{\bD}}(\bu)\right) d\bu \right\rvert \\
&\le \frac{1}{(2\pi)^r} \int_{[-\pi, \pi)^r} \left\lvert \widehat{P_{\bD}}(\bu) - \widehat{Q_{\bD}}(\bu) \right\rvert d\bu,
\end{align*}
where the inequality uses the triangle inequality.

Thus, to bound $|P_{\bD}(x) - Q_{\bD}(x)|$, it suffices to bound the point-wise difference $\left| \widehat{P_{\bD}}(\bu) - \widehat{Q_{\bD}}(\bu) \right|$. 
Letting $P_i = \Pr[X = i]$ and defining the wrapped fractional part function $\fracpart_{2\pi}(x) := 2\pi \cdot \fracpart\left(\frac{x}{2\pi}\right) \in [-\pi, \pi)$, we can write:
\begin{align*}
\widehat{P_{\bD}}(\bu) &= \EEx{\bz \sim P_{\bD}}{e^{-i \langle \bu, \bz \rangle}} = \EEx{\bx \sim P^n}{e^{-i \langle \bu, \bD \bx \rangle}} \\
&= \prod_{j \in [n]} \sum_{k \ge 0} P_k \cdot \cos\left(k \cdot \left\langle \bu, \bD^{(j)} \right\rangle \right) \\
&= \prod_{j \in [n]} \sum_{k \ge 0} P_k \cdot \cos\left(k \cdot \fracpart_{2\pi}(\langle \bu, \bD^{(j)} \rangle)\right),
\end{align*}
where the third equation uses the symmetry of $P$ (i.e., $P(t) = P(-t)$), allowing us to express the Fourier transform in terms of cosines.
Applying the Taylor expansion of the cosine function:
\[\cos(x) = \sum_{k=0}^\infty \frac{(-1)^k x^{2k}}{(2k)!},\]
we obtain:
\begin{align*}
\widehat{P_{\bD}}(\bu) &= \prod_{j \in [n]} \sum_{k \ge 0} M_P(2k) \cdot \frac{\left(\fracpart_{2\pi}(\langle \bu, \bD^{(j)} \rangle)\right)^{2k}}{(2k)!} \cdot (-1)^k, \\
\widehat{Q_{\bD}}(\bu) &= \prod_{j \in [n]} \sum_{k \ge 0} M_Q(2k) \cdot \frac{\left(\fracpart_{2\pi}(\langle \bu, \bD^{(j)} \rangle)\right)^{2k}}{(2k)!} \cdot (-1)^k,
\end{align*}
where $M_P(2k)$ and $M_Q(2k)$ denote the $2k$-th moments of $P$ and $Q$, respectively.
To bound the difference in Fourier transforms, \cite{GribelyukLWYZ24} leverages the two main properties previously established:
\begin{enumerate}
\item 
\textbf{Bounded fractional parts:} 
The matrix $\bD$ satisfies a spreading property: for any vector $y \in \mathbb{R}^r$ and any coordinate $j \in [n]$, we have
\[|\fracpart((y^\top \bD)_j)|^2 \le \frac{1}{s} \cdot \left\| \fracpart(y^\top \bD) \right\|_2^2.\]
Thus, if there exists any $j$ such that $|\fracpart_{2\pi}(\langle \bu, \bD^{(j)} \rangle)| \ge \frac{1}{K}$ (for a suitable threshold), then the overall $L_2$ norm of the vector $\left(\fracpart_{2\pi}(\langle \bu, \bD^{(j)} \rangle)\right)_{j\in[n]}$ is large enough to ensure that the corresponding terms in the product decay rapidly. 
\item 
\textbf{Moment matching:} 
On the other hand, if all coordinates satisfy $|\fracpart_{2\pi}(\langle \bu, \bD^{(j)} \rangle)| < \frac{1}{K}$, then we can rely on the fact that $M_P(2k) = M_Q(2k)$ for all $k \le K/2$, so the first $K/2$ terms in each Taylor expansion match exactly. 
The remaining higher-order terms can be shown to contribute negligibly, due to the smallness of the input.
\end{enumerate}
For full details of the bounds in both cases, see \secref{sec:fzero:full:analysis}.

Finally, since $\bD \in \mathbb{Z}^{r \times n}$ with polynomially bounded entries, the support of both $P_{\bD}$ and $Q_{\bD}$ is of size at most $n^{\O{r}}$. 
Therefore, once we have a uniform point-wise bound on $|\widehat{P_{\bD}}(\bu) - \widehat{Q_{\bD}}(\bu)|$, we can conclude the desired total variation bound via a union bound over the support:
\[\TVD((P_{\bD}, Q_{\bD}) \le \frac{1}{\poly(n)},\]
for appropriate choices of parameters $K$ and $s$.
This concludes the technical overview for the attack. 

\subsection{Pre-processing the Sketching Matrix}
\seclab{sec:fzero:preprocess}
In this section, we formally introduce the pre-processing procedure and justify the main properties resulting from the subsequent matrix.  
Suppose we use a sketching matrix $\bA$ as a basis for our adversarially robust streaming algorithm for $L_0$ estimation. 
We first consider a conceptual pre-processing of the sketching matrix $\bA$ into \textit{sparse} part and a \textit{dense} part, which consists has disjoint support on the columns. 
Observe that this pre-processing step only increases the power of the streaming algorithm, since it only allows the algorithm to observe potentially more entries of the input vector $\bx^{(t)}$. 
For a real number $x$, we define the function $\fracpart(x) = x-\mathsf{int}(x) \in (-\frac{1}{2}, \frac{1}{2}]$, where $\mathsf{int}(x)$ is the closest integer number to $x$. 
Similarly, for a vector $\bx\in\mathbb{R}^n$, let $\fracpart(\bx)\in\mathbb{R}^n$ be the coordinate-wise fractional parts of $\bx$, i.e., $\fracpart(\bx)_j=\fracpart(x_j)$. 
Formally, the matrix $A'$ formed after pre-processing $A$ will satisfy the following several key properties, where we use $\fracpart(x)$ to denote the fractional part of a real number $x$. 
\begin{lemma}
\lemlab{lem:pre}
\cite{GribelyukLWYZ24}
Let $\calA$ be a streaming algorithm that uses a sketching matrix $\bA \in \mathbb{Z}^{r \times n}$. 
Then there is a pre-processing procedure that produces a new matrix $\bA' \in \mathbb{Z}^{r' \times n}$ for $r' = \O{rs\log n}$ such that:
\begin{enumerate}
    \item The matrix $\bA'$ has the form $\begin{bmatrix}
        \bD \\
        \bS
    \end{bmatrix}$ where the matrices $\bD$ and $\bS$ are column-disjoint.
    \item For all $\by\in\mathbb{R}^{r}$ and $j\in[n]$, we have $|\fracpart(\by^\top \bD)_j|^2 \le \frac{1}{s}\cdot\|\fracpart(\by^\top \bD)\|_2^2$.  
    \item Each row and column of $\bS$ has at most one non-zero entry.
\end{enumerate} 
Moreover, we can assume without loss of generality that the algorithm $\calA$ uses sketching matrix $\bA'$ instead of $\bA$.
\end{lemma}
\begin{proof}
Consider the following iterative procedure. 
Begin with the sketching matrix $\bA$. 
For each time $t$, let $\bD^{t - 1}$ be the first $r$ rows of $\bA^{(t - 1)}$. 
We then identify a column $j_t\in[n]$ such that there exists $\by^\top$ such that
\[\left\lvert\fracpart(\by^\top \bD^{(t-1)})_{j_t}\right\rvert^2 > \frac{1}{s}\cdot\|\ \fracpart(\by^\top \bD^{(t-1)})\|_2^2.\]
We then zero out the $j_t$-th column of $\bA^{t - 1}$ and add a new row (the elementary vector) $\be_{j_t}$ to the matrix $\bA^{(t - 1)}$. 
We denote this new resulting matrix by $\bA^{(t)}$. 

Suppose that the above procedure ends in the iteration $T$. 
From \lemref{lem:remove:s:heavy:frac}, which we shall prove, we have $T = \O{rs \log n}$.
    
Let $\bD = \bD^{(T)}$ and $\bS$ be the remaining rows of $\bA^{(T)}$.
Hence, $\bA' = \begin{bmatrix}
\bD \\
\bS
\end{bmatrix}$ has at most $r + T = \O{rs \log n}$ columns. 
Moreover, it is clear from the iterative procedure that the matrices $\bD$ and $\bS$ are column-disjoint. 
Furthermore, by construction of the added elementary rows in the iterative procedure, each row and column of $\bS$ has at most one non-zero entry.

At this point, it remains to show why we can assume that the streaming algorithm $\calA$ uses the sketching matrix $\bA'$, rather than $\bA$. 
To this end, suppose that the algorithm $\calA$ uses the sketching matrix $\bA$ and a post-processing composition function $f$ on $\bA\bx$. 
Now, consider a different post-processing estimator $g$ that takes input $\bA'\bx$ and first inverts the row operations that we apply on $\bA$ to achieve $\bA'$. 
Thus, $g$ transforms $\bA'\bx$ to $\bA\bx$ and then outputs the value $f(\bA\bx)$. 
From the definition of $g$, we have $g(\bA'\bx)=f(\bA\bx)$ for every input vector $\bx$, and thus we can assume that $\calA$ has the form $g(\bA'\bx)$ without loss of generality.
\end{proof}
\noindent
It remains to bound the number of added rows. 
\begin{lemma}
\lemlab{lem:random:vec:info:frac}
\cite{GribelyukLWYZ24}
Let $\bA\in \mathbb{Z}^{r\times n}$ be a fixed matrix, let $s>1$ be a parameter, and let $c$ be a sufficiently small constant. 
Let $\bx\in\{-1, 0, 1\}^n$ be a random vector, such that each coordinate is chosen independently, so that with probability $1 - \frac{2c}{s}$, $x_i = 0$, and with probability $\frac{2c}{s}$, $x_i = 1$ or $-1$ with equal probability. 
Suppose there exists $\by\in\mathbb{R}^r$ and $j\in[n]$ such that for $\left\lvert\fracpart((\by^\top \bA)_j)\right\rvert^2\ge\frac{1}{s}\cdot\|\fracpart(\by^\top \bA)\|_2^2$. 
Then 
\[I(\bA\bx; x_j)=\Omega\left(\frac{1}{s}\right).\] 
\end{lemma}
\begin{proof}
By the data-processing inequality, we have
\[I(\bA\bx; x_j)\ge I(\by^\top \bA\bx; x_j)\ge I(\fracpart(\by^\top \bA\bx); x_j).\]
Thus, it suffices to prove
\[I(\fracpart(\by^\top \bA\bx); x_j)=\Omega\left(\frac{1}{s}\right).\]
To that end, let the vector $\ba=\by^\top \bA\in\mathbb{R}^n$. 
Let $\bx^{(1)},\ldots,\bx^{(t)}$ be $t=\O{\log s}$ different vectors whose coordinates are sampled randomly from $\{-1, 0, 1\}$ from the distribution in the lemma statement but the $j$-th coordinate is the same in all $t$ vectors, i.e., there is a single draw for the $j$-th coordinate, which is then propagated to all $t$ vectors. 
Then, the marginal distributions are the same for each $\bx^{(k)}$. 
Hence, $I(\fracpart(\langle \ba,\bx\rangle); x_j) = I(\fracpart(\langle \ba,\bx^{(k)})\rangle; x_j)$. 

We claim that we can determine the value of $x_j$ with probability at least $1-\frac{1}{\poly(s)}$ by looking at $t = \O{\log s}$ samples $\fracpart(\langle \ba,\bx^{(k)})\rangle$, where $k\in[t]$. 
In particular, we claim that
\[I(\fracpart(\langle \ba,\bx^{(1)}\rangle_,\ldots,\fracpart(\langle \ba,\bx^{(t)}\rangle); x_j)=\Omega\left(\frac{\log s}{s}\right).\]

\paragraph{Mutual information from independent instances.}
Firstly, observe that if $\|\fracpart(\ba)\|_2^2>s$, then $\frac{1}{s}\cdot\|\fracpart(\ba)\|_2^2>1$. 
Thus, there cannot exist a vector $\by$ that induces $\ba=\by^\top \bA$ such that $|\fracpart(a_j)|^2\ge\frac{1}{s}\cdot\|\fracpart(\ba)\|_2^2$. 
Therefore, it suffices to consider $\|\fracpart(\ba)\|_2^2\le s$. 

We next consider a fixed instance $\bx=\bx^{(k)}$. 
Let $S$ be the set of indices such that $x_i \ne 0$ and $i \ne j$. 
Then 
\[\Ex{\|\fracpart(\ba_S)\|_2^2}\le \frac{2c}{s} \|\fracpart(\ba)\|_2^2.\]
Hence for a sufficiently small constant $c$, then Markov's inequality implies 
\[\PPr{\|\fracpart(\ba_S)\|_2^2\le\frac{1}{200s} \|\fracpart(\ba)\|_2^2}\ge0.99.\]
Conditioning on this event, then by Markov's inequality, we also have
\[\PPr{\sum_{i \in S} \left( \fracpart(a_i) \cdot x_i \right)^2 \le \frac{1}{10s} \|\fracpart(\ba)\|_2^2}\ge 0.9,\]
since $\Ex{x_i}=0$ and $x_i^2\le 1$. 
Hence with probability at least $0.89$,
\[\sum_{i \in S} \left( \fracpart(a_i) \cdot x_i \right)^2 \le \frac{1}{10s} \|\fracpart(\ba)\|_2^2 \le \frac{1}{10}.\]
In this case, we have $\fracpart(\langle\ba,\bx\rangle) = \fracpart(\fracpart(a_j)\cdot x_j + \alpha)$, for $\alpha = \sum_{i \in S} \fracpart(a_i) \cdot x_i \le \frac{1}{3\sqrt{s}}\|\fracpart(\ba)\|_2 \le \frac{1}{3} \left| \fracpart(a_j)\right|$.

Conditioned on these events, suppose $x_j=0$. 
In this case, we have $|\fracpart(\langle \ba,\bx\rangle)| \le \frac{1}{3} \left| \fracpart(a_j)\right|$. 
Otherwise if $x_j \ne 0$, then 
\[|\fracpart(\langle a,x\rangle)| \ge \frac{2}{3}  \left| \fracpart(a_j)\right|,\]
so that the sign of $\fracpart(\langle \ba,\bx\rangle)$ is the same as the sign of $x_j$. 
Therefore, we can determine the value of $x_j$ by looking at the value of $\fracpart(\langle \ba,\bx\rangle)$.

The above procedure succeeds with probability at least $0.98$. 
To further boost the success probability, we can instead look at the majority of the outputs by $\O{\log s}$ independent instances 
\[(\fracpart(\langle \ba,\bx^{(1)}\rangle_,\ldots,\fracpart(\langle \ba,\bx^{(t)}\rangle) \;, \]
so that the failure probability is at most $\frac{1}{\poly(s)}$. 
Thus, since $x_j$ is nonzero with probability $\Omega\left(\frac{1}{s}\right)$, then
\[I(\fracpart(\langle \ba,\bx^{(1)}\rangle_,\ldots,\fracpart(\langle \ba,\bx^{(t)}\rangle); x_j)=\Omega\left(\frac{\log s}{s}\right),\]
as claimed.

\paragraph{Mutual information from a single instance.}
It remains to analyze the mutual information from a single vector sampled from the given probability distribution. 
By the chain rule for mutual information, i.e., \thmref{thm:chain:information},  
\begin{align*}
I&(\fracpart(\langle \ba,\bx^{(1)}\rangle),\ldots,\fracpart(\langle \ba,\bx^{(t)}\rangle); x_j)\\
&=\sum_{k=1}^t I(\fracpart(\langle \ba,\bx^{(k)}\rangle); x_j\,\mid\,\fracpart(\langle \ba,\bx^{(1)}\rangle),\ldots,\fracpart(\langle \ba,\bx^{(k-1)}\rangle)).   
\end{align*}
Since each of the draws $\bx^{(1)},\ldots,\bx^{(t)}$ are independent, conditioned on $x_j$, then $\fracpart(\langle \ba,\bx^{(k)}\rangle)$ is independent of $\fracpart(\langle \ba,\bx^{(1)}\rangle,\ldots\langle \ba,\bx^{(k-1)}\rangle)$ conditioned on $x_j$. 
Hence, 
\begin{align*}
I(\fracpart(\langle \ba,\bx^{(1)}\rangle),\ldots,\fracpart(\langle \ba,\bx^{(t)}\rangle); x_j)&\le\sum_{k=1}^t I(\fracpart(\langle \ba,\bx^{(k)}\rangle); x_j)\\
&=\sum_{k=1}^t I(\fracpart(\langle \ba,\bx\rangle);x_j).    
\end{align*}
Since $t=\O{\log s}$, then 
\[I(\fracpart(\by^\top \bA\bx;x_j))=I(\fracpart(\langle \ba,\bx\rangle);x_j)=\Omega\left(\frac{\log s}{st}\right)=\Omega\left(\frac{1}{s}\right).\]
\end{proof}

\begin{lemma}
\lemlab{lem:remove:s:heavy:frac}
\cite{GribelyukLWYZ24}
Let $\bA\in \mathbb{Z}^{r\times n}$ be a fixed matrix and let $s>1$ be a parameter. 
There exists a pre-processing procedure that takes input $\bA$ and outputs a matrix $\bA'\in\mathbb{Z}^{r\times n}$ that zeros out at most $\O{rs\log n\log s}$ columns of $\bA$. 
Moreover for all $\by\in\mathbb{R}^{r}$ and $j\in[n]$, we have $|\fracpart(\by^\top \bA')_j|^2 \le\frac{1}{s}\cdot\|\fracpart(\by^\top \bA')\|_2^2$. 
\end{lemma}
\begin{proof}
Consider the following iterative process. 
Let $\bA^{(0)}=\bA$ and let $\bx\in\{-1,0,1\}^n$ be drawn from the same distribution defined in \lemref{lem:random:vec:info:frac}. 
Hence, $\bA\bx$ can be completely encoded using $\O{r\log n}$ bits, since $\bA$ has integer entries with magnitude at most $\O{\log n}$ bits. 
In the $t$-th step of the process, we identify a column $j_t\in[n]$ such that $|\fracpart(\by^\top \bA^{(t-1)})_j|^2 > \frac{1}{s}\cdot\|\ \fracpart(\by^\top \bA^{(t-1)})\|_2^2$ and then we set $\bA^{(t)}$ to be the matrix $\bA^{(t-1)}$ that zeros out the identified column. 

Suppose the process terminates in $T$ rounds for some $T\le n$. 
By the chain rule for mutual information, i.e., \thmref{thm:chain:information}, 
\[I(\bA^{(T)}\bx; x_{j_1}, x_{j_2}, \cdots, x_{j_T})=\sum_{t=1}^T I(\bA^{(T)}\bx; x_{j_t}\,\mid\,x_{j_{t + 1}}, x_{j_{t + 2}}, \cdots x_{j_{T}}).\]
Observe that given the matrix $\bA^{(T)}$ and $j_{t + 1}, j_{t + 2}, \cdots, j_T$, we can recover $\bA^{(t)}$ exactly. 
Hence, by a similar approach to that in \lemref{lem:random:vec:info:frac}, we have $I(\bA^{(T)}\bx; x_{j_t}\,\mid\,x_{j_{t + 1}}, x_{j_{t + 2}}, \cdots x_{j_{T}}) \ge \O{\frac{1}{s \log s}}$ from 
\[|\fracpart(y^\top \bA^{(t-1)})_j|^2 > \frac{1}{s}\cdot\|\ \fracpart(y^\top \bA^{(t-1)})\|_2^2.\]
Since $\bA\bx$ can be encoded using $\O{r\log n}$ bits, then we must have 
\[I(\bA^{(T)}\bx; x_{j_1}, x_{j_2}, \cdots, x_{j_T}) \le\O{r \log n}.\]
Therefore, $Cr \log n \ge\frac{T}{s \log s}$ for some constant $C$, which implies $T = \O{rs \log n \log s}$, as desired.
\end{proof}

\subsection{Fingerprinting Attack on Sparse Part}
\seclab{sec:fzero:fingerprinting}
In this section, we show correctness of the fingerprinting attack on the sparse part $\bS$ of the matrix $\bA$, where we assume $\bA$ has already undergone the pre-processing procedure from \secref{sec:fzero:preprocess}. 
We split the analysis into discussions for soundness and correctness of the attack. 

Let $P_{a,b}$ denote the probability distribution with support $[a,b]$ and probability density function $\mu(p)=\frac{C_{a,b}}{\sqrt{p(1-p)}}$, for a normalizing constant $C_{a,b}$. 
For $p\in[0,1]$, let $\phi^p:\{\pm1\}\to\mathbb{R}$ be defined by $\phi^0(c)=\phi^1(c)=0$ for $c\in\{\pm1\}$. 
Moreover, let $\phi^p(1)=-\sqrt{\frac{p}{1-p}}$ and $\phi^p(-1)=\sqrt{\frac{1-p}{p}}$ for $p\in(0,1)$. 
By construction, $\phi^p(c)$ has mean $0$ and variance $1$ when $\PPr{c=-1}=p$ and $\PPr{c=1}=1-p$. 
We describe the attack on the sparse matrix in full in \figref{fig:1}.

\begin{figure}[!htb]
\begin{mdframed}
\begin{algorithmic}
Let $\alpha$ and $\beta$ be the parameters from \lemref{lem:moment:match}
\newline
Let $\calD$ be the family of distributions from \lemref{lem:moment:match}, with $K = \O{r \log n}$
\newline\noindent
$h \gets \O{rs \log n} = \O{r^4 \log^3 n}$, $\sigma\gets\O{h \log(n)}$, $\ell\gets\O{h}\cdot\sigma$, $c \gets \O{1}$
\newline \noindent
For a vector $\bv\in\mathbb{R}^n$ and a set $J\subseteq[n]$, let $\bz_J(\bv) $ denote the vector $\bv$ with all entries in $J$ set to zero, i.e., $v_i=0$ for all $i \in J$.
\newline Let $\calA$ be an instance of the $L_0$ gap-norm algorithm. 
\newline\noindent
Initialize $s_i^0=0$ for all $i\in[n]$
\newline \noindent 
For $j\in[\ell]$:
\newline \indent
Sample $\bu^1, \cdots, \bu^c \sim D_\alpha^n$ and $\bv^1, \cdots, \bv^c \sim D_{\beta}^n$ 
\newline \indent
If $\calA$ fails with constant probability on either $\bz_{I^{j - 1}}(\bu^i)$ or $\bz_{I^{j - 1}}(\bv^i)$:
\newline \indent \indent Output the corresponding distribution as the attack.
\newline\indent
Sample $p^j\sim P_{\alpha,\beta}$ and $\bv^j\sim D_{p^{j}}^n$
\newline\indent
For all $i\in[n]$, set $c_i^j=1$ if $v_i^j\neq 0$ and $c_i^j=-1$ otherwise if $v_i^j=0$
\newline\indent
Query $\bz_{I^{j - 1}}(\bv^j)\in\mathbb{Z}^n$ and receive $a^j = \mathcal{A}(\bz_{I^{j - 1}}(\bv^j))\in\{\pm1\}$ as the output 
\newline\indent
For $i\in[n]$, update the correlation scores $s_i^j\gets s_i^{j-1}+a^j\cdot\phi^{p^j}(c_i^j)$
\newline\indent
Set $I^j\gets I^{j - 1} \cup \{i\in[n]\,\mid\, s_i^j>\sigma\}$ and $\calS^{j + 1} \gets \calS \setminus I^j$    
\end{algorithmic}
\end{mdframed}   
\caption{Construction of our attack}
\figlab{fig:1}
\end{figure} 

\subsubsection{Soundness}
\seclab{sec:soundness}
In this section, we prove soundness of the fingerprinting attack, i.e., that the attack will not erroneously identify coordinates that are not contained within $\calS$. 
We first show a structural property about an exponential function on a random variable $v$ chosen from the distribution $D_p$. 
\begin{lemma}
\lemlab{lem:one:moment:gen:subg}
\cite{SteinkeU15,GribelyukLWYZ24}
For $p\in[\alpha, \beta]$, let $\tau=\min(\alpha,1 - \beta)$ and $t\in\left[-\frac{\sqrt{\tau}}{2},\frac{\sqrt{\tau}}{2}\right]$. 
Let $a\in[-1,1]$ be fixed and let $c=1$ if $v$ is nonzero and $c=-1$ if $v$ is zero. 
Then 
\[\EEx{v\sim D_p}{e^{at\phi^p(c)}}\le e^{t^2}.\]
\end{lemma}
\begin{proof}
This statement and corresponding proof are almost identical to Lemma 2.4 in \cite{SteinkeU15}, with the difference that $v$ is drawn from a different probability distribution, i.e., $v\sim D_p$ to handle the boundaries of $[\alpha,\beta]$, instead of the probability distribution in \cite{SteinkeU15}. 

Since $\PPPr{v\sim D_p}{v=0}=p$, then we have $\EEx{\bv\sim D_p}{\phi^p(c)}=p\cdot\phi^p(-1)+(1-p)\cdot\phi^p(1)$. 
Due to the setting of $\phi^p(1)=-\sqrt{\frac{p}{1-p}}$ and $\phi^p(-1)=\sqrt{\frac{1-p}{p}}$ for $p\in(0,1)$, then we have $\EEx{v\sim D_p}{\phi^p(c)}=0$. 
By a similar calculation, we have $\EEx{\bv\sim D_p}{(\phi^p(c))^2}=1$. 
Since $\alpha,\beta\in[0,1]$, then $\tau\le 1$ and so for $c\in\{\pm1\}$, we have $|\phi^p(c)|\le\frac{1}{\sqrt{\tau}}$. 
Hence for $t\in\left[-\frac{\sqrt{\tau}}{2},\frac{\sqrt{\tau}}{2}\right]$, we have $|\phi^p(c)\cdot t|\le\frac{1}{2}$. 

We have $e^x\le 1+x+x^2$ for $x\in\left[-\frac{1}{2},\frac{1}{2}\right]$. 
Thus, from the Taylor expansion of $e^{t^2}$, we have for $a\in[-1,1]$,
\[\EEx{\bv\sim D_p}{e^{at\phi^p(c)}}\le1+at\cdot\EEx{\bv\sim D_p}{\phi^p(c)}+a^2t^2\cdot\EEx{v\sim D_p}{(\phi^p(c))^2}\le 1+t^2\le e^{t^2}.\] 
\end{proof}
\FloatBarrier
Next, we show that provide tail bounds on the sum of the linear combinations of the $\phi$ functions. 
\begin{lemma}
\lemlab{lem:sum:moment:gen:subg}
\cite{SteinkeU15,GribelyukLWYZ24}
Let $p^1,\ldots,p^m\in[\alpha,\beta]$ and $\bv_i\sim D_{p^j}$. 
Let $a^1,\ldots,a^m\in[-1,1]$ be fixed and $\tau=\min(\alpha, 1 - \beta)$. 
For all $i\in[n]$, let $c_i=1$ if $v_i^j\neq 0$ and $c_i^j=-1$ otherwise if $v_i^j=0$. 
Then for all $\lambda\ge 0$,
\[\PPr{\sum_{j\in[m]}a^j\phi^{p^j}(c_i^j)\ge\lambda}\le e^{-\lambda^2/4m}+e^{-\sqrt{\tau}\lambda/4}.\]
\end{lemma}
\begin{proof}
This statement and corresponding proof are almost identical to Lemma 2.5 in \cite{SteinkeU15}, with the difference that we instead define $\tau=\min(\alpha, 1 - \beta)$ due to the range of $p\in[\alpha,\beta]$ as the probability of $D_p$ drawing a zero.  
Since $a^1,\ldots,a^m\in[-1,1]$, then for all $t\in\left[-\frac{\sqrt{\tau}}{2},\frac{\sqrt{\tau}}{2}\right]$, we have by \lemref{lem:one:moment:gen:subg},  
\[\EEx{\bv}{e^{t\sum_{i\in[m]}a^j\phi^{p^j}(c_i^j)}}\le\prod_{j\in[m]}\EEx{v_i^j\sim D_{p^j}}{e^{ta^j\phi^{p^j}(c_i^j)}}\le e^{t^2m}.\]
By Markov's inequality, 
\[\PPr{\sum_{i\in[m]}a^j\phi^{p^j}(c_i^j)\ge\lambda}\le\frac{\Ex{e^{t\sum_{i\in[m]}a^j\phi^{p^j}(c_i^j)}}}{e^{t\lambda}}\le e^{t^2m-t\lambda}.\]
Let $t=\min\left(\frac{\sqrt{\tau}}{2},\frac{\lambda}{2m}\right)$. 
Then for $\lambda\in\left[0,m\sqrt{\tau}\right]$, we have $t=\frac{\lambda}{2m}$, so that
\[\PPr{\sum_{i\in[m]}a^j\phi^{p^j}(c_i^j)\ge\lambda}\le e^{-\lambda^2/4m}.\]
Similarly, for $\lambda\ge m\sqrt{\tau}$, we have
\[\PPr{\sum_{i\in[m]}a^j\phi^{p^j}(c_i^j)\ge\lambda}\le e^{\tau m/4-\sqrt{\tau}\lambda/2}\le e^{-\frac{\sqrt{\tau}\lambda}{4}}.\]
Together, these two inequalities imply the desired claim. 
\end{proof}
We require the following concentration inequality upper bounding the magnitude of the supremum of a sequence of partial sums. 
\begin{theorem}[Etemadi's inequality]
\cite{etemadi1985some}
\thmlab{thm:etemadi}
Let $X_1,\ldots,X_n$ be independent random variables. 
For all $k\in[n]$, let $S_k=\sum_{i=1}^k X_i$ be the $k$-th partial sum of the sequence $X_1,\ldots,X_n$. 
Then for all $\lambda>0$,
\[\PPr{\max_{k\in[n]}|S_k|>4\lambda}\le4\cdot\max_{k\in[n]}\PPr{|S_k|>\lambda}.\]
\end{theorem}
We next show that each item not in the sparse set $\calS$ will not be reported with high probability. 
\begin{lemma}[Individual soundness]
\lemlab{lem:ind:sound} 
\cite{SteinkeU15,GribelyukLWYZ24}
For each $i\in[n]\setminus\calS$, we have
\[\PPr{i\in I^\ell}\le \frac{1}{n^2}.\]
\end{lemma}
\begin{proof}
The proof is similar to Proposition 2.7 in \cite{SteinkeU15}.
Let $i\in[n]\setminus\calS$ be a fixed index. 
Since $i \notin \calS$, then the adversary does not see $c_i^j$. 
Hence, we can assume without the loss of generality that the outputs $a^j$ are fixed and $v_i^j$ and then $c_i^j$ are subsequently drawn. 
Then for every $j\in[\ell]$, we have by \lemref{lem:sum:moment:gen:subg}, 
\begin{align*}  
\PPr{s_i^j>\frac{\sigma}{4}} &=\PPr{\sum_{k\in[j]} a^k\phi^{p_k}(c_i^k)>\frac{\sigma}{4}} \le e^{-\frac{\sigma^2}{64\ell}}+e^{-\sigma\sqrt{\tau}/16}
\end{align*}
In other words, the probability that there is some index $i\in[n]\setminus\calS$ with high correlation score is small. 
Similarly, we upper bound the probability that there is some index $i\in[n]\setminus\calS$ with high anti-correlation score. 
Namely, for every $j\in[\ell]$, we have by \lemref{lem:sum:moment:gen:subg}, 
\begin{align*}
\PPr{s_i^j<-\frac{\sigma}{4}} &=\PPr{\sum_{k\in[j]} a^k\phi^{p_k}(c_i^k)<-\frac{\sigma}{4}} \le e^{-\frac{\sigma^2}{64\ell}}+e^{-\sigma\sqrt{\tau}/16}
\end{align*}
Hence we can use Etemadi's inequality, c.f., \thmref{thm:etemadi}, to upper bound the probability that $i$ is assigned to $I^j$ because some partial sum has large magnitude:
\begin{align*}
\PPr{i\in I^j}&\le\PPr{\max_{t\in[j]}|s_i^t|>\sigma}\\
&\le4\max_{t\in[j]}\PPr{|s_i^t|>\frac{\sigma}{4}}\\
&\le8(e^{-\frac{\sigma^2}{64\ell}}+e^{-\sigma\sqrt{\tau}/16}) \le \frac{1}{n^2}.   
\end{align*} 
\end{proof}
We then have soundness from essentially taking a union bound over all $i\in[n]\setminus\calS$. 
\begin{lemma}[Soundness] 
\lemlab{lem:soundness}
\cite{SteinkeU15,GribelyukLWYZ24}
\[\PPr{|I^\ell\setminus\calS|\ge 1}\le \frac{1}{n}.\]
\end{lemma}
\begin{proof}
We use the same proof as Theorem 2.8 in \cite{SteinkeU15}. 
Namely, for $i\in[n]\setminus\calS$, let $Y_i$ denote the indicator random variable for the event $i\in I^\ell\setminus\calS$. 
By \lemref{lem:ind:sound}, it follows that $\Ex{Y_i}\le \frac{1}{n^2}$ for all $i\in[n]\setminus\calS$. 
Thus by Markov's inequality,
\[\PPr{|I^\ell\setminus\calS|\ge 1}\le\Ex{\sum_{i\in[n]\setminus\calS}Y_i}\le \frac{1}{n^2}(n-r)\le \frac{1}{n}.\]
\end{proof}
Finally, we show that the sum of the scores cannot drastically increase between rounds. 
\begin{lemma}
\lemlab{lem:soundness:bound}
\cite{SteinkeU15,GribelyukLWYZ24}
Let $\tau=\min(\alpha,\beta)$ and let $I^{\ell+1}=[n]$.  
For each $i\in[n]$, let $j_i\in[\ell+1]$ be the first $j$ such that $i\in I^j$. 
Then for $J\subset[n]$,
\[\PPr{\sum_{i\in J}(s_i^\ell-s_i^{j_i-1})>\lambda}\le e^{-\frac{\lambda^2}{4|J|\ell}}+e^{-\sqrt{\tau}\lambda/4}.\]
\end{lemma}
\begin{proof}
The proof follows Lemma 2.10 in \cite{SteinkeU15}. 
Let $\mathbbm{1}$ denote the standard indicator function. 
Note that
\[\sum_{i\in J}(s_i^\ell-s_i^{j_i-1})=\sum_{i\in J}\sum_{j\in[\ell]}\mathbbm{1}[j\ge j_i] a^j\phi^{p_j}(c_i^j).\]
Recall that the fingerprinting code zeros out the $i$-th coordinate after time $j_t - 1$. 
Thus we can again take the view that the outputs $a^j$ are fixed and then the terms $\phi^{p_j}(c_i^j)$ are subsequently drawn for $j \ge j_t$. 
Then by \lemref{lem:sum:moment:gen:subg}, we have
\[\PPr{\sum_{i\in J}(s_i^\ell-s_i^{j_i-1})>\lambda}\le e^{-\frac{\lambda^2}{4|J|\ell}}+e^{-\sqrt{\tau}\lambda/4},\]
as claimed.
\end{proof}

\subsubsection{Completeness}
\seclab{sec:completeness}
In this section, we prove completeness of the fingerprinting attack, i.e., that the attack will correctly identify coordinates that are contained within $\calS$, the set of non-zero indices of columns in $S$. 
Note that this is necessary in conjunction with \secref{sec:soundness} to show correctness of the fingerprinting attack because even in light of \secref{sec:completeness}, it could be that the fingerprinting attack does not report any coordinates in each round. 

\paragraph{Fourier analysis.}
We first show a structural property informally stating that the rate at which the average of $f(\bv)$ changes with $p$ is exactly given by how much $f(\bv)$ correlates with the zero vs. nonzero pattern in $\bv$.
We use $h:=|\calS|$ to denote the size of $\calS$. 
\begin{lemma}
\lemlab{lem:corr:fourier}
\cite{SteinkeU15,GribelyukLWYZ24}
Let $f:\mathbb{R}^h\to\mathbb{R}$ and let $g:[0,1]\to\mathbb{R}$ be defined so that $g(p)=\EEx{v_1,\ldots,v_h\sim D_p}{f(\bv)}$. 
For all $i\in[h]$, let $c_i=1$ if $v_i$ is nonzero and $c_i=-1$ if $v_i$ is zero. 
Then for any $p\in[\alpha,\beta]$,
\[\EEx{v_1,\ldots,v_h\sim D_p}{f(\bv)\cdot\sum_{i\in[h]}\phi^p(c_i)}=g'(p)\sqrt{p(1-p)}.\]
\end{lemma}
\begin{proof}
The analysis is similar to Lemma 2.11 of \cite{SteinkeU15}, but since the probability distribution $D_p$ is different, then there is also a different corresponding score function $\phi$. 

For $p\in(0,1)$ and $T\subset[h]$, we define the functions $\phi_T^p:\{\pm1\}^h\to\mathbb{R}$ to form an orthonormal basis with respect to the product distribution with bias $p$, by setting $\phi_T^p(\bc)=\prod_{i\in T}\phi^p(c_i)$. 
Then for all $T,U\subset[h]$, we have
\[\EEx{v_1,\ldots,v_h\sim D_p}{\phi_T^p(\bc)\cdot\phi_U^p(\bc)}=\begin{cases}1\qquad&T=U\\
0\qquad &T\neq U,\end{cases}\]
where for all $i\in[h]$, $c_i=1$ if $v_i$ is nonzero and $c_i=-1$ if $v_i$ is zero. 
Let $c(\bv)$ denote this mapping from $\bv$ to $\bc$. 
Then we can decompose $f$ by
\[f(\bv)=\sum_{T\subset[h]}\widehat{f}^p(T)\cdot\phi_T^p(c(\bv)),\]
using the Fourier coefficients
\[\widehat{f}^p(T)=\EEx{v_1,\ldots,v_h\sim D_p}{f(\bv)\cdot\phi_T^p(c(\bv))},\]
for all $T\subset[h]$. 
\cite{SteinkeU15} observes that this representation is a generalization of Fourier analysis to biased distributions~\cite{ODonnell14}. 

Now, for $p,q\in(0,1)$, we can write $g(q)$ in its Fourier representation by
\begin{align*}
g(q)&=\EEx{v_1,\ldots,v_h\sim D_q}{f(\bv)}\\
&=\sum_{T\subset[h]}\widehat{f}^p(T)\cdot\EEx{v_1,\ldots,v_h\sim D_q}{\phi_T^p(c(\bv))},
\end{align*}
using the linearity of expectation. 
Now since for $v_i\sim D_q$, the probability that $v_i$ is zero (and thus $c_i$ is $-1$) is $q$, then we have
\begin{align*}
g(q)&=\sum_{T\subset[h]}\widehat{f}^p(T)\cdot\prod_{i\in T}\EEx{v_i\sim D_q}{\phi_T^p(c_i(v_i))}\\
&=\sum_{T\subset[h]}\widehat{f}^p(T)\cdot\left(q\cdot\sqrt{\frac{1-p}{p}}+(1-q)\cdot\sqrt{\frac{p}{1-p}}\right)^{|T|}. 
\end{align*}
Taking the derivative, we have
\begin{align*}
g'(q)=\sum_{T\subset[h], T\neq\emptyset}\Bigg(\widehat{f}^p(T)\cdot|T|\cdot&\left(q\cdot\sqrt{\frac{1-p}{p}}+(1-q)\cdot\sqrt{\frac{p}{1-p}}\right)^{|T|-1}\\
&\cdot\left(\sqrt{\frac{1-p}{p}}+\sqrt{\frac{p}{1-p}}\right)\Bigg)
\end{align*}
and specifically for $q=p$, 
\[g'(p)=\sum_{i\in[h]}\widehat{f}^p(\{i\})\cdot\left(\sqrt{\frac{1-p}{p}}+\sqrt{\frac{p}{1-p}}\right).\]
Since $\widehat{f}^p(\{i\})=\EEx{v_1,\ldots,v_h\sim D_p}{f(\bv)\cdot\phi^p(c_i(v_i))}$, then by linearity of expectation,
\begin{align*}
\EEx{v_1,\ldots,v_h\sim D_p}{f(\bv)\cdot\sum_{i\in[h]}\phi^p(c_i(v_i))}&=\sum_{i\in[h]}\widehat{f}^p(\{i\})\\
&=\frac{g'(p)}{\sqrt{\frac{1-p}{p}}+\sqrt{\frac{p}{1-p}}}\\
&=g'(p)\cdot\sqrt{p(1-p)}.
\end{align*}
\end{proof}
We next show that on average over $p$, the connection between $f(\bv)$ and the zero/nonzero pattern gives us a meaningful fraction of how much $g(p)$ changes from $\alpha$ to $\beta$. 
In particular, note that $g(\beta)$ will be large and $g(\alpha)$ will be small in our setting for the $F_0$ gap problem. 
\begin{lemma}
\lemlab{lem:corr:gap}
\cite{SteinkeU15,GribelyukLWYZ24}
Let $f:\mathbb{R}^h\to\mathbb{R}$ and let $g:[0,1]\to\mathbb{R}$ be defined so that $g(p)=\EEx{v_1,\ldots,v_h\sim D_p}{f(\bv)}$. 
For all $i\in[h]$, let $c_i=1$ if $v_i$ is nonzero and $c_i=-1$ if $v_i$ is zero. 
Then there exists a constant $\zeta>0$ such that
\[\EEx{p\sim P_{\alpha,\beta}}{\EEx{v_1,\ldots,v_h\sim D_p}{f(\bv)\cdot\sum_{i\in[h]}\phi^p(c_i)}}\ge\zeta\cdot(g(\beta)-g(\alpha)).\]
\end{lemma}
\begin{proof}
The proof follows the same outline as Proposition 2.12 in \cite{SteinkeU15}. 
We define the probability distribution $P_{\alpha,\beta}$ to have probability density function $\mu(p)=\frac{C_{\alpha,\beta}}{\sqrt{p(1-p)}}$, with support exactly on the interval $[\alpha,\beta]$. 
By \lemref{lem:corr:fourier}, 
\begin{align*}
\EEx{p\sim P_{\alpha,\beta}}{\EEx{v_1,\ldots,v_h\sim D_p}{f(\bv)\cdot\sum_{i\in[h]}\phi^p(c_i)}}&=\EEx{p\sim P_{\alpha,\beta}}{g'(p)\cdot\sqrt{p(1-p)}}\\
&=\int_{\alpha}^\beta g'(p)\sqrt{p(1-p)}\cdot\mu(p)\,dp\\
&=C_{\alpha,\beta}\cdot\int_{\alpha}^\beta g'(p)\,dp\\
&=C_{\alpha,\beta}\cdot(g(\beta)-g(\alpha)).
\end{align*}
The desired claim then follows from setting $C_{\alpha,\beta}=\zeta$. 
\end{proof}

\paragraph{Concentration.}
We now show that if the algorithm has small error and the function only looks at certain coordinates, then the difference in expected outputs for $p = \beta$ and $p = \alpha$ is large. 
In particular, this crystallizes the guarantee of \lemref{lem:corr:gap} from an abstract function to our fingerprinting attack $\calA$. 
\begin{lemma}
\lemlab{lem:function:gap}
\cite{GribelyukLWYZ24}
Let $c$ be a sufficiently small universal constant. 
Suppose that at the $j$-th round, the algorithm $\calA$ has error probability $\delta_\alpha^j, \delta_\beta^j \le c$ over the input distribution $\bz_{I^{j - 1}}(\bu)$ and $\bz_{I^{j - 1}}(\bv)$ where $\bu \sim D_\alpha^n, \bv \sim D_\beta^n$. 
Moreover, suppose that the function $f^{j}: \mathbb{R}^h \to \mathbb{R}$ only depends on the interaction up to round $j - 1$ and $f^j(\bv)$ is only decided by the coordinates of $\bv$ in $\mathcal{S}^j$. 
Specifically, suppose $f^j(v^j_{\mathcal{S}^j}) = a^j$ where $\mathcal{S}^{j} = \mathcal{S} \setminus I^{j - 1}$. 
Then for $g(p)=\EEx{v_1,\ldots,v_h\sim D_p}{f(\bv)}$, we have $g^j(\beta) - g^j(\alpha) \ge 2 - \eta$ for some $\eta = \O{1}$.
\end{lemma}
\begin{proof}
By assumption, we have
\[\EEx{v_1, \ldots, v_n \sim D_\alpha}{\mathcal{A}(\bz_{I^{j - 1}}(\bv))} \le -(1 - 2c)\]
and 
\[\EEx{v_1, \ldots, v_n \sim D_\beta}{\mathcal{A}(\bz_{I^{j - 1}}(\bv))} \ge 1 - 2c.\]
Since the coordinates of $\bv$ in $I^{j-1}$ have been zeroed out in the $j$-th iteration, then without loss of generality, the output $a^j$ can be represented by the value of a function $f^j$ that is only decided by the coordinates on $\mathcal{S}^j$. 
Thus, the assumption that $\delta_\alpha^j, \delta_\beta^j \le c$ implies
\[g^j(\beta) - g^j(\alpha) = \EEx{v_1,\ldots,v_h\sim D_\beta}{f(\bv)} - \EEx{v_1,\ldots,v_h\sim D_\alpha}{f(\bv)} \ge 2 - \eta,\]
for some $\eta = \O{1}$.
\end{proof}

For $p\sim P_{\alpha,\beta}$ and $v_1,\ldots,v_h\sim D_p$, let $\xi_{\alpha,\beta}(f)=f(\bv)\cdot\sum_{i \in [h]}\phi^p(c_i)$, where for all $i\in[h]$, we set $c_i=1$ for $v_i$ is nonzero and we set $c_i=-1$ otherwise, i.e., if $v_i$ is zero. 
We now show closeness of the expectations, essentially allowing us to convert between exponential functions and linear functions. 
\begin{lemma}
\lemlab{lem:exp:bound}
\cite{SteinkeU15,GribelyukLWYZ24}
Let $f:\mathbb{R}^h\to\{\pm1\}$, $\tau=\min(\alpha, 1 - \beta)$, and $t\in\left[-\frac{\sqrt{\tau}}{8},\frac{\sqrt{\tau}}{8}\right]$. 
Then for $C=\frac{64e^{h\tau/4}}{\tau}$, we have
\[\Ex{e^{t\xi_{\alpha,\beta}(f)-\Ex{t\xi_{\alpha,\beta}(f)}}}\le e^{Ct^2}.\]
\end{lemma}
\begin{proof}
The proof is identical to Proposition 2.15 in \cite{SteinkeU15}. 
Let $Y = \sum_{i\in[h]}\phi^p(c_i)$. 
Since $\bv\sim D_p^h$, then each coordinate $v_i$ is independently drawn. 
Then by \lemref{lem:one:moment:gen:subg} and independence, 
\begin{align*}
\Ex{e^{tY}} = \Ex{e^{t \sum_{i\in[h]}\phi^p(c_i)}} = \left(\EEx{\bv \sim D_p} {e^{t\phi^p(\bc)}}\right)^h\le e^{{t^2 h}},
\end{align*}
for $t \in \left[-\frac{\sqrt{\tau}}{8},\frac{\sqrt{\tau}}{8}\right]$. 
Let $t \in \{\pm \frac{\sqrt{\tau}}{8}\}$ satisfy 
\[\sum_{k = 0}^{\infty} \frac{t^{2k + 1}}{(2k + 1)!} \Ex{Y^{2k + 1}} \ge 0.\]
By dropping all the positive terms, we have for all $j\ge 1$,
\begin{align*}
0 \le \Ex{Y^{2j}} &\le \frac{(2j)!}{t^{2j}} \sum_{k = 0}^{\infty} \frac{t^k}{k!} \Ex{Y^k} \\
& = \frac{(2j)!}{t^{2j}} \Ex {e^{tY}} \\
&\le \frac{(2j)!}{t^{2j}} e^{t^2 h } \\
&= \frac{4^j (2j)!}{\tau^j} e^{\tau h/4} \;.
\end{align*}
Hence, we have bounded the even moment of $Y$. 
To bound the magnitude of the odd terms, let $k = 2j + 1 \ge 3$. 
Then by Cauchy-Schwartz, 
\begin{align*}
\Ex{|Y|^k} &\le \sqrt{\Ex{Y^{2j}} \cdot \Ex{Y^{2j + 2}}} \\
&\le \sqrt{\frac{4^j (2j)!}{\tau^j}e^{h\tau /4}\frac{4^{j + 1} (2j + 2)!}{\tau^{j + 1}} e^{h\tau /4}} \\
&= \frac{4^{k / 2} k! }{\tau^{k / 2}} e^{h\tau / 4} \sqrt{\frac{k + 1}{k}}. 
\end{align*}
Because $|f(\bc) | \le 1$, then it follows that $\Ex{|f(\bc) \cdot Y|^k} \le \Ex{|Y|^k} \le 2\cdot\frac{4^{k / 2} k! }{\tau^{k / 2}} e^{h\tau / 4}$. 

Now for $t \in \left[-\frac{\sqrt{\tau}}{8},\frac{\sqrt{\tau}}{8}\right]$,
\begin{align*}
\Ex{e^{t\xi_{\alpha,\beta}(f)}} & \le 1 + t\Ex{\xi_{\alpha,\beta}(f)} + \sum_{k = 2}^{\infty} \frac{|t|^k}{k!} \Ex{|\xi_{\alpha,\beta}(f)|^k} \\
& \le 1 + t\Ex{\xi_{\alpha,\beta}(f)} + \sum_{k = 2}^{\infty} \frac{|t|^k}{k!} \frac{2 \cdot 4^{k / 2}k!e^{h\tau / 4}}{\tau^{k / 2}} \\
&= 1 + t\Ex{\xi_{\alpha,\beta}(f)} + 2 e^{h\tau/4}\sum_{k = 2}^{\infty} \left(\frac{\sqrt{4}t}{\sqrt{\tau}}\right)^k \\
& \le 1 + t\Ex{\xi_{\alpha,\beta}(f)} + 2 e^{h\tau/4}\sum_{k = 2}^{\infty} \left(\frac{\sqrt{4}t}{\sqrt{\tau}}\right)^2 (\sqrt{4})^{-(k - 2)} \\
& \le 1 + t\Ex{\xi_{\alpha,\beta}(f)} + 64 e^{h\tau/4} \frac{t^2}{\tau} \\
& \le e^{t\Ex{\xi_{\alpha,\beta}(f)} + Ct^2} \;.
\end{align*}
\end{proof}
We recall the following formulation of the Azuma-Doob martingale concentration inequality. 
\begin{theorem}[Azuma-Doob Inequality, Theorem 2.16 in \cite{SteinkeU15}]
\thmlab{thm:azuma:doob}
Let $X_1,\ldots,X_m\in\mathbb{R}$, $\mu_1,\ldots,\mu_m\in\mathbb{R}$, and $\calU_0,\ldots,\calU_m\in\Omega$ be random variables, such that for all $i\in[m]$:
\begin{itemize}
\item
$X_i$ and $\calU_{i-1}$ are determined by $\calU_i$
\item 
$\mu_i$ is determined by $\calU_{i-1}$.
\end{itemize}
Let $c>0$ and $C\in\mathbb{R}$ be parameters and suppose that for all $i\in[m]$, $u\in\Omega$, and $t\in[-c,c]$, we have
\[\Ex{e^{t(X_i-\mu_i)}\,\mid\calU_{i-1}=u}\le e^{Ct^2}.\]
Then for $\lambda\in[0,2Cmc]$, we have
\[\PPr{\left\lvert\sum_{i\in[m]}(X_i-\mu_i)\right\rvert\ge\lambda}\le2e^{-\frac{\lambda^2}{4Cm}},\]
and for $\lambda\ge2Cmc$, we have
\[\PPr{\left\lvert\sum_{i\in[m]}(X_i-\mu_i)\right\rvert\ge\lambda}\le2e^{-\frac{-c\lambda}{2}}.\]
\end{theorem}

\paragraph{Lower bounding the correlation.}
Given the above statements, we finally lower bound the correlation. 
We first show that there is a substantial gap in the overall sum of the scores of the coordinates in $\calS$, compared to those not in $\calS$. 
\begin{lemma}
\lemlab{lem:completeness:gap}
\cite{GribelyukLWYZ24}
Let $\zeta$ be the constant from \lemref{lem:corr:gap}, let $c$ be a sufficiently small universal constant, and let $\tau=\min(\alpha, 1 - \beta)$.  
Suppose that for every $j \in [\ell]$, the algorithm $\mathcal{A}$ has error probability $\delta_\alpha^j, \delta_\beta^j \le c$ in the $j$-th round, over the distribution $z_{I^{j - 1}}(D_\alpha^n), z_{I^{j - 1}}(D_\beta^n)$, respectively, where $z_{I^{j - 1}}$ means we zero out the coordinates in $I^{j - 1}$.
Then for any $\lambda\in\left[0,\frac{16\ell}{\sqrt{\tau}}\right]$,
\[\PPr{\sum_{i\in\calS} s_i^\ell<2\ell\zeta(1 -\eta) -\lambda}<2e^{-\frac{\lambda^2\tau}{256\ell}}.\]
\end{lemma}
\begin{proof}
For each $j \in [\ell]$, from the discussion of~\lemref{lem:function:gap}, there exists a function $f^j: \mathbb{R}^h \to \{\pm 1\}$ that only depends on the interaction up to round $j - 1$ and satisfies $f^j(v^j_{S^j}) = a^j$. 
For $j\in[\ell]$, we define the random variable 
\[X_j = f^j(v^j_{S^j}) \sum_{i\in[h]}\phi^p(c^j_i) \sim  \xi_{\alpha,\beta}(f^j) \;,\]
where $\sim$ denotes that $X_j$ has the same distribution as $\xi_{\alpha,\beta}(f^j)$. 
Therefore, 
\[\sum_{i \in \mathcal{S}} s^\ell_i = \sum_{j \in [\ell]} X_j \;.\]
From \lemref{lem:corr:gap} and \lemref{lem:function:gap}, it follows that
\[\mu_j = \Ex{X_j} \ge 2 \zeta (1 - \eta),\]
for all $f^j$. 
Then, from \lemref{lem:exp:bound} we have, 
\[\Ex{e^{t(X^j - \mu_j)}} = \Ex{e^{t\xi_{\alpha,\beta}(f^j)-\Ex{t\xi_{\alpha,\beta}(f^j)}}} \le e^{Ct^2} \;.\]
Define $\mathcal{U}_j = (f^1,p^1,v^1, \cdots, f^j, p^j,v^j,f^{j+1})$. 
Now $X_1,\ldots, X_\ell$, $\mu_1,\ldots, \mu_\ell$, and $\mathcal{U}_1, \ldots, \mathcal{U}_\ell$ satisfy the conditions of \thmref{thm:azuma:doob} for $c=\frac{\sqrt{\tau}}{8}$ and $C=\frac{64e^{h\tau/4}}{\tau}$, since $\tau\le\frac{1}{4}n$. 
Now for $\lambda \in [0, 2Cmc] = \left[0, \frac{15\ell}{\sqrt{\tau}}\right]$, we have from \thmref{thm:azuma:doob} that 
\begin{align*}
\PPr{\sum_{i\in\calS} s_i^\ell<2\ell\zeta(1 -\eta) -\lambda}&\le \PPr{\left|\sum_i X_i - \mu_i\right|}\\
&\le 2e^{-\lambda^2 / 4Cm}\\
&<2e^{-\frac{\lambda^2\tau}{256\ell}}. 
\end{align*}
\end{proof}
Next, we show that once a coordinate has been accused of being in $\calS$, then its score will not drastically increase in subsequent rounds of the fingerprinting attack. 
\begin{lemma}
\lemlab{lem:completeness:upper}
\cite{GribelyukLWYZ24}
Let $\tau=\min(\alpha, 1-\beta)$. 
Then for all $\lambda>0$,
\[\PPr{\sum_{i\in\calS}s_i^\ell>\lambda+h\sigma+\frac{h}{\sqrt{\tau}}}\le e^{-\frac{\lambda^2}{4h\ell}}+e^{-\sqrt{\tau}\lambda/4}.\]
\end{lemma}
\begin{proof}
The proof closely follows that of Lemma 2.18 in \cite{SteinkeU15}. 
For each $i \in [n]$, let $j_i$ be as in \lemref{lem:soundness:bound}, so that $i \notin \mathcal{S}^{j_i}$ and $i \in \mathcal{S}^{j_{i - 1}}$, where we define $\mathcal{S}^{\ell + 1} = \emptyset$ and $\mathcal{S}^0 = [n]$. 
By the definition of $j_i$, it follows that for all $i \in \mathcal{S}$, we have $s_i^{j_i - 2} \le \sigma$. 
Thus,
\[\sum_{i \in \mathcal{S}} s_i^{j_i - 1} = \sum_{i \in \mathcal{S}} s_i^{j_i - 2} + a^{j_i - 1} \phi^{j_i - 1}(c_i^{j_i - 1}) \le \sum_{i \in \mathcal{S}}\left(\sigma + \frac{1}{\sqrt{\tau}}\right) \le h \sigma + \frac{h}{\sqrt{\tau}} \;.\]
By \lemref{lem:soundness:bound}, 
\[\PPr{\sum_{i\in S}(s_i^\ell-s_i^{j_i-1})>\lambda}\le e^{-\frac{\lambda^2}{4h\ell}}+e^{-\sqrt{\tau}\lambda/4} \;, \]
from which the desired claim follows. 
\end{proof}
Finally, we show completeness of the fingerprinting attack over multiple rounds. 
\begin{lemma}[Completeness]
\lemlab{lem:completeness}
\cite{GribelyukLWYZ24}
With high constant probability, at the end of $\ell$ rounds, the attack finds a distribution on $\mathbb{Z}^n$ such that the algorithm $\mathcal{A}$ fails with constant probability on an input sampled from this distribution.
\end{lemma}
\begin{proof}
Suppose that at some round $j \in [\ell]$ we have $\max\left(\delta_\alpha^j, \delta_\beta^j\right) = \Omega(1)$. 
Then by standard concentration inequalities, we have with probability at least $0.99$, we can use a constant number of samples to find this distribution $z_{I^{j - 1}}(D_\alpha)$ or $z_{I^{j - 1}}(D_\beta)$ on which the algorithm $\mathcal{A}$ fails.

Thus it remains to consider the other case, where $\delta_\alpha^j, \delta_\beta^j \le c$ over the distribution $D_\alpha^n, D_\beta^n$ for all $j \in [\ell]$. 
By setting $\lambda = \O{\ell}$, we have from \lemref{lem:completeness:gap} that
\[\sum_{i \in \mathcal{S}} s_i^\ell \ge 2 \ell \zeta (1 - \eta) - \lambda = \Omega(\ell),\]
with probability at least $1  - 2\exp(-\Omega(\ell))$. 
On the other hand, we have from \lemref{lem:completeness:upper} that
\[\sum_{i\in\calS}s_i^\ell<\lambda+h\sigma+\frac{h}{\sqrt{\tau}} \le 3 h \sigma\]
with probability at least $1 - 2\exp(-\Omega(\sigma))$ by setting $\lambda = h \sigma$. 
However, this is a contradiction when $\ell \ge C\cdot h \sigma$, for a sufficiently large constant $C$.
\end{proof}

\subsubsection{Final Guarantees for Fingerprinting Attack}
We can now put together the completeness and soundness guarantees for our full guarantees for the fingerprinting attack. 
In particular, this attack must still succeed even with possible interference from the dense part $\bD$ of the sketch matrix $\bA$. 
\begin{theorem}
\thmlab{thm:main-theorem}
\cite{GribelyukLWYZ24}
Let $\mathcal{A}$ be a linear streaming algorithm that solves the $(\alpha + c, \beta - c)$-gap $F_0$ problem for some constants $\alpha, \beta,$ and $c$, where the sketching matrix $\bA \in \mathbb{Z}^{r \times n}$ satisfies $r \ll n$. 
Suppose $\mathcal{A}$ uses any estimator $f: \mathbb{Z}^{r \times n} \to \{-1, +1\}$ and outputs $f(\bA, \bA \bx)$ for each input query $\bx$.

Then, there exists a randomized adaptive attack that, with high constant probability, after making at most $\tO{r^{8}}$ queries to $\mathcal{A}$, constructs a distribution $D$ over $\mathbb{Z}^n$ on which $\mathcal{A}$ fails with constant probability. Furthermore, this attack runs in time polynomial in $r$.
\end{theorem}
\begin{proof}
First, without loss of generality, assume $n = \poly(r)$ since we can restrict queries to the first $\poly(r)$ coordinates and set the rest to zero, effectively focusing on the first $\poly(r)$ columns of the sketching matrix $\bA$.

Next, we prove the correctness of our attack. Suppose the algorithm $\mathcal{A}$ uses estimator $f$ and at time $t$ we sample $\bx \sim D_{p^{t}}$. Consider an algorithm $\mathcal{A}'$ using the same $f$, but receiving input
\[\begin{bmatrix}
\bD \bx' \\
\bS \bx_{\mathcal{S}}
\end{bmatrix},\]
where $\bx' \sim D_{\gamma}^{|D|}$ for fixed $\gamma \in [\alpha, \beta]$, independent of $\bx$. Since $\bD \bx$ and $\bS \bx$ are independent given $p^t$, by \lemref{lem:distribution:tvd} the total variation distance between $
\begin{bmatrix}
\bD \bx^{(t)}_D \\
\bS \bx^{(t)}_{\mathcal{S}}
\end{bmatrix}$ and $\begin{bmatrix}
\bD \bx' \\
\bS \bx^{(t)}_{\mathcal{S}}
\end{bmatrix}$
is at most $\frac{1}{\poly(n)}$ for each $t$. 
Thus,
\[\TVD\left(\{\mathcal{A}(\bx^{(t)})\}_{t=1}^\ell, \{\mathcal{A}'(\bx^{(t)})\}_{t=1}^\ell\right)\leq \ell \cdot \frac{1}{\poly(n)} = \frac{1}{\poly(n)}.\]
Therefore, it suffices to show that interacting with $\mathcal{A}'$ allows us to find an attack distribution on which $\mathcal{A}'$ fails with high constant probability. 
Note that $\mathcal{A}'$ depends only on $\bx_{\mathcal{S}}$. 
By \lemref{lem:soundness}, with probability at least $1 - \frac{1}{n}$, no indices outside $\mathcal{S}$ are falsely identified. 
Moreover, \lemref{lem:completeness} ensures that with high constant probability, the attack identifies some (or all) coordinates in $\mathcal{S}$ and produces a distribution on which $\mathcal{A}'$ fails. 
From this, it follows that $\mathcal{A}$ also fails on this distribution with constant probability. 
Taking a union bound on these events, our attack finds a hard query distribution $\bq$ where $\mathcal{A}$ fails with constant probability.

Finally, regarding complexity, each of the $\ell$ iterations makes $\O{1}$ queries, so total queries are $\O{\ell} = \O{r^8 \log^7 n} = \tO{r^8}$. 
Since only accumulated scores $s_i^t$ are maintained at each iteration $t \in [\ell]$ and $n = \poly(r)$, total runtime is $\O{\ell n} = \poly(r)$.
\end{proof}

\subsection{Moment Matching on Dense Part}
In this section, we give the construction of the hard distribution family that is used for moment matching to attack the dense part of the linear sketch. 

\subsubsection{Distribution via Moment Matching}
We shall utilize the existence of a polynomial with the following guarantees:
\begin{lemma}[Claim 1 of \cite{LarsenWY20}]
\lemlab{lem:cheby:at:zero}
For every $\eps\in\left(2^{-\O{R}},1\right)$, there exists a univariate polynomial $Q$ of degree at most $R-\Omega\left(\sqrt{R\log\frac{1}{\eps}}\right)$ such that
\[|Q(0)|>\eps\cdot\sum_{i=0}^R\left\lvert\binom{R}{i}\cdot Q(i)\right\rvert=\eps.\]
Moreover, for all non-negative integers $t \leq\O{\sqrt{R \log\frac{1}{\eps}}}$, this polynomial $Q$ satisfies 
\[\sum_{i=0}^R(-1)^i\binom{R}{i}\cdot Q(i)\cdot i^t=0.\]
\end{lemma}

\begin{lemma}
\lemlab{lem:moment:match}
\cite{GribelyukLWYZ24}
For any $K>0$, there exist constants $0\le\alpha<\beta\le 1$ and a family $\calD = \{D_p\}_{p\in[\alpha,\beta]}$ of probability distributions with support on $\{-R,\ldots,R\}$ where $R = \O{K^2}$ such that:
\begin{enumerate}
\item
For all $D_p\in\calD$ with $p\in[\alpha,\beta]$, we have $D_p(0)=p$ and $D_p(1) = \Omega(1)$.
\item 
For all $D_p$ with $p\in[\alpha,\beta]$ and for all $X\in[R]$, we have $D_p(X)=D_p(-X)$, i.e., $D_p$ is a symmetric distribution. 
\item 
For all $p,q\in[\alpha,\beta]$, we have $\EEx{X\sim D_p}{X^k}=\EEx{X\sim D_q}{X^k}$ for all $k\in[K]$.
\end{enumerate}
\end{lemma}
\begin{proof}
By taking $R = \Theta(K^2)$ and $\eps = 1/4$ in \lemref{lem:cheby:at:zero}, there exists a univariate polynomial $Q$ of degree at most $R-\Omega\left(\sqrt{R }\right)$ such that
\[|Q(0)|>\frac{1}{4}\cdot\sum_{i=0}^R\left\lvert(-1)^i\binom{R}{i}\cdot Q(i)\right\rvert,\]
and, for every non-negative integer $t\in[K]$,
\[
\sum_{i=0}^R(-1)^i\binom{R}{i}\cdot Q(i)\cdot i^t=0.
\]
For all $i\in[R]$, let $u(i)=(-1)^i\binom{R}{i}\cdot Q(i)$ 
Let $U=\sum_{i\in[R]}|u(i)|$ and suppose without loss of generality that $Q(0)>0$, so that $u(0)>0$ and $u(0)>\frac{1}{4}\cdot U$. 
Moreover, since $\sum_{i\in[R]}u(i)=0$, then by triangle inequality, $u(0)\le\frac{1}{2}\cdot U$. 

Let $\alpha=\left\lvert\frac{u(0)}{2U}\right\rvert$ and $\beta=2\left\lvert\frac{u(0)}{2U}\right\rvert$. 
Furthermore, let
\[B(i)=\begin{cases}
\left\lvert\frac{u(0)}{2U}\right\rvert,\qquad&i=0\\
\frac{1}{2}\left(\frac{1}{2}+\left\lvert\frac{u(1)}{2U}\right\rvert\right)&i=\pm1\\
\frac{1}{2}\left(\left\lvert\frac{u(i)}{2U}\right\rvert\right),\qquad&|i|\in\{2,\ldots,R\}
\end{cases}\]
Then for any $p\in[\alpha,\beta]$, we define 
\[D_p(0)=B(0)+\left(\frac{p}{\alpha}-1\right)\cdot\frac{u(0)}{2U},\]
and
\[D_p(i)=B(i)+\left(\frac{p}{\alpha}-1\right)\cdot\frac{u(i)}{4U},\] for all $i$ with $|i|\in\{1,\ldots,R\}$. 

We first show that $D_p(i)$ is a valid probability distribution. 
Since $\sum_{i=0}^R|u(i)|=U$, then $\sum_{i=0}^R\frac{|u(i)|}{2U}=\frac{1}{2}$. 
Hence,
\[\sum_{i:|i|\in\{0,1,\ldots,R\}}B(i)=\frac{1}{2} + \sum_{j=0}^R\frac{|u(j)|}{2U}=1.\]
Moreover, we have $|u(i)|\le\frac{U}{2}$ for all $i\in[R]$. 
Thus, $B(i)\in[0,1]$ for all $i\in[R]$, which implies that $B$ is a probability distribution. 
Moreover, $\sum_{i = 0}^R \frac{u(i)}{U}=0$. 
Hence, 
\begin{align*}
\sum_{i:|i|\in\{0,1,\ldots,R\}}D_p(i)&=\left(\sum_{i:|i|\in\{0,1,\ldots,R\}}B(i)\right)+\left(\sum_{i:i\in\{0,1,\ldots,R\}}\left(\frac{p}{\alpha}-1\right)\frac{u(i)}{2U}\right)\\
&=\sum_{i:|i|\in\{0,1,\ldots,R\}}B(i)=1.
\end{align*}
Since $\sum_i\frac{|u(i)|}{2U}=\frac{1}{2}$, then it follows that $\frac{|u(i)|}{2U}\le\frac{1}{2}$ for all $i\in[R]$. 
Furthermore, observe that for $p\in[\alpha,\beta]$ with $\alpha=\left\lvert\frac{u(0)}{2U}\right\rvert$ and $\beta=2\left\lvert\frac{u(0)}{2U}\right\rvert$, then $\left(\frac{p}{\alpha}-1\right)\in[0,1]$. 
Hence, $D_p(i)\in[0,1]$ for all $i\in\{-R,\ldots,-1,0,1,\ldots,R\}$. 
Therefore, $D_p$ is a valid probability distribution. 

We now show that the desired properties hold. 
By construction, 
\begin{align*}
D_p(0)&=\left\lvert\frac{u(0)}{2U}\right\rvert+\left(\frac{p}{\alpha}-1\right)\cdot\frac{u(0)}{2U}\\
&=\left\lvert\frac{u(0)}{2U}\right\rvert+\left(\frac{2Up}{|u(0)|}-1\right)\cdot\frac{u(0)}{2U}\\
&=p,
\end{align*}
since $u(0)>0$ by assumption. 
Thus, the first part of the claim follows. 

By construction, we have $D_p$ is symmetric distribution for all $p\in[\alpha,\beta]$, and so the second part of the claim also holds. 

Hence, it remains to prove the third part of the claim. 
Let $p<q$ be fixed, with $p,q\in[\alpha,\beta]$. 
Since $D_p$ and $D_q$ are symmetric distributions, then by the definition of expectation, $\EEx{X\sim D_p}{X^j}=\EEx{X\sim D_q}{X^j}$ if and only if $\sum_{X\in[R]}X^j\cdot(D_p(X)-D_q(X))=0$. 
For each $X\in[R]$, we have $D_p(X)-D_q(X)=\frac{q-p}{\alpha}\cdot\frac{u(X)}{2U}$ by construction. 
Since $u(X)=(-1)^X\binom{R}{X}\cdot Q(X)$, then it suffices to show that $\sum_{X\in[R]}X^j\cdot(-1)^X\binom{R}{X}\cdot Q(X)=0$. 
This is true by the choice of the polynomial $Q$ from \lemref{lem:cheby:at:zero}. 
Therefore, the third part of the claim follows.

As an alternative view, observe that since $D_p$ and $D_q$ are symmetric distributions, then their odd moments are all $0$. 
Thus it remains to match their even moments. 
To that end, we can define a matrix $M\in\mathbb{R}^{K\times R}$ be the following transposition of a Vandermonde matrix:
\[M=\begin{bmatrix}1&1&1&\ldots&1\\
1&4&9&\ldots&R^2\\
1&16&81&\ldots&R^4\\
\vdots&\vdots&\vdots&\ddots&\vdots\\
1&2^{2K}&3^{2K}&\ldots&R^{2K}
\end{bmatrix}.\]
Note that $\EEx{X\sim D_p}{X^{2j}}$ is the $j$-th row of the matrix-vector product $Mv$, where $v_i=2\cdot D_p(i)$. 
Similarly, $\EEx{X\sim D_q}{X^{2j}}$ is the $j$-th row of the matrix-vector product $Mv'$, where $v'_i=2\cdot D_q(i)$. 
Therefore, $\EEx{X\sim D_p}{X^{2j}}=\EEx{X\sim D_q}{X^{2j}}$ if and only if $Mv-Mv'=0^K$, i.e., the all zeros vector of length $K$, or equivalently, that $v-v'$ is in the kernel of $\bM$. 
Now, the $j$-th entry of $Mv-Mv'$ is precisely $2\sum_{X\in[R]}X^j\cdot(D_p(X)-D_q(X))=0$. 
We then proceed as before.
\end{proof}

\subsubsection{Bounding the Total Variation Distance}
We first prove the following structural inequality.
\begin{lemma}
\lemlab{lem:prod:sum:tele}
\cite{GribelyukLWYZ24}
Suppose that for all $i\in[n]$, we have $|a_i+\delta_i|\le 1$. 
Then
\[\left\lvert\prod_{i\in[n]}(a_i+\delta_i)-\prod_{i\in[n]} a_i\right\rvert\le\sum_{i\in[n]}|\delta_i|\cdot e^{\sum_{j\in[n]}|\delta_j|}.\]
\end{lemma}
\begin{proof}
Observe that since $|a_j+\delta_j|\le 1$ for all $j\in[n]$, then
\begin{align*}
\left\lvert\prod_{j<i}(a_j+\delta_j)\prod_{j\ge i}a_j-\prod_{j<i+1}(a_j+\delta_j)\prod_{j\ge i+1}a_j\right\rvert&=|\delta_i|\cdot\prod_{j>i}|a_j+\delta_j|\prod_{j\ge i+1}|a_j|\\
=|\delta_i|\cdot\prod_{j\ge i+1}|a_j|.
\end{align*}
Since $|a_j+\delta_j|\le 1$ for all $j\in[n]$, then by triangle inequality, we have $|a_j|\le1+|\delta_j|$ for all $j\in[n]$. 
Hence,
\begin{align*}
\left\lvert\prod_{j<i}(a_j+\delta_j)\prod_{j\ge i}a_j-\prod_{j<i+1}(a_j+\delta_j)\prod_{j\ge i+1}a_j\right\rvert&\le|\delta_i|\cdot\prod_{j\ge i+1}(1+|\delta_j|)\\
&\le|\delta_i|\prod_{j\in[n]}e^{|\delta_i|}\\
&\le|\delta_i|\cdot e^{\sum_j|\delta_j|}.
\end{align*}
Note that we have
\begin{align*}
\left\lvert\prod_{i\in[n]}(a_i+\delta_i)-\prod_{i\in[n]} a_i\right\rvert=\sum_{i=1}^n\left\lvert\prod_{j<i}(a_j+\delta_j)\prod_{j\ge i}a_j-\prod_{j<i+1}(a_j+\delta_j)\prod_{j\ge i+1}a_j\right\rvert.
\end{align*}
Thus,
\begin{align*}
\left\lvert\prod_{i\in[n]}(a_i+\delta_i)-\prod_{i\in[n]} a_i\right\rvert\le\sum_{i\in[n]}|\delta_i|\cdot e^{\sum_{j\in[n]}|\delta_j|}.
\end{align*}
\end{proof}

\begin{lemma}
\lemlab{lem:distribution:tvd}
\cite{GribelyukLWYZ24}
Let, $p,p'\in [\alpha, \beta]$ be fixed. 
Let $P=D_p$ and $Q=D_{p'}$ be the corresponding pair of probability distributions defined in \lemref{lem:moment:match}. 
Let $P^n$ and $Q^n$ be the probability distributions of vectors of dimension $n$, with each entry drawn independently from $P$ and $Q$, respectively.  
Let $D\in \mathbb{Z}^{r\times n}$ have entries bounded by $[-\poly(n), \poly(n)]$ and suppose 
\[
|\fracpart(y^\top D)_j|^2 \le \frac{1}{s}\cdot\|\fracpart(y^\top D)\|_2^2, 
\]
for all $y\in\mathbb{R}^{r}$ and $j\in[n]$. 
Let $P_{D}$ and $Q_{D}$ be the probability distributions of $Dx$ and $Dx'$ for $x\sim P^n$ and $x'\sim Q^n$ respectively. 
Let $K$ and $R$ be the parameters from \lemref{lem:moment:match} and $s$ be the parameter from \lemref{lem:remove:s:heavy:frac} with $s = \Omega(R^{5/2})$. 
Then the total variation distance between $P_{\bD}$ and $Q_{\bD}$ is at most $n^{\O{r}}\left(n\cdot e^{-\Omega(K)}+e^{-\Omega(K)}\right)$. 
In particular, for $K =\Omega(r\log n)$ and $s = \O{(r\log n)^3}$, we have
\[\TVD(P_{D}(x), Q_{D}(x)) \le \frac{1}{\poly(n)}.\]
\end{lemma}
\begin{proof}
For any fixed $u\in[-\pi,\pi]^r$ and $z=Dx$, the corresponding Fourier coefficient is defined as
\begin{align*}
\widehat{P_{D}}(u)=\EEx{z\sim P_{D}}{e^{-\langle u,z\rangle i}}=\EEx{z\sim P_{D}}{e^{-\langle u^\top Dx\rangle i}}.
\end{align*}
Note that $Dx=\sum_{j\in[n]}D^{(j)}x_j$, where $D^{(j)}$ denotes the $j$-th column of $D$. 
For all $a\in[R]$, let $P_a:=\PPPr{X\sim P}{X=a}$. 
Since each coordinate of $x$ is drawn independently from $P$, it follows that
\begin{align*}
\widehat{P_{D}}&(u)=\prod_{j\in[n]}\Ex{e^{-u^\top D^{(j)}x_j i}}\\
&=\prod_{j\in[n]}\sum_{m\in\{-R,\ldots,-1,0,1,\ldots,R\}}P_m\cdot\left(\cos(\langle u,D^{(j)}\rangle m)+i\cdot\sin(\langle u,D^{(j)}\rangle m)\right).
\end{align*}
Since $P_i=P_{-i}$, then we have
\[\widehat{P_{D}}(u)=\prod_{j\in[n]}\left(P_0+2\sum_{m>0}P_m\cdot\left(\cos(\langle u,D^{(j)}\rangle m)+i\cdot\sin(\langle u,D^{(j)}\rangle m)\right)\right).\]
We define the functions $\fracpart(x)=x-\mathsf{int}(x)\in\left[-\frac{1}{2},\frac{1}{2}\right)$ and $\fracpart_{2\pi}(x)=2\pi\cdot\fracpart\left(\frac{x}{2\pi}\right)\in[-\pi,\pi)$, so that $\cos(m\theta)=\cos\left(m\cdot\fracpart_{2\pi}(\theta)\right)$. 
Hence, we can write
\[\widehat{P_{D}}(u)=\prod_{j\in[n]}\left(P_0+2\sum_{m>0}P_m\cdot\cos\left(m\cdot\fracpart_{2\pi}(\langle u,D^{(j)}\rangle)\right)\right).\]
By utilizing the Taylor expansion $\cos(x)=1-\frac{x^2}{2!}+\frac{x^4}{4!}-\frac{x^6}{6!}+\ldots$, we have
\[\widehat{P_{D}}(u)=\prod_{j\in[n]}\left(P_0+2\sum_{m>0}P_m\cdot\sum_{k\ge 0}\frac{\left(m\cdot\fracpart_{2\pi}(\langle u,D^{(j)}\rangle)\right)^{2k}}{(2k)!}\cdot(-1)^k\right).\]
Since $\cos(x)$ is well-defined, then the summation is absolutely convergent. 
Therefore,
\[\widehat{P_{D}}(u)=\prod_{j\in[n]}\left(P_0+2\sum_{k\ge 0}\left(\sum_{m>0}P_m\cdot m^{2k}\right)\cdot\frac{\left(\fracpart_{2\pi}(\langle u,D^{(j)}\rangle)\right)^{2k}}{(2k)!}\cdot(-1)^k\right).\]
Let 
\begin{align*}
M_P(2k)&=\left(\sum_{m\in\{-R,\ldots,-1,0,1,\ldots,R\}}P_m\cdot m^{2k}\right),\\
M_Q(2k)&=\left(\sum_{m\in\{-R,\ldots,-1,0,1,\ldots,R\}}Q_m\cdot m^{2k}\right)
\end{align*}
be the $2k$-moments of $P$ and $Q$, respectively. 
Hence,
\[\widehat{P_{D}}(u)=\prod_{j\in[n]}\sum_{k\ge 0}M_P(2k)\cdot\frac{\left(\fracpart_{2\pi}(\langle u,D^{(j)}\rangle)\right)^{2k}}{(2k)!}\cdot(-1)^k,\]
and similarly,
\[\widehat{Q_{D}}(u)=\prod_{j\in[n]}\sum_{k\ge 0}M_Q(2k)\cdot\frac{\left(\fracpart_{2\pi}(\langle u,D^{(j)}\rangle)\right)^{2k}}{(2k)!}\cdot(-1)^k.\]
We claim $|\widehat{P_{D}}(u)-\widehat{Q_{D}}(u)|\le n\cdot e^{-\Omega(K)}+e^{-\Omega(K)}$ for all $u\in[-\pi,\pi]^n$. 
To that end, consider a fixed $u\in[-\pi,\pi]^n$. 
Either there exists $j\in[n]$ such that $|\fracpart_{2\pi}(\langle u,D^{(j)}\rangle)|>\frac{1}{4K}$ or for all $j\in[n]$, we have $|\fracpart_{2\pi}(\langle u,D^{(j)}\rangle)|\le\frac{1}{4K}$. 
We analyze these cases separately. 

For the first case, suppose there exists $j\in[n]$ such that 
\[|\fracpart_{2\pi}(\langle u,D^{(j)}\rangle)|>\frac{1}{4K}.\] 
Let $\iota\left(\frac{u}{2\pi}\right)_j:=\fracpart_{2\pi}\langle u,D^{(j)}\rangle$, so that $|\iota\left(\frac{u}{2\pi}\right)_j|>\frac{1}{4K}$ by assumption. 
By the definition $\fracpart_{2\pi}(x)=2\pi\cdot\fracpart\left(\frac{x}{2\pi}\right)\in[-\pi,\pi)$, we have
\[\iota\left(\frac{u}{2\pi}\right)_j^2=\left\lvert4\pi^2\cdot\fracpart\left(\left\langle\frac{u}{2\pi},D^{(j)}\right\rangle\right)^2\right\rvert>\frac{1}{16K^2}.\] 
Since we have $|\fracpart(y^\top D)_j|^2<\frac{1}{s}\cdot\|\fracpart(y^\top D)\|_2^2$ for all vectors $y\in\mathbb{R}^r$, then it follows that 
\[\left\|\iota\left(\frac{u}{2\pi}\right)\right\|_2^2\ge\frac{s}{(16K^2) 4\pi^2} = \frac{K}{64 \pi^2}.\]
by setting $s = \O{K^3}$. 
From before, we have
\begin{align*}
|\widehat{P_{D}}(u)|&=\left\lvert\prod_{j\in[n]}\left(P_0+2\sum_{m>0}P_m\cdot\cos\left(m\cdot\fracpart_{2\pi}(\langle u,D^{(j)}\rangle)\right)\right)\right\rvert\\
&=\prod_{j\in[n]}\left\lvert P_0+2\sum_{m>0}P_m\cdot\cos\left(m\cdot\iota\left(\frac{u}{2\pi}\right)_j\cdot 2\pi\right)\right\rvert. 
\end{align*}
Since $P_0=1-\sum_{m>0} P_m$ and $P_1=\Omega(1)$, then
\begin{align*}
|\widehat{P_{D}}(u)|&\le\prod_{j\in[n]}\left\lvert1-2P_1\left(1-\cos\left(\iota\left(\frac{u}{2\pi}\right)_j\cdot 2\pi\right)\right)\right\rvert.
\end{align*}
By the Taylor expansion $\cos(x)=1-\frac{x^2}{2!}+\frac{x^4}{4!}-\frac{x^6}{6!}+\ldots$ and the inequality $1-x\le e^{-x}$, we have
\begin{align*}
|\widehat{P_{D}}(u)|&\le\prod_{j\in[n]} e^{-\Omega\left(\left(\iota\left(\frac{u}{2\pi}\right)_j\right)^2\right)}. 
\end{align*}
Therefore,
\begin{align*}
|\widehat{P_{D}}(u)|&\le e^{-\Omega\left(\left\|\iota\left(\frac{u}{2\pi}\right)\right\|_2^2\right)}\le e^{-\Omega(K)},
\end{align*}
and by similar reasoning, $|\widehat{Q_{D}}(u)|\le e^{-\Omega(K)}$. 
Hence, in this case, we have that 
\[|\widehat{P_{D}}(u)-\widehat{Q_{D}}(u)|\le e^{-\Omega(K)},\]
by triangle inequality. 

For the other case, suppose that for all $j\in[n]$, we have 
\[|\fracpart_{2\pi}(\langle u,D^{(j)}\rangle)|\le\frac{1}{4K}.\] 
From before, we have
\[\widehat{P_{D}}(u)=\prod_{j\in[n]}\sum_{k\ge 0}M_P(2k)\cdot\frac{\left(\fracpart_{2\pi}(\langle u,D^{(j)}\rangle)\right)^{2k}}{(2k)!}\cdot(-1)^k.\]
Recall that by \lemref{lem:moment:match}, we have $R = \O{K^2}$. 
For all $j\in[n]$, we have $|\fracpart_{2\pi}(\langle u,D^{(j)}\rangle)|\le\frac{1}{4K}$. 
Hence, we have
\begin{align*}
\Bigg|\sum_{k\ge K/2}M_P(2k)&\cdot\frac{\left(\fracpart_{2\pi}(\langle u,D^{(j)}\rangle)\right)^{2k}}{(2k)!}\cdot(-1)^k\Bigg|\\
&\le\sum_{k>K/2}R^{2k}\cdot\frac{1}{(2k)!}\cdot \left(\frac{1}{16K^2}\right)^k.
\end{align*}
We can then use Stirling's approximation to upper bound the higher moments as follows:
\begin{align*}
\Bigg|\sum_{k\ge K/2}M_P(2k)&\cdot\frac{\left(\fracpart_{2\pi}(\langle u,D^{(j)}\rangle)\right)^{2k}}{(2k)!}\cdot(-1)^k\Bigg|\\
&\le\frac{K^{4K}}{(2K)^{2K}/e^{2K}\cdot \sqrt{4\pi K} \cdot (16)^K} \cdot \frac{1}{K^{2K}} \le e^{-\Omega(K)}.
\end{align*}
Next, we apply \lemref{lem:prod:sum:tele} with $a_j=\sum_{k\le K/2}M_P(2k)$ and $\delta_j=\sum_{k>K/2}M_P(2k)$ so that
\[\left\lvert\widehat{P_{D}}(u)-\prod_{j\in[n]}\sum_{k\le K/2}M_P(2k)\cdot\frac{\left(\fracpart_{2\pi}(\langle u,D^{(j)}\rangle)\right)^{2k}}{(2k)!}\cdot(-1)^k\right\rvert\le n\cdot e^{-\Omega(K)}.\]
Similarly,
\[\left\lvert\widehat{Q_{D}}(u)-\prod_{j\in[n]}\sum_{k\le K/2}M_Q(2k)\cdot\frac{\left(\fracpart_{2\pi}(\langle u,D^{(j)}\rangle)\right)^{2k}}{(2k)!}\cdot(-1)^k\right\rvert\le n\cdot e^{-\Omega(K)}.\]
Furthermore, we have $M_Q(2k)=M_Q(2k)$ for $k\le K/2$. 
Hence, by triangle inequality, we have $|\widehat{P_{D}}(u)-\widehat{Q_{D}}(u)|\le n\cdot e^{-\Omega(K)}$. 

Combining both cases, for all $u\in[-\pi,\pi]^n$, we have $|\widehat{P_{D}}(u)-\widehat{Q_{D}}(u)|\le n\cdot e^{-\Omega(K)}+e^{-\Omega(K)}$, as desired. 
Thus,
\begin{align*}
|P_{D}(x)-Q_{D}(x)|&=\left\lvert\frac{1}{(2\pi)^r}\int_{[-\pi,\pi)^r}e^{i\langle u,x\rangle}\left(\widehat{P_{D}}(u)-\widehat{Q_{D}}(u)\right)\,du\right\rvert\\
&\le n\cdot e^{-\Omega(K)}+e^{-\Omega(K)}.
\end{align*}
Finally, we observe that since $D\in \mathbb{Z}^{r\times n}$ with entries bounded in $[-\poly(n), \poly(n)]$, then $P_{D}(x)$ and $Q_{D}(x)$ only have support on a set of size $n^{\O{r}}$. 
Thus, we have that 
\[\TVD(P_{D}(x), Q_{D}(x)) \le n^{\O{r}}\left(n\cdot e^{-\Omega(K)}+e^{-\Omega(K)}\right).\]
Now, observe that in the proof of \lemref{lem:distribution:tvd}, we have $s = \O{K^3}$. 
Hence, by setting $K = r \log n$, we see that $s = \O{(r\log n)^3}$. 
For these choices of parameters of $s, K$, it follows that
\[\TVD(P_{D}(x), Q_{D}(x)) \le \frac{1}{\poly(n)},\]
as desired.
\end{proof}

\subsection{Analysis of Attack on \texorpdfstring{$F_0$}{F0} Estimation}
\seclab{sec:fzero:full:analysis}
\begin{theorem}
\thmlab{thm:f0:main}
\cite{GribelyukLWYZ24}
Let $\calA$ be a streaming algorithm that uses a sketching matrix $A \in \mathbb{Z}^{r \times n}$ with $r\ll n$ and a post-processing function $f$ to solve the $(\alpha + c, \beta - c)$-gap $F_0$ gap norm problem with some constant $\alpha, \beta$ and $c$. 
    
Then, there exists a randomized algorithm that makes an adaptive sequence of queries to $\calA$, and with high constant probability, generates a distribution $D$ on $\mathbb{Z}^n$ such that $\calA$ fails on $D$ with constant probability. 
Moreover, this adaptive attack algorithm makes at most $\tO{r^8}$ queries and runs in $\poly(r)$ time. 
\end{theorem}
\begin{proof}
We first observe that we can assume $n=\poly(r)$ without loss of generality, since we can always query on the first $\poly(r)$ coordinates and set the remaining $\poly(r)$ coordinates of the query vector $\bx$ to $0$. 
This corresponds to attacking the first $\poly(r)$ columns of the sketching matrix $\bA$.
    
Now, suppose that the algorithm $\calA$ uses post-processing function $f(\bA,\bA\bx)$ for each query $\bx$. 
Suppose we sample $\bx \sim D_{p^{t}}$ at time $t$ and consider instead an algorithm $\calA'$ that uses the same estimator $f$, but instead takes the input $\begin{bmatrix}
\bD \bx'\\
\bS \bx_{\calS},
\end{bmatrix}$
where $\bx' \sim \bD_{\gamma}^{|\bD|}$ for a fixed $\gamma \in [\alpha, \beta]$ that is independent of input $\bx$, and $\calS$ is a small disjoint set of coordinates.  
Observe that $\bD\bx'$ and $\bS\bx_{\calS}$ are independent conditioned on $p^t$. 
Hence by \lemref{lem:distribution:tvd}, for each iteration $t$, the total variation distance between $\begin{bmatrix}
\bD \bx^{(t)}_{\calD} \\
\bS \bx^{(t)}_{\calS}
\end{bmatrix}$
and 
$\begin{bmatrix}
\bD \bx' \\
\bS \bx^{(t)}_{\calS}
\end{bmatrix}$
is at most $\frac{1}{\poly(n)}$, where we use $\calD$ to denote the coordinates outside of $\calS$. 
Therefore, 
\[\TVD\left(\{\calA(\bx^{(t)})\}_{t = 1, 2, \cdots, \ell},\{\calA'(\bx^{(t)})\}_{t = 1, 2, \cdots, \ell}\right) \le \ell \cdot \frac{1}{\poly(n)} = \frac{1}{\poly(n)}.\]
Thus, it suffices to show that we can find the attack distribution on which $\calA'$ fails with high constant probability just by interacting with $\calA'$.  
Importantly, $\calA'$ only uses $x_{\calS}$ as input to its computation. 
By \lemref{lem:soundness}, we never falsely accuse any index $i \notin\calS$, with probability at least $1 - \frac{1}{n}$. 
Moreover, by \lemref{lem:completeness}, the attack correctly identifies (some, or all) coordinates $i \in\calS$ and outputs a distribution on which $\calA'$ fails, with high constant probability. 
From the total variation distance bounds in the above discussion, it follows that $\calA$ must also fail on this distribution with high constant probability. 
Conditioning on these events and taking a union bound, it follows that the attack find a query distribution on which $\calA$ fails with high constant probability.

It remains to analyze the query complexity and time complexity of the attack. 
In each of the $\ell$ iterations, the attack makes $\O{1}$ queries. 
Hence, the total number of queries is $\O{\ell} = \O{r^8 \log^7 n} = \tO{r^{8}}$. 

The attack only maintains the accumulated correlation scores $s_i^t$ in each iteration $t \in [\ell]$. 
Thus, the total runtime of the attack is $\O{\ell n} = \poly(r)$, since without loss of generality, it suffices to assume $n=\poly(r)$.
\end{proof}

\section{Dense-Sparse Decomposition}
In this section, we show a more general positive bound by \cite{Ben-EliezerEO22}, which demonstrates that we can do better than the $\tO{\sqrt{m}}$ space algorithm that results from applying the differential privacy technique from \secref{sec:dp:framework:isolate}.
As previously discussed, techniques that rely solely on the flip number $\lambda$ inherently require $\tilde{\Omega}(\sqrt{\lambda})$ space, yielding no significant improvement for moment estimation problems where $\lambda = \Theta(m)$. 
Instead, the approach by \cite{Ben-EliezerEO22} tracks the frequency moments by switching between two distinct regimes. 
In the \emph{sparse regime}, the current frequency vector is explicitly maintained using standard sparse recovery methods. 
In the \emph{dense regime}, the previous differential privacy techniques are utilized. 
This hybrid approach allows \cite{Ben-EliezerEO22} to surpass the previously known $\tilde{\Theta}(m^{1/2})$ space barrier for fixed $p$. 
Specifically, \cite{Ben-EliezerEO22} achieves a space complexity of $\tilde{O}(m^{2/5})$ for estimating $F_2$, and $\tilde{O}(m^{1/3})$ for estimating $F_0$, the number of distinct elements.

To achieve this, \cite{Ben-EliezerEO22} exploits a crucial structural property specific to $F_p$-moment estimation: the $p$-th moment cannot undergo significant changes unless the frequency vector is either small or highly sparse. 
Based on this observation, \cite{Ben-EliezerEO22} introduces a switching strategy that maintains the algorithm in either a \emph{sparse} or \emph{dense} state, according to a threshold $T$. 
If the frequency vector contains at most $T$ non-zero entries, it is deemed sparse; if it exceeds $4T$ non-zeros, it is considered dense. 
Intermediate cases may be temporarily classified either way.

In the sparse regime, \cite{Ben-EliezerEO22} adopts a simple strategy of explicitly storing the frequency vector using a sparse representation, incurring only $\O{T}$ space. 
This allows exact computation of the $F_p$ moment. 
In the dense regime, \cite{Ben-EliezerEO22} applies the robust DP algorithm from \secref{sec:dp} introduced by \cite{HassidimKMMS22}, which uses differential privacy not to protect input data, but to preserve the internal randomness of a collection of $k$ estimators. 
This framework introduces a $\tilde{O}(\sqrt{\lambda})$ overhead due to repeated estimator queries.

The key observation by \cite{Ben-EliezerEO22} is that for dense vectors, the $F_p$ moment cannot change too quickly. 
For example, when $p = 0$ or $p = 1$, $\Omega(T)$ updates are needed to change the moment by a constant factor; for $p = 2$, at least $\Omega(\sqrt{T})$ updates are required. 
Consequently, the effective flip number in the dense regime is much smaller than $\lambda$, and we can reduce the number of estimator queries accordingly, significantly decreasing the overall space complexity.

Finally, we must handle regime transitions carefully. 
When moving from sparse to dense, the frequency vector is known exactly and can be used to initialize the dense regime. 
Conversely, when transitioning from dense to sparse, \cite{Ben-EliezerEO22} applies the sparse recovery techniques of \thmref{thm:sparse:recovery} to reconstruct the frequency vector. 
To detect when this transition should occur, the robust distinct element counting algorithm of \thmref{thm:lzero:est} is run in parallel; this subroutine only needs to be queried every $\Omega(T)$ updates. 
The algorithm appears in full in  \algref{alg:robust:lp}. 

We utilize the following streaming algorithm for sparse recovery, which outputs all the coordinates of an underlying frequency vector if it is sparse. 
\begin{theorem}
\thmlab{thm:sparse:recovery}
\cite{GilbertSTV07}
There exists a deterministic algorithm $\SparseRecover$ that recovers a $k$-sparse frequency vector defined by a turnstile stream of length $n$. 
The algorithm uses $k\cdot\polylog(n)$ bits of space. 
\end{theorem}
We use the following algorithm for $F_p$ estimation for $p\in(0,2]$, though either \thmref{thm:strong:F2} or \thmref{thm:strong:Fp:smallp} could also suffice. 
\begin{theorem}
\thmlab{thm:fp:est}
\cite{KaneNW10a}
For $p\in(0,2]$, there exists an insertion-deletion streaming algorithm $\FPEst$ that outputs a $(1+\eps)$-approximation to $L_p$ with probability at least $1-\delta$ while using $\O{\log\frac{1}{\delta}\left(\frac{1}{\eps^2}\log m+\log\log n\right)}$ bits of space.
\end{theorem}
We restate the guarantees of \thmref{thm:Fp:bigp} for $F_p$ estimation for $p>2$:
\begin{theorem}
\thmlab{thm:fp:est:big}
\cite{GangulyW18}
For $p>2$, there exists an insertion-deletion streaming algorithm $\FPEst$ that outputs a $(1+\eps)$-approximation to $L_p$ with probability at least $1-\delta$ while using $\O{\frac{1}{\eps^2}n^{1-2/p}\log^2 n\log\frac{1}{\delta}}$ bits of space.
\end{theorem}
Similarly, we use the following algorithm for $F_0$ estimation, though \thmref{thm:strong:F0} would also suffice. 
\begin{theorem}
\thmlab{thm:lzero:est}
\cite{KaneNW10b}
There exists an insertion-deletion streaming algorithm $\LZeroEst$ that uses 
\[\O{\frac{1}{\eps^2}\log n\log\frac{1}{\delta}\left(\log\frac{1}{\eps}+\log\log m\right)}\]
bits of space, and with probability at least $1-\delta$, outputs a $(1+\eps)$-approximation to $L_0$.
\end{theorem}

\begin{algorithm}[!htb]
\caption{Adversarially robust $L_p$-estimation}
\alglab{alg:robust:lp}
\begin{algorithmic}[1]
\Require{Turnstile stream of length $m$ for a frequency vector of dimension $n$, input parameter $T$}
\Ensure{Adversarially robust $L_p$-estimation}
\State{$\ell\gets\O{\eps T}$ for $p\in[0,1]$, $\ell\gets\O{\eps^{1+1/p}T^{1/p}}$ for $p>1$, $\FLAG\gets\SPARSE$}
\State{Initialize $\SparseRecover$ with sparsity $4T$}
\State{Initialize $\LZeroEst$ with accuracy $2$ robust to $b:=\frac{m}{\ell}$ queries}
\State{Initialize $\FPEst$ with accuracy $\O{\eps}$ robust to $b$ queries}
\For{each block of $\ell$ updates}
\State{Update $\LZeroEst$, $\SparseRecover$, and $\FPEst$}
\If{$\FLAG=\SPARSE$ at the beginning of the block}
\State{Let $\bg$ be the vector output by $\SparseRecover$}
\State{$\widehat{G}\gets\|\bg\|_p^p$}
\State{\Return $\widehat{G}$}
\Else
\State{Let $\widehat{G}$ be the output of $\FPEst$ at the beginning of the block}
\State{\Return $\widehat{G}$}
\EndIf
\State{Let $Z$ be the output of robust $\LZeroEst$}
\If{$Z>2T$}
\State{$\FLAG\gets\DENSE$}
\Else
\State{$\FLAG\gets\SPARSE$}
\EndIf
\EndFor
\end{algorithmic}
\end{algorithm}

We first recall the following two structural properties about the number of updates needed to change the $L_p$ norm of a $k$-dense vector $\bx$. 
The first statement is for $p\in[0,1]$:
\begin{lemma}[Lemma 16 in~\cite{Ben-EliezerEO22}]
\lemlab{lem:moments:close:small:p}
Let $p\in[0,1]$, $\eps\in(0,1)$, and $k$ be a positive integer. 
Suppose $\bx,\bx'\in\mathbb{Z}^n$ are vectors such that $\bx$ is $k$-dense and $\|\bx-\bx'\|_1\le\eps k$. 
Then $\|\bx'\|_p^p$ is a $(1+\eps)$-approximation to $\|\bx\|_p^p$. 
\end{lemma}
The second statement is for $p\ge 1$:
\begin{lemma}[Lemma 19 in~\cite{Ben-EliezerEO22}]
\lemlab{lem:moments:close:big:p}
Let $p\ge 1$, $\eps\in(0,1)$, and $k$ be a positive integer. 
Suppose $\bx,\bx'\in\mathbb{Z}^n$ are vectors such that $\bx$ is $k$-dense and $\|\bx-\bx'\|_1\le\frac{\eps}{8p}\cdot\left(\frac{\eps m}{4}\right)^{1/p}$. 
Then $\|\bx'\|_p^p$ is a $(1+\eps)$-approximation to $\|\bx\|_p^p$. 
\end{lemma}
We now show correctness of the dense-sparse decomposition in \algref{alg:robust:lp}. 
\begin{lemma}
\lemlab{lem:dense:sparse:correct}
\cite{Ben-EliezerEO22}
With high probability, the output $Z$ by \algref{alg:robust:lp} at each time $t\in[m]$ is a $(1+\eps)$-approximation for $\|\bx\|_p^p$, where $\bx$ is the frequency vector at time $t$. 
\end{lemma}
\begin{proof}
Let $t\in[m]$ be a fixed time. 
Observe that since each block of length $\ell$ makes a single query to $\FPEst$ and $\LZeroEst$, the total number of queries is at most $b=\frac{m}{\ell}$ and thus it suffices for these subroutines to be robust to $b$ such adaptive queries. 
Let $\calE$ be the event that all instances of the algorithms $\LZeroEst$, $\FPEst$, and $\SparseRecover$ succeed at all times $1,\ldots,t-1$ prior to $t$. 
By \thmref{thm:robust:dp}, \thmref{thm:fp:est}, and \thmref{thm:lzero:est}, we have that $\LZeroEst$ and $\FPEst$ succeed with high probability at all times across $b$ adaptive queries. 
By \thmref{thm:sparse:recovery}, we have that $\SparseRecover$ is deterministic and thus succeeds. 
Therefore, $\calE$ holds with high probability, i.e., $\PPr{\calE}\ge1-\frac{1}{\poly(n)}$. 
Let $t_0$ be the beginning of the block that contains time $t$. 

Conditioned on $\calE$, we have that the flag is set to $\SPARSE$ at $t_0$ only if there are at most $2T$ nonzero coordinates at time $t_0$, by the correctness of $\LZeroEst$ at time $t_0$.  
Over the course of a block of length $\ell\le T$, at most $\ell\le T$ more coordinates in the frequency vector can be nonzero. 
Hence, there can be at most $2T+\ell\le 3T$ nonzero coordinates at time $t$ if the flag is set to $\SPARSE$ at time $t_0$. 
By the correctness of $\SparseRecover$ for recovering all coordinates of a $4T$-sparse vector, we can exactly identify the nonzero coordinates, thereby recovering $\bx$ exactly, i.e., the vector $\bg$ returned by $\SparseRecover$ is identical to $\bx$. 
Therefore, $Z=\|\bg\|_p^p=\|\bx\|_p^p$. 

In the other case, the flag at time $t_0$, the beginning of the block of time $t$, must have been $\DENSE$. 
In this case, there must have been at least $\frac{1}{2}\cdot T$ nonzero coordinates by the correctness of $\LZeroEst$ at time $t_0$. 
Let $\bh$ be the frequency vector at time $t_0$, so that $\widehat{G}$ is a $\left(1+\frac{\eps}{10}\right)$-approximation to $\|\bh\|_p^p$. 
Since the block has at most $\ell=\O{\eps\cdot m^{c/p}}$ coordinates, then $\bx$ is at least $\frac{1}{4}\cdot T$-dense. 
Hence by \lemref{lem:moments:close:small:p} for $p\in[0,1]$ and \lemref{lem:moments:close:big:p} for $p\ge 1$, $\|\bh\|_p^p$ is a $(1+\eps)$-approximation to $\|\bx\|_p^p$. 

Therefore, in both cases, $Z$ is a $(1+\eps)$-approximation of $\|\bx\|_p^p$, as desired. 
The desired claim then follows from a union bound over all $t\in[m]$. 
\end{proof}
Next, we analyze the space complexity of the dense-sparse decomposition algorithm in \algref{alg:robust:lp}. 
\begin{lemma}
\lemlab{lem:dense:sparse:space}
\cite{Ben-EliezerEO22}
The space complexity of \algref{alg:robust:lp} is 
\begin{itemize}
\item 
$\tO{\frac{1}{\eps^{5/3}}\cdot m^{1/3}}$ bits of space for $p\in[0,1]$, 
\item 
$\tO{\frac{1}{\eps^{(5p+1)/(2p+1)}}\cdot m^{p/(2p+1)}}$ bits of space for $p\in[1,2]$,
\item 
and $\tO{\frac{1}{\eps^{(5p+1)/(2p+1)}}\cdot m^{p/(2p+1)}\cdot n^{1-5/(2p+1)}}$ bits of space for $p>2$.
\end{itemize}
\end{lemma}
\begin{proof}
By \thmref{thm:robust:dp}, we require $\tO{\sqrt{\frac{m}{\ell}}}$ instances of $\LZeroEst$ and $\FPEst$. 
By \thmref{thm:lzero:est}, each independent instance of $\LZeroEst$ uses $\O{\frac{1}{\eps^2}\log^2(nm)}$ bits of space. 
By \thmref{thm:sparse:recovery}, $\SparseRecover$ recovering $\O{T}$ coordinates requires $T\cdot\polylog(m)$ bits of space for $\log m=\Theta(\log n)$. 

For $p\in[0,1]$, we set $\ell=\O{\eps T}$. 
By \thmref{thm:fp:est}, each instance of $\FPEst$ uses $\O{\frac{1}{\eps^2}\log^2(nm)}$ bits of space. 
Therefore, by setting $T=\O{\eps^{-5/3}m^{1/3}}$, the total space is 
\[\tO{\frac{1}{\eps^{5/3}}\cdot m^{1/3}}.\]

For $p\in[1,2]$, we set $\ell=\O{\eps^{1+1/p}T^{1/p}}$. 
By \thmref{thm:fp:est}, each instance of $\FPEst$ uses $\O{\frac{1}{\eps^2}\log^2(nm)}$ bits of space. 
Therefore, by balancing $T$ appropriately, the total space is 
\[\tO{\frac{1}{\eps^{(5p+1)/(2p+1)}}\cdot m^{p/(2p+1)}}.\]

Finally, for constant $p>2$, we again set $\ell=\O{\eps^{1+1/p}T^{1/p}}$. 
By \thmref{thm:fp:est:big}, each instance of $\FPEst$ uses $\tO{\frac{1}{\eps^2} n^{1-2/p}}$ bits of space. 
Hence, by balancing $T$ appropriately, the total space is 
\[\tO{\frac{1}{\eps^{(5p+1)/(2p+1)}}\cdot m^{p/(2p+1)}\cdot n^{1-5/(2p+1)}}.\]
\end{proof}
The full guarantees then follow from combining \lemref{lem:dense:sparse:correct} and \lemref{lem:dense:sparse:space}. 
\begin{theorem}
\cite{Ben-EliezerEO22}
There exists an algorithm that with high probability, outputs a $(1+\eps)$-approximation for $\|\bx\|_p^p$, where $\bx$ is the frequency vector at time $t$ at each time $t\in[m]$. 
The algorithm uses 
\begin{itemize}
\item
$\tO{\frac{1}{\eps^{5/3}}\cdot m^{1/3}}$ bits of space for $p\in[0,1]$, 
\item
$\tO{\frac{1}{\eps^{(5p+1)/(2p+1)}}\cdot m^{p/(2p+1)}}$ bits of space for $p\in[1,2]$, 
\item
$\tO{\frac{1}{\eps^{(5p+1)/(2p+1)}}\cdot m^{p/(2p+1)}\cdot n^{1-5/(2p+1)}}$ bits of space for $p>2$.
\end{itemize}
\end{theorem}
Thus the dense-sparse decomposition achieves adversarial robustness on turnstile streams using more efficient space than the $\tO{\sqrt{m}}$ space algorithm that results from applying the differential privacy technique from \secref{sec:dp}. 
Finally, we remark by using heavy-hitter algorithms to further track large updates to a small number of coordinates, \cite{WoodruffZ24} is able to further decrease the flip number of the dense regime and therefore achieve slightly better space dependencies in $m$. 
Quantitatively, the improvement of \cite{WoodruffZ24} is mild, e.g., roughly $\tO{m^{0.373}}$ rather than the bound of $\tO{m^{0.375}}$ by \cite{Ben-EliezerEO22} for $F_p$ moment estimation for $p=1.5$. 
However, these results nevertheless demonstrate that the dense-sparse decomposition does not serve as a natural barrier. 

\chapter{White-Box Model}
\chaplab{chap:white:box}

\begin{chapterbannerbox}
\centering
Even if an algorithm's entire data structure is revealed to an adversary, cryptographic tools can still provide robustness.  
\end{chapterbannerbox}
\vspace{0.4in}

In this chapter, we focus on the \emph{white-box} adversarial model, where an adversary $\Adv$ again has repeated interactions with an algorithm $\Alg$ through a data stream that represents the queries of $\Adv$. 
As in the black-box setting defined in \chapref{chap:black:box}, the adversary $\Adv$ has access to the previous outputs of the algorithm $\Alg$. 
Additionally, the adversary $\Adv$ now also has the internal state of the algorithm $\Alg$ at each time step over the duration of the stream. 

Formally, the white-box adversarial model can be defined through a two-player game between a streaming algorithm $\Alg$ and a source $\Adv$ of adaptive or adversarial input to $\Alg$. 
As in the black-box setting, a query function $\calQ$ is fixed before the game begins. 
Subsequently, the game proceeds over $m$ rounds, so that in the $t$-th round:
\begin{enumerate}
\item
$\Adv$ computes an update $s_t$ for the stream, which possibly depends on all previous stream updates and observations by $\Adv$. 
\item
$\Alg$ updates its internal data structures $\calD_t$ with $s_t$, possibly drawing a fresh batch $R_t$ of random bits, and outputs a response $Z_t$.
\item
$\Adv$ observes the response $Z_t$, the data structures $\calD_t$, and the random bits $R_t$.
\end{enumerate}
The algorithm $\Alg$ is permitted space sublinear in the size of the input $m$ and only a single pass over the stream. 
As before, the goal of the adversary $\Adv$ is to compel an incorrect response $Z_t$ to the query $\calQ$ at some time $t\in[m]$ throughout the stream through its choices of $s_1,\ldots,s_m$. 

\paragraph{Applications of white-box adversaries.}
The white-box adversarial model captures situations in which an adversary can observe and adapt to the internal state of an algorithm, going beyond what is possible in the black-box setting. 
This added flexibility allows the model to reflect a range of realistic scenarios in which future inputs are influenced by the algorithm’s own behavior.

A canonical example comes from dynamic algorithms. 
Here, the goal is to maintain a data structure that produces correct answers at every time step $t \in [m]$ as updates $u_1,\ldots,u_m$ arrive one by one, while keeping the overall running time or worst-case update time small. 
In some applications, space is also a limiting factor and the data structure must use space sublinear in the input size, though this is not always required. 
In this setting it is common to consider adaptive adversaries~\cite{Chan10,Wajc20,ChanH21,RoghaniSW22} that choose each update after inspecting the current state of the data structure, naturally fitting the white-box adversarial model.

More broadly, an algorithm’s internal state may play an active role in shaping future inputs. 
To see this, consider a distributed streaming scenario in which a central coordinator seeks to compute statistics over data generated by many remote users. 
To save space, the coordinator may avoid storing all updates, and to reduce communication costs it may share parts of its internal state $S$, e.g., random seeds or sketch parameters, with the users. 
The users may then rely on this shared state in their own local processes, which in turn influences how subsequent data is generated. 
In this case, the data stream depends directly on components of the coordinator’s internal state, a dependence that is naturally captured by the white-box model. 
If some users are malicious and deliberately exploit knowledge of $S$ to cause the coordinator to fail, the resulting inputs are both state-dependent and adversarial. 
Similar considerations apply when the coordinator’s state is stored in the cloud and is visible to all users.

Related ideas also appear in the pan-private streaming model~\cite{DworkNPRY10}, which allows the internal state of an algorithm to be partially or fully exposed. 
This model is motivated by applications involving distributed data curators such as hospitals, government agencies, or large technology companies. 
As noted in~\cite{DworkNPRY10}, a data curator may be pressured to allow data to be used in ways that go beyond its original purpose, potentially influencing the distribution of future inputs. 
In a similar spirit,~\cite{MirMNW11} studies problems such as distinct counting and heavy-hitter detection when the internal state of the algorithm is revealed, motivated by scenarios in which an insider manipulates traffic patterns while probing a system that tracks user statistics. 
While the focus there is on privacy, these examples also highlight the relevance of a white-box adversarial perspective, in which future inputs need not be independent of previously revealed internal information.

White-box adversaries have also become central in the study of adversarial robustness in machine learning. 
Many of the most effective attacks rely on access to a model’s parameters and training weights in order to craft inputs that cause misclassification while remaining close to a given example~\cite{BiggioCMNSLGR13,SzegedyZSBEGF14,GoodfellowSS14}. 
Subsequent work has shown that such attacks can produce perturbations that are almost imperceptible to humans, both in digital images~\cite{SzegedyZSBEGF14,HuangPGDA17} and in physical settings~\cite{KurakinGB17a,SharifBBR16,AthalyeEIK18}. 
This line of work has led to a substantial literature on designing learning algorithms that are robust to white-box attacks, e.g.,~\cite{IlyasEM18,MadryMSTV18,SchmidtSTTM18,TramerKPGBM18,CubukZMVL18,KurakinGB17,LiuCLS17}.

Finally, white-box considerations also arise in persistent data structures, whose purpose is to provide access to past versions of a data structure while keeping space and query time under control~\cite{DriscollSST89,FiatK03,Kaplan04}. 
In collaborative systems that provide version control over shared repositories, the underlying persistent data structures may be visible to all users. 
Updates made by different collaborators may then depend explicitly on previously stored states, again fitting naturally within the white-box adversarial framework.

\section{Lower Bounds}
A natural question is whether there exist robust streaming algorithms against white-box adversaries for standard problems in the streaming model, such as $L_p$ norm estimation or distinct element estimation. 
Somewhat surprisingly, there exists a general technique to prove lower bounds for randomized white-box robust algorithms through lower bounds for deterministic two-player communication problems. 
Informally, we say a streaming algorithm solves a two-player communication game if there is a reduction from the communication problem to the streaming problem. 
The players encode their private inputs as parts of a single data stream and jointly simulate the streaming algorithm on this stream. 
The memory state of the streaming algorithm is passed between the players, and its final output is used to recover the value of the communication game. 
Under this reduction, the space used by the streaming algorithm corresponds directly to the communication cost of the protocol. 
\begin{restatable}{theorem}{thmtwopinf}
\thmlab{thm:two:p}
Suppose there exists a randomized white-box robust streaming algorithm that uses $S(n,\eps)$ space and solves a two-player communication game with $S(n,\eps)$ bits of communication with some constant probability $p\in\left(\frac{1}{2},1\right]$. 
Then there exists a deterministic protocol for the two-player communication game that uses $S(n,\eps)$ bits of communication and succeeds on at least a $p$ fraction of all possible inputs. 
\end{restatable}
\begin{proof}
Suppose there exists a randomized white-box robust streaming algorithm $\calA$ that uses $S(n,\eps)$ space and solves a two-player communication game $\calP$ with $S(n,\eps)$ bits of communication with some constant probability $p\in\left(\frac{1}{2},1\right]$. 
We say a random string $R$ chosen by the first player is \emph{good}, if $R$ results in a two-player protocol with the correct answer, over all possible inputs to the second player and at least a $p$ fraction of the possible random strings chosen by the second player. 
By the definition of white-box robustness, the algorithm succeeds with probability at least $p$ even if the second player's input is chosen adaptively by an adversary who observes the internal state of the algorithm. 
Thus, the expected success probability over the first player's random string $R$, evaluated against the worst-case second-player input, is at least $p$. 
An averaging argument then implies that for any local input of the first player, there exists at least one good random string $R$.

Now, the first player can enumerate over all possible inputs to the second player as well as all possible random strings and then select a good random string to generate a state $\calS$ of the algorithm $\calA$ on their local input, which uses $S(n,\eps)$ bits to store.  
The first player then sends the state $\calS$ of the algorithm to the second player, using $S(n,\eps)$ bits of communication. 
Because the chosen state $\calS$ guarantees a success probability of at least $p$ over the second player's randomness for every possible pair of inputs, an averaging argument over all inputs implies there exists a globally fixed random string for the second player that achieves the correct answer on at least a $p$ fraction of all possible inputs. 
The second player can then update the algorithm with their input and a fixed string. 
Consequently, the white-box robust algorithm results in a deterministic protocol that solves the one-way two-player communication game with $S(n,\eps)$ bits of communication that succeeds on at least a $p$-fraction of all inputs.
\end{proof}

We demonstrate \thmref{thm:two:p} by establishing a lower bound for $F_p$ moment estimation, achieved through a reduction from the following version of the Gap Equality problem:

\begin{definition}[Gap Equality Problem]
In the deterministic Gap Equality problem $\detgapeq_n$, Alice receives a vector $\bu\in\{0,1\}^n$ with $\|\bu\|_0=\frac{n}{2}$ and Bob receives a vector $\bv\in\{0,1\}^n$ with $\|\bv\|_0=\frac{n}{2}$. 
Given the promise that either $\bu=\bv$ or $\HAM(\bu,\bv)\ge\frac{n}{10}$, the two players must perform a deterministic protocol to identify which case characterizes the input.  
\end{definition}
When the communication is not required to be deterministic, it is possible to use $\O{1}$ bits of communication and succeed with probability at least $\frac{2}{3}$, by hashing each player's input into a universe of constant size, such as through Karp-Rabin fingerprints~\cite{KarpR87}. 
However, the deterministic Gap Equality problem $\detgapeq_n$ has communication complexity $\Omega(n)$:
\begin{theorem}
\thmlab{thm:detgapeq:cc}
\cite{BuhrmanCW98,BlaisBM12}
The communication complexity of the deterministic Gap Equality problem $\detgapeq_n$ is $\Omega(n)$. 
\end{theorem}
We now use the deterministic Gap Equality problem to show lower bounds for white-box robust streaming algorithms for $F_p$ moment estimation. 
Our lower bounds are parameterized by the number $k$ of private bits the algorithm may utilize, noting that $k=0$ corresponds to the standard white-box model previously discussed. 
\begin{theorem}
\thmlab{thm:fp:lb}
\cite{AjtaiBJSSWZ22}
For any $p\ge 0$ with $p\neq 1$, there exists a constant $C_p>1$ such that any white-box robust algorithm that outputs a $C_p$-multiplicative approximation to the $F_p$ moment estimation problem with probability at least $\frac{9}{10}$ using $k$ hidden private bits must use $\Omega\left(\frac{n}{2^k}\right)$ bits of space.
\end{theorem}
\begin{proof}
Consider an instance of $\detgapeq_n$, so that Alice and Bob are given vectors $\bu,\bv\in\{0,1\}^n$ with $\|\bu\|_0=\|\bv\|_0=\frac{n}{2}$. 
Now if $\|\bu+\bv\|_0\ge\frac{n}{2}+\frac{n}{10}$ so that the Hamming distance between $u$ and $v$ is at least $\frac{n}{10}$, then there exists a constant $C_p>1$ such
\[C_p\cdot\|2\bu\|_p\le\|\bu+\bv\|_p\]
for $p\in[0,1)$ and
\[C_p\cdot|\bu+\bv\|_p\le\|2\bu\|_p\]
for $p>1$. 
Let $\gamma$ be the hidden constant in \thmref{thm:detgapeq:cc}, so that any deterministic protocol that solves $\detgapeq_n$ requires at least $\gamma n$ bits of communication. 

Suppose by way of contradiction, there exists a white-box robust streaming algorithm $\calA$ that outputs a $C_p$-multiplicative approximation to the $F_p$ moment estimation problem, with probability at least $\frac{9}{10}$, while using $k$ hidden private bits and $\frac{\gamma}{100}\frac{n}{2^k}$ bits of space. 
Given the input vector $\bu$ for $\detgapeq_n$, Alice creates a stream $S$ that induces the frequency vector $\bu$ by inserting each element $i\in[n]$ for which $u_i=1$ into the stream $S$. 

Since Alice and Bob must solve $\detgapeq_n$ deterministically, Alice and Bob must utilize the $k$ hidden private bits given to $\calA$ in a deterministic manner. 
To that end, Alice runs a separate instance of $\calA$ on $S$ for each of the $2^k$ possible fixings of the $k$ random bits. 
For each $i\in[2^k]$ corresponding to the $i$-th fixing of the $k$ hidden bits, Alice deterministically chooses a sequence $R_i$ of public random bits such that $\calA$ is correct for at least $\frac{9}{10}$ fraction of the possible values of $\bv$, if such a sequence exists. 
Otherwise if such a sequence does not exist, then Alice instead sets the sequence $R_i$ to be the all zeros sequence. 
For each $i\in[2^k]$, Alice then sets the $i$-th fixing of the deterministic fixing $R_i$ as the public random bits input to $\calA$.
Alice then runs the algorithm $\calA$ with public random bits $R_i$ on the input $\bu$ to create a state $\sigma_i(\bu)$. 
Finally, Alice sends the set of states $\sigma_1(\bu),\ldots,\sigma_{2^k}(\bu)$ to Bob. 

Bob then continues updating the stream in a manner that is consistent with the frequency vector $\bv$, so that Bob inserts element $i\in[n]$ into the stream $S$ if $v_i=1$. 
Thus, the underlying frequency vector induced by the stream corresponds to $\bu+\bv$. 
For each $i\in[2^k]$, Bob takes the state $\sigma_i(u)$ passed from Alice and continues running $\calA$ on the stream and queries the algorithm. 
Since a $C_p$-approximation to $\|\bu+\bv\|_p$ distinguishes whether $\bu=\bv$ or $\HAM(\bu,\bv)\ge\frac{n}{10}$, then for each $i\in[2^k]$, Bob can determine whether the $i$-th instance of $\calA$ suggests $\bu=\bv$ or $\HAM(\bu,\bv)\ge\frac{n}{10}$. 

By the guarantee that $\calA$ is white-box robust with probability at least $\frac{9}{10}$, then at least $\frac{9}{10}$ fraction of the possible fixings of the random bits will also correspond to a correct output for \emph{all possible values} of $\bv$, across at least $\frac{9}{10}$ fraction of the additional possible public random bits used by the algorithm. 
Consequently, at least $\frac{9}{10}$ fraction of the states $\sigma_i(\bu)$ sent by Alice will succeed for all possible values of $\bv$, across $i\in[2^k]$. 
Therefore, at least $\frac{9}{10}$ fraction of the $2^k$ outputs by Bob will be correct, allowing the protocol to distinguish whether $\bu=\bv$ or $\HAM(\bu,\bv)\ge\frac{n}{10}$. 

From our assumption, each instance of $\calA$ uses $\frac{\gamma}{100}\frac{n}{2^k}$ bits of space.  
Thus the states $\sigma_i(\bu)$ sent by Alice use $\frac{\gamma n}{100}$ total communication across the states $i\in[2^k]$. 
However, by definition of $\gamma$ and \thmref{thm:detgapeq:cc}, any deterministic protocol that solves $\detgapeq_n$ requires at least $\gamma n$ bits of communication, which is a contradiction. 
Thus, it follows that $\calA$ must use $\Omega\left(\frac{n}{2^k}\right)$ space.
\end{proof}
We remark that the lower bound of \thmref{thm:fp:lb} relies on the derandomization of two-party communication protocols, and this technique seems to be inherently limited to establishing lower bounds only within a specific constant factor $C_p$. 
\cite{EfremenkoKSZ26} strengthened this result to show that estimating the $F_p$ moment to within any constant factor in the white-box model requires $\Omega(n)$ memory. 
In a similar vein, unconditional white-box streaming lower bounds have been established for other classical problems, such as approximating the length of the longest increasing subsequence (LIS) \cite{GalKSY26}.

\section{Upper Bounds}

\subsection{\texorpdfstring{$F_0$}{F0} Estimation Against Computationally-Bounded White-Box Adversaries}
In this section, we present the streaming algorithm by \cite{AjtaiBJSSWZ22} for estimating the $F_0$ value that is robust against white-box adversaries. 
As established by the lower bound in \thmref{thm:fp:lb}, it is impossible to approximate the $F_0$ value of a frequency vector $\bx$ to an arbitrarily small constant in sublinear space. 
However, perhaps unexpectedly, it is still possible to achieve a multiplicative approximation within a factor of $n^{\eps}$ for any small constant $\eps > 0$ provided the adversary is \emph{computationally bounded}. 
This assumption is in line with standard practices in cryptography. 
Our notion of a computationally bounded adversary will be based on the hardness of the Short Integer Solution (SIS) problem, a well-known assumption from lattice-based cryptography, defined as follows:

\begin{definition}[Short Integer Solution (SIS) Problem]
\deflab{def:SIS}
Let $n,m,q$ be integers and let $\beta >0$. 
Given a uniformly random matrix $\bA \in \mathbb{Z}^{n \times m}_q$, the short integer solution (SIS) problem is to find a nonzero integer vector $\bz \in \mathbb{Z}^m$ such that $\bA\bz\equiv\mathbf{0}^n\bmod q$ and $\|\bz\|_2 \le \beta$.
\end{definition}
We remark that in \defref{def:SIS}, the parameters can be chosen so that there does exist such a nonzero integer vector $\bz \in \mathbb{Z}^m$ such that $\bA\bz\equiv\mathbf{0}^n\bmod q$ and $\|\bz\|_2 \le \beta$, by a simple Pigeonhole argument, so the defined task is not vacuously impossible. 

Since the foundational work of Ajtai~\cite{Ajtai96}, it has been known that the SIS problem exhibits average-case to worst-case hardness. 
Specifically, for suitable choices of parameters, solving an ``average'' case instance of SIS is at least as difficult as approximating certain key lattice problems in the worst case.

\begin{theorem}
\thmlab{thm:SIS_hardness}
\cite{MicciancioP13}
Let $n$ be an integer, and suppose $m$, $\beta$, and $q$ are polynomially bounded in $n$, with the additional condition that $q \geq n \cdot \beta$. 
Then, solving the SIS problem with parameters $n, m, q, \beta$ and with non-negligible success probability is at least as hard as approximating the Shortest Vector Problem (SVP) within a factor $\gamma$, where $\gamma \in \poly(n)$.
\end{theorem}

In the cryptographic literature, lattice-based cryptographic constructions are typically designed under the assumption that approximating lattice problems within a factor $\gamma < 2^{o(n \log \log n / \log n)}$ is computationally hard. 
The best known polynomial-time algorithms, such as the LLL algorithm \cite{Vaikuntanathan15}, only achieve an approximation factor of $\gamma = 2^{\O{n \log \log n / \log n}}$. 
Any improvement beyond this barrier to a significantly smaller $\gamma$ would constitute a major breakthrough in the field. 
Based on this, we adopt the following computational hardness assumption, which implies that breaking our algorithms would require progress that challenges fundamental cryptographic assumptions:

\begin{assumption}
\assumlab{assumption:lattice}
No polynomial-time algorithm can approximate the Shortest Vector Problem (SVP) in $n$ dimensions within a factor of $\gamma = 2^{o(n (\log \log n) / \log n)}$.
\end{assumption}

The $F_0$ estimation algorithm by \cite{AjtaiBJSSWZ22} only requires computational hardness for significantly smaller values of $\gamma$. 
The algorithm proceeds as follows. 
First, partition the universe $[n]$ into $n^{1-\eps}$ contiguous blocks, each consisting of $n^{\eps}$ coordinates. 
For each block, maintain a separate vector by applying updates using a fixed sketching matrix $\bA$, which is derived from the hardness of the SIS problem. 
At the end of the stream, the algorithm outputs an estimate equal to the number of these $n^{1-\eps}$ sketches that are nonzero. 
Importantly, the same matrix $\bA$ is reused across all blocks, as described in more detail in \algref{alg:l0:ub}. 

\begin{algorithm}[!htb]
\caption{$F_0$ Estimation Algorithm on Turnstile Streams from Computationally-Bounded White-Box Adversaries}
\alglab{alg:l0:ub}
\begin{algorithmic}[1]
\Require{Universe size $n$, accuracy parameter $\eps$, and a stream of updates $u_1, u_2, \ldots$, where each $u_t \in [n]$ denotes an update to a coordinate of the frequency vector $\bx$}
\Ensure{An $n^\eps$-multiplicative approximation of $\|\bx\|_0$}
\State{Let $\bA \in \mathbb{Z}^{n^{c\eps} \times n^\eps}_q$ be a uniformly random matrix, where $q = \poly(n)$ and $0 < c < \frac{1}{2}$}
\State{Partition $[n]$ into $n^{1 - \eps}$ consecutive chunks, each of size $n^\eps$}
\State{Initialize $n^{1 - \eps}$ sketch vectors in $\mathbb{Z}^{n^{c\eps}}_q$, all initially $\vec{0}$, one per chunk}
\For{each update $u_t$ in the stream}
\State{Determine the chunk index $i$ and local coordinate $k$ within that chunk corresponding to $u_t$}
\State{Update the $i$-th sketch vector by adding $u_t \cdot \bA_k$, where $\bA_k$ is the $k$-th column of $\bA$}
\EndFor
\State{\Return the number of nonzero sketch vectors at the end of the stream}
\end{algorithmic}
\end{algorithm}

We now argue that if the final frequency vector $\bx$ satisfies $\|\bx\|_{\infty} \le \poly(n)$, then our algorithm provides an $n^\eps$-multiplicative approximation for the $F_0$ value. 
Additionally, the algorithm can be made more space-efficient under the \emph{random oracle model} introduced by Bellare and Rogaway~\cite{BellareR93}.
In the random oracle model, both the algorithm and the adversary have access to a public random function. 
Each query to this oracle returns a uniformly random output from a fixed domain, and repeated queries with the same input yield the same output. 
This model is widely used in cryptographic design~\cite{BellareR93, BellareR96, CanettiGH2004, KoblitzM2015}, and in practice, hash functions like SHA256 can serve as heuristic implementations of such oracles.
We remark that the random oracle assumption was also removed by \cite{FengJW24}, who also gave algorithms with faster update time. 
Instead of the random oracle model, \cite{FengJW24} leverages a suitably chosen fully homomorphic encryption (FHE) scheme satisfying mild structural properties to build the relevant hash functions.
These structural properties are currently satisfied by most schemes based on LWE and Ring-LWE.
At a high level, this construction bears resemblance to classical approaches for building collision-resistant functions from private information retrieval (PIR) schemes, where the PIR is instantiated using fully homomorphic encryption. 

We emphasize that to apply \thmref{thm:SIS_hardness}, it suffices for the approximation factor $\gamma$ to be $\poly(n)$. 
Moreover, our algorithm remains valid in the \textbf{turnstile} streaming setting, where updates to the coordinates of $\bx$ may be both positive and negative. 
This is because \thmref{thm:SIS_hardness} only requires $\|\bx\|_{\infty} \le \poly(n)$, not a bound on the signs of the entries.

\begin{restatable}{theorem}{thmlzeroub}
\thmlab{thm:lzero:ub}
\cite{AjtaiBJSSWZ22}
Let $c \in \left(0,\frac{1}{2}\right)$, and suppose that \assumref{assumption:lattice} holds. 
Then there exists a streaming algorithm on turnstile streams generated by polynomial time-bounded white box adversaries, which outputs an $n^{\eps}$-multiplicative approximation to the $F_0$ value in the turnstile streaming model.
Moreover, the algorithm has a space complexity of $\tilde{\mathcal{O}}(n^{1-\eps + c\eps} + n^{(1+c)\eps})$. 
In the random oracle model, the space usage improves to $\tilde{\mathcal{O}}(n^{1-\eps + c\eps})$.
\end{restatable}
\begin{proof}
Consider \algref{alg:l0:ub}. 
From \thmref{thm:SIS_hardness} and \assumref{assumption:lattice}, no polynomial-time adversary can find a nonzero integer vector $\bz$ with $\|\bz\|_{\infty} \le \poly(n)$ such that $\bA\bz = 0$. 
Therefore, if a sketch vector (for a given chunk) equals zero, it implies that all coordinates within that chunk have zero frequency in $\bx$ at the end of the stream. 
Conversely, if the sketch vector is nonzero, at least one coordinate in the corresponding chunk must be nonzero in $\bx$. 
Since each chunk contains at most $n^\eps$ coordinates, this yields a multiplicative approximation of at most $n^\eps$, as desired.

For the space complexity, observe that the matrix $\bA$ does not need to be stored explicitly if we adopt the random oracle model; its columns can be generated on demand via oracle queries. 
Hence, the primary space usage arises from maintaining $n^{1-\eps}$ vectors of dimension $n^{c\eps}$, one for each chunk of size $n^\eps$.
\end{proof}

\subsection{Vector Recovery}

We can further work with the SIS problem without going through the SVP problem. 
To this end, we remark the best known algorithm for solving the SVP within a $\poly(n)$ approximation is due to \cite{AggarwalDRS15} and runs in $\tilde{\mathcal{O}}(2^n)$ time.

\begin{assumption}
\assumlab{asm:strong}
Given $n \in \mathbb{N}$, there exist parameters $m, \beta, q \le \poly(n)$ with $q \ge n \cdot \beta$, so that no adversary running in time $o(2^n)$ can solve the $\mathsf{SIS}_{n, m, q, \beta}$ problem with non-negligible probability.
\end{assumption}

We now describe the first of multiple algorithms that follow a unified two-stage framework. 
The framework runs two procedures in parallel: (i) a detection algorithm that determines whether the input belongs to a structured class, such as being sparse, low rank, or both, and (ii) a deterministic recovery algorithm that efficiently reconstructs the input once membership in this class is certified. 
The detection procedure relies on the assumed hardness of the SIS problem to withstand adaptive, white-box adversaries, while the recovery procedure is fully deterministic and efficient in both time and space, ensuring robustness once the structure is verified. 
In this section, we will apply the framework to vector and matrix recovery problems, but in subsequent sections, we will also apply the framework to robust PCA, tensor recovery, and other problems as well. 

\begin{lemma} 
\lemlab{lem:fw23:vec}
\cite{FengW23}
Suppose \assumref{asm:strong} holds and let $\bA \in \mathbb{Z}_q^{n \times m}$ be drawn uniformly at random, where $q, m, \beta \in \poly(n)$ and $q \geq n \cdot \beta$. 
If a vector $\bx \in \mathbb{Z}^m_\beta$ is produced by an adversary running in time $o(2^n)$, then with probability at least $1 - \negl(n)$ over the choice of $\bA$, there does not exist a $k$-sparse vector $\by \in \mathbb{Z}^m_\beta$ such that $\bx \not\equiv \by \pmod{q}$ but $\bA\bx \equiv \bA\by \pmod{q}$, for any $k \in o\left(\frac{n}{\log n}\right)$.
\end{lemma}
\begin{proof}
If an adversary were able to find a vector $\by \in \mathbb{Z}^m_\beta$ such that $\bx \not\equiv \by \pmod{q}$ but $\bA\bx \equiv \bA\by \pmod{q}$, then it could compute $(\bx - \by) \bmod q$, yielding a non-zero, short vector in the kernel of $\bA$, thereby solving the SIS problem. 
Since the entries of $\by$ are bounded by $q$, the total number of $k$-sparse vectors in $\mathbb{Z}^m_\beta$ is at most $\O{q^k \cdot \binom{m}{k}} \leq \poly(n)^k$, meaning an adversary could exhaustively search over all such vectors in time $\poly(n)^k$. 
Consequently, if $k \in o\left(\frac{n}{\log n}\right)$, then the probability of such a $k$-sparse $\by$ existing must be negligible in $n$, as otherwise an $o(2^n)$-time adversary could find it with non-negligible probability, contradicting the assumed hardness of SIS.
\end{proof}

We remark that given a random matrix $\bA \in \mathbb{Z}^{n \times m}_q$, if both vectors $\bx$ and $\by$ are $k$-sparse with bounded entries, an information-theoretic union bound implies that, with high probability, any distinct $\bx \neq \by$ will satisfy $\bA\bx \neq \bA\by$. 
However, there can exist a binary vector $\bx$ that is not $k$-sparse and a $k$-sparse vector $\by$ with bounded entries such that $\bA\bx = \bA\by$. 
In these scenarios, the SIS assumption is required to argue that it is computationally hard for an adversary to find such an $\bx$ and deceive the algorithm.

\begin{lemma} 
\lemlab{lem:fw23:mat}
\cite{FengW23}
Suppose \assumref{asm:strong} holds and consider a uniformly random matrix $\bA \in \mathbb{Z}^{n \times m}_q$ with parameters $q, m, \beta \in \poly(n)$ satisfying $q \geq n \cdot \beta$. 
If an $o(2^{n})$-time adversary generates a matrix $\bX \in \mathbb{Z}^{\sqrt{m} \times \sqrt{m}}_\beta$, then with probability at least $1 - \negl(n)$, there is no matrix $\bY \in \mathbb{Z}^{\sqrt{m} \times \sqrt{m}}_\beta$ of rank at most $k$ such that $\bX \neq \bY \mod q$ and $\bA\bx = \bA\by \mod q$, where $\bx$ and $\by$ denote the vectorized forms of $\bX$ and $\bY$, respectively, and $k \in o\left(\frac{n}{\sqrt{m} \log n}\right)$.
\end{lemma}
\begin{proof} 
Following the reasoning in the proof of \lemref{lem:fw23:vec}, an adversary can exhaustively search through all matrices $\bY \in \mathbb{Z}^{\sqrt{m} \times \sqrt{m}}_\beta$ with $\rank(\bY) \leq k$ in time $\poly(n)^{\sqrt{m}k}$. 
This is due to the fact that there are $\mathcal{O}\binom{\sqrt{m}}{k}$ ways to select the positions of the linearly independent columns of $\bY$, and for each such selection, there are $\poly(n)^{\sqrt{m}k}$ possible values for these columns since $\beta \in \poly(n)$. 
The remaining columns are linear combinations of these independent columns. 
Given that there are $\poly(n)^k$ possible coefficient combinations for each dependent column and $(\sqrt{m} - k)$ such columns, the total number of choices for dependent columns is $\poly(n)^{(\sqrt{m}-k)k}$. 
Consequently, the total number of candidate matrices is bounded by $\poly(n)^{\sqrt{m}k}$. 
For $k \in o\left(\frac{n}{\sqrt{m}\log n}\right)$, an $o(2^{n})$-time adversary can enumerate all candidates. 
Hence, under \assumref{asm:strong}, with probability $1-\negl(n)$, no such matrix $\bY$ exists; otherwise, an adversary could solve the SIS problem by computing $(\bx - \by) \mod q$ given $\bX$ and $\bY$.
\end{proof}

\subsection{\texorpdfstring{$k$}{k}-Sparse Recovery Algorithm}

\begin{theorem}
\cite{FengW23}
Suppose \assumref{asm:strong} holds. 
Given a parameter $k \in \Theta\left(\frac{n^{c}}{\log n}\right)$ with any constant $c > 0$, and given an input vector of length $n$ whose integer entries are bounded by $\poly(n)$, there exists a streaming algorithm that is secure against $o(n^k)$ time-bounded white-box adversaries. 
This algorithm can decide whether the input vector is $k$-sparse, and if so, reconstructs the $k$-sparse vector while using only $\tilde{\mathcal{O}}(k)$ bits of space within the random oracle model.
\end{theorem}
\begin{proof}
Consider \algref{alg:fw23:sparse-recovery}, which identifies and reconstructs a $k$-sparse vector using only $\tilde{\mathcal{O}}(k)$ bits of space. 
The input to the algorithm is a stream of integer updates to an underlying vector, where each coordinate is assumed to be bounded by $\beta \in \poly(n)$ at all times. 
Consequently, we can treat all updates as being modulo $q$ for $q, \beta \in \poly(n)$ and $q \geq n \cdot \beta$.

When the input vector $\bx$ is $k$-sparse, \lemref{lem:fw23:vec} ensures that for a uniformly random sketching matrix $\bA$, if $\bA\by = \bA\bx \bmod q$, then it must be that $\by = \bx$. 
Therefore, in this case, \algref{alg:fw23:sparse-recovery} successfully recovers $\bx$ by exhaustively checking all possible $k$-sparse vectors. 
In contrast, if the input vector has sparsity greater than $k$, \lemref{lem:fw23:vec} guarantees that the post-processing step will not find any $k$-sparse vector $\by$ such that $\bA\by = \bv \bmod q$. Hence, the algorithm correctly outputs \texttt{None}.
In the random oracle model, the columns of the random matrix $\bA$ can be generated on demand. 
Thus, the algorithm only needs to maintain a sketch vector of length $f(k) \cdot \log n$ with entries bounded by $\poly(n)$, leading to a total space complexity of $\tilde{\mathcal{O}}(k)$ bits.
\end{proof}

\begin{algorithm}
\caption{Sparse-Recovery($n$, $m$, $k$)}
\alglab{alg:fw23:sparse-recovery}
\begin{algorithmic}
\Require{A stream of $m$ updates $u_t$ to a vector of length $n$, with entries in $[-\beta, \beta]$ for $\beta \in \poly(n)$ and a modulus $q \in \poly(n)$ with $q \gg \beta$}
\Ensure{The frequency vector if it is $k$-sparse or \texttt{None} otherwise}
\State{Let $f(k) \in \omega(k) \cap \tO{k}$}
\State{Sample a uniformly random matrix $\bA \in \mathbb{Z}_q^{f(k) \cdot \log n \times n}$}
\State{Initialize a vector $\bv \in \mathbb{Z}_q^{f(k) \cdot \log n}$ with all zeros}
\For{each update $u_t\in[n]$ with $t \in [m]$}
\State{Let $i$ be the coordinate index updated by $u_t$}
\State{Update $\bv \gets \bv + u_t \cdot \bA_i$, where $\bA_i$ is the $i^{\text{th}}$ column of $\bA$}
\EndFor
\For{each $k$-sparse vector $\by \in \mathbb{Z}^n$ with entries in $[-\beta, \beta]$}
\If{$\bA\by \equiv \bv \mod q$}
\State{\Return $\by$}
\EndIf
\EndFor
\State{\Return \texttt{None}}
\end{algorithmic}
\end{algorithm}

We remark that any algorithm for $k$-sparse recovery that uses approximately $k$ words of space and aims to be robust against white-box adversaries must assume that the adversary is limited to at most $n^k$ time. 
This is because an algorithm with $k$ words of memory can only maintain at most $n^k$ distinct internal states. 
Consequently, there exists some $k$-sparse vector $x$ and some $k'$ sparse vector $\bx'$ with $k' > k$ that results in the same internal state as $\bx$. 
An adversary with $n^k$ time could efficiently find such a pair $(\bx, \bx')$. 
If the adversary inserts either $\bx$ or $\bx'$ into the stream, followed by $-\bx$, the algorithm would be unable to distinguish whether the true input is the zero vector or $\bx' - \bx$. 
Therefore, the above algorithm achieves near-optimality by using only $\tilde{\mathcal{O}}(k)$ bits of space, assuming the adversary is limited to $o(2^{k \log n}) = o(n^k)$ time.

\paragraph{Fast recovery.}
On the other hand, \algref{alg:fw23:sparse-recovery} performs a brute-force enumeration over all possible $k$-sparse vectors in the post-processing stage, which is time-inefficient. 
To achieve a faster version of $k$-sparse recovery, it suffices to run in parallel an existing deterministic fast recovery scheme for $k$-sparse inputs. 

\begin{theorem}
\thmlab{thm:det:ksparse}
\cite{jafarpour2011deterministic}
There exists a deterministic sparse recovery scheme that uses $\tO{k}$ bits of space and outputs a $k$-sparse vector defined by a data stream on a universe of size $n$. 
The algorithm uses $\poly(n)$ post-processing time. 
\end{theorem}

The algorithm in \thmref{thm:det:ksparse} can recover the input vector if it is $k$-sparse, even against white-box adversaries, since the algorithm is deterministic. 
However, the algorithm fails when the input vector is not $k$-sparse. 
Therefore, we can run \algref{alg:fw23:sparse-recovery} and the algorithm of \thmref{thm:det:ksparse} in parallel. 
If the deterministic $k$-sparse recovery scheme outputs a vector $\by^*$, then it suffices to check whether $\bA\by^*$ matches the vector $\by$ stored by \algref{alg:fw23:sparse-recovery}. 
Putting these together, we have:
\begin{theorem}
\thmlab{thm:fw23:ksparse:fast}
\cite{FengW23}
Suppose \assumref{asm:strong} holds. 
Given any constant $c > 0$ and parameter $k \in \Theta\left(\frac{n^c}{\log n}\right)$, there exists a streaming algorithm that operates in $\poly(n)$ time and uses only $\tilde{\mathcal{O}}(k)$ bits of space in the random oracle model. 
Given an input vector of length $n$ defined by a turnstile stream generated by a white-box adversary with $o(n^k)$ runtime, the algorithm determines whether the input is $k$-sparse and, if so, correctly recovers the $k$-sparse vector.
\end{theorem}

\paragraph{Applications to $F_0$ estimation.}
By leveraging our $k$-sparse recovery algorithm as a subroutine, we can design an efficient algorithm for estimating the $F_0$ value of vectors defined by white-box adversaries with bounded runtime. 
Namely, we set $k=n^{1-\eps}$ to provide an $n^\eps$-approximation to the $F_0$ value of a vector, under the assumption that the vector's entries are bounded by $\poly(n)$.
Now, if the $k$-sparse algorithm returns \texttt{None}, then the $F_0$ value is larger than $n^{1-\eps}$. 
However, since the vector has $n$ coordinates, then the $F_0$ value is certainly at most $n$, and thus we can simply output $n^{1-\eps}$ as an $n^{\eps}$-approximation to the $F_0$ value.  
Otherwise, the vector has $F_0$ value at most $n^{1-\eps}$, in which case the $k$-sparse algorithm will recover it exactly. 

\begin{theorem}
\cite{FengW23}
Suppose \assumref{asm:strong} holds. 
Then given any constant $\eps < 1$, there exists a streaming algorithm that, in the random oracle model, computes a multiplicative $n^{\eps}$-approximation to the $F_0$ value of a vector of length $n$, using $\tilde{\mathcal{O}}(n^{1-\eps})$ bits of space and $\poly(n)$ time. 
Moreover, the algorithm is robust against white-box adversaries running in $o(n^{n^{1-\eps}})$ time.
\end{theorem}
\begin{proof}
Let $\bv$ be the underlying vector. 
Observe that if $\|\bv\|_0\ge n^{1-\eps}$, then the $k$-sparse algorithm will return \texttt{None} for $k=n^{1-\eps}$, in which case our algorithm will output $n^{1-\eps}$. 
Since $\|\bv\|_0\le n$, then $n^{1-\eps}$ is an $n^\eps$-approximation to $\|\bv\|_0$. 
On the other hand, if $\|\bv\|_0\le n^{1-\eps}$ then the $k$-sparse algorithm will recover it exactly. 
The space and runtime guarantees follow from the guarantees of the $k$-sparse algorithm in \thmref{thm:fw23:ksparse:fast}, with the setting of $k=n^{1-\eps}$. 
\end{proof}

\subsection{Matrix Recovery}

\subsubsection{Low-Rank Matrix Recovery}
Beyond sparse vector recovery, these techniques also extend to low-rank matrix recovery. 
We introduce a white-box adversarially robust algorithm by \cite{FengW23} for this task that is both time and space-efficient.
Analogous to the method used in the $k$-sparse vector setting, the approach ensures fast update time while maintaining correctness when the input matrix exceeds the target rank by simultaneously maintaining two sketches: one constructed using a uniformly random matrix to detect when the input rank is too high for recovery, and another designed to enable recovery when the input is indeed of low rank.

\begin{theorem}
\cite{RechtFP10}
\thmlab{thm:iso}
Let $\alpha = \mathcal{O}(nk \log n)$ and define $\bA\in\mathbb{R}^{\alpha\times n^2}$ as a random matrix, where each entry is drawn independently from a symmetric Bernoulli distribution:
\[A_{i,j} = 
\begin{cases}
\sqrt{\frac{1}{\alpha}} & \text{with probability } \frac{1}{2} \\
-\sqrt{\frac{1}{\alpha}} & \text{with probability } \frac{1}{2}
\end{cases}\]
Interpret $\bA$ as a linear transformation $\mathcal{A}: \mathbb{R}^{n \times n} \rightarrow \mathbb{R}^\alpha$ acting on a matrix $\bX \in \mathbb{R}^{n \times n}$ via vectorization, i.e., $\mathcal{A}(\bX) = \bA\bx$ where $\bx = \text{vec}(\bX)$. 
Then, for any matrix $\bX_0 \in \mathbb{R}^{n \times n}$ of rank $r$, where $1 \leq r \leq \min\left(k,\frac{n}{2}\right)$, and given $\bb = \mathcal{A}(\bX_0)$, it holds with high probability that $\bX_0$ is the unique rank-$r$ matrix satisfying $\mathcal{A}(\bX) = \bb$. 
Furthermore, $\bX_0$ can be exactly recovered as the solution to the following convex optimization problem:
\[\arg\min_{\bX} \lVert \bX \rVert_* \quad \text{subject to } \mathcal{A}(\bX) = \bb\]
where $\|\bX\|_*$ denotes the nuclear norm of $\bX$, i.e., the sum of the singular values: $\|\bX\|_* = \sum_i \sigma_i(\bX)$.
\end{theorem}

\begin{algorithm}
\caption{Recover-Matrix($n$, $m$, $k$)}
\alglab{alg:m-rec}
\begin{algorithmic}[1]
\Require{$m$ integer updates $u_t$ to an $n \times n$ matrix with entries bounded by $\beta \in \poly(n)$; a modulus $q \in \poly(n)$ such that $q \gg \beta$.}
\Ensure{Either a rank-$k$ matrix recovery or \textsc{None}.}
\State{Let $f(k)$ be a function in $\omega(k)$ and $\tilde{\mathcal{O}}(k)$.}
\State{Initialize a uniformly random matrix $\bH \in \mathbb{Z}_q^{f(k) \cdot n \log n \times n^2}$.}
\State{Let $\bA \in \mathbb{Z}_q^{\alpha \times n^2}$ be as in \thmref{thm:iso} scaled accordingly.}
\State{Initialize zero vectors $\bv, \bw \in \mathbb{Z}_q^{f(k) \cdot n \log n}$.}
\For{each update $u_t$ with $t \in [m]$}
\State{Let $i$ be the vectorized index of the update.}
\State{Update $\bv \gets \bv + u_t \cdot \bH_i$.}
\State{Update $\bw \gets \bw + u_t \cdot \bA_i$.}
\EndFor
\State{Let $\bX_0 \gets \arg\min_{\bX} \lVert \bX \rVert_*$ subject to $\bA \cdot \mathsf{vec}(\bX) = \bw$.}
\If{$\rank(\bX_0) \leq k$ \textbf{and} $\bX_0 \in \mathbb{Z}_\beta^{n \times n}$ \textbf{and} $\bH \cdot \mathsf{vec}(\bX_0) = \bv \mod q$}
\State{\Return $\bX_0$}
\Else
\State{\Return \textsc{None}}
\EndIf
\end{algorithmic}
\end{algorithm}

We state our main theorem for matrix recovery:

\begin{theorem}
\cite{FengW23}
Suppose \assumref{asm:strong} holds. 
Then given any integer parameter $k$, there exists a streaming algorithm that takes an input matrix generated by a white-box adversary with $o(n^{nk})$ runtime and either correctly reports that the input matrix has rank exceeding $k$, or successfully recovers the matrix if its rank is at most $k$. 
The algorithm uses $\tilde{\mathcal{O}}(nk)$ bits of space and runs in $\poly(n)$ time within the random oracle model.
\end{theorem}
\begin{proof}
Consider \algref{alg:m-rec}. 
Given an input matrix $\bX \in \mathbb{Z}^{n \times n}_\beta$ with $\beta \in \poly(n)$ and a modulus $q \geq n \cdot \beta$, if $\rank(\bX) \leq k$, then by the uniqueness guarantee of \thmref{thm:iso}, $\bX$ can be recovered by solving a convex program, and the sketch $\bv$ computed via $\bH$ matches the product $\bH \cdot \mathrm{vec}(\bX)$. 
Conversely, if $\rank(\bX) > k$, then by \lemref{lem:fw23:mat} under the SIS assumption, no other low-rank matrix $\bY$ satisfies $\bH \cdot \mathrm{vec}(\bY) = \bH \cdot \mathrm{vec}(\bX) \mod q$, so the algorithm correctly returns \texttt{None}.

In the random oracle model, the sketching matrices $\bH$ and $\bA$ can be generated on demand, allowing the algorithm to maintain only two vectors of length $\tilde{\mathcal{O}}(nk)$ with entries bounded by $\poly(n)$, resulting in a total space usage of $\tilde{\mathcal{O}}(nk)$ bits. 
Recovery involves solving a convex program using the ellipsoid method and verifying the result, both of which can be done in $\poly(n)$ time.
\end{proof}

We again remark the near-optimality of the low-rank matrix recovery algorithm: using approximately $\tilde{\mathcal{O}}(nk)$ bits of space, any algorithm that is robust against white-box adversaries must assume that the adversary is bounded to $n^{nk}$ time. 
Without this time constraint, an adversary could identify two matrices $\bX \neq \bX'$ such that the algorithm reaches the same internal state on both, and $\rank(\bX' - \bX) > k$. 
By inserting $\bX$ followed by $-\bX$, or $\bX'$ followed by $-\bX$, the input would result in either the zero matrix or $\bX' - \bX$, which the algorithm could not distinguish. 
Thus, the previous algorithm, which uses $\tilde{\mathcal{O}}(nk)$ space and assumes an adversary with runtime $o(2^{nk \log n}) = o(n^{nk})$, is essentially optimal under these constraints.

\paragraph{Applications to rank decision problem.}
The low-rank matrix recovery algorithm can be applied to a number of other problems on data streams. 
For example, consider the following definition of the rank decision problem:

\begin{definition}[Rank Decision Problem]
Given an integer $k\ge 0$, and a matrix $\bA\in\mathbb{R}^{n\times n}$, determine whether the rank of $\bA$ is larger than $k$.    
\end{definition}

Then the following holds by running \algref{alg:m-rec} with parameter $k$.
\begin{theorem}
\cite{FengW23}
Suppose \assumref{asm:strong} holds. 
Then given any integer parameter $k$, there exists a streaming algorithm that, in the random oracle model, solves the rank decision problem using $\tilde{\mathcal{O}}(nk)$ bits of space and $\poly(n)$ time. 
Moreover, the algorithm is robust against white-box adversaries running in time $o(n^{nk})$.
\end{theorem}

\paragraph{Applications to graph matching.}
Next, we consider the maximum matching problem, defined as follows:
\begin{definition}[Maximum Matching Problem]
Given an undirected graph $G = (V, E)$ defined by a sequence of insertions and deletions to edges in a stream, the maximum matching problem is to identify a maximum cardinality set of vertex disjoint edges in $G$. 
\end{definition}

\begin{theorem}
\thmlab{thm:wbox:max:match}
\cite{FengW23}
Suppose \assumref{asm:strong} holds and suppose an integer upper bound $k'$ on the maximum matching size is known. 
Then there exists a streaming algorithm that, in the random oracle model, computes a maximum matching using $\tilde{\mathcal{O}}(nk')$ bits of space and $\poly(n)$ time. 
The algorithm is robust against white-box adversaries with running time bounded by $o(n^{2nk'})$.
\end{theorem}
\begin{proof}
To recover a maximum matching, we use the fact that the rank of the $n \times n$ Tutte matrix $\bA$ associated with a graph $G$ is equal to twice the size of a maximum matching in $G$. 
Specifically, $A_{i,j} = 0$ if there is no edge between vertices $i$ and $j$, and otherwise, $A_{i,j} = x_{i,j}$ and $A_{j,i} = -x_{i,j}$ for distinct indeterminates $x_{i,j}$. 
The rank of $\bA$ is defined as the maximum rank over all real assignments to the $x_{i,j}$. 
However, unlike traditional approaches, e.g., Sections 4.2.1 and 4.2.2 in~\cite{CheungKL13}, we cannot assign the $x_{i,j}$ values randomly during the stream in the white-box setting, as an adversary with access to the algorithm's internal state could exploit this.

To overcome this, we deterministically replace each $x_{i,j}$ with $1$ as edges arrive in the stream, resulting in a fixed matrix $\bA'$. 
Since $\bA'$ is a special case of $\bA$, its rank is at most that of $\bA$, which is bounded above by twice the maximum matching size. 
We then apply our low-rank matrix recovery algorithm with parameter $k = 2k'$, where $k'$ is an upper bound on the size of the maximum matching. 
If the rank of $\bA'$ exceeds $2k'$, it implies the true matching size exceeds $k'$, and we terminate and report this. 
Otherwise, we recover $\bA'$ successfully. 
Once the stream ends, the nonzero entries in $\bA'$ correspond exactly to the edges in $G$, allowing us to reconstruct $\bA$ and thereby the original graph $G$. 
We can then apply any offline algorithm to compute a maximum matching.
\end{proof}

We remark that while \thmref{thm:wbox:max:match} provides an efficient algorithm for maximum matching when the white-box adversary is computationally bounded, \cite{EfremenkoKSZ26} established an unconditional $\Omega(n)$ memory lower bound for estimating the maximum matching size against information-theoretic (unbounded) white-box adversaries.

\subsection{Extension to Robust PCA and Tensors}
A common approach in the previous algorithms is to run two procedures in parallel: (1) an algorithm that identifies whether the input belongs to a restricted class of inputs, such as those that are sparse, low rank, or both, and (2) a deterministic algorithm that efficiently recovers the input if it is confirmed to be from that class. 
The correctness of the first algorithm depends on the hardness of the SIS problem, while the second algorithm is any deterministic method that is efficient in time and space, making it robust against white-box adversaries.
\cite{FengW23} observes that this framework can also be applied to robust PCA and tensors. 

\subsubsection{Robust Principal Component Analysis}
\seclab{sec:fw23:rpca}
The problem of Robust Principal Component Analysis is defined as follows:

\begin{definition}[Robust Principal Component Analysis]
Consider a data matrix $\bM \in \mathbb{Z}_q^{n \times n}$, with $q \ge \poly(n)$, which admits a decomposition $\bM=\bL+\bS$, where $\bL \in \mathbb{Z}_q^{n \times n}$ has rank at most $k$, i.e., $\rank(\bL) \le k$, and $\bS \in \mathbb{Z}_q^{n \times n}$ is a sparse matrix containing no more than $r$ non-zero entries.  
The goal of the robust principal component analysis (RPCA) problem is to recover the components $\bL$ and $\bS$.
\end{definition}

\cite{FengW23} notes that the following lemma can be derived using the hardness of the SIS problem.

\begin{lemma}
\lemlab{lem:rpca}
\cite{FengW23}
Suppose \assumref{asm:strong} holds and consider a uniformly random matrix $\bA \in \mathbb{Z}_q^{n \times m}$ with parameters $q, m, \beta \in \poly(n)$ satisfying $q \geq n \cdot \beta$. 
Suppose an adversary running in time $o(2^n)$ produces a matrix $\bX \in \mathbb{Z}_\beta^{\sqrt{m} \times \sqrt{m}}$.
Then, with probability at least $1 - n\cdot\negl(n)$, there do not exist matrices $\bL, \bS \in \mathbb{Z}_\beta^{\sqrt{m} \times \sqrt{m}}$ such that $\rank(\bL) \leq k$ and $\mathsf{nnz}(\bS) \leq r$ (where $\mathsf{nnz}(\bS)$ denotes the number of non-zero entries in $\bS$), for which
\[\bX \neq \bL + \bS \mod q \quad \text{and} \quad \bA \bx = \bA(\mathbf{\ell} + \bs) \mod q,\]
where $\bx, \mathbf{\ell}, \bs$ are the vectorizations of $\bX, \bL, \bS$ respectively, and with
\[k \in o\left(\frac{n - r \log n}{\sqrt{m} \log n}\right).\]
\end{lemma}
\begin{proof} 
Following a similar approach to the proof of \lemref{lem:fw23:mat}, an adversary can enumerate all pairs of matrices $\bL, \bS \in \mathbb{Z}_\beta^{\sqrt{m} \times \sqrt{m}}$ where $\rank(\bL) \leq k$ and $\mathsf{nnz}(\bS) \leq r$ within time $\poly(n)^{r + k\sqrt{m}}$. 
As established in \lemref{lem:fw23:mat}, there are $\poly(n)^{\sqrt{m}k}$ possible candidates for $\bL$. 
For the sparse matrix $\bS$, the number of ways to select the positions of its $r$ non-zero entries is $\binom{m}{r} \in \poly(n)^r$, with each non-zero value chosen from a set of size $\poly(n)$. 
Consequently, the total number of candidate pairs $(\bL, \bS)$ is bounded by $\poly(n)^{r + k \sqrt{m}}$.

If $k \in o\left(\frac{n - r \log n}{\sqrt{m} \log n}\right)$, then there exists an adversary running in time $o(2^n)$ capable of enumerating all such pairs. 
Therefore, under \assumref{asm:strong}, with probability $1-n\cdot\negl(n)$, no such matrices $\bL$ and $\bS$ exist. 
Otherwise, given $\bL$ and $\bS$, an adversary could solve the SIS problem by outputting $(\bx - \mathbf{\ell} - \bs) \mod q$.
\end{proof}

Similar to the matrix recovery procedure, \cite{FengW23} observes that we can execute a compressed sensing-based approach for RPCA in parallel to enable efficient recovery. 
This approach quickly approximates a unique decomposition of the input into a low-rank and a sparse matrix, under the assumption that such a decomposition exists.

\begin{theorem}
\cite{TannerV20}
\thmlab{thm:iso-rs}
Let $\alpha = \O{(nk + r)\cdot \log n}$, and define a random matrix $\bA \in \mathbb{R}^{\alpha \times n^2}$ whose entries are sampled independently from a symmetric Bernoulli distribution:
\[A_{i,j} = \begin{cases} 
\sqrt{\frac{1}{\alpha}} & \text{with probability } \frac{1}{2}, \\
-\sqrt{\frac{1}{\alpha}} & \text{with probability } \frac{1}{2}.
\end{cases}\]
We interpret $\bA$ as a linear operator $\mathcal{A}: \mathbb{R}^{n \times n} \rightarrow \mathbb{R}^{\alpha}$ acting on the vectorization $\bx$ of a matrix $\bX \in \mathbb{R}^{n \times n}$ via matrix multiplication $\bA\bx = \mathcal{A}(\bX)$. 
Given a vector $\bb = \mathcal{A}(\bL_0 + \bS_0)$, with high probability the matrices $\bL_0$ and $\bS_0$ are the unique pair satisfying $\mathcal{A}(\bL + \bS) = \bb$ such that $\rank(\bL_0) \leq k$ and $\nnz(\bS_0) \leq r$. 

Moreover, this pair can be efficiently approximated by solving the following semidefinite program:
\[\min_{\bL, \bS} \left(\|\bL \|_* + \sqrt{\frac{2r}{k}} \cdot \| \bS \|_1 \right) \quad \text{subject to} \quad \|\mathcal{A}(\bL + \bS) - \bb \|_2 \leq \eps,\]
which yields matrices $\bL, \bS$ such that $\|(\bL + \bS) - (\bL_0 + \bS_0) \|_F \leq 42\eps$. 

Here, $\|\cdot\|_*$ denotes the nuclear norm, i.e., the sum of the singular values: $\|\bM\|_* = \sum_i \sigma_i(\bM)$, and $\|\cdot\|_1$ denotes the matrix $1$-norm, defined as the maximum absolute column sum: $\|\bM \|_1 = \max_{0 \leq j \leq n} \sum_{i=1}^n |M_{i,j}|$.
\end{theorem}

We remark that to guarantee exact recovery for a stream of updates that are integers bounded in magnitude by $\poly(n)$, we can set the error parameter $\eps \leq \frac{1}{\poly(n)}$ and then round the entries of the output to integers.
Then using the previous statements, \cite{FengW23} achieves the following guarantees for robust PCA:

\begin{theorem}
\thmlab{thm:rpca-1}
\cite{FengW23}
Suppose \assumref{asm:strong} holds. 
Then given parameters $r, k > 0$, there exists a streaming algorithm that is robust against white-box adversaries running in $o(n^{nk + r})$ time, which decides whether an $n \times n$ input matrix can be written as the sum of a rank-$k$ matrix and a matrix with at most $r$ non-zero entries. 
If such a decomposition exists, the algorithm finds it using $\tilde{\mathcal{O}}(nk + r)$ bits of space and runs in $\poly(n)$ time in the random oracle model.
\end{theorem}
\begin{proof}
Let $f(k)$ be a function in $\omega(k)$ and $\Tilde{\mathcal{O}}(k)$. 
Consider the algorithm where we generate a uniformly random matrix $H \in \mathbb{Z}^{(f(k)\cdot n + r)\log n \times n^2}_{q}$ for $q \in \poly(n)$, a fast recovery matrix $A\in\mathbb{R}^{(nk+r)\log n \times n^2}$ as specified in \thmref{thm:iso-rs}, and initialize zero vectors $\bv, \bw$ of length $(f(k)\cdot n + r)\log n$. 
Then for each update $u_t$ with $t\in [m]$, we update $\bv$ by adding $\bu_t\cdot \bH_i$ to it, and update $\bw$ by adding $\bu_t \cdot \bA_i$ to it, where $i$ corresponds to the vectorized index of the update, and where $\bH_i, \bA_i$ are the $i$-th columns of $\bH, \bA$, respectively. 
Let $\bL_0$ and $\bS_0$ be the solutions to the semidefinite program $\argmin_{\bL, \bS}(\|\bL\|_* + \sqrt{\frac{2r}{k}}\cdot\|\bS\|_1)$ subject to $\|\mathcal{A} (\bL+\bS) -\bb \|_2 \leq \frac{1}{\poly(n)}$. 
If $\rank(\bL_0) \leq k$, $\nnz(\bS_0) \leq r$, $\bL_0, \bS_0 \in \mathbb{Z}^{ n \times n}_{\beta}$, and $\bH\cdot(\mathbf{\ell}_0+\bs_0)\equiv \bv\mod q$, then the algorithm returns $\bL_0$ and $\bS_0$. 
Otherwise, the algorithm reports \texttt{None}. 
Here, $\mathbf{\ell}_0$ and $\bs_0$ denote the vectorizations of $\bL_0$ and $\bS_0$, respectively. 

Given an input matrix $\bX_0 = \bL_0 + \bS_0 \in \mathbb{Z}^{n \times n}_\beta$ with $\rank(\bL_0) \leq k$ and $\nnz(\bS_0) \leq r$, \thmref{thm:iso-rs} guarantees the uniqueness of the decomposition, and $\bL_0, \bS_0$ can be recovered by solving a semidefinite program. 
Moreover, the sketch $\bv$ is preserved, as $\bH(\bL_0 + \bS_0) = \bH\bx_0$. 

Conversely, if the input $\bX$ cannot be decomposed into a low-rank and sparse pair, then by \lemref{lem:rpca} and under the SIS hardness assumption, there does not exist a pair $\bL', \bS'$ with $\bX \neq \bL' + S'$ such that $\bH(\mathbf{\ell}'+\bs') = \bv = \bH\bx \mod q$, where $\mathbf{\ell}', \bs', \bx$ are the vectorizations of $\bL', \bS'$, and $\bX$, respectively. 
Hence, in this case, the algorithm correctly reports \texttt{None}. 

Both random matrices $\bH$ and $\bA$ can be generated on-the-fly in the random oracle model. 
Thus, the algorithm only needs to store two sketch vectors of length $\tilde{\mathcal{O}}(nk + r)$ with entries bounded by $\poly(n)$, using a total of $\tilde{\mathcal{O}}(nk + r)$ bits of space. 
Solving the semidefinite program and verifying the solution against the sketch takes $\poly(n)$ time, resulting in overall $\poly(n)$ runtime.
\end{proof}

\subsubsection{Tensor Recovery}
\seclab{sec:fw23:tensor}
Similar to the vector and matrix recovery algorithms, \cite{FengW23} also considers an algorithm that recovers tensors with low CANDECOMP/PARAFAC (CP) rank.
Throughout this section, we use $\otimes$ to denote the outer product of two vectors. 
Observe that a rank-$1$ tensor can be built in $\mathbb{Z}^{n_1 \times n_2 \times \cdots \times n_d}$ by taking the outer product $x_1 \otimes x_2 \otimes \cdots \otimes x_d$ where $x_i \in \mathbb{Z}^{n_i}$. 

\begin{definition}[CP-rank]
Let $\bX \in \mathbb{Z}^{n_1 \times \ldots \times n_d}_q$ be a tensor over $\mathbb{Z}_q$. 
Suppose $\bX$ can be written as a sum of $r$ rank-1 tensors:
\[\bX = \sum_{i=1}^r (\bx_{i,1} \otimes \bx_{i,2} \otimes \cdots \otimes \bx_{i,d}),\]
where each $\bx_{i,j} \in \mathbb{Z}^{n_j}_q$. 
The tensor rank of $\bX$ is defined as the minimum number $r$ for which such a decomposition exists.
\end{definition}
\cite{FengW23} shows the following hardness result based on the SIS problem. 
\begin{lemma}
\lemlab{lem:fw23:tensor}
\cite{FengW23}
Suppose \assumref{asm:strong} holds and let $\bA \in \mathbb{Z}_q^{n \times m}$ be a uniformly random matrix, where $q, m, \beta \in \poly(n)$ and $q \geq n \cdot \beta$. 
Suppose a tensor $\bX \in \mathbb{Z}^{n_1 \times \cdots \times n_d}_\beta$ with $\prod n_i = m$ is generated by an $o(2^n)$-time adversary. 
Then, with probability at least $1 - \negl(n)$, there does not exist a tensor $\bY \in \mathbb{Z}^{n_1 \times \cdots \times n_d}_\beta$ such that $\rank(\bY) \leq k$, $\bX \not\equiv \bY \mod q$, and $\bA\bx \equiv \bA\by \mod q$, where $\bx$ and $\by$ are the vectorizations of $\bX$ and $\bY$, respectively, and $k \in o\left(\frac{n}{(n_1 + \cdots + n_d)\log n}\right)$.
\end{lemma}
\begin{proof} 
As in the proof of \lemref{lem:fw23:mat}, an adversary can exhaustively search over all low-rank tensors $\bY \in \mathbb{Z}^{n_1 \times \cdots \times n_d}_\beta$ with $\rank(\bY) \leq k$ in $\poly(n)^{k(n_1 + \cdots + n_d)}$ time. 
For each factor $\bx_{i,j}$ in a rank-1 decomposition, there are $\poly(n)^{n_j}$ possible values, resulting in $\poly(n)^{n_1 + \cdots + n_d}$ distinct rank-1 tensors. 
Selecting $k$ such tensors to construct a rank-$k$ tensor yields a total of $\poly(n)^{k(n_1 + \cdots + n_d)}$ candidates.

When $k \in o\left(\frac{n}{(n_1 + \cdots + n_d)\log n}\right)$, an $o(2^n)$-time adversary can enumerate all such candidates. 
Hence, under \assumref{asm:strong}, with overwhelming probability no such tensor $\bY$ exists; otherwise, the adversary could solve the SIS problem by outputting $(\bx - \by) \mod q$, where $\bx$ and $\by$ are the vectorizations of $\bX$ and $\bY$, respectively.
\end{proof}

As with the vector and matrix recovery problems, \cite{FengW23} notes that it is possible to run the following fast low-rank tensor estimation scheme in parallel with the tensor recovery algorithm. 

\begin{theorem}
\cite{GrotheerLMNQ19}
\thmlab{thm:iso-tensor}
Let the measurement operator $\mathcal{A}: \mathbb{R}^{n_1 \times \cdots \times n_d} \rightarrow \mathbb{R}^{k(n_1 + \cdots + n_d)\log n}$ have entries that are i.i.d. Gaussian random variables with appropriate normalization. 
Given the measurement $\bb = \mathcal{A}(\bX)$ for a tensor $\bX \in \mathbb{R}^{n_1 \times \cdots \times n_d}$, there exists an algorithm that, with high probability, recovers an estimate $\bX_0 \in \mathbb{R}^{n_1 \times \cdots \times n_d}$ satisfying $\|\bX_0 - \bX \|_F \leq \frac{1}{\poly(n)}$, where $n = \prod_{i=1}^d n_i$. 
The algorithm runs in $\poly(n)$ time.
\end{theorem}

We remark that to maintain a proper linear sketch with bounded precision, it suffices to round the Gaussian random variables to additive integer multiples of $\frac{1}{\poly(n)}$. 
This discretization only changes the norm of the measurement by at most additive $\frac{1}{\poly(n)}$ and thus does not asymptotically change the result in \thmref{thm:iso-tensor}. 
The required precision for the rounded Gaussian random variables can be achieved using uniformly random bits generated by a random oracle, as described in~\cite{Karney16}.

By executing the algorithm from \thmref{thm:iso-tensor} in the streaming setting, we only need to maintain and update a measurement vector of length $\tilde{\mathcal{O}}(k(n_1 + \cdots + n_d))$, whose entries are bounded by $\poly(n)$. 
This results in an overall space complexity of $\tilde{\mathcal{O}}(k(n_1 + \cdots + n_d))$ bits. 

\begin{theorem}
\thmlab{thm:fw23:tensor}
\cite{FengW23}
Suppose \assumref{asm:strong} holds. 
Given an input tensor $\bX \in \mathbb{Z}^{n_1 \times \cdots \times n_d}$ and a parameter $k$ satisfying $k \in \Theta\left(\frac{n^c}{(n_1 + \cdots + n_d)\log n}\right)$ for $n = \prod_{i=1}^d n_i$ and some constant $c > 0$, there exists a streaming algorithm that is resilient against $o(n^{k(n_1 + \cdots + n_d)})$-time white-box adversaries. 
This algorithm determines whether $X$ has CP rank at most $k$, and if so, it recovers $\bX$ using $\tilde{\mathcal{O}}(k(n_1 + \cdots + n_d))$ bits of space and runs in $\poly(n)$ time in the random oracle model.
\end{theorem}
\begin{proof}
The algorithm is similar to that described in the proof of \thmref{thm:rpca-1}. 
We similarly maintain a sketching matrix $\bH$ to check whether the CP rank is at most $k$ and a fast recovery matrix $\bA$ corresponding to \thmref{thm:iso-tensor} to recover the decomposition if the CP rank is indeed at most $k$. 

For any input tensor $\bX \in \mathbb{Z}^{n_1 \times \cdots \times n_d}_\beta$ with $\rank(\bX) \leq k$, \thmref{thm:iso-tensor} guarantees that $\bX$ can be successfully reconstructed and the resulting tensor satisfies $\bH\bx = \bv$, where $\bx$ is the vectorization of $\bX$. 
Conversely, if $\rank(\bX) > k$, then by \lemref{lem:fw23:tensor} and the SIS hardness assumption, no tensor $\bY$ of rank at most $k$ exists such that $\bX \neq \bY$ and $\bH\by = \bv = \bH\bx \mod q$. 
Hence, in this case, the algorithm returns \texttt{None}.

The random matrix $\bH$ can be generated on demand using the random oracle model. 
Thus, the recovery process only needs to maintain a sketch vector of length $\tilde{\mathcal{O}}(k(n_1 + \cdots + n_d))$, with each entry bounded by $\poly(n)$, resulting in total space usage of $\tilde{\mathcal{O}}(k(n_1 + \cdots + n_d))$. 
Additionally, the fast recovery scheme also requires $\tilde{\mathcal{O}}(k(n_1 + \cdots + n_d))$ bits of space. 
Both recovering the rank-$k$ decomposition and checking the output against the sketch vector take $\poly(n)$ time, resulting in an overall runtime of $\poly(n)$.
\end{proof}

\chapter{Robust Algorithms for Turnstile Streams}
\chaplab{chap:turnstile:algs}

\begin{tallchapterbannerbox}
\centering
In some cases, adversarially robust algorithms for turnstile streams with minimal space overhead are possible!
\end{tallchapterbannerbox}
\vspace{0.4in}

As a number of adaptive attacks in \chapref{chap:turnstile} focus on linear sketches, one might ask whether there exist adaptive attacks on general streaming algorithms. 
In this chapter, we present a result of \cite{GribelyukLWYZ26} that surprisingly shows that for many problems that satisfy an approximate notion of triangle inequality, there exists an adversarially robust streaming algorithm that can handle both insertions and deletions. 
In particular, \cite{GribelyukLWYZ26} showed that there exists an adversarially-robust algorithm that uses $\poly\left(\frac{1}{\eps}, \log n\right)$ space for $(1+\eps)$-approximation for $F_2$ moment estimation on turnstile streams of length $m=\poly(n)$ with a universe of size $n$. 
We first give a technical overview of the various approaches in \secref{sec:turnstile:algs:overview}. 
We then describe the robust $F_2$ estimation algorithm of \cite{GribelyukLWYZ26} in \secref{sec:turnstile:algs:ftwo} and the robust framework for many functions that satisfy approximate triangle inequality in \secref{sec:turnstile:algs:triangle}. 
Finally, we present the robust algorithm of \cite{GribelyukLWYZ26}, for $L_2$ heavy-hitters on insertion-deletion streams in \secref{sec:turnstile:hh}. 
Throughout the section, we follow the presentation of \cite{GribelyukLWYZ26}. 

\section{Technical Overview}
\seclab{sec:turnstile:algs:overview}
In this section, we provide a technical overview of the main approach of \cite{GribelyukLWYZ26}. 
Recall from the discussion in the previous chapters that, given a data stream inducing a frequency vector $\bx\in\mathbb{R}^n$, a standard approach is to sample a random sketching matrix $\bA\in\mathbb{R}^{r\times n}$ and maintain the sketch $\bA\bx$ as the stream evolves. 
The matrix is chosen so that there exists a recovery function $g$ for which $g(\bA\bx)$ serves as an estimator for the target quantity on $\bx$, for instance $g(\bA\bx)\approx\|\bx\|_p$. 
Because $\bA\bx$ consists of only $r$ entries, the algorithm typically stores just $\O{r\log n}$ bits over the course of the stream, and in many settings one can take $r\ll n$. 
Recall that as a concrete example, constant-factor approximations to the $L_p$ norm are achievable with $r=\polylog(n)$ when $p\le 2$, and with $r=\O{n^{1-2/p}}$ when $p>2$.

However, the lower bounds by \cite{HardtW13,GribelyukLWYZ24,GribelyukLWYZ25} presented in \chapref{chap:turnstile} show that any linear sketch that is adversarially robust to insertion-deletion streams must have dimension $\Omega(n)$. 
A key issue is that linear sketches draw all their randomness once, at initialization, after which the algorithm behaves deterministically.
An adaptive adversary can exploit this by issuing $\poly(n)$ queries to progressively uncover the initial random choices, and then create an adversarial stream that ``defeats'' the fixed sketching matrix. 
This strategy essentially describes the approaches by \cite{HardtW13,GribelyukLWYZ24,GribelyukLWYZ25}  presented in \chapref{chap:turnstile}. 
Consequently, overcoming these lower bounds appears to require injecting some sort of fresh randomness during the execution of the streaming algorithm, rather than committing all randomness upfront. 
Our approach accomplishes this while maintaining small space by introducing new, independent sketches partway through the stream.

To formalize this idea, suppose the frequency vector $\bx$ decomposes as $\bx=\bz+\bq$, where $\bz$ represents the contribution of all updates up to some time $t$, and $\bq$ captures the updates that occur after time $t$. 
In other words, $\bz$ corresponds to a prefix of the stream, while $\bq$ corresponds to the remaining suffix. 
Now consider a sketching matrix $\bB$ that is initialized at time $t+1$, so that the algorithm maintains $\bB\bq$. 
If $\bB$ is a random sketch designed for estimating the $F_2$ moment, then knowing $\bB\bz$ would allow us to recover $\bB\bx=\bB\bz+\bB\bq$, and hence approximate $\|\bx\|_2^2$. 
Of course, $\bB\bz$ is not directly available, since $\bB$ is only created after the updates defining $\bz$ have already occurred.
In what follows, we show that either $\|\bx\|_2^2$ can still be estimated accurately without access to $\bB\bz$, or the algorithm can incrementally recover sufficient information about $\bB\bz$ to proceed.

\subsection{Technical Overview for \texorpdfstring{$F_2$}{F2} Moment Estimation}
We will first introduce the $F_2$ moment because its inner product structure provides a clean and easily digestible exposition. 
However, we emphasize from the outset that the core algorithmic ideas are not limited solely to $F_2$. 
As we will detail later in \secref{sec:tech:overview:approx:triangle} and \secref{sec:turnstile:algs:triangle}, a similar approach extends naturally to a broad class of norms and symmetric functions satisfying an approximate triangle inequality. 
There is, however, one crucial mathematical distinction regarding the approximation guarantees. 
$F_2$ is unique in that the geometric properties of Hilbert spaces allow our algorithm to achieve a $(1+\eps)$-approximation, while general functions that rely solely on looser triangle inequalities give $\O{1}$-approximations. 
Keeping this distinction in mind, we now detail the $F_2$ construction.

Now suppose there is an additional random sketch matrix $\bA$ for $F_2$ moment estimation that is initialized at the start of the stream, allowing the algorithm to maintain $\bA\bz$ and later continue tracking $\bA\bx$ after time $t$.
In this setting, the information that the algorithm has about the prefix $\bz$ is captured entirely by the sketch $\bA\bz$.

If the suffix vector $\bq$ points in a random direction that is independent of $\bz$, then it is unrealistic to expect to extract meaningful information about $\bB\bz$ from $\bq$.
Crucially, this is also the regime in which such information is unnecessary for estimating $\|\bx\|_2^2$.
Indeed, in high dimensions random vectors are nearly orthogonal, and thus we have
\[\|\bz+\bq\|_2^2\approx \|\bz\|_2^2+\|\bq\|_2^2,\]
which can be estimated separately using $\bA\bz$ and $\bB\bq$.

On the other hand, if $\|\bz\|_2^2+\|\bq\|_2^2$ provides a poor approximation to $\|\bx\|_2^2$, then, since
\[\|\bx\|_2^2=\|\bz+\bq\|_2^2=\|\bz\|_2^2+2\langle\bz,\bq\rangle+\|\bq\|_2^2,\]
it must be the case that $|\langle\bz,\bq\rangle|$ is somewhat large relative to both $\|\bz\|_2^2$ and $\|\bq\|_2^2$. 
This situation can only arise when $\bz$ exhibits non-trivial alignment with $\bq$, in which case the suffix $\bq$ potentially reveals information about $\bz$.

Using the sketch $\bA\bx$, we can test whether $\|\bz\|_2^2+\|\bq\|_2^2$ is an inaccurate estimate.
Depending on the outcome, the algorithm either outputs a reliable approximation to $\|\bx\|_2^2$ or extracts additional information about $\bz$.
This intuition motivates a simplified version of our algorithm, which we describe next.

\paragraph{Estimator, corrector, learner framework.}
We consider the following simplified setting:
\begin{enumerate}
\item 
Initially, we are given the sketch $\bA\bz$, after which vectors $\bq$ arrive in a stream; for each $\bq$, we are able to maintain both $\bA\bq$ and $\bB\bq$.
\item 
Upon receiving each $\bq$, the goal is to output an estimate of $\|\bz+\bq\|_2^2$.
\item 
We focus only on protecting the sketch $\bA$ from an adversary; that is, we assume that norm estimates obtained from $\bB$ are always accurate (for the moment, one may think of $\bB$ as the identity matrix).
\end{enumerate}
Our simplified algorithm for $F_2$ moment estimation in this setting is built around three key components:
\begin{enumerate}
\item 
An \emph{estimator}, which produces an estimate of $\|\bz+\bq\|_2^2$.
\item
A \emph{learner}, whose purpose is to gradually acquire information about $\bz$ in order to improve future estimates.
Whenever the estimator fails, the learner extracts information about $\bz$ from the current query $\bq$.
\item
A \emph{corrector}, which detects and informs the learner/estimator when the estimate is incorrect. 
\end{enumerate}
The corrector maintains the sketch $\bA\bx$ and uses it to identify incorrect estimates. 
The learner maintains a vector $\bz'$, formed as a linear combination of the queries $\bq$ on which the algorithm previously erred. 
The learner explicitly stores only the sketches $\bA\bz'$ and $\bB\bz'$. 
The estimator maintains both $\bA\bz$ and $\bB\bq$. 
Given a query $\bq$, the estimator outputs $\|\bz-\bz'\|_2^2+\|\bz'+\bq\|_2^2$ as an estimate of $\|\bx\|_2^2=\|\bz+\bq\|_2^2$, where the first term is approximated using $\bA\bz-\bA\bz'$ and the second term is approximated using $\bB\bz'+\bB\bq$.

As a consistency check, observe that if $\bz'=\mathbf{0}^n$, meaning no information about $\bz$ has been learned, the estimator reduces to $\|\bz\|_2^2+\|\bq\|_2^2$. 
At the other extreme, if $\bz'=\bz$, so that $\bz$ has been fully learned, the estimator exactly recovers $\|\bz+\bq\|_2^2$. 
Finally, whenever the estimator is detected to be incorrect, the algorithm simply outputs the estimate provided by the corrector for that query, based on $\bA\bx$.

\paragraph{Robustness and progress.}
Using standard minimax arguments, it is enough to consider a deterministic adversary. 
Within our framework, one potential vulnerability is that the corrector relies on $\bA\bx$ to detect incorrect estimates. 
Each such interaction could reveal some information about $\bA$, so after multiple interactions, an adversary might be able to exploit their knowledge of $\bA$ to ``fool'' the corrector. 

To handle this, suppose that the estimator makes errors at most $L$ times. 
We then apply the key idea from the bounded computation paths technique of \cite{Ben-EliezerJWY22}: there are at most $\binom{m}{L}$ possible sets of times at which the corrector can flag errors over a stream of length $m$. 
Consequently, there are only $\binom{m}{L}$ adaptive input streams that a deterministic adversary can generate based on the information revealed by the corrector. 
By setting the failure probability of the corrector to $\delta = \left(\frac{1}{\poly(n)}\right)^L\cdot\binom{m}{L}^{-1}$, then by a union bound argument, the corrector will be robust against all possible streams from the adversary. 

Since the space of streaming algorithms typically scales with $\log\frac{1}{\delta}$, the resulting space complexity grows roughly linearly with $L$. 
We also use the estimate of the corrector at most $L$ times as output, and $\bA$ is used at most $L$ times to estimate $\|\bz-\bz'\|_2^2$ after each update to $\bz'$. 
A sketch of size $L$ can safely be used $\O{L}$ times even in the presence of an adaptive adversary (for example, by assigning a different block of rows to each query). 
Therefore, the main remaining task is to bound the total number $L$ of times the estimator produces an incorrect output.

To do this, consider each instance when the estimator is inaccurate. 
In such a case, $\bz-\bz'$ cannot be nearly orthogonal to $\bz'+\bq$. 
By updating $\bz'$ towards $-\bq$ (or away from $-\bq$), the distance to $\bz$ decreases significantly, meaning that $\bz'$ becomes a better approximation of $\bz$. 
Formally, we use $\|\bz-\bz'\|_2^2$ as our ``progress measure''. 
An incorrect estimate of $\|\bz + \bq \|_2^2$ then implies 
\[\left\lvert \|\bz-\bz'\|_2^2 + \|\bz'+\bq\|_2^2 - \|\bz+\bq\|_2^2 \right\rvert \ge \eps \cdot \|\bz+\bq\|_2^2.\] 
Expanding both sides gives 
\[| \langle \bz - \bz', \bz' + \bq \rangle | \ge \frac{\eps}{2} \|\bz + \bq\|_2^2 = \frac{\eps}{2} \left( \|\bz - \bz'\|_2^2 + \|\bq + \bz'\|_2^2 + 2\langle \bz - \bz', \bq + \bz' \rangle \right).\] 
Without loss of generality, suppose $\langle \bz - \bz', \bz' + \bq \rangle \ge \frac{\eps}{2} \|\bz + \bq\|_2^2$ (the other case is similar). 
In particular, this implies 
\[\langle \bz - \bz', \bz' + \bq \rangle \ge \eps \cdot \|\bz - \bz'\|_2 \cdot \|\bq + \bz'\|_2.\] 
When the algorithm errs on query $\bq$, we update $\bz'$ to $\bz'' = \bz' + \alpha(\bq + \bz')$ with a carefully chosen $\alpha$. 
Note that in the sketch space, this can be done by scaling the sketch of $\bz'$ by $(1+\alpha)$, followed by adding a multiple of the sketch of $\bq$. 
Then
\begin{align*}
\|\bz - \bz''\|_2^2 &= \|\bz - \bz' - \alpha(\bq + \bz')\|_2^2 \\
&= \|\bz - \bz'\|_2^2 + \alpha^2 \|\bq + \bz'\|_2^2 - 2\alpha \langle \bz - \bz', \bq + \bz'\rangle \\
&\le \|\bz - \bz'\|_2^2 + \alpha^2 \|\bq + \bz'\|_2^2 - 2\alpha \cdot \eps \|\bz - \bz'\|_2 \cdot \|\bq + \bz'\|_2.
\end{align*} 
By setting the step size $\alpha = \frac{\eps \cdot \|\bz - \bz'\|_2}{\|\bq + \bz'\|_2}$, we obtain
\[\|\bz - \bz''\|_2^2 \le (1-\eps^2) \|\bz - \bz'\|_2^2.\] 
Importantly, this step size $\alpha$ can be approximated up to a $(1+\eps)$-multiplicative factor using robust $L_2$ sketches with high probability. 
Since $\|\bz\|_2^2 \le \poly(n)$ for a stream of length $m = \poly(n)$, it follows that $\bz'$ will be updated at most $\O{\frac{1}{\eps^2} \log n}$ times, giving an upper bound on $L$. 
Hence, the algorithm solves the simplified setting using only $\poly\left(\frac{1}{\eps}, \log n\right)$ rows in $\bA$.

\paragraph{Recursive estimator.}
To extend the previous approach to a full streaming algorithm, the main challenge is that while the sketching matrix $\bB$ allows access to $\bB\bq$ and lets us gradually learn $\bB\bz$, it still needs to correctly answer a large number of queries involving $\bB\bq$ throughout the stream. 
In other words, $\bB$ itself must be robust against adaptive queries that appear along the way to $\bq$. 
To handle this, we apply recursion over the stream.

Recall that our estimator is $\|\bz-\bz'\|_2^2+\|\bz'+\bq\|_2^2$, and we maintain both $\bA\bz'$ and $\bB\bz'$ while updating $\bz'$. 
The first term, $\|\bz-\bz'\|_2^2$, is estimated using $\bA\bz-\bA\bz'$. 
The key insight is that estimating the second term, $\|\bz'+\bq\|_2^2$, is essentially the same problem on a shorter stream: given a (rounded) sketch $\bB\bz'$ for an unknown $\bz'$, estimate $\|\bz'+\bq\|_2^2$ for $\bq$ arriving in a stream. 
This enables a recursive approach for estimating the second term.

\paragraph{Two levels and beyond.}
To illustrate the idea, consider a two-level construction in which we decompose the query as $\bq = \bq_1 + \bq_2$, so that
\[\|\bz' + \bq\|_2^2 = \|\bz' + \bq_1 + \bq_2\|_2^2.\]
We use one sketching matrix $\bB_1$ to maintain $\bB_1(\bz' + \bq_1)$ and another sketching matrix $\bB_2$ to maintain $\bB_2 \bq_2$. 
At the second level, we additionally introduce a learner whose task is to approximate $\bz' + \bq_1$ throughout the stream updates corresponding to $\bq_2$. 

Concretely, this learner maintains a vector $\bu$, formed as a linear combination of those queries on which the second-level estimator is inaccurate when using 
\[\|\bz' + \bq_1 - \bu\|_2^2 + \|\bu + \bq_2\|_2^2\]
as a proxy for $\|\bz' + \bq_1 + \bq_2\|_2^2$. 
By the preceding intuition, this proxy fails precisely when the inner product $\langle \bz' + \bq_1, \bq_2 \rangle$ is large, indicating that the procedure handling $\bq_2$ has implicitly learned information about $\bz' + \bq_1$ from the updates in $\bq_2$.

To detect such inaccuracies, we again couple the learner and the estimator at the second level with a corrector for $\|\bz' + \bq_1 + \bq_2\|_2^2$. 
One subtlety is that the iterate $\bz'$ may change during the stream of updates to $\bq_2$, since the first-level estimator might fail on some query $\bq$. 
To accommodate this, we initiate a new block at the second level whenever $\bz'$ is modified. 
In other words, instead of fixing two blocks corresponding to streams of length $\frac{m}{2}$, we form blocks dynamically—either when the stream length reaches a multiple of $\frac{m}{2}$ or when $\bz'$ is updated to $\bz' + \alpha(\bq + \bz')$ due to an inaccurate query $\bq$. 
Since we have already bounded the total number of updates to $\bz'$ by $L = \O{\frac{1}{\eps^2}\log n}$, it follows that the total number of second-level blocks is at most $\O{\frac{1}{\eps^2}\log n}$.

More generally, let $B = \O{\frac{1}{\eps^2}\log n}$. 
We further limit each recursive estimator for $\|\bz'+\bq\|_2^2$ to run only for a $\frac{1}{B}$ fraction of the steps before deliberately terminating it and starting a new instance. 
We emphasize that this forced termination is an important mechanism that enforces the boundaries of a $B$-ary tree, so that the full algorithm has at most $H = \log_B m = \O{\log n}$ levels of recursion, which avoids an exponential blow-up in the overall space. 
Since at most $\poly(B,H) = \poly\left(\frac{1}{\eps},\log n\right)$ sketches are maintained simultaneously, the algorithm achieves the desired space complexity. 
By starting with $\bz = \mathbf{0}^n$ at the beginning of the stream, the algorithm outputs a correct estimate of $\|\bx\|_2^2$ at all times.

\paragraph{Comparison to other techniques.}
It is instructive to describe the differences between our approach with various other strategies. 
First, we contrast our strategy with the difference estimators presented in \secref{sec:diff:est}. 
Difference estimators operate on value-based granularity, partitioning the stream based on dyadic increases in the function's value and allocating lower-accuracy sketches to later blocks. 
This is highly efficient for insertion-only streams due to the small flip number. 
However, turnstile streams can fluctuate wildly in value, potentially resulting in large flip numbers. 
Our framework overcomes this by utilizing time-based recursive decomposition, dividing the stream into blocks based on timestamps and reusing memory by aggressively discarding subroutines once their specific time block expires, avoiding polynomial space dependency. 

Furthermore, the learner only updates the iterator $\bz'$ when the corrector detects a significant error from the estimator. 
This shares a conceptual connection with the ``update-until-converge'' phenomenon found in the Private Multiplicative Weights (MWU) method from the differential privacy literature~\cite{HardtR10,HardtLM12}. 
In Private MWU, a synthetic dataset is updated only when a DP query mechanism reveals it is inaccurate, and the total number of updates is bounded by a potential function (relative entropy). 
Similarly, our learner updates $\bz'$ only when the estimator errs. 
Using $\|\bz - \bz'\|_2^2$ as a potential function, we can upper bound the number of updates to $\poly\left(\frac{1}{\eps},\log n\right)$. 
In both paradigms, upper bounding the number of updates minimizes the exposure of internal randomness, which is the crucial for maintaining robustness and privacy.

\subsection{Technical Overview for Approximate Triangle Inequality}
\seclab{sec:tech:overview:approx:triangle}
As a natural extension of our $F_2$ algorithm, we can design an algorithm to robustly estimate functions $\calF$ that satisfy an approximate triangle inequality and have non-robust turnstile sketches. 
Concrete examples of such functions, which we formally discuss in \secref{sec:turnstile:algs:triangle}, include the $F_p$ frequency moments, i.e., $F(x) = \|x\|_p^p$, distinct elements, symmetric norms, and robust $M$-estimators and loss functions commonly used in machine learning, such as the Pseudo-Huber and Cauchy losses. 
Notice that for any function $\calF$ satisfying the triangle inequality and $\calF(\bv)=\calF(-\bv)$ for all $\bv\in\mathbb{R}^n$, if we aim for an $\alpha$-approximation with a sufficiently large constant $\alpha>3$, we can use 
\[\calF(\bz-\bz')+\calF(\bz'+\bq)\] 
as an estimator for $\calF(\bz+\bq)$. 
By the $\beta$-approximate triangle inequality, this provides a lower bound of $\frac{1}{\beta}\cdot\calF(\bz+\bq)$. 
The estimator only fails when
\[\calF(\bz-\bz')+\calF(\bz'+\bq)\ge \alpha \cdot \calF(\bz+\bq).\] 
In this case, the triangle inequality gives 
\[\calF(\bz+\bq)+\calF(\bz-\bz')\ge \calF(\bz'+\bq),\] 
and therefore 
\[\calF(\bz+\bq)+2\calF(\bz-\bz')\ge \alpha\cdot \calF(\bz+\bq),\] 
which implies
\[\calF(\bz+\bq)\le \frac{2}{\alpha-1}\cdot \calF(\bz-\bz').\] 
Now, if we set $\bz''$ to be the vector in the span of $\bq$ and $\bz'$ that is closest to $\bz$ under $\calF$, we get
\[\calF(\bz-\bz'')\le \calF(\bz+\bq)\le \frac{2}{\alpha-1}\cdot \calF(\bz-\bz'),\] 
demonstrating progress. 
Thus, the learner can iteratively update $\bz'$, and within $\O{\log n}$ steps, $\bz'$ converges to $\bz$. 

A similar reasoning applies for functions $\calF$ that only satisfy the approximate triangle inequality. 
Using this framework with a corrector, an estimator, and a learner, combined with recursion, we can design a robust algorithm that outputs a $\Theta(1)$-approximation to $\calF$, using $n^{1/C}$ levels of recursion for some constant $C>1$.

\section{Turnstile Streaming Algorithm for \texorpdfstring{$F_2$}{F2} Moment Estimation}
\seclab{sec:turnstile:algs:ftwo}
In this section, we provide our robust algorithm for $F_2$ moment estimation on turnstile streams of length $m=\poly(n)$ with a universe of size $n$. 
Our algorithm uses $\poly\left(\frac{1}{\eps},\log n\right)$ space for $(1+\eps)$-approximation, as compared to the optimal algorithms that use $\O{\frac{1}{\eps^2}\log^2 n}$ bits of space on non-adaptive streams~\cite{AlonMS99}. 

First, we recall the following guarantees of the classical AMS algorithm:
\begin{theorem}[AMS sketch for estimating $F_2$]
\cite{AlonMS99}
\thmlab{thm:ams}
Given an accuracy parameter $\eps\in(0,1)$ and a failure probability parameter $\delta\in(0,1)$, let $k=\O{\eps^{-2}\log\frac{1}{\delta}}$. 
There exists a turnstile streaming algorithm $\AMS$ that maintains a linear sketch $\bA\bx$, where $\bA\in\mathbb{R}^{k\times n}$ is sampled from an explicit distribution over random matrices, uses $\O{k\log n}$ bits of space, and outputs, with probability at least $1-\delta$, a $(1+\eps)$-approximation to the second frequency moment $F_2$ for streams of length $m=\poly(n)$.
\end{theorem}

\begin{figure*}[!htb]
\begin{mdframed}
\textbf{Algorithm}:
\begin{enumerate}
\item
Maintain a hierarchical decomposition of the stream into a tree, where level $1$ consists of individual updates and each internal node has branching factor $B$.
\item
A node at level $i$ is created either after observing $B$ consecutive nodes from level $i-1$ in the stream, or when the iterate node at the parent level $i+1$ is updated.
\item
For every block $C_{i,j}$, apply an associated sketching matrix $\bB_{i,j}$ to summarize relevant vectors, e.g., the frequency vector of the block and any iterate vectors required by subroutines.
\item
Upon processing an update, let $H$ denote the current height of the tree, and maintain a sketch matrix $\bB_{H+1}$ at the root level.
\item
Return $\EstLevel(H+1,\bB_{H+1}\bz)$, where $\bz$ denotes the frequency vector of the entire stream.
\end{enumerate}
\end{mdframed}
\caption{Adversarially robust $F_2$ norm estimation algorithm on turnstile streams from \cite{GribelyukLWYZ26}}
\figlab{fig:alg:ltwo}
\end{figure*}

\paragraph{Algorithm overview.}
The algorithm organizes the stream into a hierarchical tree structure. 
At the base level $1$, the algorithm partitions the stream into blocks consisting of $B$ individual updates, while each higher level aggregates $B$ nodes from the level below into a single block. 
Every block $C_{i,j}$ is summarized via an associated sketching matrix $\bB_{i,j}$, which is used to maintain the frequency vector of the block, as well as any auxiliary iterate vectors. 
As the stream evolves, these sketches are recursively combined and passed upward, culminating in a root-level sketch $\bB_{H+1}$ that represents the entire stream. 
An estimate of the $F_2$ moment is obtained by invoking the subroutine $\EstLevel$ at the highest level of this recursive construction, which aggregates information from lower levels to yield an adversarially robust estimate. 
This hierarchical organization ensures that each sketch is subjected to only a bounded number of adaptive queries.  
With a choice of block size $B=\O{\eps^{-2}\log n}$ and tree height $H=\O{\log n}$, the algorithm uses $\poly(1/\eps,\log n)$ space while retaining robustness against adaptive updates. 
The full algorithm is given in \figref{fig:alg:ltwo}, with a schematic depiction shown in \figref{fig:alg:tree}.

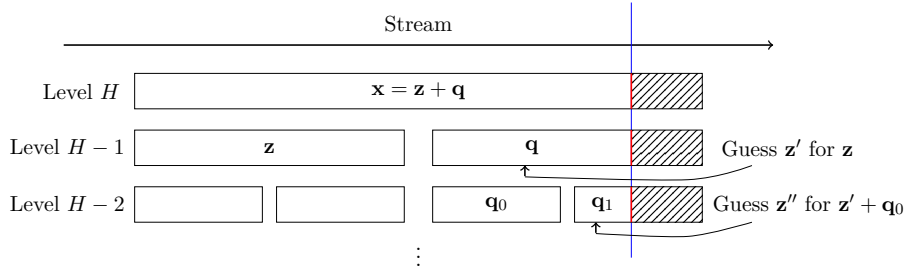
\begin{figure*}[!thb]
\scalebox{0.75}{
\centering
\begin{tikzpicture}[scale=1.25]
\draw(-1,4.2)[thick,->] -- (9,4.2);
\node at (4,4.5){Stream};

\draw(7,4.8)[color=blue] -- (7,1.2);
\draw(0,3.3) rectangle+(8,0.5); 
\fill[pattern=north east lines, draw=none] (7,3.3) rectangle (8,3.8);
\node at (4,3.55){$\bx=\bz+\bq$};
\draw(7,3.3)[thick,color=red] -- (7,3.8);
\node at (-0.75,3.55){Level $H$};

\draw(0,2.5) rectangle+(3.8,0.5); 
\node at (1.9,2.75){$\bz$};
\draw(4.2,2.5) rectangle+(3.8,0.5);
\fill[pattern=north east lines, draw=none] (7,2.5) rectangle (8,3);
\node at (5.6,2.75){$\bq$};
\draw(7,2.5)[thick,color=red] -- (7,3);
\node at (9.2,2.75){Guess $\bz'$ for $\bz$};
\draw plot [smooth] coordinates {(5.5,2.34) (7.8,2.3) (8.7,2.5)};
\draw[->, thick] (5.5,2.35) -- ++(0,0.15);
\node at (-0.95,2.75){Level $H-1$};

\draw(0,1.7) rectangle+(1.8,0.5); 
\draw(2,1.7) rectangle+(1.8,0.5); 
\draw(4.2,1.7) rectangle+(1.8,0.5); 
\node at (5.1,1.95){$\bq_0$};
\draw(6.2,1.7) rectangle+(1.8,0.5); 
\fill[pattern=north east lines, draw=none] (7,1.7) rectangle (8,2.2);
\node at (6.6,1.95){$\bq_1$};
\draw(7,1.7)[thick,color=red] -- (7,2.2);
\node at (9.5,1.95){Guess $\bz''$ for $\bz'+\bq_0$};
\draw plot [smooth] coordinates {(6.5,1.54) (7.8,1.5) (8.7,1.7)};
\draw[->, thick] (6.5,1.55) -- ++(0,0.15);
\node at (-0.95,1.95){Level $H-2$};

\node at (4,1.3){$\vdots$};
\end{tikzpicture}
}
\caption{Illustration of the recursive tree decomposition of the stream of \cite{GribelyukLWYZ26}, with branching factor $B=2$ and height $H$. 
At level $H-1$, the estimate for $\|\bx\|_2^2=\|\bz+\bq\|_2^2$ is given by $\|\bz-\bz'\|_2^2+\|\bz'+\bq\|_2^2$. 
At level $H-2$, the estimate for $\|\bz'+\bq\|_2^2=\|\bz'+\bq_0+\bq_1\|_2^2$ is computed as $\|\bz'+\bq_0-\bz''\|_2^2+\|\bz''+\bq_1\|_2^2$.}
\figlab{fig:alg:tree}
\end{figure*}

\paragraph{The $\EstLevel$ subroutine.}
We next describe the subroutine $\EstLevel$ and its guarantees. 
The role of $\EstLevel$ is to produce a robust estimate of the squared $L_2$ norm at a specified level~$i$ of the tree. 
For simplicity of notation, we present the description assuming that $\bq$ denotes the query vector at level~$i$; in the general execution, however, $\bq$ is further decomposed as $\bq=\bq_0+\bq_1$, where $\bq_0$ corresponds to contributions from the left siblings in the preceding block and $\bq_1$ corresponds to the contribution from the active block (namely, the blocks along the ancestor path of the current update). 
We denote by $\bz$ the portion of the stream accumulated in completed blocks of the parent node, so that the task of the subroutine is to estimate $\|\bz+\bq\|_2^2$.

\begin{algorithm}[!htb]
\caption{$\EstLevel(i,\bB_i\bz)$ for $F_2$ moment estimation \cite{GribelyukLWYZ26}}
\alglab{alg:est:level:ltwo}
\begin{algorithmic}[1]
\State{Let $\calP_i$ be the active block at level $i$}
\State{Use $\bB_i$ as the sketch matrix corresponding to $\calP_i$}
\State{}\Comment{This provides a $\left(1+\frac{\eps}{100H}\right)$-approximation for the $L_2$ norm, robust to $\poly\left(\frac{1}{\eps},\log n\right)$ adaptive queries}
\State{Let $A_i$ be a $\left(1+\frac{\eps}{100H}\right)$-approximation to $\|\bz+\bq\|_2^2$}
\State{Let $\bv$ be the current iterate obtained from $\MaintainIter(i)$}
\State{Let $\bq_0$ represent the portion of the query in the left siblings of $\calP_i$}
\If{$i\neq 1$}
    \State{Compute $P_i \gets \|\bB_i\bz+\bB_i\bq_0-\bB_i\bv\|_2^2$}
    \State{Recursively compute $Q_i \gets \EstLevel(i-1,\bB_{i-1}(\bv+\bq_1))$}
\Else
    \State{Set $P_i \gets 0$}
    \State{Define $\bq_1$ as the active portion of the query in $\calP_i$}
    \State{Compute $Q_i \gets \|\bB_1\bv+\bB_1\bq_0+\bB_1\bq_1\|_2^2$}
\EndIf
\If{$P_i+Q_i \in \left(1+\frac{\eps}{100H}\right)^{3i}\cdot A_i$}
    \State{Return $P_i + Q_i$}
\Else
    \State{Return $A_i$}
\EndIf
\end{algorithmic}
\end{algorithm}

Intuitively, the subroutine $\EstLevel$ operates by coordinating three components: an estimator, a learner, and a corrector. 
The learner maintains an iterate vector $\bv$, formed as a linear combination of past queries on which the estimator failed, thereby incrementally inferring information about $\bz$. 
Using this iterate, the estimator seeks to approximate $\|\bz+\bq\|_2^2$ by combining estimates of $\|\bz-\bv\|_2^2$ and $\|\bv+\bq\|_2^2$. 
The corrector observes the sketches $\bB_i\bz$ and $\bA\bz$ to identify when the estimator’s output deviates significantly from the true value, and then signals the learner to update $\bv$. 
Notably, the sketch $\bB_i$ is initialized once the preceding block at level $i+1$ is finalized. 
Consequently, it captures the frequency vectors contributed by all left siblings of the active block at level~$i$, and in particular maintains access to $\bB_i\bz$. 

At a high level, $\EstLevel$ decomposes the estimate into two terms: $P_i$, representing the contribution from the current block after adjustment by the learner’s iterate, and $Q_i$, which is obtained recursively from the level below. 
Concretely, $P_i$ approximates $\|\bz-\bv\|_2^2$, whereas $Q_i$ approximates $\|\bv+\bq\|_2^2$. 
When the combined estimate $P_i+Q_i$ roughly agrees with an independent estimate $A_i$ of the target quantity $\|\bz+\bq\|_2^2$, the algorithm returns $P_i+Q_i$; otherwise, it falls back to outputting $A_i$. 
This decision rule guarantees that each sketch is exposed to only a bounded number of adaptive queries, while the recursive aggregation across levels yields a robust and accurate estimate of the $F_2$ moment over the entire stream. 
This completes the description of the estimator–corrector interaction, which is formalized in \algref{alg:est:level:ltwo}. 

\paragraph{The $\MaintainIter$ subroutine.}
We now turn to the learner, implemented by the subroutine $\MaintainIter$, whose task is to update the iterate $\bv$ within the currently active block so as to approximate $\bz$. 
At the beginning of each block, the iterate $\bv$ is set to zero, reflecting the absence of information about the accumulated frequency vector $\bz$ from earlier blocks. 
As the stream evolves, the estimator provides tentative estimates to the $F_2$ moment. 
If such an estimate is inconsistent with the reference value $A_i$, the corrector signals that the estimator has failed. 
In response, $\MaintainIter$ updates $\bv$ by adding a small, appropriately scaled multiple of the current query vector, in the direction that reduces the estimation error.  
Through this iterative process, the learner progressively accumulates information about $\bz$ over the duration of the block, while operating exclusively within the sketched space. 
We give the full details of the algorithm in \algref{alg:mainiter:ltwo}, where we emphasize that all computations occur in the sketch space, i.e., under the image of the sketch matrix $\bB_i$ at level~$i$, since the algorithm never has direct access to the vectors $\bv$ and $\bq$.

\begin{algorithm}[!htb]
\caption{$\MaintainIter(i)$ for $F_2$ moment estimation \cite{GribelyukLWYZ26}}
\alglab{alg:mainiter:ltwo}
\begin{algorithmic}[1]
\State{At the start of the active block at level $i$, initialize the iterate $\bv=\mathbf{0}^n$}
\State{$\bq \gets \bq_0+\bq_1$}
\If{$P_i+Q_i \notin \left(1+\frac{\eps}{100H}\right)^i \cdot A_i$}
\State{Set $\sigma \gets \mathrm{sign}\left(A_i-(P_i+Q_i)\right)$}
\State{Compute $\alpha \gets \frac{1}{4}\sqrt{\frac{P_i}{Q_i}} \cdot \sigma \cdot \frac{\eps}{100H}$}
\State{Update the iterate as $\bv \gets \bv + \alpha(\bv+\bq)$}
\EndIf
\State{\Return $\bv$, truncated to precision $\frac{1}{\poly(n)}$}
\Comment{All operations performed in the sketch space}
\end{algorithmic}
\end{algorithm}

The main idea is to argue that the learner can recover $\bz$ after only a small number of updates. 
Each update occurs exactly when the estimator makes a mistake, and thus each such event effectively ``charges'' one additional adaptive interaction to the corrector. 
We upper bound the total number of these updates in \lemref{lem:bounded:iterations:ltwo}, using the following reasoning. 
Let $\bz'$ denote the value of the iterate $\bv$ just before an update. 
If the estimator outputs an inaccurate estimate of $\|\bz+\bq\|_2^2$, then the residual $\bz-\bz'$ must be aligned with the direction $\bq+\bz'$, since an incorrect estimate implies
\[|\langle \bz - \bz', \bq + \bz' \rangle| \ge \eps \cdot \|\bz - \bz'\|_2 \cdot \|\bq + \bz'\|_2.\]
In this case, the learner updates the iterate to $\bz''=\bz'+\alpha(\bq+\bz')$, where the step size $\alpha$ is chosen proportional to the residual magnitude and its sign is given by $\sigma=\mathrm{sign}(\langle \bz-\bz',\,\bq+\bz'\rangle)$. 
Viewing $\|\bz-\bz'\|_2^2$ as a potential function, one can show that
\[\|\bz - \bz''\|_2^2 \le (1-\eps^2) \|\bz - \bz'\|_2^2,\]
so each update makes noticeable progress toward learning $\bz$. 
Moreover, the step size $\alpha$ can be estimated to within a $(1\pm\eps)$ factor from the estimates $P_i$ and $Q_i$ with high probability. 
Since $\|\bz\|_2^2 \le \poly(n)$, this geometric decrease implies that the learner performs at most $\O{\frac{1}{\eps^2} \log n}$ iterations in total. 

\paragraph{Algorithm analysis.}
We now argue the guarantees of the algorithm. 
For simplicity of presentation, we focus on the case in which $\calP_i$ is the leftmost child of its parent. 
In this setting, $\bq_0=\mathbf{0}^n$, and hence $\bz+\bq_0=\bz$, so it is sufficient for the iterate $\bv$ to approximate $\bz$. 
In the general case, however, the requirement is that $\bv$ should approximate $\bz+\bq_0$.

We begin by showing that $P_i$ computed by $\EstLevel$ provides an accurate estimate of $\|\bz+\bq_0-\bv\|_2^2$.
\begin{lemma}
\lemlab{lem:est:proj}
\cite{GribelyukLWYZ26}
Let $\bv$ be the fixed iterate vector at a specific point in the stream, as defined in \algref{alg:est:level:ltwo}, conditioned on all previous updates. 
Then with high probability, 
\[\|\bz+\bq_0-\bv\|_2^2\le P_i\le\left(1+\frac{\eps}{100H}\right)\cdot\|\bz+\bq_0-\bv\|_2^2.\]
\end{lemma}
\begin{proof}
Since $\bB_i$ is a sketch matrix that provides a $\left(1+\frac{\eps}{100H}\right)$-multiplicative approximation, the bounds on $P_i$ follow directly from the guarantees of the $\AMS$ algorithm in \thmref{thm:ams}.
\end{proof}

Next, we argue that $Q_i$ computed by $\EstLevel$ is an accurate estimation of $\|\bv+\bq_1\|_2^2$. 
As a result, the sum $P_i+Q_i$ provides an accurate estimation of $\|\bz+\bq_0-\bv\|_2^2+\|\bv+\bq_1\|_2^2$.
\begin{lemma}
\lemlab{lem:estlvl}
\cite{GribelyukLWYZ26}
For each $i\in[H]$, with high probability the output $P_i+Q_i$ of $\EstLevel(i,\bB_i\bz)$ satisfies
\begin{align*}
\|\bz+\bq_0-\bv\|_2^2+\|\bv+\bq_1\|_2^2&\le P_i+Q_i\\
&\hspace{-0.3in}\le\left(1+\frac{\eps}{100H}\right)^{3i-2}\cdot\left(\|\bz+\bq_0-\bv\|_2^2+\|\bv+\bq_1\|_2^2\right).
\end{align*}
\end{lemma}
\begin{proof}
Assume that $\EstLevel(i-1,\bB_i(\bv+\bq_1))$ produces $Q_i$ satisfying 
\[\|\bv+\bq_1\|_2^2\le Q_i\le\left(1+\frac{\eps}{100H}\right)^{3(i-1)-2}\cdot\|\bv+\bq_1\|_2^2.\] 
Then, applying \lemref{lem:est:proj}, it follows that with high probability, the output $P_i+Q_i$ of $\EstLevel(i,\bB_{i+1}\bz)$ satisfies
\begin{align*}
\|\bz+\bq_0-\bv\|_2^2+\|\bv+\bq_1\|_2^2&\le P_i+Q_i\\
&\hspace{-0.3in}\le\left(1+\frac{\eps}{100H}\right)^{3i-2}\cdot\left(\|\bz+\bq_0-\bv\|_2^2+\|\bv+\bq_1\|_2^2\right).
\end{align*}
This argument serves as the inductive step for $i\in[H]$, so it remains to check the base case $i=1$. 
At this level, we have $Q_1=\|\bB_1\bv+\bB_1\bq_0+\bB_1\bq_1\|_2^2$. 
By the correctness of $\bB_1$, it holds that 
\[\|\bv+\bq_0+\bq_1\|_2^2\le Q_1\le\left(1+\frac{\eps}{100H}\right)\cdot\|\bv+\bq_0+\bq_1\|_2^2,\]
with high probability. 
Since $3i-2\ge 1$ for $i=1$, the base case
\[\|\bv+\bq_0+\bq_1\|_2^2\le Q_1\le\left(1+\frac{\eps}{100H}\right)^{3i-2}\cdot\|\bv+\bq_0+\bq_1\|_2^2\]
also holds with high probability. 
Finally, applying a union bound completes the induction, establishing the claim for all $i\in[H]$.
\end{proof}

Next, we argue that, conditioning on the correctness of the sketches at each node, the algorithm maintains an invariant: the estimator at every level $i$ remains correct, because the corrector at that level flags any inaccurate estimates. 
\begin{invariant}
\invarlab{invar:output:acc:ptwo}
\cite{GribelyukLWYZ26}
For any level $i\in[H]$, let $\calP$ denote the active block at level $i$, and let $\bB_{i+1}$ denote the sketch matrix of the parent of $\calP$. 
Let $\bq$ represent the portion of the query within $\calP$, and let $\bz$ denote the contribution from previous blocks. 
Then, with high probability,
\[\|\bz+\bq\|_2^2\le\EstLevel(i,\bB_i\bz)\le\left(1+\frac{\eps}{100H}\right)^{3i+1}\cdot\|\bz+\bq\|_2^2.\]
\end{invariant}
\begin{proof}
First, consider the case where the current iterate $\bv$ is not updated to $\bv + \alpha(\bq + \bv)$ for the current query $\bq$. 
By the design of the algorithm, we then have
\[\calA_i(\bz+\bq)\le P_i+Q_i\le\left(1+\frac{\eps}{100H}\right)^{3i}\cdot\calA_i(\bz+\bq).\] 
Using the correctness guarantee of $\calA_i$, it follows that
\[\|\bz+\bq\|_2^2\le\calA_i(\bz+\bq)\le\left(1+\frac{\eps}{100H}\right)\cdot\|\bz+\bq\|_2^2.\] 
Combining these inequalities gives
\[\|\bz+\bq\|_2^2\le\EstLevel(i,\bB_i\bz)\le\left(1+\frac{\eps}{100H}\right)^{3i+1}\cdot\|\bz+\bq\|_2^2,\] 
so the invariant holds when $\bv$ is not updated.

Alternatively, if the estimator is incorrect, we update the iterate to $\bv' = \bv + \alpha(\bq + \bv)$ and return $A_i$, which is a $\left(1+\frac{\eps}{100H}\right)$-approximation of $\|\bz+\bq\|_2^2$. 
Therefore, the invariant holds for all levels $i\in[H]$.
\end{proof}

We now upper bound the total number of updates to the iterate at any level $i\in[H]$ over the course of the stream. 

\begin{lemma}[Bounded iterations]
\lemlab{lem:bounded:iterations:ltwo}
\cite{GribelyukLWYZ26}
Let $L_i$ denote the number of updates to $\bv$ at level $i$ and let $\eta = \Theta\left(\frac{\eps}{H}\right)$. 
With high probability, $L_i=\O{\frac{1}{\eta^2}\log n}$ for all levels $i \in [H]$ and all times in the stream, assuming $m \le \poly(n)$.
\end{lemma}
\begin{proof}
Fix a level $i\in[H]$ and a specific time in the stream, and let $\calP_i$ and $\calP_{i+1}$ be the active blocks at levels $i$ and $i+1$, respectively. 
Let $\bz$ denote the frequency vector corresponding to all updates prior to $\calP_{i+1}$. 
Conditioned on the accuracy of $\EstLevel$ from \lemref{lem:estlvl}, we have
\begin{align*}
\|\bz - \bv\|_2^2 + \|\bv + \bq \|_2^2 \leq \EstLevel(i, \bB_{i+1}\bz) \leq (1+\eta) \cdot \left( \|\bz - \bv\|_2^2 + \|\bv + \bq \|_2^2\right).
\end{align*}
An update occurs only if the estimate is inaccurate, meaning
\[\left\lvert\|\bz-\bv\|_2^2+\|\bv+\bq\|_2^2-\|\bz+\bq\|_2^2\right\rvert\ge\eta\cdot\|\bz+\bq\|_2^2.\] 
Expanding the left-hand side gives
\begin{align*}
|\langle \bz - \bv, \bv + \bq \rangle| &\geq \frac{\eta}{2} \|\bz + \bq\|_2^2\\ 
&= \frac{\eta}{2}\left(\|\bz - \bv\|_2^2 + 2 \langle \bz - \bv, \bv + \bq \rangle + \|\bv + \bq \|_2^2 \right).
\end{align*}
Suppose $\langle \bz -\bv, \bv + \bq \rangle \geq \frac{\eta}{2} \|\bz + \bq\|_2^2$. 
Rearranging, we get
\begin{align*}
(1-\eta) \cdot\langle \bz - \bv, \bv + \bq \rangle &\geq \frac{\eta}{2} \left(\|\bz - \bv\|_2^2 + \|\bv + \bq\|_2^2 \right) \\ 
& \geq \eta \|\bz - \bv\|_2\cdot \|\bv + \bq\|_2. 
\end{align*}
Thus, $\langle \bz - \bv, \bv + \bq \rangle \geq \eta \|\bz - \bv\|_2\cdot\|\bv + \bq\|_2$, and $\MaintainIter$ updates $\bv$ to $\bv' = \bv + \alpha(\bv + \bq)$ for some $\alpha \in [-1,1]$. 

To bound the total number of updates, we track the progress measure $\|\bz - \bv \|_2^2 - \|\bz - \bv'\|_2^2$. 
First, note
\begin{align*}
\|\bz - \bv'\|_2^2 &= \|\bz - \bv - \alpha(\bv+\bq)\|_2^2 \\
&= \|\bz - \bv\|_2^2 - 2\alpha \langle \bz - \bv, \bv + \bq \rangle + \alpha^2 \|\bv + \bq\|_2^2.
\end{align*}
Hence,
\begin{align*}
\|\bz - \bv\|_2^2 - \|\bz - \bv'\|_2^2 &\geq 2\alpha \langle \bz - \bv, \bv + \bq \rangle - \alpha^2 \|\bv + \bq\|_2^2 \\ 
&\geq 2 \alpha \cdot \eta \|\bz - \bv\|_2 \cdot\|\bv + \bq \|_2 - \alpha^2 \|\bv + \bq\|_2^2.
\end{align*}
For any $\alpha = \Theta(\eta) \cdot \frac{\|\bz - \bv\|_2}{\|\bv + \bq\|_2}$ with a constant in $\left[\frac{1}{10},1\right]$, we obtain
\begin{align*}
\|\bz - \bv\|_2^2 - \|\bz - \bv'\|_2^2 \geq \frac{\eta^2}{100} \cdot \|\bz - \bv\|_2^2.
\end{align*}
Similarly, if $\langle \bz - \bv, \bv + \bq \rangle \leq -\frac{\eta}{2} \|\bz + \bq\|_2^2$, then
\begin{align*}
\langle \bz - \bv, \bv + \bq \rangle &\leq -\frac{\eta}{2} \left(\|\bz - \bv\|_2^2 + 2 \langle \bz - \bv, \bv + \bq \rangle + \|\bv + \bq \|_2^2 \right),
\end{align*}
so that $(1+\eta) \cdot \langle \bz - \bv, \bv + \bq \rangle \leq  -\eta \|\bz - \bv\|_2 \|\bq + \bv\|_2$, i.e.,
\[- \langle \bz - \bv, \bv + \bq \rangle \geq  \frac{\eta}{2} \|\bz - \bv\|_2 \|\bq + \bv\|_2.\] 
Again, considering the progress measure $\|\bz - \bv\|_2^2 - \|\bz - \bv'\|_2^2$, we have
\begin{align*}
\|\bz - \bv\|_2^2 - \|\bz - \bv'\|_2^2 &\geq 2\alpha \langle \bz - \bv, \bv + \bq \rangle - \alpha^2 \|\bq + \bv\|_2^2,
\end{align*}
and for $\alpha = -\Theta(\eta) \cdot \frac{\|\bz - \bv\|_2}{\|\bq + \bv\|_2}$ with a constant in $\left[\frac{1}{10},1\right)$, we have
\[\|\bz - \bv\|_2^2 - \|\bz - \bv' \|_2^2 \geq \frac{\eta^2}{100} \|\bz - \bv\|_2^2.\]
Thus, in both cases, $\|\bz - \bv'\|_2^2$ decreases by a factor of $(1-\O{\eta^2})$. 
Since $\|\bz\|_2^2 \le \poly(n)$, this holds even when the image of $\bv'$ is truncated to precision $\frac{1}{\poly(n)}$. 
Starting from $\bv = \mathbf{0}^n$ and noting $\|\bz\|_2^2 \le \poly(n)$, it follows that $\bv$ is updated at most $\O{\frac{1}{\eta^2} \log n}$ times.
\end{proof}

Finally, we prove that our algorithm uses polylogarithmic space and robustly outputs a $(1+\eps)$-approximation of the $F_2$ moment on insertion-deletion streams. 

\begin{restatable}{theorem}{thmltwo}
\thmlab{thm:ltwo}
\cite{GribelyukLWYZ26}
Given accuracy parameter $\eps\in(0,1)$, there exists an adversarially robust turnstile streaming algorithm for a stream of length $m=\poly(n)$ that uses $\poly\left(\frac{1}{\eps},\log n\right)$ bits of space and with high probability, outputs a $(1+\eps)$-approximation to the $F_2$ moment at all times for the underlying frequency vector over a universe of size $n$.
\end{restatable}
\begin{proof}
First, note that a randomized adversary can be viewed as a distribution over internal randomness, and the probability of a successful attack is upper bounded by the maximum success probability over all sequences. 
Hence by fixing the adversary's private random tape, it suffices to consider a deterministic adversary without loss of generality. 

We begin by tracking the adaptive interactions with the linear sketches used by the corrector. 
At each step, $\MaintainIter$ queries each $F_2$ linear sketch $\bB_i$ to determine whether to update the iterate $\bv$ based on the current query $\bq$. 
We can encode the sketch transcript as a sequence of symbols, either $\bot$ or $\top$, where $\bot$ indicates no update and $\top$ triggers an update. 
Suppose that $\top$ occurs at most $L$ times in a single block; since there are $B$ blocks in the level below, the corrector will be used to flag at most $B \cdot L$ queries overall. 

Each time the iterate is updated also triggers the sketch at level $i-1$ to maintain $\bB_{i-1}\bv'$, where $\bv' = \bv + \eps(\bq + \bv)$ is the updated iterate, and releases the estimate produced by $\bB_i$ without revealing any other information about $\bB_i$. 
Additionally, the linear sketch outputs an estimate for the $F_2$ moment at the start of each of the $B$ blocks, encoded in $C_1\log n$ bits for some constant $C_1>0$.  
The entries of the truncated image of $\bv$ are multiples of precision $\frac{1}{\poly(n)}$ and bounded in magnitude by $\poly(n)$, as otherwise $\bv$ could not be a good approximation to $\bz$, violating the progress analysis in \lemref{lem:bounded:iterations:ltwo}. 
Thus, the total number of possible output sequences from $\bB_i$ is at most
\[\binom{m}{B \cdot L}\cdot \left(2^{C_1\log n}\right)^{B \cdot L + B}.\]
As we assume without loss of generality that the adversary is deterministic, any input stream corresponds to one of these sequences. 
By setting the total failure probability $\delta$ so that
\[\delta \le \frac{1}{(nm)^3} \cdot \left(\binom{m}{B \cdot L}\cdot \left(2^{C_1\log n}\right)^{B \cdot L + B}\right)^{-1},\]
then by a union bound over all possible input streams that the deterministic adversary can produce, it follows that the sketch $\bB_i$ is correct with probability $1-\frac{1}{(nm)^3}$.  
For $m=\poly(n)$, it thus suffices to take 
\[\log\frac{1}{\delta} = \O{(B\cdot L + B)\cdot \log n}.\]
This establishes robustness, after which correctness then follows from \invarref{invar:output:acc:ptwo}.

Next, we upper bound the number of adaptive interactions with the estimator, which consists of the two components that output $P_i$ and $Q_i$. 
While the correctness of $Q_i$ follows from recursion, we use a bounded computation paths argument for $P_i$. 
Each sketch matrix $\bB_i$ has entries rounded to $\O{\log n}$ bits, so each $P_i$ output can be encoded in $\O{\log n}$ bits. 
The number of adaptive interactions with $\bB_i$ is at most $L$, the number of updates to the iterate $\bv$. 
Hence, the total number of computation paths is at most 
\[\binom{m}{L}\cdot \left(2^{C_2 \log n}\right)^L\]
for some constant $C_2>0$, so it again suffices to set 
\[\log\frac{1}{\delta} = \O{(L+B)\cdot \log n}.\]
Every adaptive query to $P_i$ corresponds to a different iterate $\bv$, and all other queries occur between updates and are correct conditioned on the correctness of the value of $\bv$. 

\paragraph{Space complexity.}
Finally, we analyze the space complexity.  
For a fixed tree, at each level $i \in [H]$, we maintain $B$ sketching matrices for the $B$ nodes in level $i+1$. 
Each sketch uses accuracy $(1+\eta)$ for $\eta = \frac{\eps}{100H}$ and failure probability $\delta$, with 
\[\log\frac{1}{\delta} = \O{(B \cdot L + B)\cdot\log n}.\]
By \lemref{lem:bounded:iterations:ltwo}, $L \le \frac{C_3}{\eta^2} \log n$ for some constant $C_3>0$ with high probability, conditioned on the correctness of the sketch. 
We need $B^H \ge m$ and $B \ge L$, so it suffices to set 
\[B = \O{\frac{1}{\eta^2} \log n}.\]
By \thmref{thm:ams}, each linear sketch has dimension 
\[\tO{\frac{1}{\eta^2} \log \frac{1}{\delta}}.\]
Since 
\[\log\frac{1}{\delta} = \O{(B \cdot L + B)\cdot\log n} = \O{\frac{1}{\eta^4}\log^3 n},\]
this gives sketch dimension 
\[\tO{\frac{1}{\eta^6}\log^3 n},\]
and each sketch uses 
\[\tO{\frac{1}{\eta^6}\log^4 n}\]
bits.  

There are $B=\O{\frac{1}{\eta^2}\log n}$ sketches per level across $H=\O{\log n}$ levels in the recursion. 
However, it suffices to maintain only one active sketch per level at any given time. 
Thus, the total space is 
\[\O{\frac{1}{\eta^6}\log^5 n} = \tO{\frac{1}{\eps^6}\log^{11} n}\]
bits of space.  
\end{proof}

\begin{remark}
Instead of the computation paths argument presented above, we could alternatively guarantee that every sketch $\bB_i$ is robust against $B\cdot L + B= \O{B^2}$ adaptive interactions by utilizing the differential privacy framework of \cite{HassidimKMMS20} within each block $C_{i,j}$. 
According to \thmref{thm:frame:dp}, this framework requires a space complexity of 
\[\O{\frac{1}{\eta^2} \log\left(\frac{1}{\delta}\right) \log n \cdot \sqrt{(B \cdot L + B) \cdot \log \left(\frac{1}{\delta}\right)} \cdot \log\left(\frac{m}{\eta \delta}\right)}\]
per sketch $\bB_i$. 
Because we only maintain one active sketch at any given time across each of the $H = \O{\log n}$ levels, substituting $B = \frac{1}{\eta^2} \log n$, $\eta = \frac{\eps}{H}$, and $\delta = \frac{1}{(nm)^3}$ for $m \leq \poly(n)$ results in an overall space footprint for our algorithm of
\[\O{\frac{1}{\eps^4} \cdot \log^{9.5} n + \frac{1}{\eps^4} \log^{8.5} n \cdot \log\left(\frac{\log n}{\eps}\right)}.\]
\end{remark}

\section{Approximate Triangle Inequality}
\seclab{sec:turnstile:algs:triangle}
In this section, we generalize our robust algorithm for $F_2$ estimation to a wider class of functions $\calF$ that satisfy an approximate triangle inequality.  
Recall that for some constant $\beta>0$, a function $\calF$ obeys a $\beta$-approximate triangle inequality if  
\[\calF(\bx-\bz)\le\beta\cdot(\calF(\bx-\by)+\calF(\by-\bz)),\]  
for all vectors $\bx,\by,\bz\in\mathbb{R}^n$.  

As before, the framework uses an estimator, a corrector, and a learner, so that the algorithm maintains an iterate $\bz'$ approximating the previous stream vector $\bz$.  
Each incoming query $\bq$ is processed by computing $P_i\approx\calF(\bz-\bz')$ and $Q_i\approx\calF(\bz'+\bq)$, where $P_i$ is obtained using a non-adaptive sketch and $Q_i$ via a recursive procedure.  
The estimator then outputs the sum $P_i+Q_i$, which by the triangle inequality is at least $\calF(\bz+\bq)$.  
If this sum significantly overestimates $\calF(\bz+\bq)$, then, by the approximate triangle inequality, the iterate can be updated to a new vector $\bz''=-\bq$ in a way that provably reduces the distance to the true vector under $\calF$.  
Specifically, \lemref{lem:precond:tri} formalizes how such a precondition triggers an update, ensuring that $\calF(\bz-\bz'')$ is a constant factor smaller than $\calF(\bz-\bz')$, reflecting progress toward learning $\bz$.  
Consequently, \lemref{lem:bounded:iterations:tri} shows that the algorithm performs at most $\O{\log n}$ updates per level.  
Finally, we employ the same recursive structure to bound the number of adaptive interactions at each level.  
The key subroutine $\EstLevel$ is described in \algref{alg:est:level:tri}, and it is embedded within the same tree structure as in \figref{fig:alg:tree}.  

We first argue that when the estimator produces an inaccurate output on a query, there is significant progress: $\calF(\bz-\bz'')$ becomes a constant factor smaller than $\calF(\bz-\bz')$, where $\bz'$ and $\bz''$ are the iterates before and after the query.  
Since $\calF$ satisfies a $\beta$-approximate triangle inequality, we have  
\[\calF(\bz - \bz') + \calF(\bz' + \bq) \ge \frac{1}{\beta} \cdot \calF(\bz + \bq).\]

\begin{lemma}[Precondition triggering]
\lemlab{lem:precond:tri}
\cite{GribelyukLWYZ26}
Let $\calA$ be an algorithm that outputs a $\kappa$-approximation to a symmetric function $\calF$ satisfying the $\beta$-triangle inequality, and define $Z=\calA(\bz-\bz')+\calA(\bz'+\bq)$.  
Suppose $Z>\kappa^2\cdot\calA(\bz+\bq)$.  
Then after setting $\bz''=-\bq$, with high probability, we have
\[\calF(\bz-\bz'')\le\frac{\beta+1}{\kappa-\beta}\cdot\calF(\bz-\bz').\]
\end{lemma}
\begin{proof}
We first condition on the correctness of the subroutines. 
Then with high probability, we have:
\begin{enumerate}
\item
$Z\in\left[\calF(\bz-\bz')+\calF(\bz'+\bq),\kappa\cdot(\calF(\bz-\bz')+\calF(\bz'+\bq))\right]$
\item
$\calA(\bz+\bq)\in\left(\calF(\bz+\bq),\kappa\cdot\calF(\bz+\bq)\right]$
\end{enumerate}
Hence, if $Z>\kappa^2\cdot\calA(\bz+\bq)$, we obtain  
\[\kappa\cdot(\calF(\bz-\bz')+\calF(\bz'+\bq))\ge Z>\kappa^2\cdot\calA(\bz+\bq)\ge\kappa^2\cdot\calF(\bz+\bq),\]  
which implies  
\[\calF(\bz-\bz')+\calF(\bz'+\bq)\ge\kappa\cdot\calF(\bz+\bq).\]

Using the $\beta$-approximate triangle inequality of $\calF$, we also have $\beta\cdot(\calF(\bz+\bq)+\calF(\bz'-\bz))\ge\calF(\bz'+\bq)$. 
Since $\calF$ is symmetric, then $\calF(\bz'-\bz)=\calF(\bz-\bz')$, so that
\[(\beta+1)\cdot\calF(\bz-\bz')\ge(\kappa-\beta)\cdot\calF(\bz+\bq).\]

Therefore, updating the iterate from $\bz'$ to $\bz''=-\bq$ gives
\[\calF(\bz-(-\bq))\le\frac{\beta+1}{\kappa-\beta}\cdot\calF(\bz-\bz'),\]  
which establishes our notion of progress.
\end{proof}
Consequently, the number of times the iterate $\bv$ approximating $\bz$ can be updated is at most $\O{\log n}$.  

\begin{lemma}[Bounded iterations]
\lemlab{lem:bounded:iterations:tri}
\cite{GribelyukLWYZ26}
Suppose $\calF$ is a function that, on a stream of length $m=\poly(n)$, takes values in $\left[\frac{1}{\poly(m)},\poly(m)\right]$.  
Let $L_i$ denote the number of updates to the iterate $\bz'$ at level $i$.  
Then with high probability, $L_i\le\O{\log n}$ throughout the stream for all $i\in[H]$. 
\end{lemma}
\begin{proof}
Fix a level $i \in [H]$ and a point in the stream, and let $\calP_i$ be the active block at level $i$.  
Let $\bq_1$ be the active portion of the query in $\calP_i$, and $\bq_0$ be the contribution from left siblings of $\calP_i$, so that $\bq = \bq_0 + \bq_1$.  
Let $\bz$ represent the current state of the stream, and $\bz'$ be the previous iterate maintained by $\EstLevel$.

Conditioned on the correctness of the subroutines in \algref{alg:est:level:tri}, if the iterate $\bv$ is not updated, we have $P_i + Q_i \le \kappa^i \cdot A_i$, where $P_i$ and $Q_i$ estimate $\calF(\bz + \bq_0 - \bz')$ and $\calF(\bz' + \bq_1)$, respectively, and $A_i$ is a $\kappa^{4i}$-approximation to $\calF(\bz + \bq)$.  
If the iterate $\bv$ is updated, then the precondition of \lemref{lem:precond:tri} triggers, replacing $\bz'$ with $\bz'' = -\bq_1$.  
By \lemref{lem:precond:tri}, this ensures progress:
\[\calF(\bz - \bz'') = \calF(\bz + \bq) \le \frac{\beta+1}{\kappa-\beta} \cdot \calF(\bz - \bz').\]

Thus, each update to $\bz'$ guarantees a constant-factor reduction in $\calF(\bz - \bz')$, provided $\kappa>2\beta+1$.  
Since initially $\calF(\bz) \le \poly(n)$ and the minimum value of $\calF$ is at least $\frac{1}{\poly(n)}$, it follows that the iterate $\bv$ can be updated at most $L_i \le \O{\log n}$ times per level $i$.  
Taking a union bound over all $H$ levels gives the claimed high-probability bound.
\end{proof}

\begin{algorithm}[!htb]
\caption{$\EstLevel(i,\bB_i\bz)$ for function $\calF$ with $\beta$-approximate triangle inequality}
\alglab{alg:est:level:tri}
\begin{algorithmic}[1]
\State{Let $\calP_i$ be the active blocks at level $i$}
\State{Set $\bq_1$ as the active query portion in $\calP_i$}
\State{Set $\bq_0$ as the query contribution from the left siblings of $\calP_i$}
\State{$\bq\gets\bq_0+\bq_1$}
\State{Use $\bB_i$ as the sketch matrix for $\calP_i$}
\State{}\Comment{$\kappa$-approximation for function $\calF$, robust to $\tO{n^{1/C}}$ adaptive queries}
\State{Let $A_i$ be a $\kappa^{4i}$-approximation to $\calF(\bz+\bq)$}
\State{Retrieve $\bv$ as the previous iterate}
\State{Compute $P_i$ as an estimate of $\calF(\bz+\bq_0-\bv)$ using $\bB_i$}
\If{$i\neq 1$}
\State{Compute $Q_i \gets \EstLevel(i-1,\bB_{i-1}(\bv+\bq_1))$ recursively}
\Else
\State{Compute $Q_i$ as an estimate of $\calF(\bv+\bq_1)$ using $\bB_1$}
\EndIf
\If{$P_i+Q_i\le\kappa^i\cdot A_i$}
\State{\Return $P_i + Q_i$}
\Else
\State{$\bv\gets-\bq_1$}
\Comment{Performed in the sketch space}
\State{\Return $A_i$}
\EndIf
\end{algorithmic}
\end{algorithm}

We now give the full guarantees of our algorithm for estimation of functions that satisfy $\beta$-approximate triangle inequality. 
\begin{restatable}{theorem}{thmtri}
\thmlab{thm:tri}
\cite{GribelyukLWYZ26}
Suppose $\calF$ is a function over a stream of length $m=\poly(n)$, taking values in the range $\left[\frac{1}{\poly(m)},\poly(m)\right]$, and satisfies the $\beta$-approximate triangle inequality. 
Let $\kappa>2\beta+1$ be a fixed constant, and assume there exists a non-adaptive turnstile streaming algorithm that uses $S(n)\cdot\log\frac{1}{\delta}$ bits of space and, with probability at least $1-\delta$, outputs a $\kappa$-approximation to $\calF$. 
Then, for any constant $C>1$, there is an adversarially robust insertion-deletion streaming algorithm on a stream of length $m$ that uses $\tO{n^{1/C}}\cdot S(n)$ bits of space and, with high probability, maintains a $\kappa^{\O{C}}$-approximation to $\calF$ at all times.
\end{restatable}
\begin{proof}
As before, we start by reducing to the case of a deterministic adversary. 
Any randomized adversary is just a distribution over private random tapes, and its success probability is upper bounded by the maximum over deterministic sequences. 
Thus by fixing the adversary's private random tape, it suffices to consider only deterministic adversaries.

Next, we examine the interactions with the sketches used in \algref{alg:est:level:tri}. 
At each step, $\EstLevel$ queries the sketch at level $i$ to decide whether to update the iterate $\bv$. 
We can view the sketch outputs as symbols: $\bot$ if no update is performed, and $\top$ if the iterate is updated. 
By \lemref{lem:bounded:iterations:tri}, the number of $\top$ events at level $i$ is at most $L_i = \O{\log n}$ with high probability.  

Let $B$ denote the number of blocks in each level. 
Each update at level $i$ triggers a query to level $i-1$, but does not leak further information about level $i$. 
Additionally, at the start of each block, the sketch outputs a $\kappa^i$-approximation encoded in $\O{1}$ words. 
Therefore, the total number of possible output transcripts at level $i$ is upper bounded by 
\[\binom{m}{B \cdot L_i} \cdot \left(2^{C_1 \log n}\right)^{B \cdot L_i + B},\]
where $C_1>0$ is a constant bounding the number of bits per output.

To maintain a constant-factor approximation, we require $\kappa^{\O{H}} = \O{1}$, which implies $H = \O{1}$. 
Specifically, for any constant $C>1$, there exists $H = \O{1}$ such that the bottom-level block sizes are at most $n^{1/C}$. 
Moreover, each output from the estimator or from $P_i$ uses at most $\O{\log n}$ bits. 
Hence, at the bottom level, the number of possible transcripts is bounded by 
\[\left(2^{C_1 \log n}\right)^{n^{1/C}+L},\]
while at other levels it is upper bounded by 
\[\binom{m}{B \cdot L}\cdot \left(2^{C_1 \log n}\right)^{B \cdot L + B},\]
since there are $L$ adaptive interactions with the corrector for each of the $B$ blocks in the level below. 
Choosing $B = \O{n^{1/C}}$, it suffices to set the failure probability of each sketch so that 
\[\log\frac{1}{\delta} \le \O{n^{1/C}} \cdot \log m\]
for some constant $C>1$. 

Conditioned on all sketches being correct, the iterate updates always make progress by \lemref{lem:precond:tri}, reducing $\calF(\bz-\bv)$ by a constant factor each time. 
Since $\calF$ lies in $\left[\frac{1}{\poly(m)},\poly(m)\right]$, this ensures that each level performs at most $\O{\log n}$ updates, consistent with $L_i$.

To analyze the space complexity, note that each sketch requires 
\[\log\frac{1}{\delta} \le \O{n^{1/C}}\cdot\log m.\]
Given that each sketch occupies $S(n)\cdot \log\frac{1}{\delta}$ space, each sketch uses $\tO{n^{1/C}}\cdot S(n)$ space. 
Although we set $B = \O{n^{1/C}}$, at any given time it suffices to maintain only a single active sketch per level. 
Thus, the total space per level is $\tO{n^{1/C}}\cdot S(n)$. 
Since there are $H = \O{1}$ levels, the overall space bound follows.

Finally, correctness is guaranteed by the approximate triangle inequality argument: the iterate updates always ensure the output remains a $\kappa^{\O{C}}$-approximation to $\calF$ at all times. 
Furthermore, while the $\beta$-approximate triangle inequality only guarantees a lower bound of $\frac{1}{\beta}\calF(\bz+\bq)$ at each recursive step, across the $H$ levels of recursion this lower bound degrades by at most a factor of $\beta^H$; because $H = \mathcal{O}(1)$ and $\beta$ is a constant, this degradation is absorbed into the final $\kappa^{\O{C}}$ two-sided approximation factor.
Thus, the algorithm is adversarially robust and achieves the stated space and approximation guarantees.
\end{proof}

\paragraph{Applications.}
We now give several examples of functions that obey the approximate triangle inequality. 
Our first example is $\mathcal{F}(\bx) = \|\bx\|_p^p$ for $0 \le p \le 1$.
\begin{lemma}
\lemlab{lem:xp:small}
\cite{GribelyukLWYZ26}
For any $\bx, \by \in \mathbb{R}^n$ and $p\in(0,1]$, we have $\|\bx + \by\|_p^p \le \|\bx\|_p^p + \|\by\|_p^p$. 
Moreover, we have $\|\bx + \by\|_0 \le \|\bx\|_0 + \|\by\|_0$. 
\end{lemma}
\begin{proof}
By a standard argument, we have $\|\bx + \by\|_0 \le \|\bx\|_0 + \|\by\|_0$. 
Recall that for $p\in(0,1]$, the inequality $|x + y|^p \le |x|^p + |y|^p$ holds. 
Applying this coordinate-wise gives
\[\|\bx + \by\|_p^p = \sum_{i = 1}^n |x_i + y_i|^p \le \sum_{i = 1}^n \left(|x_i|^p + |y_i|^p\right)\le \|\bx\|_p^p + \|\by\|_p^p.\]
\end{proof}
Next, we consider $\mathcal{F}(\bx) = \|\bx\|_p^p$ for $p \ge 1$.
\begin{lemma}
\lemlab{lem:xp:big}
\cite{GribelyukLWYZ26}
For any $\bx, \by \in \mathbb{R}^n$ and $p\ge 1$, we have $\|\bx + \by\|_p^p \le 2^p\cdot(\|\bx\|_p^p + \|\by\|_p^p)$.
\end{lemma}
\begin{proof}
Recall that for $p\ge1$, the inequality $|x + y|^p \le 2^p(|x|^p + |y|^p)$ holds. 
Applying this coordinate-wise gives
\[\|\bx + \by\|_p^p = \sum_{i = 1}^n |x_i + y_i|^p \le \sum_{i = 1}^n 2^p\cdot\left(|x_i|^p + |y_i|^p\right)\le 2^p\cdot(\|\bx\|_p^p + \|\by\|_p^p).\]
\end{proof}
It is known how to achieve $(1+\eps)$-approximation for $F_p$ estimation for all $p$ on non-adaptive turnstile streams~\cite{AlonMS99,IndykW05,Indyk06,Li08,KaneNW10a,KaneNW10b,AndoniKO11,Ganguly11,KaneNPW11,GangulyW18}. 

We also recall the following definition of a symmetric norm:
\begin{definition}[Symmetric Norm]
A norm $\|\cdot\|$ on $\mathbb{R}^n$ is called a \emph{symmetric norm} if it is invariant under permutations and sign changes of coordinates. 
That is, for every $\bx \in \mathbb{R}^n$, every permutation $\pi$ of $\{1,\ldots,n\}$, and every choice of signs $\sigma_i \in \{-1,1\}$, we have
\[\|\bx\| = \|(\sigma_1 x_{\pi(1)}, \sigma_2 x_{\pi(2)}, \ldots, \sigma_n x_{\pi(n)})\|.\]
Equivalently, $\|\bx\|$ depends only on the multi-set of absolute values $\{|x_1|,\ldots,|x_n|\}$.
\end{definition}
It is known how to achieve $(1+\eps)$-approximation to all symmetric norms on non-adaptive turnstile streams~\cite{BlasiokBCKY17,BravermanWZ21,BravermanMWZ23}. 
Additionally, because all norms must satisfy the triangle inequality, symmetric norms must also satisfy the triangle inequality. 
Hence, we can apply our framework for robust estimation on turnstile streams. 

Next, recall that a Bernstein function satisfies the following definition: 
\begin{definition}[Bernstein function, e.g., \cite{schilling2012bernstein}]
\deflab{def:bernstein-function}
A function $f:(0,\infty)\to[0,\infty)$ is said to be a \emph{Bernstein function} if it is infinitely differentiable on $(0,\infty)$ and its first derivative $f'$ is completely monotone. 
Formally, this means that for every $n\in\mathbb{N}_0$ and all $x>0$,
\[(-1)^n \cdot f^{(n+1)}(x) \ge 0.\]
\end{definition}

We now turn to functions $\mathcal{F}$ of the form $\mathcal{F}(\bx) = \sum_{i = 1}^n g(x_i)$, where $g$ satisfies $g(t) = f(t^2)$ for some Bernstein function $f$.
In particular, we list several representative choices of the function $g$, all of which are commonly encountered in robust statistics. 
\begin{itemize}
\item 
(Pseudo-Huber loss): $g_{\tau}(x) = \tau\left(\sqrt{1 + \left(x / \tau\right)^2} - 1\right)$
\item 
(Cauchy/Lorentzian loss): $g_{\tau}(x) = \log\left(1 + x^2 / \tau\right)$
\item 
(Generalized Charbonnier loss): $g_{\tau}(x) = \left(1 + x^2 / \tau\right)^\beta - 1$ for $0 < \beta \le 1$
\item 
(Welsch/Leclerc loss): $g_{\tau}(x) = 1 - e^{-x^2 / \tau}$
\item 
(Geman-McClure loss): $g_{\tau}(x) = \frac{x^2}{x^2 + \tau}$
\end{itemize}

To establish the approximate triangle inequality, following the same high-level approach as in~\lemref{lem:xp:small} and \lemref{lem:xp:big}, it is enough to analyze the one-dimensional setting.

\begin{lemma}
\cite{GribelyukLWYZ26}
Assume $g(t) = f\left(t^2\right)$ for some Bernstein function $f$, with $g(0) = 0$. 
Then for all $a, b \in \mathbb{R}$, we have
\[g(a + b) \le 2\left(g(a) + g(b)\right).\]
\end{lemma}
\begin{proof}
From the definition of a Bernstein function, the derivative $f'(x)$ is non-negative but non-increasing. 
We also have $f(0)=0$ since $g(0)=0$. 
Thus, $f$ is sub-additive on $[0,\infty)$, so $f(a+b)\le f(a)+f(b)$ for all $a,b\ge 0$. 
Hence, we have $f(2a)\le 2f(a)$ for all $a\ge 0$. 

Consequently, for any $0 \le a \le b$, we have
\[(a+b)^2 \le 2 a^2 + 2 b^2,\]
so that 
\begin{align*}
g(a+b) &= f((a+b)^2) \\
&\le f(2 a^2 + 2 b^2) \\
&\le f(2 a^2) + f(2 b^2) \\
&\le 2(f(a^2) + f(b^2)) = 2(g(a) + g(b)),
\end{align*}
as desired. 
\end{proof}

To apply our framework, we additionally require a (non-robust) streaming algorithm for estimating $\mathcal{F}(\bx)$. 
In particular, the work of~\cite{BravermanCWY16} establishes the following zero-one law characterizing functions of the form $g$.
\begin{lemma}[Zero-one law for normal functions, \cite{BravermanCWY16}]
\lemlab{lem:zero-one-law}
Let $g: \mathbb{Z}_{n \ge 0} \to \mathbb{R}$ be a function. 
There exists a one-pass turnstile streaming algorithm that provides a $(1 \pm \eps)$-approximation to
\[\|\bx\|_g = \sum_{i = 1}^n g(x_i)\]
using sub-polynomial space if and only if $g$ is slow-dropping, slow-jumping, and predictable, where:
\begin{enumerate}
\item 
A function $g \in \mathcal{G}$ is \emph{slow-dropping} if there exists a sub-polynomial function $h$ such that for all $y \in \mathbb{N}$ and $x < y$, we have $g(x) \le g(y)\, h(y)$.
\item 
A function $g \in \mathcal{G}$ is \emph{slow-jumping} if there exists a sub-polynomial function $h$ such that for any $x < y$,
\[g(y) \le \lfloor y/x \rfloor^{2}\, h\left(\lfloor y/x \rfloor\, x\right)\, g(x).\]
\item 
A function $g \in \mathcal{G}$ is \emph{predictable} if for every sub-polynomial $\eps > 0$, there exists a sub-polynomial function $h$ such that for all $x \in \mathbb{N}$ and $y \in [1,\, x/h(x))$, either $|g(x+y) - g(x)| \le \eps(x)\, g(x)$ or $g(y) \ge g(x)/h(x)$.
\end{enumerate}
In particular, when $h$ is polylogarithmic, the corresponding streaming algorithm uses only polylogarithmic space.
\end{lemma}

\begin{lemma}
\lemlab{lem:zero-one-law-h}
\cite{GribelyukLWYZ26}
Assume $f:(0,\infty)\to[0,\infty)$ is a Bernstein function and define $g(x) = f\left(x^2\right)$. Then:
\begin{enumerate}
\item 
$g$ is slow-dropping with $h(x) = 1$.
\item 
$g$ is slow-jumping with $h(x) = 4$.
\item 
$g$ is predictable with $h(x) = \lceil 3/\eps(x) \rceil$.
\end{enumerate}
\end{lemma}
\begin{proof}
(1) Since $f$ is a Bernstein function, we have $f' \ge 0$, implying that $f$ is non-decreasing. 
As a consequence, $g(x) = f\left(x^2\right)$ is also non-decreasing, and thus for $x < y$, we obtain $g(x) \le g(y)$.

(2) Because $f$ is a Bernstein function, its derivative $f'$ exists and is non-increasing. 
Define
\[s(t) := \frac{f(t)-f(0)}{t} = \frac{1}{t}\int_0^t f'(u)\,du, \qquad t>0.\]
Since $f'$ is non-increasing, the average $s(t)$ is likewise non-increasing in $t$, which implies $s(y) \le s(x)$ for $0 < x < y$. Therefore,
\[\frac{f(y)}{y} = \frac{f(0)}{y} + s(y) \le \frac{f(0)}{y} + s(x) \le \frac{f(0)}{x} + s(x) = \frac{f(x)}{x}.\]
This shows that $\frac{f(y)}{y} \le \frac{f(x)}{x}$, and hence
\[\frac{g(y)}{g(x)} = \frac{f\left(y^2\right)}{f\left(x^2\right)} \le \frac{y^2}{x^2},\]
which gives
\[g\left(y^2\right) \le \frac{y^2}{x^2} \cdot g\left(x^2\right).\]
It follows that $g$ satisfies the slow-jumping condition with $h(x)=4$.

(3) Since $f$ is concave, non-decreasing, and satisfies $f(0)\ge 0$, we may use the standard slope bound
\[f'(t) \le \frac{f(t)}{t} \qquad (t>0).\]
Let $g(x)=f\left(x^2\right)$ and define $\Delta := (x+y)^2 - x^2 = 2xy + y^2$. 
Then
\begin{align*}
\frac{|g(x+y)-g(x)|}{g(x)} &= \frac{f\left(x^2+\Delta\right)-f\left(x^2\right)}{f\left(x^2\right)} \\
&\le \frac{f'(x^2)}{f(x^2)}\,\Delta \\
& \le \frac{\Delta}{x^2} = \frac{2y}{x} + \left(\frac{y}{x}\right)^2.
\end{align*}
For $y < x/h(x) \le \left(\eps(x)/3\right)x$, we have $\frac{y}{x} \le \eps(x)/3$, and therefore
\[\frac{2y}{x} + \left(\frac{y}{x}\right)^2 \le \frac{2}{3}\eps(x) + \frac{1}{9}\eps(x)^2 \le \frac{7}{9}\eps(x)
< \eps(x),\]
where we used $\eps(x)\le 1$. Hence,
\[\frac{|g(x+y)-g(x)|}{g(x)} \le \eps(x),\]
which shows that $g$ is predictable with $h(x) = \lceil 3/\eps(x) \rceil$.
\end{proof}

Combining \lemref{lem:zero-one-law} and \lemref{lem:zero-one-law-h}, we immediately obtain the following consequence.

\begin{lemma}
\cite{GribelyukLWYZ26}
Let $g: \mathbb{Z}_{n \ge 0} \to \mathbb{R}$ be defined as $g(x) = f\left(x^2\right)$ for some Bernstein function $f$. 
Then there exists a one-pass (non-robust) turnstile streaming algorithm that, with high probability, outputs a $(1 \pm \eps)$-approximation to $\|\bx\|_g = \sum_{i = 1}^n g(x_i)$ using space $\poly\left(\frac{1}{\eps}, \log n\right)$.
\end{lemma}

\section{\texorpdfstring{$L_2$}{L2} Heavy-Hitters}
\seclab{sec:turnstile:hh}
Next, we present a natural extension of our robust $F_2$ estimation algorithms to adversarially robustly compute the $L_2$ heavy hitters of the frequency vector at every point in the stream. 
Recall that an algorithm solves the $L_2$ heavy hitters problem if it outputs every coordinate $i$ such that 
\[|x_i| \ge \eps \|\bx\|_2,\]
and does not output any coordinate $i$ satisfying 
\[|x_i| \le \frac{\eps}{2}\|\bx\|_2.\]
As a direct corollary of our robust $L_2$ heavy hitters result, we also obtain an adversarially robust algorithm for recovering the $L_p$ heavy hitters for all $p \le 2$.

Suppose without loss of generality that $x_i > 0$; the case $x_i < 0$ follows symmetrically. 
The main idea is to first compute a robust approximation of the $L_2$ norm so that
\[\|\bx\|_2 \le X \le \left(1 + \frac{\eps}{100}\right)\|\bx\|_2.\]
To identify heavy hitters, we then \emph{deterministically} iterate over each coordinate $i \in [n]$ of the frequency vector, add mass $\frac{1}{2}\eps \cdot X \cdot \be_i$ to coordinate $i$, and recompute the approximate $L_2$ norm, denoted $S_i$. 
If the approximate $F_2$ value increases by at least 
\[1.15 \cdot \eps^2 \cdot \|\bx\|_2^2,\]
we report $i$ as a heavy hitter and continue to the next coordinate. 
The full procedure is formalized in \figref{fig:alg:heavy:hitter}.

\begin{figure*}[!htb]
\begin{mdframed}
\textbf{Algorithm}:
\begin{enumerate}
\item 
Let $\calA$ be a robust $L_2$ norm estimation algorithm with accuracy $(1+\O{\eps^2})$.
\item 
Let $\bx^{(t)}$ denote the frequency vector at time $t$, and obtain an estimate $X$ such that
\[
\|\bx^{(t)}\|_2 \le X \le \left(1+\frac{\eps^2}{100}\right)\|\bx^{(t)}\|_2.
\]
\item
Initialize $H_t \gets \emptyset$.
\item 
For each $i \in [n]$:
\begin{enumerate}
\item 
Set $\bv_i \gets \frac{1}{2}\eps \cdot X \cdot \be_i$.
\item 
Compute $S_i \gets \calA(\bx^{(t)} + \bv_i)$ and $T_i \gets \calA(\bx^{(t)} - \bv_i)$.
\item 
If $S_i^2 - X^2 \ge 1.15\,\eps^2 X^2$ or $T_i^2 - X^2 \ge 1.15\,\eps^2 X^2$, update $H_t \gets H_t \cup \{i\}$.
\item 
Restore coordinate $i$ by inserting $-\bv_i$ or $\bv_i$, as appropriate.
\end{enumerate}
\item 
Return $H_t$ as the set of heavy hitters at time $t$.
\end{enumerate}
\end{mdframed}
\caption{Algorithm for adversarially robust heavy hitters in insertion-deletion streams.}
\figlab{fig:alg:heavy:hitter}
\end{figure*}

To establish correctness of the algorithm, we first argue that perturbing a coordinate by roughly $\frac{\eps}{2}\|\bx\|_2$ produces noticeably different changes in the $F_2$ moment depending on whether that coordinate is heavy or not. 
In particular, adding this mass to a coordinate $i \in [n]$ that is $\eps$-heavy causes the total $F_2$ moment to increase by at least some quantity, whereas adding the same mass to a coordinate that is at most $\frac{\eps}{2}$-heavy increases the $F_2$ moment by strictly less. 
These two increases are separated by a constant gap, and a sufficiently accurate $F_2$ approximation can therefore distinguish the two scenarios, certifying whether $i$ is at least $\eps$-heavy or at most $\frac{\eps}{2}$-heavy.

\begin{lemma}
\lemlab{lem:heavy:hitters}
\cite{GribelyukLWYZ26}
Let $\bx \in \mathbb{R}^n$ be a frequency vector and suppose 
\[\|\bx\|_2 \le X \le \left(1+\frac{\eps^2}{100}\right)\|\bx\|_2.\]
Fix $i \in [n]$, and let $Z$ be a $\left(1+\O{\eps^2}\right)$-approximation to 
\[\left\|\bx + \frac{1}{2}\eps \cdot X \cdot \be_i\right\|_2^2.\]
Then with high probability:
\begin{enumerate}
\item 
If $x_i \ge \eps \|\bx\|_2$, then 
\[Z - X^2 > 1.15 \cdot \eps^2 \cdot X^2.\]
\item
If $x_i \le \frac{\eps}{2}\|\bx\|_2$, then 
\[Z - X^2 < 1.15 \cdot \eps^2 \cdot X^2.\]
\end{enumerate}
\end{lemma}
\begin{proof}
Let $\bx_{-i}$ denote the vector obtained from $\bx$ by zeroing out the $i$-th coordinate, i.e., 
\[\bx_{-i} := \bx - x_i \be_i.\]
Then
\[\|\bx\|_2^2 = \|\bx_{-i}\|_2^2 + x_i^2\]
and
\[\left\|\bx + \frac{1}{2}\eps X \be_i\right\|_2^2 = \|\bx_{-i}\|_2^2 + \left(x_i + \frac{1}{2}\eps X\right)^2.\]
Therefore,
\begin{align*}
\left\|\bx + \frac{1}{2}\eps X \be_i\right\|_2^2 - \|\bx\|_2^2
&= \left(x_i + \frac{1}{2}\eps X\right)^2 - x_i^2.
\end{align*}

If $x_i \le \frac{\eps}{2}\|\bx\|_2$, then using $X \le 1.05\|\bx\|_2$ for sufficiently small $\eps$, we obtain
\[\left(x_i + \frac{1}{2}\eps X\right)^2 - x_i^2\le (1.05^2 - 0.5^2)\,\eps^2\,\|\bx\|_2^2.\]
On the other hand, if $x_i \ge \eps\|\bx\|_2$, then by convexity of the quadratic function,
\[\left(x_i + \frac{1}{2}\eps X\right)^2 - x_i^2
\ge (1.5^2 - 1)\,\eps^2\,\|\bx\|_2^2.\]

There is thus a constant multiplicative gap between the two cases. 
Consequently, a $\left(1+\O{\eps^2}\right)$-approximation $Z$ to 
\[\left\|\bx + \frac{1}{2}\eps X \be_i\right\|_2^2\]
is sufficient to distinguish them. 
In particular, for sufficiently small constant $\eps \in (0,1)$, the threshold $1.15 \cdot \eps^2 X^2$ separates the two regimes, completing the proof.
\end{proof}
Finally, we justify the correctness of our adversarially robust heavy-hitter algorithm on insertion-deletion streams. 
\begin{restatable}{theorem}{thmhh}
\label{thm:hh}
\cite{GribelyukLWYZ26}
Given any $\eps \in (0,1)$, there exists an adversarially robust insertion-deletion streaming algorithm on a stream of length $m$ that solves the $L_2$ heavy-hitters problem at all times, for the underlying frequency vector of universe size $n$. 
For $m = \poly(n)$, the algorithm uses $\poly\left(\frac{1}{\eps}, \log n \right)$ bits of space.
\end{restatable}
\begin{proof}
The algorithm relies on maintaining a robust $F_2$ moment estimator and issues only deterministic queries when identifying the heavy hitters at each time step. 
Therefore, adversarial robustness follows immediately from \thmref{thm:ltwo}. 

Moreover, by \lemref{lem:heavy:hitters}, any $\left(1 + \frac{\eps^2}{100}\right)$-approximation to the $F_2$ moment suffices to separate the case $x_i \ge \eps \|\bx\|_2$ from the case $x_i \le \frac{\eps}{2}\|\bx\|_2$. 
Consequently, the overall space complexity is inherited directly from the bounds established in the aforementioned theorem.

Finally, for $p \le 2$, every $\eps$-$L_p$ heavy hitter is also an $\eps$-$L_2$ heavy hitter. 
Indeed, if $|x_i| \ge \eps \|\bx\|_p$, then
\[|x_i|^2 \ge \eps^2 \|\bx\|_p^2 \ge \eps^2 \|\bx\|_2^2,\]
where the last inequality holds because $p \le 2$. 
Thus, identifying $\eps$-$L_p$ heavy hitters reduces to running the $\eps$-$L_2$ heavy hitter algorithm, though we require a robust $L_p$ estimation algorithm to reject items with frequency less than $\frac{\eps}{2}\|\bx\|_p$. 
For the purposes of presentation, we omit this discussion and instead refer to \cite{GribelyukLWYZ26b}. 
\end{proof}

\chapter{Conclusion and Future Directions}
\chaplab{chap:conclusion}

This monograph has studied the essential and rapidly developing area of adversarial robustness in streaming algorithms.  
Throughout this exploration, we have navigated the complexities introduced by dynamic, adaptive, and even malicious data inputs, and surveyed a broad array of theoretical techniques and algorithmic strategies designed to address these challenges.  
From the subtle manipulations of a black-box observer to the omniscient insights of a white-box adversary, the discussion has made clear that traditional models assuming fixed, non-adaptive streams often fall short when faced with the demands of practical, real-world systems.

We began by framing the core challenges within the streaming model, emphasizing that even seemingly benign adaptive behavior without adversarial intent can disrupt the correctness of standard algorithms.  
Using concrete examples, we showcased how foundational techniques like basic random sampling and the classic Alon-Matias-Szegedy (AMS) sketch can become unexpectedly fragile when subjected to thoughtfully designed adaptive inputs.

Within the black-box adversarial model for insertion-only streams, we presented a suite of general techniques for strengthening algorithmic robustness. 
A key insight came from the notion of the \emph{flip number}, which captures how frequently a function’s output can change over time.  
We introduced sketch-switching to carefully hide internal randomness, thereby converting non-robust algorithms into robust ones.  
Additionally, the use of difference estimators led to algorithms with nearly optimal space guarantees for core problems like $F_p$ estimation, distinct elements, and entropy, revealing that, in many insertion-only settings, the ``price'' of adversarial robustness can be surprisingly minimal. 
We also presented bounded computation paths as an approach that uses an abundance of randomness to handle all possible inputs, even from an adaptive adversary. 

Our exploration continued by uncovering deep theoretical connections between adversarial robustness and established areas in computer science. 
We showed how techniques from differential privacy, originally designed to protect individual data, can be repurposed as a powerful tool for algorithmic robustness by deliberately limiting what internal information an algorithm reveals.  
Complementarily, tools from adaptive data analysis offered a rigorous lens through which to formalize separation results, revealing that certain tasks fundamentally require more computational effort in adaptive settings than in oblivious ones.

The landscape shifted considerably in the more complex setting of turnstile streams, where both insertions and deletions are allowed. 
This generalization breaks many of the structural properties that made insertion-only models more tractable.  
Nevertheless, techniques like the dense-sparse decomposition offered improved space bounds for some problems by dynamically switching between algorithmic regimes. 
On the other hand, we also established strong lower bounds and powerful attacks on linear sketches. 
These results explicitly exposed inherent limitations, demonstrating that even popular sketching methods for problems such as $F_p$ estimation and $F_0$ estimation can be undermined by adaptive strategies in this more challenging model. 

We continued our exploration with arguably the most challenging setting: the white-box adversarial model, where adversaries have full access to an algorithm’s internal state and random bits. 
We showed that unfortunately, many problems do not admit sublinear-space algorithms in this setting. 
On a positive note, we showed that strong guarantees can still be achieved against computationally bounded adversaries by drawing on modern cryptographic hardness assumptions, most notably the Short Integer Solution (SIS) problem.  
We introduced streaming algorithms for sparse vectors, low-rank matrices, and tensors that remain robust under white-box attacks—capable of both verifying structural consistency and recovering the true input. 
These techniques have immediate applications to problems such as $F_0$ estimation, rank decision, graph matching, and robust principal component analysis.

Finally, we showed that robust $F_2$ moment estimation can be achieved on turnstile streams in space polylogarithmic in the universe size $n$. 
This demonstrates that there is hope for designing robust turnstile streaming algorithms beyond traditional linear sketches. 
An important direction for future work is to identify additional problems that admit similarly robust algorithms through fundamentally different techniques, such as hierarchical decompositions, recursive estimators, or learning-based approaches. 
Perhaps the central open question in adversarially robust streaming is now to determine which turnstile problems admit similarly robust streaming algorithms through fundamentally new techniques, and which problems are inherently impossible even beyond the linear sketch paradigm.

\subsection*{Additional Related Work}
In this section, we highlight several other recent developments related to adversarial robustness. 

\paragraph{Graph coloring.} 
We briefly discussed graph coloring in \secref{sec:separation:graph:coloring} as an example of a separation between non-adaptive and adversarial insertion-only streams. 
In particular, we showed that $\Omega(\Delta^2)$ colors are necessary to color a graph by an algorithm using $\tO{n}$ space for adversarial insertion-only streams, whereas there exists an algorithm that uses $\tO{n}$ space to color a graph with $(\Delta+1)$ colors for oblivious (non-adaptive) streams. 
On the positive side, \cite{ChakrabartiGS22} presented a semi-streaming algorithm, i.e., space $\tO{n}$, for adversarial insertion-only streams using $\O{\Delta^3}$ colors, which was subsequently improved to $\O{\Delta^{5/2}}$ colors by \cite{AssadiCGS23}. 

\paragraph{Online learning with experts.}
Online learning with adversarial input has been considered in a number of other settings, e.g., \cite{AlonBDMNY21}. 
In the memory-bounded online learning with experts problem, an algorithm selects one of $n$ experts on each of $T$ rounds, receiving the losses of all experts and incurring the loss of the selected expert after each round. 
The goal is to minimize regret, defined as the cumulative loss of the algorithm minus that of the best expert in hindsight. 
In the streaming model, the algorithm must make decisions using memory sublinear in $n$ and $T$, e.g., the algorithm cannot store full loss history or weight vectors of losses for each expert. 
\cite{PengR23} showed that $\polylog(nT)$ memory suffices to achieve $\tO{\sqrt{nT}}$ regret in the oblivious setting, but $\tilde{\Omega}(\sqrt{n})$ memory is necessary to achieve $o(T)$ regret in the adaptive setting, where the losses of each expert on each round can be dependent on previous outcomes. 
For sufficiently large $n$ and $T$, \cite{PengR23} gave an algorithm using $S$ space with regret $\tO{\max\left(\frac{\sqrt{n}T}{S},\sqrt{\frac{nT}{S}}\right)}$ against adaptive adversaries, with high probability, while \cite{WoodruffZZ23b} gave an algorithm using $\tO{\frac{n}{R\sqrt{T}}}$ space for regret $R$ when the best expert makes $\O{R^2 T \log^2 n}$ mistakes. 

\paragraph{Robust data structures.} 
Rather than focusing on the streaming model, a number of recent works have instead focused on attaining data structures robust against a number of adaptive queries~\cite{CherapanamjeriN20,CohenLNSSS22,FleischhackerLS22,NaorO22,MenuhinN22,CherapanamjeriSWZZZ23,KapralovMS25,BogdanovRVV26,FengFLSWZ25}, and conversely, designing attacks using a number of adaptive queries against existing non-adaptive data structures~\cite{CohenNSS23,CohenNSSS24}. 
For example, \cite{CohenLNSSS22} showed that the aforementioned $\countsketch$ data structure, c.f., \thmref{thm:countsketch}, is not robust in the sense that a sketch with size $k$ can be attacked using $\tO{k}$ queries, but with an appropriate modification, can be made robust to $m$ rounds of adaptive queries using space roughly $\sqrt{m}\cdot k$ sketch size. 
Similarly, \cite{CohenNSS23} gave an attack on certain classes of data structures based on hashing, using $\tO{k^2}$ adaptive queries to break a sketch of size $k$, while more recently, \cite{AhmadianC24,CohenNSSS24} gave attacks on cardinality-based estimators, such as those used for distinct element estimation. 
Specifically, \cite{AhmadianC24} achieved an adaptive attack using $\O{k}$ queries on cardinality sketches based on the \texttt{HyperLogLog} paradigm~\cite{FlajoletFGM07}, which randomly prioritizes keys in the universe and then keeps the lowest priorities of keys that are in the dataset to estimate the overall number of distinct elements. 
Their attack works by sampling elements into random sets and then querying the items with lowest scores, correlating with items with lower priorities, thus biasing the random sketch. 
The attack requires the ability to remove previously inserted elements to isolate the influence of individual keys, and thus applies only in insertion-deletion, i.e., turnstile models.
Additionally, \cite{CohenNSSS24} gave an attack on any union-composable sketching map using $\tO{k^4}$ adaptive queries, as well as a tight $\tO{k^2}$ bound for monotone maps, such as MinHash, statistical queries, and Boolean linear sketches. 
Moreover, they prove that linear sketching maps over $\mathbb{R}$ and finite fields $\mathbb{F}_p$ can also be broken with $\tO{k^2}$ adaptive queries, strengthening the prior polynomial bounds by \cite{GribelyukLWYZ24}, though their techniques do not extend to integer sketches, i.e., linear sketches with integer-valued entries and integer-valued inputs. 
On the positive side, \cite{CohenSS25} circumvented the quadratic barrier by introducing fine-grained robustness guarantees, given an estimator that can support an exponential number of adaptive queries, provided that each individual element appears in at most $\tO{k^2}$ adaptive queries. 
This shifts the viewpoint of adaptivity from the overall number of queries to how often each element is involved, allowing for greater robustness in realistic workloads and opening new directions for building fine-grained sketching algorithms.

\paragraph{Dynamic algorithms.}
In the dynamic model, updates to an underlying setting arrive sequentially. 
However, the focus is sometimes on fast update time rather than sublinear space. 
This setting often considers an adaptive adversary~\cite{Chan10,NanongkaiS17,Wajc20,ChanH21,BeimelKMNSS22,BernsteinBGNSS022,BhattacharyaSS22,RoghaniSW22,BateniEFHJMW23,AlokhinaB24,BravermanDPSW25} that generates the updates $u_1,\ldots,u_m$ upon seeing not only the output of the algorithm after the previous update, i.e., a black-box adversary, but also the entire data structure maintained by the algorithm after the previous update, i.e., a white-box adversary. 

\paragraph{Pseudo-deterministic algorithms.}
Pseudo-deterministic algorithms~\cite{GatG11,GoldreichGR13,GrossmanL19,GoldwasserGMW20,BravermanK0S23,GrossmanGS23} are randomized algorithms that, despite using internal randomness, produce the same output with high probability on each execution for a given input. 
Formally, a randomized algorithm $\calA$ is pseudo-deterministic if for every input $x$, there exists an output $y$ such that $\PPr{\calA(x)=y}\ge1-\eps$, for some small error parameter $\eps$. 
These algorithms offer a compelling middle ground between deterministic and randomized computation, retaining many of the efficiency benefits of randomness while yielding predictable outputs.

In adversarial settings, pseudo-determinism can enhance robustness by reducing the variability that an adversary can exploit. 
For instance, in online or streaming environments where an adversary may adapt to algorithmic randomness, pseudo-deterministic algorithms limit the adversary's power by narrowing the range of observable outcomes. 
As such, they provide a principled way to achieve robust and reproducible behavior even under adaptive or adversarial input sequences.

\paragraph{Sketches for fast iterative algorithms.}
Adversarial robustness plays an important role in the design of iterative algorithms, especially in continuous optimization methods that repeatedly rely on sketching techniques or fast data structures as subroutines, e.g., \cite{ChenKLPGS25}. 
In these settings, the outputs of one iteration, such as gradient estimates or compressed representations, are reused to guide future updates, creating a natural feedback loop within the algorithm. This interaction can allow an adaptive adversary to exploit information leaked by earlier iterations and influence later inputs, potentially harming accuracy or convergence. 
As a result, subroutines that are reliable in a single-shot setting may become fragile when used iteratively, which highlights the need for sketching and data-structural primitives that remain robust under adaptive and repeated use.

\paragraph{Adversarially robust distributed streaming.}
In the \emph{distributed streaming model}, also known as \emph{distributed functional monitoring}, a central server coordinates with $k$ distributed sites, each of which receives a stream of updates. 
The goal is to track a global function over the union of all items seen so far, while minimizing the communication cost between the sites and the server. 
This model captures practical scenarios such as monitoring network traffic, sensor data aggregation, or distributed databases under bandwidth constraints.

\cite{XiongZH23} initiated the study of adversarial robustness in the distributed streaming model, so that the stream of updates may be based on previous outputs produced by the algorithm. 
As in the streaming setting, typical analyses in the distributed setting often assume that the input is independent of the internal randomness used by the distributed protocol~\cite{LinSWXZ26}. 
\cite{XiongZH23,CohenSS26} showed that for the \emph{count tracking} problem, where the server must approximate the total count of items across all sites at all times, it is possible to achieve a $(1+\eps)$-approximation to the count at all times in a distributed stream of total length $n$, using $\tO{k+\frac{\sqrt{k}}{\eps}\log n}$ bits of communication. 

\subsection*{Future Directions and Open Problems}
While this monograph consolidates significant advancements in adversarial robustness for streaming algorithms, it also illuminates a rich landscape of open problems and promising research avenues:

\begin{enumerate}
\item 
\textbf{Space complexity gaps in turnstile $F_p$ estimation:} 
Although the results in \secref{sec:turnstile:algs:ftwo} make initial progress toward understanding the landscape of robust $F_p$ moment estimation on turnstile streams, it should be noted that the dependencies on the accuracy parameter $\eps$ are suboptimal. 
Furthermore, the framework initially only achieved constant-factor approximations for $F_p$ estimation when $p \neq 2$. 
However, very recently, \cite{GribelyukLWYZ26b} generalized the $L_2$ estimation algorithm to successfully achieve a $(1+\eps)$-approximation for $L_p$ estimation for all $p \in [0, 2)$ on turnstile streams. 
This rapid progress is a strong indication that this is a highly active area of research. 
Nonetheless, optimizing the space dependencies, as well as extending these $(1+\eps)$-approximation guarantees to $L_p$ estimation for $p>2$, remain substantial open challenges. 
In particular, oblivious algorithms can achieve $(1+\eps)$-approximation for $F_p$ estimation for all ranges of $p$ in sublinear space, while the best known robust algorithms exhibit a polynomial dependence on stream length for $p>2$. 
Thus, a prominent open problem remains the substantial gap in space complexity for general $F_p$ estimation in the turnstile model: 
{
\renewcommand{\thetheorem}{\thechapter.\arabic{theorem}}
\renewcommand{\thequestion}{\thechapter.\arabic{theorem}}
\begin{question}
Is this an inherent separation for general $F_p$ moment estimation (or more general functions) in turnstile streams in the black-box model, or do more efficient robust algorithms exist? 
\end{question}
}
In particular, it is natural to study whether there is a separation between oblivious and adversarially robust turnstile streaming algorithms beyond scalar frequency moments to tasks with richer structural and geometric complexities. 
For example, the landscape of adversarially robust streaming remains largely unexplored for fundamental optimization objectives within numerical linear algebra, dynamic graph processing, and computational geometry. 
Determining whether the core sketching primitives in these broad domains fundamentally require additional space overheads against adaptive turnstile updates, or if they admit highly space-efficient robust algorithms, provides a rich and critical avenue for future research. 
Further investigations into novel techniques that transcend the limitations of current frameworks are warranted. 

\item
\textbf{Capabilities and limitations of cryptographic techniques against computationally-bounded white-box adversaries:}
The current white-box algorithms for sparse and low-rank recovery often provide exact reconstruction. 
In many realistic scenarios, the underlying data might only be \textit{approximately} sparse or low-rank. 
Developing provably robust algorithms for \textit{approximate} recovery under white-box adversaries poses a significant theoretical and practical challenge. 
{
\renewcommand{\thetheorem}{\thechapter.\arabic{theorem}}
\renewcommand{\thequestion}{\thechapter.\arabic{theorem}}
\begin{question}
Can we achieve robust algorithms for approximate recovery under computationally-bounded white-box adversaries?
\end{question}
}
More generally, lattice-based cryptographic assumptions are the predominant approach to generate sublinear space streaming algorithms for input generated by white-box adversaries using polynomial runtime. 
Thus a natural question is:
{
\renewcommand{\thetheorem}{\thechapter.\arabic{theorem}}
\renewcommand{\thequestion}{\thechapter.\arabic{theorem}}
\begin{question}
What are the limitations of cryptographic assumptions for constructing sublinear-space streaming algorithms that are adversarially robust against polynomial-time white-box adversaries?
\end{question}
}
\item 
\textbf{Refining adversary models and interactions:} 
The analysis of robust algorithms against adversaries with more fine-grained computational power, e.g., polynomial time within specific complexity classes, or more nuanced observational capabilities, e.g., delayed observations, partial visibility of randomness, offers fertile ground for research. 
Understanding the interplay between robustification techniques and communication constraints in multi-party streaming environments is also crucial. 
In particular, for a number of settings the source of adversarial input is the algorithm itself, due to repeated interactions causing a feedback loop. 
In this case, the input is adaptive, but not necessarily adversarial, e.g., a user purchasing items online might continuously react to recommendation systems but does not explicitly devote memory to remember much beyond the last few recommendations. 
Then the ``adversary'' for an efficient streaming algorithm might also be bounded in space or time. 
Hence, one could ask:
{
\renewcommand{\thetheorem}{\thechapter.\arabic{theorem}}
\renewcommand{\thequestion}{\thechapter.\arabic{theorem}}
\begin{question}
What are the tradeoffs when the source of adaptive input is itself bounded in space or time?
\end{question}
}
Recently, \cite{Ben-EliezerOS26} initiated the study of this direction by formalizing robust streaming against low-memory adversaries. 
\item
\textbf{Robustness in the distributed streaming model:} 
Recently initiated by \cite{XiongZH23}, the adversarially robust distributed streaming model~\cite{CohenSS26,LinSWXZ26} focuses on minimizing the amount of communication to accurately estimate a statistic on an evolving dataset distributed across multiple sites. 
The non-adaptive setting has been extensively studied, achieving near-optimality for a number of problems such as norm estimation, distinct element estimation, entropy estimation, and heavy hitters~\cite{DilmanR02,BabcockO03,CormodeGMR05,KeralapuraCR06,ArackaparambilBC09,CormodeMY11,TirthapuraW11,CormodeMYZ12,WoodruffZ12,YiZ13,ChenZ17,HuangYZ19,WuGZ20}. 
\cite{XiongZH23,CohenSS26} showed that for the simple count tracking (or counting) problem, achieving robustness in this model requires almost no overhead for insertion-only streams. 
An interesting open direction is:
{
\renewcommand{\thetheorem}{\thechapter.\arabic{theorem}}
\renewcommand{\thequestion}{\thechapter.\arabic{theorem}}
\begin{question}
Do central problems in the distributed streaming model require communication overhead to achieve adversarial robustness?
\end{question}
}
\item
\textbf{Optimality for other problem classes:} 
While near-optimal bounds have been achieved for several problems in the black-box insertion-only model, determining the precise optimal overhead for robustification across a broader range of streaming problems, particularly those with complex output structures or for randomized algorithms not based on sketching, remains an active area.
{
\renewcommand{\thetheorem}{\thechapter.\arabic{theorem}}
\renewcommand{\thequestion}{\thechapter.\arabic{theorem}}
\begin{question}
Beyond the problems discussed in this monograph, which problems admit space-efficient streaming algorithms in the adversarially robust black-box model?
To what extent can we characterize problems that admit separations between the adversarial and non-adaptive models? 
\end{question}
}
\item 
\textbf{Broader applicability of robustification frameworks:} 
Can the principles underpinning \emph{difference estimators} or \emph{differential privacy-based frameworks} be generalized and applied effectively to a wider array of streaming problems, such as geometric computations, graph stream algorithms beyond coloring, or other machine learning tasks? 
Exploring the extension of the \emph{flip number} concept to higher-dimensional outputs, e.g., matrices, coresets, could yield new robustification paradigms.
\end{enumerate}

This monograph serves as a testament to the vibrant and rapidly evolving field of adversarial robustness in streaming algorithms. 
By bridging theoretical computer science foundations with real-world applications, the discussion illuminates both the profound challenges and the creative solutions that empower robust computation in an increasingly complex and adversarial digital landscape. 
The pursuit of reliable algorithmic systems remains a cornerstone of dependable computing, and the avenues for future discovery are as vast as the data streams themselves.

\chapter*{Acknowledgments} 
We would like to thank our coauthors Mikl\'os Ajtai, Omri Ben-Eliezer, Vladimir Braverman, Itai Dinur, Ying Feng, Elena Gribelyuk, Moritz Hardt, Avinatan Hassidim, Aayush Jain, Rajesh Jayaram, T. S. Jayram, Honghao Lin, Yossi Matias, Mariano Schain, Sandeep Silwal, Uri Stemmer, Alec Sun, Eylon Yogev, Huacheng Yu for contributions to the technical content. 
The results in \secref{sec:attack:specific:turnstile} are unpublished and discovered in joint work with Honghao Lin. 
We also would like to thank Edith Cohen, Uri Stemmer, and anonymous reviewers for detailed feedback on previous versions of this monograph. 

David P. Woodruff is supported in part by the Office of Naval Research award number N000142112647, and a Simons Investigator Award. 
Samson Zhou is supported in part by NSF CCF-2335411. 
Samson Zhou gratefully acknowledges funding provided by the Oak Ridge Associated Universities (ORAU) Ralph E. Powe Junior Faculty Enhancement Award. 

\bibliography{references}
\bibliographystyle{alpha}

\end{document}